%% file: main.tex
\documentclass[
11pt, 
english,
singlespacing,
liststotoc 
headsepline, 
]{MastersDoctoralThesis} 

\usepackage[utf8]{inputenc} 
\usepackage[T1]{fontenc} 

\usepackage{xcolor}
\definecolor{Sapienza}{rgb}{0.51,0.14,0.20}
\usepackage{hyperref}
\usepackage{siunitx} 
\usepackage{enumitem}
\usepackage{epigraph} 
\usepackage{subcaption}

\usepackage{empheq} 
\usepackage{fontawesome}
\usepackage{amsthm}
\usepackage{amssymb} 
\usepackage{slashed}
\usepackage{multirow}
\usepackage{scalerel} 
\usepackage{wrapfig}  
\usepackage{physics} 
\usepackage{cancel} 
\usepackage{mathrsfs}
\usepackage[most]{tcolorbox}
\tcbuselibrary{listingsutf8}
\tcbset{
    mybox/.style={enhanced jigsaw,
        colback=gray!20,
        colframe=gray!75,
        boxrule=1mm,
        breakable,
        pad at break*=1mm
    }
}

\DeclareSIUnit \parsec{pc}

\newcommand{\M}{M^2_{pl}}
\newcommand{\ie}{\textit{i.e.}\,}

\renewcommand\({\left(}
\renewcommand\){\right)}
\renewcommand\[{\left[}
\renewcommand\]{\right]}
\newcommand\gev{\text{ GeV}}
\newcommand\mev{\text{ MeV}}
\newcommand\kev{\text{ KeV}}
\renewcommand\ev{\text{ eV}}
\newcommand\mpc{\text{ Mpc}}

\newcommand\Mpl{M_{\rm pl}}
\newcommand\neff{N_{\rm eff}}
\newcommand{\Hl}{{\scaleto{H}{4pt}}}
\newcommand{\Vl}{{\scaleto{V}{4pt}}}
\newcommand{\na}{\boldsymbol{\nabla}}
\newcommand{\eh}{\epsilon_\Hl}
\newcommand{\etah}{\eta_\Hl}
\newcommand{\epsv}{\epsilon_\Vl}
\newcommand{\etav}{\eta_\Vl}
\newcommand{\msolar}{\textup{M}_\odot}

\newcommand\eq[1]{Eq.~\eqref{eq:#1}}
\newcommand\sect[1]{Sec.~\ref{sec:#1}}
\newcommand\chap[1]{Ch.~\ref{ch:#1}}
\newcommand\appx[1]{Appx.~\ref{app:#1}}
\newcommand\fig[1]{Fig.~\ref{fig:#1}}
\newcommand\tab[1]{Tab.~\ref{tab:#1}}
\newcommand{\mscr}[1]{\mathscr{#1}}

\newcommand{\mcal}[1]{\mathcal{#1}}

\usepackage{mathpazo} 

\usepackage[backend=bibtex,natbib=true,sorting=none,style=numeric-comp]{biblatex}

\usepackage[autostyle=true]{csquotes} 

\usepackage{titlesec}
\usepackage{lmodern}
\usepackage{lipsum}

\thesistitle{Probing new physics in the era of precision cosmology} 
\supervisor{\textcolor[rgb]{0.51,0.14,0.20}{Prof. Alessandro Melchiorri}} 
\examiner{} 
\degree{Doctor of Philosophy} 
\author{\textcolor[rgb]{0.51,0.14,0.20}{Matteo Forconi}} 
\addresses{} 
\subject{} 
\university{\textcolor[rgb]{0.51,0.14,0.20}{La Sapienza, University of Rome}} 
\department{\textcolor[rgb]{0.51,0.14,0.20}{Physics Department}} 
\group{{Astronomy, Astrophysics and Space Science}} 
\faculty{\textcolor[rgb]{0.51,0.14,0.20}{Faculty of Mathematical, Natural and Physical Sciences}} 
\newcommand{\coordname}{\textcolor[rgb]{0.51,0.14,0.20}{Prof. Francesco Piacentini}} 

\AtBeginDocument{
\hypersetup{pdftitle=\ttitle} 
\hypersetup{pdfauthor=\authorname} 
\hypersetup{pdfkeywords=\keywordnames} 
}

\begin{document}

\frontmatter 

\pagestyle{plain} 


\begin{titlepage}
\begin{center}
\begin{minipage}[t]{1.0\textwidth}
    \centering
    \includegraphics[width=0.9\textwidth]{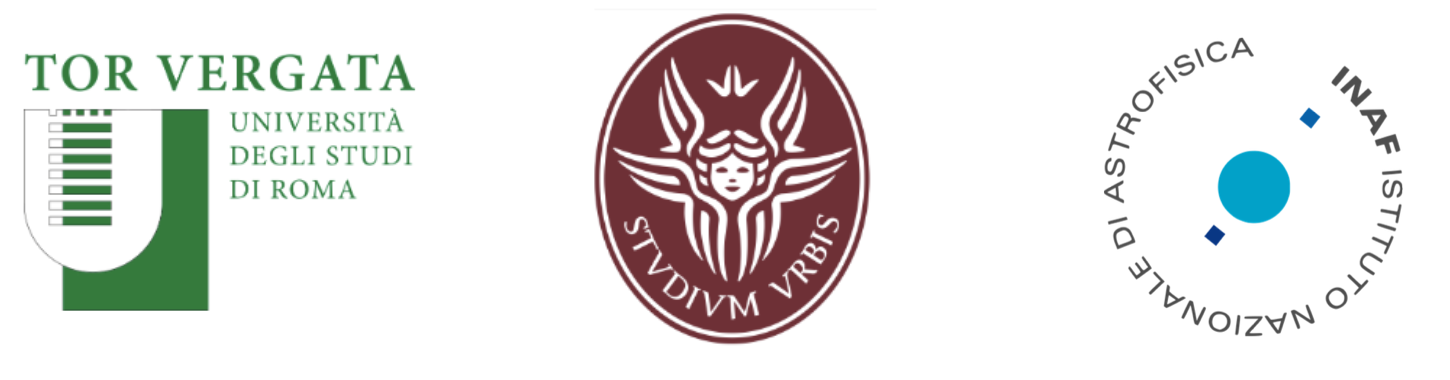} 
\end{minipage}

\vspace*{.05\textheight}
{\scshape\large \textcolor[rgb]{0.51,0.14,0.20}{Doctoral Thesis submitted in fulfilment }\par}\vspace{0.25cm}
{\scshape\large \textcolor[rgb]{0.51,0.14,0.20}{of the requirements for the degree of}\par}\vspace{0.25cm} 
{\scshape\large \textcolor[rgb]{0.51,0.14,0.20}{Doctor of Philosophy}\par}\vspace{0.25cm}
{\scshape\large \textcolor[rgb]{0.51,0.14,0.20}{in}\par}\vspace{0.5cm}
{\scshape\LARGE \textcolor[rgb]{0.51,0.14,0.20}{Astronomy, Astrophysics and Space Science}\par}\vspace{0.5cm}
{\scshape\large \textcolor[rgb]{0.51,0.14,0.20}{XXXVII Cycle}\par}\vspace{0.5cm}
\HRule \\[0.4cm] 
{\scshape\huge \ttitle\par}\vspace{0.4cm} 
\HRule \\[1.5cm] 

\begin{minipage}[t]{0.4\textwidth}
\begin{flushleft} \large
\emph{PhD Candidate}\\
{\authorname} 
\end{flushleft}
\end{minipage}
\begin{minipage}[t]{0.4\textwidth}
\begin{flushright} \large
\emph{Supervisor} \\
\supname 
\end{flushright}
\end{minipage}\\[1cm]

\begin{minipage}[t]{0.4\textwidth}
\begin{flushleft} \large
\emph{}\\
{} 
\end{flushleft}
\end{minipage}
\begin{minipage}[r]{0.4\textwidth}
\begin{flushright} \large
\emph{Coordinator} \\
\coordname 
\end{flushright}
\end{minipage}\\[1cm]
\vspace{0.85cm}

\large 
\textsc{La Sapienza, Univeristy of Rome}\\[0.3cm] 
\textsc{Tor Vergata, University of Rome}\\[0.3cm] 
\textsc{INAF, National Institute for Astrophysics}\\[0.3cm] 
\vfill

{\large \textit{Academic Year 2023/2024}}\\[4cm] 

\vfill
\end{center}
\end{titlepage}



\vspace*{0.2\textheight} 
\noindent\hfill
\begin{minipage}{0.9\textwidth} 
    \raggedleft
    
    \textit{\large That you are here - that life exists and identity,} \\
    \textit{\large That the powerful play goes on, and you may contribute a verse.} \\
    \vspace*{2mm}
    \hfill{\large -- Walt Whitman, O Me! O Life!}
\end{minipage}


\begin{abstract}
\addchaptertocentry{\abstractname} 
\begin{center}
    \Large \textbf{Abstract}
\end{center}

\vspace{0.5em} 
    \noindent\rule{\textwidth}{0.5pt} 
    \vspace{0.5em} 
\begin{center}

{\noindent \large Over the past decades, advancements in observational cosmology have introduced us in an era of precision cosmology, dramatically enhancing our understanding of the Universe's history as well as bringing new tensions to light. Observations of the Cosmic Microwave Background, large-scale structure, and distant galaxies have provided unprecedented insights into the processes that shaped our Universe. This PhD thesis contributes to this research by exploring how these cosmic observables can be leveraged to constrain new physics beyond the standard $\Lambda$-Cold Dark Matter model.}

\end{center}
\vspace{0.5em} 
    \noindent\rule{\textwidth}{0.5pt} 
    \vspace{0.5em} 

\end{abstract}
\newpage
\section*{\Huge Scientific Production}
\addcontentsline{toc}{chapter}{Scientific Production}

{\large This thesis PhD is based on the following papers (ordered by date) that I authored or co-authored during my PhD}

    \vspace{1em}
    \noindent\rule{\textwidth}{1.5pt}
    \vspace{1em}
\begin{itemize}
\begin{large}
    \item[\textbf{Ref.~\cite{Forconi:1}}] \textbf{Matteo Forconi}, William Giarè, Eleonora Di Valentino, and Alessandro Melchiorri, "\textit{Cosmological Constraints on Slow Roll Inflation: an update}", \href{https://journals.aps.org/prd/abstract/10.1103/PhysRevD.104.103528}{\color{Sapienza}{Phys. Rev. D 104, 103528}}
    
    \item[\textbf{Ref.~\cite{Forconi:2}}] William Giarè, \textbf{Matteo Forconi}, Eleonora Di Valentino, and Alessandro Melchiorri, "\textit{Towards a reliable calculation of relic radiation from primordial gravitational waves}", \href{https://academic.oup.com/mnras/article/520/2/1757/7005230?login=true}{\color{Sapienza}{Mon. Not. Roy. Astron. Soc. 520, 2}}

    \item[\textbf{Ref.~\cite{Forconi:3}}] \textbf{Matteo Forconi}, Ruchika, Alessandro Melchiorri, Olga Mena and Nicola Menci, "\textit{Do the Early Galaxies observed by JWST disagree with Planck's CMB polarization measurements?}", \href{https://inspirehep.net/files/a574d7b16a76cfd972aca6df2b5b6108}{\color{Sapienza}{JCAP 10 (2023) 012}}

    \item[\textbf{Ref.~\cite{Forconi:4}}] \textbf{Matteo Forconi}, Eleonora DI Valentino, Alessandro Melchiorri, and Supriya Pan, "\textit{Possible Impact of Non-Gaussianities on Cosmological Constraints in Neutrino Physics}", \href{https://journals.aps.org/prd/abstract/10.1103/PhysRevD.109.123532}{\color{Sapienza}{Phys.Rev.D 109 (2024) 12, 123532}}

    \item[\textbf{Ref.~\cite{Forconi:5}}] \textbf{Matteo Forconi}, William Giar\'e, Olga Mena, Ruchika, Eleonora DI Valentino, Alessandro Melchiorri, and Rafael C. Nunes, "\textit{A double take on early and interacting dark energy from JWST}", \href{https://iopscience.iop.org/article/10.1088/1475-7516/2024/05/097}{\color{Sapienza}{JCAP 05 (2024) 097}}
\end{large}
\end{itemize}
\vspace{1em}
\noindent\rule{\textwidth}{1.5pt}
\vspace{1em}

{\large \noindent The discussion on how they fit the storyline is outlined in the subsequent \hyperref[Overview]{overview} while a brief summary of all the major results is displayed in the \hyperref[conclusionnn]{conclusions}. Works currently in progress have been omitted as well as other forthcoming projects that do not fit the overall topic.}
\newpage

\section*{\Huge Overview}
\addcontentsline{toc}{chapter}{Overview}
\phantomsection
\label{Overview}

{\large In this thesis I study how important cosmological observations, either in Early- or Late-time Universe, are able to set constraints in models that go beyond our standard description of the Universe. Here, I briefly summarize the structure of the work to guide the interested reader. My scientific productions inspired \sect{bBNPGW}~\cite{Forconi:2}, \sect{CMBPGW}~\cite{Forconi:1} , \sect{PSOGFNSN}~\cite{Forconi:4}, \sect{PolJWST} and \sect{JDE}~\cite{Forconi:3,Forconi:5} }

\vspace{1em}
\noindent\rule{\textwidth}{1.5pt}
\vspace{1em}

{\large \begin{enumerate}[label=\large\arabic*)]
    \item \underline{In \textbf{ \large Chapter \ref{fhgge}}} I provide a review of the Unperturbed Universe in the Hot Big Bang Theory. It is organized as follows
    \begin{enumerate}[label=\Large$\bullet$]
        \item \underline{In \textit{\textbf{Section \ref{fhsd}}}} I introduce the spacetime geometry of our Universe. In order to do so, I present the theory of General Relativity and write down the Einstein equations. Then, based on the symmetry of our Universe, I build the Friedmann-Lema\^{i}tre-Robertson-Walker metric. 
        \item \underline{In \textit{\textbf{Section \ref{sec:DynamoSassari}}}} I solve the Einstein equations for our spacetime, deriving the equations of motion that relate the dynamics of the Universe to its matter and energy content.
        \item \underline{In \textit{\textbf{Section \ref{entropyandstatistics}}}} I collect the basic equations that describe the thermodynamics in an expanding Universe and then I review the most important steps of the thermal history.
    \end{enumerate}
     \item \underline{In \textbf{ \large Chapter \ref{Chapter2}}} I study the small scale structure of the Universe, exploring the cosmological perturbations and following their linear evolution. In order to do so
    \begin{enumerate}[label=\Large$\bullet$]
        \item \underline{In \textit{\textbf{Section \ref{sec:fieldEquation}}}} I split the metrics according to the perturbations symmetry proprieties and I introduce the concept of gauge theory.
        \item \underline{In \textit{\textbf{Section \ref{sec:setensor}}}} I derive the form of the perturbed stress-energy tensor.
        \item \underline{In \textit{\textbf{Section \ref{sec:Evo}}}} I perturb the Einstein equations and study the dynamics of density perturbations, introducing the adiabaitc modes.
        \item \underline{In \textit{\textbf{Section \ref{frera}}}} I study the dynamics of scalar perturbations along the different cosmological epochs using the linearized Theory developed in the previous subsection.
        \item \underline{In \textit{\textbf{Section \ref{sec:GW}}}} I describe the dynamical evolution of tensor perturbations in en expanding Universe introducing the concept of metric waves as gravitational waves.
    \end{enumerate}
    \item \underline{In \textbf{ \large Chapter \ref{ch:CMBCHAPTER}}} I introduce perhaps the most important cosmological observable: the Cosmic Microwave Background. I describe the physics of temperature anisotropies and polarization, connecting the small irregularities observed in the CMB with the physics of the Early Universe. 
    \begin{enumerate}[label=\Large$\bullet$]
        \item \underline{In \textit{\textbf{Section \ref{TEMEE}}}} I introduce the temperature multipoles expansion to better study the CMB .
        \item \underline{In \textit{\textbf{Section \ref{sec:TA}}}} I review the theory of CMB temperature anisotropies, discussing different physical mechanisms able to produce primary and secondary anisotropies and the respective signatures left in the angular power spectrum.
        \item \underline{In \textit{\textbf{Section \ref{sec:CapitoloPoloDOPO}}}} I review the theory of CMB polarization, discussing in details different physical mechanisms able to produce them.
    \end{enumerate}
    \item \underline{In \textbf{ \large Chapter \ref{ch:InfloBinglo}}} I introduce the theory of cosmological inflation, showing how an early epoch of fast accelerated expansion can solve many fine-tuning problems with the initial conditions.
    \begin{enumerate}[label=\Large$\bullet$]
        \item \underline{In \textit{\textbf{Section \ref{androi}}}} I introduce the shortcomings of the Standard Big Bang Theory and how a condition of an early accelerated phase is the solution.
        \item \underline{In \textit{\textbf{Section \ref{oshoah}}}} I characterize the simplest dynamical models of inflation that involve a scalar field and a sufficiently flat potential to allow a phase of slow-roll evolution.
        \item \underline{In \textit{\textbf{Section \ref{sec:perturbationss}}}} I show that inflation provides an elegant mechanism able to generate the primordial scalar and tensor perturbations. I perform a detailed and complete calculation in quantum field theory, deriving the expressions for the spectra of scalar perturbations in an almost de-Sitter spacetime.
        \item \underline{In \textit{\textbf{Appendix \ref{sjsdjk}}}} I present two examples of quantization method
    \end{enumerate}
    \item \underline{In \textbf{ \large Chapter \ref{iphone}}} I introduce the idea of an Effective Field Theory and I apply it to the Inflationary framework 
    \begin{enumerate}[label=\Large$\bullet$]
        \item \underline{In \textit{\textbf{Section \ref{sec:GoldTheorem}}}} I introduce the concept of spontaneous symmetry breaking. I compute the derivation for the Goldstone boson and study the Goldstone model as an effective theory
        \item \underline{In \textit{\textbf{Section \ref{sec:HiggsModel}}}} I study the Higgs model and the Higgs mechanism with the gauge field theories.
        \item \underline{In \textit{\textbf{Section \ref{sec:GoldsotneBos}}}} I present the inflationary theory as a theory of a Goldstone boson as well as I introduce the effective action of inflation.
        \item \underline{In \textit{\textbf{Section \ref{sec:HFLEQ}}}} I conclude the chapter presenting the interesting mechanism to tackle the dynamics of the parameters of effective action that uses the Hubble Flow Equations.
    \end{enumerate}
    \item \underline{In \textbf{ \large Chapter \ref{ch:PGWs}}} I present the results obtained during my PhD in the field of inflation and Primordial Gravitational Waves.
    \begin{enumerate}[label=\Large$\bullet$]
        \item \underline{In \textit{\textbf{Section \ref{sec:TensoTensoO}}}} I study in detail the tensor perturbations and I derive the expressions for the spectra of tensor perturbations in an almost de-Sitter spacetime.
        \item \underline{In \textit{\textbf{Section \ref{sec:PIE}}}} I present the propagation of primordial gravitational waves. 
        \item \underline{In \textit{\textbf{Section \ref{sec:bBNPGW}}}} I study the implications of constraining the extra radiation from primordial gravitational waves using the BBN as observable by means of both a stochastic approach and the Hubble flow equations. The study is based on the work~\textbf{Ref.~\cite{Forconi:2}}
        \item \underline{In \textit{\textbf{Section \ref{sec:CMBPGW}}}} I provide an updated review of the observational constraints on the standard slow roll paradigm of inflation based on~\textbf{Ref.~\cite{Forconi:1}}
        \item \underline{In \textit{\textbf{Appendix \ref{ewjkq}}}} I briefly sketch the idea of gravitons
    \end{enumerate}
    \item \underline{In \textbf{ \large Chapter \ref{ch:NGty}}} I study the implications of a non-negligible contribution of non-Gaussianity to the neutrino sector.
    \begin{enumerate}[label=\Large$\bullet$]
        \item \underline{In \textit{\textbf{Section \ref{eq:Ahan}}}} I present the three-point function and the main features of the shape function
        \item \underline{In \textit{\textbf{Section \ref{sec:QSFIFIA}}}} I introduce the Quasi-single field inflationary models and how they have a boosted tirspectrum
        \item \underline{In \textit{\textbf{Section \ref{sec:PSOGFNSN}}}} I study the costraining power of the Super-$\Lambda$CDM model on the neutrino sector. It is based on~\textbf{Ref.~\cite{Forconi:4}}
    \end{enumerate}
    \item \underline{In \textbf{ \large Chapter \ref{ch:Chahahah}}} I present the non linear evolution of perturbation and study the possible implications of observing galaxy too massive to be observed in the standard cosmological model
    \begin{enumerate}[label=\Large$\bullet$]
        \item \underline{In \textit{\textbf{Section \ref{sec:NLC}}}} I introduce the concept of non-linear clustering by means of the spherical-collapse. I then present the window and mass function 
        \item \underline{In \textit{\textbf{Section \ref{sec:PolJWST}}}} I study the JWST observations in light of a possible systematic in Planck polarization, based on \textbf{Ref.~\cite{Forconi:3}}.
        \item \underline{In \textit{\textbf{Section \ref{sec:JDE}}}} I study the JWST observations within two extension of the standard model: Early Dark Energy and Interactive Dark Energy. This section is based on \textbf{Ref.~\cite{Forconi:5}}
    \end{enumerate}
\end{enumerate}}
{\large \underline{In \textit{\textbf{Appendix \ref{app:DATa}}}} I present some of the statistical tool to perform data analysis in cosmology.}

\vspace{1em}
\noindent\rule{\textwidth}{1.5pt}
\vspace{1em}

\newpage\section*{\Huge Conventions}
\addcontentsline{toc}{chapter}{Conventions}
\input{Appendices_More/Conventions}

\tableofcontents 

\listoffigures 

\listoftables 







\mainmatter 

\pagestyle{thesis} 


\renewcommand{\thechapter}{\arabic{chapter}}
\titleformat{\chapter}[display]
  {\bfseries\Huge\centering}
  {\rule{\textwidth}{1pt}\\\MakeUppercase{\chaptertitlename} \Huge\thechapter}
  {2ex}
  {\titlerule\vspace{2ex}\filcenter}

\renewcommand{\thesection}{\arabic{chapter}.\arabic{section}}
\titleformat{\section}[block]
  {\Large\bfseries}
  {\thesection}{.5em}{\titlerule[1pt]\\[.8ex]\bfseries}

\titleformat{\subsection}
  {\large\bfseries}
  {\thesubsection}
  {1em}
  {\bfseries\underline}

\titleformat{\subsubsection}
  {\large\bfseries\itshape}
  {\thesubsection}
  {}
  {}

\setcounter{secnumdepth}{2}

\renewcommand{\theequation}{\arabic{chapter}.\arabic{equation}}
\renewcommand{\thefigure}{\arabic{chapter}.\arabic{figure}}
\renewcommand{\thetable}{\arabic{chapter}.\arabic{table}}
\newcommand{\Ne}{\textit{e}-folds}
\newcommand{\ncl}{95\% CL\,}
\newcommand{\scl}{68\% CL\,}
\newcommand{\nnu}{N_{\rm eff}}
\newcommand{\meff}{m^{\rm eff}_{\nu,\rm sterile}}
\newcommand{\mef}{m_{\rm eff}}
\newcommand{\mnu}{\Sigma m_\nu}
\newcommand{\slcdm}{Super-$\Lambda$CDM\,}
\newcommand{\lcdm}{\Lambda\mathrm{CDM}}

\newcommand{\GZ}[1]{{\color{violet} [GZ: #1]}}
\renewcommand{\d}{\text{d}}

\input{Chapters/Unperturbed}
\input{Chapters/Perturbations}

\input{Chapters/CMB}
\input{Chapters/Inflation}

\input{Chapters/EFT}
\input{Chapters/PGW}
\input{Chapters/Non-Gaussianity}
\input{Chapters/NonLinear}

\titleformat{\chapter}[display]
  {\bfseries\Huge}
  {\rule{\textwidth}{1pt}\\\MakeUppercase{\chaptertitlename} \Huge\thechapter}{}{}

\chapter*{Conclusions}
\addcontentsline{toc}{chapter}{Conclusions}
\markboth{Conclusions}{Conclusions}
\input{Chapters/Conclusions}

\chapter*{Acknowledgements}
\addcontentsline{toc}{chapter}{Acknowledgements}
\markboth{Acknowledgements}{Acknowledgements}
\input{Appendices_More/Acknowledgment}

\titleformat{\chapter}[display]
  {\bfseries\Huge\centering}
  {\rule{\textwidth}{1pt}\\\MakeUppercase{\chaptertitlename} \Huge\thechapter}
  {2ex}
  {\titlerule\vspace{2ex}\filcenter}
\appendix 

\renewcommand{\thechapter}{\Alph{chapter}} 
\renewcommand{\thesection}{\Alph{chapter}.\arabic{section}}
\renewcommand{\theequation}{\thechapter.\arabic{equation}} 

\include{Appendices_More/HarmonicOscillator}

\include{Appendices_More/Gravitons}

\include{Appendices_More/DataAnalysis}


\printbibliography[heading=bibintoc]


\end{document}

%% file: Appendices_More/Conventions.tex
Even though sometimes I will keep the fundamental constants explicit in the equations, in this work I will largely adopt the so-called natural units: $c=\hbar=k_B=1$. A quantity with the units $\text{kg}^{\alpha}\text{m}^\beta\text{s}^\gamma$ in natural units becomes
 \begin{equation}
     \(\frac{E}{c^2}\)^\alpha\(\frac{\hbar c}{E}\)^\beta\(\frac{\hbar}{E}\)^\gamma=E^{\alpha-\beta-\gamma}\hbar^{\beta+\gamma}c^{\beta-2\alpha}
 \end{equation}
 which means that everything is measured in powers of units of energy. For example
 \begin{equation}
     [\text{Energy}]=[\text{mass}]=[\text{time}]^{-1}=[\text{length}]^{-1}=[\text{pressure}]^4=[\text{energy density}]^4.
 \end{equation}
Useful conversion are

\begin{gather*}
    1kg=5.610\times 10^{26}\gev c^{-2}\\
    1m=5.068\times 10^{15}\gev^{-1}(\hbar c)\\
    1s=1.519\times10^{24}\gev^{-1}\hbar\\
    1 K= 8.6\times 10^{-14} \gev\\
    1 \mpc =3.08\times 10^{22}m=1.56\times 10^{38}\gev^{-1}
\end{gather*}

while useful physical constants are

\begin{gather*}
    \text{speed of light in vacuum}\equiv c=2.9979\times 10^8ms^{-1}\\
    \text{Planck's constant}\equiv \hbar = 1.05457 \times 10^{-34}Js\\
    \text{Electron volt}\equiv 1\ev=1.6022\times 10^{-19}J\\
    \text{Boltzmann's constant}\equiv k_B=1.380\times 10^{-23}JK^{-1}\\
    \text{Newton's constant}\equiv G=6.674\times 10^{-11}m^3kg^{-1}s^{-2}\\
    \text{Planck mass}\equiv m_{\rm pl}\equiv \sqrt{\hbar c/G}=1.220\times 10^{19}\gev\\
    \text{Reduced Planck mass}\equiv M_{\rm pl}\equiv \sqrt{\hbar c/8\pi G}=2.435\times 10^{18}\gev\\
    \text{Planck length}\equiv l_{\rm pl}\equiv\sqrt{\hbar G/c^3}=1.616\times 10^{-35}m\\
    \text{Planck time}\equiv t_{\rm pl}\equiv\sqrt{\hbar G/c^5}=5.391\times 10^{-44}s
\end{gather*}

and in particle physics

\begin{gather*}
    \text{Electron mass}\equiv m_e =0.510\mev\\
    \text{Proton mass}\equiv m_p =938\mev\\
    \text{Neutron mass}\equiv m_n =939\mev\\
    \text{Proton-neutron mass difference}\equiv Q =1.293\mev\\
    \text{Neutron lifetime}\equiv \tau_n =879 s\\
    \text{Deuteron mass}\equiv m_D =1875\mev\\
    \text{Thomson cross section mass}\equiv \sigma_T =0.665\times 10^{-28}m^2\\
    \text{Fermi's constant}\equiv G_F=1.116\times10^{-5}\gev^{-2}.
\end{gather*}

 Some useful Cosmological parameters, using the values from the best fit of Planck 2018 using 6-parameter $\Lambda$CDM model, are

\begin{gather*}
    \text{CMB Temperature today}\equiv T_0 = 2.725 K\\
    \text{Number density of photon}\equiv n_\gamma = 410.7cm^{-3}\\
    \text{CMB density}\equiv \Omega_\gamma = 5.38\times10^{-5}\\
    \text{Number density of baryons}\equiv n_b 2.515\times 10^{-7}cm^{-3}\\
    \text{Baryon density}\equiv \Omega_b = 0.0493\\
    \text{Dark matter density}\equiv \Omega_c = 0.265\\
    \text{Matter density}\equiv \Omega_m = 0.315\\
    \text{Dark energy density}\equiv \Omega_\Lambda = 0.685\\
    \text{Hubble expansion rate}\equiv H_0 = 100h\,\text{km} s^{-1}\mpc^{-1}\\
    \text{Hubble parameter}\equiv h = 0.674\\
    \text{Critical density}\equiv \rho_{\rm crit} = 8.532\times 10^{-30}gcm^{-3}\\
\end{gather*}

Some important epochs in the universe are

\begin{gather*}
    \text{BBN}\rightarrow z_{BBN}=4\times 10^8 (3')\\
    \text{Electron-positron annihilation}\rightarrow z_{ee^+}=2\times 10^9 (6s)\\
    \text{Neutrino Decoupling}\rightarrow z_{\nu dec}=6\times 10^9 (1s)\\
    \text{Matter-radiation}\rightarrow z_{eq}=3402\\
    \text{Recombination}\rightarrow z_{rec}=1270\\
    \text{Photon decoupling}\rightarrow z_{\star}=1089\\
    \text{Last-scattering}\rightarrow z_{\scaleto{LSS}{4pt}}=1089\\
    \text{Baryon decoupling}\rightarrow z_{bdec}=1060\\
    \text{Half reionization}\rightarrow z_{re}=7.7\\
\end{gather*}

\subsubsection{General Relativity}
In this work, I adopt the signature $\(-,+,+,+\)$ for the metric tensor. I recall that the signature of a metric tensor is defined as the number (counted with multiplicity) of positive, negative and zero eigenvalues of the real symmetric matrix associated to the metric tensor with respect to a basis. Here, the $-$ is associated to the time dimension, and the $\(+,+,+\)$ to the space and physical dimension. With this choice, the line element of a flat maximally symmetric Minkowsky spacetime reads
\begin{equation}
    ds^2=-cdt^2+dx^2+dy^2+dz^2.
\end{equation}
and the three-dimensional Euclidean sub-space admits a positive scalar product. The interval between timelike separated events (i.e. the interval between a given event and the set of points that are inside its past and future light cone) is negative ($\Delta s^2<0$), while the interval between spacelike events (i.e. the interval between a given event and the set of points that are outside its past and future light cone) is positive ($\Delta s^2>0$). Finally, light-like events ($\Delta s^2=0$) define the limit between the two cases.

Below, some useful definition for working in the General Relativity. The comma represent the normal derivative whereas the semicolumn the covariant derivative. 

\begin{itemize}
    \item The covariant derivative for a covariant vector $v^\alpha_{;\beta}=v^\alpha_{,\beta}+\Gamma^\alpha_{\beta\gamma}v^\gamma$\\
    \item Christoffel's symbols
\begin{equation*}
    \Gamma^\mu_{\alpha \beta}\equiv \frac{1}{2}g^{\mu \nu}\left[g_{\alpha \nu,\beta}+g_{\beta \nu, \alpha}-g_{\alpha \beta,\nu}\right].
\end{equation*}\\
\item The geodesic equation
\begin{equation*}
    \frac{d^2x^\alpha}{d\tau^2}+\Gamma^\alpha_{\mu\nu}\frac{dx^\mu}{d\tau}\frac{dx^\nu}{d\tau}=0.
\end{equation*}\\
\item The Riemann tensor 
\begin{equation*}
    R^\alpha_{\beta\mu\nu}\equiv\Gamma^\alpha_{\beta\nu,\mu}-\Gamma^\alpha_{\beta\mu,\nu}-\Gamma^\alpha_{\kappa\nu}\Gamma^\kappa_{\beta\mu}+\Gamma^\alpha_{\kappa\mu}\Gamma^\kappa_{\beta\nu}.
\end{equation*}
\item The Ricci tensor 
\begin{equation*}
    R_{\mu \nu}\equiv\Gamma^\alpha_{\mu \nu,\alpha}-\Gamma^\alpha_{\mu \alpha,\nu}+\Gamma^\alpha_{\beta \alpha}\Gamma^\beta_{\mu \nu}+\Gamma^\alpha_{\beta \nu}\Gamma^\beta_{\mu \alpha}.
\end{equation*}
\item The scalar curvature 
\begin{equation*}
    R\equiv g^{\mu\nu}R_{\mu\nu}.
\end{equation*}
\item The Strong Principle of Equivalence: 
\textit{At every spacetime point in an arbitrary gravitational field, it is possible to choose a locally inertial frame (LIF) such that, within a sufficiently small region of the point in question, the laws of nature take the same form as in an unaccelerated Cartesian coordinate system in the absence of gravitation.}
\item The Principle of General Covariance: \textit{A physical equation holds in a general gravitational field if two conditions are met:}
\begin{enumerate}
    \item \textit{The equation holds in the absence of gravitation; namely, it agrees with the laws of special relativity.}
    \item \textit{The equation is generally covariant; that is, it preserves its form under a general coordinate transformation $x\rightarrow x'$.}
\end{enumerate}
\end{itemize}

%% file: Chapters/Unperturbed.tex
\chapter{The Unperturbed Universe} 
\label{fhgge}
In this chapter, we embark on an exploration of the Universe's large-scale structure, focusing initially on the theory of general relativity, which provides the framework for all our subsequent discussions. Then, we define the cosmological spacetime and its metric, laying the foundation for understanding cosmic dynamics on large scales. After establishing these geometric concepts, we proceed to examine the dynamics and thermodynamics of cosmic expansion, including the key steps in the Universe's thermal evolution. We will ignore the possible presence of an inflationary period, even though it is no secret that the standard model has multiple shortcomings that the inflation paradigm is able to resolve. The inflationary theory will be widely discussed from \chap{InfloBinglo}.

\section{Spacetime geometry}
\label{fhsd}
Modern Cosmology is based on Einstein’s Theory of General Relativity~\cite{Einstein:1916vd,Misner:1973prb,Carroll:2004st}. Gravity is \textit{universal}: given the same initial velocity, all bodies follow the same trajectory in a gravitational field, regardless of their internal constitution. That is, we can describe its effects in terms of curved geometry. Let us take the general action
\begin{equation}
    \boxed{S=\frac{c^4}{16\pi G}S_H+S_M}\,
    \label{Hilbertaction}
\end{equation}
where $S_H=\int{\sqrt{-g}Rd^nx}$ is the Hilbert action and $S_M$ is an additional term that describes matter. Under small variations of the metric, $S_H$ gives a term proportional to $\delta R_{\mu\nu}$ that assumes the form of an integral of a covariant divergence of a vector. We can set it to zero by means of the Stokes theorem.\footnote{If $V^\mu$ is a vector field over a region $\Sigma$ with boundary $\partial\Sigma$, Stokes's theorem is $\int_\Sigma{V^\mu_{;\mu}\sqrt{\lvert g\rvert}d^nx}=\int_{\partial\Sigma}{n_\mu V^\mu\sqrt{\lvert \gamma\rvert}d^{n-1}x}$, where $n_\mu$ is normal to $\partial\Sigma$ and $\gamma_{ij}$ is the metric on $\partial\Sigma$.}. For $S_M$, we can define the stress-energy tensor as
\begin{equation}
    T_{\mu\nu}=-2\frac{1}{\sqrt{-g}}\frac{\delta S_M}{\delta g^{\mu\nu}}\,.
    \label{eq:Vinted}
\end{equation}
Being aware that the critical points of an action are the classical solutions, we get the Einstein's equations
\begin{equation}
    \boxed{G_{\mu\nu}=R_{\mu\nu}-\frac 12 Rg_{\mu\nu}=\frac{8\pi G}{c^4}T_{\mu\nu}.}
    \label{eq:EinsteinEquations}
\end{equation}
with $G_{\mu\nu}$ the \textit{Einstein Tensor}. It is possible to show (see Appendix E in \cite{Wald:GR}) that using the \textit{Noether's theorem}, which associates a conservation law for every symmetry of the Lagrangian, the invariance of $S_M$  under diffeomorphisms implies 
\begin{equation}
    \boxed{T^{\mu\nu}_{\,\,\,\,;\mu}=0}\,.
    \label{eq:Tmunukio}
\end{equation}

Since $T_{\mu\nu}$ is a symmetric two-index tensor, as well as $G_{\mu\nu}$, we are supposed to have ten independent equations, precisely the number of independent components of the metric tensor. However, the \textit{Bianchi identity}\footnote{The general form of the Bianchi identity is $\nabla_{[\lambda}R_{\rho\sigma]\mu\nu}=\nabla_\lambda R_{\rho\sigma\mu\nu}+\nabla_\rho R_{\sigma\lambda\mu\nu}+\nabla_\sigma R_{\lambda\rho\mu\nu}=0$. Contracting twice we get $\nabla^\mu R_{\rho\mu}-\nabla_\rho R+\nabla^\nu R_{\rho\nu}=0$. They can also be written as $G_{\mu\nu;\mu}=0$.} gives us four constraints on $R_{\mu\nu}$. That is, there are only six truly independent equations\footnote{This is true because, given a solution to the Einstein equations in one coordinate system, it remains so in every other coordinate system. The four functions, which bring us from $x^\mu$ to a generic $x^{\mu'}$, are the four unphysical degrees of freedom in $g_{\mu\nu}$}. 

The stress-energy tensor is the source of the gravitational theory. Therefore, we are interested in the actual value of $T_{\mu\nu}$ and not just its difference between states ( e.g the motion of a particle with potential $V(x)$ is the same as that with $V(x)+V_0$, for any constant $V_0$). This leads us to an energy density characteristic of the empty space: the \textit{vacuum energy}. Such energy has no preferred direction and through the Principle of General Covariance, can be written as a constant times the metric tensor. Therefore, we can write the Einstein equations as
\begin{equation}
    \boxed{R_{\mu\nu}-\frac12 Rg_{\mu\nu}=\frac{8\pi G}{c^4}T_{\mu\nu}-\Lambda g_{\mu\nu}}\,,
    \label{eq:EinsteinLambda}
\end{equation}
where we have defined the \textit{cosmological constant}~\cite{Carroll:2000fy} as 
\begin{equation}
    \Lambda=\frac{8\pi G}{c^4}\rho_{vac}.
    \label{eq:Lambdarho}
\end{equation}

These equations are essential to define the dynamics within our Universe, which will be outlined in \sect{DynamoSassari}.

\subsection{Metric}
\label{sec:Metronomo}
In spite of the presence of a wide range of astrophysical objects, such as stars, black holes, galaxies, and clusters of galaxies, the Universe appears the same everywhere on large scales (length greater than \SI{100}{Mpc}). In particular, according to the \textit{Cosmological Principle}, on large
scale the Universe is \textit{isotropic} (space looks the same in any direction) and \textit{homogeneous} (the metric is the same throughout the manifold). The isotropy of the Universe can be probed by angular diameter distances, lensing distortion, and transverse velocities. Assuming the \textit{Copernican Principle}, the observations for the isotropy of the Universe can lead to its homogeneity~\cite{Maartens:Homogeneous}. The homogeneity and isotropy imply that a space is \textit{maximally symmetric}~\cite{Carroll:2004st}. 

\begin{tcolorbox}[mybox]
A $n$-dimensional manifold is said to be \textit{maximally symmetric} if it has $\frac 1 2n(n+1)$ \textit{Killing vectors}. The Killing vectors are associated with conserved quantities, which may be hidden by an unsuitable coordinate choice. The \textit{Killing equation} is $\xi_{\alpha;\beta}+\xi_{\beta;\alpha}=0$. If this equation admits a solution, the spacetime has a symmetry.  For example, the $\mathbb{R}^n$ space with flat Euclidean metric has two types of isometries: the translations and the rotations. If we take a point $P$, we have $n$ independent axes along which it can be moved; thus, there are $n$ translations. In addition, we can rotate one of the $n$ axes into $n-1$ other axes and, avoiding double counting, we have $\frac 1 2n(n-1)$ independent rotations. Eventually, we obtain $\frac 1 2n(n+1)$ independent symmetries, which correspond to as many Killing vectors. 

Taking a generic point $P$ in a maximally symmetric space, we can define a local inertial frame (where $g_{\mu\nu}=\eta_{\mu\nu}$). The isotropy of space demands that it must not be possible to distinguish between different directions by their curvature, that is, the curvature tensor $R_{\mu\nu\rho\sigma}$ must remain unchanged under any coordinate rotation. Since $\eta_{\mu\nu}$ is the only quantity unchanged under rotations, the curvature tensor must be some combination of unit tensors:
\begin{equation}
    R_{\mu\nu\rho\sigma}=k\left(\eta_{\mu\rho}\eta_{\nu\sigma}\right)+k_1\left(\eta_{\mu\sigma}\eta_{\nu\rho}\right)+k_2\left(\eta_{\mu\nu}\eta_{\sigma\rho}\right).
\end{equation}
The asymmetry relation $R_{\mu\nu\rho\sigma}=-R_{\mu\nu\sigma\rho}$ requires $k_1=k$ and $k_2=0$. Then, we can contract both members twice using the fact that for a $n$ dimensional
manifold $\eta^{\mu\rho}\eta_{\mu\rho}=n$ and
$\eta^{\mu\rho}\eta_{\nu\rho}=\delta^\mu_\nu$. Since it is a tensorial equation, it holds in any
maximally symmetric space and with any coordinate
system, thanks to the principle of general covariance. Eventually, we obtain
\begin{equation}
    \boxed{
    R_{\mu\nu\rho\sigma}=k\left[g_{\mu\rho}g_{\nu\sigma}-g_{\mu\sigma}g_{\nu\rho}\right] \quad \text{with} \quad k=\frac{R}{n(n-1)}
    }\,.
\end{equation}
\end{tcolorbox}

Maximally symmetric spaces are characterized by the dimension of the manifold, the signature of the metric, and by $R$. In our case we fix four dimensions and the signature, therefore maximally symmetric universes are only characterized by the sign of $k$~\cite{Hawking:1973uf}:
\begin{itemize}
    \item \textbf{$k=0$}. It corresponds to the Minkowski spacetime with the metric
    \begin{equation*}
        ds^2=-dt^2+dx^2+dy^2+dz^2.
    \end{equation*}
    \item \textbf{$k>0$}. A maximally symmetric spacetime with positive curvature is called \textit{de Sitter space}. If we embed a hyperboloid in a five-dimensional Minkowski space and introduce a suitable set of coordinates on the hyperboloid, we obtain the metric
    \begin{equation*}
        ds^2=-dt^2+\alpha^2\cosh[2](t/\alpha)\left[d\chi^2+\sin[2](\chi)(d\theta^2+\sin[2](\theta)d\phi^2)\right]\,.
    \end{equation*}
    \item \textbf{$k<0$}. The negative curvature describes an \textit{anti-de Sitter space}. We embed a hyperboloid in a fictitious five-dimensional flat manifold (the metric has two negative signs) and, with a suitable choice of coordinates, we get the metric
    \begin{equation*}
        ds^2=-\alpha^2\left(-\cosh[2](\rho) dt+d\rho^2+\sinh[2](\rho) d\Omega^2\right).
    \end{equation*}
    There are no closed time-like curves in this space. 
\end{itemize}

If our Universe had been a maximally symmetric Universe we would have only vacuum energy. This is because in a four-dimensional manifold, we get
\begin{equation}
    R_{\mu\nu}=3kg_{\mu\nu}, \quad R=12k\longrightarrow  G_{\mu\nu}=-3kg_{\mu\nu},
\end{equation}
which implies
\begin{equation}
    T_{\mu\nu}=-\frac{3k}{8\pi G}g_{\mu\nu}\,.
    \label{Tt}
\end{equation}

We shall not think of our spacetime as maximally symmetric. Instead, we should consider the Universe to be $\mathbb{R}\times \Sigma$ , where $\mathbb{R}$ is the time dimension, whereas $\Sigma$ is the spatial part; only $\Sigma$ ought to be maximally symmetric. We cannot extend the homogeneity and isotropy to time because the Universe is evolving. Therefore, we may write more explicitly the Cosmological Principle: On large scale the Universe is \textit{spatially} homogeneous and isotropic. 

\subsubsection{Friedmann-Lema\^{i}tre-Robertson-Walker metric}
To introduce the \textit{Friedmann-Lema\^{i}tre-Robertson-Walker metric} (FLRW metric henceforth), we consider the previously mentioned foliation $\mathbb{R}\times \Sigma$. Time intervals among slices do not depend on the position because of homogeneity, therefore we can choose a time coordinate such that it agrees with the proper time of all observers; this leads us to $g_{00}=-1$. Moreover, to preserve homogeneity, $g_{ij}$ depends on time only via a common  factor $R(t)$ known as the \textit{scale factor}. Furthermore, we have $g_{0i}=0$ because of isotropy, in fact a non-vanishing values would have introduced a preferred direction. The metric takes the form
\begin{equation}
    ds^2=-dt^2+R^2(t)\left[\frac{dr^2}{1-kr^2}+r^2d\Omega^2\right].
    \label{eq:FRWmetricwithR}
\end{equation}
The spatial coordinate $r$ is called \textit{comoving coordinate}\footnote{Such coordinates consider the uniform motion of the expansion of the
Universe. Namely, we can think that each galaxy carries the spatial coordinates with itself, with the result that intervals between any two galaxies remain constant, and the expansion of the Universe results in the change of the metric tensor. Only in comoving coordinate isotropy is preserved.}.

\begin{tcolorbox}[mybox]
To obtain \eq{FRWmetricwithR} we have used the fact that the maximally symmetric metric is also spherically symmetric. It implies
\begin{equation}
    d\sigma^2=e^{2\beta(r)}dr^2+r^2d\Omega^2\,,
\end{equation}
with $ds^2=-dt^2+R^2(t)d\sigma^2$. Then, we compute the $rr$ component of the Ricci tensor for this metric and obtain
\begin{equation}
    R_{\rm rr}=\frac2r\partial_r\beta(r)\,.
\end{equation}
Because of maximal symmetry we have
\begin{equation}
    R_{ij}=2k\gamma_{ij} \quad ^{(3)}R=6k
    \label{3R}
\end{equation}
where $(3)$ indicates that the quantity is associated with the three-metric. Putting them together yields
\begin{equation}
    \frac{2}{r}\partial_r\beta(r)=2ke^{2\beta(r)}
\end{equation}
whose solution is
\begin{equation}
    \beta(r)=-\frac 1 2 \ln(1-kr^2)\,.
\end{equation}
\end{tcolorbox}

Since $R=6k$, the value of $k$ sets the curvature of the spatial surface. It is common to impose a normalization such that we have $k=1,-1,0$, and the physical size of the manifold is absorbed into the scale factor $R(t)$. $R(t)$ in this case has the dimension of a length, whereas the radial coordinate is dimensionless. Since the metric is invariant under the simultaneous transformations
\begin{equation*}
   \left\{ \begin{array}{l}
    R(t)\rightarrow \lambda^{-1}R(t)\\
    r\rightarrow \lambda r\\
    k\rightarrow \lambda^{-2}k
    \end{array}\right.,
\end{equation*}
we can change our coordinates in a way that the scale factor becomes dimensionless; we have just to perform the previous transformations with $\lambda=R_0$ the physical curvature scale today. Now $k$ can take any value and the metric reads\footnote{It has to be noted that there is a $c^2$ factor in front of $dt^2$. But we are in natural units.}
\begin{equation}
\boxed{
    ds^2=-dt^2+a^2(t)\left[\frac{dr^2}{1-kr^2}+r^2d\Omega^2\right] 
    }\,.
    \label{eq:FRWmetricwitha}
\end{equation}
Henceforth, when we mention the scale factor, we are referring to the dimensionless version $a(t)$. The sign of $k$ determines the nature of the spacetime: $k=0$ corresponds to a flat Universe, $k>0$ to a closed Universe and $k<0$ to an open Universe. 

The comoving coordinate $r$ in \eq{FRWmetricwitha} is not unique, as we have seen moving from \eq{FRWmetricwithR} to \eq{FRWmetricwitha}. Therefore, we need to define the \textit{physical coordinate} $r_{\text{phys}}=a(t)r$. To better understand the difference, let us compute the physical velocity of a galaxy
\begin{equation}
    \textbf{v}_{phys}\equiv\frac{da}{dt}\textbf{r}+a(t)\frac{d\textbf{r}}{dt}\equiv H\textbf{r}_{phys}+\textbf{v}_{pec}
    \label{eq:vphys}
\end{equation}
where we have introduced the \textit{Hubble parameter}
\begin{equation}
    \boxed{H\equiv \frac{\dot{a}}{{a}}}\,.
    \label{eq:Hubble}
\end{equation}
In \eq{vphys} we have two terms. One is the \textit{Hubble flow} $H\textbf{r}_{phys}$, which describes the velocity of the galaxy from the expansion of the space. The second one, is the \textit{peculiar velocity} that describes the motion of the galaxy with respect to a cosmological rest frame. 

If we introduce a new angular variable depending on the value of $k$
\begin{equation}
    r=S(\chi)=\left\{\begin{array}{ccl}
    \sin{\chi} & \text{for} & k=1,\\
    \chi & \text{for} & k=0,\\
    \sinh{\chi} & \text{for} & k=-1,\\
    \end{array}\right.
\end{equation}
with $d\chi=\frac{dr}{\sqrt{1-kr^2}}$, together with the \textit{conformal time}
\begin{equation}
    \boxed{\eta=\int{\frac{dt}{a(t)}}}\,,
    \label{Conformaltime}
\end{equation} 
we obtain
\begin{equation}
    \boxed{
    ds^2=a^2(\eta)\left[-d\eta^2+d\chi^2+S^2(\chi)(d^2\theta+\sin^2{\theta}d^2\phi)\right]}\,.
    \label{eq:FRWmetricwithtau}
\end{equation}
The latter metric has the same form for $k=0$ as the one of a uniformly expanding Minkowski space. 

Once we have fixed the FLRW metric, we can find the Christoffel symbols and compute the geodesic equation. If we define the \textit{four-momentum} as $P^\mu=(E,p)$ for a massless particle and we focus only on the zeroth component, after some computations, and remembering that $P^\mu P_\mu=0$, we end up with
\begin{equation}
    \frac{1}{E}\frac{dE}{dt}=-\frac{\dot{a}}{a}
\end{equation}
which implies that the energy of massless particles decays with the expansion of the Universe $E\propto a^{-1}$. This result is of great importance as allows us to introduce a fundamental concept to describe our Universe: the \textit{redshift}. The energy of a photon is inversely proportional to its wavelength $\lambda$. Hence, as the Universe expand, light emitted in the past sees its wavelength elongated and this phenomenon is called redshift. If a light emitted at time $t_1$ is observed at a later time $t_0$, $\lambda_0$ would be
\begin{equation}
    \lambda_0=\frac{a(t_0)}{a(t_1)}\lambda_1\,.
\end{equation}
Conventionally, the redshift $z$ is defined as the fractional shift in the wavelength. If we set $t_0$ to the present time and $a(t_0)=1$ we can write
\begin{equation}
    \boxed{1+z=\frac{1}{a(t_1)}}
    \label{redshift}
\end{equation}
which in general implies that, if an event take place at $z=6$, for example, the Universe at that time was one-seventh its current size.

\section{Dynamics of the Universe}
\label{sec:DynamoSassari}

We have introduced in \eq{Vinted} the stress-energy tensor, which represent the source of the gravitational field in \eq{EinsteinEquations}. Thus, it is essential, in order to give a complete picture of the Universe dynamics, to fix a model for our stress-energy tensor. We should give a convenient definition of its components:
\begin{center}
       \fbox{ \textit{T}$^{\mu\nu} \equiv$ \textit{Flux of $\mu$ momentum across a surface of constant x$^{\nu}$}}\,.
\end{center}
With this definition and recalling that the $0th$ momentum corresponds to the energy, it is easy to see that $T^{00}$ corresponds to the energy density, $T^{0i}$ is the flux of energy across $ith$ surface, $T^{i0}$ is the $ith$ momentum density and $T^{ij}$ the flux of $ith$ momentum across $jth$ surface. Additionally $T^{0i}=T^{i0}$ because the energy flux is the density of energy times the speed it flows at, which, in turn, is equal to the density of mass times the speed it is moving at (the density momentum), due to the equivalence between energy and mass. $T^{\mu\nu}$ is symmetric.

Our assumption of homogeneity and isotropy of space imposes that also the stress-energy tensor should be homogeneous and isotropic; such a tensor is that of a perfect fluid. A perfect fluid in the comoving reference frame has no viscosity and no heat conduction (a generalization of the \textit{ideal gas} in thermodynamics). Eventually, in a local inertial comoving frame, the stress-energy tensor for a perfect fluid takes the form
\begin{equation}
T^{\mu\nu}= \left(
\begin{array}{cccc}
-\rho &0 &0&0 \\
0 &p &0&0 \\
0 &0&p&0 \\
0 &0 &0&p \\
 \end{array} \right)
 \label{TmunuLif}
\end{equation}
where $\rho$ is the proper energy density of the fluid and $p$ is its pressure ($p>0$) or strain ($p<0$). It can also be written as
\begin{equation}
\boxed{
    T_{\mu\nu}=(\rho+p)u_\mu u_\nu+pg_{\mu \nu}
    }\,.
    \label{eq:Tmunu}
\end{equation}
and, as already mentioned, the conservation laws of energy and momentum are expressed with \eq{Tmunukio}.

Now, we are ready to study the Einstein equations to extract more information about the scale factor $a(t)$. Since the only non-zero terms in $T_{\mu \nu}$ are along the diagonal, we expect four independent equations. Furthermore, considering the fact that the pressure components are equal, the number of independent equations is reduced to only two. If we take the $00$ component of \eq{EinsteinLambda} we obtain the \textit{first Friedmann equation}
\begin{equation}
\boxed{
\left(\frac{\dot{a}}{a}\right)^2=\frac{8\pi G}{ 3}\rho-\frac{k}{a^2}+\frac{\Lambda}{3}
}\,.
\label{eq:IFriedmann}
\end{equation}
whereas with $\{i,j\}$ we have the \textit{second Friedmann equation}
\begin{equation}
\boxed{
\frac{\ddot{a}}{a}=-\frac{4\pi G}{3}(\rho + 3p)+\frac{\Lambda}{3}
}\,.
\label{eq:IIFriedmann}
\end{equation}

Using the conservation law of $T_{\mu\nu}$ in \eq{Tmunukio} we can derive a third important equation, the \textit{continuity equation}:
\begin{equation}
    \boxed{\dot{\rho}+3H(\rho+p)=0}\,.
    \label{eq:ContinuityEq1}
\end{equation}
As it can also be obtained by differentiating the Friedmann equations, it is not independent from~\eq{IFriedmann} and~\eq{IIFriedmann}. As a consequence, to study the dynamics of the various components of the Universe, \ie the three unknowns $\[\rho,p,a(t)\]$, we are still one equation short. This necessity of a further information is fulfilled by the \textit{equation of state} which is a relation between the pressure and the energy density of the different compontents of our Universe. Since we have considered the Universe as filled with perfect fluids, it is customary to assume a \textit{barotropic} equation, namely a linear relationship in the form
\begin{equation}
    \boxed{p=w\rho}\,,
    \label{eq:equationofstate}
\end{equation}
where $w$ is the \textit{equation of state parameter} and it is taken constant for each component. The evolution of the energy density can be computed by integrating the conintuity equation \eq{ContinuityEq1} substituing the prassure with the relation in \eq{equationofstate}. We obtain
\begin{equation}
    \boxed{\rho\propto a^{-3(1+w)}}\,.
    \label{rhoa}
\end{equation}

We should now find the value of $w$ for the components of the Universe. The non-relativistic matter is characterised by $v\ll c$ and by a negligible kinetic energy $kT\ll m$. Therefore, $\rho\sim m\gg kT\sim p$, the pressure is negligible and we get $w=0$ and thus
\begin{equation}
    \rho\propto a^{-3}\,.
\end{equation}
The energy density is diluted by the expansion of the Universe simply because the volume of a region in space increases as $V\propto a^3$ while the energy within that region stays constant. For radiation (or relativistic matter in general), the pressure is $1/3$ of $\rho$ therefore
\begin{equation}
    \rho\propto a^{-4}\,.
\end{equation}
Radiation is not only diluted by the expansion, but also the energy of a single particle is additionally redshifted ($E\propto 1/a$). Additionally, we have seen that in our Universe we define also a cosmological constant (see \eq{Lambdarho}) that is a natural candidate for dark energy. It is characterised by a negative pressure $p=-\rho$ ($w=-1$) so
\begin{equation}
    \rho\propto a^{0}\,,
\end{equation}
which implies constant energy density. As the Universe expands,  the energy is not diluted which means that the energy increases in proportion to the volume. For this reason it is often called \textit{vacuum energy}. In general we have a decelerating Universe as long as $w>-1/3$. A generic component with $w$ less than the deceleration limit, namely $w\leq-1/3$, is called \textit{Dark Energy}. Lastly, it is easy to see that the curvature term can be assimilated with a fluid with an equation of state parameter $w=-1/3$.

The Hubble parameter that we have defined in \eq{Hubble} allows us to define the \textit{critical} energy density $\rho_c(t)$ and the \textit{density parameter} for a generic component:
\begin{equation}
    \boxed{\Omega\equiv \frac{\rho(t)}{\rho_c(t)} \quad \text{with} \quad \rho_c(t)\equiv \frac{3H^2}{8\pi G}}\,.
    \label{eq:densityparameter}
\end{equation}
We may also define a density parameter associated with the curvature term
\begin{equation}
    \boxed{\Omega_k=-\frac{k}{H^2a^2}}
    \label{eq:curvatureparameter}
\end{equation}
and thus from \eq{IFriedmann} follows that $\Omega_k=0$ corresponds to a spatially flat universe and $\Omega_k>0$, $\Omega_k<0$ to an open and closed universe respectively.

Obviously the Universe contains more than one species (baryons, neutrinos, dark matter, dark energy, ...), and for this reason, the energy density and pressure shall be decomposed as a sum of all components and the density parameter becomes:
\begin{equation}
    \Omega=\sum_i\Omega_i,\quad \Omega_i\equiv\frac{\rho_i(t)}{\rho_c(t)}\qquad i=\{matter, radiation, \Lambda,k\}
    \label{eq:Omega}
\end{equation}

An important notation that would be adopted from now on is to denote the present time as $t_0$ (in general the subscript zero indicates that the quantity is referred to the present time) and we normalize the scale factor such that henceforth $a_0=a(t_0)\equiv 1$. As a consequence $H_0=\dot{a}_0$ is the present day Hubble parameter. The present density parameter for each species $i$ allows us to write the first Friedmann equation as
\begin{equation}
    \left(\frac{H}{H_0}\right)^2=\sum_i\Omega_{0i}a^{-3(1+w_i)}\qquad i=\{matter, radiation, \Lambda,k\
\end{equation}
where we have used the fact that the curvature can be thought of as a fluid with $w_k=-1/3$. The second Friedmann equation can be written as:
\begin{equation}
    \boxed{\frac{1}{H_0^2}\frac{\d^2a_0}{\d t^2}=-\frac 1 2 \sum_i\Omega_{0i}(1+3w_i)a^{-3(1+w_i)}}
\end{equation}
again with $i=\{matter, radiation, \Lambda,k\}$ which defines the condition for the accelerated expansion today.

If we are interested on the evolution law of the scale parameter we can combine \eq{IFriedmann} with \ref{rhoa}. For a simple time dependence, we ought to consider a flat universe (\ie $k=0$) which is also the most relevant case considering that observations suggest $\Omega_k\sim0$ \cite{Planck:2018vyg}. We end up with two different cases:
\begin{itemize}
    \item $w\neq-1$. The Hubble rate equation implies $\dot{a}\propto a^{-(1+3w)/2}$. Integrating we obtain 
    \begin{equation}
        a(t)\propto t^{\frac{2}{3(1+w)}}.
        \label{powerlaw}
    \end{equation}
    From this expression, we can find the evolution law in terms of conformal time 
    \begin{equation}
    \left\{\begin{array}{ll}
        a(\eta)\propto\eta^{\frac{2}{1+3w}}&\text{with}\quad w\neq -\frac 13\\
        
        a(\eta)\propto e^\eta &\text{with}\quad w=-\frac13
        \end{array}\right. .
    \end{equation}\\
    \item  $w=-1$. Here, we have $\dot{a}\propto a$, hence, by integrating we get
    \begin{equation}
        a(t)\propto e^{Ht}
        \label{deSitterstage}
    \end{equation}
    and the conformal time dependency of the scale factor is
    \begin{equation}
        a(\eta)\propto -\frac 1 {H\eta}, \quad \text{for} \quad \eta<0.
        \label{DeSitterTime}
    \end{equation}
\end{itemize}
The behaviour for the energy density, the equation of state parameter and the scale factor for the specific components is summarized in \tab{Table}
\begin{table}[h!]
\centering
\begin{tabular}{ |c|c|c|c|c|c|c| }
\hline
  & $w$ & $\rho$ & $H$ & $\Omega-1$& $a(t)$ & $a(\eta)$ \\ 
  \hline\hline
 RD & $\frac 13$ & $a^{-4}$&$a^{-2}$&$a^2$&$t^{\frac 12}$&$\eta$ \\  
 MD& $0$ & $a^{-3}$&$a^{-\frac 32}$&$a$&$t^{\frac 23}$&$\eta^2$ \\ 
 $\Lambda$ & $-1$ & $\rho_{\Lambda}$&$\frac{\Lambda}{3}$&$e^{-2Ht}$&$e^{Ht}$&$-\frac{1}{H\eta}$ \\ 
 \hline
\end{tabular}
\belowcaptionskip=5pt
\caption[ Evolution laws]{\small Some evolution laws for a universe dominated by a fluid with equation of state parameter $w$. RD stands for radiation-dominated while MD means matter-dominated.}
\label{tab:Table}
\end{table}

\section{Thermodynamics}
\label{entropyandstatistics}
We now describe briefly the thermal history~\cite{Weinberg:1977ji,Dodelson:2003ft,Kolb:1988aj} of our Universe. The early Universe was a hot and dense gas, and a precise deterministic description of each particle is not possible. Instead, a statistical description is more suitable. For this reason, in \sect{stato} an introduction to the basic concepts in statistically thermodynamics is given. Then in \sect{ThermalHistory} are presented briefly some of the main thermal history events. Hereafter, the reduced Planck mass is introduced with $\M=\hbar c/(8\pi G)$.

\subsection{Statistical Mechanics}
\label{sec:stato}
The probability to find a particle with $3$-momentum within $[\textbf{p},\textbf{p}+d\textbf{p}]$ in $[\textbf{x},\textbf{x}+d\textbf{x}]$ is
\begin{equation}
    P=f(\textbf{x},\textbf{p},T,t)d\textbf{p}d\textbf{x}.
    \label{eq:P}
\end{equation}
Given the fact that we have a homogeneous Universe, we can exclude the dependence in \textbf{x} as well as $t$ if we restrict ourselves to species in equilibrium. In \eq{P} we have introduced $f$, which is the distribution function in phase space defined as
\begin{equation}
    f(\textbf{p})=\frac{1}{e^\frac{(E(\textbf{p})-\mu)}{kT}\pm1} \text{\, \,  with \,\,} E^2(\textbf{p})=|\textbf{p}|^2+m^2
\end{equation}
where the plus sign refers to \textit{Fermi-Dirac} species whereas the minus to \textit{Bose-Einstein} ones. With the distribution function, we can define the number density $n$, the energy density $\rho$ and the pressure $p$~\cite{Baumann:2022mni,Mukhanov:2005sc,Weinberg:2008zzc} as
\begin{equation}
        n =\frac{g}{(2\pi)^3}\int{f(\textbf{p})d\textbf{p}},\quad\rho =\frac{g}{(2\pi)^3}\int{E(\textbf{p})f(\textbf{p})d\textbf{p}} ,\quad
        p=\frac{g}{(2\pi)^3}\int \frac{|\textbf{p}|^2}{3E(\textbf{p})}f(\textbf{p})d\textbf{p},
        \label{eq:nrhop}
\end{equation}
where $g$ represents the internal degrees of freedom of the gas of particles and $g/(2\pi)^3$ is the denisty of states (remember we are in natural units $\hbar\rightarrow h/(2\pi)$). At early times, the chemical potentials of all particles is much smaller than the temperature and therefore can be neglected from now on.

To find an explicit expression for $n$ and $\rho$ we need to numerically evaluate the integrals. However, we can find an analytical solution if we consider the temperature away from the mass threshold. In the relativistic limit, the temperature $T$ is much larger than the particle mass $T\gg m$, and the integral gives a contribution equal to $2\zeta(3)$ for bosons  leading to 
\begin{equation}
    n=\frac{\zeta(3)}{\pi^2}g_BT^3,\quad \rho=\frac{\pi^2}{30} g_BT^4
    \label{eq:bosonsnandrho}
\end{equation}
and $\frac23\zeta(3)$ for fermions,
\begin{equation}
n=\frac{3}{4}\frac{\zeta(3)}{\pi^2}g_FT^3\quad \rho=\frac{7}{8}\frac{\pi^2}{30}g_FT^4\,.
\label{eq:fermionsnandrho}
\end{equation}
where the subscripts B and F stand for \textit{Bose} and \textit{Fermi} species. It is easy to check that being relativistic particles we have $p=E$ and we restore the known result $p=\frac13\rho$. Knowing that the observed temperature for CMB photons is $T_0=2.725K$~\cite{Planck:2018nkj}, we can find that, today, we have approximately $410\,\text{photons cm}^{-3}$ with an energy density of $\Omega_{\gamma}h^2\approx 2.5\times 10^{-5}$\,.

\begin{tcolorbox}[mybox]
On the other hand, we have another analytical solutions which is coming from massive particles ($T\ll m$). In this scenario we do not have a distinction with bose and fermion species. Instead, we have the same exponential suppression, leading to a density number of
\begin{equation}
    n=g\left(\frac{mT}{2\pi}\right)^{\frac32}e^{-\frac{m}{T}}\,.
    \label{eq:nnonrel}
\end{equation}
The energy density is computed through the expansion of the energy E
\begin{equation}
    \rho\approx mn+\frac32 nT
\end{equation}
and the pressure $P$ gets the natural expression of an ideal gas law. Given the non-relativistic limit, we have that the gas is a pressureless fluid (matter).  
\end{tcolorbox}

As we have already mentioned, there are multiple components in the Universe. For this reason, it is instructive to fix a specific temperature, $T=T_\gamma$ where $T_\gamma$ is the photon temperature as photons are the dominant specie and their temperature well approximate the one of the Universe, and define the effective number of relativistic degrees of freedom $g_\star(T)$\footnote{we have include only relativistic species that gives the greatest contribution} as
\begin{equation}
    g_\star(T)\equiv \sum_{i=b}g_i\left(\frac{T_i}{T}\right)^4+\frac78\sum_{i=f}g_i\left(\frac{T_i}{T}\right)^4
    \label{eq:gstar}
\end{equation}
in order to write
\begin{equation}
    \rho=\frac{\pi^2}{30}g_\star(T)T^4\,.
\end{equation}

The last important quantity we want to introduce, of primary importance due to the fact it is conserved during the evolution of the Universe, is the entropy density $\mathbf{s}$. From the first law of thermodynamics we can find
\begin{equation}
    s=\frac{\rho+P}{T}\quad\text{and}\quad\frac{\d s}{\d T}=\frac1T\frac{\d \rho}{\d T}\,.
\end{equation}
From the latter expression, using the continuity equation, we see that $\d (sa^3)/\d t=0$ and therefore $s\propto a^{-3}$. Hence, $s$ is conserved in equilibrium. 

As we have done for $\rho$, we introduce the total entropy density
\begin{equation}
    s=\left(\frac{2\pi^2}{45}\right)q_{\star}T^3
\end{equation}
with 
\begin{equation}
    q_{\star}(T)=\sum_{i=b}g_i\left(\frac{T_i}{T}\right)^3+\frac{7}{8}\sum_{i=f}g_i\left(\frac{T_i}{T}\right)^3.
    \label{eq:qstaar}
\end{equation}
It is interesting to note that, in thermal equilibrium $g_{\star}(T)=q_\star(T)$~\cite{Schwarz:2003du}.

\subsection{Thermal History}
\label{sec:ThermalHistory}
Now  that some of the mean thermodynamics features are highlighted, some important steps in the thermal history of our Universe are presented. 

\paragraph{\textit{Neutrino Decoupling}}
Being the neutrinos the most weakly interacting particles, they decoupled first from the thermal plasma at $T\sim 1\mev$. The interaction rate per particle is~\cite{Dolgov:2002ab,Wong:2002fa,Abazajian:2002qx} 
\begin{equation}
    \boxed{\Gamma\equiv n\sigma\lvert v\rvert}\,,
\end{equation}
where $n$ is the number density of the target particles, $\sigma$ is the cross section and $v$ is the relative velocity. For weak scale interaction we can approximate $\sigma\approx G_F^2T^2$ with $G_F\approx1.2\times10^{-5}\gev^{-2}$ the Fermi's constant. Recalling the results in \eq{fermionsnandrho} we see that $\Gamma\propto G_F^2T^5$. On the other hand, from \eq{fermionsnandrho} and \eq{bosonsnandrho}, at early time (radiation dominated Universe) we can see that: 
\begin{equation}
    H^2=\frac{\rho}{3\Mpl^2}\approx\frac{\pi^2}{90}g_\star \frac{T^4}{\Mpl^2}\,.
\end{equation}
Therefore, we can find
\begin{equation}
    \frac{\Gamma}{H}\approx\( \frac{T}{1 \mev} \)
\end{equation}
which implies that the interaction rate become smaller than the Hubble rate around $T\sim 1\mev$. However, despite being decoupled from the primordial plasma, neutrinos temperature still decrease as $T_\nu\propto a^{-1}$, maintaining $T_\nu=T_\gamma$. But that is not for long. 

\paragraph{\textit{Electron-Positron Annihilation}}
As soon as the temperature drops below the electron mass (namely $T\sim 0.5\mev$), we can have pair annihilation between electron and positron, resulting into two photons. This injection of energy does not affect the neutrinos which are now decoupled. Therefore from now on $T_\nu\neq T_\gamma$. We can quantify these difference by computing the effective number of degrees of freedom entropy\footnote{the annihilation of $e^+$ and $e^-$ occurs adiabatically therefore entropy in comoving volume is conserved $q_\star(T)a^3T^3=const$.} defined in \eq{qstaar}. Before the energy injection from the pair annihilation into the $\gamma$ plasma, electron and positrons contribute to $q_\star(T)$ with a factor of $\frac72$ and therefore, we have $q_{\star}=\frac{11}2$. After the annihilation, we reduce to $q_{\star}=2$. Since $T_\nu=T_\gamma$ was holding before annihilation and $aT_\nu$ remains the same, we end up with 
\begin{equation}
    T_\nu=\(\frac4{11}\)^{\frac13}T_\gamma .
\end{equation}
However, when computing effective number of relativistic degrees of freedom either for energy density \eq{gstar} and entropy \eq{qstaar}, we introduce the parameter $\neff$, called \textit{effective number of relativistic species}~\cite{Mangano:2001iu,Bennett:2019ewm,Mangano:2005cc} that takes into account the non instantaneous deocoupling of neutrinos (in which case we would have $\neff=3$) and the consequently injection of some energy. For example, we have the energy density of radiation $\rho_r$ written as the sum of photons and neutrinos
\begin{equation}
    \rho_r=\[1+\frac78\(\frac{4}{11}\)^{\frac43}\neff\]\rho_\gamma
    \label{eq:Neff}
\end{equation}
with $\neff=3.046$~\citep{Mangano:2005cc,2016JCAP...07..051D,Akita:2020szl,Froustey:2020mcq,Bennett:2020zkv,Planck:2018vyg}
Knowing today's photon temperature,  we can compute $T_{\nu 0}=1.95K$. Also, we have $112\,\text{neutrinos cm}^{-2}$ and $\Omega_{\nu}h^2\approx 1.7\times10^{-5}$ for massless neutrinos. On the other hand, neutrino oscillations experiments set a lower bound to the sum of neutrinos masses $\Sigma m_{\nu,i}>0.06 \ev$ (see e.g.~\cite{Lesgourgues:2006nd}) that leads to
\begin{equation}
    \Omega_\nu h^2\approx\frac{\Sigma m_{\nu,i}}{94\ev} .
\end{equation}

\paragraph{\textit{Big Bang Nucleosynthesis}}
Between $1$ to $300s$ after the Big Bang, or between $0.1\mev< T<50\kev$, light elements are formed via the \textit{Big Bang Nucleosynthesis}~\cite{Fields:2006ga,Steigman:1998vy,Sarkar:1995dd,Burles:2000zk,Alpher:1948ve,Cooke:2024nqz}. Until $T\sim 0.1\mev$, protons and neutrons are in equilibrium and we can write the ratio of the number density as
\begin{equation}
    \(\frac{n_n}{n_p}\)_{\rm eq}=e^{-\frac QT}
\end{equation}
where we simplified the masses in the prefactor approximating them as equal, but preserving them in the exponent, defining $Q\equiv m_n-m_p=1.30 \mev$. Hence, the number of neutrons starts to drop when $T<1\mev$ until it reaches a constant value of about $X^\infty_n\sim 0.15$ with $X_n$ the neutron fraction $X_n\equiv n_n/(n_n+n_p)$. 
\begin{tcolorbox}[mybox]
    To get the value $X^\infty_n$ we must introduce the \textit{Boltzmann equation}:
    \begin{equation}
        \frac{dn_i}{dt}+3\frac{\dot{a}}{a}n_i=C[\{n_j\}]
        \label{eq:Boltzmann}
    \end{equation}
     where $C[\{n_j\}]$ is a collision term. It gives the evolution of the number density of a particle species $i$. If we suppose that particles $1$ and $2$ interact producing $3$ and $4$, the latter equation can also be written as
     \begin{equation}
         \frac1{a^3}\frac{d(n_1 a^3)}{dt}=-\langle\sigma v\rangle\[n_1n_2-\(\frac{n_1n_2}{n_3n_4}\)_{\rm eq}n_3n_4\]\,.
         \label{eq:Boltzmannexp}
     \end{equation}
     In our case of study, we have protons coupled to neutrons towards the $\beta$-decay and inverse $\beta$-decay. We can obtain the evolution of the neutron number density from \eq{Boltzmannexp}
     \begin{equation}
         \frac{dX_n}{dt}=-\Gamma_n\[X_n-(1-X_n)e^{-\frac{Q}{T}}\]
     \end{equation}
     and the total rate $\Gamma_n$~\cite{Dodelson:2003ft} is given by
     \begin{equation}
         \Gamma_n(x)=\frac{255}{\tau_n}\frac{12+6x+x^2}{x^5}
     \end{equation}
     where $\tau_n=886.7\pm0.8\,s$ is the neutron lifetime. On the other hand, we are in a radiation dominated Universe, which gives us the Hubble rate
     \begin{equation}
         H(x)=\frac\pi 3\sqrt{g_\star}{{10}}\frac{Q^2}{\Mpl}\frac{1}{x^2}=\frac{H_1}{x^2}
     \end{equation}
     with $H \sim 1.13\,s^{-1}$. Now we want to find the freeze-out abundance. We can see that $dx/dt=xH$ knowing that $T\propto a^{-1}$ and consequently
     \begin{equation}
         \frac{d X_n}{dx}=\frac{\Gamma_n(x)}{H_1}x\[e^{-x}-X_n(1+e^{-x})\]
     \end{equation}
     and from that we can numerically solve it and find the already mentioned result $X^\infty_n\sim 0.15$
\end{tcolorbox}
To that freeze-out abundance, we need to add the dependence on the lifetime of the neutron which, at temperatures below $0.2\,\,mev$ becomes relevant
\begin{equation}
    \boxed{X_n(t)=X_n^{\infty}e^{-\frac{t}{\tau_n}}}
\end{equation}

\begin{figure}[h!]
\centering
\includegraphics[width=0.9\textwidth]{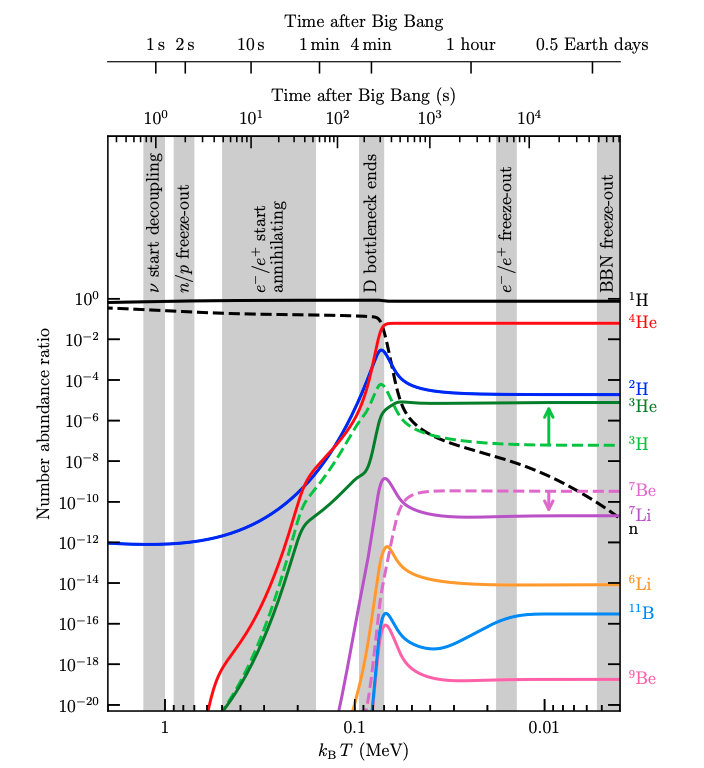}
\caption[The time evolution of the primordial nuclides]{\small The time evolution of the primordial nuclides.  The neutrons and protons are in equilibrium at the earliest times, until the weak interaction rates fall below the expansion rate of the Universe and the relative abundance of neutrons and protons freezes out. Free neutrons (dashed black curve) start decaying, and continue to decay throughout and beyond the period of BBN, until there are no free neutrons remaining. Meanwhile, the lightest nuclides, particularly deuterions (dark blue curve), start forming in appreciable abundance, at which point $4He$ can establish its nuclear statistical equilibrium abundance (red curve). Most of the primordial nuclides freeze-out around 8 hours after the Big Bang. Figure taken from~\cite{Cooke:2024nqz}.}
\label{fig:BBNCOOK}
\end{figure}

At this point, protons and neutrons combine to form the first nucleus: deuterium D. Other heavier nuclei need sufficient number of deuterium to be available. This phenomenon is called the \textit{deuterium bottleneck} (see \fig{BBNCOOK}).
The abundance of deuterium at equilibrium can be written as
\begin{equation}
    \(\frac{n_D}{n_p}\)_{\rm eq}=\frac34 n_n^{\rm eq}\(\frac{4\pi}{m_p T}\)^{\frac32}e^{\frac{B_D}{T}}
\end{equation}
with $B_D$ the binding energy of deuterium $B_D=2.22\,\mev$. If we approximate
\begin{equation}
    n_n\sim n_b=\eta n_\gamma=\eta\times\frac{2\zeta(3)}{\pi^2}T^3
\end{equation}
with $\eta=n_b/n_\gamma\approx 6\times 10^{-10}$~\cite{Fields:2019pfx,Cyburt:2015mya,Pitrou:2018cgg,Foley:2017xex} so small that to compensate the temperature has to drop at round $T_{\rm nuc}\sim0.06\mev$ to get a deuterium fraction comparable to the protons. 

Once deuterium is available, helium start to form. Being the binding $B_{\rm He}>B_D$, the formation of helium is favored. The reactions are
\begin{gather*}
    D+p^+\longleftrightarrow ^3He+\gamma\\
    D+D\longleftrightarrow ^3H+p^+\\
    D+D\longleftrightarrow ^3He+n\\
    ^3H+p^+\longleftrightarrow ^3He+n\\
    ^3H+D\longleftrightarrow ^4He+n\\
    ^3He+D\longleftrightarrow ^4He+p^+\\
\end{gather*}
and we can see that all the neutrons at $t_{nuc}\sim250\,s$ goes int $^4He$. We can define the mass fraction of helium as
\begin{equation}
    \boxed{Y_p=\frac{4n_{\rm He}}{n_H}=\frac{4n_n/2}{n_n+n_p}\sim 0.25}
\end{equation}
more exact equation can be found in~\cite{Pitrou:2018cgg}.

Regarding the other light elements, only a few amount of beryllium and lithium are created and almost no nuclei beyond a mass number of $8$ as they are not stable. An example evoultion of nuclei formation is given in \fig{BBNCOOK}.

\paragraph{\textit{Recombination}}
Until now, the temperature was high enough to make it impossible for the electrons to be bound into atoms. Electrons and protons are tightly coupled via Coulomb scattering while free electrons and photons reach the thermal equilibrium as long as their Thomson scattering rate is larger than the expansion rate. But, with the temperature steadily decreasing, we enter the epoch called \textit{recombination epoch}. The atoms are being formed and photons eventually decoupled. First and foremost, it has to be underlined that helium formation, which completes when $T\sim 0.47\ev$ happened before hydrogen formation. Moreover, it takes place before photon decoupling, therefore it does not have a great impact on the photons we observe today as the Universe was still optically thick. Therefore, we can ignore it while discussing the hydrogen recombination. Also, we will assume that the number density of free electrons is equal to that of free protons: $n_e=n_p$. We now define the fractional ionization
 \begin{equation}
     X\equiv\frac{n_p}{n_p+n_H}=\frac{n_e}{n_{b}},
 \end{equation}
 whose domain is $X\in [0;1]$, where $X=0$ corresponds to an universe neutral, whereas $X=1$ to a completely ionized one. If a photon has an energy greater than the ionization energy of hydrogen ($Q=\SI{13.6}{\electronvolt}$), the photoionization of the atom can occur. The formation of hydrogen atoms occurs via the reaction 
\begin{equation}
    e^{-}+p^{+}\longleftrightarrow H+\gamma
    \label{eq:Hreaction}
\end{equation}
 where the right to left relation is the \textit{radiative recombination}. Trivially, $X$ depends on the balance between these two reactions.
 
At equilibrium, with the temperature less than the mass of hydrogen, electron and proton, we can write the \textit{Saha equation}
\begin{equation}
    \boxed{\(\frac{1-X_e}{X^2_{e}}\)_{\rm eq}=\frac{2\zeta(3)}{\pi^2}\eta\(\frac{2\pi T}{m_e}\)^{\frac32}e^{\frac{E_I}{T}}}
\label{eq:Saha}
\end{equation}
with $X_e$ the free-electron fraction and $E_I$ the ionization energy of hydrogen.
\begin{tcolorbox}
    The Saha equation can be derived using the definition in \eq{nrhop} in the limit \eq{nnonrel} because $T\ll m$ with $m$ the mass of the components in \eq{Hreaction}. Also, it is important to take into account the fact that, at equilibrium, the chemical potential of both members in the reaction should be equal, with $\mu_\gamma=0$. Hence, we can write
    \begin{equation}
        \(\frac{n_H}{n_en_p}\)_{\rm eq}=\frac{g_H}{g_eg_p}\(\frac{m_H}{m_em_p}\frac{2\pi}{T}\)^{\frac32}e^{\frac{(m_p+m_e-m_H)}{T}}\,.
    \end{equation}
    If we now set $g_p=g_e=2$ and $g_H=4$, $n_e=n_p$ because of the neutral charge of our Universe, and we approximate $m_H\sim m_p$ (only) in the prefactor we have
    \begin{equation}
        \(\frac{n_H}{n_e^2}\)_{\rm eq}=\(\frac{2\pi}{m_e T}\)^{\frac32}e^{\frac{E_I}{T}}\,.
    \end{equation}
    Lastly, introducing the free-electron fraction $X_e\equiv n_e/(n_e+n_H)$ and neglecting the small amount of helium atoms ($n_b\approx n_e+n_H$), we recover the Saha equation in \eq{Saha}
\end{tcolorbox}
The recombination temperature is defined as the temperature at which $X_e=0.5$ and for $\eta\approx 6\times 10^{-10}$ we have $T\approx 0.32\ev$ ($z_{\rm rec}\approx 1270$). It is important to highlight that the temperature has to drop well below the binding energy of the hydorgen because of $\eta$: rare high energy photons are still sufficient in number to ionize hydrogen atoms. Shortly after, the photons decouple. This happens because there are no more free electrons, due to recombination process. In general the interaction rate of photons depend on Thomson scattering ($\sigma_T\approx 2\times 10^{-3}\mev^{-2}$) and the number density of electrons, which decreases until we reach $\Gamma_\gamma(T_{\rm dec)}\approx H(T_{\rm dec})$. Keeping in mind that we are in a matter dominated Universe and $n_bX_e(T_{\rm dec})=n_e$ we find\footnote{This number is given by adjusting our findings with a more precise treatment of this process. Otherwise we would have get $T_{dec}\approx 0.27\ev$} $T_{dec}\approx 0.25\ev$ ($z_{\rm dec}\approx 1090$). The fact that $T_{rec}\neq T_{dec}$ implies that a large degree of neutrality is necessary before the Universe becomes transparent to photons ($X_e(T_{rec}\approx 0.5>X_e(T_{dec}\approx 0.001$).

The decoupling of photons introduce one last important concept in the recombination epoch: the moment of \textit{last-scattering} (LS). It corresponds to the moment when the scattering between photons and electrons ends and the photons are free to travel towards us. We can define the \textit{optical depth} $\tau$ as
\begin{equation}
    \tau(t)=\sigma_T\int^{t_0}_{t}n_e(t')dt'
    \label{eq:taureio}
\end{equation}
and the probability that a photon did not scatter off an electron between redshift $z$ and redshift  $z_0$ as 
\begin{equation}
    P(z,z_0)=e^{-\tau(z,z_0)}\,.
    \label{eq:probnoscatt}
\end{equation}
As a consequence, the probability that a photon scattered last at certain redshift $z$ is given by the \textit{visibility function}
\begin{equation}
    g(z)=\frac{d\tau}{dz}e^{-\tau}\,.
\end{equation}
Being the visibility function the product between the optical depth, which is high at early time, and its derivative, which is small at late time because the density of free electrons is small, it has a peculiar peak feature. The maximum of $g$ is at $z_{\star}=1080$ and it defines our moment of last scattering~\cite{Baumann:2022mni}. The photons emitted at the LS surface defines the \textit{Cosmic Microwave Background}.

\begin{figure}[h!]
\centering
\includegraphics[width=0.7\textwidth]{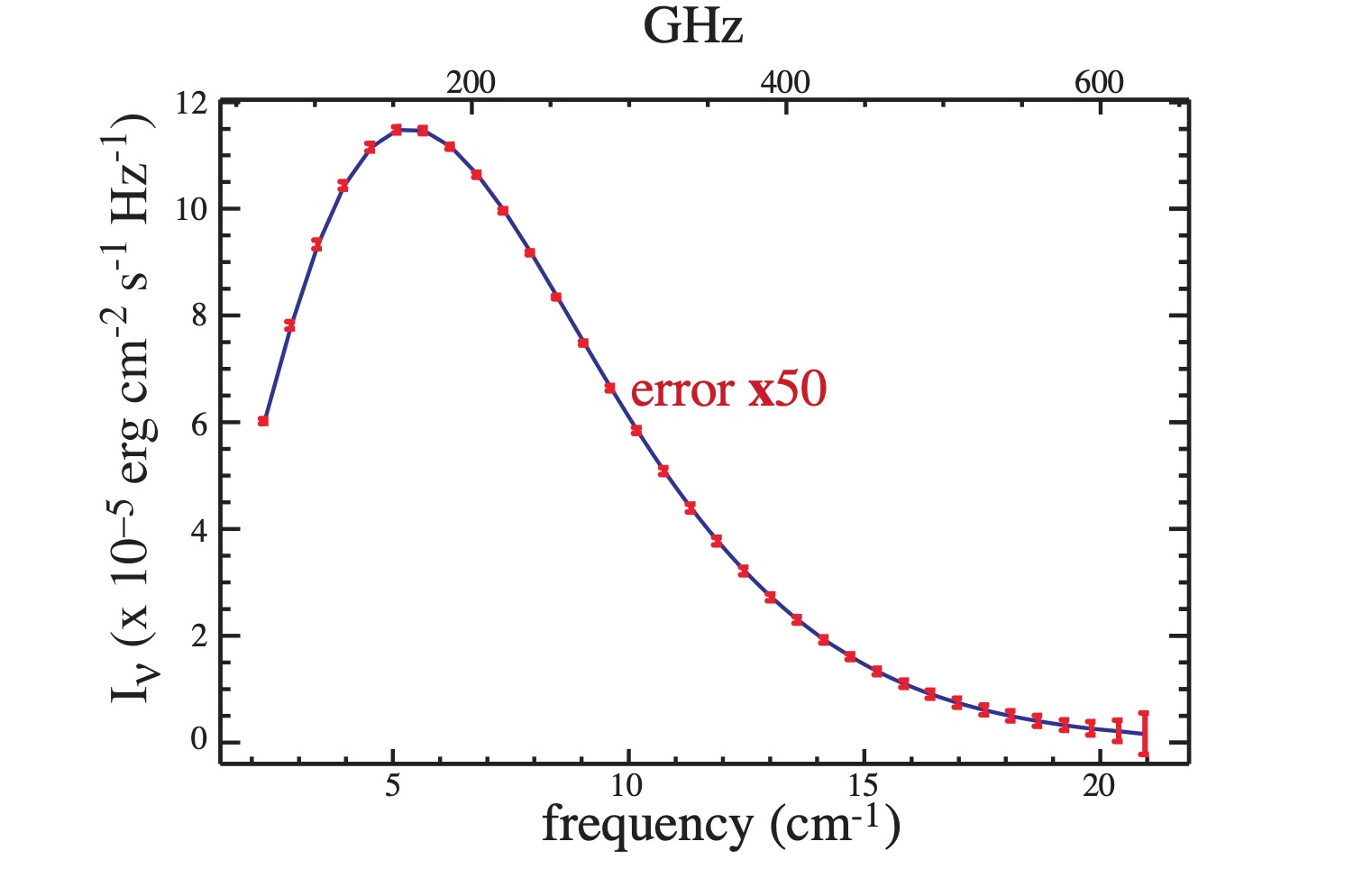}
  \caption[Black-body Cosmic Microwave Background spectrum] {\small We can see the remarkable agreement between the Cosmic Microwave Background spectrum measured by COBE~\cite{Wright:1993re,Mather:1993ij} and the theoretical black-body spectrum. Image taken from~\cite{Hu:2008hd}.}
  \label{fig:BBspectrum}
\end{figure}

The number density of photons at temperature $T$, in thermal equilibrium, with a frequency between $\nu$ and $\nu+d\nu$, is given by the black-body spectrum~\cite{Fixsen:1996nj,Peebles:1991ch}\,:
\begin{equation}
    n_T(\nu)d\nu=\frac{8\pi\nu^2d\nu}{e^{\frac{2\pi\hbar\nu}{k_BT}}-1}\,,
    \label{eq:BBnot}
\end{equation}
with $k_B$ the Boltzmann's constant which, from now on, we are going to set $k_B=1$;  we have also restored $\hbar$ in this equation for completeness. Once the scattering rate drops below $H$, the radiation expands freely, but its spectrum keeps the same form. However, the frequency changes with the scale factor as well as the number density. Both contributions are cancelled when they are inserted in \eq{BBnot}, except in the exponential. We have
\begin{equation}
        n_{T(t)}(\nu)d\nu=\frac{8\pi\nu^2d\nu}{e^{\frac{2\pi\nu}{T(t)}}-1},
    \label{BBody}
\end{equation}
where we have absorbed the scale factor in the temperature term, $T(t)=T(t_{\rm LS})a(t_{\rm LS})/a(t)$, with $t_L$ the time of last scattering. The observed black-body spectrum is shown in \fig{BBspectrum}.

Since our CMB photons are consistent with a Planck distribution, the energy density is given by
\begin{equation}
    \rho_\gamma=\int_0^\infty{2\pi\nu n(\nu)d\nu}=\alpha T^4,
    \label{eq:rhorad}
\end{equation}
where $\alpha$ is the radiation energy constant
\begin{equation}
    \alpha=\frac{8\pi^5k_B^4}{15h^3c^3}.
\end{equation}
Thus, the current energy density of CMB photons is
\begin{equation}
    \rho_{\gamma0}=\alpha T^4_{\gamma0}=\SI{4.64e-34}{\gram\per\cubic\centi\metre}.
\end{equation}
where we have used the present temperature of microwave background photons. The energy per photon is$\approx \SI{6.34e-4}{\electronvolt}$. To comparison, the baryon's energy density is approximately 900 times larger. But on the other hand, the CMB photons outnumber the baryons
\begin{equation}
    n_{\gamma0}=410 \frac{\text{photons}}{cm^3},\quad n_{b0}=\SI{1.123e-5}\Omega_Bh^2\frac{\text{nucleons}}{cm^3}.
\end{equation}
In fact, as we have previously seen, the baryon-to-photon ratio is of order $10^{-10}$. Since both $n_{\gamma}$ and $n_B$ scale as $a(t)^{-3}$, $\eta$ has been the same, at least during the whole period where the Universe became transparent to photons.

%% file: Chapters/Perturbations.tex
\chapter{Perturbation Theory} 
\label{Chapter2}
In this chapter, we explore the theory of perturbations, which plays a crucial role in understanding the formation of structures in the Universe. We begin by introducing the formalism for decomposing perturbations in spacetime and the concept of gauge freedom. It requires careful choices, and we will explore different gauge conditions that simplify the equations governing perturbations.

We will then focus on the dynamics of perturbations, beginning with the definition of the perturbed stress-energy tensor and the perturbed Einstein equations, which lead to the growth of inhomogeneities. We will also introduce tensor perturbations, which describe gravitational waves. This chapter lays the foundation for the analysis of cosmic microwave background anisotropies, which will be explored in \chap{CMBCHAPTER}. The process of producing these perturbations is outlined in \chap{InfloBinglo}.

\section{Metric decomposition}
\label{sec:fieldEquation}

Let us now perturb the FLRW metric at the first order
\begin{equation}
   \boxed{ g_{\mu\nu}=\bar{g}_{\mu\nu}+h_{\mu\nu},\qquad h_{\mu\nu}\ll \bar{g}_{\mu\nu}}\,,
   \label{eq:perturbeggmunu}
\end{equation}
where the bar indicates the unperturbed quantity and $h_{\mu\nu}$ is a small perturbation. We want to decompose the metric perturbation tensor $h_{\mu\nu}$ following the symmetry properties of the background metric. Under spatial rotations, we have that 
\renewcommand{\labelitemi}{$\bullet$}
\begin{itemize}
    \item The $h_{00}$ component behaves like a 3-scalar, which we indicate with $\Psi$, and it is called \textit{lapse}.
    \item The $h_{i0}$ component, as well as $\boldsymbol{h_{0i}}$, behaves as a 3-vector, $w_i$, which we  call \textit{shift}.
    \item The $h_{ij}$ component forms a 2-index spatial symmetric tensor. We will call it $E_{ij}$.
\end{itemize}
So, we can write the metric as~\cite{Baumann:2022mni,Kodama:1984ziu,Bernardeau:2001qr,Durrer:2004fx,Nakamura:2004rm} 
\begin{equation}
    ds^2=a^2(\eta)\[-(1+2\Psi)d\eta^2+2w_id\eta dx^i+(\delta_{ij}+2E_{ij})dx^idx^j\]\,,
    \label{eq:perturbedMetric}
\end{equation}
where we have introduced 2 factors and signs to simplify later expressions. 

At this point we perform a SVT decomposition and we find that
\begin{gather}
    w_i=\partial_i B+ \hat{B}_i,\\
    E_{ij}=-\Phi\delta_{ij}+\partial_{\langle i}\partial_{j\rangle}E+\partial_{(i}\hat{E}_{j)}+\hat{E}_{ij}
\end{gather}
where
\begin{equation}
    \partial_{\langle i}\partial_{j \rangle}E\equiv\(\partial_i\partial_j-\frac13\delta_{ij}\nabla^2\)E\quad \partial_{(i}\hat{E}_{j)}\equiv \frac12\(\partial_i\hat{E}_j+\partial_j\hat{E}_i\)
\end{equation}
with the hatted quantities divergenceless. The term $\Phi$ contains the trace of the spatial perturbation. 
\begin{tcolorbox}[mybox]
The SVT decomposition derived from the following theorem 
\newtheorem*{remark}{Helmotz Theorem}
\begin{remark}
Let $\boldsymbol{F}(\boldsymbol{r})$ be any continuous vector field with continuous first partial derivatives. Then $\boldsymbol{F}(\boldsymbol{r})$ can be uniquely expressed in terms of the negative gradient of a scalar potential $\phi(\boldsymbol{r})$ and of the curl of a vector potential $\boldsymbol{A}(\boldsymbol{r})$.
\end{remark}
Which simply means
\begin{equation}
    \boldsymbol{F}(\boldsymbol{r})=-\boldsymbol{\nabla}\phi(\boldsymbol{r})+\boldsymbol{\nabla}\times\boldsymbol{A}(\boldsymbol{r})=\boldsymbol{F}_{(l)}(\boldsymbol{r})+\boldsymbol{F}_{(t)}(\boldsymbol{r}),
\end{equation}
where the subscripts $l$ and $t$ stand for \textit{longitudinal} and \textit{transverse}, respectively. If we move in Fourier space, we have a monochromatic plane wave described by the scalar function $Q(\boldsymbol r)=exp(i\boldsymbol{k}\vdot\boldsymbol{r})$. The longitudinal vector can be constructed as $\boldsymbol{F}_{(l)}\equiv \boldsymbol{k}Q(\boldsymbol x)$. Similarly, for a transverse vector we have $\boldsymbol{F}_{(t)}\equiv \boldsymbol{n}Q(\boldsymbol x)$ with $\boldsymbol{n}$ the normal vector, \ie $\boldsymbol{n}\vdot\boldsymbol{k}=0$. By definition, these conditions imply 
\begin{itemize}
  \item Longitudinal and irrotational component
  \begin{equation}
      \boldsymbol{\nabla}\times\boldsymbol{F}_{(l)}(\boldsymbol{r})=0\rightarrow \boldsymbol{k}\parallel \boldsymbol{F}_{(l)}(\boldsymbol{r}).
    \end{equation}
 \item Transverse and divergenceless component
 \begin{equation}
     \boldsymbol{\nabla}\vdot\boldsymbol{F}_{(t)}(\boldsymbol{r})=0\rightarrow \boldsymbol{k}\,\bot\, \boldsymbol{F}_{(t)}(\boldsymbol{r}).
      \end{equation}
\end{itemize}
We can apply analogous considerations to tensors:
\begin{itemize}
\item Doubly longitudinal tensor
\begin{equation}       
T^{(ll)}_{ij}=k_ik_jQ(\boldsymbol{x}).
\end{equation}
 \item Singly longitudinal tensor (traceless and divergenceless)
 \begin{equation}
     T^{(lt)j}_{i}=k_in^jQ(\boldsymbol{x}).
\end{equation}
 \item Doubly transverse tensor (divergenceless)
 \begin{equation}
T^{(tt)i}_j=n^in_jQ(\boldsymbol{x}).
 \end{equation}
\end{itemize}

\end{tcolorbox}

And with these four scalars ($\Psi$, $B$, $\Phi$ and $E$), two vectors ($\hat{B}_i$ and $\hat{E}_i$) and one tensor $\hat{E}_{ij}$ perturbations, we can write the three metrics~\cite{Weinberg:2008zzc}
\begin{equation}
    \boxed{ds^2=a^2(\eta)\[-(1+2\Psi)d\eta^2+2B_{,i}d\eta dx^i+\left[(1-2\Phi)\delta_{ij}+2\partial_{\langle i}\partial_{j\rangle}E\right]dx^idx^j\]}\,,
\end{equation}
\begin{equation}
        \boxed{ds^2=a^2(\eta)\[-d\eta^2+2\hat{B}_id\eta dx^i+\left(\delta_{ij}+2E_{(i,j)}\right)dx^idx^j\]}\,
\end{equation}
and
\begin{equation}
        \boxed{ds^2=a^2(\eta)\[-d\eta^2+\left(\delta_{ij}+E_{ij}\right)dx^idx^j\]}\,.
        \label{eq:Tens}
\end{equation}
With this decomposition we can work with scalar, vector and tensor Einstein equations independently, without worrying of any possible modes coupling.
    
\begin{tcolorbox}[mybox]
Moving to the Fourier space, we are able to give a simple proof of the fact that scalar, vector, and tensor variables obey differential equations that in linear theory are decoupled from each other.

Let us define a general perturbation $P(t,\boldsymbol{x})$ in the Fourier space as
\begin{equation}
    P(t,\boldsymbol{k})=\int{d^3xP(t,\boldsymbol{x})e^{-i\boldsymbol{k}\vdot\boldsymbol{x}}}.
\end{equation}
We have two important properties derived from the maximal symmetry of the spatial metric:
\newtheorem*{remarks}{Translation invariance}
\begin{remarks}
The translation invariance implies that different Fourier modes evolve independently.
\end{remarks}
\begin{proof}
We consider the evolution of $N$ perturbations  $P_i(t,\boldsymbol{k})$, with $i=1,...,N$, from $t_1$ to $t_2$, using the transfer matrix $T_{ij}(t_2,t_1,\boldsymbol{k},\boldsymbol{q})$. We will allow, in theory, the modes to mix themselves during the evolution:
\begin{equation}
     P_i(t_2,\boldsymbol{k})=\sum_{j=1}^N\int{d\boldsymbol{q}T_{ij}(t_2,t_1,\boldsymbol{k},\boldsymbol{q})P_j(t_1,\boldsymbol{q})}.
     \label{eq:sumperturbation}
\end{equation}
We now perform a translation: 
\begin{equation}
    x^{i'}=x^i+\varepsilon^i,\qquad \varepsilon^i=const.
\end{equation}
From the translational invariance, we have that equations of motion must be the same in both coordinate systems. Together with the relation
\begin{equation}
    P'_i(t,\boldsymbol{k})=e^{-ik_j\varepsilon^j}P_i(t,\boldsymbol{k}),
\end{equation}
we arrive at the conclusion that reads
\begin{equation}
    T_{ij}(t_2,t_1,\boldsymbol{k},\boldsymbol{q})=T_{ij}(t_2,t_1,\boldsymbol{k},\boldsymbol{q})e^{i(q_j-k_j)\varepsilon^j}.
\end{equation}
The latter equation holds for any $\varepsilon^i$. This implies that either $\boldsymbol{q}=\boldsymbol{k}$ or the transfer matrix is equal to zero. Hence, $P_i(t_2,\boldsymbol{k})$ depends only on the initial $\boldsymbol{k}$. There is no coupling.
\end{proof}
Taken an angle $\varphi$, we can perform rotations around the wave vector: $\boldsymbol{k}$ 
\begin{equation}
    P(t,\boldsymbol{k})\rightarrow e^{im\varphi}P(t,\boldsymbol{k}).
\end{equation}
It is possible to use the \textit{helicity} $m$ to define the type of perturbations. For instance, a 3-scalar perturbation has helicity $0$. Besides, 3-vector perturbations are characterised by $m=\pm1$, whereas the tensor perturbations have $m=\pm2$. We are now ready to present the second property:
\newtheorem*{remarkss}{Rotational invariance}
\begin{remarkss}
Rotational invariance implies that scalar, vector, and tensors perturbations evolve independently.
\end{remarkss}
\begin{proof}
To the perturbation in the form \eq{sumperturbation}, we apply the property of linear evolution and obtain
\begin{equation}
    P_i(t_2,\boldsymbol{k})=\sum_{j=1}^NT_{ij}(t_2,t_1,\boldsymbol{k})P_j(t_1,\boldsymbol{k}).
\end{equation}
Then, finding that a given perturbation under rotations transforms as 
\begin{equation}
    P'_i(t,\boldsymbol{k})=e^{im_i\varphi}P_i(t,\boldsymbol{k}),
\end{equation}
and taking into consideration the rotational invariance, we have the following result:
\begin{equation}
    T_{ij}(t_2,t_1,\boldsymbol{k},)=T_{ij}(t_2,t_1,\boldsymbol{k})e^{i(m_i-m_j)\varphi}.
\end{equation}
Holding for every angle $\varphi$, it follows that either $m_i=m_j$ or $T_{ij}(t_2,t_1,\boldsymbol{k})=0$. Since the helicity defines the type of perturbation, the evolution does not mix modes of different helicity.
\end{proof}
 For a proof in the real space, see Appendix B~\cite{Kodama:1984ziu}. 
\end{tcolorbox}

\subsection{Gauge choice}
\label{GaugeChoice}

Ten independent functions define our metric perturbations: four for the 3-scalar perturbations, four for the 3-vector perturbations (counting one constraint for each vector), and two for the tensor perturbations (our symmetric tensor has four constraints). This is in agreement with the result
\begin{figure}[h!]
\centering
\includegraphics[width=0.9\textwidth]{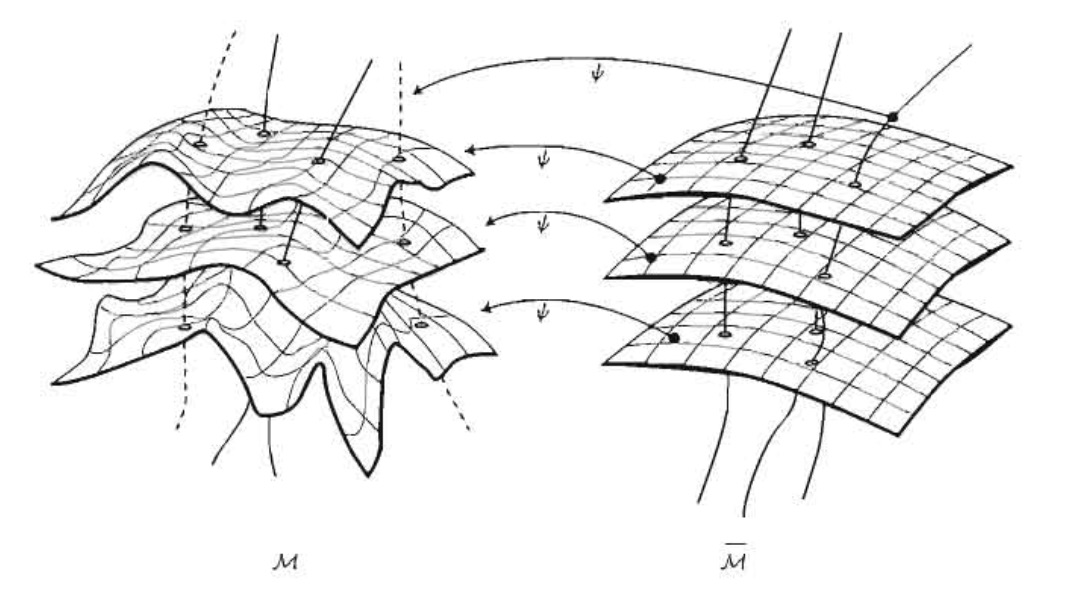}
  \caption[Gauge choice]{\small $\mathcal{M}$ represents the perturbed spacetime whilst $\bar{\mathcal{M}}$ is the background spacetime. The image is taken from \cite{Peter:2013avv}}
  \label{fig:Transformations}
\end{figure}
we have obtained in \sect{Metronomo}, where our metric possess $\frac12 n(n+1)$ Killing vectors. As a matter of fact, only six of these degrees of freedom represent physical quantities;
the reason is that we can choose the four coordinates used to describe the perturbations. Let us go back to the unperturbed metric; in that case, we had space-like hypersurfaces of constant time $t$, which are the \textit{slicing} of the four-dimensional spacetime, and the time-like worldlines orthogonal to the slicing, which define the \textit{threading}. It was a special coordinate choice that permitted us to consider comoving observers who could see the Universe as homogeneous, and its expansion isotropic. Moreover, the constant-time surfaces have constant spatial curvature due to their maximally symmetric property. In other words, the coordinate system is fixed by the symmetry's properties of the background. With a perturbed Universe, the slicing and the threading are not unique anymore; we do not have a preferred coordinate system. This freedom, called \textit{gauge freedom}, leads to the appearance of fictitious perturbation modes (see \fig{Transformations}). These fictitious modes do not describe any real inhomogeneities. Instead, they reflect only the properties of the coordinate system considered. For instance, let us take the energy density of an unperturbed homogeneous isotropic universe, $\rho(\boldsymbol{x},t) = \rho(t)$. We now decide to use a time coordinate, $\widetilde{t}$, related to the old time $t$ via $\widetilde{t}=t+\delta t(\boldsymbol{x}, t)$. Then, the energy density on the hypersurface with constant $\widetilde{t}$ depends, in general, on the spatial coordinates $\boldsymbol{x}$. Assuming that $\delta t \ll t$, we have
\begin{equation}
    \rho(t)=\rho(\widetilde{t}-\delta t(\boldsymbol x,t))\simeq\rho(\widetilde{t})-\frac{\partial \rho}{\partial t}\delta t=\underbrace{\rho(\widetilde{t})}_{\hidewidth\text{Background energy density}\hidewidth}+\overbrace{\delta\rho(\boldsymbol x,\widetilde{t})}^{\hidewidth\text{Linear perturbation}\hidewidth}.
\end{equation}
The perturbation is non-physical and due to the choice of the new time entirely. Namely, perturbing the coordinates, we obtain fictitious perturbations. Besides, we can also have the opposite problem: if we choose the hypersurfaces of constant time to be the same as the hypersurfaces of constant energy, we obtain $\delta\rho=0$ despite the presence of real inhomogeneities. 

As we have seen in \eq{perturbeggmunu}, perturbations are defined as the difference between the real value that a certain quantity assumes on the real physical spacetime, and the value it assumes on the unperturbed background. However, to make this comparison, we need to compute them in the same spacetime point, even though the values live in two different geometries. Hence, we need to specify a map that allows us to link the same point between two different spacetimes. This correspondence is called a \textit{gauge choice} and defines a threading and a slicing for our spacetime. Changing the map means performing a \textit{gauge transformation}. 

\subsubsection{Gauge transformations}
Gauge transformations can be either active, where we slightly alter the manifold, or passive, where we do not alter the manifold and we just change the coordinate system. From now on, we will focus on the passive transformations. Using this latter point of view, we compare a quantity at a point $P$, with a new quantity at the point $P'$, whose new coordinates have the same value as $P$ had in the old coordinate system, \ie\, $\widetilde{x}^\mu(P')=x^\mu(P)$.

Let us now consider the coordinate transformation
\begin{equation}
    x^\mu\rightarrow \widetilde{x}^\mu=x^\mu+\varepsilon^\mu,\qquad \varepsilon^\mu\ll 1.
    \label{eq:newcoordinate}
\end{equation}
The metric tensor in the new coordinate system is
\begin{equation}
    \widetilde{g}_{\mu\nu}(\widetilde{x}^\gamma)=\frac{\partial x^\sigma}{\partial\widetilde{x}^\mu}\frac{\partial x^\rho}{\partial\widetilde{x}^\nu}g_{\sigma\rho}(x^\gamma)=(\delta_{\sigma\mu}-\varepsilon^\sigma_{,\mu})(\delta_{\rho\nu}-\varepsilon^\rho_{,\nu})g_{\sigma\rho}\approx \bar{g}_{\mu\nu}(x^\gamma)+h_{\mu\nu}-\bar{g}_{\mu\rho}\varepsilon^\rho_{,\nu}-\bar{g}_{\sigma\nu}\varepsilon^\sigma_{,\mu},
    \label{eq:newcoordinatess}
\end{equation}
where we have kept only terms linear in $h$ and $\varepsilon$. Furthermore, we have
\begin{equation}
    \widetilde{g}_{\mu\nu}(\widetilde{x}^\gamma)=\bar{g}_{\mu\nu}(\widetilde{x}^\gamma)+\widetilde{h}_{\mu\nu},
\end{equation}
where we have split the metric into the background and perturbation part. Taking into account that
\begin{equation}
    \bar{g}_{\mu\nu}(x^\gamma)\approx\bar{g}_{\mu\nu}(\widetilde{x}^\gamma)-\bar{g}_{\mu\nu,\sigma}\varepsilon^\sigma,
\end{equation}
we eventually obtain the following transformation law:
\begin{equation}
    \boxed{h_{\mu\nu}\rightarrow  \widetilde{h}_{\mu\nu}=h_{\mu\nu}-\bar{g}_{\mu\nu,\sigma}\varepsilon^\sigma-\bar{g}_{\mu\sigma}\varepsilon^\sigma_{,\nu}-\bar{g}_{\sigma\nu}\varepsilon^\sigma_{,\mu}=h_{\mu\nu}-\pounds_\varepsilon\bar{g}_{\mu\nu}}\,.
    \label{eq:transflaw}
\end{equation}
In \eq{transflaw}, we have used the definition of the Lie derivative with respect to the vector field $\varepsilon^\mu$, applied to a covariant tensor\footnote{The Lie derivative corresponds to the change determined by an observer who goes from a point $P$, with coordinate $x^\mu$, in the direction of $\varepsilon^\mu$ to the infinitesimally neighbouring point $\widetilde{P}$, with coordinates $\widetilde{x}^\mu$, defined in \eq{newcoordinate}. The Lie derivative of a scalar $\varphi$ is $\pounds_\varepsilon\varphi=\varepsilon^\lambda\varphi_{,\lambda}$, whereas for a covariant vector $v_\mu$ we have $\pounds_\varepsilon v_\mu=v_{\mu,\alpha}\varepsilon^\alpha+v_\alpha\varepsilon^\alpha_{,\mu}$. See e.g. Ref. \cite{Stephani:2004ud}}. Afterwards, we should proceed to apply SVT decomposition to the spatial part of the vector field $\varepsilon^\mu\equiv(\varepsilon^0,\varepsilon^i)$
\begin{equation}
\varepsilon^i=\partial^i\varepsilon+\hat{\varepsilon}^i,\qquad \partial_i\varepsilon^{i}=0.
\end{equation}
This decomposition results in perturbations in the scalar and the vector parts of the metric tensor and stress-energy tensor, but not in the tensor parts. The tensor perturbations are \textit{gauge-invariant}. Applying this result to~\eq{perturbedMetric}, we get
\begin{gather}
    \widetilde{h}_{00}=h_{00}-\bar{g}_{00,0}\varepsilon^0-2\bar{g}_{00}\varepsilon^0_{,0}=h_{00}+2aa'\varepsilon^0+2a^2\varepsilon^{0'},\\
    \widetilde{h}_{0i}=h_{0i}-\cancel{\bar{g}_{0i,\sigma}\varepsilon^\sigma}-\bar{g}_{00}\varepsilon^0_{,i}-\bar{g}_{i j}\varepsilon^j_{,0}=h_{0i}+a^2\varepsilon^0_{,i}-a^2\varepsilon'_{,i}-\hat{\varepsilon}^{i'},\\
    \widetilde{h}_{ij}=h_{ij}-g_{ij,0}\varepsilon^0-g_{ii}\varepsilon^i_{,j}-g_{jj}\varepsilon^j_{,i}=h_{ij}+2aa'\delta_{ij}\varepsilon^0-2a^2\varepsilon^0_{,ij}+a^2(\hat{\varepsilon}^i_{,j}+\hat{\varepsilon}^j_{,i}).
\end{gather}
From these expressions, we obtain the significant transformations\footnote{In the last equation we have add and subtracted $\frac23\delta_{ij}a^2\nabla^2\varepsilon^0$}:
\begin{multline}
    \widetilde{\Psi}=\Psi-\varepsilon^{0'}-\mathcal{H}\varepsilon^0,\quad \widetilde{\Phi}=\Phi-\mathcal{H}\varepsilon^0-\frac13\nabla^2\varepsilon,\quad \widetilde{B}=B+\varepsilon^0-\varepsilon',\quad
    \widetilde{E}=E-\varepsilon,\\
    \widetilde{\hat{B}}_i=\hat{B}_i-\hat{\varepsilon}_i',\quad \widetilde{\hat{E}}_i=\hat{E}_i-\hat{\varepsilon}_i\\
    \widetilde{\hat{E}}_{ij}=\hat{E}_{ij}\\
    \label{eq:Transformationgauge}
\end{multline}

\subsubsection{Gauge fixing}
To get rid of these spurious degree of freedoms we need to fix a gauge, namely, we need to impose four conditions. The \textit{Poisson gauge} is defined by the conditions
\begin{equation}
\partial^iw_i=\partial^iE_{ij}=0.
\end{equation}
If we additionally restrict ourselves to the condition
\begin{equation}
E=B=0,
\label{eq:Newtoniangauge}
\end{equation}
with a suitable choice of the gauge functions $\varepsilon^0$ and $\varepsilon$, we obtain what is known in the literature as the \textit{Newtonian gauge}. These latter conditions can be applied only if the stress-energy tensor contains no vector or tensor parts and there are no free gravitational waves so that only the scalar metric perturbations are present. While this condition may apply, in principle, in the linear regime, non-linear density fluctuations generally induce vector and tensor modes even if none were present initially. One benefit of working in this gauge is that the metric tensor is diagonal. Besides, $\Psi$ plays the role of the gravitational potential in the Newtonian limit, and consequently it has a simple physical interpretation. In general, this is not a valid gauge condition, and it should really be called a \textit{restricted gauge}. The Poisson gauge, by contrast, allows all physical degrees of freedom present in the metric.

From \eq{Transformationgauge}, it is possible to construct gauge-invariant variables, the so-called \textit{Bardeen variables}~\cite{Bardeen:1980kt} which are the real spacetime perturbations
\begin{equation}
    \Psi^B\equiv \Psi+\mathcal{H}(B-E')+(B-E')',\quad\Phi^B\equiv\Phi+\frac13\nabla^2E-\mathcal{H}(B-E')
    \label{eq:Bardeen}
\end{equation}
where we should add $\hat{\Phi}^B_{i}\equiv\hat{B}_i-\hat{E}_i'$ but since, as we will see, the vector perturbations do not contribute to structure formation, we can omit that\footnote{Imposing the Newtonian gauge in~\eq{Newtoniangauge}, we have the simple relation
\begin{equation}
\Psi_B=\Psi,\qquad \Phi_B=\Phi\,.
\end{equation}}. Given that these variables do not change under coordinate transformations, if they both are equal to zero, then the metric perturbations are fictitious and can be removed. Another important gauge is the \textit{Synchronous gauge}~\cite{Lifshitz:1945du}, where we set
\begin{equation}
    \Psi=w_i=0
\end{equation}
Hence, we have eliminated two scalar fields and one vector field with a suitable choice of $\varepsilon^0$, $\varepsilon$, and $\hat{\varepsilon}^i$. However, we have not eliminated all the gauge freedom. In this gauge, there exists a set of \textit{fundamental observers}: they are comoving observers who fall freely without changing their spatial coordinates. Each of the fundamental observers carries a clock reading a conformal time $\eta$ and a fixed spatial coordinate label $x^i$. The residual gauge freedom arises from the possibility to choose the initial settings of the clocks and the initial coordinates for such observers. We ought to underline that this synchronous gauge has a flaw in the non-linear perturbation theory: the trajectory of two fundamental observers may intersect. In that case, two different sets of $x^\mu$ label the same spacetime event. If we just focus on the scalar perturbations, it is easy to convert the Newtonian gauge fixing into the synchronous one, and vice-versa \cite{Ma:1995ey}. Another gauge worth mentioning as it is useful when computing inflationary perturbations is the \textit{Spatially flat gauge} with
\begin{equation}
    \Phi=E=0\,.
    \label{eq:SPAtially}
\end{equation}

\section{Stress-energy tensor perturbations}
\label{sec:setensor}
Next, we shall investigate the stress-energy perturbations. First of all, let us write it in the explicitly covariant form
\begin{equation}
        T^{\mu\nu}=(\rho+p)u^\mu u^\nu+pg^{\mu\nu},
\end{equation}
where $u^\mu$ is the relative four-velocity between the fluid and the observer. For a comoving observer we have $u^\mu=(\frac{1}{a},0,0,0)$ and $u_\mu u^\nu=-1$ . Now, we need to consider the form of $T_{\mu\nu}$ for an imperfect fluid, \ie we have to introduce shear and bulk viscosity (the isotropic stress generated when an imperfect fluid is rapidly compressed or expanded), thermal conduction and other physical processes. To do that, we add a tensor $\Sigma_{\mu\nu}$, called \textit{anisotropic stress}, which we can require to be traceless and flow orthogonal ($\Sigma^\mu_\nu u^\nu=0$) without any loss of generality. Moreover, we define $u^\mu$ so that the heat conduction is included in the term $\rho u^\mu$, $p$ includes any bulk viscosity and the shear viscosity is in the tensor $\Sigma_{\mu\nu}$. Hence, we have
\begin{equation}
    T^{\mu\nu}=(\rho+p)u^\mu u^\nu+pg^{\mu\nu}+\Sigma^{\mu\nu}.
    \label{eq:imperfecttmunu}
\end{equation}
At this point we can write $\rho=\bar{\rho}+\delta{\rho}$ where the bar indicates the background quantity. The explicit form of the four-velocity in the case of the metric in \eq{perturbedMetric} is
\begin{equation}
    u^\mu=a^{-1}(1-\Psi,v^i),\quad u_\mu=a(-(1+\Psi),v_i+w_i).
    \label{eq:4velocity}
\end{equation}
\begin{tcolorbox}[mybox]
To find the form in \eq{4velocity} we start from the idea that $\delta(g_{\mu\nu}u^\mu u^\nu)=0$. It means, at linear order
\begin{equation}
    \delta g_{\mu\nu}\bar{u}^\mu \bar{u}^\nu+2\delta u^\mu \bar{u}^\nu=0
\end{equation}
with the bar indicating the unperturbed 4-velocity. From the latter equation, we can find $\delta u^0=-\Psi a^{-1}$ and we can define $\delta u^i=v^i a^{-1}$ where $v^i$ is the \textit{bulk velocity} defined as $dx^i/d\eta$. Using the fact that $u_0=g_{00}u^0+g_{0i}u^0$ and $u_i=g_{0i}u^i+g_{ij}u^j$, neglecting the second order terms, we recover the expressions in \eq{4velocity}. From the fact that the spatial derivative of $\Psi$ is non-zero, we infer that $\Psi$ is space-dependent: the proper time interval $a(\eta)(1+\Psi)d\eta$ depends on the position, namely, the clocks run at different rates in different places. Whereas, requiring a non-vanishing $w_i$ implies that an observer in $x^i=const$ sees a clock at $x^i+dx^i=const$ running faster by an amount $w_idx^i$. With these definitions it is straightforward to interpret $\Psi$  as the \textit{gravitational redshift} and $w_i$ as the \textit{dragging of inertial frames}. 
\end{tcolorbox}

That been said, we can now explicitly write the perturbations for the stress-energy tensor with mixed upper and lower indices for simplicity:
\begin{equation}
    \delta T^\mu_{\nu}=(\delta \rho + \delta p)\bar{u}^\mu\bar{u}_\nu+(\bar{\rho}+\bar{p})(\delta u^\mu \bar{u}_\nu+\delta u_\nu \bar{u}^\mu)-\delta p\delta^\mu_\nu-\Sigma^\mu_\nu.
\end{equation}
Therefore, we obtain
\begin{gather}
    T^0_0\equiv -(\bar{\rho}+\delta\rho),\\
    T^i_0\equiv-q^i,\\
    \label{eq:TIO}
    T^i_j\equiv(\bar{p}+\delta p)\delta^i_j+\Sigma^i_j
\end{gather}
with $q^i\equiv(\bar{\rho}+\bar{p})v^i$ the \textit{momentum density}. We can again apply the SVT decomposition to the vector $v_i=\partial_i v+\hat{v}_i$ and similarly to $q_i$ and $\Sigma_{ij}=\partial_{\langle i}\partial_{j\rangle}\Sigma+\partial_{(i}\hat{\Sigma}_{j)}+\hat{\Sigma}_{ij}$ remembering that  $\Sigma_{ij}$ is defined traceless. 

\subsubsection{Gauge transformation for the stress-energy tensor.} Before studying the scalar, vector and tensor modes separately, we should stress that our field equations will be invariant only if we perform the same gauge transformation to the stress-energy tensor. The transformation law for the stress-energy tensor has the same form as \eq{transflaw}:
\begin{equation}
    \delta\widetilde{T}_{\mu\nu}= \delta T_{\mu\nu}-\pounds_{\varepsilon}\bar{T}_{\mu\nu}\,.
\end{equation}
However, for sake of simplicity, having the tensor written in mixed indices, with respect to the matrix transformation in \eq{newcoordinatess} we now have
\begin{equation}
    T^\mu_\nu(x)=\frac{\partial x^\mu}{\partial \widetilde{x}^\alpha}\frac{\partial \widetilde{x}^\beta}{\partial x^\nu}\widetilde{T}^\alpha_\beta(\widetilde{x})
\end{equation}
which implies
\begin{equation}
    \delta\widetilde{\rho}=\delta \rho-\bar{\rho}'\varepsilon^0,\quad \delta \widetilde{p}=\delta p-\bar{p}'\varepsilon^0,\quad
    \widetilde{q}_i=q_{i}+(\bar{\rho}+\bar{p})\varepsilon'_i,\quad\widetilde{v}_i=v_i+\varepsilon'_i,\quad \widetilde{\Sigma}_{ij}=\Sigma_{ij}.
    \label{eq:gaugetensor}
\end{equation}

We have seen that gauge transformations lead to non-physical quantities. Therefore, to get rid of these degrees of freedom we can either define a specific gauge, like the \textit{uniform density gauge} ($\delta\rho=0$) or the \textit{comoving gauge} ($v+B=0$), or introduce gauge-invariant variables. Specifically, we have 
\begin{equation}
\boxed{
    \Delta\equiv\delta+\frac{\bar{\rho}}{\bar{\rho}}'(v+B)}
\end{equation}
where $\delta$ is the \textit{density contrast} $\delta\rho/\rho$, and the more important \textit{curvature perturbations}
\begin{equation}
\boxed{\zeta=\Phi+\frac13\nabla^2E+\mathcal{H}\frac{\delta \rho}{\bar{\rho}'},\\}
\label{eq:zetafirst}
\end{equation}
and 
\begin{equation}
    \boxed{\mathcal{R}=\Phi+\frac13\nabla^2E-\mathcal{H}(v+B),\\}\,.
    \label{eq:mathcalR}
\end{equation}

In the Newtonian gauge \eq{zetafirst} can also be written as 
\begin{equation}
    \zeta=\Psi+\frac13\frac{\delta\rho}{\bar{\rho}+\bar{p}}=\Psi+\frac13\frac{\delta}{3(1+w)}
    \label{eq:mathcalzeta}
\end{equation}
where we used the continuity equation in \eq{ContinuityEq1} and the equation of state (see \eq{equationofstate})

\begin{tcolorbox}[mybox]
    These three gauge-invariant perturbations are related via the equation~\cite{Baumann:2009ds}
    \begin{equation}
        \zeta=\mathcal{R}+\frac{\mathcal{H}}{\bar{\rho}'}\bar{\rho}\Delta\,.
    \end{equation}

A few more words should be spent regarding these variables. For $\Delta$ we can write the generalize Poisson equation
\begin{equation}
    \nabla^2\Phi^B=4\pi Ga^2\bar{\rho}\Delta
    \label{eq:Poissonhorizon}
\end{equation}
with $\Phi^B$ defined in \eq{Bardeen}. Therefore, it is easy to see if we move in Fourier space, that on superhorizon scales $k\ll\mathcal{H}$ $\zeta=\mathcal{R}$. As we will see in \chap{InfloBinglo}, inhomogeneities that we observe today originates from quantum fluctuations during the inflationary era. From the end of inflation until relatively near the present, all observable perturbations were outside the horizon, namely,  $k\ll \mathcal{H}$. That is, for perturbations outside the horizon, $\zeta$ is conserved if $\mathcal{R}$ is. It is possible to demonstrate \cite{Weinberg:2003sw, Weinberg:2008zzc} that whatever the constituents of the Universe are, there is always an adiabatic solution of the field equations for which $\zeta$ and $\mathcal{R}$ are conserved outside the horizon. So, if cosmological fluctuations are described by such a solution during inflation, it will be so until they re-enter the horizon. Moreover, this conservation implies that, if scalar perturbations are adiabatic at the end of inflation, later processes such as reheating cannot generate entropic perturbations \cite{Weinberg:2004kr}. On top of that, since the scalar modes have two independent adiabatic solutions, if these equations have no more than two independent solutions in general, any scalar perturbations must be adiabatic \cite{Weinberg:2008zzc}. This is the case for inflation generated by a single scalar field~\cite{Salopek:1990jq,Lyth:2004gb,Cheung:2007sv}. Therefore, the scalar perturbations produced during inflation induced by a single scalar field, are adiabatic and remain so until they re-enter the horizon. We will discuss in the following sections what are the adiabatic solutions.
\end{tcolorbox}

\section{Perturbed dynamics}
\label{sec:Evo}
If we want to define the evolution of the components of the Universe, we can use our results in \sect{setensor} and apply them into the conservation equation \eq{Tmunukio}. But, we need one last ingredient. From the definition of covariant derivative, we need the perturbed Christoffel's symbol 
\begin{equation}
    \delta\Gamma^\mu_{\alpha\beta}=\frac 12\bar{g}^{\mu\nu}\left[-2h_{\nu\sigma}\bar{\Gamma}^\sigma_{\alpha\beta}+h_{\nu\alpha,\beta}+h_{\nu\beta,\alpha}-h_{\beta\alpha,\nu}\right]\
    \label{eq:Christoff}
\end{equation}
where we have neglected higher-order perturbations and we have used the relation\footnote{The perturbation of the inverse of a general matrix $M$ is $\delta M^{-1}=-M^{-1}(\delta M)M^{-1}$.}
\begin{equation}
    h^{\mu\nu}=-\bar{g}^{\mu\rho}\bar{g}^{\nu\sigma}h_{\rho\sigma}.
\end{equation}

As previously anticipated, we will fix the gauge to the Newtonian gauge presented in \eq{Newtoniangauge}. By explicitly computing the covariant derivative using the perturbed $\Gamma$s and $T^\mu_{\nu}$, if we set $\nu=0$ we can study the evolution of the density perturbations
\begin{equation}
    \delta\rho'=\overbrace{-3\mathcal{H}(\delta\rho+\delta p)}^{\text{expansion}}-\underbrace{\partial_i q^i}_{\text{fluid}}+\overbrace{3\Phi'(\bar{\rho}+\bar{p})}^{\text{relativistic}}
\end{equation}
with the first term that takes into account the background expansion, the second the local fluid flow and the third term is a relativistic effect. Using the definition of density contrast, we can write
\begin{equation}
    \delta'=-\(1+\frac{\bar{p}}{\bar{\rho}}\)(\partial_i v^i-3\Phi')-3\mathcal{H}\(\frac{\delta p}{\delta \rho}-\frac{\bar{p}}{\bar{\rho}}\)\delta
\end{equation}
or, using $w=\bar{p}/\bar{\rho}$,
\begin{equation}
    \boxed{\delta'=-\(1+w\)(\partial_i v^i-3\Phi')-3\mathcal{H}w\(\frac{\delta p}{p}-\delta\)}\,.
    \label{eq:continuityrel}
\end{equation}
Now, if we set $\nu=i$ we get the relativistic form of the Euler equation
\begin{equation}
    \boxed{v'_i=-\(\mathcal{H}+\frac{\bar{p}'}{\bar{\rho}+\bar{p}}\)v_i-\frac{1}{\bar{\rho}+\bar{p}}(\partial_i\delta p+\partial^j\Sigma_{ij})-\partial_i\Psi}\,.
    \label{eq:eulerrel}
\end{equation}

If we now combine the time derivative of \eq{continuityrel} with the divergence of \eq{eulerrel}, we can obtain the special case for matter and radiation\footnote{For non-relativistic fluid we have $p_m=0$ and $\Sigma_m^{ij}=0$ and for relativistic fluid we have $p_r=\frac13\rho_r$ and $\Sigma_r^{ij}=0$}
\begin{gather}
\delta''_m+\mathcal{H}\delta'_m=\nabla^2\Psi+3(\Phi''+\mathcal{H}\Phi')\,,
\label{eq:mattercontinuity}\\
\delta''_r+\frac13\nabla^2\delta_r=\frac43\nabla^2\Psi+4\Phi''
\label{eq:radiationcontinuity}
\end{gather}
where we note that radiation perturbations do not experience the Hubble friction but feel an additional pressure-induced force.

\subsection{Einstein Equations}
The evolution of the fluids \eq{continuityrel} and \eq{eulerrel} is dependent on perturbation of the metric $\Phi$ and $\Psi$. Therefore, to close the system of equations we need to solve the perturbed Einstein equations. In order to do so, we need the perturbed Ricci tensor
\begin{equation}
    \delta R_{\mu\nu}=\delta\Gamma^\sigma_{\mu\nu,\sigma}-\delta\Gamma^\sigma_{\mu\sigma,\nu}+\delta\Gamma^\eta_{\mu\rho}\bar{\Gamma}^\rho_{\nu\eta}+\delta\Gamma^\rho_{\nu\eta}\bar{\Gamma}^\eta_{\mu\rho}-\delta\Gamma^\eta_{\mu\nu}\bar{\Gamma}^\rho_{\rho\eta}-\delta\Gamma^\rho_{\rho\eta}\bar{\Gamma}^\eta_{\mu\nu}\,.
    \label{eq:PerturbationR}
\end{equation}
and the Ricci scalar
\begin{equation}
    \delta R=\delta g^{\mu\nu}R_{\mu\nu}+g^{\mu\nu}\delta R_{\mu\nu}\,.
\end{equation}
The components of the Einstein equations with mixed lower and upper index $G^\mu_\nu=8\pi G T^\mu_\nu$ are the following
\begin{equation}
    \delta G^0_0\longrightarrow\boxed{\nabla^2\Phi-3\mathcal{H}(\Phi'+\mathcal{H}\Psi)=4\pi G a^2\delta \rho}\,,
    \label{eq:einstein00}
\end{equation}
\begin{equation}
    \delta G^0_i\longrightarrow\boxed{\Phi'+\mathcal{H}\Psi=-4\pi Ga^2q}\,,
    \label{eq:einstein0i}
\end{equation}
\begin{equation}
    \delta G^i_j\xrightarrow[]{\textit{tracefree}}\boxed{k^2(\Phi-\Psi)=8\pi Ga^2(\bar{\rho}+\bar{p})\Sigma}\,,
    \label{eq:einsteinij}
\end{equation}
and
\begin{equation}
    \delta G^i_i\xrightarrow[]{\textit{trace}}\boxed{\Phi''+\frac13\nabla^2(\Psi-\Phi)+(2\mathcal{H}'+\mathcal{H}^2)\Psi+\mathcal{H}\Psi'+2\mathcal{H}\Phi'=4\pi G a^2\delta p}\,,
    \label{eq:einsteinii}
\end{equation}
where the last equation, if we assume $\Phi\approx\Psi$, can be simplified to
\begin{equation}
    \boxed{\Phi''+3\mathcal{H}\Phi'+(2\mathcal{H}'+\mathcal{H}^2)\Phi=4\pi G a^2\delta p}\,.
    \label{eq:einsteiniisimply}
\end{equation}

\begin{tcolorbox}
    A few considerations about the perturbed Einstein equations. From \eq{einstein00} we can see that, inside the Hubble radius ($k\gg\mathcal{H}$), the second term in the left-hand side is negligible with respect to the second, therefore we end up with the Poisson equation in the Newtonian limit. In \eq{einstein0i}  we moved in Fourier space with $\Sigma_{ij}= -(\bar{\rho}+\bar{p})\hat{k}_{\langle i}\hat{k}_{ j \rangle}\Sigma$. Dark matter and baryons can be described as perfect fluids with negligible anisotropic stress. Photons have an anisotropic stress only in the matter-dominated epoch, when their contribution is negligible. Neutrinos are the only species with anisotropic-stress but whose effect is relatively small~\cite{Baumann:2022mni} and we can ignore it unless otherwise stated. Therefore, this implies that $\Psi\approx \Phi$. This assumption is used to derived \eq{einsteinii}. Combining \eq{einsteinij} with \eq{einstein00} we can obtain the general Poisson equation \eq{Poissonhorizon} without any scale restriction. Lastly, \eq{einsteiniisimply} is a closed equation for $\Phi$ if we write $\delta p=c_s^2\delta\rho$, with $c_s^2$ the sound speed, and use the generalized Poisson equation for $\Phi$.
\end{tcolorbox}

Having defined the perturbed Einstein tensor, we can use \eq{einstein0i} in the Newtonian gauge to rewrite \eq{mathcalR}
\begin{equation}
    \mathcal{R}=\Phi+\frac{2}{3(1+w)}\(\frac{\Phi'}{\mathcal{H}}+\Phi\)\xrightarrow[]{k\ll\mathcal{H}}\mathcal{R}=\frac{5+3w}{3+3w}\Phi.
    \label{eq:mathcalRw}
\end{equation}
If we use \eq{einstein00} in \eq{mathcalzeta} we can find the relation
\begin{equation}
    \zeta-\mathcal{R}=\frac{2}{9(1+w)}\frac{\nabla^2}{a^2H^2}\Phi
        \label{eq:mathcalRZw}
\end{equation}
which, at large-scale gives the already found result $\zeta\sim\mathcal{R}$.

\subsubsection{Adiabatic mode}
On superhorizon scales  ($k\ll \mathcal{H}$) we see from \eq{continuityrel} that for matter and radiation we have
\begin{equation}
    \delta_m'=3\Phi'\quad\text{and}\quad \delta_r'=4\Phi'\,.
    \label{eq:adiabaticic}
\end{equation}
If we integrate, we end up with an additional integration constant. Specifically, the different components of the Universe are related via the relation
\begin{gather}
    \delta_\gamma=4\Phi+S_\gamma\,,\\
    \delta_\nu=\delta_\gamma+S_\nu\,,\\
    \delta_c=\frac34\delta_\nu+S_c\,,\\
    \delta_b=\frac34\delta_\gamma+S_b\,,\\
\end{gather}
The $S$ parameters are called \textit{isocuravture modes}. We define the \textit{adiabatic mode} with 
\begin{equation}
    S_\nu=S_c=S_b=0\,,\quad S_\gamma\neq0\,.
    \label{eq:adiabatic}
\end{equation}
In the same large scale regime, we see that \eq{einstein00} loses the first term. Moreover, at early times we are in the radiation epoch, so we have only photons and neutrinos contribution. Also, from the Friedmann equation \eq{IFriedmann} we have $4\pi Ga^2(\bar{\rho}_\gamma+\bar{\rho}_\nu)=3\mathcal{H}^2/2$ and $\mathcal{H}=1/\eta$ and thus
\begin{equation}
    \eta\Phi'+\Phi=-\frac12\delta_\gamma\,.
\end{equation}
Taking a time derivative and using \eq{adiabaticic}, we have
\begin{equation}
    \eta\Phi''+4\Phi'=0
\end{equation}
with one decaying ($\Phi\propto \eta^{-3}$) and one growing ($\Phi=const$) mode. Focusing on the latter we find $\delta_\gamma=-2\Phi_i$ so $S_\gamma=-6\Phi_i$ and therefore
\begin{equation}
\delta_\gamma=\delta_\nu=\frac43\delta_c=\frac43\delta_b=-2\Phi_i
\label{eq:initialPhi}
\end{equation}
which implies we need just the value of the primordial potential $\Phi_i$ to specify the initial condition of all fluctuations. Adiabatic perturbations are a natural prediction for most of the simple inflationary models~\cite{Hawking:1982cz,Bardeen:1983qw,Guth:1982ec,Starobinsky:1982ee}. 

\begin{tcolorbox}[mybox]
    Adiabatic perturbations are generated thorugh a local shift in time $\pi(\eta,\textbf{x})$ which is identical for all background quantities. This implies that $\delta \rho(\eta,\textbf{x})=\bar{\rho}(\eta,\textbf{x})\pi(\eta,\textbf{x})$ and similar definition for the induced pressure perturbation. We ends up with
    \begin{equation}
        \delta p_a=c^2_{s,a}\delta\rho_a
    \end{equation}
    where a perturbation is defined adiabatic if $c^2_s=\delta p/\delta\rho$, we used the definition of the sound horizon $c^2_s=\bar{p}'/\bar{\rho}'$ and we have called $a$ a generic species. So $\pi$ induce adiabatic perturbations. Now we can also see, using the continuity equation \eq{ContinuityEq1} in conformal time, that
    \begin{equation}
        \frac{\delta_a}{1+w_a}=\frac{\delta_b}{1+w_b}
        \label{eq:adiabaticpert}
    \end{equation}
    which implies $\delta_m=\frac34\delta_r$. Hence, these are adiabatic modes. In general, primordial perturbations can be decomposed into adiabatic modes and isocurvature. Isocurvature perturbations~\cite{He:2015msa,Liddle:1999pr,Bucher:1999re} take into account the deviation from \eq{adiabaticpert} because, for two species $a$ and $b$, are defined as
    \begin{equation}
        S_{ab}\equiv\frac{\delta_a}{1+w_a}-\frac{\delta_b}{1+w_b}\,.
    \end{equation}
    They would  leave an imprint on the temperature and polarization power spectra of the CMB~\cite{Langlois:2000ar,Bucher:2004an,Bucher:2000kb} but with the current observations, we have no evidences of their existence~\cite{Planck:2019evm,Planck:2019nip}\,.
\end{tcolorbox}

\section{Growth of perturbations}
\label{frera}
We have now everything in order to study the evolution of perturbations. We will work in the Newtonian gauge and assume adiabatic perturbations. An important threshold throughout the the section is the wavevector at the matter-radiation equality of a mode that is just entering the horizon: $k_{eq}=\mathcal{H}_{eq}$. If $k>k_{eq}$ the modes enter the horizon during the radiation era whereas $k>k_{eq}$ experience a MD universe. 

\paragraph{\textit{Primordial potential}}
Let us start to compute the evolution of $\Phi$. \eq{einsteinii} can be written as
\begin{equation}
    \Phi''+3(1+w)\mathcal{H}\Phi'-w\nabla^2\Phi=0\,.
\end{equation}
In matter domination ($w=0$, $a\propto\eta^2$) the solutions are
\begin{equation}
\boxed{\Phi(a,\textbf{k})=C_1(\textbf{k})+C_2(\textbf{k})a^{-\frac52}}\,.
\label{eq:phimd}
\end{equation}
In radiation domination ($w=\frac13$, $a\propto\eta$) we have \begin{equation}
    \Phi(\eta,\textbf{k})=2\mathcal{R}_i\frac{\sin{\varphi}-\varphi\cos{\varphi}}{\varphi^3}
\end{equation}
where $\varphi=k\eta/\sqrt{3}$ and the normalization is found using \eq{mathcalRw}, and is valid on all scales. The solutions for the limits $\varphi\ll 1$ and $\varphi\gg 1$ are
\begin{equation}
    \boxed{\Phi(\eta,\textbf{k})\approx\left\{ \begin{array}{l}
    \frac23\mathcal{R}_i\qquad\qquad\quad\text{(superhorizon)}\\
    -18\mathcal{R}_i\frac{\cos(\varphi)}{\varphi^2}\qquad\text{(subhorizon)}
    \end{array}\right.}\,.
    \label{eq:phird}
\end{equation}
We can conclude that the potential is constant outside the horizon. We have only a drop of $\frac23\times\frac53$ during the switch from radiation to matter domination, coming from \eq{mathcalRw}. If $k>k_{eq}$ the modes enter during radiation era, therefore decreases as $a^{-2}$ and as soon as we enter the matter dominated era, $\Phi$ stays constant because \eq{phimd} is valid at all scales.

\paragraph{\textit{Matter}} To study the matter density perturbations evolution in a radiation dominated universe, we take \eq{mattercontinuity} and consider that the $\Phi=\Phi_r+\Phi_m$ where, as we have seen in \eq{phird} the radiation contribution rapidly oscillates while the matter contribution is constant (see \eq{phimd}). However, the matter effectively only evolves to the time-averaged potential  therefore $\delta_m$ is only sourced by $\Phi_m$~\cite{Weinberg:2008zzc}. Using the Poisson equation and approximating $\Phi''=\Phi'\sim0$, we eventually arrive at the so-called \textit{Meszaros equation}~\cite{Meszaros:1974tb}
\begin{equation}
    \frac{d^2\delta_m}{dy^2}+\frac{2+3y}{2y(1+y)}\frac{d\delta_m}{dy}-\frac{3}{2y(1+y)}\delta_m=0
    \label{eq:Mexi}
\end{equation}
where $y=a/a_{eq}$. Therefore, when radiation dominates the Universe we have $y\ll 1$
\begin{equation}
    \boxed{\delta_m\propto
    1+\frac23y}
    \label{eq:MZSE}
\end{equation}
which grows logarithmically. Moreover, if we call $a_{\rm hor}$ the scale factor when the perturbation crosses the horizon, the growth factor is 
\begin{equation}
    \frac{\delta_m(a_{eq})}{\delta_m(a_{\rm hor})}=\frac{1+\frac32}{1+\frac32\frac{a_{\rm hor}}{a_{eq}}}\leq \frac52
\end{equation}
which is the \textit{Meszaros effect}: Dark matter fluctuations do not grow much within the horizon before the equivalence epoch. After the equivalence, matter is dominant and we can solve the Poisson equations for the comoving density contrast $\Delta_m$, finding
\begin{equation}
\boxed{\Delta_m(a,\textbf{k})=-\frac{k^2\Phi}{4\pi Ga^2\bar{\rho}_m}}\,.
\label{eq:mmdom}
\end{equation}
Therefore, we need to divide the solutions \eq{phimd} by $a$ ( $\bar{\rho}_ma^2\propto a^{-1}$), finding that $\Delta_m$ evolves as $a$. It has to be remembered that $\delta_m$ outside the horizon is constant\footnote{we are using adiabatic mode and $\delta_m\propto\Phi$} but as soon as the modes enter the horizon, $\Delta_m\approx\delta_m\propto a$.

\paragraph{\textit{Radiations}} In a radiation dominated universe, we can use again the Poisson equation and find
\begin{equation}
    \boxed{\Delta_r(\eta,\textbf{k})=-\frac23(k\eta)^2\Phi(k,\textbf{k})}
    \label{eq:radevo}
\end{equation}
which implies, looking at \eq{phird} that $\Delta_r\propto a^2$ outside the horizon and the modes inside the horizon $\delta_r$ oscillates without damping. On the other hand, when matter is the principal component, we can look at \eq{radiationcontinuity} which, being the potential constant on all scales, recovers the form of an harmonic oscillation equation whose solution is oscillatory.

\begin{tcolorbox}[mybox]
Baryons are tightly coupled to radiation well after $z_{eq}$, and they cannot grow as they oscillate with the radiation component (with additional damping effects like \textit{ Silk damping}~\cite{Silk:1967kq}). The fractional contribution to the photon perturbations is often characterized by the dimensionless parameter $R$
\begin{equation}
    R\equiv\frac34\frac{\bar{\rho}_b}{\bar{\rho}_\gamma}\,.
\end{equation}
With adiabatic perturbations, we have constant entropy per baryon $\delta n_\gamma/n_\gamma=\delta n_b/ n_b$. It is possible to show that the anisotropy in temperature of the last scattering surface photons gives an upper limit to the baryonic perturbations 
\begin{equation}
    \frac{\delta T}{T}=\frac13\delta_b
\end{equation}
with $\frac{\delta T}{T}\sim10^{-5}$. Therefore the baryonic perturbations cannot grow enough to form the structures we observe today. We need the dark matter. Even though it also grows linearly with $a$, being a matter component, the dark matter is not coupled with the radiation and it is able to grow earlier in the past without any constraint. Taking \eq{mattercontinuity} and taking into account that $\bar{\rho}_m\delta_m=\Sigma_i\bar{\rho}_i\delta_i$ with $i=c,b$, where $c$ stands for cold dark matter we can write
\begin{equation}
    \delta''_b+\mathcal{H}\delta_m'-4\pi Ga^2\bar{\rho}_c\delta_c\approx 0\,.
\end{equation}
we can write $\delta_c\sim Da$ with $D$ a constant factor. If we change the derivation with respect to time to a derivation with respect to the scale factor, we find the solution in the form $\delta_b=Ba+C$. We can easily see that $B=D$ and, imposing $\delta_b(a_{d})\approx 0$ with $a_d$ the time of decopling from radiation (\ie the baryonic perturbations do not grow until they decouple from radiation), $C=-Da_d$. Thus, the solution reads
\begin{equation}
    \delta_b=\delta_c\(1-\frac{a_d}{a}\)\,.
\end{equation}
Looking at the equation, it is possible to see that once the baryonic matter decouples from radiation, it goes through an accelerated phase, the \textit{catching-up} phase. Namely, it is attracted to the potential wells of the dark matter halos, which are already formed.
\end{tcolorbox}

\section{Tensor perturbations}
\label{sec:GW} 
Gravitational waves (GW) are \textit{metric waves}. Hence, when they propagate, the geometry changes in time. We want to focus on the derivation of gravitational waves from a simple perturbation of the flat metric and see if it is gauge invariant~\cite{Ferrari:2020nzo}. In the end, we will show how gravitational waves affect a circle of test particles. In \chap{CMBCHAPTER} we will see how it will match the signature in the polarization and in \chap{PGWs} we will study the phenomenology of the gravitational waves coming from inflationary perturbations.

\subsection{Flat time perturbation}
The metric that results from a perturbation of the flat spacetime is
\begin{equation}
    g_{\mu\nu}=\eta_{\mu\nu}+h_{\mu\nu},\quad \lvert h_{\mu\nu}\rvert\ll 1,
    \label{eq:flatperturbed}
\end{equation}
before computing the perturbed Einstein equations, it is better to use a different form of them. We can contract both side of \eq{EinsteinEquations} obtaining $R=-kT$ with $k=8\pi G/c^4$. Next, we introduce the \textit{Source tensor}
\begin{equation}
    S_{\mu\nu}\equiv T_{\mu\nu}-\frac12g_{\mu\nu}T
\end{equation}
and write \eq{EinsteinEquations} in the form
\begin{equation}
    R_{\mu\nu}=k\(T_{\mu\nu}-\frac12Tg_{\mu\nu}\)\,.
\end{equation}
To compute the perturbed Einstein equations, we need \eq{Christoff}, \eq{PerturbationR} and
\begin{equation}
    \delta S_{\mu\nu}=\delta T_{\mu\nu}-\frac12\bar{g}_{\mu\nu}\delta T-\frac12h_{\mu\nu}\bar{T}
\end{equation}
for a generic perturbation~\eq{perturbeggmunu}. However, now we are working with $\bar{g}_{\mu\nu}=\eta_{\mu\nu}$, \ie $\bar{\Gamma}=0$ and we obtain
\begin{gather}
\delta\Gamma^\lambda_{\mu\nu}=\frac12\eta^{\lambda\rho}[h_{\rho\nu,\mu}+h_{\rho\mu,\nu}-h_{\mu\nu,\rho}]+\mathcal{O}(h^2),
    \label{eq:Gflat}\\
    \delta R_{\mu\nu}=\delta\Gamma^\alpha_{\mu\nu,\alpha}-\delta\Gamma^\alpha_{\mu\alpha,\nu}+\mathcal{O}(h^2),
    \label{eq:Rflat}\\
    \delta S_{\mu\nu}=\delta T_{\mu\nu}-\frac12\eta_{\mu\nu}\delta T+\mathcal{O}(h^2).
    \label{eq:Sflat}
\end{gather}
If we use \eq{Gflat} to write explicitly \eq{Rflat}, we obtain
\begin{equation}
\begin{split}
    \delta R_{\mu\nu}=\frac12[-\eta^{\alpha k}h_{\mu\nu,\alpha k}+\eta^{\alpha k}(h_{k\nu,\mu\alpha}-h_{k\alpha,\mu\nu}+h_{\mu\alpha,k\nu})]\\=\frac12[-\Box_F h_{\mu\nu}+(h^\alpha_{\nu,\mu\alpha}-h_{,\mu\nu}+h^k_{\mu,k\nu})],
    \end{split}
\end{equation}
where we have used the d'Alambert operator (\ie $\Box_F=-\frac1{c^2}\frac{\partial^2}{\partial t^2}+\mathbf{\nabla}^2$), and the subscript $F$ to stress that we are in a flat space. Eventually, the perturbed Einstein equation becomes 
\begin{equation}
    \Box_Fh_{\mu\nu}-(h^\alpha_{\nu,\mu\alpha}+h^k_{\mu,k\nu}-h_{,\mu\nu})=-16\pi G(\delta T_{\mu\nu}-\frac 12\eta_{\mu\nu}\delta T).
    \label{eq:n}
\end{equation}
We know, from Sec. \ref{GaugeChoice}, that the solution of Einstein's field equations is not uniquely determined, as we can always perform a coordinate transformation and obtain a solution to the equation again. Here, we are limited to the only transformations that preserve the weak field limit. We perform an infinitesimal transformation following \eq{newcoordinate}, and \eq{newcoordinatess} becomes
\begin{equation}
    \widetilde{g}_{\mu\nu}=\eta_{\mu\nu}+h_{\mu\nu}-\varepsilon_{\mu,\nu}-\varepsilon_{\nu,\mu}+\mathcal{O}(h^2).
\end{equation}
Using this gauge freedom, we decide to work in the  \textit{harmonic gauge} \ie imposing $g^{\mu\nu}\Gamma^\lambda_{\mu\nu}=0$. This condition is equivalent to say
\begin{equation}
    h^\mu_{\nu,\mu}=\frac12 h_{,\nu}
    \label{eq:HarmonicGauge}
\end{equation}
up to first order in $h_{\mu\nu}$. The latter equation makes the term in brackets in \eq{n} vanishes and the Einstein equation turns into a simple wave equation, supplemented by the condition of the harmonic gauge:
\begin{equation}
    \left\{\begin{array}{l}
    \Box_Fh_{\mu\nu}=-16\pi G(\delta T_{\mu\nu}-\frac 12\eta_{\mu\nu}\delta T)\\
    h^\mu_{\nu,\mu}=\frac12 h_{,\nu}
    \end{array}\right. .
    \label{eq:Einsteinsystem}
\end{equation}
We can now introduce the tensor
\begin{equation}
    \bar{h}_{\mu\nu}\equiv h_{\mu\nu}-\frac12\eta_{\mu\nu}h
    \label{eq:hbarr}
\end{equation}
which simplifies \eq{Einsteinsystem} into\footnote{To obtain this result, we should multiply the first equation by the inverse of the flat metric: $\eta^{\mu\nu}\Box_F h_{\mu\nu}=\Box_Fh=-16\pi G(T-2T)$, and combine it with the original equation.}
\begin{equation}
    \left\{\begin{array}{l}
    \Box_F\bar{h}_{\mu\nu}=-16\pi G\delta T_{\mu\nu}\\
    \bar{h}^\mu_{\nu,\mu}=0,
    \end{array}\right.\,.
    \label{eq:WaveEq}
\end{equation}
This is exactly the wave equation with a source term $\delta T_{\mu\nu}$. A perturbation of a flat spacetime propagates as a wave travelling at the speed of light. The solution of \eq{WaveEq} can be written in terms of retarded potentials~\cite{Ferrari:2020nzo}
\begin{equation}
    \bar{h}_{\mu\nu}(t,\boldsymbol{x})=\frac{4G}{c^4}\int{\frac{T_{\mu\nu}(t-\frac{\lvert\boldsymbol{x-x'}\rvert}{c},\boldsymbol{x}')}{\lvert\boldsymbol{x-x'}\rvert}d^3x'},
\end{equation}
where the integral is over the past light cone of the event $(t,\boldsymbol{x})$. In the solution, we have restored $c$ for completeness, and we have removed the $\delta$ term considering $T$ as a quantity of order $h$. 
\begin{tcolorbox}[mybox]
    The simplest solution of \eq{WaveEq}, with $T=0$, is a monochromatic plane wave
\begin{equation}
    \bar{h}_{\mu\nu}=\Re{A_{\mu\nu}e^{ik_\alpha x^\alpha}},
    \label{eq:solutio}
\end{equation}
with $A_{\mu\nu}$ the wave amplitude or polarization tensor, and $\boldsymbol{k}$ the wave vector. The equation of the wavefront is 
\begin{equation}
    k_\alpha x^\alpha=const.
\end{equation}
Inserting the solution in \eq{solutio} in the wave equation in vacuum yields
\begin{equation}
    \Box_F\bar{h}_{\mu\nu}=\eta^{\alpha\beta}\bar{h}_{\mu\nu,\alpha\beta}=-\eta^{\alpha\beta}k_\alpha k_\beta A_{\mu\nu}e^{i k_\lambda x^\lambda}=0\rightarrow \eta^{\alpha\beta}k_\alpha k_\beta=0.
\end{equation}
Therefore, $\boldsymbol{k}$ is a null vector. From the harmonic gauge, we get
\begin{equation}
    \bar{h}^\mu_{\nu,\mu}=iA^\mu_\nu k_\mu e^{ik^\alpha x_\alpha}=0\rightarrow k_\mu A^\mu_\nu=0.
\end{equation}
Namely, the wave vector is orthogonal to the polarization tensor $A_{\mu\nu}$: it is a transverse wave. 
\end{tcolorbox}

\paragraph{\textit{Gauge freedom}} We want to understand how many of the ten components of $h_{\mu\nu}$ have a real physical meaning. Let us consider a progressive wave propagating along the $x$ direction:
\begin{equation}
    \bar{h}_{\mu\nu}=\bar{h}_{\mu\nu}(f),\quad f=t-\lvert \boldsymbol{x}\rvert.
\end{equation}
Since it is independent of $y$ and $z$, we have from \eq{HarmonicGauge} the following relation:
\begin{equation}
    \bar{h}^t_{\nu,t}+\bar{h}^x_{\nu,x}=0.
\end{equation}
Being
\begin{equation}
    \frac{\partial\bar{h}^\mu_{\nu}}{\partial t}=\frac{\partial\bar{h}^\mu_\nu}{\partial f},\quad \frac{\partial\bar{h}^\mu_{\nu}}{\partial x}=-\frac{\partial\bar{h}^\mu_\nu}{\partial f},
\end{equation}
we get
\begin{equation}
    \frac{\partial}{\partial f}[\bar{h}^t_\nu-\bar{h}^x_\nu]=0.
\end{equation}
Integrating the last equation and setting to zero the constant of integration, as we are interested only in the time-dependent part, we obtain
\begin{equation}
    \bar{h}^t_t=\bar{h}^x_t,\quad\bar{h}^t_x=\bar{h}^x_x,\quad\bar{h}^t_y=\bar{h}^x_y,\quad\bar{h}^t_z=\bar{h}^x_z.
    \label{eq:h=h}
\end{equation}

The harmonic gauge does not completely fix the gauge; if we make an infinitesimal coordinate transformation, $x^{\mu'}=x^\mu+\varepsilon^\mu$, the new tensor $\bar{h}'_{\mu\nu}$ still satisfies the wave equation, as long as $\Box_F\varepsilon^\mu=0$. That said, we can use the four functions $\varepsilon^\mu$ to set to zero the following four quantities:
\begin{equation}
\bar{h}^t_x=\bar{h}^t_y=\bar{h}^t_z=\bar{h}^y_y+\bar{h}^z_z=0.
\end{equation}
This result leads us to more simplifications using \eq{h=h}. In fact, we have:
\begin{equation}
    \bar{h}^x_x=\bar{h}^x_y=\bar{h}^x_z=\bar{h}^t_t=0.
\end{equation}
Moreover, taking into account all the vanishing terms, we get
\begin{equation}
    0=\bar{h}^\mu_\mu=h^\mu_\mu-2h^\mu_\mu.
\end{equation}
Namely, in this gauge, called \textit{TT-gauge}, $h_{\mu\nu}$ and $\bar{h}_{\mu\nu}$ coincide and are traceless:
\begin{equation}
    h_{\mu\nu}=\begin{pmatrix}
    0&0&0&0\\
    0&0&0&0\\
    0&0&h_{yy}&h_{yz}\\
    0&0&h_{yz}&-h_{yy}\\
    \end{pmatrix}.
    \label{eq:tttt}
\end{equation}
We now have only two degrees of freedom which correspond to the two possible polarization states. The components of $h_{\mu\nu}$ are different from zero only in the plane orthogonal to the direction of propagation (transverse). That is way it is called T(transverse)T(traceless)-gauge. 

\subsection{Circle of test particles}
\label{sec:Circe}
A single particle is not sufficient to detect gravitational waves. To see why, we should consider a particle at rest in flat spacetime and pin the origin of the inertial frame to it. Additionally, we have an incoming gravitational wave in the TT-gauge propagate along $\hat{x}$. At $t=0$ the particle is at rest and, the geodesic equation of the curved spacetime generated by the wave, reduces to
\begin{equation}
    \left(\frac{d U^\alpha}{d\tau}\right)_{t=0}=-\frac12\eta^{\alpha\beta}[h_{\beta0,0}+h_{0\beta,0}-h_{00,\beta}].
\end{equation}
However, in the TT-gauge the right member vanishes. It implies that the particle is not accelerated and remains at a constant coordinate position, irrespective of the wave presence. Thus, we should consider more than one particle to detect gravitational waves. Let us assume two particles initially at rest and a TT-gravitational wave that reaches them at $t=0$. We have:
\begin{equation}
    ds^2=(\eta_{\mu\nu}+h^{TT}_{\mu\nu})dx^\mu dx^\nu.
\end{equation}
The single particle remains at the same coordinate position despite the passing of the wave as well as their coordinate separation $\delta x^\mu=x^\mu_B-x^\mu_A$. Nevertheless, it is not the case for the proper distance. If the particles are on the $y$-axes, we calculate
\begin{equation}
    \Delta l=\int{ds}=\int^{y_B}_{y_A}{\lvert 1+h^{TT}_{yy}(f)\rvert^{\frac12}dy}\neq constant.
\end{equation}
We should now attach a local inertial frame to the particle A, so that $x^\mu_A=(t_A,0,0,0)$ and $\delta x^i=x^i_B$ (we have omitted the prime of the new coordinates for simplicity). The separation vector satisfies the equation of geodesic deviation. Evaluating it for particle A, recalling that we are working in the TT-gauge, we get
\begin{equation}
    \frac{d^2}{dt^2}\delta x^\lambda=\frac12\eta^{\lambda i}\frac{\partial^2 h^{TT}_{im}}{\partial t^2}\delta x^m,
    \label{eq:geoGW}
\end{equation}
\begin{table}
\centering
\begin{tabular}{ |c|c|c| } 
\hline
 &\multicolumn{2}{| c |}{$A_\times=0$ and $A_+\neq 0$ }\\
\hline
&A&B\\
\hline
\multirow{2}{2em}{\centering\Large $\frac{\pi}{2}$}&$y=y_0$ & $y=0$ \\ 
& $z=0$ & $z=z_0$\\
\hline
\multirow{2}{2em}{\centering\large$>\frac{\pi}{2}$}&$y=y_0[1+A_+\cos{\omega(t-x)}]$ & $y=0$ \\ 
& $z=0$ & $z=z_0[1-A_+\cos{\omega(t-x)}]$\\ 
\hline
\multirow{2}{3em}{\centering\Large$\pi$}&$y=y_0[1-A_+]$ & $y=0$ \\ 
& $z=0$ & $z=z_0[1+A_+]$\\ 
\hline\multirow{2}{3em}{\centering\Large$\frac{3\pi}{4}$}&$y=y_0$ & $y=0$ \\ 
& $z=0$ & $z=z_0$\\ 
\hline
\multirow{2}{3em}{\centering\large$2\pi$}&$y=y_0[1+A_+]$ & $y=0$ \\ 
& $z=0$ & $z=z_0[1-A_+]$\\ 
\hline
\end{tabular}
\belowcaptionskip=5pt
\caption{\small Displacement of two particles with $A_\times=0$}
\label{tab:x0}
\end{table}
\begin{table}
\centering
\begin{tabular}{ |c|c|c|c|c| } 
\hline
 &\multicolumn{4}{| c |}{$A_+=0$ and $A_\times\neq 0$ }\\
\hline
&A&B&C&D\\
\hline
\multirow{2}{2em}{\centering\Large $\frac{\pi}{2}$}&y=r&y=-r&y=-r&y=r\\
&z=r&z=r&z=-r&z=-r\\
\hline
\multirow{2}{2em}{\centering\Large ${\pi}$}&$y=r[1-A_\times]$&$y=-r[1+A_\times]$&$y=-r[1-A_\times]$&$y=r[1+A_\times]$\\
&$z=r[1-A_\times]$&$z=r[1+A_\times]$&$z=-r[1-A_\times]$&$z=-r[1+A_\times]$\\
\hline
\multirow{2}{2em}{\centering\Large $\frac{3\pi}{4}$}&$y=r$&$y=-r$&$y=-r$&$y=r$\\
&$z=r$&$z=r$&$z=-r$&$z=-r$\\
\hline
\multirow{2}{2em}{\centering\Large ${2\pi}$}&$y=r[1+A_\times]$&$y=-r[1-A_\times]$&$y=-r[1+A_\times]$&$y=r[1-A_\times]$\\
&$z=r[1+A_\times]$&$z=r[1-A_\times]$&$z=-r[1+A_\times]$&$z=-r[1-A_\times]$\\
\hline
\end{tabular}
\belowcaptionskip=5pt
\caption{\small Displacement of two particles with $A_+=0$}
\label{tab:x1}
\end{table}
with $i,m=2,3$. For $t\leq 0$, we have assumed the particles at rest, therefore, their separation is constant, $\delta x_0^j=const$. When the wave arrives, the relative position of the particles will change infinitesimally because of the smallness of $h_{\mu\nu}$ perturbation. Thus
\begin{equation}
    \delta x^j(t)=\delta x^j_0+\delta x_1^j(t),\quad t>0\quad\text{and}\quad\delta x_0^j=const,
\end{equation}
with $\delta x_1^j(t)$ a small perturbation with respect to the initial position $\delta x_0^j$. Substituting this result in \eq{geoGW} and retaining only terms of order $h$, we infer
\begin{equation}
    \frac{d^2}{dt^2}\delta x^j=\frac12\eta^{j i}\frac{\partial^2 h^{TT}_{ik}}{\partial t^2}\delta x^k_0,
\end{equation}
whose integration yields
\begin{equation}
    \delta x^j=\delta x_0^j+\frac12\eta^{ji}h^{TT}_{ik}\delta x^k_0.
    \label{eq:pospos}
\end{equation}
\begin{figure}[h!]
\centering
\includegraphics[width=0.9\textwidth]{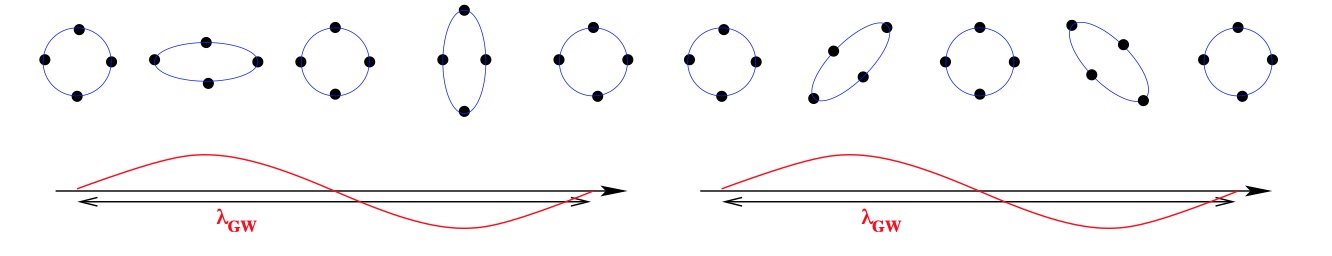}
\caption[GWs influence for a ring of particles]{\small Point particles, arranged to form a ring, move due to the interaction with a gravitational wave propagating in the direction perpendicular to the plane of the ring. The left panel refers to a wave with $A_+$ polarization, while the right panel with $A_\times$ polarization. The image is adopted from \cite{Buonanno:2007yg}.}
\label{fig:Ale}
\end{figure}
Next, we want to study the polarization of the wave. Firstly, we consider a plane wave whose non-vanishing components are
\begin{gather}
    h^{TT}_{yy}=-h^{TT}_{zz}=2\Re{A_+e^{i\omega(t-x)}},\\
    h^{TT}_{zy}=-h^{TT}_{zy}=2\Re{A_\times e^{i\omega(t-x)}}
\end{gather}
with $k=(\omega,\boldsymbol{k})$. Secondly, we consider particle A in $(0,y_0,0)$ and B in $(0,0,z_0)$. If $A_\times=0$ we have  
\begin{equation}
    h_{yy}=-h_{zz}=2A_+\cos{\omega(t-x)},\quad h_{yz}=h_{zy}=0.
\end{equation}
Whereas, if $A_+=0$, we get
\begin{equation}
    h_{yy}=-h_{zz}=0,\quad h_{yz}=h_{zy}=2A_\times\cos{\omega(t-x)}.
\end{equation}
From \eq{pospos} and assuming that, at $t=0$, $\omega=\pi/2$, we have the pattern showed in \tab{x0}.
On the other hand, to study the case $A_\times\neq0$, we consider four particles (named A,B,C,D), displaced at the four vertex of a square with length $2r$. At $t=0$, we fix $\omega=\pi/2$, whilst, for $t>0$, we have
\begin{gather}
    y=y_0+z_0A_\times\cos{\omega(t-x)},\\
    z=z_0+y_0A_\times\cos{\omega(t-x)}.
\end{gather}
and the different displacement are showed in \tab{x1}.

If we consider a ring of particles, the pattern is shown in \fig{Ale}. We can define $A_+$ and $A_\times$ as the \textit{polarization amplitudes} of the wave. Linearly polarized waves have only one of the two amplitudes different from zero.

%% file: Chapters/CMB.tex
\chapter{Cosmic Microwave Background}
\label{ch:CMBCHAPTER}

\begin{figure}[htp]
\centering
\includegraphics[width=0.7 \textwidth]{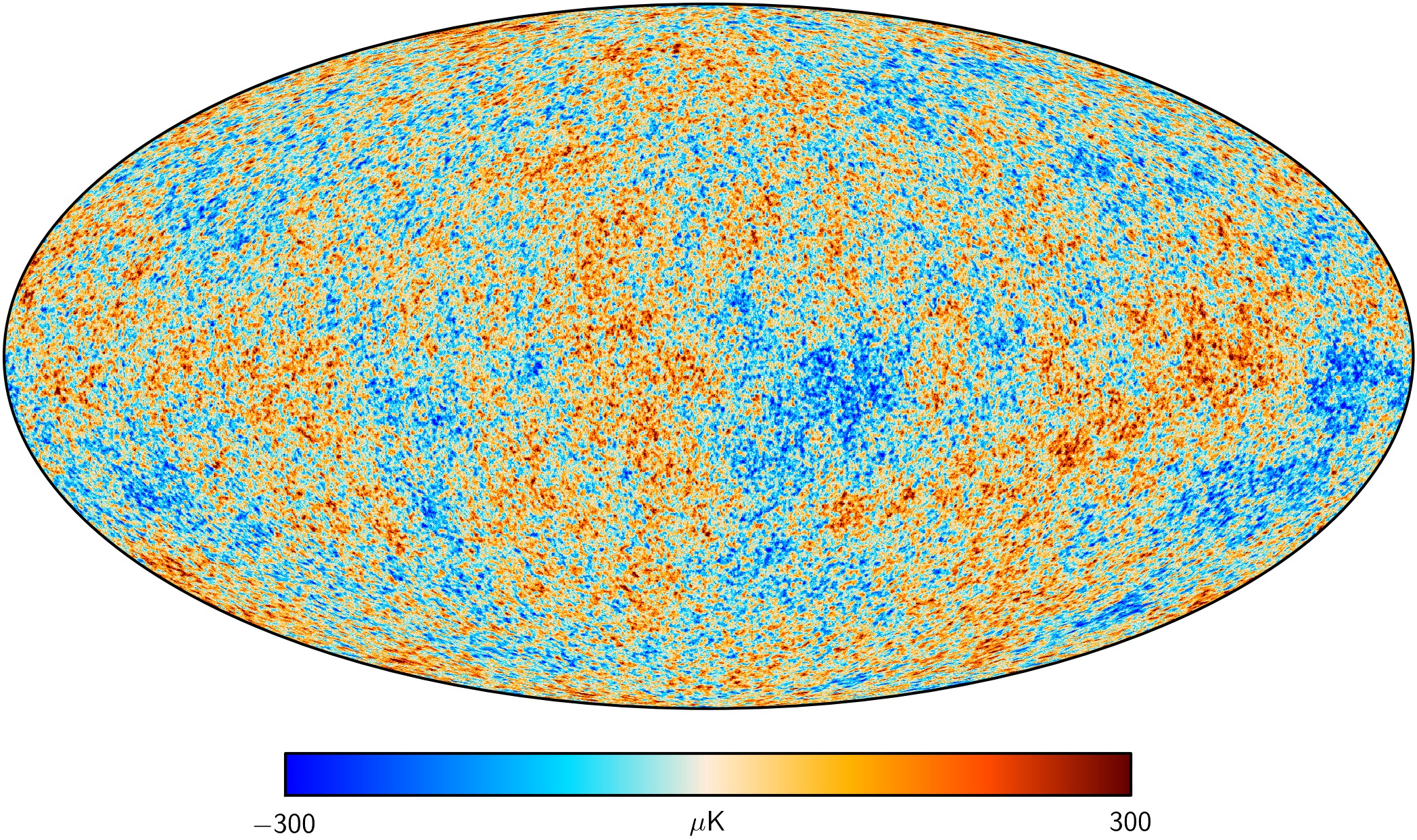}
\caption[CMB map]{\small It is the map captured by Planck observations. It shows tiny temperature fluctuations that correspond to regions of slightly different densities, representing the seeds of all future structures: the stars and galaxies we see today. The image is taken from the \href{https://www.cosmos.esa.int/web/planck/picture-gallery}{\textit{Planck ESA} website}.}
\label{fig:Map}
\end{figure}

The Cosmic Microwave Background (CMB) is a critical observable that provides a snapshot of the early Universe at the time of photon decoupling, approximately $380.000$ years after the Big Bang. Spatial variations in the CMB temperature at recombination are seen as temperature anisotropies today~\cite{Hu:2001bc,Hu:1995kot,Hu:2008hd,Lyth:2004gb,Durrer:2020fza,Hu:1994uz,Weinberg:2008zzc,Baumann:2022mni,Pan:2016zla, Dodelson:2003ft}. As soon as photons decouple, they fall out of thermal equilibrium and develop different anisotropy, perturbation in the temperature of the black-body distribution, for each of the two polarization states. Most of its structure is associated with \textit{acoustic oscillations} of the photon-baryon plasma on $\sim\ang{1}$ scales. The angular variations in temperature that we observe today, see \fig{Map}, are a snapshot of the local properties of relic photons at redshift $z\sim 1090$ (see \sect{ThermalHistory}) that must be related to primordial perturbations (see \chap{InfloBinglo}).

We have already studied that, due to its pressure, radiation does not grow via gravitational instability (see \eq{radevo}). On the other hand, matter inhomogeneities grow (see \eq{mmdom}); as a result, the Universe is now organized in very non-linear structures with a clumpy matter distribution (see \chap{Chahahah}). Nevertheless, both inhomogeneities in the matter and anisotropies in the CMB originate from the same source (see \chap{InfloBinglo}). It has to be stressed out that, since the photon distribution is close to uniformity, perturbations are small and CMB anisotropies fall almost entirely under linear perturbation theory. 

\section{Temperature multipole expansion}
\label{TEMEE}
If we call $T(\hat{n})$ the measured CMB temperature in a direction $\hat{n}$ and $T_0$ the average CMB temperature today, we can define the \textit{brightness function} as $\Theta\equiv(T(\hat{n})-T_{0})/T_0$. These fluctuations are observed on a spherical surface (the LS surface), so we can expand the temperature fluctuations in terms of spherical harmonics $Y_{\ell m}(\hat{n})$, which constitute the complete orthonormalized set of functions on a unit sphere. The expansion reads:
\begin{equation}
    \Theta(\hat{n})=\sum_{\ell m}\Theta_{\ell m}Y_{\ell m}(\hat{n})\,.
    \label{eq:Bright}
\end{equation}
The $\Theta_{\ell m}$ are called \textit{multipole moments} and the spherical harmonics are defined as
\begin{equation}
    Y_{lm}=\left[\frac{2\ell+1}{4\pi}\frac{(\ell -m)!}{(\ell +m)!}\right]P^m_\ell(\cos{\theta})e^{im\phi}\,,\quad P^m_\ell(x)=(-1)^\ell \frac{(1-x^2)^m}{2!\ell !}\frac{\d^{\ell +m}}{\d x^{\ell +m}}(1-x^2)^\ell 
\end{equation}
and $\theta$ and $\phi$ are the spherical coordinates that identify the direction $\hat{n}$, and $P^m_\ell(x)$ is the associated Legendre function. We shall list two important relations. The first one is
\begin{equation}
    \sum_mY^*_{\ell m}(\theta_1,\phi_1)Y_{\ell m}(\theta_2,\phi_2)=\frac{2\ell+1}{4\pi}P_\ell(\cos{\theta_{12}}),
    \label{eq:po}
\end{equation}
with $\theta_{12}$ the angle between the directions $1$ and $2$. The second one is the expansion of a plane wave of unit amplitude: 
\begin{equation}
e^{i\boldsymbol{k}\vdot\boldsymbol{x}}=\sum_\ell(2l+1)i^\ell j_\ell(kx)P_\ell(\hat{\boldsymbol{x}}\vdot\hat{\boldsymbol{k}}),
\label{eq:pi}
\end{equation}
where $j_\ell(x)$ are the spherical Bessel functions of integral order which satisfy the condition:
\begin{equation}
    \int_0^\infty{j^2_\ell(x)d\ln{x}}=\frac{1}{2\ell(\ell +1)}\,.
    \label{bessel}
\end{equation}
As the fluctuations are statistically isotropic, we should define the \textit{angular power spectrum} $C_\ell$ as the two-point correlation function of the multipole moments
\begin{equation}
\boxed{\langle\Theta_{\ell m}^*\Theta_{\ell 'm'}\rangle=\delta_{\ell \ell'}\delta_{mm'}C_\ell}\,,
    \label{eq:corre}
\end{equation}
that is 
\begin{equation}
    C_\ell=\langle \Theta_{\ell m}^*\Theta_{\ell m}\rangle=\langle\lvert\Theta_{\ell m}\rvert^2\rangle.
    \label{eq:expectvalue}
\end{equation}
The angular power spectrum in the harmonic space is equivalent to the one in real space. In fact  
\begin{equation}
    C(\theta)\equiv\langle \Theta(\hat{n})\Theta(\hat{n}')\rangle=\sum_\ell\frac{2\ell+1}{4\pi}C_\ell P_\ell(\cos{\theta}),
    \label{eq:angularpsreal}
\end{equation}
where $\theta$ is the angle between $\hat{n}$ and $\hat{n}'$, and $P_\ell(x)$ represents the Legendre polynomial with $m=0$. The moments of the angular power spectrum appear as coefficients in the expansion of $C(\theta)$. Using the orthogonality of the Legendre polynomials, we can write
\begin{equation}
C_\ell=2\pi\int^1_{-1}d\cos{\theta}C(\theta)P_\ell(\cos{\theta}).
\end{equation}
The information contained in the $C_\ell$ is completely equivalent to that of $C(\theta)$.
\begin{tcolorbox}[mybox]
Why should we expect \eq{angularpsreal} to depend only on the difference in the directions? In the Universe, there are non-deterministic phenomena. As a consequence, we cannot predict exactly the observed Universe. For instance, we are not able to know the exact positions of all galaxies in the Virgo cluster; instead, we shall use a statistical approach. It should be stressed that the Universe is just one statistical realisation of all random fields of cosmological interest. All the possible universes or realisations, make up the ensemble. Given a random field $f(\boldsymbol{x})$,  we call $f_n(\boldsymbol{x})$ the specific realisation of the ensemble. Let us now see the propriety of the \textit{two-point correlation function}
\begin{equation}
    \xi(x_1,x_2)\equiv \langle f(\boldsymbol{x_1}f(\boldsymbol{x_2})\rangle=\sum_n P_nf_n(\boldsymbol{x}_1)f_n(\boldsymbol{x}_2)\xRightarrow[limit]{continuous}\int{p[\hat{f}(\boldsymbol{x})]f(\boldsymbol{x}_1)f(\boldsymbol{x}_2)df}.
    \label{Coppa}
\end{equation}
where, in the continuous limit, we have used a functional integral over the random field configurations. This result can be simplified if we apply the cosmological principle. First, the homogeneity of the field implies that $p[\hat{f}(\boldsymbol{x})]$ must be invariant under translations. Thus, the two-point correlation function has to be a function of $(\boldsymbol{x_1}-\boldsymbol{x_2})$. Second, the probability must be invariant under rotations due to the condition of isotropy: the two-point function does not depend on the direction of the vectors. Therefore, for a homogeneous and isotropic field, we have
\begin{equation}
    \xi(\boldsymbol{x}_1,\boldsymbol{x}_2)=\xi(\lvert \boldsymbol{x}_1-\boldsymbol{x}_2\rvert)=\xi(r).
\end{equation}
If we move to the Fourier space, our random field is
\begin{equation}
    f(\boldsymbol{x})=\frac{1}{(2\pi)^3}\int{\hat{f}(\boldsymbol{k})e^{i\boldsymbol{k}\vdot\boldsymbol{x}}d\boldsymbol{k}}.
    \label{eq:Fourier}
\end{equation}
and remembering that the three-dimensional delta function is given by
\begin{equation}
    \delta^{(3)}_D(k)=\frac{1}{(2\pi)^3}\int{e^{i\boldsymbol{k}\vdot\boldsymbol{x}}d\boldsymbol{x}},
\end{equation}
we can now rewrite the correlation function as
\begin{equation}
    \xi(r)=\frac{1}{(2\pi)^6}\int{d\boldsymbol{k}d\boldsymbol{k}'\langle\hat{f}(\boldsymbol{k})\hat{f}(\boldsymbol{k}')\rangle e^{i\boldsymbol{k}\vdot(\boldsymbol{x}+\boldsymbol{r})}e^{i\boldsymbol{k}'\vdot\boldsymbol{x}}}.
\end{equation}
At this point we define the important quantity called \textit{power spectrum}, $P(k)$, of the random field $f(\boldsymbol{x})$:
\begin{equation}
     \langle\hat{f}(\boldsymbol{k})\hat{f}(\boldsymbol{k}')\rangle=(2\pi)^3P(k)\delta^{(3)}_D(\boldsymbol{k}+\boldsymbol{k}').
     \label{twopoin}
\end{equation}
Thus, we have
\begin{equation}
    \boxed{\xi(r)=\int{P(k)e^{i\boldsymbol{k}\vdot\boldsymbol{r}}d\boldsymbol{k}}}\,.
    \label{eq:xiR}
\end{equation}
The power spectrum can be expressed also as
\begin{equation}
   \boxed{ P(k)\equiv \int{d\boldsymbol{r}\xi(r)e^{-i\boldsymbol{k}\vdot\boldsymbol{r}}}}
\end{equation}
and in a dimensionless form
\begin{equation}
   \boxed{ \Delta^2(k)\equiv \frac{k^3}{2\pi^2}P(k)}\,.
   \label{eq:dimensionlessPS}
\end{equation}
\end{tcolorbox}

\begin{figure}[h!]
\centering
\includegraphics[width=0.9\textwidth]{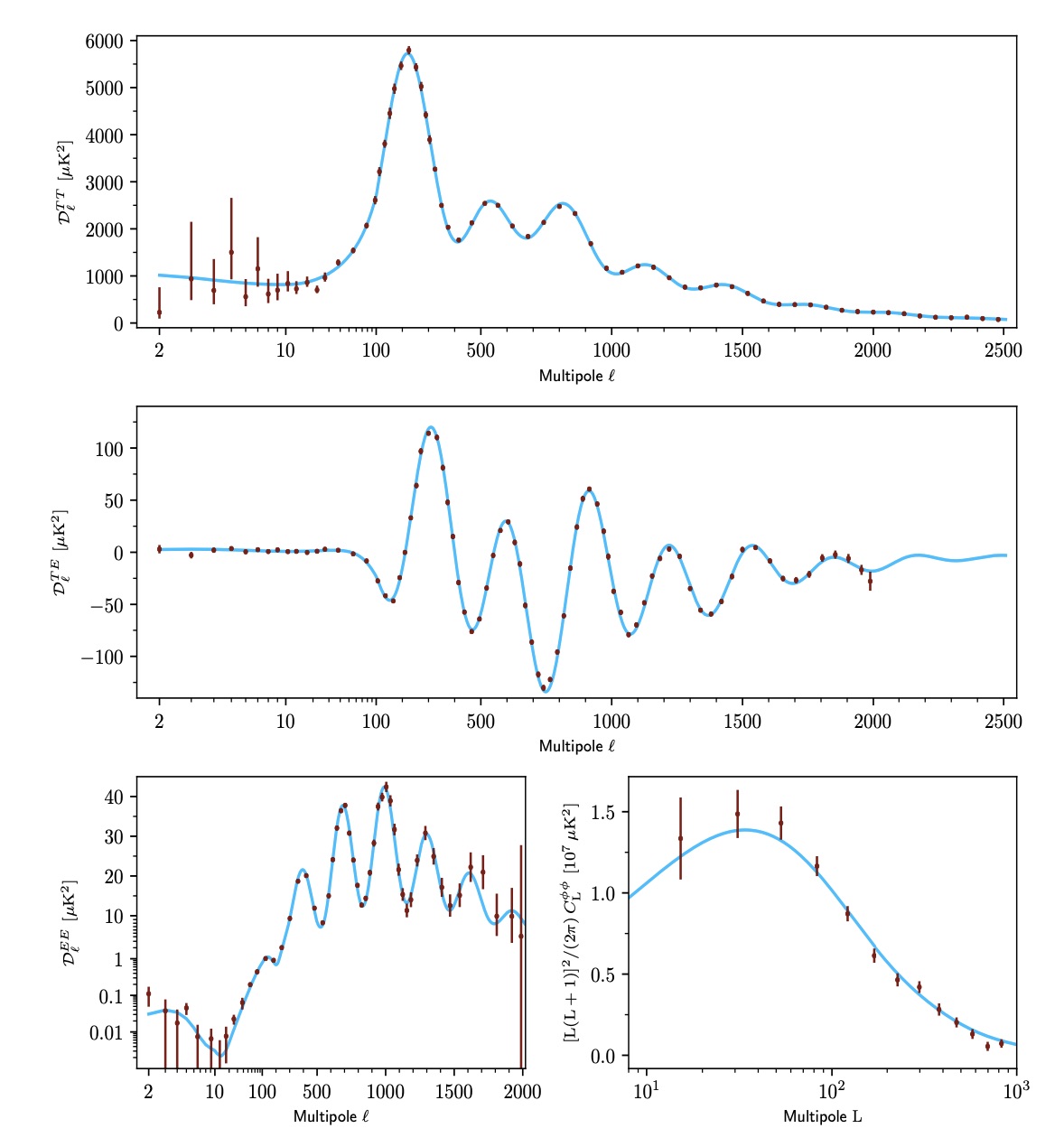}
\caption[CMB spectrum]{\small $\mathcal{D}$ is our $\Delta_T^2$ defined in \eq{AngularPower}. In the upper panel, the square root of the temperature power spectrum \eq{AngularPower}, recently measured by Planck \cite{Planck:2018nkj}. The continuous line represents the predicted behaviour for our standard cosmological model. The main features are the five peaks and the damping tail. At low $\ell$, the cosmic variance \eq{CosmicVariance} becomes dominant. In the middle panel we have the TE spectrum, that, as we can see, has good constraining power. Whereas in the lower left-hand side panel we have the EE spectrum which is noise-contaminated. In the right-hand side panel, we have the lensing angular power spectrum. Figure taken from~\cite{Planck:2018nkj}.}
\label{fig:PlotAngular}
\end{figure}

If the fluctuations are Gaussian, the power spectrum will contain all the statistical information of the field because every other correlation function is expressed in terms of \eq{corre}.
\begin{tcolorbox}[mybox]
    If the random filed is gaussian, the other correlators up to fourth order are
    \begin{gather}
        \langle f(\textbf{k})\rangle =0,\\
        \langle f(\textbf{k}_1)f(\textbf{k}_2)f(\textbf{k}_3)\rangle=0,
        \end{gather}
        \begin{multline}
        \langle f(\textbf{k}_1)f(\textbf{k}_2)f(\textbf{k}_3)f(\textbf{k}_4)\rangle=(2\pi)^6\delta^3_{D}(\textbf{k}_1+\textbf{k}_2)\delta^3_{D}(\textbf{k}_3+\textbf{k}_4)\times\\
        P(k_1)P(k_3)+\,\text{2 permutations}\,.
    \end{multline}
    Therefore, for gaussian random fields, all the information are contained in the power spectrum $P$.
\end{tcolorbox}
As long as we consider approximately instantaneous recombination, the angular temperature fluctuation is simply a projection of the spatial temperature fluctuation
\begin{equation}
\Theta(\hat{n})=\int{dD\Theta(\boldsymbol{x})\delta(D-D_{LS})},
    \label{eq:proj}
\end{equation}
where $D=\int{dz/H}$ is the comoving distance, and $D_{LS}$ denotes the distance from the last scattering surface. In a flat geometry, the spatial temperature fluctuations can be described by Fourier modes (as in \eq{Fourier}) and consequently (see \eq{xiR} and \eq{dimensionlessPS})
\begin{equation}
    \langle \Theta(\boldsymbol{x})\Theta(\boldsymbol{x})\rangle\equiv \int{d\ln{k}\Delta_T^2(k)}.
\end{equation}
If we put \eq{proj} in \eq{Bright} and we bear in mind that the exponential can be expanded as in \eq{pi}, we are able to relate the angular and spatial spectrum:
\begin{equation}
    \langle \Theta^*_{\ell m}\Theta_{\ell'm'}\rangle=\delta_{\ell \ell'}\delta_{mm'}4\pi\int{d\ln{k}j_l^2(kD_{LS})\Delta^2_T(k)=\delta_{\ell \ell'}\delta_{mm'}C_\ell}.
    \label{eq:Hello}
\end{equation}
Since observations suggest a nearly scale-invariant spatial power spectrum, we can take out $\Delta^2_T$. Furthermore, as the Bessel functions are characterised by peaks at $\ell $, we can substitute $kD_{LS}\approx \ell $, and Eq. (\ref{bessel}) yields 
\begin{equation}
    \Delta^2_T\equiv\frac{\ell(\ell+1)}{2\pi}C_\ell T_0^2\,.
    \label{eq:AngularPower}
\end{equation}
The CMB power spectrum $\Delta^2_T$,  with the characteristic acoustic, is plotted in \fig{PlotAngular}. 

As our Universe is just one statistical realisation of all cosmological random fields, we have only one CMB sky to observe. Thus, we expect large statistical fluctuations. From the assumption of Gaussian temperature fluctuations, we get that also $\Theta_{\ell m}$ are Gaussian variables (see e.g. Appendix C \cite{Gorbunov:2011zzc}); in the perfect case according to which our detection covers all the sky, we are able to determine $2\ell+1$ statistically independent $\Theta_{\ell m}$ for a given $\ell $. 
\begin{tcolorbox}[mybox]
    We now shall state three crucial assumptions in cosmology:
\begin{itemize}
    \item[\textbf{1.}] The part of the Universe that we can observe is a fair sample of the whole.
\end{itemize}
As we still have the problem that the Universe is just one realisation of the ensemble, we need for a second hypothesis:
\begin{itemize}
    \item[\textbf{2.}] \textit{The hypothesis of ergodicity.} Averaging over many realisations is equivalent to averaging over a large volume. 
\end{itemize}
We are left to understand how large this volume must be. This brings us to the third assumption:
\begin{itemize}
    \item[\textbf{3.}] \textit{The fair sample hypothesis.} Samples from well separated part of the Universe are independent realisations of the same physical process. Furthermore, in the observable part of the Universe, there are enough independent samples to be representative of the statistical ensemble.
\end{itemize}
These assumptions leave us free to take in a given volume the cosmological fields we are interested in, consider them as a realisation of the statistical process and eventually averaging over a large volume.
\end{tcolorbox}
This limitation brings into the conversation an additional quantity, the \textit{cosmic variance}:
\begin{equation}
    \sigma_\ell=\sqrt{\frac{\langle(C_\ell^{obs}-C_\ell)^2\rangle}{C_\ell^2}}.
    \label{eq:cosmicvariance}
\end{equation}
$C_\ell^{obs}$ is the variable we obtain when averaging over the $2\ell+1$ measured $\Theta_{\ell m}$ and is equal to $C_\ell^{obs}=(2\ell+1)^{-1}\sum_m{\lvert\Theta_{\ell m}\rvert^2}$. In other words, $C_\ell^{obs}$ is the \textit{estimator} for $C_\ell$\footnote{Holds the unbiased condition $\langle\hat{C}_\ell\rangle=C_\ell$ but it has non-zero variance} whereas $C_\ell$ is the statistical expectation value and is given by \eq{expectvalue}. The numerator of \eq{cosmicvariance} yields:
\begin{equation}
    \langle(C_\ell^{obs}-C_\ell)^2\rangle=\frac{1}{(2\ell+1)}\sum_{mm'}{(\langle\lvert\Theta_{\ell m}\rvert^2\lvert\Theta_{\ell m'}\rvert^2\rangle-\langle\lvert\Theta_{\ell m}\rvert^2\rangle\langle\lvert\Theta_{\ell m'}\rvert^2\rangle)};
    \label{p}
\end{equation}
the second term on the right-hand side is simply $C_\ell^2$. In the first term we apply Wick's theorem\footnote{The theorem states that the $2n$-point correlation function of a set of Gaussian variables is given by the sum of all the possible $2$-point correlation functions that can be formed from it.} that allows us to write 
\begin{equation}
\langle\lvert\Theta_{\ell m}\rvert^2\lvert\Theta_{\ell m'}\rvert^2\rangle=C_\ell^2+\delta_{mm'}C_\ell^2+\delta_{m-m'}C_\ell^2.
\end{equation}
Summation over $m$ and $m'$ gives a factor of $2$ and, therefore
\begin{equation}
   \boxed{ \sigma_l=\sqrt{\frac{2}{2\ell+1}}}\,.
   \label{eq:CosmicVariance}
\end{equation}
This is the absolutely minimal error that can be achieved from one sky \cite{Durrer:2020fza}.

\subsubsection{Monopole and dipole}
Since the energy density of the CMB photons is $\rho_\gamma\propto T^4$, from \eq{rhorad}, we can calculate the energy contrast:
\begin{equation}
    \delta_\gamma=\frac{\delta \rho_\gamma}{\rho_\gamma}=4\frac{\delta T}{T}=4\Theta_{00}.
    \label{eq:rhoT}
\end{equation}
It implies that the \textit{monopole} ($\ell =0$) is proportional to the energy density contrast\footnote{Remember that $Y_{00}$ is a constant}. We should ignore this term in \eq{Bright} because we can only measure the CMB radiation from one point, and we cannot distinguish whether we detect the average value or not. The monopole can always be removed if we redefine the background temperature. 

Now, let us call $n_\gamma(\boldsymbol{p})$ the density of CMB photons in phase space and $N_\gamma(\boldsymbol{p})d\boldsymbol{p}$ the number of photons of each polarization per unit spatial volume in a momentum-space volume centred at $\boldsymbol p$. From \eq{BBnot}, we find
\begin{equation}
   n_\gamma(\boldsymbol{p})=\frac{1}{8\pi^3}\frac{1}{e^{\frac{\lvert \boldsymbol{p}\rvert}{T}}-1},
\end{equation}
with $\lvert \boldsymbol{p}\rvert=2\pi\nu$. This is the density detected by an observer at rest in the radiation background. From Liouville's theorem and the Boltzmann equation, both the phase volume and the number of photons are conserved. Hence, $n_\gamma$ is a scalar, \ie it is invariant under Lorentz transformation
\begin{equation}
    n'_\gamma(\boldsymbol{p}')=n_\gamma(\boldsymbol{p}).
\end{equation}
Let us now take into account the velocity, $\beta$, of our reference frame (Earth). With $\boldsymbol{p'}$ the photon momentum measured by the terrestrial observer, we get
\begin{equation}
    \lvert \boldsymbol{p}\rvert =\gamma(1+\beta\cos{\theta})\lvert \boldsymbol{p}'\rvert,
\end{equation}
with $\gamma\equiv (1-\beta^2)^{-1/2}$ and $\theta$ the angle between the photon and the Earth's velocity. Thus, we arrive at the result 
\begin{equation}
     n'_\gamma(\boldsymbol{p}')=\frac{1}{8\pi^3}\frac{1}{e^{\frac{\lvert \boldsymbol{p}'\rvert}{T'}}-1},
\end{equation}
where the temperature $T'$ has become a function of $\theta$
\begin{equation}
    T'=\frac{T}{\gamma(1+\beta\cos{\theta})}.
    \label{eq:Tprime}
\end{equation}
As we can see, the temperature changes according to the direction of our observations: photons that are coming from the direction towards which the Earth is moving have the maximal value, whilst photons that are moving in the same direction have the least apparent temperature. As $\beta$ is small (of order $10^{-3}$) and $\gamma$ is of order $1$, we can expand \eq{Tprime} in powers of $\beta$. The temperature shift reads
\begin{equation}
    \delta T\equiv T'-T=T\left[-\frac{\beta^2}{6}-\beta P_1(\cos{\theta})+\frac{2\beta^2}{3}P_2(\cos{\theta})+...\right],
    \label{eq:dipole}
\end{equation}
where we have used
\begin{equation}
    P_0(x)=1,\quad P_1(x)=x\quad\text{and}\quad P_2(x)=\frac12(3x^2-1).
\end{equation}
Due to the smallness of $\beta$, in \eq{dipole} the temperature shift is primarily a dipole. Therefore, this Doppler effect will be neglected.
\begin{tcolorbox}[mybox]
If we have a single photon with a given polarization, its state is completely characterised by its position $x^i(\tau)$, and 3-momentum $p^i(\tau)$ in fact, the 4-momentum satisfies $g^{\mu\nu}p_\mu p_\nu=0$, thus, the energy $p_0$ is defined once known the spatial components of $p_\mu$. The phase volume element is defined simply as
\begin{equation}
    d\boldsymbol{x}d\boldsymbol{p}.
\end{equation}
Under coordinate transformations
\begin{equation}
d\widetilde{\boldsymbol{x}}d\widetilde{\boldsymbol{p}}=Jd\boldsymbol{x}d\boldsymbol{p},
\end{equation}
it is possible to show that $J=1$, namely, the phase volume is an invariant. 
An important consequence is given by the \textit{Louville's theorem} which states that the phase volume of the Hamiltonian system is conserved along the trajectory of the particle. If we have an ensemble of non-interacting identical particles, $dN$ defines the number of particles per phase volume. We introduce the distribution function $f$:
\begin{equation}
    dN=f(x^i,p_j,t)d\boldsymbol{x}d\boldsymbol{p}.
\end{equation}
 From Liouville's theorem, the phase volume is conserved and $f$ is a scalar. The Boltzmann equation (see \eq{Boltzmann})is
 \begin{equation}
     \frac{df(x^i(\tau),p_i(\tau),\tau)}{d\tau}\equiv \frac{\partial f}{\partial \tau}+\frac{dx^i}{d\tau}\frac{\partial f}{\partial x^i}+\frac{dp_i}{d\tau}\frac{\partial f}{\partial p_i}=C[f].
     \label{Boltzmann}
 \end{equation}
For a collisionless fluid, $C[f]=0$ and the particle number is conserved within the phase volume. The derivative $dx^i/d\tau$ and $dp_i/d\tau$ are calculated along the geodesics.

\end{tcolorbox}

\section{Temperature anisotropies}
\label{sec:TA}
To characterize the temperature anisotropies, we can ignore at first the gravitational contribution and simply consider the Thomson scattering between photons and electrons\footnote{The mean free path near recombination is $\lambda_C\sim 2.5Mpc$ which is two orders of magnitude smaller than the horizon at recombination} during epochs before recombination; it implies a conservation of photon number density in an expanding Universe. The continuity equation combined with the Euler equation
becomes an harmonic equation whose solution is a standing wave for each $k$
\begin{equation}
\Theta(\eta)=\Theta(0)\cos{(kr_s)}+\frac1{kc_s}\dot{\Theta}(0)\sin{(kr_s)}
\label{eq:acusticpeak}
\end{equation}
where we have used the fact that $n_\gamma\propto a^{-3}$. In adiabatic condition we have $c^2_s=1/3$ and a vanishing initial velocity perturbation\footnote{the $\sin$ term vanishes}. $r_s$ is the \textit{sound horizon} defined as
\begin{equation}
    r_s\equiv\int{c_sd\eta}.
    \label{eq:soundhorizon}
\end{equation}
These oscillations in real space represent the heating and cooling of the photon-baryon fluid that is compressed and rarefied by acoustic waves which continue to oscillate until recombination. As the photons’ decoupling is a fast process, the pattern of these waves is impressed in the last scattering surface and is seen by the observer as the \textit{acoustic peaks} in the temperature anisotropy.  Both maxima and minima contributions are peaks in the spectrum as the power spectrum is quadratic in the fluctuations. The fundamental scale of the extrema is $k_f=\pi/r_{s\star}$ with the following peaks at $k_n=nk_f$. The angular scale of the acoustic peaks is defined as $\theta_{s\star}\equiv r_{s\star}/D_{A\star}$ where $D_{A\star}$ is the comoving angular diameter distance to the last-scattering surface which, for a flat universe is simply $\eta_0-\eta_\star$. From \eq{Hello} we know that $\ell_n\sim k_nD_A$ due to the bessel function behavior. It is important to note that modes with wavelengths bigger than $r_{s\star}$ do not propagate. As the sound horizon at decoupling is roughly $145\mpc$, sound waves affect the CMB for $\ell>100$. 
If we take into account the perturbations in the bulk velocity of electrons at recombination, we see that photons acquire a Doppler shift that leads to a contribution to temperature perturbation as (see Appendix B~\cite{Baumann:2022mni})
\begin{equation}
    \left.\frac{\delta T}{T}\right\rvert_{\rm Doppler}=-\hat{\textbf{n}}\vdot \textbf{v}_\gamma=\Theta(0)\sin({kr_s})\,
\end{equation}
where we have used \eq{acusticpeak} in adiabatic condition. Despite not having a dependence on the wavevector\footnote{if we add in quadrature the two contributions of the anisotropies we are simply left with $\Theta(0)$ so it is a scale invariant contribution} it carries an angular dependence. In fact, when the observer is looking perpendicular to $\textbf{v}\lvert\lvert\,\textbf{k}$ whereas, in that direction we have the most contribution~\cite{Hu:1997hp}. This term, the \textit{Doppler term}, contributes the most on small scales as on superhorizon scales the baryon velocity vanishes. Its effect is to reduces the contrast between the peaks and the throughs in the spectrum, being out of phase with the acustic oscillations.

\subsubsection{Gravitational forces}  
So far, we have defined the charateristic acustic oscillations of the photon-baryon fluid in the thight-coupling approximation and adiabatic conditions. The next step is to take into account the gravitational forces and their contribution to the CMB signatures. If we take \eq{continuityrel} and \eq{eulerrel}, we can use \eq{rhoT} and the fact that $p_\gamma=1/3\rho_\gamma$ to write in Fourier space
\begin{gather}
    \Theta'=-\frac13kv_\gamma+\Phi',\label{eq:continuitytheta}\\
    v_\gamma'=k(\Theta+\Psi)\label{eq:eulertheta}
\end{gather}
where we neglected the anisotropic stress (\ie from \eq{einstein0i} $\Phi\approx\Psi$). We then take the derivative of \eq{continuitytheta},  and  we end up with 
\begin{equation}
    \Theta''+\Psi''+c_s^2k^2(\Theta+\Psi)=0\,.
    \label{eq:Thetadouble}
\end{equation}
whose solution is 
\begin{equation}
    \frac{\delta T}{T}(\eta)=\left.\frac{\delta T}{T}\right\rvert_{\rm ini}\cos{(kr_s)}
\end{equation}
with 
\begin{equation}
    \frac{\delta T}{T}(\eta)=\Theta+\Psi\,,
    \label{eq:Sachs}
\end{equation}
the \textit{Sachs Wolfe term}~\cite{Sachs:1967er,White:1997vi}. It represents oscilations around a displaced minimum. The gravitational fall compresses the fluid until resistance from photons pressure reverses the motion. It correlates the the photon density fluctuations with the induced temperature perturbations arising from gravitational redshift of the photons. From \eq{Sachs} we can see that the CMB photons have \textit{intrinsic} temperature perturbations $\Theta$, originated from earlier epochs than recombination (acoustic oscillations), and temperature fluctuations which arise from the energy lost when the photon climbed out of a potential well. In other words, $\delta T/T$ can be defined as the \textit{effective temperature}. In our Newtonian gauge, with adiabatic initial condition, we know that $\Psi$ is constant in a MD universe and \eq{initialPhi} holds. Hence, 
\begin{equation}
    \Theta+\Psi=\frac13\Psi=\frac15\mathcal{R}_i\,.
    \label{eq:15R}
\end{equation}
Negative $\Psi$ corresponds to an overdensity ($\delta_\gamma<0$) and thus a loss of energy from the climbing of the potential wells: overdense regions are cold spots in the effective temperature ($\delta T/T<0$). From \eq{mathcalRw} and \eq{mathcalRZw} we can see that $\mathcal{R}=\frac35\Psi$. The solution of \eq{Thetadouble} with the initial displacement therefore is
\begin{equation}
    [\Theta+\Psi](\eta)=\frac13\Psi\cos{(ks)}.
    \label{eq:SolutionPhi}
\end{equation}
If we take \eq{Hello} and apply to \eq{Sachs} using \eq{15R} we find
\begin{equation}
    C^{SW}_\ell=\frac{4\pi}{25}\int d\ln{\ell }\Delta_\mathcal{R}^2(k)j^2_\ell(kD_{LS}).
\end{equation}
If we have a power spectrum in a power law form $\Delta^2_\mathcal{R}(k)=A_s(k/k_0)^{n_s-1}$, we find that $C^{SW}_\ell\propto A_s$ so the amplitude of the large-scale CMB spectrum is a direct measure of the amplitude of the primordial fluctuations. Additionally, for a scale invariant spectrum ($n_s=1$) we have 
\begin{equation}
    \frac{\ell(\ell+1)}{2\pi}C^{SW}_\ell=\frac{A_s}{25}
\end{equation}
which implies a non dependence of multiploles for the scale-invariant power spectrum. At low multipoles $\ell<100$ we have the \textit{Sachs Wolfe Plateau} (see \fig{PlotAngular}). This comes from the fact that we have at recombination superhorizon scales, so no acoustic oscillations are present and $\Psi$ is constant.

The dependence on space of $\Psi$ gives us the displacements of the potential wells and therefore the Sachs-Wolf term. On the other hand, however, we need to take into account the fact that the CMB photons that we observe contains information also about the journey they undergo to reach us. In fact, if the metric potentials along the line-of-sight change with time, the net effect of photons moving in an overdense region would not cancel out. This is the so called \textit{Integrated Sachs-Wolfe}~\cite{Planck:2015fcm,Dupe:2010zs,Manzotti:2014kta,} (ISW) effect. The contribution is given by 
\begin{equation}
    \int d\eta(\Phi'+\Psi')
\end{equation} 
where the integral is computed from the time of decoupling until today\footnote{the contribution is obtained by computing the geodesic equation for photons with a perturbed metric in Newtonian gauge}. If we neglect the anisotropic stress, we end up with a factor $2$. 

\paragraph{\textit{Early ISW}} We have seen that $\Psi$ is constant during matter domination. However, the transition epoch from radiation to matter dominance is not instantaneous and, for a short period after decoupling, the gravitational potential still slightly changes in (conformal) time. In this case we have the \textit{early} ISW effect, which increase height of the first peaks, with the maximum contribution to the first one. For modes already inside the horizon, CMB photons are tightly coupled to electrons therefore, in the integral there is the additional contribution of the probability of non scattering (see \eq{probnoscatt}) which is small enough that the ISW effect is negligible. For this reason we have a boost on the first peak: $k\ll k_{LS}$ enter during matter domination whereas $k\gg k_{LS}$ have negligible contributions. The effect on the angular power spectrum is suppressed by the factor
\begin{equation}
    \frac{\rho^2_r(\eta_{LS})}{\rho^2_m(\eta_{LS})}=\(\frac{1+z_{LS}}{1+z_{eq}}\)^2
\end{equation}
Therefore, an increase of the amount of radiation during this epoch will shift the matter-radiation equality ($z_{eq}$) and result in a larger amplitude of the early ISW effect. 

\paragraph{\textit{Late ISW}} Similarly, modes that where outside the horizon and enter only at late time, after the matter-dark energy equality, experience again a non constant $\Psi$, leading to the \textit{late} ISW effect on low multipole (the rise in the SW plateau that can be seen in \fig{PlotAngular}) being the only contribution at scales of our present horizon $\ell \sim 1$ as the one for shorter wavelength is averaged away. If we go to second order, we have the \textit{Rees-Sciama effect}~\cite{Rees:1968zza}. It is due to the nonlinear collapse of initial perturbations and the ensuing structure formation, which can cause the gravitational potential to change in time even in a flat, matter dominated Universe, modifying the energy of CMB photons as they cross nonlinear structures. This effect dominates at small scales where the nonlinear nature of structure formation becomes relevant. Increasing the density of dark energy shift the dark matter-dark energy equality at earlier times increasing the power on low $\ell$.  

One last remark is concerning the decaying of $\Psi$ when the mode enter the horizon during radiation era. At horizon crossing we have the compression of the fluid as well. This entails that during the counteraction of the pressure for the fluid, the potential has decayed and does not fight against it. We can no longer consider the photon-baryon fluid to be oscillating in a fixed gravitational potential well and the amplitude of oscillations increases. This is the \textit{radiation driving} and affects only modes that enter the sound horizon during the radiation dominated epoch. We can therefore increase the peaks heights by reducing the matter density and shifting the time of matter-radiation equality. 

\subsubsection{Baryon influence}
At this point, an additional piece that we need to better understand the CMB pattern is baryons dynamic which, until now, has been neglected. Since they are tightly coupled with photons, the baryons share the same bulk velocity $v_b\simeq v_\gamma$ and can be considered as a single fluid. Only the momentum density is conserved
\begin{equation}
\textbf{q}=\frac43(1+R)\bar{\rho}_\gamma\textbf{v}_\gamma
\end{equation}
with 
\begin{equation}
    R=\frac{(\rho_b+p_b)v_b}{(\rho_\gamma+p_\gamma)v_\gamma}=\frac34 \frac{\rho_b}{\rho_\gamma}=0.6\(\frac{\Omega_bh^2}{0.02}\)\left(\frac{a}{10^{-3}}\right).
    \label{eq:Bary}
\end{equation}
We can see that our approximation is valid up to the recombination epoch where $R$ becomes of order unity, and subsequently the baryonic effect starts to appear in the oscillations just as they are frozen-in. The continuity equation remains unaltered but the Euler equation now changes so that the solution \eq{SolutionPhi} now is~\cite{Hu:1995en,Hu:2002aa}
\begin{equation}
    [\Theta+(1+R)\Psi](\eta)=\frac13(1+3R)\Psi\cos{(kr_s)}.
    \label{eq:Ferr}
\end{equation}
As we can notice, in \eq{Ferr} we have increased the amplitude of oscillation by a factor $1+3R$ as well as shifted the equilibrium point of oscillation:
\begin{equation}
    [\Theta+\Psi](\eta)=\frac13(1+3R)\Psi(0)\cos{(kr_s)}-R\Psi(0).
    \label{eq:zerofhi}
\end{equation}
The zero point shift in \eq{zerofhi} breaks the symmetry of the oscillations. Specifically, the baryon load enhance the odd peaks. In fact, the amplitude of the even peaks is $\frac13\Psi(1+3R-3R)$ hence remains unaltered with respect to the non-baryon case, while the odd peaks amplitude increases as $(1+6R)$. In other words, the presence of baryons drags the fluid farther inside the gravitational field inducing a greater compression. Baryons contribute to the effective mass of the fluid but not to the pressure. All peaks from compression are enhanced and their position is shifted as well because the sound horizon $r_{s\star}$ is the integral of the sound velocity \eq{soundhorizon}; with the baryon load, the sound speed decreases 
\begin{equation}
    c^2_s=\frac{1}{3(1+R)}
    \label{eq:csRcs}
\end{equation}
which, in turns, brings to a change in the angular scale of the sound horizon $\theta_s$, that determines the peak locations. This effect is degenerate with other parameters that changes the distance to last-scattering ($D_{LS}$) like the dark matter density or dark energy density. The latter parameter has a geometic degeneracy with other parameters such as curvature\footnote{The dark matter density also changes the peaks height therefore it is not degenerate}.

\subsubsection{Damping}

An additional feature of the CMB power spectrum is given when we consider imperfections to the photon-baryon fluid. That is, we drop the assumption of tight-coupling approximation. If we do so, we need to consider also the continuity equation in \eq{continuitytheta} for the baryons 
\begin{equation}
    \delta_b'=-kv_b+3\Phi'
\end{equation}
and modified Euler equations. We want to take into account the exchanges of momentum by the scattering via the \textit{drag term} $\Gamma(v_\gamma-v_b)$, with $\Gamma$ the scattering rate and $v_\gamma-v_b$ the \textit{slip velocity} and we want to consider the anisotropic stress $\Sigma_\gamma$ coming from radiation viscosity. Eventually, we have
\begin{gather}
    v'_\gamma+\frac14k\delta_\gamma-\frac23k\Sigma_\gamma+k\Phi=-\Gamma(v_\gamma-v_b),\\
    v'_b+\mathcal{H}\delta_b+k\Phi=\frac{\Gamma}{R}(v_\gamma-v_b)
\end{gather}
where the right-hand side of both equations are fixed by the fact that the combined momentum must be conserved and $R$ was previously defined in \eq{Bary}. From the Boltzmann equation~\cite{Kaiser:1983}
\begin{equation}
    \Sigma_\gamma\approx2A_vv_\gamma\frac{k}{\Gamma}
\end{equation}
with $A_v=16/15$.
Eventually, the oscillation equation acquires a dissipation term
\begin{equation}
    \frac{k^2c_s^2}{\Gamma}[A_v+A_h]\Theta'
\end{equation}
where $A_h=R^2/(1+R)$ is the heat conduction term: relative motion between the photons and baryons also damp oscillations. The solution of the damped oscillator equation in the adiabatic approximation gives an exponential damping of the oscillation amplitude $e^{-(k/k_D)^2}$. For small and large values of R, we find that $k_D^2\sim6\Gamma\eta^{-1}$. We can write the dissipation scale as~\cite{Hu:2004kn}
\begin{equation}
    \frac{\lambda_D}{\mpc}\approx 64.5\(\frac{\Omega_mh^2}{0.14}\)^{-0.278}\(\frac{\Omega_bh^2}{0.024}\)
\end{equation}
This implies tht CMB fluctuations are expected to be damped for $\ell>1400$. This is valid also for the \textit{Landau damping} due to the finite width of the visibility function~\cite{Baumann:2022mni}. It comes automatically that a change on the baryon density and total matter density changes the damping as they regulate the photon free path and the moment of recombination respectively.

\subsubsection{Secondaries anisotropy}
We have already encountered a secondary isotropy when we introduced the gravitational forces in the picture. In fact, the time dependence of the gravitational potential gives an integrated contribution (the ISW) which is a secondary anisotropy. Here we present a few more. 

\paragraph{\textit{Weak lensing}}

The \textit{weak lensing}~\cite{Okamoto:2003zw,Durrer:2020fza,Lewis:2006fu} indicates the small deflection angles of order of arcminutes\footnote{On the other hand the \textit{strong} lensing happens for example when lines of sight cross, giving caustics, multiple images and points of infinite magnification and shear} which is a secondary signal and represents the deflection of CMB photons due to galaxies and clusters of galaxies. If we consider the CMB temperature in a point $\hat{n}$ in the sky $T(\hat{n})$, after a deflection by a small angle $\alpha$, we receive the temperature from the direction $\hat{n}'=\hat{n}+\alpha$. To lowest order we have $\delta T=\alpha\vdot\nabla_{\hat{n}}T(\hat{n})$. As the angular dependence of the temperature is of first-order, as well as the deflection angle, this effect is second-order. The distortion of light is typically of a few arcmins and the result is the smearing of the oscillations of the CMB angular power spectrum at small scales. However, it is difficult to resolve as lensing by transverse gradients does not change
the frequency of the photons and they maintain the same blackbody spectrum as the unlensed CMB, and multi-frequency observations cannot be used to separate the lensing signal.

\paragraph{\textit{Sunyaev-Zeldovich effect}} Clusters and galaxies not only deflect the CMB photons, but produce hot gas ($T\sim 10^8K$) that affect the energy of photons via Compton scattering. This is the so called \textit{Sunyaev-Zeldovich effect}~\cite{Rephaeli:1995sr,Carlstrom:2002na,Zeldovich:1969ff}. Electrons in the gas have high temperature so the photons scattered are more energetic. This effect distort the blackbody spectrum\footnote{It does not alter the number of photon, so it is just a transfer of a fraction of photons in the spectrum} but is not significant as the optical depth is really small so the majority of photons from the CMB never scatter at all in a given cluster. If we take into account the velocity of the cluster, we end up with the \textit{kinematic Sunyaev-Zeldovich effect} but is one order of magnitude smaller than the thermal one~\cite{1972CoASP...4..173S}. 

\paragraph{\textit{Reionization damping}} At late times, the Universe reionized again due to the ultraviolet light from first stars, and CMB photons have a small probability to be scattered off free electrons, reaching the observer from a different direction with respect to the initial one giving the effect of \textit{reionization damping} ( for example see ~\cite{Knox:2003ch,Qin:2020xrg,Giare:2023ejv}). Its signature on the CMB angular spectrum is an overall suppression $C_\ell\longleftarrow C_\ell e^{-\tau}$ where $\tau$ is the optical depth defined as 
\begin{equation}
    \tau\equiv\int_{\eta_{\rm reio}}^{\eta_{today}}d\widetilde{\eta}n_e\sigma_Ta
\end{equation}
This effect occurs only on scales smaller than the horizon at recombination ($\ell>10$) and is degenerate with the amplitude $A_s$ of the initial fluctuations. Consequently, with temperature anisotropies we can only constrain $A_se^{-\tau}$. On the other hand, the reionization signal dominates the position and the height of the peaks in the polarization spectra (see \sect{CapitoloPoloDOPO}) at multipoles $\ell \lesssim 10$. Hence, it is worth noting that the constraints on $\tau$ come basically from polarization and not from anisotropies (see a related discussion on possible systematics in \sect{PolJWST}).

\section{Polarization}
\label{sec:CapitoloPoloDOPO}
The CMB is polarized. Different sources of temperature fluctuations, \ie scalar, vector and tensor, gives different patterns in the polarization. We start by introducing briefly some useful relations to describe the polarization. Afterwards, we will see how the CMB is polarized, and which are the patterns given by different sources. 

\subsection{Stokes parameters}
Consider a generic electromagnetic wave arriving at the observer's position along the  direction $\hat{z}$
\begin{equation}
    \boldsymbol{E}=E_x\hat{x}+E_y\hat{y}.
\end{equation}
The \textit{Stokes parameters} are defined as
\begin{gather}
    I=\lvert E_x\rvert^2+\lvert E_y\rvert^2,\quad Q=\lvert E_x\rvert^2-\lvert E_y\rvert^2,\\
    U=(E_x^*E_y+E_y^*E_x)=2\Re{E^*_xE_y}\quad\text{and}\quad V=2\Im{E^*_xE_y}.
\end{gather}
$I$ is the intensity of the electromagnetic wave, in our case it is proportional to the energy density of the CMB photons, thus, we have $\Theta=\delta I/I$. The parameters $Q$ and $U$ describe the linear polarization of the radiation. Specifically, $Q$ represents the difference between the intensity along the $x$ and $y$-axis, which means polarization along the $x$($y$)-axis have $Q>0$ ($Q<0$). $U$ measures the amount of linear polarization at $\ang{45}$ with respect $x$-axis. The Stokes parameter $V$ describes circular polarization. Introducing the Pauli matrices
\begin{equation*}
    \sigma^{(0)}=\begin{pmatrix}
    1&0\\
    0&1\\
    \end{pmatrix},\quad
    \sigma^{(1)}=\begin{pmatrix}
    0&1\\
    1&0\\
    \end{pmatrix},\quad
    \sigma^{(2)}=\begin{pmatrix}
    0&-i\\
    i&0\\
    \end{pmatrix},\quad\text{and}\quad
    \sigma^{(3)}=\begin{pmatrix}
    1&0\\
    0&-1\\
    \end{pmatrix},
\end{equation*}
we can proceed to define the \textit{polarization tensor}
\begin{equation}
    \boxed{P=\frac12[I\sigma^{(0)}+U\sigma^{(1)}+V\sigma^{(2)}+Q\sigma^{(3)}]=\frac12I\sigma^{(0)}+\mathcal{P}}\,,
\end{equation}
with $\mathcal{P}$ a symmetric traceless matrix. If we rotate the transverse plane by an angle $\alpha$ around $\hat{n}$, we obtain
\begin{equation}
    \begin{array}{c}
\hat{x}'=\cos{\alpha}\hat{x}+\sin{\alpha}\hat{y}\\
    \hat{y}'=\cos{\alpha}\\hat{y}-\sin{\alpha}\hat{x}
    \end{array}\rightarrow
    \begin{array}{c}
    E_x'=E_x\cos{\alpha}-E_y\sin{\alpha}\\
    E_y'=E_y\cos{\alpha}+E_x\sin{\alpha}
    \end{array},
\end{equation}
which implies
\begin{equation}
    I'=I,\quad V'=V,\quad Q'=Q\cos{2\alpha}-U\sin{2\alpha}\quad\text{and}\quad U'=U\cos{2\alpha}+Q\sin{2\alpha}.
    \label{eq:QUtrans}
\end{equation}
The parameters $I$ and $V$ are physical observables that do not depend on the coordinate system, whereas $Q$ and $U$ are not. Since circular polarization is not produced in the early Universe we can set $V\equiv 0$ from now on. We can define two invariant quantities under rotations from $\mathcal{P}$: $\partial_a\partial_b\mathcal{P}_{ab}$ and $\epsilon_{ac}\partial_b\partial_c\mathcal{P}_{ab}$ with $\epsilon_{ab}$ an antisymmetric tensor. Using Stokes parameter we get
\begin{gather}
\nabla^2E\equiv(\partial^2_x-\partial^2_y)Q+2\partial_x\partial_yU\\
\nabla^2B\equiv(\partial^2_x-\partial^2_y)U-2\partial_x\partial_yQ
\end{gather}
where we have introduced the E and B-modes. These transformations can also be written, in terms of $Q$ and $U$, as
\begin{equation}
    Q'\pm iU'=e^{\pm2 i\alpha}(Q\pm iU).
    \label{eq:Q+U}
\end{equation}
The quantity in (\ref{eq:Q+U}) transforms like 2-spin variables with a magnetic quantum number $\pm2$ under rotations around $\hat{n}$. It depends not only on the direction $\hat{n}$, but also on the orientation of $\hat{x}$. For instance, a rotation of $\ang{45}$ turns $U$ into $Q$ and $Q$ into $-U$.  It is convenient to expand it in spin-weighted spherical harmonics $_sY_{lm}$ \cite{Kamionkowski:1996ks,Zaldarriaga:1996xe}
\begin{equation}
Q(\hat{n})\pm iU(\hat{n})=\sum_{\ell ,m}a_{(\pm2,lm)}\, _{\pm2}Y_{\ell m}(\hat{n}),
\label{eq:QUSTOKES}
\end{equation}
with the spin-weighted harmonics given in terms of rotation matrices as (see Appendix A of \cite{Lyth:2004gb})
\begin{equation}
_sY_{\ell m}(\theta,\phi)=\left(\frac{2\ell+1}{4\pi}\right)^{\frac12}\mathcal{D}^\ell_{-s,m}(\phi,\theta,0).
\end{equation}
Such functions satisfy the relations:
\begin{gather}
    \sum_{\ell m}\,_sY^*_{\ell m}(\hat{n})_sY_{\ell m}(\hat{n}')=\delta(\phi-\phi')\delta(\cos{\alpha}-\cos{\alpha'}),\\
    Y_{\ell m}=\,_0Y_{\ell m}.
\end{gather}
It is possible to show that the terms $a_{\pm2,\ell m}$ transforms under rotations in the same way as $\Theta_{\ell m}$. We define the \textit{polarization multipoles}, $E_{\ell m}$ and $B_{\ell m}$, as
\begin{equation}
    a_{\pm2,\ell m}=E_{\ell m}\pm iB_{\ell m}.
\end{equation}
Under parity transformations we have $Q(\hat{n})\rightarrow Q(-\hat{n})$, while $U(\hat{n})\rightarrow -U(-\hat{n})$. It is possible to show~\cite{Weinberg:2008zzc} that the space inversion changes $a'_{\pm2,\ell m}=(-1)^\ell a^*_{\pm2,\ell -m}$. Consequently, $E_{\ell m}\rightarrow(-1)^\ell E_{\ell m}$ and $B_{\ell m}\rightarrow-(-1)^\ell B_{\ell m}$. The polarization multipoles are the coefficients for the E and B-modes
\begin{equation}
    E(\hat{n})=\sum_{\ell m}E_{\ell m}Y_{\ell m}(\hat{n}),\quad B(\hat{n})=\sum_{\ell ,m}B_{\ell m}Y_{\ell m}(\hat{n}),
\end{equation}
and use them instead of the spin-2 Stokes parameters in \eq{QUSTOKES}. The $E$-mode polarization field is characterised by its curl-free property whereas the $B$-mode is divergence-free.\footnote{That is why they are called E and B, reflecting the property of the electric and magnetic field: $\mathbf{\nabla}\times\boldsymbol{E}=0$ and $\mathbf{\nabla}\vdot\boldsymbol{B}=0$.} The symmetries of temperature and polarization anisotropies allow the following correlators between pairs of multipoles:
\begin{equation}
\boxed{\langle\Theta^*_{\ell m}E_{\ell 'm'}\rangle=C_\ell^{TE}\delta_{\ell\ell'}\delta_{mm'}}\,,\quad\boxed{\langle E^*_{\ell m}E_{\ell 'm'}\rangle=C_\ell^{EE}\delta_{\ell\ell'}\delta_{mm'}}
     \label{eq:EETE}
\end{equation}
and 
\begin{equation}
    \boxed{\langle B^*_{\ell m}B_{\ell 'm'}\rangle=C_\ell^{BB}\delta_{\ell\ell'}\delta_{mm'}}
    \label{eq:BB}
\end{equation}
together with \eq{corre}. The correlators $EB$ and $TB$ vanish for symmetry reasons. It has been proven \cite{Kamionkowski:1996ks,Zaldarriaga:1996xe} that scalar perturbations create only $E$ mode, vector perturbations create mainly $B$ modes, and tensor perturbations create both $E$ and $B$ modes. 

\subsection{Thomson scattering}
The CMB photons are polarized due to scattering. In fact Thomson scattering induces a linear polarization in the scattered radiation. The incoming radiation with polarization $\hat{\epsilon}$ shakes an electron in the same direction $\hat{\epsilon}$ (the direction of the electric field vector of the incoming radiation) and causes it to radiate with outgoing polarization parallel to that direction. Polarization direction must be transverse to the direction of propagation. In general, we have
\begin{equation}
    \frac{d\sigma_T}{d\Omega}\propto \lvert \hat{\epsilon}\vdot\hat{\epsilon}'\rvert^2.
\end{equation}
which underline that only the component of the incoming polarization that is orthogonal to the direction of the outgoing radiation is trasmitted. If the Universe were completely isotropic, the net outgoing polarization would be zero. We need a quadrupole temperature anisotropy to generate linear polarization from Thomson scattering. Let us assume that we have an incident ray in the $\hat{x}$ direction. 
\begin{figure}[h!]
\centering
    \includegraphics[width=0.8\textwidth]{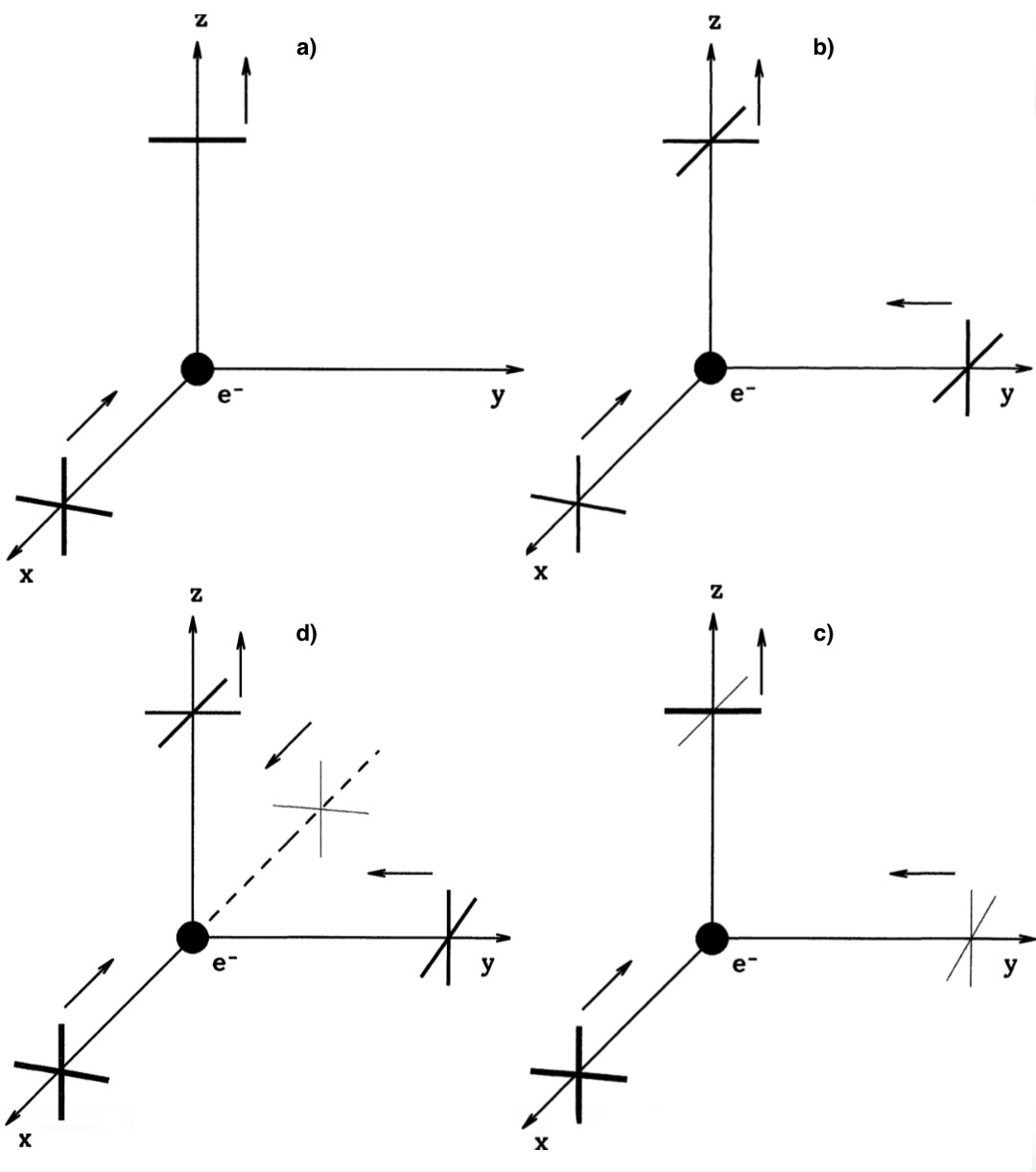}
  \caption[Thomson scattering]{\small In the isotropic case, we can see that one incoming ray (a) gives a linear polarization, but it is cancelled by another ray from a transverse direction (b). Now we consider anisotropies;  thick (thin) lines represent hotter (colder) regions with respect to the average temperature represented by a medium line. A simple dipole pattern (c) results in an unpolarized outgoing ray thus we need a quadrupole (d). The graphs are adopted from \cite{Dodelson:2003ft}.}
  \label{fig:Scatter}
\end{figure}
If it is unpolarized, it has equal intensity along the vector basis of the transverse plane. It scatters an electron and gets deflected into the $\hat{z}$ direction. That is, only the intensity along the $y$-axis is transmitted. The result is outgoing polarization in the $\hat{y}$ direction. If we assume another ray coming from the $\hat{y}$ direction with equal intensity, the outgoing ray is unpolarized. Namely, generalizing this example, we see that the isotropic radiation cannot produce polarization. So, we now assume a dipole pattern in the $x$-axis. Here, the outgoing intensity will be neither colder nor hotter than the average temperature. Therefore, if there is incoming radiation along $\hat{y}$, the result will be again unpolarized. Hence, we should consider the incoming radiation with a non-zero quadrupole: the intensity of the radiation coming from the $\hat{x}$-axis is larger than the ones coming from the $\hat{y}$-axis therefore the radiation will be polarized in the $\hat{y}$-direction. In \fig{Scatter}, our examples are shown. Since the Thomson scattering produces polarization, we shall focus on the epoch before electrons and photons have completely decoupled from each other. However, in this epoch, the quadrupole is almost negligible. Thus, we expect polarization from the standard decoupling epoch to be smaller than anisotropies. That said, the electron scattered oscillates with dipole moment $\boldsymbol{d}(t)=-e\boldsymbol{r}(t)$. The electric field of the outgoing radiation at position $\boldsymbol{r}=r\hat{n}'$ is 
\begin{equation}
    \boldsymbol{E}(\boldsymbol{r},t)=\frac{[\ddot{\boldsymbol{d}}(t-r)\times\hat{n}']\times\hat{n}'}{4\pi r}, 
\end{equation}
whose components are 
\begin{equation}
    E'_x=\frac{\alpha}{m_er}E_x\cos{\theta},\quad E'_y=\frac{\alpha}{m_er}E_y,
\end{equation}
with $\alpha$ the fine structure constant, and $\theta$ the angle formed by the scattered direction with respect to the incident direction. In terms of the Stokes parameters, we get
\begin{gather}
    I'=\frac{3\sigma_T}{8\pi r^2}[2(\cos^2{\theta}+1)I+(\cos^2{\theta}-1)Q_++(\cos^2{\theta}-1)Q_-],\\
    Q'_{\pm}=\frac{3\sigma_T}{8\pi r^2}[2(\cos^2{\theta}-1)I++(\cos^2{\theta}\pm1)^2Q_++(\cos^2{\theta}\mp1)^2Q_-],
\end{gather}
with $\sigma_T\equiv(8\pi/3)(\alpha/m_e)^2$. 

\subsubsection{Patterns of polarization}
Regarding the multipole decomposition of the radiation field into spherical harmonics $Y_{\ell m}(\theta,\phi)$, the five quadrupole moments are represented by $\ell =2$, $m=0,\pm1,\pm2$. The orthogonality of the spherical harmonics guarantees that no other moment can generate polarization from Thomson scattering. Let us now distinguish the tree sources: scalar, vector, and tensor perturbations~\cite{Hu:1997hv}.
\begin{figure}[h!]
\centering
\includegraphics[width=0.9\textwidth]{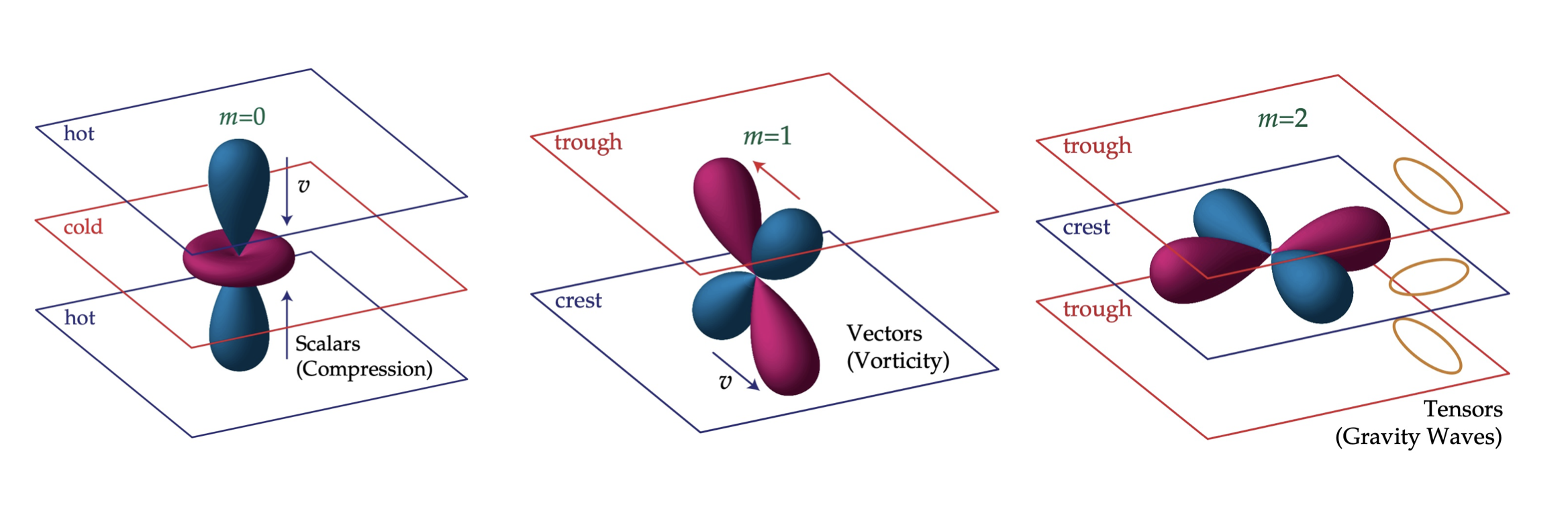}
\caption[Polarization pattern]{\small \textit{Left}: The flows from hot (blue) regions into cold (red) generate the azimuthally symmetric pattern $m=0$. \textit{Center}: Since in vector modes $\boldsymbol{v}\bot\boldsymbol{k}$, the Doppler effect generates a quadrupole pattern associated with $m=1$. \textit{Right}: GW stretches the space, changing a circle of test particles into an ellipse, we associate it with $m=2$ (see \fig{Ale}). The pictures are taken from \cite{Hu:1997hv}.}
\label{fig:m0m1m2}
\end{figure}

\paragraph{\textit{Scalar}} As we have already seen, gradients in the effective temperature create flows from hot to cold effective temperature (remember that overdense regions are cold because photons must climb out of the potential wells). Thus, we have a bulk flow. Considering the components of the temperature pattern seen by an observer located in a trough of a plane wave, we have $\boldsymbol{v}\parallel\boldsymbol{k}$, that implies the flow is irrotational. Photons located to the crests flow into the trough from the $\pm\hat{k}$ directions, while cold photons surround the observer in the plane. The pattern seen in a trough has a structure corresponding to $Y^0_2\propto 3\cos^2{\theta}-1$, see the left-hand side panel of \fig{m0m1m2}.
    
\paragraph{\textit{Vector}} For the sake of completeness, we will see also the vector modes, even if they decay straight away. The velocity is orthogonal with respect to $\boldsymbol{k}$, and it is the opposite in crests and troughs. The radiation field at these extrema has a dipole pattern due to the Doppler shift from bulk motion; the net effect is a quadrupole pattern similar to $Y^{\pm1}_{2}\propto\sin{\theta}\cos{\theta}e^{\pm i\phi}$, see the center panel of \fig{m0m1m2}.

\paragraph{\textit{Tensor}} Lastly, we have tensor fluctuations which, we shall remember, can be thought as gravitational waves. A gravitational wave perturbation represents a quadrupolar stretching of space in the plane perturbation (see \sect{Circe}). The pattern is similar to $Y^{\pm2}_{2}\propto \sin^2{\theta}e^{\pm i\phi}$ and is represented in the right-hand side panel in \fig{m0m1m2}. The quadrupole anisotropy imprinted by the gravitational waves contributes to the $BB$ power \eq{BB}.
\vspace{2mm}

We have obtained exactly the SVT decomposition described in \sect{fieldEquation}. More details can be found in~\cite{Kamionkowski:1996ks,Zaldarriaga:1996xe,Dodelson:2003ft,Hu:1997hp,Hu:1997hv,Kosowsky:1994cy,Melchiorri:1996tn} and, for details about the quadrupole pattern derived from gravitational waves, see \cite{Polnarev:1985}. 

%% file: Chapters/Inflation.tex
\chapter{Inflationary Theory}
\label{ch:InfloBinglo}
The Standard Scenario provides a reliable and tested account of the history of the Universe from at least as early as the time of nucleosynthesis (see~\sect{ThermalHistory}), until today. However, even though the model is remarkably successful, there is still an important set of inconsistencies which need to be addressed, all of which can be reduced to a common root: with only radiation and matter dominating in the past, only a high improbable fine-tuned conditions can give to the Universe what it needs to become as we observe it today.  Since we are not writing a novel, there is no point to make the readers wait with bated breaths for an answer\footnote{which could be already be found on the title of this chapter}. The way to solve these shortcomings is really simple: we just require
\begin{equation}
\boxed{
    \ddot{a}>0 
    }\,.
    \label{eq:acceleration}
\end{equation}
This condition defines what is called a stage of \textit{inflation}~\cite{Guth:1980zm,Guth:2004tw,Starobinsky:1980te}. More specifically, it is defined as \textit{a stage of accelerated expansion of the Universe when gravity acts as a repulsive force}. It was firstly introduced by Guth in 1980~\cite{Guth:1980zm} and soon enough many models were proposed to such an extent that inflation starts to denote a paradigm more than a simple theory. 

In what follows, we present the shortcomings of the Standard Bug Bang Theory and how the inflationary condition \eq{acceleration} provides a solution. We then present the key equations governing inflation, and explore the dynamics of the inflaton field, including the slow-roll approximation. In later sections, we study how inflation plants the primordial seeds of perturbations and outline the phenomenology used to constrain the inflationary epoch with current observational data.

\section{Where the Big Bang Theory comes short}
\label{androi}
The Big Bang theory that was outlined in the previous chapters is incomplete unless we call into action an extremely fine-tuned and anthropocentric initial conditions. Nevertheless, with the addition of another ingredient, inflation, we are able to naturally explain many of the shortcomings of the theory. 

\paragraph{\textit{Flatness problem}} The first key feature which needs any sort of hand-imposed condition if we are leaving the theory as it is, is the geometry of our Universe. From \eq{curvatureparameter}, we can notice that both in a MD and RD Universe, the curvature contribution to the total energy budget increases with the scale factor. In turns, if we go backwards in time, the quantity $\lvert\Omega(t)-1\rvert$ decreases, where $\Omega$ is defined in \eq{Omega}. Thus, a nearly flat universe today, implies an extreme fine-tuning of $\Omega$ in the early Universe, which must be unbelievably close to $1$. Let us change viewpoint, and start from an early epoch. The comoving Hubble radius, $\chi_\Hl(\eta)=(aH)^{-1}$ increases and $\lvert \Omega-1\rvert$ diverges: $\Omega=1$ is an unstable fixed point. Any small fluctuations in primordial times should have driven $\Omega$ exponentially far from unity. Concretely, since today measurements set the tight constraint  $\lvert\Omega_{k,0}\rvert<0.005$, at early epochs the constraint becomes more stringent. For example, at BBN we find that the deviation from flatness is $\lvert \Omega(t)-1\rvert_{BBN} \approx \mathcal{O}(10^{-16})$. To better understand why this issue is related to a fine-tuned problem, we take the definition of critical density \eq{densityparameter}, and write
\begin{equation}
    \lvert \Omega-1\rvert=\left\lvert \frac{\rho-\rho_c}{\rho_c}\right\rvert.
\end{equation}
This result implies that, if the density of the Universe had been initially greater than $\rho_c$ by a small amount, e.g. $10^{-55}\rho_c$, the Universe would have collapsed long ago. On the other hand, if the initial value had been smaller by the same amount, the present energy density would have been so low that life could not have existed. Only an extremely fine-tuning to flatness could explain the present value. This is the reason why this problem is often dubbed as \textit{flatness problem}.

\begin{tcolorbox}[mybox]
We can redirect the problem of flatness to an entropy problem. The second law of thermodynamics (\ie $dS \geq 0$) holds for the entire Universe \cite{Penrose:Road}: the entropy of the Universe does not decrease with time. The present horizon contains a total entropy of 
\begin{equation}
S_U=V_Us=\frac{4}{3}\pi H_0^{-3}s\simeq 10^{90}.
 \label{eq:Nowentropy}
 \end{equation}
Under the hypothesis of adiabaticity, we have that 
\begin{equation}
    \lvert \Omega-1\rvert=\frac{\M}{S_U^{\frac{2}{3}}T_{Pl}^2}.
\end{equation}
 With the value we have found in \eq{Nowentropy}, we have at Planck epoch
\begin{equation}
    |\Omega-1|_{\rm Planck}\sim S_U^{-\frac{2}{3}}\approx 10^{-60}.
\end{equation}
To understand whether our hypothesis of adiabatic evolution holds or not, we ought to go back to the recombination epoch. Here, the entropy of the photons emitted from the black-body radiation of the CMB is $S_{CMB}\sim\sim 10^{90}$. The entropy of the photons in the Universe remains constant as the Universe expands because it is proportional to the number of photons. The number of photons in the volume remains constant and so does the entropy \cite{Lineweaver:entropy,Lineweaver:entropy2}. Thus, far back, at $t_{CMB}$, our hypothesis works. Even though the adiabatic expansion does not conflict with the second law of thermodynamics, one would have indeed expected the entropy to be of the order of unity at the Planckian epoch, when the horizon was of the order of the Planckian length~\cite{Lineweaver:entropy2}. In fact, according to the \textit{Past Hypothesis}: \textit{The initial macrostate of the Universe has very low thermodynamic entropy.}  (see e.g.~\cite{pittphilsci8894}). 
\end{tcolorbox}

\paragraph{\textit{Causality problem}}The fact that the Hubble radius increase in a standard-component dominated universe brings another problem related to the homogeneous distribution of the temperature anisotropies observed in the CMB. Specifically, the comoving particle horizon and the comoving Hubble horizon are related by a logarithmic integral $\chi_p(\eta)=\int{\chi_\Hl(\eta)d\ln{a}}$ and are roughly of the same order. This result inevitably entails that the regions of the Universe which cross the particle horizon every instant have never been in causal contact before. This comes from the fine distinction between these two quantities: the Hubble horizon separates regions that cannot communicate with each other \textit{now}. On the other hand, the particle horizon separates regions which could \textit{never} have exchanged information with one another. To test whether the CMB could have been in casual contact, let us call $\lambda_H(t_{\rm LS})$ the length corresponding to our present Hubble radius at the time of last scattering. We can write
\begin{equation}
    \lambda_H(t_{\rm LS})=R_H(t_0)\left(\frac{a_{\rm LS}}{a_0}\right)=R_H(t_0)\left(\frac{T_0}{T_{\rm LS}}\right)
\end{equation}
where $R_H(t_0)=\chi_\Hl a$. Conversely, the Hubble radius at $t_{\rm LS}$ is
\begin{equation}
    R_H(t_{\rm LS})=R_H(t_0)\left(\frac{T_{0}}{T_{LS}}\right)^{\frac{3}{2}}
\end{equation}
having considered that we are working in a MD period. Comparing the volumes related to these two scales yields
\begin{equation}
    \frac{\lambda^3_H(t_{\rm LS})}{R^3_H(t_{LS})}=\left(\frac{T_{\rm LS}}{T_0}\right)^{\frac{3}{2}}\approx 10^6.
\end{equation}
Namely, there are $\sim 10^6$ causally disjoint regions within the volume that now corresponds to our horizon. This is difficult to reconcile with the fact that all CMB photons are emitted with a precise black-body distribution. The probability that the expansion of the Universe has begun simultaneously at random is close to $e^{-10^{90}}$. 
\begin{tcolorbox}[mybox]
A closely related shortcoming is for our initial assumption of a Universe spatially homogeneous and isotropic. Any induced isotropy in an expanding Universe must be local and cannot lead to a global property because regions separated by a distance greater than the particle horizon cannot influence each other. Therefore, an interesting question is to understand which are the conditions that brought the Universe to satisfy the cosmological principle. that the early Universe was in a chaotic state with inhomogeneities and anisotropies of all kinds~\cite{Dewitt:1968lxx}. However, various dissipation processes like the effect of neutrino viscosity, damped out nearly all of these imperfections by the time the temperature had fallen to about $\SI{e10}{K}$, leaving only the very small amounts that we observe today. But this just shift the initial condition problem because these processes need specific value of initial anisotropy (see e.g. \cite{Stewart:NotNeutrinoDump}). That is, not all initial conditions would lead to a universe similar to the one we observe today. On the other hand, if in the space of all initial data there were an open set which produces approximately homogeneous and isotropic universes, we would be satisfied. Such an open set exists, but, unfortunately, it is possible to prove that there is no intersection with the subspace of homogeneous initial conditions \cite{Collins:1972tf}. Namely, only inhomogeneous initial conditions give rise to models that approach homogeneity and isotropy. It appears really counterintuitive: inhomogeneities are expected to produce anisotropy rather than isotropy. Besides, FLRW models are unstable, since any open neighbour of the FLRW initial data will contain initial data that does not lead to isotropy.
\end{tcolorbox}
But the CMB is not only homogenous, it is also correlated. Let us test the possibility of that to happen with some back-of-the-envelope calculations. The comoving distance of the last scattering surface from us is $\eta_0-\eta_{\rm LS}$. If we neglect curvature effects, a general comoving scale $\lambda$ is projected on the last-scattering surface sky on an angular scale equal to
\begin{equation}
    \theta\simeq\frac{\lambda}{(\eta_0-\eta_{\rm LS})}.
\end{equation}
Now, we want to identify $\lambda$ as the comoving sound horizon at the time of last-scattering, $\lambda\sim c_s\eta_{LS}$ where $c_s=1/\sqrt{3}$ is the sound velocity at which photons propagate in the plasma at the time (we neglect the baryon load in \eq{csRcs}). We get
\begin{equation}
    \theta\simeq c_s\frac{\eta_{\rm LS}}{(\eta_0-\eta_{\rm LS})}\simeq c_s\frac{\eta_{\rm LS}}{\eta_0},
\end{equation}
where we have used the fact that $\eta_0\gg\eta_{\rm LS}$. Using \tab{Table}, we find 
\begin{equation}
    \theta_H\simeq c_s\left(\frac{T_0}{T_{ls}}\right)^\frac{1}{2}\sim 1^o,
\end{equation}
which corresponds approximately to $\ell\sim 200$, the fundamental multipole for acoustic oscillations that we found in \sect{TA}. The angle $\theta_H$ is the limit inside which two photons are in causal contact. Anisotropies for points outside this limit should be random and big. However, observations of the CMB anisotropies are homogeneous, $\frac{\Delta T}{T_0}\sim 10^{-5}$, and correlated, over the entire surface, far beyond $\theta_H$. We are forced to suppose a fine-tuning of thousands initial conditions to explain homogeneity in the Early Universe

\paragraph{\textit{Monopoles}} Within the standard Big Bang Theory, we expect to find an abundance of monopoles that can compromise the existence of our Univere.  According to the GUTs, for $E\gg E_{GUT}\sim 10^{16}$, all interactions except gravity have equal strength; namely,  electromagnetic, weak, and strong interactions can be unified into a single force~\cite{Georgi:1974sy}. As the temperature drops, this symmetry is broken, and it occurs independently in all causally disconnected regions of the Universe. The field responsible for such symmetry breaking can assume different values in different regions. Consequently, these regions are separated by domain walls~\cite{Zeldovich:1974uw} whose energy density is so high that they should not exist in our observable part of the Universe or they would have led to unacceptable consequences. This is the Kibble mechanism for defect generation~\cite{Kibble:1976sj}. Other kinds of defects can be produced by this mechanism. The most important ones are the magnetic monopoles. Consider a point between four domains. It is possible that the field cannot continuously interpolate between all four domains without vanishing in the middle, and a monopole or an anti-monopole (a monopole with negative magnetic charge) is formed. It is possible to assume that roughly one monopole or anti-monopole is produced per domain, or per particle horizon~\cite{Kibble:1976sj}. Because the particle horizon was very short when the GUT phase transition would have occurred, this leads to a very high density of monopoles and anti-monopoles. In fact, today the number density of magnetic monopoles would be comparable to the number density of protons and neutrons \cite{Preskill:1979zi,Preskill:1984gd}. With a back-of-the-envelope calculation, the density number of monopoles would be 
\begin{equation}
    n_M\sim \frac{1}{2ct_{GUT}}\sim \SI{e82}{\per\cubic \metre}.
\end{equation}
Such a big abundance would have caused catastrophic consequences: because their masses are $\sim 10^{16}$ times that of the proton and the energy density $15$ orders of magnitude higher than the critical density, it would have brought the Universe to the MD phase earlier, $t_{eq}\sim 10^{-16}$, and would have collapsed long ago. A simple but detailed introduction to magnetic monopoles can be found in \cite{Rajantie:2012xh}.

\begin{tcolorbox}[mybox]
One of the reason why the GUT theories are considered a valid piece of the energy history of our Universe is because of the baryon asymmetry problem: nearly all the Universe is matter and not antimatter and  baryons are way scarcer than photons ($\eta \sim 10^{-10}$). An effortless solution could be assuming extremely fine-tuned initial conditions that precisely give the form of the Universe we observe today. More elaborate answers involve the so-called \textit{Anthropic Principle}: \textit{our Universe has these properties simply because otherwise, life would have been impossible and these questions could not have even been asked} \cite{Dicke:1961zz,Carter:1974zz}. It certainly gives a metaphysical outlook to our problems. However, it does not explain the small baryons-photons ratio nor the uniformity of all properties of the Universe. Namely, life could have arisen even if the favourable conditions had existed only in a region of the size of the solar system. Moreover, there is the implicit assumption that either universes are constantly created or there exist many different universes, and life begins only in those which are hospitable. A major breakthrough in finding a more \textit{physical} solution to the baryon asymmetry problem was achieved at first by Sakharov~\cite{Sakharov:1967dj}. He discovered that a solution could be found in theories that consider non-equilibrium processes with C and CP violations in the early Universe. Such violations imply the non-conservation for the baryon number. This discovery opens the door to the Grand Unified Theories (GUTs) that, inexorably, bring a new series of puzzles. 
\end{tcolorbox}

\subsubsection{Where the acceleration comes as a solution}

Let us now see what happen to the aforementioned shortcomings once a period of accelerated expansion is introduced. First and foremost, we want to define the dynamics. Recalling the second Friedmann equation \eq{IIFriedmann}, we can note that we are bound to require a violation of the strong energy condition
\begin{equation}
    \ddot{a}>0\leftrightarrow p<-\frac{\rho}{3}
    \label{eq:acc}
\end{equation}
to obtain acceleration. Neither a radiation nor a matter-dominated phase satisfies the condition. However, we have seen that the cosmological constant $\Lambda$ is characterised by $p_\Lambda=-\rho_\Lambda$ and an exponential growth of the scale factor. Similarly, we can assume such extreme conditions (which nonetheless satisfies \eq{acc}) for our period of inflation which results in a constant Hubble rate $H_I$ and a constant energy density $\rho_I$. The solution of the first Friedmann equation yields \eq{IFriedmann}
\begin{equation}
    a=a_Ie^{H_I(t-t_I)},
\end{equation}
where $t_I$ is the time at which inflation starts. The period where these conditions holds is called \textit{de Sitter stage}. We shall bear in mind our requirement for a graceful exit which inexorably exclude our de Sitter stage. To obtain that, we need the Hubble parameter to vary in time. 
\begin{tcolorbox}[mybox]
    Nevertheless, we want to see how erroneous is to approximate an inflationary period to a de Sitter stage. In order to achieve a graceful exit, $\ddot{a}$ should become negative as well as the derivative of the Hubble rate because
\begin{equation}
    \frac{\ddot{a}}{a}=H^2+\dot{H}.
    \label{eq:infacc}
\end{equation}
Precisely, the graceful exit occurs when $\lvert \dot{H}\rvert$ becomes of order $H^2$. That is, the ratio $\lvert \dot{H}\rvert/H^2$ grows towards the end of inflation. Assuming that
\begin{equation}
    \lvert \ddot{H}\rvert<2H\dot{H}, 
\end{equation}
we obtain a rough estimate for the duration of inflation:
\begin{equation}
    t_f\sim\frac{H_i}{\lvert \dot{H}_i\rvert}. 
    \label{eq:tfinal}
\end{equation}
When $t\sim t_f$ the right-hand side of \eq{infacc} changes sign and the Universe begins to decelerate. At this point two facts come to play. The fluctuations in the CMB are of order $10^{-5}$, namely, the factor $\dot{a}_i/\dot{a}_0$ shall be at most of the same order. In addition, the observable Universe implies that $\dot{a}_f/\dot{a}_0>10^{28}$. Thus, we have
\begin{equation}
   \frac{\dot{a_i}}{\dot{a}_0}=\frac{\dot{a_i}}{\dot{a}_f}\frac{\dot{a}_f}{\dot{a}_0}=\frac{a_i}{a_f}\frac{H_i}{H_f}\frac{\dot{a}_f}{\dot{a}_0}<10^{-5}\rightarrow \frac{a_f}{a_i}>10^{33}\frac{H_i}{H_f}.
\end{equation}
If we assume that $\lvert \dot{H}_i\rvert\ll H_i^2$, and neglect the change of the Hubble parameter ($H_i/H_f\sim1$), we roughly have
\begin{equation}
    \frac{a_f}{a_i}\sim e^{H_it_f}\sim e^{\frac{H^2_i}{\lvert \dot{H}_i\rvert}}>10^{33}
\end{equation}
where we have used the estimate \eq{tfinal}. We can rewrite the latter equation as
\begin{equation}
    \frac{\lvert \dot{H}_i\rvert}{H_i^2}<\frac{1}{75}.
\end{equation}
Combining \eq{IFriedmann} and \eq{ContinuityEq1}, we get
\begin{equation}
    \frac{(\rho+p)_i}{\rho_i}<0.01
    \label{eq:percentage}
\end{equation}
which implies that the deviation from our assumption $p\approx-\rho$ is less than $1\%$ \cite{Mukhanov:2005sc}. Therefore, our de Sitter stage is a good approximation for the initial stage of inflation. Inflation ends when $\rho+p\sim\rho$.
\end{tcolorbox}

A shrinking Hubble radius break the parallelism with the particle horizon. Therefore, any physical length $\lambda$, which is inside the horizon today does not mean anymore that it has never been in casual contact in the past but quite the opposite: this scale has already entered the Hubble horizon in the past, and perturbations from microphysical processes were established. This \textit{earlier talk} could easily explain the homogeneity and isotropy of the CMB. Obviously, it depends on the length of the scale. If we ask that the present horizon, $H_0^{-1}$, was within the Hubble radius during inflation, we can resolve our horizon puzzle. Specifically, we get
\begin{equation}
    \lambda_{H_0}(t_I)=H_0^{-1}\left(\frac{a_{t_f}}{a_{t_0}}\right)\left (\frac{a_{t_I}}{a_{t_f}}\right)=H_0^{-1}\left(\frac{T_0}{T_f}\right)e^{-N}\lesssim H_I^{-1},
    \label{eq:efoldsHorizon}
\end{equation}
where $i$ and $f$ stands for beginning and end of inflation, respectively. With \eq{efoldsHorizon}, apart from a logarithmic dependence, we find that $N\gtrsim 60$ where $N$ is the number of \Ne. Moreover, during inflation
\begin{equation}
\frac{\lvert \Omega-1\rvert_{t=t_f}}{\lvert\Omega-1\rvert_{t=t_I}}=\left(\frac{a_I}{a_f}\right)^2=e^{-2N}.
\label{eq:streatching}
\end{equation}
Hence, the solution $\Omega=1$ from an unstable fixed point becomes an \textit{attractor} during inflation: \textit{inflation drives the Universe towards flatness}.  Since we identify the beginning of the radiation epoch with the end of inflation, we find that $N\approx 60$ solves our flatness problem. Does it mean that the Universe is flat because of inflation? No, it does not. What inflation does is to \textit{stretch} the Universe locally by magnifying the radius of curvature but it does not change the nature of the spacetime: if it is globally open or closed, it will remain so. 

\begin{tcolorbox}[mybox]
A similar value of \Ne~is obtained when we try to solve the entropy problem. To solve this issue we need to introduce a non-adiabatic period where the large entropy observed is produced. Such a period does not correspond to the accelerating phase of inflation. Instead, the condition is satisfied during the phase transition after inflation, which leads the Universe to the radiation-dominated phase. After inflation, the entropy in a comoving volume is conserved, thus
\begin{equation*}
    S_f=S_0,
\end{equation*}
where $S_f$ is the entropy at the end of the inflationary period. On the other hand, during the non-adiabatic phase, we can postulate to have
\begin{equation}
    S_f=A^3S_I,
\end{equation}
with $A$ a numerical factor. From the present value of $S_0$ it is trivial to assume $A\sim 10^{30}$ and, since $S\sim (aT)^3$, we have
\begin{equation}
    \left( \frac{a_f}{a_I} \right)=e^N\approx 10^{30}\left( \frac{T_I}{T_f} \right),
\end{equation}
from which we obtain $N\sim 60$ aside from the logarithmic factor of the ratio between temperatures.
\end{tcolorbox}

The absence of monopoles, which is against the prediction of the Standard Model, could be solved more trivially. If the defects were created before inflation, they were consequently diluted to a maximum of one monopole per visible Universe \cite{Guth:1980zm}: one monopole per $\SI{e60}{ Mpc^3}$. All other particles would be diluted too. However, if we consider a scalar field as the origin of inflation, it then decays into the current particles and not into the monopoles, which are generated only from the break of the symmetry at the end of the GUT. 

\begin{tcolorbox}[mybox]
We do not need to employ the dynamics of the de Sitter stage to takcle the issue of the initial condition problem. First of all, we assume that the initial energy density fluctuations are of order unity on scales equal to the initial Hubble horizon. Our assumption implies
\begin{equation}
   \left.\frac{\delta \rho}{\rho}\right\rvert_{t_i}\sim\frac{1}{\rho}\frac{\lvert \nabla \rho\rvert}{a_i}H_i^{-1}=\frac{\lvert\nabla\rho\rvert}{\rho}\frac{1}{\dot{a}_i}\sim \mathcal{O}(1).
\end{equation}
At a later time $t$, if we assume that the ratio of the spatial derivative of $\rho$ to $\rho$ does not change substantially during the expansion, we have
\begin{equation}
   \left.\frac{\delta \rho}{\rho}\right\rvert_{t}\sim\frac{1}{\rho}\frac{\lvert \nabla \rho\rvert}{a(t)}H(t)^{-1}\sim \mathcal{O}(1)\frac{\dot{a}_i}{\dot{a}(t)}.
   \label{eq:densityexp}
\end{equation}
What \eq{densityexp} tells us is that an accelerated expansion ($\dot{a}(t)>\dot{a}_i$) dilutes large initial inhomogeneities. A patch of size $H^{-1}$ becomes more and more homogeneous because the physical scale size of the perturbations grows faster than the Hubble horizon, while the perturbations amplitude does not change substantially. The term \textit{inflation} fairly captures the effect of devaluation of initial inhomogeneities. Therefore, we do not need a subspace of initial homogeneous conditions: inhomogeneities can now \textit{intuitively} produce isotropy and homogeneity. We are free to ignore what has happened before inflationary epochs because it is irrelevant: the effect of the exponential expansion rules out any initial condition trace from which inflation itself started. 
\end{tcolorbox}

\section{Inflationary framework}
\label{oshoah}
The original model of inflation, now called \textit{old inflation} was presented in Guth's original work \cite{Guth:1980zm}, where we have the scalar field $\phi$, called \textit{inflaton}, trapped in the metastable state called false vacuum. During this stage, the energy density of the state dominates the Universe and triggers the exponential expansion. The potential has two minima degenerate but, with the expansion of the Universe, the second minimum drops lower and becomes an absolute local minimum called true vacuum (remember the inflaton is still in the false vacuum due to the potential barrier which traps it). The phase transition occurs with a tunneling effect which takes place independently on each causal region of the Universe. Since the inflationary epoch is driven as long as the inflaton is trapped, we need a high potential barrier to solve the standard model's puzzles. However, this old scenario was soon abandoned becaue of the so-called \textit{bubbles}~\cite{Kleban:2011pg}, caused by the randomness of the triggering of the phase.

\begin{tcolorbox}[mybox]
The \textit{bubbles} are regions of true vacuum (where the phase transition occurs), which rapidly expanded into the background of false vacuum (where the scalar field is still trapped)\cite{Hawking:1982ga}. The main issue, that also brought to look for a different inflationary scenario, is that the latent heat released in the phase transition is wound up into the bubble walls and is released during a coalescence event. This energy reheats the Universe (which underwent the supercooling phase) and leads to the present contents. Nevertheless, since the thermalization can happen only undergoing many collisions, the Universe becomes greatly inhomogeneous and anisotropic. A further reason to abandon such a model is the following: can these bubbles fill the whole Universe? At first, Guth thought they would have merged, but soon he understood that since the background continued to inflate, the bubble walls cannot merge and the bubbles do not percolate. Moreover, the bubbles continue to be created and inflation never ends. We have no graceful exit. Even if we ask for a high collision rate, we will end up with too many topological defects. We should highlight that theoretical developments in string theory, in the last few years, have brought new interest on the physics of individual bubble universes and on collisions between bubbles (see for example \cite{Kleban:2011pg,Hook:2019pbh} and the bibliography therein). Neighbouring bubbles can leave an observational signature in our Universe if a collision takes place \cite{Kleban:2011pg,Freivogel:2014hca,Chang:2008gj}. For instance, a bubble can influence the local physical conditions in our early Universe even though it never enters inside our visual horizon: it is inside our primordial particle horizon. This could lead to a circle in the CMB sky.
\end{tcolorbox}

This scenario has been replaced by the \textit{new inflation}~\cite{Linde:1981mu,Linde:1990flp}. Where the phase transition this time is smooth, therefore of the second-order. The de Sitter stage does not occur while the inflaton is trapped in the false vacuum. Instead, it takes place when the field is slow rolling towards the true vacuum.  With the new inflation, we drop the requirement of a high potential barrier and replace it with the essential plateau where the inflaton undergoes a \textit{slow-roll} stage. In such a stage, the kinetic energy of the scalar field is negligible with respect to the constant potential energy on the plateau. The de Sitter phase stops as soon as the inflaton falls into the absolute local minimum. The slow-rolling evolution can be initiated by a tunnelling effect but not necessarily. Again, the phase transition occurs by forming bubbles, but for low temperature the potential barrier is very small, and so the scalar field in the interior of the bubble starts with $\phi$ nearly zero. Our observable Universe is supposed to
be a small region inside one bubble. However, even if we have resolved the graceful exit problem, we are now dealing with more stringent conditions~\cite{Steinhardt:1984jj}. Here, the inflaton, at some early time, has to take a value at which the potential $V$ is large but almost flat. Moreover, both the old and the new theory assumed $\phi$ to be in thermal equilibrium with other matter fields before the onset of inflation. It is an unacceptable requirement since the quantum fluctuations of the scalar field, give rise to density perturbations larger than the one observed, unless the inflaton is weakly coupled; that is, there is no thermal equilibrium. 

\begin{tcolorbox}[mybox]
A \textit{phase transition} represents the transition between a disordered phase, characterised by a certain symmetry, and an ordered phase with a smaller degree of symmetry. Such a phase transition is defined by an order parameter that we will call $\Phi$, whose value is zero in the disordered phase. For instance, in ferromagnetic substances, we have $\Phi\equiv\mathcal{M}$, where $\mathcal{M}$ is the net magnetisation. As long as temperature is above the Curie temperature, we have a disordered system with $\mathcal{M}=0$. Once the temperature drops, a magnetisation appears in the Weiss domains and its direction in each domain breaks the rotational symmetry possessed by the disordered phase. However, the lowering of the degree of symmetry of the system takes place even though the Hamiltonian conserves the same degree of symmetry: $\mathcal{M}$ can, in theory, assume any direction. If we take into account all the possibilities,  we still have a homogeneous and isotropic state. However, as soon as a small fluctuation will pick one preferred solution out from such a degenerate state, we will lose the degree of symmetry.We can also think about classical mechanics: the equation $\dot{\boldsymbol{v}}=0$ has both translation and rotational symmetries. The solutions $\boldsymbol{r}=\boldsymbol{r}_0+\boldsymbol{v}_0t$, as long as we do not choose the initial conditions, form a set with the same symmetries. A phase transition can be caused by external influences of sufficient intensity (\textit{induced} symmetry breaking) whilst if the phase transition comes from a gradual change of the parameters of the system, it is called \textit{spontaneous} symmetry breaking. Let us consider the free energy of the system
\begin{equation}
    F=U-TS,
    \label{eq:freeenergy}
\end{equation}
with $U$ the internal energy. The equilibrium state of a system is characterised by the minimum of $F$. If $T=0$, $F$ coincides with the internal energy; otherwise, whatever is the form of $U$, an increase of entropy (disorder) is favourable. When we have a phase transition, the free energy is a function of $\Phi$ and must respect the symmetry of the Hamiltonian of the system. For example, we shall take a Hamiltonian with a reflection symmetry which is broken by the appearance of an order parameter. If $\Phi$ is not too large, we have
\begin{equation}
    F(\Phi)\simeq F_0+\alpha\Phi^2+\beta\Phi^4,
    \label{eq:TaylorF}
\end{equation}
where $\alpha$ and $\beta$ depend on some parameters of the system. In the case in which both the factors are positive, we have a curve with one minimum ($\Phi=0$) that corresponds to the disordered state. Nevertheless, if $\alpha$ turns negative, the curve will be modified: two minima appear ($\Phi_m=\pm(-\alpha/(2\beta))^\frac12$), which are two equal probable ordered states, and the previous minimum becomes a maximum. Namely, the disordered state is unstable and any small external perturbation will \textit{choose} one minimum in the sense that makes one of them deeper or nudges the system towards it. In the latter case, we have achieved a spontaneous symmetry breaking. If the system is only described by the temperature, we can write $\alpha(T-T_c)$, where $T_c$ is some critical temperature below which $\alpha$ becomes negative. When the temperature grows towards the critical value, $\Phi$ decreases slowly and the difference $\Delta F$ between $T>T_c$ and $T<T_c$ at $T\simeq T_c$ is infinitesimal. This is called second-order phase transition. On the other hand, if the order parameter appears or disappears rapidly and $\Delta F$ is finite (and called latent heat) at temperature values near the critical one, we are in a first-order phase transition. We can have such a transition if we add to \eq{TaylorF} an extra term $\lambda(\Phi^2)^{\frac32}$ with $\lambda<0$. $F$ acquires two new minima which, at $T_c$, are equal to $F(0)=F_0$. When the temperature drops, the system is trapped in the disordered state ($\Phi=0$); this is the phenomenon of supercooling. As soon as the system is perturbed or the temperature drops low enough, the system rapidly evolves from    the metastable equilibrium (when $T<T_c$) into stable equilibrium, and liberates the latent heat. The system in the ordered state is then heated up to temperature of order $T_c$ by the release of $\Delta F$. This is the \textit{reheating}.
\end{tcolorbox}

\begin{figure}[h!]
\centering
\includegraphics[width=0.9\textwidth]{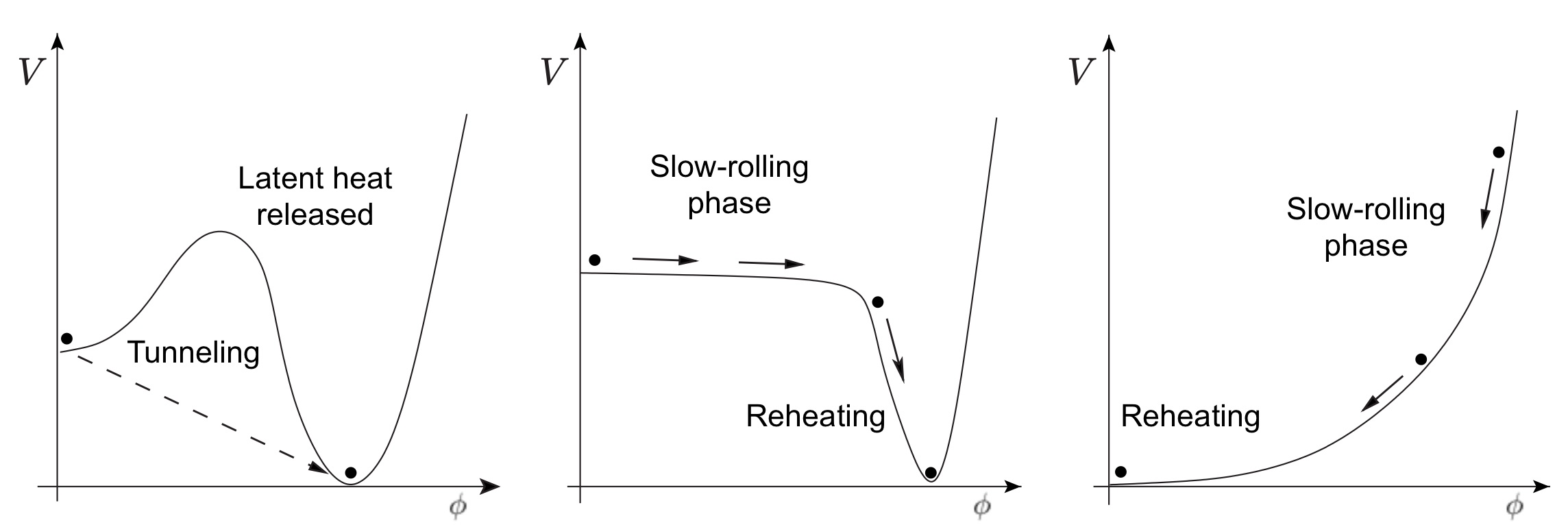}
\caption[First inflationary scenarios]{\small On the left we have the potential form according to the Old scenario, where the de Sitter stage lasts as long as the inflaton is trapped in the false vacuum. In the center, we have the New scenario: the de Sitter stage takes place while the inflaton slowly rolls down the plateau of the potential. On the right-hand side panel, we have the Chaotic scenario. No particular initial conditions are required. It is enough that the inflaton undergoes a slow-rolling phase in a small region which subsequently expands and hosts our Universe.}
\label{fig:Potentials}
\end{figure}

To overcome these shortcomings, the \textit{Chaotic scenario} has been proposed~\cite{Linde:1983gd} where, instead of having high-temperature phase transitions, we have a chaotically distributed inflaton. The basic idea behind this scenario is simply that the assumption according to which the field $\phi$ lies at a minimum (or flat maximum) of its  potential is no longer needed. Instead, it is only necessary to study the evolution of $\phi$ for a variety of fairly natural initial conditions, and check to see whether inflation could start or not. We do not need a symmetry breaking anymore and we no longer try to obtain a globally homogeneous and isotropic universe: it is sufficient to focus on a small region which, after an exponential expansion, becomes greater than the observable Universe. In \fig{Potentials}, we can see the different types of potential in these theories. From the left-hand side panel to the centre one, we move from first to second-order phase transition and introduced a slow-rolling phase. From the panel in the centre to the right one, we eliminate the initial flat potential. 

\subsection{Inflaton dynamics}
\label{ArOd}
The natural candidate for driving inflation is the inflaton $\phi$. We take the simplest scenario of the inflaton minimally coupled to gravity that is governed by the action
\begin{equation}
    S=\int d^4x \sqrt{-g} [\frac{\M}{2}R+\mathcal{L}_\phi]=S_{H}+S_{\phi}
    \label{eq:ActionInflatonMinimallycoupled}
\end{equation}
with
\begin{equation}
   \mathcal{L}_\phi=\frac 12 \partial^\mu\phi\partial_\mu\phi-V(\phi)
   \label{eq:Linflaton}
\end{equation}
and where $S_H$ is the Hilbert action and $V(\phi)$ is the potential of the scalar field which describes its self-interactions. In order to obtain the dynamic of the inflaton, we shall vary the action with respect to the metric $g^{\mu\nu}$ and the field $\phi$. For the Hilbert action the former gives the Einstein equations (see\eq{EinsteinEquations}) and the latter is zero. Hence, we are left to compute
\begin{equation}
     \frac{\delta S_\phi}{\delta g^{\mu\nu}}=\int{d^4x\frac{\sqrt{-g}}{2}\left[2\frac{\delta\mathcal{L}_\phi}{\delta g^{\mu\nu}}-g_{\mu\nu}\mathcal{L}_\phi\right]}.
\end{equation}
Putting all together, we arrive at the expression
\begin{equation}
    R_{\mu\nu}-\frac12 g_{\mu\nu} R+2\frac{\delta\mathcal{L}_\phi}{\delta g^{\mu\nu}}-g_{\mu\nu}\mathcal{L}_\phi=0.
\end{equation}
Similary as we have done for the Einstein equations, we define the stress-energy tensor for the inflaton (or more generally a scalar field minimally coupled to gravity) as
\begin{equation}
    T^{(\phi)}_{\mu\nu}=g_{\mu\nu}\mathcal{L}_\phi-2\frac{\delta \mathcal{L}_\phi}{\delta g^{\mu\nu}}
    \label{eq:StressInflaton}
\end{equation}
or explicitly 
\begin{equation}
T^{(\phi)}_{\mu \nu}=\partial_\mu \phi \partial_\nu \phi + g_{\mu \nu}\left(\frac12\partial^\sigma\phi\partial_\sigma\phi-V(\phi)\right)\,.
\label{eq:TforInflaton}
\end{equation}
The variation of the action with respect to $\phi$ yields
\begin{multline}
    \frac{\delta S_\phi}{\delta \phi}=\int{d^4x\sqrt{-g}\left[\frac12\partial_\mu\delta\phi\partial^\mu\phi+\frac12\partial_\mu\phi\partial^\mu\delta\phi-\frac{\delta V(\phi)}{\delta \phi}\delta\phi\right]}\\
    =\int{d^4x\sqrt{-g}\left[\partial_\mu\phi\partial^\mu\delta\phi-\frac{\delta V(\phi)}{\delta \phi}\delta\phi\right]}\\
    =\int{d^4x\left[ \partial_\mu(\sqrt{-g}\delta\phi\partial^\mu\phi)-\delta\phi\partial_\mu(\sqrt{-g}\partial^\mu\phi)-\sqrt{-g}\frac{\delta V(\phi)}{\delta \phi}\delta\phi\right]}.
\end{multline}
For the variational principle, $\delta \phi$ vanishes at the extrema; thus, the first term, which is a surface term, is zero. It follows
\begin{equation}
    \frac{\delta S_\phi}{\delta \phi}=-\int{d^4x\sqrt{-g}\left[\frac{1}{\sqrt{-g}}\partial_\mu(\sqrt{-g}\partial^\mu\phi)+\frac{\delta V(\phi)}{\delta \phi}\right]\delta \phi}=0.
\end{equation}
Hence, 
\begin{equation}
    \frac{1}{\sqrt{-g}}\partial_\mu(\sqrt{-g}\partial^\mu\phi)+\frac{\delta V(\phi)}{\delta \phi}.
    \label{eq:hen}
\end{equation}
To find the equation of motion for the inflaton, we ought to explicitly write the terms in \eq{hen}. Considering that, as we have seen, inflation stretches the spacetime and the fact that the observational consequences arise from its ending phase, we can use the flat FLRW metric as a successful approximation. Knowing that $\sqrt{-g}=a^3(t)$, we eventually obtain the equation of motion
\begin{equation}
    \ddot{\phi}+3H\dot{\phi}-\frac{\nabla^2 \phi}{a^2} + V'(\phi)=0\,,
    \label{eq:EOMInflatonGrad}
\end{equation}
which is nothing but the \textit{Klein-Gordon} equation with a source term. Henceforth, we will consider a homogeneous field, \ie there is no dependence of space in the inflaton and the spatial derivatives are zero. This assumption brings us to
\begin{equation}
\boxed{
    \ddot{\phi}+3H\dot{\phi}+ \frac{\d V(\phi)}{\d \phi}=0
    }
    \label{eq:EOMInflaton}
\end{equation}
We are allowed to neglect the spatial derivatives for the so-called inflationary \textit{non-hair} theorem, namely, the fact that initial inhomogeneities are diluted as shown in \eq{densityexp}. If we take \eq{TforInflaton} and the expression for the stress-energy tensor in \eq{Tmunu}, we obtain
\begin{equation}
\boxed{
    \rho_\phi=\frac{1}{2}\dot{\phi}^2+V(\phi)
    }
    \label{eq:densityinflaton}
\end{equation}
and
\begin{equation}
\boxed{
    p_\phi=\frac{1}{2}\dot{\phi}^2-V(\phi)
    }\,,
    \label{eq:pressureinflaton}
\end{equation}
always neglecting the gradient term. From these two definitions, we are able to get the equation of state parameter $w$:
\begin{equation}
   \boxed{ w_\phi\equiv \frac{p_\phi}{\rho_\phi}=\frac{\frac{1}{2}\dot{\phi}^2-V(\phi)}{\frac{1}{2}\dot{\phi}^2+V(\phi)}}\,.
   \label{eq:EOSInfl}
\end{equation}
Since we can write
\begin{equation}
    p_\phi=-\rho_\phi+\dot{\phi}^2,
\end{equation}
the deviation from the de Sitter stage is entirely characterised by the kinetic energy and can be obtained if $\dot{\phi}^2\ll V(\phi)$.

\subsection{Slow-Roll approximation}
\label{sec:SRAP}
Let us now delve deeper into the slow-rolling phase~\cite{Lyth:2009zz,Mukhanov:2005sc,Dodelson:2003ft,Weinberg:2008zzc}. First of all, it is important to parameterize these \textit{slowness} of the crucial features. If we consider the variation of the Hubble radius
\begin{equation}
    \frac{1}{aH}=-\frac{\dot{a}H+a\dot{H}}{(aH)^2}=-\frac1{a}(1-\eh)
\end{equation}
therefore, we have introduced the first \textit{Hubble slow-roll parameter} (HSR)
\begin{equation}
    \boxed{\eh\equiv-\frac{\dot{H}}{H^2}=-\frac{d\ln{H}}{dN}}
    \label{eq:ISRpar}
\end{equation}
where $dN=d\ln{a}=Hdt$. This parameter is associated to the behavior of the Hubble radius, which is essential to regulate if we want to solve the puzzles of the standard theory. $\eh<1$ implies a shrinking Hubble radius. Furthermore, the more $\eh$ is small the better this stage of inflation resemble the de Sitter stage, with a constant $H$. Nonetheless, as we have previously stated the Hubble parameter has to vary in time to allow a graceful exit. For this reason, we introduce a second dimensionless parameter which quantifies our requirement to $\eh$ to be small for a sufficiently large number of \Ne~(remember we need at least 60 \Ne)
\begin{equation}
    \boxed{\kappa\equiv\frac{d\ln{\eh}}{dN}=\frac{\dot{\eh}}{H\eh}}
    \label{eq:IISRpar}
\end{equation}
which implies we need $\lvert\kappa\rvert<1$. 

We want to compute now the dynamics of the inflaton in this slow-rolling regime. With the definition of the energy density \eq{densityinflaton} we can write the first Friedmann equation and combine it with the already defined equation of motion \eq{EOMInflaton}, finding
\begin{equation}
    \dot{H}=-\frac12\frac{\dot{\phi}^2}{\M}\,
    \label{eq:Ip}
\end{equation}
or, dividing by $\dot{\phi}$ we can get rid of the time dependency and obtain
\begin{gather}
    H(\phi)_{,\phi}=-\frac1{2\M}\dot{\phi}
    \label{eq:IpHJ}\\
    H(\phi)_{,\phi}^{2}-\frac3{2\M}H(\phi)^2=-\frac1{2M_{\rm pl}^4}V(\phi)\,.
    \label{eq:HJ}
\end{gather}
This new set of equations is called the \textit{Hamilton-Jacobi equations} where we have eliminated the time dependence in the Friedmann equation to better show how the equation of motion and the Friedmann equations are coupled~\cite{Salopek:1990jq,Lidsey:1991zp,Muslimov:1990be,Kinney:1997ne}. We can also writhe the HSR parameters with derivative of $\phi$ noting that
\begin{equation}
    dN\equiv d\ln{a}= Hdt =\frac{H}{\dot{\phi}}d\phi=-\frac{H}{2\M H_{,\phi}}d\phi
    \label{eq:NefoldHJ}
\end{equation}

\begin{tcolorbox}[mybox]
The set of equations \eq{IpHJ} and \eq{HJ} allow to generate a collection of exact inflationary solutions~\cite{Lidsey:1991zp,Lidsey:1991dz}. Once the form of $H(\phi)$ is chosen, \eq{HJ} allows to find the potential for which the exact solution applies; subsequently, \eq{IpHJ} converts the $\phi$ dependence to be converted into time-dependence, to get $H(t)$. However, the solution of \eq{HJ} depends on the choice of initial conditions for the field $\phi$. We can try to check the sensitivity to the initial conditions because, to have robust results, we need that the late time solution is independent of the choice of the initial conditions~\cite{Salopek:1990jq}. To see it, we should assume that $H_0(\phi)$ is the slow-roll solution and $\delta H$ a small perturbation such that the general solution is $H(\phi)=H_0(\phi)+\delta H(\phi)$. Putting this solution in \eq{HJ} and linearizing the result, we find that $\delta H$ is a solution of
\begin{equation}
    H_{0,\phi}\delta H_{,\phi}\simeq \frac{3}{2\M}H_0\delta H
\end{equation}
and has the general form
\begin{equation}
    \delta H=\delta H(\phi_i)\text{exp}\left[\frac3{2\M}\int^{\phi}_{\phi_i}\frac{H_0}{H_{0,\phi}}d\phi\right],
\end{equation}
where $\delta H(\phi_i)$ represents a different choice of initial condition. Using \eq{NefoldHJ}, we can write
\begin{equation}
    \boxed{\delta H=\delta H(\phi_i)\text{exp}[-3N(\phi)]}\,.
\end{equation}
Thus, the effect of specific initial conditions is exponentially erased and has no observable effect. All possible inflationary trajectories, \ie solutions of \eq{HJ}, will quickly converge to a common attractor solution if they are sufficiently close to each other. In the case which $\dot{\phi}$ reverses its sign, as long as the perturbation is insufficient to knock the scalar field over the maximum in the potential, the perturbed solution will inevitably reverse and pass through the initial value $\phi_i$ again and can be treated as a perturbation with a fixed sign. We should underline that all solutions are attractors for one another and converge asymptotically.~\cite{Liddle:1994dx}
\end{tcolorbox}

Substituting \eq{Ip} into \eq{ISRpar} the slow-rolling stage arises when the kinetic term is negligible with respect to the potential, that is why it is called slow-roll inflation. If we want a negligible kinetic term, we need to ensure that the velocity of the inflaton does not increase. To regulate this, we introduce the second HSR parameter\footnote{Sometimes $\kappa$ in \eq{IISRpar} is defined as second slow-roll parameter (see e.g.~\cite{Baumann:2022mni,Schwarz:2001vv}). However, these two parameters are related because $\kappa=2(\eh-\etah)$ therefore the condition on $\etah$ induce automatically our requirement on $\kappa$ and vice-versa.}
\begin{equation}
    \boxed{\etah\equiv -\frac{\ddot{\phi}}{H\dot{\phi}}=\frac{d\ln{\dot{H}}}{d\ln{a}}}
\end{equation}
whose smallness implies that the kinetic term is subdominant and the friction term in \eq{EOMInflaton} dominates. We have now parameterized the slow-rolling phase and quantified the necessary conditions. Now, we want to see how these affects the dynamics. From \eq{Ip} and the definition \eq{ISRpar}, we see that with the smallness of $\eh$ we require $\dot{\phi}^2\ll V(\phi)$ that leads to $H^2\approx V/(3\M)$. This is now telling us that, in this approximation, the behavior of the Hubble radius is determined by the potential. And with the further condition $\etah\ll 1$, we can see that $3H\dot{\phi}\approx-V_{,\phi}$ with the comma indicating the derivative. From \eq{pressureinflaton} and \eq{densityinflaton} we can see how these conditions yield to the approximate relation $p_\phi\approx-\rho_\phi$ wanted, with the inflaton behaving as a vacuum-like matter and gives rise to almost exponential expansion.

\paragraph{\textit{Potential slow-roll parameters}} What is interesting to see, from this approximation, is that we can now determine whether a potential can induce the slow roll-inflation, computing the \textit{potential slow-roll parameters} (PSR)
\begin{equation}
    \boxed{\epsv\equiv\frac{\M}{2}\(\frac{V_{,\phi}}{V}\)^2\,\,,\,\,\eta_\Vl\equiv\M\frac{V_{,\phi\phi}}{V}}
    \label{eq:evetav}
\end{equation}
with $\eh\approx\epsv$ and $\eh+\etah\approx\etav$. There is an important distinction between the Hubble and potential slow-roll parameters:  the smallness of the PSR parameters is a necessary condition, but not a sufficient one. With their smallness we can restrict the form of the potential, not the properties of dynamic solutions. The value of $\dot{\phi}$ that defines the kinetic term is not yet specified; it could, be as large as one wants, regardless of the smallness of the PSR parameters. Consequently, we need to add a further assumption: the scalar field evolves to approach an asymptotic attractor solution, determined by
\begin{equation}
    \dot{\phi}\simeq -\frac{V_{,\phi}}{3H},
    \label{eq:attractor}
\end{equation}
which is the same as neglecting the acceleration of the inflaton field in \eq{EOMInflaton}. The HSR and PSR are related through
\begin{gather}
    \epsilon_\Vl=\epsilon_\Hl\left(\frac{3-\eta_\Hl}{3-\epsilon_\Hl}\right)^2,\label{eq:EpsilonHV}\\
    \eta_\Vl=\frac{3\epsilon_\Hl+3\eta_\Hl-\eta_\Hl^2-\xi^2_\Hl}{3-\epsilon_\Hl}\label{eq:EtaHV},
\end{gather}
with $\xi_\Hl=\,^2\beta_\Hl$  and $^n\beta_\Hl$ is the HSR parameter hierarchy
\begin{equation}
    ^n\beta_\Hl\equiv\left[\prod_{i=0}^{n}{\left(-\frac{d\ln{H^{(i)}}}{d\ln{a}}\right)}\right]^\frac1n=2\M\left(\frac{(H_{,\phi})^{n-1}H^{(n+1)}}{H^n}\right)^{\frac1n}
    \label{eq:Hierarchy}
\end{equation}
where the superscript $(n)$ indicates the $n$th derivative with respect to $\phi$. The parameter $\eh$ is defined separately, but we associate it to $^0\beta_H$, while $\etah\equiv\,^1\beta_H$. We can do the same for PSR parameters, defining
\begin{equation}
    ^n\beta_\Vl\equiv\M\left(\frac{d\ln{V}}{d\phi}\right)\left[\prod_{i=0}^{n}{\left(\frac{d\ln{V^{(i)}}}{d\phi}\right)}\right]^\frac1n=\M\left(\frac{(V')^{n-1}V^{(n+1)}}{V^n}\right)^{\frac1n}.
    \label{eq:eVhierarchy}
\end{equation}
The relations between these parameters at higher order can be found in~\cite{Liddle:1994dx}. One last remark, before, moving to the next section, is that, using the PSR parameters, we can generate an analytic solution to fourth-order in the potential $V(\phi)$
\begin{equation}
    H^2(\phi)=\frac {\M}{3}V(\phi)\left[1+\frac 1 3\epsilon_\Vl-\frac 1 3\epsilon_\Vl^2+\frac 2 9 \epsilon_\Vl\eta_\Vl+\frac {25} {27} \epsilon_\Vl^3+\frac 5 {27} \epsilon_\Vl\eta_\Vl^2-\frac {26}{27} \epsilon_\Vl^2\eta_\Vl+\frac 2 {27} \epsilon_\Vl\xi_\Vl^2 +\mathcal{O}_4\right].
    \label{eq:expansionH}
\end{equation}
We have generated an analytic solution for inflation in the potential $V(\phi)$, that is accurate up to fourth-order in slow-roll parameters. For further details see \cite{Liddle:1994dx,Schwarz:2001vv,Leach:2002ar}. For observational constraints to PSR and HSR parameters up to fourth-order, see \chap{PGWs}

\subsection{Duration of inflation}

We saw how it is also important for the inflationary period to last long enough to solve the shortcomings of the Standard Scenario. And this is usually quantized by the number of \Ne. In the slow-roll scenario, they have a great significance and we need to impose a constraint on the minimum value possible. In order to compute that, we need to make some assumptions. First of all, we need to schematize into intervals the history of our Universe and we are assuming sudden transition between epochs. We further assume a universe MD during the phase of reheating\footnote{The theory of reheating is essential to reconcile inflation with the Hot Big Bang scenario: the Universe is left at low temperature while we require at least $T\simeq1\,\text{MeV}$ to start BBN. Besides, as we have already mentioned, all the particles, as well as the monopoles, are diluted due to the Universe rapid expansion; we need to convert the energy stored in the scalar
field to, ultimately, relativistic products that lead to a radiation-dominated scenario in thermal
equilibrium from which the Standard evolution could begin. Details on this epoch could be found in~\cite{Kofman:1994rk,Kofman:1997pt,Kofman:1997yn,Bassett:2005xm}} after inflation. The four intervals are
we need to approximately divide the evolution of the Universe into four intervals:
\begin{equation*}
    \text{Inflation}\rightarrow  a_{end} \rightarrow \text{Reheating} \rightarrow a_{reh}
    \rightarrow \text{RD epoch}\rightarrow a_{eq} \rightarrow \text{MD epoch}\rightarrow a_{0},
\end{equation*}
The label \textit{end} stands for the end of the inflationary period, while \textit{eq} labels the equivalence period when the Universe goes from a RD to a MD epoch. This separation yields
\begin{equation}
    \frac{a_kH_k}{a_0H_0}=\frac{a_k}{a_{end}}\frac{a_{end}}{a_{reh}}\frac{a_{reh}}{a_{eq}}\frac{a_{eq}}{a_0}\frac{H_k}{H_{eq}}\frac{H_{eq}}{H_0}
\end{equation}
where $a_kH_k$ represent the  hubble radius  equalt to certain comoving scale $k$. Knowing the proportionality between the energy density and the scale factor for MD and RD epochs, we have
\begin{equation}
    \frac{k}{a_0H_0}=e^{-\Delta N_k}\left(\frac{\rho_{reh}}{\rho_{end}}\right)^{\frac{1}{3}}\left(\frac{\rho_{eq}}{\rho_{reh}}\right)^{\frac{1}{4}}\left(\frac{H_k}{H_{eq}}\right)\left(\frac{a_{eq}H_{eq}}{a_0H_0}\right).
\end{equation}
$\Delta N_k$ is the number of \Ne~before the end of inflation at which our scale k is equal to the Hubble radius. Considering our slow-roll approximations, we can write $H_k^2\simeq \frac{1}{3\M}V_k$ and, performing a logarithm to both members, we get~\cite{Liddle:2000cg,Liddle:2003as,Liddle:1993fq,Martin:2013tda,Kinney:2021nje}
\begin{equation}
    \Delta N_k=-\ln{\frac{k}{a_0H_0}}+\frac{1}{3}\ln{\frac{\rho_{reh}}{\rho_{end}}}+\frac{1}{4}\ln{\frac{\rho_{eq}}{\rho_{reh}}}+\ln{\sqrt{\frac{V_k}{3\M}}\frac{1}{H_{eq}}}+\ln{219\Omega_0h}.
    \label{eq:deltaN}
\end{equation}
At this point, we need to make further approximations to limit $N$. First and foremost, we are interest in the current horizon scale with $k\rightarrow k_{\rm hor}=a_0H_0$. Then, we assume that even during the last part of inflation $V_k=\rho_{\rm end}$ holds. Finally, the we impose an instantaneous reheating ($\rho_{\rm reh}=\rho_{\rm end}$). Therefore~\cite{Liddle:2003as}
\begin{equation}
    N^{max}_{hor}=\frac{1}{4}\ln{\frac{\rho_{eq}}{V_{hor}}}+\ln{\sqrt{\frac{V_{hor}}{3\M}}\frac{1}{H_{eq}}}+\ln{219\Omega_0h}. 
\end{equation}
The substitution of the known quantities leads us to the relation
\begin{equation}
N^{max}_{hor}=68.5+\frac14\ln{\frac{V_{hor}}{M_{\rm pl}^4}}.
\end{equation}
\newpage
\section{Inflationary perturbations}
\label{sec:perturbationss}
We now want to study the mechanism on how the primordial seeds for perturbations are planted~\cite{Guth:1985ya,Starobinsky:1992ts, Starobinsky:1979ty,Starobinsky:1983zz,Mukhanov:1981xt,Mukhanov:1990me,Mukhanov:2013tua,Mukhanov:1982nu,Bardeen:1980kt,Bardeen:1983qw,Adams:1992bn,Bartolo:2001rt,Choe:2004zg,Gordon:2000hv,Jackson:2013vka,Martin:2002vn,Adams:2001vc}. As we have seen in the previous sections, the early Universe is made \textit{nearly} uniform by a primordial inflationary stage. Nevertheless, the fact that we have not used the word \textit{perfectly} is due to the smallness of the \textit{seed} perturbations which are present at that stage. They are vital as they evolve until they give rise to the Large Scale Structure of the Universe. They explain the CMB anisotropies at angular scales larger than $\ang{1}$ because, otherwise the microphysical processes are prevented. The aim of this section is to understand how such seeds are produced. Small fluctuations of the inflaton field are related to fluctuations of the spacetime metric since the gravity talks to any component of the Universe. Such a contact leads to perturbations of the curvature $\mathcal{R}$ \eq{mathcalR}. Fluctuations are created on all length scales and since the comoving Hubble radius shrinks exponentially all fluctuations exit the horizon. We have already mentioned that $\mathcal{R}$ remains constant outside the horizon. Hence, its amplitude is not affected by the unknown physics shortly after inflation and can be considered as classical. Ultimately, these fluctuations re-enter the horizon at some epoch (radiation-dominated or matter-dominated) and produce matter and temperature perturbations via the Poisson equation. Then, they will start growing.

\subsection{Mukhanov-Sasaki Equation}
We  have assumed that the inflaton field dominates the energy density of the Universe. This means that a perturbation on $\phi$ implies a perturbation on the stress-energy tensor. Consequently, the perturbation of $T_{\mu\nu}$ implies a perturbation on the metric tensor, via the Einstein equations. But a perturbation in the metric tensor, induce a perturbation on $\phi$ via the Klein-Gordon equation. Let us compute the relevant quantities. 

First of all, we want to set ourselves in the spatially flat gauge \eq{SPAtially}, where $\Phi=E=0$. In this gauge, we have
\begin{equation}
    ds^2=a^2(\eta)[-(1+2\Psi)d\eta^2+2\partial_i Bd\eta dx^i+\delta_{ij}dx^idx^j]
\end{equation}
and
\begin{equation}
    g_{\mu\nu}=a(\eta)^2\begin{pmatrix}
    -(1+2\Psi)&\partial_i B\\
    \partial_i B& 0\\
    \end{pmatrix},\quad g^{\mu\nu}=a(\eta)^2\begin{pmatrix}
    -1+2\Psi&\partial^i B\\
    \partial^i B& 0\\
    \end{pmatrix}
\end{equation}
and we remember that 
\begin{equation}
    \delta\sqrt{-g}=-\frac{\delta g}{2\sqrt{-g}},\quad \delta g=g g^{\mu\nu}\delta g_{\mu\nu}.
\end{equation}
With this information and the introduction of the perturbed field $\phi=\phi_0+\delta\phi$ we can, from \eq{hen}, obtain 
\begin{equation}
    \delta \phi''+2\mathcal{H}\delta\phi'-\nabla^2\delta\phi=(\Psi'+\nabla^2 B)\phi_0'-2a^2V_{,\phi}\Psi-a^2V_{,\phi\phi}\delta\phi.
\label{eq:inflatonperturbedeq}
\end{equation}
Moreover, from the perturbed Einstein tensor, we can write the following set of equations~\cite{Riotto:2002yw,Baumann:2022mni}
\begin{gather}
    \Psi=\eh\frac{\mathcal{H}}{\phi_0'}\delta\phi,\label{eq:infEn}\\
    \nabla^2B=-\eh\frac{\mathcal{H}}{\phi_0'}\(\delta\phi'+(\eta_\Hl-\eh)\mathcal{H}\delta\phi\).
    \label{eq:infEn2}
\end{gather}
From these equations it is easy to see that for the HSR parameters go to zero, the mixing with the inflaton flucutations vanishes. With \eq{infEn} and \eq{infEn2}, we can remove the metric perturbations and \eq{inflatonperturbedeq} reduce to 
\begin{equation}
    \delta \phi''+2\mathcal{H}\delta\phi'-\nabla^2\delta\phi=\[2\varepsilon_\Hl(3+\varepsilon_\Hl-2\eta_\Hl)-\frac{a(\eta)^2V_{,\phi\phi}}{\mathcal{H}}\]\mathcal{H}^2\delta\phi.
    \label{eq:preMS}
\end{equation}
If we perform another derivative to the equation of motion $\ddot{\phi}+2H\dot{\phi}=-V_{,\phi}$ and knowing that
\begin{equation}
    \frac{\dddot{\phi}}{H^2\dot{\phi}}=-\frac{\eta_\Hl'}{\mathcal{H}}+3\varepsilon_\Hl+\eta_\Hl^2+\eta_\Hl\varepsilon_\Hl
\end{equation}
we can write 
\begin{equation}
    \delta \phi''+2\mathcal{H}\delta\phi'-\nabla^2\delta\phi=\[(\eh-\eta_\Hl)(3+2\eh-\eta_\Hl)-\frac{\eta_\Hl'}{\mathcal{H}}\]\mathcal{H}^2\delta\phi\,.
\end{equation}
If we define $f\equiv a\delta\phi$, and see that $a''/a=2(-\varepsilon_\Hl)\mathcal{H}^2$ we find
\begin{equation}
    \delta \phi''+2\mathcal{H}\delta\phi'=\frac1a\(f''-(2-\eh)\mathcal{H}^2f\)
\end{equation}
and inserting it into \eq{preMS}, moving into Fourier space, we eventually obtain the \textit{Mukhanov-Sasaki equation}~\cite{Sasaki:1986hm,Mukhanov:2005sc,Baumann:2022mni,Riotto:2002yw}
\begin{equation}
    \boxed{f''+\(k^2-\frac{z''}{z}\)f=0}
    \label{eq:MS}
\end{equation}
where $z\equiv a\phi'_0/\mathcal{H}$ is the \textit{Mukhanov variable}. In this equation we have represented the coupling between the inflaton and matter perturbations, without any approximations. This is so beautiful because \eq{MS} is the equation of a harmonic oscillator, with a time-dependent frequency
\begin{equation}
    \omega^2(\eta,k)\equiv k^2-\frac{z''}{z}
\end{equation}
and it is something we know how to deal with. During slow-roll inflation, we can notice that, being $H$ and $\phi'_0$ approximately constant, $z''/z\approx 2\mathcal{H}^2$. Let us study qualitatively two different limits \cite{Kinney:2009vz,Baumann:2022mni}:
\begin{enumerate}
    \item The \textit{Short wavelength limit}, \ie $k\gg \lvert z{''}/z\rvert$ is valid at early times because all modes are inside the horizon. In this case, the equation of motion in \eq{MS} becomes that for a conformally Minkowski Klein-Gordon field
    \begin{equation}
        f''+k^2f=0,
    \end{equation}
    whose solution is 
\begin{equation}
    f(\eta)=\frac{1}{\sqrt{2k}}(A_ke^{-ik\eta}+B_ke^{ik\eta})
    \label{eq:RotTor}
\end{equation}
which is that of a harmonic oscillator with a fixed frequency $\omega=k$.
    \item The \textit{Long wavelength limit} is characterized by \ie $k\ll \lvert z{''}/z\rvert$. In the infrared limit, the modes are all outside the horizon and \eq{MS} reads
    \begin{equation}
    f''-\frac{z''}{z}f=0.
    \label{eq:fredy}
    \end{equation}
    \eq{fredy} has two solutions
    \begin{gather}
        f\propto z\rightarrow \delta \phi = const\\
        f\propto z^{-2}\rightarrow \delta \phi = a^{-1}.
        \label{eq:Freeze}
    \end{gather}
    It is important to note that these scales are microscopic which, in the proper time framework, are stretched outside the constant Hubble radius, and becomes macroscopic fluctuations. 
\end{enumerate}
In the considered gauge, from \eq{mathcalR} we have 
\begin{equation}
    \mathcal{R}=-\mathcal{H}(v+B)
\end{equation}
and if we compute the perturbed $\delta T^i_0$ for the scalar field, and confronting it with \eq{TIO}, we see that $v+B=-\delta\phi/\phi'_0$ and therefore
\begin{equation}
\mathcal{R}=\frac{\mathcal{H}}{\phi_0'}\delta\phi=\frac{f}{z}.
\label{eq:Rflz}
\end{equation}
From \eq{Freeze}, we see that on superhorizon scales \eq{Rflz} is constant; as consequence, we can restrict ourselves to study the UV limit, which is fortunate since we know how to quantize fields in the Minkowski space. 

\subsection{Quantization of Primordial Fluctuations}
In general, whichever the system is, there are some fundamental steps to make for quantization. We now try to list them, in the trivial case of a system with one particle. In one dimension, if we have a particle of mass $m$ moving in a time-dependent potential $V(x,t)$, the steps for canonical quantization are the following: 
\begin{enumerate}
    \item Firstly, we define the Lagrangian and the action
    \begin{equation}
        L=\frac12 m\dot{x}^2-V(x,t),\quad S=\int{dt L}
        \label{eq:HOAction}
    \end{equation}
    from which we can derive the equation of motion through the variation principle
    \begin{equation}
        \frac{\delta S}{\delta x}=0\rightarrow m\ddot{x}=-\partial_x V(x,t).
        \label{eq:HOequationofmotion}
    \end{equation}
    \item Afterwards, we define the momentum conjugate to $x$,
    \begin{equation}
        p\equiv\frac{\partial L}{\partial \dot{x}}=m\dot{x},
        \label{eq:HOmomentum}
    \end{equation}
    which agrees with the standard notion of the particle's momentum $p=mv$.
    \item We now replace the classical variables, $x$ and $p$, with the operators $\hat{x}$ and $\hat{p}$ and subsequently impose the canonical commutation relation
    \begin{equation}
        [\hat{x},\hat{p}]=\hat{x}\hat{p}-\hat{p}\hat{x}=i\hbar.
        \label{eq:HOcanonical}
    \end{equation}
    \item Next, we should choose a formalism: the Schr\"{o}dinger's one, where the state is time-dependent whilst the operators are not, or the Heisenberg picture where is valid the opposite\footnote{there is also a third well-known formalism, the Dirac representation, which is a mix of the two}. We choose the latter. It is trivial to see that, since the operators in the Heisenberg representation are time dependent, the commutation relation \eq{HOcanonical} should  hold at all times. In fact, the equation of motion implies that if it holds at one initial time, it will hold at all times. We can rewrite it in the form 
    \begin{equation}
        [x(t),\dot{x}(t)]=\frac{i\hbar}{m}.
        \label{eq:HOcanonical2}
    \end{equation}
\end{enumerate}

An instructive description of the quantization process for a simple one-dimensional harmonic oscillator is given in \appx{HO1D}. A more advanced quantization method is given in \appx{qsfcs} where a massive scalar field is quantizied in curved space time.

\subsection{Inflaton Quantization}
To promote our field $f$ in \eq{MS}, we have seen in \eq{HOmomentum} that we should define the conjugate momentum
\begin{equation}
    \pi\equiv\frac{\partial\mathcal{L}}{\partial f'}
    \label{eq:conjugatepi}
\end{equation}
for which is valid the commutator relation \eq{HOcanonical}. Then, in Fourier space we define the single time-independent non-Hermitian operator 
\begin{equation}
    f_{\boldsymbol{k}}\rightarrow f_k(\eta)\hat{a}_{\boldsymbol{k}}+f^*_{k}(\eta)\hat{a}^\dagger_{-\boldsymbol{k}},
    \label{eq:Promoting2}
\end{equation}
that satisfies the canonical commutation relation
\begin{equation}
    [\hat{a}_{\boldsymbol{k}},\hat{a}^\dagger_{\boldsymbol{k}'}]=(2\pi)^3\delta(\boldsymbol{k}-\boldsymbol{k}')
    \label{eq:Communist}
\end{equation}
only if the mode functions are normalized as follows:
\begin{equation}
    \langle f_k, f_k\rangle\equiv \frac{i}{\hbar}(f^*_kf'_k-f^{*'}_kf_k)=1.
    \label{eq:CondCond}
\end{equation}
We have provided just one boundary condition on the solutions of the Mukhanov-Sasaki equation in \eq{MS} and we need to fix the second one from the \textit{vacuum selection}.

\subsubsection{Vacuum Selection}
The action of creation and annihilation operators is
\begin{gather}
\hat{a}^\dagger_{\boldsymbol{k}}\ket{n(\boldsymbol{k})}=\sqrt{n+1}\ket{n(\boldsymbol{k})+1}\\
\hat{a}_{\boldsymbol{k}}\ket{n(\boldsymbol{k})}=\sqrt{n}\ket{n(\boldsymbol{k})-1}.
\end{gather}
The ground state is just $a_{\boldsymbol{k}}\ket{0}=0$. It is natural to define the vacuum state as the ground state at the beginning of inflation. But we have seen that this produce a simplification to \eq{MS} reducing it to the simple harmonic oscillator. Namely, we are in the short wavelength limit in \eq{RotTor}; at the beginning of inflation all modes are well inside the Hubble horizon. We can write
\begin{equation}
    \lim_{k\eta \to -\infty}f_k(\eta)=\frac1{\sqrt{2k}}e^{ik\eta}
    \label{eq:condition}
\end{equation}
because in this regime, $\omega(k)\longrightarrow k$. At this point, \eq{CondCond} and \eq{condition} completely fix the mode functions on all scales and we can define vacuum by for all modes.

Let us take the classical wave equation that comes from our subhorizon limit
\begin{equation}
    f''-\nabla^2f=0.
    \label{eq:WaveWave}
\end{equation}
We decompose it into Fourier modes $f_k$
\begin{equation}
    f=\int{d\boldsymbol{k}[f_{\boldsymbol{k}}(\eta)a_{\boldsymbol{k}}e^{i\boldsymbol{k}\vdot\boldsymbol{x}}+f^*_{\boldsymbol{k}}(\eta)a^*_{-\boldsymbol{k}}e^{-i\boldsymbol{k}\vdot\boldsymbol{x}}]}
    \label{eq:zerofluct}
\end{equation}
that consequently satisfies the differential equation
\begin{equation}
    f''_k+k^2f_k=0
\end{equation}
whose solution is the mode function
\begin{equation}
    f_k\propto e^{-i\omega_k \eta},\quad\text{where}\quad \omega_k^2-k^2=0.
    \label{eq:GiraIlMondo}
\end{equation}
Next, promoting the Fourier coefficients into creation and annihilation operators, we discover that the solution has an interesting symmetry. If we define a new mode function which is the rotation of \eq{GiraIlMondo}, we find
\begin{equation}
    f_k=A(k)e^{-i\omega \eta+i\boldsymbol{k}\vdot\boldsymbol{x}}+B(k)e^{-i\omega \eta-i\boldsymbol{k}\vdot\boldsymbol{x}},
\end{equation}
which is still a valid solution of \eq{WaveWave}. However, we can write
\begin{equation}
    f=\int{d\boldsymbol{k}[\hat{b}_{\boldsymbol{k}}e^{-i\omega \eta+i\boldsymbol{k}\vdot\boldsymbol{x}}+\hat{b}^\dagger_{\boldsymbol{k}}e^{+i\omega \eta-\boldsymbol{k}\vdot\boldsymbol{x}}]},
\end{equation}
where
\begin{equation}
    \hat{b}_{\boldsymbol{k}}=A(k)\hat{a}_{\boldsymbol{k}}+B^*(k)\hat{a}^\dagger_{\boldsymbol{k}}.
    \label{eq:BbB}
\end{equation}
It is completely equivalent to our original solution as the new operators \eq{BbB} satisfy the same commutation relation \eq{Communist}. It is possible to demonstrate that we are consequently able to put a condition on the coefficients $A$ and $B$:
\begin{equation}
    \lvert A\rvert^2-\lvert B\rvert^2=1.
\end{equation}
Here, we are at the crux of the matter; considering that
\begin{equation}
    \hat{a}_{\boldsymbol{k}}\ket{0_a}=0 \leftrightarrow \hat{b}_{\boldsymbol{k}}\ket{0_b}=0,
\end{equation}
we have two vacuum states which are not the same. If we rotate the operator, the previous vacuum state will contain particles and a previous state which had particles will become the vacuum. How do we tell which is the physical state? We need to focus on which spacetime the field is living in, e.g. in special relativistic quantum field theory, the true vacuum is the zero-particle state seen by an inertial observer. We also want to require the Lorentz invariance for our physical vacuum. In this way, we are able to fix the vacuum and unambiguously define our creation and annihilation operators\footnote{ An accelerated observer in the Minkowski vacuum will think that the space is full of particles, a phenomenon that is known as the \textit{Unruh effect} \cite{Unruh:1976db}}. In our case of FLRW spacetime, our inertial observer is an observer at rest in comoving coordinates. In the ultraviolet limit, a FLRW spacetime is asymptotically Minkowskian (see the previous section). Hence, we choose the vacuum field that corresponds to the usual Minkowski vacuum in that limit (see e.g. Chapter 3 in \cite{Birrell:1982ix}), which means, from \eq{RotTor}, that
\begin{equation}
    f_k(\eta)\propto e^{-ik\eta}\rightarrow A_k=1, B_k=0.
\end{equation}
This is known as the \textit{Bunch-Davies} vacuum \cite{Bunch:1978yq}. 

\subsection{Phenomenology}
If we apply the slow-roll approximation to \eq{MS}, $z''/z\sim 2/\eta^2$ we end up with
\begin{equation}
    \boxed{f''_\textbf{k}+\(k^2-\frac{2}{\eta^2}\)f_\textbf{k}}
    \label{eq:Rising}
\end{equation}
whose solution, with \eq{condition} is \textit{Bunch-Davies mode function}
\begin{equation}
    \boxed{f_k(\eta)=\frac1{\sqrt{2k}}\(1-\frac{i}{k\eta}\)e^{-ik\eta}}\,.
\label{eq:BDmf}
\end{equation}
The operator $\hat{f}$ in \eq{zerofluct} has zero expectation value. But the variance does not vanishes
\begin{equation}
    \langle\lvert\hat{f}\rvert^2\rangle=\int{\frac{d^3k}{(2\pi)^3}}\int{\frac{d^3k'}{(2\pi)^3}f_k(\eta)f^\star_{k'}(\eta)\bra{0}[\hat{a}_\textbf{k},\hat{a}^\dagger_{-\textbf{k}}\ket{0}]}=\int d\ln{k}\frac{k^3}{2\pi^2}\lvert f_k(\eta)\rvert^2
\end{equation}
from which comes the interesting relation 
\begin{equation}
    \Delta^2_f(k,\eta)\equiv\frac{k^3}{2\pi^2}\lvert f_k(\eta)\rvert^2.
\end{equation}
If we use the Bunch-Davis mode function presented in \eq{BDmf} and explicitly write the definition of $f$, we end up with
\begin{equation}
    \Delta^2_{\delta\phi}(k,\eta)=\(\frac{H}{2\pi}\)^2\[1+(k\eta)^2\]
    \label{eq:PSphi}
\end{equation}
where on superhorizon limit, we have a scale-invariant spectrum, but, as previously stated, there is a small time dependence on the Hubble rate during inflation. Now we are able to connect the missing dots. The primordial curvature perturbatins are related via \eq{Rflz}. Therefore, we get
\begin{equation}
    \boxed{\Delta^2_\mathcal{R}=\left.\frac{1}{8\pi^2\varepsilon_\Hl}\frac{H^2}{\M}\right\rvert_{k=aH}}
    \label{eq:PSscalar}
\end{equation}
The time-dependence is captured by the slow-roll parameter. We define the scalar spectral index as
\begin{equation}
    \boxed{n_s-1\equiv \frac{d\ln{\Delta^2_\mathcal{R}(k)}}{d\ln{k}}}
    \label{eq:nsns}
\end{equation}
which, at horizon crossing ($k=aH$) can be written as
\begin{equation}
    \boxed{n_s-1=-2\epsilon_\Hl-\eta_\Hl}
    \label{eq:nsnsepsilon}
\end{equation}
and if we call 
\begin{equation}
     \boxed{A_s\equiv\frac{1}{8\pi^2\varepsilon_{\Hl\star}}\frac{H_\star^2}{\M}}
\end{equation}
we have
\begin{equation}
    \boxed{\Delta^2_\mathcal{R}(k)=A_s\(\frac{k}{k_\star}\)^{n_s-1}}\,.
    \label{eq:PSfirst}
\end{equation}
We can also introduce the next-to-leading order generalization with a scale dependency of the spectral index, by including its running
\begin{equation}
    \boxed{\alpha_s\equiv\frac{\d n_s}{\d\ln{k}}}\,.
    \label{eq:runningss}
\end{equation}
Eventually, using \eq{PSfirst} and the definitions \eq{runningss} and \eq{nsns}, we can write
\begin{equation}
    \boxed{\ln \Delta_{\rm s}(k)=\ln(A_{\rm s}) + (n_{\rm s}-1)\,\ln(k/k_{\star}) + \alpha_{\rm s}\,\ln^2(k/k_{\star})}
\label{eq:PLscalarexpansion}
\end{equation}

\subsubsection{Transfer Function}
We have computed the power spectra of the primordial scalar, $\mathcal{R}$, and similar calulations are done for tensor fluctuations, $h$ (see \chap{PGWs}). In particular, we have evaluated them at horizon exit because for greater wavelength they froze until they re-enter. 
\begin{figure}[htp]
\centering
\includegraphics[width=0.9\textwidth]{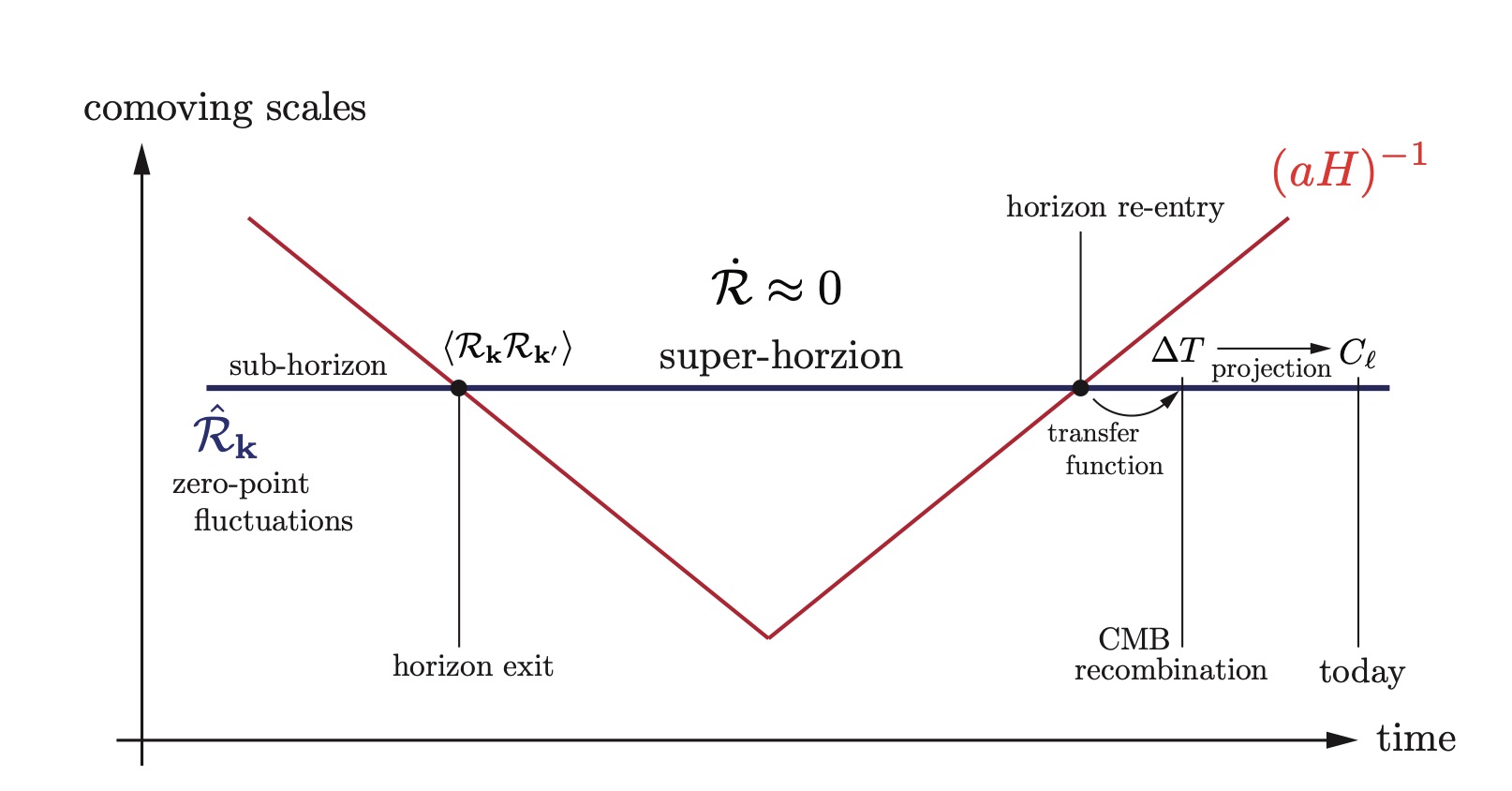}
\caption[Evolution of perturbations]{\small This picture underlines the evolution of perturbations in the inflationary Universe. On superhorizon scale, they are approximately constant. After horizon reentry, they evolve into quantities that we observe, e.g. the anisotropies in the CMB. To relate the primordial and observed quantities we need a transfer function. Image taken from~\cite{Baumann:2009ds}.}
\label{fig:TransferFunction}
\end{figure}
As shown in \fig{TransferFunction}, we want to relate the previous results to the present observables, like the CMB. In order to do that, we need to introduce the \textit{transfer function} $T_\mathcal{Q}$: a relation between $\mathcal{R}$ fluctuations at a generic time $\eta_\star$ of horizon exit and a generic $\mathcal{Q}$ fluctuations at some later time. The latter corresponds to what is really measured in an experiment. In particular
\begin{equation}
   \boxed{\mathcal{Q}_{\boldsymbol{k}}(\eta)=T_{\mathcal{Q}}(k,\eta,\eta_\star)\mathcal{R}_{\boldsymbol{k}}(\eta_\star)}\,.\label{eq:TrF}
\end{equation}
According to \eq{Hello}, we can write 
\begin{equation}
    \boxed{\Theta_{\ell m}=4\pi (-i)^\ell\int{\frac{d\boldsymbol{k}}{(2\pi)^3}T_{T\ell}(k)\mathcal{R}_{\boldsymbol{k}}Y_{\ell m}(\hat{k})}}\,.
\end{equation}
Now, substituting in \eq{po} using the relation \eq{po}, we eventually get
\begin{equation}
    C_\ell^{TT}=\frac{2}{\pi}\int{d\ln{k}\Delta_{\mathcal{R}}(k)T_{T\ell}(k)T_{T\ell}(k)}.
\end{equation}
The transfer function generally has to be computed numerically using Boltzmann-codes such as \texttt{CAMB}~\cite{Lewis:1999bs,Howlett:2012mh} or \texttt{CLASS}~\cite{Blas:2011rf}. We are greatly interested in large-scale CMB as the modes were still outside the horizon at recombination and consequently the spectrum has not been affected by subhorizon evolution and is plainly the geometric projection of the primordial spectrum from recombination to us, today. In this regime, $T_{T\ell}$ is simply a Bessel function
\begin{equation}
    T_{T\ell}(k)=\frac13 j_\ell(k[ \eta_0-\eta_{rec}]).
\end{equation}
Thus, we have
\begin{equation}
    C_\ell^{TT}=\frac{2}{9\pi}\int{d\ln{k}\Delta_{\mathcal{R}}(k)j^2_{\ell}(k[\eta_0-\eta_{rec}])}.
\end{equation}
The Bessel function is peaked at $k[\eta_0-\eta_{rec}]\approx \ell$ and so it acts like a $\delta$-function. We have~\cite{Dodelson:2003ft}
\begin{equation}
    \ell(\ell+1)C_\ell^{TT}\propto \left.\Delta_{\mathcal{R}}^2(k)\right\rvert_{k\approx \ell/(\eta_0-\eta_{rec})}\propto \ell^{n_s-1}.
\end{equation}

Similarly, we can define the power spectrum of $E$ \eq{EETE} and $B$ modes \eq{BB}
\begin{gather}
C_\ell^{EE}\approx (4\pi)^2\int{d\ln{k}\Delta_{\mathcal{R}}(k)T_{E\ell}(k)T_{E\ell}(k)},\\
C_\ell^{TE}\approx (4\pi)^2\int{d\ln{k}\Delta_{\mathcal{R}}(k)T_{T\ell}(k)T_{E\ell}(k)},\\
C_\ell^{BB}\approx (4\pi)^2\int{d\ln{k}\Delta_{h}(k)T_{B\ell}(k)T_{B\ell}(k)}.
\end{gather}
As we have already mentioned, the quadrupole anisotropy imprinted by the gravitational waves contributes to the $BB$ power; the other power spectrums are dominated by inflationary scalar modes. In conclusion, if we call $\Delta(k)=\{\Delta_{\mathcal{R}}(k),\Delta_h(k)\}$ we can write
\begin{equation}
    \boxed{C_\ell^{XY}=\frac{2}{\pi}\int{d\ln{k}\Delta(k)T_{Xl}(k)T_{Yl}(k)}\label{CXYT}}\,,
\end{equation}
where $X,Y=T,E,B$. Generally speaking, the transfer function beyond the large-scale approximation can be written as the line-of-sight integral~\cite{Dodelson:2003ft}
\begin{equation}
    T_{X\ell}(k)=\int_0^{\eta_0}d\eta S_X(k,\eta)P_{X\ell}(k[\eta_0-\eta])
\end{equation}
where we have factorized $T$ with a source term, $S_X(k,\eta)$, and geometric projection factors $P_{X\ell}$ which are combination of Bessel functions.

\paragraph{\textit{Large Scale Structure}} 
To study inhomogeneities of the Large Scale Structure (LSS), we use a different transfer function. We can roughly assume the following function~\cite{Vittorio:2017foh}
\begin{equation}
    T_{\delta_m}(k)\approx\left\{ \begin{array}{cl}
    1& k<k_{eq}\\
    \left(\frac{k_{eq}}{k}\right)^2& k>k_{eq}
    \end{array}\right.
    \label{eq:TransMat}
\end{equation}
where $\delta_m$ is the matter perturbations that evolves according to \eq{MZSE} and \eq{mmdom}. However, a more accurate fitting function for the matter transfer function is the following \cite{Bardeen:1985tr,Eisenstein:1997ik}
\begin{equation}
    T_{\delta_m}(q)=\frac{\ln(1+2.34q)}{2.34q}[1+3.89q+(1.61q)^2+(5.46q)^3+(6.71q)^4]^{-\frac14},
\end{equation}
where
\begin{equation}
    q=\frac{k}{\Gamma h}Mpc^{-1},\quad \Gamma\equiv \Omega h e^{-\Omega_b-\sqrt{2h}\frac{\Omega_b}{\Omega}},
\end{equation}
with $\Gamma$ called the shape parameter. Exact transfer functions can be computed numerically using, again, numerical codes. That said, what is relevant is the fact that \textit{exists} a transfer function which can relate the dark matter power spectrum $P_m(k,z)$ to the inflationary spectrum~\cite{Baumann:2009ds}
\begin{equation}
    \boxed{P_m(k\eta)=\frac{4}{25}\left(\frac{k}{aH}\right)^4T^2_{\delta_m}(k,\eta)P_{\mathcal{R}}(k)}\,,
    \label{eq:TransferLSS}
\end{equation}
where for convention we have factored out the numerical and $k$ factors from the transfer function. The dark matter is only directly observed through gravitational lensing. On the other hand, we are able to observe luminous or baryonic matter. The assumption
\begin{equation}
    \delta_g(x)=\delta(x)
\end{equation}
where $g$ stands for galaxies, is generally wrong because the galaxies distribution does not follow the dark matter's one. Thus, we should introduce a new degree of freedom in the relation, the \textit{bias factor}
\begin{equation}
    \delta_g(x)=b\delta(x).
\end{equation}
The bias factor could be a long density function or a simple constant. It describes our ill-understood physics of galaxy formation. Nevertheless, we are more interested in the power spectrum relation
\begin{equation}
   \boxed{P_{\delta_g}=b^2 P_m}
\end{equation}
which should be paired with the relation in \eq{TransferLSS}.

%% file: Chapters/EFT.tex
\chapter{Effective Field Theory of Inflation}
\label{iphone}
Nature comes to us in many scales and phenomena involving distinct scales can be analysed by considering one relevant scale at a time. For instance, one does not worry about the size of planets when studying the motions in the Solar system, as well as we can ignore the presence of gluons and quarks inside a proton when the hydrogen spectrum is studied. The \textit{Effective Field Theory} (see e.g.~\cite{Skiba:2010xn,Luty:2005sn,Kaplan:2005es}) (EFT) occurs from such scale separation in quantum field theories. The effects of large, or short, energy scales are suppressed by powers of the ratio of scales in the problem so we end up with the lowest dimension operators compatible with the underlying symmetries. In what follows, we outline the basics of EFT

\begin{itemize}
\item Usually, the heavy fields are integrated out by performing path integral over the heavy degrees of freedom\footnote{Heavy degrees correspond, for example, to particles that cannot be produced on shell at the energies available to the experiment interest.}. As a result, we have an \textit{effective action} for light degrees of freedom. We can write it by defining the Lagrangian density
\begin{equation}
    \mathcal{L}_{eff}(\phi_L)=\mathcal{L}_{\Delta\leq4}+\sum_ic_i\frac{\mathcal{O}_i(\phi_L)}{\Lambda^{\Delta_i-4}},
    \label{eq:Leef}
\end{equation}
where the first term contains the finite number of renormalizable terms of dimension four or less, whilst the sum is over the non-renormalizable terms. $\Delta_i$ are the dimensions of the operators $\mathcal{O}_i$. These operators are made of the light degrees of freedom and are local in spacetime; for example in a relativistic scalar theory with the symmetry $\phi_L\rightarrow-\phi_L$, they take the form $\phi_L^{4+2n}$ or $(\partial_\mu\phi_L)^2\phi_L^{2n}$ and so forth. However, we shall note that the infinite sum reduces to a finite number of terms because, at the end, we need to reproduce experiments to finite accuracy. In \eq{Leef}, we have used the same suppressing scale $\Lambda$ which is an oversimplification. $\Lambda$ is usually called the cut-off of the effective field theory. 
\item We can use the EFT when the full theory is unknown. When it decouples at scale $\Lambda$, the physics manifests itself at low energies as an effective Lagrangian of the form \eq{Leef}. Besides, if the symmetries that survive at low energies are known, then the operator $\mathcal{O}_i(x)$ in \eq{Leef} must respect those symmetries. Thus, we can write the effective Lagrangian  with the most general set of operators consistent with the symmetries and we take into account the UV physics in a completely model-independent way. 
\item We should be able to predict the magnitudes of the different operators $\mathcal{O}_i$, using the \textit{ratios of energy scales}. The simplest way is by conducting a dimensional analysis in natural units, where $[m]=[l]^{-1}$. Since the action is dimensionless, the Lagrangian has dimension $4$. The dimensions of the fields are determined from the kinetic energies because in weakly interactive theories these terms always dominate.\footnote{For example the kinetic term for a scalar field $\partial_\mu\phi\partial^\mu\phi$ implies that $\phi$ has dimension $1$, whereas in the fermion case, $i\bar{\psi}\slashed{\partial}\psi$, $\psi$ has dimension $\frac32$. Always implying four-dimensional spacetime.}. For processes at scale $E$, we estimate dimensionally the magnitude of a given term in the action in \eq{Leef} as
\begin{equation}
    c_i\left(\frac{E}{\Lambda}\right)^{\Delta_i-D},
\end{equation}
where we have generalized the result to $D$ dimensions. Operators with large dimensions are therefore suppressed at $E\ll \Lambda$. 
\item Once we are at energies below $\Lambda$, the behaviour of the different operators is determined by their dimension. In four-dimensional spacetime, if $\Delta>4$, we have \textit{irrelevant operators} that can be neglected because suppressed by powers of $E/\Lambda$ at low energies. Conversely, $\Delta<4$ indicates all the \textit{relevant operators} which become more important at lower energies. Finally, $\Delta=4$ is the dimension of \textit{marginal operators} which are equally important at all energy scales. The EFT which contains only relevant and marginal operators is called \textit{renormalizable}.
\item Lastly, we should take into consideration the symmetries of the theory. They can put a further constraint on the infinite sum in \eq{Leef}. The gauge symmetries and global symmetries possessed in the UV-theory are inherited by the low energy theory. However, global symmetries can be spontaneously broken with the consequent appearance of massless particles (we will see better in the following sections)  which are light degrees of freedom. 
\end{itemize}

In this chapter, our goal is to formulate inflation as an example of a spontaneous symmetry breaking theory, through an effective field theory. From \sect{GoldTheorem} we study the phenomenon of spontaneous symmetry breaking and see how the Goldstone boson arises in both global and gauge symmetries. At last, in \sect{GoldsotneBos} we will apply the results to our case of interest: cosmology, where the broken symmetry is the time translation invariance of the spacetime.

\section{Spontaneous symmetry breaking}
\label{sec:GoldTheorem}
A global continuous symmetry implies a Noether current $J_\mu(x)$ which is conserved ($\partial_\mu J^\mu=0$) on the equation of motion. It is possible to demonstrate that the conserved charge
\begin{equation}
    Q=\int{d\boldsymbol{x}J_0(\boldsymbol{x})}
\end{equation}
 generates the symmetry transformations \cite{Schwartz:2014sze,Weinberg:1995mt,Itzykson:1980rh,Peskin:1995ev}. Namely, we can write
\begin{equation}
    \left[Q_a,Q_b\right]_{x^0=y^0}=f_{abc}Q_c,
    \label{eq:formgroup}
\end{equation}
where $f_{abc}$ are the group structure constants. This implies that the $Q^a$ are an algebra representation of the group $G$, reducible. Hence, an infinitesimal transformation can be written as
\begin{equation}
    U=e^{ig^aQ^a}\approx I+ig^aQ^a,
\end{equation}
which, applied to the ground state, gives
\begin{equation}
    U\ket{0}=I\ket{0}+ig^aQ^a\ket{0}.
\end{equation}
From the last result we can see that if $Q^a\ket{0}=0$, the vacuum is invariant. We can now introduce the \textit{Goldstone's Theorem}:
\emph{If the vacuum is not invariant under a group of symmetry transformations, we should expect the existence of massless particles.}
In other words, whenever there is a continuously broken symmetry\footnote{If we have a degenerate ground state, as we have seen in \eq{TaylorF}, the corresponding eigenstates are not invariant under the symmetry transformations of the Lagrangian. Once we select one of the degenerate states, the ground state no longer shares the symmetry of the system. When such a choice is made, the spontaneous symmetry breaking occurs.}, a massless scalar particle appears for each generator of the symmetry that is broken \cite{Goldstone:1961eq,Goldstone:1962es,Nambu:1961tp}; it is the \textit{Nambu-Goldstone boson} or simply \textit{Goldstone boson} . A symmetry is spontaneously broken when the ground state no longer shares the symmetries of $\mathcal{L}$.

\subsection{The Goldstone boson is massless} 
Let us suppose that our system is invariant under a symmetry group $G$. We have shown that the conserved charges $Q^a$ are also a representation of the group. To show that the vacuum state is not invariant, we just have to prove that  $Q^a\ket{0}\neq0$. Suppose we have $Q^a=Q^{a\dagger}$, we find
\begin{equation}
    \bra{0}Q^aQ^a\ket{0}=\int{d\boldsymbol{x}d\boldsymbol{y}\bra{0}J_0^a(x^0,\boldsymbol{x})J_0^a(y^0,\boldsymbol{y})\ket{0}},
\end{equation}
which diverges for the translational invariance. Hence, we need to define the non invariance of the vacuum in other terms. Introducing a set of scalar operators which reproduce the fundamental commutator of the group,  $\{O_i(x)\}$,
\begin{equation}
    \comm{Q^a}{O_i(x)}=i T^a_{ij}O_j(x),
    \label{eq:parameters}
\end{equation}
 where $T^a$ is a matrix belonging to an irreducible representation of $G$ (it has not to be fundamental). Without any loss of generality, we can set $x=0$ and do the expectation value in the vacuum of \eq{parameters}. The non-invariance is now defined as 
 \begin{equation}
     \bra{0}\comm{Q^a}{O_i(0)}\ket{0}=i T^a_{ij}\bra{0}O_j(0)\ket{0}\neq 0,
     \label{eq:kkknunu}
 \end{equation}
which is a better definition than the previous one because an irreducible representation implies
\begin{equation}
    T^a_{ij}v_j=0 \rightarrow v_j=0, \forall j.
\end{equation}
Thus, it is sufficient that one of the expected values $\bra{0}O_j(0)\ket{0}$ is different from zero, to imply that $Q^a$ does not annihilate the vacuum and consequently is not invariant
\begin{equation}
    \bra{0}Q^aO_i(0)\ket{0}-\bra{0}O_i(0)Q^a\ket{0}\neq 0.
    \label{eq:noninvariance}
\end{equation}
The condition of spontaneous symmetry breaking is that we have at least one $\bra{0}O_j(0)\ket{0}\neq 0$. To begin with, we take the first term of the commutator in \eq{noninvariance}:
\begin{equation}
\bra{0}Q^aO_i(0)\ket{0}=\int_{\mathbb{R}^3}{\dd[3]{x}\bra{0}J_0^a(0,\textbf{x})O_i(0)\ket{0}}.
\end{equation}
Now we calculate the generic element
\begin{equation}
    \bra{0}J_\mu^a(\textbf{x})O_i(0)\ket{0}
\end{equation} 
and we will set $\mu=0$ at the end of the calculations. Introducing a sum over all the eigenstates of the momentum $p_\mu$ and applying the translation operators, we get
\begin{equation}
   \sum_n \bra{0}e^{i\textbf{p}\vdot\textbf{x}}J_\mu^a(0)e^{-i\textbf{p}\vdot\textbf{x}}\ket{n}\bra{n}O_i(0)\ket{0}=\int_{\mathbb{R}^4}{\dd[4]{q}e^{-i\textbf{q}\vdot\textbf{x}}\widetilde{\rho}^a_{\mu i}(q)},
\end{equation}
where
\begin{equation}
    \widetilde{\rho}^a_{\mu i}(q)\equiv\sum_n \bra{0}J_\mu^a(0\ket{n}\bra{n}O_i(0)\ket{0}\delta^{(4)}(q-p_n).
    \label{eq:rhotilde}
\end{equation}
The function \eq{rhotilde} has the support within the physical states spectrum since $\widetilde{\rho}^a_{\mu i}(q)\neq 0$ only if $q=p_n$. As a consequence, we must impose $q^0\geq0$ and $q^2\geq 0$. Also, we see that from the Lorentz invariance
\begin{equation}
    \widetilde{\rho}^a_{\mu i}(\Lambda q)=\Lambda^\nu_\mu\widetilde{\rho}^a_{\nu i}(q),
\end{equation}
the index $\mu$ is brought by $q$. Thus, we can recast the definition \eq{rhotilde} and obtain
\begin{equation}
     \tilde{\rho}^a_{\mu i}(q)=\rho_i^a(q^2)\frac{\theta(q^0)}{(2\pi^3)}q_\mu.
\end{equation}
From the current conservation, $\partial_\mu J^{a\mu}=0$, we get
\begin{equation}
    \partial_\mu\bra{0}J^{a\mu}(x)O_i(0)\ket{0}\rightarrow \partial_\mu\int{d^4q e^{-i q\vdot x}\widetilde{\rho}^{a\mu}_{ i}(q)=0}
\end{equation}
and eventually 
\begin{equation}
    \rho_i^a(q^2)q^2\frac{\theta(q^0)}{(2\pi)^3}=0.
\end{equation}
If $\rho_i^a(q^2)$ was a function, it would be equal to $0$. But, since it is a distribution, we obtain
\begin{equation}
    \rho_i^a(q^2)=c_i^a\delta(q^2).
\end{equation}
Therefore, we have
\begin{equation}
    \bra{0}J_\mu^a(\boldsymbol{x})O_i(0)\ket{0}=c_i^a\int_{\mathbb{R}^4}{\frac{\dd[4]{q}}{(2\pi)^3}q_\mu e^{-i\boldsymbol{q}\vdot\boldsymbol{x}}\delta(q^2)\theta(q^0)}.
    \label{eq:1stcommutator}
\end{equation}
Since the support of $\rho_i^a$ is the spectrum physical states and that $\rho_i^a$ is proportional to $\delta(q^2)$, there could be a state with $q^2=0$ which actually represents a massless particle. However, to be certain, we have to show that $c_i^a\neq 0$. To achieve that, we will use the \textit{CPT theorem} which states that any Lorentz-invariant local theory, is invariant under the CPT symmetry. If we call $\Theta$ the CPT operator with the following property
\begin{gather}
    (\phi,\Theta^+ \psi)=(\Theta \phi,\psi)^*,\\
    \Theta^\dagger J^\mu(\boldsymbol{x})\Theta=-J^\mu(-\boldsymbol x),\\
    \Theta^\dagger O_i(\boldsymbol x)\Theta=O(-\boldsymbol x),\\
    \Theta^\dagger\Theta=I=\Theta\Theta^\dagger,\\
    \Theta^{-1}=\Theta^\dagger,\\
    \Theta\ket{0}=\ket{0}
\end{gather}
and use the notation $\psi_\Omega=\ket{0}$, we can write
\begin{multline}
    \bra{0}J^a_\mu(\boldsymbol{x})O_i(0)\ket{0}=\left(\psi_\Omega, \Theta^\dagger\Theta J_\mu^a(\boldsymbol{x})O_i(0)\psi_\Omega\right)=
    \left(\Theta\psi_\Omega, \Theta 
    J_\mu^a(\boldsymbol{x})O_i(0)\psi_\Omega\right)^*\\
    =\left(\psi_\Omega, \Theta J_\mu^a(\boldsymbol{x})O_i(0)\psi_\Omega\right)^*
    =\left(\Theta J_\mu^a(\boldsymbol{x})\Theta^\dagger\Theta 
    O_i(0)\Theta^\dagger\Theta\psi_\Omega,\psi_\Omega\right)\\
    =-\left(J_\mu^a(-\boldsymbol{x})O_i(0)\psi_\Omega,\psi_\Omega\right)=-\left(O_i(0)\psi_\Omega,J_\mu^a(-\boldsymbol{x})\psi_\Omega\right)\\
    =-\left(\psi_\Omega,O_i(0)J_\mu^a(-\boldsymbol{x})\psi_\Omega\right)=-\bra{0}O_i(0)J^a_\mu(-\boldsymbol{x})\ket{0}.
\end{multline}
Hence,
\begin{equation}
    \bra{0}O_i(0)J_\mu^a(\boldsymbol{x})\ket{0}=-\bra{0}J_\mu^a(-\boldsymbol{x})O_i(0)\ket{0}=-c_i^a\int_{\mathbb{R}^{3,1}}{\frac{\dd[4]{q}}{(2\pi)^3}q_\mu e^{i\boldsymbol{q}\vdot\boldsymbol{x}}\delta(q^2)\theta(q^0)}.
    \label{eq:2ndcommutator}
\end{equation}
 Being \eq{1stcommutator} the complex conjugate of \eq{2ndcommutator}, the factor $c_i^a$ is an imaginary number 
 \begin{equation}
     (c_i^a)^*=-c^a_i.
      \end{equation} 
We eventually obtain 
\begin{equation}
    \bra{0}\comm{J_\mu^a(\boldsymbol{x})}{O_i(0)}\ket{0}=ic_i^a\partial_\mu \Delta_c(\boldsymbol{x};0),
\end{equation}
where
\begin{equation}
    \Delta_c(x;m)=\bra{0}\comm{\phi(x)}{\phi(0)}\ket{0}
\end{equation}
and
\begin{equation}
\bra{0}\phi(x)\phi(0)\ket{0}=\int_{\mathbb{R}^3}{\frac{\dd[3]{q}}{(2\pi)^32\omega_q}e^{-i\boldsymbol{q}\vdot \boldsymbol{x}}}\quad \text{with} \quad \omega_q=\sqrt{\lvert\boldsymbol{q}\rvert^2+m^2}
\end{equation}
for a free scalar theory. Eventually, we arrive at the relation
\begin{equation}
    \bra{0}\comm{Q^a}{O_i(0)}\ket{0}=c_i^a.
\end{equation}
Hence, recalling \eq{kkknunu}, we have a spontaneous symmetry breaking if $c_i^a\neq 0$ which implies, from the fact that \eq{1stcommutator} represents the spectrum of physical states, that there are massless particles in the theory.

\subsection{The Goldstone boson is a spinless boson}
\label{sec:NGB}
Next, we want to know what kind of particles they are. Let us indicate such particles with the state $\ket{GB}$. We have to bear in mind that in order to have such particles, the coefficient related to the massless particles ($p_n^2=0$) in the spectral function \eq{rhotilde} has to be non-vanishing
\begin{equation}
\bra{0}J_\mu^a(0)\ket{GB}\bra{GB}O_i(0)\ket{0}\neq 0,
\end{equation}
which holds only if 
\begin{gather}
    \bra{0}J_\mu^a(0)\ket{GB}\neq0, \label{eq:conditionprime}\\
    \bra{GB}O_i\ket{0}\neq0.
    \label{eq:conditionboson}
\end{gather}
The condition in \eq{conditionboson} brings us to two important results:
\begin{enumerate}
    \item Firstly, if we induce a rotation of $2\pi$ with the operator $U(2\pi)$,  we get
\begin{equation}
    \bra{0}O_i(0)\ket{GB}=\bra{0}U^\dagger(2\pi)O_i(0)U(2\pi)\ket{GB}=\bra{0}O_i(0)\left(U(2\pi)\ket{GB}\right).
    \label{eq:Boson}
\end{equation}
If these new particles were fermions, there would be a minus sign in the last member in \eq{Boson} and the term would be zero, which violates our requirement in \eq{conditionboson}. That is, these massless particles are \textit{boson}.
\item Secondly, we  define the projection of the spin $s_p=\textbf{s}\hat{p}$ along the momentum direction and we make a general rotation around $\hat{p}$. We get
\begin{equation}
    \bra{0}O_i(0)\ket{\textbf{p},s}=\bra{0}U^\dagger(\hat{p},\theta)O_i(0)U(\hat{p},\theta)\ket{\textbf{p},s}=e^{is_p\theta} \bra{0}O_i(0)\ket{\textbf{p},s}.
\end{equation}
Thus 
\begin{equation}
    e^{is_p\theta}=1,\quad \forall \theta\in [0,2\pi)\rightarrow s=0.
\end{equation}
 We have found that from a continuously breaking symmetry a massless boson arises and it is \textit{spinless}. 
\end{enumerate}

Being aware of the above results,  we focus on \eq{conditionprime}. We change the notation $\ket{GB}=\ket{\tilde{p}}$, where the tilde stands for a covariant normalized state, and write
\begin{equation}
    \bra{0}J_\mu^a(\textbf{x})\ket{\tilde{p}}=e^{-i\textbf{p}\vdot\textbf{x}}\bra{0}J_\mu^a(0)\ket{\tilde{p}}=e^{-i\textbf{p}\vdot\textbf{x}}F_\mu^a(\textbf{p}),
\end{equation}
where we have used the invariance under translations. To study the function $F_\mu^a(\textbf{p})$ we perform a Lorentz transformation
\begin{equation}
    F^{a\mu}(\Lambda \textbf{p})=\bra{0}J^{a\mu}(0)\ket{\tilde{\Lambda p}}=\bra{0}U^\dagger(\Lambda)J^{a\mu}(0)U(\Lambda)\ket{\tilde{p}}=\Lambda^\mu_\nu\bra{0}J^{a\nu}(0)\ket{\tilde{p}}.
\end{equation}
Hence, 
\begin{equation}
    F^{a\mu}(\Lambda \textbf{p})=\Lambda^\mu_\nu F^{a\nu}(\textbf{p})\rightarrow F^{a\mu}(\boldsymbol{p})=F^a(p^2)p^\mu
\end{equation}
because the index $\mu$ is brought by $\boldsymbol p$. $F^a(p^2)$ is a relativistic invariant, thus, we can write $F^a(p^2)=f^a$ and get
\begin{equation}
    \bra{0}J^{a\mu}(\boldsymbol{x})\ket{\widetilde{p}}=e^{-i\boldsymbol{p}\vdot\boldsymbol{x}}f^ap^\mu.
\end{equation}
If we take the derivative
\begin{equation}
    \partial_\mu\bra{0}J^{a\mu}(\textbf{x})\ket{\tilde{p}}=-ie^{-i\textbf{p}\vdot\textbf{x}}f^ap_\mu p^\mu
\end{equation}
and consider that the derivative must be zero, we obtain $p^2=0$ because $f^a$ cannot be zero otherwise there would be no contribution on the sum of the states. Our massless bosons are coupled to the current in the following way
\begin{equation}
\boxed{
    \bra{0}J^{a\mu}(\textbf{x})\ket{\tilde{p}}=e^{-i\textbf{p}\vdot\textbf{x}}f^ap^\mu}\,.
    \label{eq:orderparameter}
\end{equation}
The current interpolates the vacuum state and the Goldstone boson state with a strength set by the scale $f$. This identifies $f$ as the order parameter of the symmetry breaking and which will be of vital importance on the construction of our effective field theory.

\subsection{Goldstone model}
\begin{figure}
\centering
\includegraphics[scale=0.25]{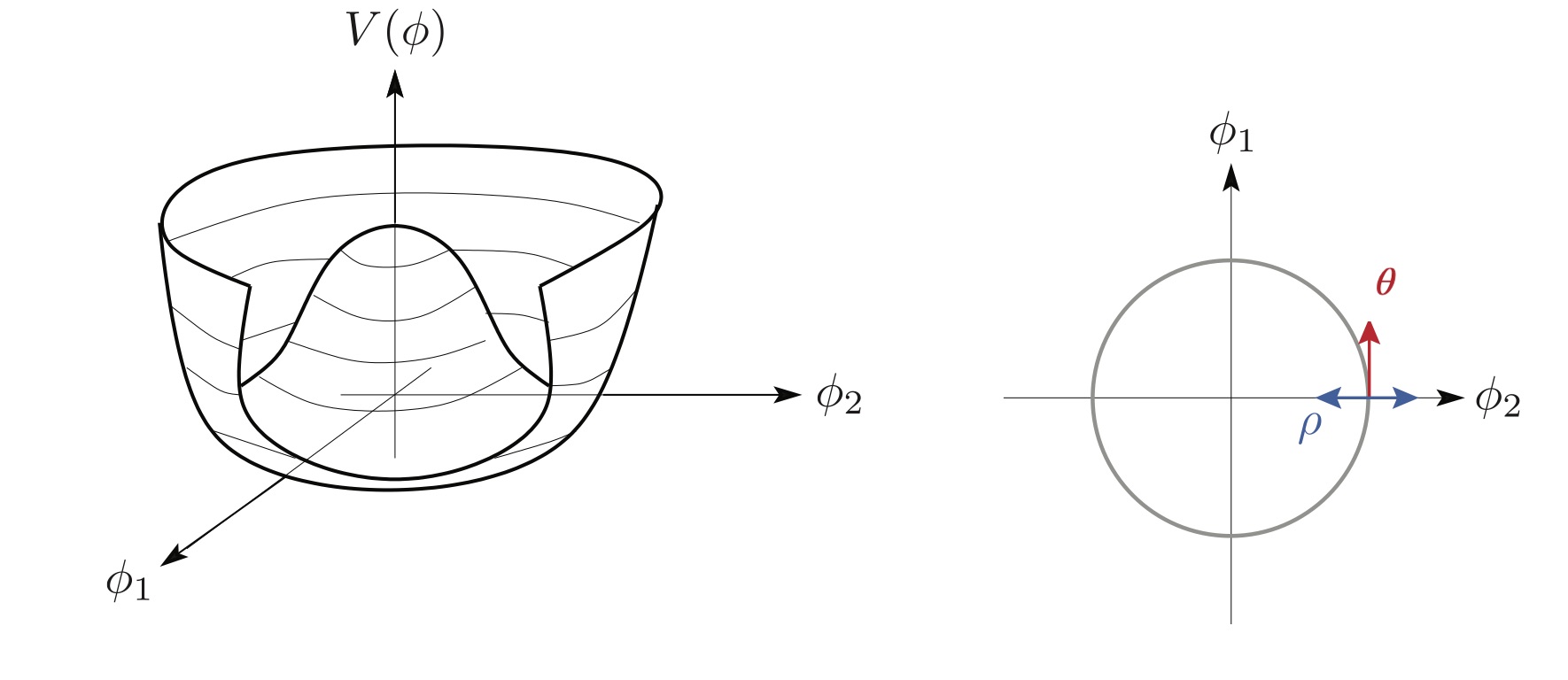}
\caption[Mexican hat potential]{\small When $m^2<0$ in \eq{Gmodel}, the potential assume the so-called \textit{Mexican hat} form, displayed in the left-hand side graph. In the right panel, we can see why the field $\rho(x)$ represents the displacement in the radial plane, whereas $\theta(x)$ represents it along the valley of minimum potential energy. Image taken from~\cite{Senatore:2016aui}.} 
\label{fig:Displacement}
\end{figure}

We now want to study the simplest example of a field theory exhibiting spontaneous symmetry breaking: the \textit{Goldstone model}~\cite{Coleman:1985rnk}. Its Lagrangian density is
\begin{equation}
    \mathcal{L}= \partial_\mu\bar{\phi}(x)\partial^\mu\phi(x)-V(\phi),
    \label{eq:Gmodel}
\end{equation}
with
\begin{equation}
    V(\phi)=m^2\bar{\phi}(x)\phi(x)+g\left(\bar{\phi}(x)\phi(x)\right)^2
\end{equation}
and
\begin{equation}
    \phi(x)=\frac 1 {\sqrt{2}}\left(\phi_1(x)+i\phi_2(x)\right), \quad \bar{\phi}(x)=\frac 1 {\sqrt{2}}\left(\phi_1(x)-i\phi_2(x)\right),
    \label{eq:phicomplex}
\end{equation}
where $\phi_{1,2}(x)$ are two real scalar fields. $\mathcal{L}$ is invariant under the group $U(1)$ of global phase transformations 
\begin{equation}
    \left\{ \begin{array}{c}
    \phi(x)'=e^{i\alpha}\phi(x)\\
    \bar{\phi}'(x)=e^{-i\alpha}\bar{\phi}(x)
    \end{array}\right.
    \label{eq:globaltransformation}
\end{equation}
with $\alpha\in \mathbb{R}$. The Hamiltonian density of the system is
\begin{equation}
    \mathcal{H}=\partial^0\bar{\phi}(x)\partial_0\phi(x)+\na\bar{\phi}(x)\na\phi(x)+V(\phi).
    \label{eq:Hgoldstone}
\end{equation}
Thus, the constant $g$ has to be positive in order to bound from below the energy of the fields. Since the derivative terms in \eq{Hgoldstone} are positive and vanish for a constant value of the fields, the minimum value of the system's energy corresponds to the value of $\phi(x)$ and $\bar{\phi}(x)$ which minimize the potential $V(\phi)$. Depending on the sign of $m^2$ we will face two different situations:
\begin{itemize}
    \item If $m^2>0$, both the terms in the potential are positive. $V(\phi)$ is a paraboloid and has an absolute minimum at $\phi(x)=\bar{\phi}(x)=0$. In this case the ground state shares the symmetry of the Lagrangian and no spontaneous symmetry breaking can occur.
    \item If $m^2<0$ the potential has the typical form that is called \textit{Mexican hat} (see \fig{Displacement}), which has a local maximum in  $\phi(x)=\bar{\phi}(x)=0$ and degenerate absolute minima which form a circle at
    \begin{equation}
        \phi(x)=\phi_0=\left(-\frac{m^2}{g}\right)^\frac12e^{i\beta},\quad \text{with}\quad \beta\in[0,2\pi]
        \label{eq:circle}
    \end{equation}
    and $\beta$ defines the direction in the complex $\phi$-plane. Once $\beta$ is chosen, the ground state is no longer invariant under the symmetry. Therefore, in this case, spontaneous symmetry breaking can occur. From \eq{globaltransformation} we see that the choice of $\beta$ is not relevant, thus we choose $\beta=0$ and the equilibrium ground state becomes real
\begin{equation}
    \phi_{eq}=\frac{m}{\sqrt g}=\frac{\bar{\rho}}{\sqrt{2}}.
\end{equation}
\end{itemize}
We will focus on the second case. Let us now introduce two real scalar fields which measure the deviations of $\phi(x)$ from the equilibrium ground state
\begin{gather}
    \phi(x)=\frac{1}{\sqrt{2}}[1+\rho(x)+i\theta(x)],\\
    \bar{\phi}(x)=\frac{1}{\sqrt{2}}[1+\rho(x)-i\theta(x)].
    \label{eq:deviationfields}
\end{gather}
The Lagrangian density is now
\begin{multline}
    \mathcal{L}=\frac12\partial^\mu\rho(x)\partial_\mu\rho(x)-\frac12(2g\bar{\rho}^2)\rho^2(x)+\frac12\partial^\mu\theta(x)\partial_\mu\theta(x)\\
    -g\bar{\rho}\rho(x)(\rho^2(x)+\theta^2(x))-\frac14g(\rho^2(x)+\theta^2(x))^2+c.
    \label{eq:laglag}
\end{multline}
where $c$ represents the constant contributes which can be neglected. Both the Lagrangian in \eq{laglag} and in \eq{Gmodel} represent the same physical result. We can write Eq. \eq{laglag} as a sum of the free Lagrangian plus an interaction term. The free Lagrangian is
\begin{equation}
    \mathcal{L}_0=\frac12\partial^\mu\rho(x)\partial_\mu\rho(x)-\frac12(2g\bar{\rho}^2)\rho^2(x)+\frac12\partial^\mu\theta(x)\partial_\mu\theta(x).
    \label{eq:Lo}
\end{equation}
Since no terms couple the fields in \eq{Lo}, they are normal coordinates. In particular, we see the similarity with the KG equation and consequently identify $\theta(x)$ and $\rho(x)$ as real Klein-Gordon fields. Those fields lead to neutral spinless particles, one with a mass $\sqrt{2g\bar{\rho}}$, $\rho(x)$, and the other is massless, $\theta(x)$, since there are no terms in $\theta^2(x)$. The latter field corresponds to our Goldstone boson. 
As shown in \fig{Displacement}, the field $\rho(x)$ represents a displacement in the radial plane ($\phi_2(x)=0$) in which the potential energy increases quadratically with $\rho(x)$. On the other hand, $\theta(x)$ is the displacement along the valley of minimum potential energy where $V(\phi)$ is constant. Therefore, the corresponding quantum excitations, the $\theta$ bosons, are massless.

\begin{tcolorbox}[mybox]
Let us generalize the previous argument introducing the breakdown of a general continuous internal symmetry (for a more detailed description see e.g.~\cite{Low:2001bw}). If we assume to have $n$ real fields $\phi$ and that the potential is invariant under a group $G$ of transformations, we have
\begin{equation}
    \phi\rightarrow\phi'=e^{i\theta^ag^a}\phi,
\end{equation}
where the $g^a$ are the generators of the group $G$ and $\theta^a$ are arbitrary real parameters. The associated infinitesimal transformations are
\begin{equation}
    \delta \phi=ig_a\delta\theta^a\phi.
\end{equation}
Since the $g_a$ are the generators of the group, they satisfy
\begin{equation}
    [g_a,g_b]=f_{abc}g_c
\end{equation}
as in \eq{formgroup}. The invariance of the Lagrangian in the form \eq{Gmodel}, implies that
\begin{equation}
    V(\phi)=V(e^{i\theta^ag^a}\phi).
\end{equation}
Besides, making the further assumption that the vacuum state, which we will call $\phi_{eq}$, is invariant under a subgroup $H$ of $G$, yields
\begin{equation}
    \delta\phi_{eq}=ig_a\theta^a\phi_{eq}=0 \quad \text{for} \quad a \in H.
    \label{eq:deltaphi0}
\end{equation}
We will denote with $U(g)$ the representation of $G$ and $U(h)$ the representation of $H$. If we now expand the potential $V(\phi)$ around the minimum, for an infinitesimal group transformation we obtain
\begin{equation}
   V(U(g)\phi_{eq})=V(\phi_{eq})+\frac 1 2m_{ij}^2\delta\phi_i\delta\phi_j+...,
   \label{eq:potentialmij}
\end{equation}
where we have defined
\begin{equation}
    m_{ij}^2\equiv \left.\frac{\partial^2 V}{\partial\phi_i\partial\phi_j}\right\rvert_{\phi=\phi_{eq}},
\end{equation}
which are the elements of a symmetric and positive matrix $M^2$. Bearing in mind that the theory is invariant for the group transformations $U(g)$, we have
\begin{equation}
    V(\phi_{eq})=V(U(g)\phi_{eq}),
    \label{eq:Vug}
\end{equation}
The second term on the right-hand side in  \eq{potentialmij} is zero, in particular
\begin{equation}
    M^2\delta\phi_{eq}=M^2[ig_a\theta^a]\phi_{eq}=0,
\end{equation}
and two different possibilities arise:
\begin{enumerate}
    \item If $g \in H$, $\delta\phi_{eq}=0$ as we have seen in \eq{deltaphi0}  and the potential is automatically invariant.
    \item If $g \in G/H$, where $G/H$ is the left coset, the vacuum is not invariant. Thus, in order to satisfy \eq{Vug}, we get that the matrix $M^2$ must have zero eigenvalues.
\end{enumerate}
If we consider the eigenvalues of $M^2$ as the mass of the bosons, we have obtained the same result as the Goldstone model: the theory has $n-n_h$ Goldstone bosons, where $n_h$ is the number of generators of the subgroup of symmetry $H$ for which the vacuum is invariant. Moreover, we have $n_h$ massive bosons in correspondence to the directions of the generators along which the symmetries are unbroken. We have obtained the generalization for the Goldstone theorem: there exists one massless Golstone boson for each broken generator. 
\end{tcolorbox}

\subsubsection{Effective theory}
If we now want to determine the EFT of the Goldstone bosons, we replace the constant transformation parameter $\theta^a$ with a spacetime-dependent parameter $\pi^a(x)$. The fields $\pi^a(x)$ parameterize massless excitations of the theory after spontaneous symmetry breaking along the $n-n_h$ directions. The remaining $n_h$ directions in field space are massive and decouple from the dynamics of the Goldstone bosons.  For this reason, we want to introduce the low-energy effective theory. We define the field
\begin{equation}
    U(x)\equiv e^{i\pi(x)\vdot T}.
    \label{eq:Ufield}
\end{equation}
At lowest order in derivative expansion, it is possible to show that the unique $G$-invariant Lagrangian is \cite{Baumann:2014nda}
\begin{equation}
    \mathcal{L}_{eff}=-\frac{f^2_\pi}4 \Tr[\partial_\mu U\partial^\mu U^{\dagger}],
    \label{eq:Generaldod}
\end{equation}
where $f_\pi$ is a parameter with the dimension of mass. Expanding the Lagrangian in powers of $\pi_c\equiv f_\pi \pi$, it reads
\begin{equation}
    \mathcal{L}_{eff}=-\frac 1 2 \partial_\mu \pi_c \vdot \partial^\mu \pi_c-\frac 1 {6f^2_\pi}\left[(\pi_c\vdot\partial_\mu\pi_c)^2-\pi_c^2(\partial_\mu\pi_c\vdot\partial^\mu\pi_c)\right]+...,
    \label{eq:Leff}
\end{equation}
where we have used the fact that $2\Tr(T^aT^b)=\delta^{ab}$. An infinite series of non-renormalizable interactions appears. The broken symmetry dictates relations among these interactions, and the couplings are determined by the single parameter $f_\pi$. These interactions are called \textit{universal}. Whereas \textit{non-universal} interactions are obtained using higher-order derivative expansion. For instance, we have
\begin{equation}
    \mathcal{L}_{eff}=-\frac{f^2_\pi}{4}\Tr{\partial_\mu U\partial^\mu U^\dagger}+c_1\left(\Tr{\partial_\mu U\partial^\mu U^\dagger}\right)^2+c_2\Tr{\partial_\mu U\partial_\nu U^\dagger}\Tr{\partial^\mu U\partial^\nu U^\dagger}+...,
\end{equation}
where $c_1$ and $c_2$ are model-dependent variables. The Noether conserved current for the effective Lagrangian \eq{Leff} is
\begin{equation}
    J^\mu=-f_\pi\partial^\mu\pi_c+...,
    \label{eq:readoff}
\end{equation}
which compared to \eq{orderparameter} leads us to the approximation $f=f_\pi+\ldots$ Namely, the important fact is that the symmetry breaking scale is restored when the right-hand side of \eq{orderparameter} vanishes. This occurs when high-order corrections cancel the leading term in $f$, which means at energies of order $f_\pi$. Consequently, the symmetry-breaking scale can be read off from the scale appearing in the Noether current for the canonically normalized field.

\section{Higgs model}
\label{sec:HiggsModel}
Goldstone's theorem appears to be clear proof that the spontaneous breakdown of continuous symmetries cannot take place in the real world, since there is no evidence in any experiment whatsoever of such goldstone bosons. However, as we will see, the gauge field theories come to the rescue. 
 Let be $\phi$ a set of fields whose dynamics is determined by $\mathcal{L}$. The Lagrangian is invariant under one parameter group transformation
 \begin{equation}
     \phi\rightarrow e^{iQ\omega}\phi.
 \end{equation}
 The associated infinitesimal transformation is
 \begin{equation}
     \delta\phi=iQ\phi\delta\omega.
 \end{equation}
 We now introduce a spacetime-dependence in $\delta\omega$ (\ie it becomes a gauge transformation). The Lagrangian \eq{Gmodel} is not invariant because of the appearance of a second term with the variation of the partial derivatives
 \begin{equation}
     \delta(\partial_\mu\phi)=iQ(\partial_\mu\phi)\delta\omega+iQ\phi\partial_\mu(\delta\omega).
 \end{equation}
As a consequence, to achieve invariance, we should introduce a gauge field $A_\mu(x)$, which transforms as follows:
 \begin{equation}
     \delta A_\mu=-\frac{1}{e}\partial_\mu(\delta\omega),
 \end{equation}
with $e$ the electric charge. Besides, we must now use the covariant derivative
\begin{equation}
    D_\mu\phi=[\partial_\mu+ieA_\mu]\phi.
\end{equation}
To make the gauge field a true dynamical variable, we must add a term involving derivatives. The simplest gauge-invariant choice is a term proportional to $(F_{\mu\nu})^2$, where
\begin{equation}
    F_{\mu\nu}=\partial_\mu A_\nu-\partial_\nu A_\mu.
\end{equation}
The Lagrangian density changes accordingly
\begin{equation}
 \mathcal{L}= \partial_\mu\bar{\phi}(x)\partial^\mu\phi(x)-m^2\bar{\phi}(x)\phi(x)-g\left(\bar{\phi}(x)\phi(x)\right)^2-\frac14F_{\mu\nu}(x)F^{\mu\nu}(x),
 \label{eq:Higgsmoel}
\end{equation}
and it defines the \textit{Higgs model}.

\subsubsection{Higgs boson}
As in the Goldstone model, if $m^2>0$ the symmetry breaking cannot occur because the state of lowest energy corresponds to both $\phi(x)$ and $A_\mu(x)$ vanishing. Nevertheless, with $m^2<0$, we obtain again a circle of minimum energy corresponding to $\phi(x)$ taking the same values as in \eq{circle}. The vector field must vanish for the vacuum to ensure Lorentz invariance. Introducing once more the real fields as \eq{deviationfields}, the Lagrangian density becomes
\begin{multline}
    \mathcal{L}=\frac12\partial^\mu\rho(x)\partial_\mu\rho(x)-\frac12(2g\bar{\rho}^2)\rho^2(x)-\frac14 F_{\mu\nu}(x)F^{\mu\nu}(x)\\
    +\frac12(e\bar{\rho})^2A_\mu(x)A^\mu(x)+\frac12\partial^\mu\theta(x)\partial_\mu\theta(x)+e\bar{\rho}A^\mu(x)\partial_\mu\theta(x)+...,
    \label{eq:higgsLag}
\end{multline}
where we have discarded the constant and interactions terms. It is easy to see in \eq{higgsLag} that the field $\rho(x)$ is a Klein-Gordon field that gives spinless bosons with mass $\sqrt{2g\bar{\rho}^2}$. However, the product term $A^\mu(x)\partial_\mu\theta(x)$ shows that both the fields are not independent normal coordinates. If we count the degrees of freedom, we see that \eq{Higgsmoel} has four degrees of freedom: two from the complex scalar fields $\phi(x)$ and $\bar{\phi}(x)$ and two from the real massless vector field $A^\mu(x)$. Conversely, \eq{higgsLag} has two real scalar fields (two degrees of freedom) and a real massive vector field (three degrees of freedom). A change of variables cannot alter the number of degrees of freedom of the system. Thus, there must be an unphysical field which we should eliminate.  

To solve this problem, we will use the \textit{unitary gauge}:
\begin{equation}
    \partial_\mu\theta(x)-eA_\mu=-eB_\mu \leftrightarrow B_\mu=A_\mu-\frac 1 e \partial_\mu \theta(x).
\end{equation}
Here, $B$ is gauge-invariant and $F_{\mu\nu}=\partial_\nu B_\mu-\partial_\mu B_\nu$. The Lagrangian consequently becomes
\begin{equation}
    \mathcal{L}=\frac 1 2 \partial_\mu\rho(x)\partial^\mu\rho(x)+\frac {e^2}2\bar{\rho}^2B_\mu(x)B^\mu(x)-\frac 1 4 F_{\mu\nu}F^{\mu\nu}+...
    \label{eq:LagrangianB}
\end{equation}
In this gauge, we see that the $\theta$ field, which is associated with the Goldstone boson, has vanished. Furthermore, a massive vector meson has appeared, associated with the $B$ field, with a mass $e\bar{\rho}$. Such a massive vector meson has three degrees of freedom: the three spin states. What happened is that the two degrees of freedom of the massless gauge field and the one of the real field $\theta(x)$ have combined together to make the three degrees of freedom of the $B$ field. The phenomenon according to which a vector boson acquires mass without destroying the gauge invariance of the Lagrangian density is known as \textit{Higgs mechanism} and the spinless massive boson associated with $\rho(x)$ is called \textit{Higgs Boson}. The field $\theta(x)$ is eliminated by the gauge invariance.

\subsubsection{Higgs mechanism in the general case}

To extend the Higgs phenomenon to a general internal symmetry group, we merely have to add extra degrees of freedom (gauge fields) to promote the whole internal symmetry group to a gauge group. Afterwards, we gauge away the degrees of freedom that would correspond to Goldstone bosons. Let us call $U(x)$ a spacetime-dependent symmetry transformation as in \eq{Ufield}. The gauge fields transform as
\begin{equation}
    A_\mu(x)\rightarrow U(x)\left(A_\mu(x)-\frac{i}{g}\partial_\mu\right)U^\dagger(x),
\end{equation}
where $g$, like $e$, is a free parameter. The low-energy effective Lagrangian is now \cite{Baumann:2014nda}
\begin{equation}
   \mathcal{L}_{eff}= -\frac14\Tr{F_{\mu\nu}^2}-\frac{f^2_\pi}{4}\Tr{D_\mu U(D^\mu U)^\dagger}+...
\end{equation}
The Goldstone bosons are described by the spacetime parameter $\pi(x)$. The quadratic Lagrangian for the Goldstone bosons and the gauge fields becomes
\begin{equation}
    \mathcal{L}^{(2)}_{eff}=-\frac14\Tr{F_{\mu\nu}^2}-\frac12(\partial_\mu\pi_c)^2-\frac12m^2A^2_\mu+m\partial_\mu\pi_c\vdot A^\mu,
    \label{eq:higgs2}
\end{equation}
where $m^2\equiv f_\pi^2g^2$. With the unitary gauge, as previously stated, we can set $\pi=0$.
Nevertheless, we can still reintroduce the Goldstone boson along with the associated gauge redundancy. This process is called \textit{Stückelberg trick} \cite{Stueckelberg:1938hvi,Ruegg:2003ps}. The advantages of the Goldstone bosons presence are that it makes the high-energy behaviour of the theory manifest. The mixing term $\partial_\mu\pi_c\vdot A^\mu$, in \eq{higgs2}, has one fewer derivative than the kinetic term. That is, the term will become irrelevant at sufficiently high energies. We can take the decoupling limit
\begin{equation}
    g\rightarrow 0,\quad m\rightarrow 0\quad \text{for}\quad f_\pi=\frac{m}{g}=const.
    \label{eq:decouplinglimit}
\end{equation}
In this limit, the mixing term can be neglected and the Goldstone boson part of the Lagrangian density reduces to \eq{Generaldod}: the local symmetry has effectively become a global symmetry. 

\section{Inflation as the theory of a Goldstone boson}
\label{sec:GoldsotneBos}
Finally, we are ready to build the effective field theory of inflation. Inflation, as we have already seen in \chap{InfloBinglo}, is a period of accelerated expansion where the Universe is quasi de Sitter. Our request for a graceful exit requires an order parameter, or physical clock, that provides us a smooth connection between the inflationary period and the hot Big Bang evolution. A natural example of such a clock is the expansion rate $H$ which decreases during inflation or the expectation value of some field $\phi$ (like the inflaton). We will opt for the latter. The presence of this field defines a preferred time slicing with different time slices labeled by distinct values of $\phi(t)$. Consequently, the gauge invariance of General Relativity, \ie the invariance under spacetime diffeomorphisms 
\begin{equation}
    x^\mu\rightarrow x'^{\mu}(x^\nu),
\end{equation}
in our inflationary scenario no longer holds. Only the spatial diffeomorphisms are, in fact, unbroken, whereas the spacetime-dependent transformation
\begin{equation}
    t\rightarrow \widetilde{t}=t+\pi(t,\boldsymbol{x})
\end{equation}
does not leave the action invariant unless $\pi$ is constant. As we have seen in \sect{GoldTheorem}, this broken symmetry implies the existence of a Goldstone excitation along the direction of the broken generator
\begin{equation}
    U(t,\boldsymbol{x})\equiv t+\pi(t,\boldsymbol{x}).
\end{equation}
The Goldstone boson is connected to a generic adiabatic perturbation of the field
\begin{equation}
    \delta\phi(t,\boldsymbol{x})\equiv \phi(t+\pi(t,\boldsymbol{x}))-\phi(t)\approx\dot{\phi}\pi(t,\boldsymbol{x})
    \label{eq:deltapsi}
\end{equation}
at linear order. It is possible to show (see Appendix C \cite{Baumann:2014nda}) that in the spatial gauge (\ie $g_{ij}\equiv a^2(t)\delta_{ij}$), all metric perturbations are related to the Goldstone mode by the Einstein equations. Here, we choose the unitary gauge ($\pi=0$) where the privileged slicing coincides with surface of constant $t$. Namely, the adiabatic fluctuations vanish and we are left with only explicit metric perturbations 
\begin{equation}
   \delta g_{ij}=a^2(t)e^{-2\mathcal{R}}\delta_{ij}.
\end{equation}
because $\mathcal{R}=-H\pi$ from \eq{Rflz} and \eq{deltapsi}. As we have previously mentioned, the Goldstone field gets \textit{eaten} by the metric. 

\subsection{Most general action in the unitary gauge}
\label{EffectiveAction}
At this point, we want to write down the most general Lagrangian in the unitary gauge; a Lagrangian written in unitary gauge is no longer invariant under the broken symmetries, while it is still invariant under the unbroken symmetries. In our case of inflation, the theory of spacetime diffeomorphisms are spontaneously broken to time-dependent spacial diffeomorphisms. Following~\cite{Baumann:2014nda,Cheung:2007sv} we have a set of rules regarding the terms we can add into this Lagrangian. Firstly, terms invariant under spacetime diffeomorphisms are allowed (polynomials of the Riemann tensor $R_{\mu\nu\rho\sigma}$ and of its covariant derivatives, contracted by a scalar) as well as generic functions in time. Moreover, we can use any four-dimensional covariant tensor with a free $0$ index and spatial indeces contracted because they are scalars under spatial diffeomorphisms\footnote{This comes from the fact that $\frac{\partial\widetilde{t}}{\partial x^\mu}=\delta^0_\mu$. So, for example, $\widetilde{g}^{00}=g^{00}$}. Furthermore, as we have fixed the slice, we can define a induced spatial metric $h_{\mu\nu}$ on it by introducing the normal vector to these surfaces 
\begin{equation}
        n_\mu=\frac{\partial_\mu\widetilde{t}}{\sqrt{-g^{\mu\nu}\partial_\mu\widetilde{t}\partial_\nu\widetilde{t}}}.
        \label{eq:nugg}
\end{equation}
As a consequence, all the tensors projected on the surfaces are allowed as well as  the induced metric and the extrinsic curvature tensor $K_{\mu\nu}$ which is defined as the covariant derivative of $n_\mu$ projected on the slices $K_{\mu\nu}\equiv h^\sigma_\mu\nabla_\sigma n_\nu$. However, the \textit{Gauss-Codazzi} relation~\cite{Wald:GR} demonstrate that using the three-dimensional Riemann tensor and the extrinsic curvature at the same time is redundant. Thus, we can exclude the former. That said, the most generic action in the unitary gauge is 
\begin{equation}
    S=\int{d^4x\sqrt{-g}\mathcal{L}(R_{\mu\nu\rho\sigma},g^{00},K_{\mu\nu},\nabla_\mu;t)},
\end{equation}
where all the free indices must be upper $0$s. Writing it explicitly 
\begin{multline}
    S=\int d^4x\sqrt{-g}\left[\frac{\M}2R-c(t)g^{00}-\Lambda(t)+\frac{1}{2!}M_2(t)^4(g^{00}+1)^2+\frac{1}{3!}M_3(t)^4(g^{00}+1)^3\right.\\
    \left.-\frac{\bar{M}_1(t)^3}{2}(g^{00}+1)\delta K^\mu_\mu-\frac{\bar{M}_2(t)^2}{2}\delta K^\mu_\mu\,^2-\frac{\bar{M}_3(t)^2}{2}\delta K^\mu_\nu\delta K^\nu_\mu+...\right]
    \label{eq:genaa}
\end{multline}
where we have omitted terms of higher order in the fluctuations or with more derivatives and
\begin{equation}
    \delta K_{\mu\nu}=K_{\mu\nu}-a^2Hh_{\mu\nu}\,,\quad \delta g^{00}=(g^{00}+1)\,.
\end{equation}
Only the first three terms in the general action \eq{genaa} contain linear perturbations around the chosen FLRW solution. Therefore, we can fix the coefficients $c(t)$ and $\Lambda(t)$ requiring a given $H(t)$ evolution. Our FLRW background gives
\begin{equation}
    g^{00}=-1,\quad R\equiv R^\mu_\mu=12H^2+6\dot{H},\quad K\equiv K^\mu_\mu=3H.
\end{equation}
We can rewrite \eq{genaa} including all the quadratic order and higher terms in $\Delta S$:
\begin{equation}
    S=\int d^4x\sqrt{-g}\left[\frac12 \M R-c(t)g^{00}-\Lambda(t)\right]+\Delta S.
    \label{eq:purr}
\end{equation}
Lastly, varying the linear terms in the above action with respect to the metric, we get the Friedmann equations
\begin{gather}
    H^2=\frac{1}{3\M}[c(t)+\Lambda(t)],\\
    \dot{H}+H^2=-\frac{1}{3\M}[2c(t)-\Lambda(t)].
\end{gather}
This set of equations implies
\begin{gather}
    \Lambda(t)=\M(3H^2+\dot{H}),\\
    c(t)=-\M\dot{H}.
\end{gather}
Eventually, the action in the form \eq{purr}, becomes
\begin{equation}
    S=\int{d^4x\sqrt{-g}\left[\frac{\M}{2}R-\M(3H^2+\dot{H})+\M\dot{H}g^{00}\right]}+\Delta S.
    \label{eq:hophop}
\end{equation}
We, therefore, conclude that \cite{Cheung:2007sv}: \textit{the unperturbed history} (of our Universe) \textit{fixes $c(t)$ and $\Lambda(t)$, while the difference among different models will be encoded into higher order terms.}

For instance, if we set all but the first three terms to zero, we get the same action for a model with a minimal kinetic term and a slow-roll potential $V(\phi)$ in the unitary gauge ($\phi=\phi(t)$). Namely, recollecting the action
\begin{equation}
   S_\phi= \int{d^4x\sqrt{-g}\left[-\frac12(\partial_\mu\phi\partial^\mu\phi)^2-V(\phi)\right]}=\int{d^4x\sqrt{-g}\left[-\frac{\dot{\phi}^2}{2}g^{00}-V(\phi(t))\right]},
\end{equation}
and using the the relation \eq{Ip} and \eq{densityinflaton}, we obtain
\begin{equation}
    S_\phi=\int{d^4x\sqrt{-g}\left[\M\dot{H}g^{00}-\M(3H^2+\dot{H})\right]}
\end{equation}
which, adding the Hilbert action, is identical to the first three term in \eq{hophop}. 

\subsection{Action for the Goldstone boson}
Now that we have a Lagrangian written in unitary gauge,  if we force on the fields a gauge transformation via the St\"uckelberg trick, gauge invariance can be restored and we can introduce the Goldstone bosons~\cite{Piazza:2013coa,Baumann:2014nda}. We perform a spacetime-dependent time parametrization $t\rightarrow\widetilde{t}=t+\pi(t,\boldsymbol{x})$ and $\boldsymbol{x}\rightarrow\boldsymbol{\widetilde{x}}=\boldsymbol{x}$. The volume element $\sqrt{-g}d^4x$ and the Ricci scalar are invariant under general four-dimensional diffeomorphisms. Time-dependent coefficients transform like 
\begin{equation}
    f(t)\rightarrow f(t+\pi)=f(t)+\dot{f}\pi+\frac12\ddot{f}\pi^2+...
\end{equation}

\begin{tcolorbox}[mybox]
    Contravariant and covariant components of any tensor transform like
\begin{gather}
    t^{\mu\nu}\rightarrow \frac{\partial x^{\mu'}}{\partial x^\alpha}\frac{\partial x^{\nu'}}{\partial x^\beta}t^{\alpha\beta}=(\delta^\mu_\alpha+\delta^\mu_0\partial_\alpha\pi)(\delta^\nu_\beta+\delta^\nu_0\partial_\beta\pi)t^{\alpha\beta}\\
     t_{\mu\nu}\rightarrow \frac{\partial x^{\alpha}}{\partial x^{\mu'}}\frac{\partial x^{\beta}}{\partial x^{\nu'}}t_{\alpha\beta}=(\delta^\alpha_\mu+\delta^\alpha_0\partial_\mu\pi)^{-1}(\delta^\beta_\nu+\delta^\beta_0\partial_\nu\pi)^{-1}t^{\alpha\beta}.
     \label{eq:transform}
\end{gather}
So, for example,
\begin{gather}
    g^{00}\rightarrow g^{00}+2\partial_\mu\pi g^{0\mu}+\partial_\mu\pi\partial_\nu\pi g^{\mu\nu},\\
    g^{0i}\rightarrow g^{0i}+\partial_\mu\pi g^{\mu i},\\
    g^{ij}\rightarrow g^{ij}.
\end{gather}
\end{tcolorbox} 
Great care must be taken with the quantities defined in the time slice surfaces because they changes as well under a coordiantes transformation~\cite{Baumann:2014nda}. Neglecting for simplicity the terms that involve the extrinsic curvature, the action now reads
\begin{multline}
    S=\int d^4x\sqrt{-g}\left[\frac12\M R-\M\left(3H^2(t+\pi)+\dot{H}(t+\pi)\right)+\right.\\
    \M\dot{H}(t+\pi)\left((1+\dot{\pi})^2g^{00}+2(1+\dot{\pi}\partial_i\pi) g^{0i}+g^{ij}\partial_i\pi\partial_j\pi\right)+\\
    \left.\sum_{n=2}^{\infty}\frac{M_n^4(t+\pi)}{n!}\left(1+g^{00}+2\partial_\mu\pi g^{0\mu}+\partial_\mu\pi\partial_\nu\pi g^{\mu\nu}\right)^n\right].
    \label{eq:LongAction}
\end{multline} 
In analogy with \eq{higgs2}, the reintroduction of the Goldstone $\pi$ is useful because at sufficiently short distances, the physics of the Goldstone can be studied by neglecting metric fluctuations. Similar to the gauge theory case, we have just to look at the mixing terms: they contain fewer derivatives than the kinetic term of $\pi$ so they can be neglected.  In the simplest case of $M_n=0$, \ie the standard slow-roll inflationary case, we have
\begin{equation}
    \M\rightarrow\infty,\quad \dot{H}\rightarrow 0,\quad\text{for}\quad \M\dot{H}=const,
\end{equation}
where we can easily identify $g=1/\M$ from the case \eq{decouplinglimit}. Hence, in this simplified scenario, the leading mixing with gravity come from $\sim \M \dot{H}\dot{\pi}\delta g^{00}$ which, after canonical normalization~\cite{Cheung:2007sv}, leads to
\begin{equation}
    E_{mix}\sim\sqrt{\eh}H
\end{equation}
If the second operator $M_2$ gets large, we have~\cite{Cheung:2007sv,Cheung:2007st,Baumann:2014nda}
\begin{equation}
     E_{mix}\sim\frac{M_2^2}{M^2_{pl}}.
     \label{eq:Emix}
\end{equation}
As a result, when $E\gg E_{mix}$, the action dramatically simplifies because we can evaluate \eq{LongAction} in the unperturbed spacetime
\begin{multline}
    S=\int d^4x\sqrt{-g}\left[\frac12\M R-\M\dot{H}\left(\dot{\pi}^2-\frac{(\partial_i\pi)^2}{a^2}\right)\right.\\
    \left.+2M_2^4\left(\dot{\pi}^2+\dot{\pi}^3-\dot{\pi}\frac{(\partial_i\pi)^2}{a^2}\right)-\frac43 M_3^4\dot{\pi}^3+...\right].
    \label{eq:decouplinglimite}
\end{multline}
which is really useful because we are interested in computing predictions for present cosmological observations, given an inflationary model. Namely, we have seen that the curvature perturbation $\mathcal{R}$ is constant out of the horizon at any order in perturbation theory. As we have said, roughly speaking, the reason for the existence of a conserved quantity is that after exiting the horizon different regions evolve exactly in the same way.  Therefore, we are reduced to calculate correlation functions just after horizon crossing. That is, we are interested in studying our Lagrangian with an IR energy cutoff of order $H$. If $E_{mix}<H$, the Lagrangian in \eq{decouplinglimite} will give correct predictions up to terms suppressed by $E_{mix}/H$. The decoupling limit $\omega \gg \omega_{mix}=E_{mix}$ is reached at relatively low frequencies and the horizon crossing is within our decoupling regime. 

\paragraph{\textit{Symmetry-Breaking scale}} We are now interested in computing the symmetry-breaking scale. As we have done in \eq{readoff}, we are able to do so by reading the normalization term from the Noether current. Since~\cite{Baumann:2014nda}
\begin{equation}
    J^\mu=-\sqrt{2\M\lvert\dot{H}\rvert}\partial^\mu\pi_c+\mathcal{O}(\pi^2_c),
\end{equation}
where $\pi_c\equiv\sqrt{2\M\lvert\dot{H}\rvert}\pi$, the normalization of the current gives us the scale
\begin{equation}
    f^4_\pi\equiv 2\M\lvert \dot{H}\rvert.
    \label{eq:fpi}
\end{equation}
In our slow-roll regime, we recover
\begin{equation}
    f^4_\pi=\dot{\phi}^2.
\end{equation}
It perfectly matches our intuition that, in the single field inflation, the time variation of the inflaton is responsible for breaking the symmetry. As a last remark, we want to see what happen if we set $M_2\neq0$. In \eq{decouplinglimite} the Goldstone spatial gradient term do not change because it is completely fixed by the background evolution ($\M \dot{H}(\partial_i\pi_i)^2$. However, the kinetic term $\dot{\pi}$ changes due to a contribution from $\delta g^{00}$
\begin{equation}
    \M\dot{H}\dot{\pi}^2\longleftarrow \(-\M\dot{H}+2M^4_2\)\dot{\pi}\,.
\end{equation}
This contribution induces a nontrivial speed of sound for the fluctuations. It happens because spatial and time kinetic terms do not share the same coefficients anymore and the resulting speed of sound $c_s$ of the $\pi$ waves is not $1$ anymore\footnote{$c_s=1$ is obtained from the Lorentz invariance, which does not hold anymore}. The quadratic action therefore is~\cite{Baumann:2014nda}
\begin{equation}
    \mathcal{L}^{(2)}_\pi=\frac{\M\lvert \dot{H}\rvert}{c_s^2}\(\dot{\pi}^2-c_s^2\frac{(\partial_i\pi)^2}{a^2}\)
\end{equation}
with
\begin{equation}
    c_s^2\equiv\frac{\M\dot{H}}{\M\dot{H}-2M^4_2}
    \label{eq:cs2eft}
\end{equation}
and the symmetry-breaking scale in \eq{fpi} is now~\cite{Baumann:2014nda}
\begin{equation}
    f^4_\pi\equiv 2\M\lvert\dot{H}\rvert c_s
\end{equation}

\section{Hubble Flow Equations in the EFT of inflation}
\label{sec:HFLEQ}
Before concluding this section, it is instructive to introduce a method to investigate the observable predictions of a very broad class of inflationary models in the most general framework. It will be used in \sect{bBNPGW} when studying the constraining power of BBN on primordial gravitational waves. The method is based on the so-called Hubble Flow Equation~\citep{Hoffman:2000ue,Kinney:2002qn,Easther:2002rw,Friedman:2006zt}. The Hubble Flow Equations were first introduced by Hoffman and Turner~\citep{Hoffman:2000ue} for the simplest single-field slow-roll case where it is straightforward to define an infinite hierarchy of slow-roll parameters that, starting from the Hubble parameter $H$ and its derivatives with respect to the field, completely specify the evolution of the main observable quantities during inflation. Since the integration of the equations yields a trajectory in slow-roll parameter space that can be ultimately interpreted as a model whose dynamics is a solution of the flow equations, solving numerically a truncated system of Hubble Flow Equations for a set of suitably defined initial conditions has been proposed as a sophisticated algorithm for generating large numbers of slow-roll inflationary models, without relying on the explicit form of the action~\citep{Kinney:2002qn}. Recently, the method has been extended to the EFT framework of inflation~\cite{Capurri:2020qgz} to include a much broader class of beyond-standard inflationary models and explore a wide variety of possible high-energy corrections to the simplest slow-roll scenario. 

Taking into account the EFT of inflation \eq{genaa} and its background block \eq{purr}, the Hubble parameter can be written as a function of inflaton field, $H(\phi)$. To switch from the time domain to field domain we can  exploit a relation betweeen the time-derivative of the field, $c(\phi)$ and $H(\phi)$ that follows from a combination of the Friedmann equation and the continuity equation 
\begin{equation}
    \frac{\text{d}\phi}{\text{d}t}=-\frac{c(\phi)}{M_{\mathrm{pl}}^{2} H^{\prime}(\phi)}
\end{equation}
where, in this section, the prime indicates a derivative with respect to the field. Using the relation above, it is easy to see that the slow roll parameter $\eh$ becomes:
\begin{equation}
    \eh=\frac{c(\phi)}{M_{\mathrm{pl}}^{2} H^{2}(\phi)}.
\end{equation}
Starting from $\eh$, we can define the higher-order slow-roll parameters by iterated derivations:
\begin{equation}
    \begin{aligned} \etah(\phi)=& \frac{c(\phi)}{M_{\mathrm{pl}}^{2}} \frac{H^{\prime \prime}(\phi)}{H(\phi) H^{\prime 2}(\phi)} \\ & \vdots \\{ }^{l} \lambda(\phi) &=\left(\frac{c(\phi)}{M_{\mathrm{pl}}^{2}}\right)^{l}\left(\frac{1}{H(\phi)}\right)^{l}\left(\frac{1}{H^{\prime}(\phi)}\right)^{l+1} \frac{d^{l+1} H(\phi)}{d \phi^{l+1}} \end{aligned}
    \label{Htower}
\end{equation}
with $l \geq 2$ and $\etah(\phi)\equiv \, ^{1}\lambda(\phi)$. The evolution of $\eh$ and the other higher-order parameters will depend on the additional unknown function $c(\phi)$. Therefore, it is important to define another set of new slow-roll parameters to describe the evolution of $c(\phi)$. The Hubble flow equations in this case start from the parameter $\theta$~\cite{Capurri:2020qgz} 
\begin{equation}
    \theta \equiv-\frac{\dot{c}}{H c}=\frac{1}{M_{\mathrm{pl}}^{2}} \frac{c^{\prime}(\phi)}{H(\phi) H^{\prime}(\phi)}
\end{equation}
and the the other higher-order parameters by taking iterated derivations 
\begin{equation}
   \begin{aligned} \kappa(\phi) &=\frac{1}{M_{\mathrm{pl}}^{2}} \frac{c^{\prime \prime}(\phi)}{H^{\prime 2}(\phi)} \\ & \vdots \\ ^{l}{\xi}(\phi) &=\left(\frac{c(\phi)}{M_{\mathrm{pl}}^{2}}\right)^{l}\left(\frac{1}{H(\phi)}\right)^{l-1}\left(\frac{1}{H^{\prime}(\phi)}\right)^{l+1} \frac{1}{c(\phi)} \frac{d^{l+1} c(\phi)}{d \phi^{l+1}} \end{aligned} 
   \label{Ctower}
\end{equation}
always with $l \geq 2$ and $\kappa(\phi)\equiv \, ^{1}\xi(\phi)$. An explicit calculation of the equations above lead to derive the generalized Hubble flow equation for the background parameters~\citep{Capurri:2020qgz} (also called Hubble tower):
\begin{equation}
    \begin{cases}
    \frac{\text{d}\eh}{\text{d}N}=\eh\,\left( \theta - 2\eh \right)\\
    
    \frac{\text{d}\eta}{\text{d}N}= \etah\,\left( \theta - \eh -2\etah\right) +\,^{2}\lambda \\
    \vdots
    \\
    \frac{\text{d} \, ^{l}\lambda }{\text{d}N} = \, ^{l}\lambda \, \left[ l\left( \theta -\eh \right) -\left(l+1\right)\etah \right] + \, ^{l+1}\lambda
    \\
    \\
    \frac{\text{d}\theta}{\text{d}N}=\eh\,\kappa \, -\theta\,\left( \eh + \etah \right)\\
    \frac{\text{d} {\kappa} }{\text{d}N} = -2\kappa\etah + \, ^{2}\xi\\
    \vdots
    \\
    \frac{\text{d} \, ^{l}\xi }{\text{d}N} = \, ^{l}\xi \left[ \left(l-1\right)\left(\theta -\eh \right) -\left(l+1\right)\etah\right] + \, ^{l+1}\xi
    \end{cases}
    \label{eq:HFE_background}
\end{equation}
The integration of this system of coupled equations completely specifies the dynamics of the background during inflation. Once the background dynamics is reconstructed by solving the system in \eq{HFE_background}, it is useful to derive a further system of equations to describe the evolution of the $M$ coefficients over that background. It is possible to do so in a quite general and elegant way by noting that for any quantity described by a generic scalar function $Q(\phi)$, one can always define a slow-roll
parameter $\epsilon_Q$ as follows: 
\begin{equation}
    \epsilon_{Q}=-\frac{\dot{Q}}{H\,Q}=\frac{1}{M_{\mathrm{pl}}^{2}}\frac{c(\phi)}{H(\phi)\,H^{\prime}(\phi)}\,\frac{Q(\phi)}{Q^{\prime}(\phi)}
    \label{eq:epsQ}
\end{equation}
In analogy to the discussion for the background parameters, it is possible to define also the higher-order parameters for the quantity $Q(\phi)$ by taking its derivatives:

\begin{equation}
\begin{aligned} 
\rho_{Q}(\phi)&=\frac{1}{M_{\mathrm{pl}}^{2}} \frac{c(\phi)}{H^{\prime\,2}(\phi)}\frac{Q^{\prime \prime}(\phi)}{Q(\phi)}\\ & \vdots \\
{}^{l}\chi_{Q}(\phi)&=\left( \frac{c(\phi)}{M_{\mathrm{pl}}^{2}} \right)^l\,\left(\frac{1}{H(\phi)}\right)^{l-1}\,\left(\frac{1}{H^{\prime}(\phi)}\right)^{l+1}\,\frac{1}{Q} \frac{\text{d}^{l+1} \,Q}{\text{d}\phi^{l+1}}
\end{aligned} 
\end{equation}
again with $l\geq 2$ and $\rho_{Q}(\phi)\equiv{}^{1}\chi_{Q}(\phi)$. The system of Hubble flow equations for $Q(\phi)$ is
\begin{equation}
\begin{cases}
\frac{\text{d} \epsilon_{Q}}{\text{d} N}= \eh{Q}\left(\theta - \eh -\etah -\epsilon_{Q} \right) + \, \eh\,\rho_{Q}\\
\frac{\text{d} \rho_{Q}}{\text{d} N}= \rho_{Q} \left(\theta - 2\etah -\epsilon_{Q}\right) + {}^{2}\chi_{Q}\\
\vdots 
\\
\frac{\text{d}  {}^{l}\chi_{Q}}{\text{d} N}= {}^{l}\chi_{Q} \left[l\theta -\left(l-1\right)\eh - \left(l+1\right)\etah -\epsilon_{Q} \right] + {}^{l+1}\chi_{Q}
\end{cases}  
\label{eq:system_Q}
\end{equation}
Solving the system implies predicting the evolution of any generic quantity $Q(\phi)$ that will depend also on the background via the slow-roll parameters $\eh$, $\etah$ and $\theta$, as expected. This means that, in principle, one can evolve all the $M$ coefficients and study different models of inflation in full generality.

%% file: Chapters/PGW.tex
\chapter{Primordial Gravitational Waves}
\label{ch:PGWs}

The inflationary scenario also predicts  the production of a background of stochastic gravitational waves~\cite{Abbott:1984fp,Grishchuk:1974ny,Starobinsky:1979ty,Rubakov:1982df,Fabbri:1983us,Starobinsky:1980te,Linde:1981mu,Vilenkin:1983xq,Ungarelli:2005qb,Guzzetti:2016mkm,Mukhanov:2005sc,Lyth:2009zz,Weinberg:2008zzc,Martin:2013tda,Riotto:2002yw,Caprini:2018mtu,Cabass:2015jwe}. As we have seen in \sect{GW}, tensor fluctuations of the metric represent the degrees of freedom of the gravitational waves. Their evolution is regulated only by the traceless spatial part of the Einstein equation~\cite{Guzzetti:2016mkm} in a more generic case. We discuss how these perturbations re-enter the Hubble horizon after inflation, forming a stochastic background of gravitational waves that carry a quasi-invariant spectrum. We also study how the contribution of the primordial gravitational waves to radiation is effectively constrained by BBN bounds as well as how much their imprint on the CMB in the form of B-mode polarization can constrain the slow-roll inflation.

\section{Tensor perturbations}
\label{sec:TensoTensoO}
In this section, we expand the de Sitter treatment in \sect{perturbationss} focusing on the tensor modes. The tensor modes, or \textit{radiative modes}, involve only two traceless and divergenceless symmetric tensors: $E_{ij}$ in \eq{Tens} and $\Sigma_{ij}$ in \eq{imperfecttmunu}. In \sect{GW} we have seen that the Radiative modes describe gravitational waves. In general, perturbing the stress-energy tensor and the Einstein tensor, we get the wave equation for gravitational radiation:
\begin{equation*}
    -\frac{1}{2\M}a^2\hat{\Sigma}_{ij}=\nabla^2E_{ij}-a^2\ddot{E}_{ij}-3a\dot{a}\dot{E}_{ij}.
    \label{eq:QuoVado}
\end{equation*}
As it was previously stated, $E_{ij}$ does not change under coordinate transformations. Thus, it already describes the gravitational waves in a gauge-invariant manner. If we decompose tensor perturbations into eigenmodes of the spatial Laplacian, $\nabla^2e_{ij}=-k^2e_{ij}$ with comoving wavenumber $k$, we can write
\begin{equation}
    E_{ij}=h(t)e^{(+,\times)}_{ij}(x),
    \label{eq:GWp}
\end{equation}
where $h(t)$ is the scalar amplitude and $+$, $\times$ denote the two possible polarization states of the gravitational waves (see \sect{GW}). Moreover, it is a good assumption not to consider the anisotropic stress. Thus, we arrive at the wave equation describing the evolution of gravitational waves in an expanding Universe:
\begin{equation}
\boxed{E''_{ij}+3\mathcal{H}E'_{ij}+k^2E_{ij}=0}\,.
    \label{eq:quoVadis}
\end{equation}
We expand the Einstein-Hilbert action to the second order of tensor fluctuations and obtain
\begin{equation}
S^{(2)}_T=\int{d^4xa^2(\eta)\left[(E'_{ij})^2-(\nabla E_{ij})^2\right]}.
\end{equation}
Up to some factors, this is the same action for a massless scalar field in the FLRW Universe. Being aware of  the result in \eq{GWp}, we promote the field to an operator and, in define the Fourier expansion
\begin{equation}
    E_{ij}=\int{\frac{d\boldsymbol{k}}{(2\pi)^3}\sum_{\lambda=+,\times}\epsilon_{ij}^\lambda(k)h_{\boldsymbol{k}}^\lambda(t)e^{i\boldsymbol{k}\vdot\boldsymbol{x}}},
    \label{eq:GWWp}
\end{equation}
and write the tensor action in the following way:
\begin{equation}
    S_T^{(2)}=\sum_\lambda\int d\tau d\boldsymbol{k}a^2[h^{\lambda'}_{\boldsymbol{k}}h^{\lambda'}_{\boldsymbol{k}}-k^2h^{\lambda}_{\boldsymbol{k}}h^{\lambda}_{\boldsymbol{k}}].
\end{equation}
In analogy with the scalar procedure, we perform the transformation
\begin{equation}
    f_{\boldsymbol{k}}^\lambda\equiv a h^\lambda_{\boldsymbol{k}}
    \label{eq:DefOfV}
\end{equation}
to get
\begin{equation}
    S^{(2)}_T=\sum_\lambda\frac12\int d\tau d\boldsymbol{k}\left[(f_{\boldsymbol{k}}^{\lambda'})^2-\left(k^2-\frac{a''}{a}\right)(f^\lambda_{\boldsymbol{k}})^2\right].
\end{equation}
The equation of motion for each mode reads
\begin{equation} f^{\lambda''}_{\boldsymbol{k}}+\left(k^2-\frac{a''}{a}\right)f^\lambda_{\boldsymbol{k}}=0.
    \label{eq:Lobello}
\end{equation}
which resembles \eq{MS}. Similar to the case for scalar perturbations, we can identify two main regimes. The subhorizon behavior with the non decaying solution (as in \eq{RotTor}) which reads $f_{\textbf{k}}(\eta)=Ae^{-ik\eta}$. From the definition of $f_{\textbf{k}}$, the amplitude of the field decreases like $a^{-1}$ as an effect of the Universe expansion. The second regime gives, similarly to \eq{Freeze}, a constant and decreasing solution. 

In order to compute accurately the power spectrum, we perform the standard quantization like in \eq{Promoting2} and the normalization in \eq{CondCond}. Once we have ensured that $\hat{a}^\lambda_{\boldsymbol{k}}$ and $\hat{a}^{\lambda\dagger}_{\boldsymbol{-k}}$ behave as the canonical creation and annihilation operators, we assume that the Universe was in the vacuum state at past infinity, \ie we select the Bunch-Davis vacuum. The following computatins are done fixing one polarization state so we can drop the $\lambda$. Using $\eh$ we can write
\begin{equation}
    \boxed{f''_k+\left[k^2-\frac{1}{\eta^2}\left(\nu^2-\frac14\right)\right]f_k=0,\quad\nu\equiv \frac32+\eh}\,.
    \label{eq:GeneralmMotor}
\end{equation}

\begin{tcolorbox}[mybox]
From the definition of $\epsilon_\Hl$ we can see that  
    \begin{equation}
    H(N+\Delta N)\simeq H(N)e^{-\epsilon_\Hl\Delta N},
\end{equation}
which implies
\begin{equation}
    \eta=\int{\frac{dN}{aH}}\simeq-\frac{1}{(1-\epsilon_\Hl)\mathcal{H}}\,.
\end{equation}
In particular
\begin{equation}
    \mathcal{H}'\simeq (1-\epsilon_\Hl)\mathcal{H}^2
\end{equation}
and eventually
\begin{equation}
    \frac{a''}{a}\equiv \mathcal{H}'+\mathcal{H}^2\simeq(2-\epsilon_\Hl)\mathcal{H}^2\simeq\frac{(2-\epsilon_\Hl)}{(1-\epsilon_\Hl)^2\eta^2}\simeq\frac{1}{\eta^2}(2+3\epsilon_\Hl)
\end{equation}
that allows us to write \eq{GeneralmMotor}
\end{tcolorbox}

It is easy to see that the exact de Sitter solution is obtained with $\nu=3/2$; namely, the slow-roll parameter $\epsilon_\Hl\neq0$ represents a linear deviation from it. The general solution of \eq{GeneralmMotor} is
\begin{equation}
    f_k(\eta)=\sqrt{-\frac{1}{\eta}}[c_1(k)H^{(1)}_\nu(-k\eta)+c_2(k)H^{(2)}_\nu(-k\eta)],
\end{equation}
where $H^{(1)}_\nu(-k\eta)$ and $H^{(2)}_\nu(-k\eta)$ are the Hankel functions of the first and second kind.

\begin{tcolorbox}[mybox]
The Hankel functions are defined as
\begin{gather}
    H^{(1)}_\nu(x)\equiv J_\nu(x)+iY_\nu(x)\\
    H^{(1)}_\nu(x)\equiv J_\nu(x)+iY_\nu(x)
\end{gather}
where $J_\nu(x)$ are the Bessel function of the first kind\footnote{Which should not be confused with $j_n$ that are the spherical Bessel functions} while $Y_\nu(x)$ are the Bessel function of the second kind, also called Neumann functions. Some properties of the Hankel functions are
\begin{gather}
    H_\nu^{(\alpha)}(x)=\frac{J_{-\nu}(x)-e^{(-1)^{\alpha}\nu\pi i}J_{\nu}(x)}{(-1)^{(\alpha+1)} i\sin{\nu\phi}},\\
    H^{(\alpha)}_{-\nu}(x)=e^{(-1)^{\alpha+1}\nu \pi i}H^{(\alpha)}_{\nu}(x),
\end{gather}
and the limits
\begin{gather} H^{(\alpha)}_\nu(x\gg1)\simeq\sqrt{\frac{2}{\pi x}}e^{(-1)^{(\alpha+1)}(x-\nu-\frac{\pi}{4})}, \label{eq:LimitHankel}\\
    H_{\nu}^{(1)}(x\ll1)\simeq \sqrt{\frac{2}{\pi}}e^{-i\frac{\pi}{2}}2^{\nu-\frac32}\frac{\Gamma(\nu)}{\Gamma(\frac32)}\frac{1}{x^{\nu}}.\label{eq:LowHankel}
\end{gather}
\end{tcolorbox}

We shall now verify that the general solution is, in fact, in agreement with our previous results. Specifically, in the UV regime, the solution must match \eq{condition}. Using the Hankel functions property (see \eq{LimitHankel}) we get
\begin{equation}
    \frac{e^{-ik\eta}}{\sqrt{2k}}=\sqrt{-\eta}\sqrt{-\frac{2}{\pi k\eta}}[c_1(k)e^{-i(k\eta+\nu+\frac{\pi}{4})}+c_2(k)e^{i(k\eta+\nu+\frac{\pi}{4})}]
\end{equation}
thus
\begin{equation}
   c_1(k)=\frac{\sqrt{\pi}}{2}e^{\frac{i}{2}(\nu+\frac12)}\quad\text{and}\quad c_2(k)=0.
\end{equation}
The exact solution then becomes
\begin{equation}
    f_k=\frac{\sqrt{\pi}}{2}e^{\frac{i}{2}(\nu+\frac12)}\sqrt{-\eta}H^{(1)}_{\nu}(-k\eta)
   \label{eq:DireS}
\end{equation}
which, depending on the scale, reduces to
\begin{equation}
   f_k\simeq \frac{e^{ik\eta}}{\sqrt{2k}}\quad\text{subhorizon}\quad k\gg aH
    \label{eq:SubSUbBUS}
\end{equation}
as it should, and, using \eq{LowHankel}, we have
\begin{equation}
    f_k\simeq e^{-i\frac{\pi}{2}(\nu-\frac12)}2^{\nu-\frac32}\frac{\Gamma(\nu)}{\Gamma(\frac32)}\frac{1}{\sqrt{2k}}(-k\eta)^{(\frac12-\nu)}\quad\text{superhorizon}\quad k\ll aH\,.
    \label{eq:SuperSuperHo}
\end{equation}
Using \eq{Lobello} together with \eq{SuperSuperHo}, we can find the amplitude of physical tensor modes $h_k$ introduced in \eq{GWWp}, in the super-Hubble scales as
\begin{equation}
    \boxed{\lvert h_k(\eta)\rvert\simeq \frac{H}{\Mpl}k^{-\frac32}\hat{g}(\epsilon_\Hl)\left(\frac{k}{aH}\right)^{-\epsilon_\Hl}\quad\text{for}\quad k\ll aH}\,
    \label{ICCROss}
\end{equation}
with
\begin{equation}
\hat{g}(\epsilon_\Hl)\equiv 2^{\eh}(1-\eh)^{(1+\eh)}\frac{\Gamma(\frac32+\eh)}{\Gamma(\frac32)}\simeq 1-(1-\ln2-\psi_0(\frac32))\eh\simeq 1-0.27\eh
\end{equation}
where $\psi_0$ is the Digamma function. Iin the limit $\eh\rightarrow 0$, we get $\hat{g}(0)=1$ and
\begin{equation}
     \lvert h_k(\eta)\rvert\simeq \frac{H}{\Mpl}k^{-\frac32}\quad\text{for}\quad\eh\rightarrow0,\quad k\ll aH.
     \label{eq:HomeAlone}
\end{equation}
which corresponds to a power spectrum assuming a de Sitter stage. 

\subsection{Tensor phenomenology}
the definition of power spectrum, its expression at super-Hubble scales is
\begin{equation}
    \Delta_h(k)\simeq \frac{1}{\pi^2}\frac{H^2}{\M}\hat{f}^2(\eh)\left(\frac{k}{aH}\right)^{-2\eh}
\end{equation}
with a small difference with the de Sitter result, as the factor $f^2(\eh)\simeq 1-0.54\eh$ is a tiny correction in the amplitude of $\sim 0.5(\eh/0.01)\%$. Evaluated at horizon crossing 
\begin{equation}
\Delta_h(k)\simeq \frac{1}{\pi^2}\left(\frac{H_\star}{\M}\right)^2\,.
\end{equation}
Since we have two polarizations, the power spectrum for tensor fluctuations is
\begin{equation}
\boxed{\Delta^2_T=2\Delta^2_h(k)=\frac{2}{\pi^2}\frac{H^2_\star}{\M}}\,.
\label{eq:PSTensor}
\end{equation}
Like we have done for the scalar perturbations, we define a spectral index that quantifies the scale dependence
\begin{equation}
\boxed{n_T\equiv\frac{d\ln{\Delta^2_T}}{d\ln{k}}}
\label{eq:ntnt}
\end{equation}
which at first order in the Hubble parameters can be written as
\begin{equation}
    \boxed{n_T=-2\eh}\,.
    \label{eq:ntepsH}
\end{equation}
and if we call
\begin{equation}
    A_T=\frac{2}{\pi^2}\frac{H^2_\star}{\M}
\end{equation}
we have a similar phenomenological parametrization as in \eq{PSfirst}, for the tensorial power spectrum 
\begin{equation}
    \boxed{\Delta^2_T(k)=A_T\left(\frac{k}{k_\star}\right)^{n_T}}
    \label{eq:PhenomenologicalTensor}\,.
\end{equation}
We can move to next-to-leading order generalization and introduce a scale dependence of the tensor tilt, by including its running
\begin{equation}
    \boxed{\alpha_T\equiv\frac{\d n_T}{\d\ln{k}}}\,.
    \label{eq:running}
\end{equation}
In the inflationary scenario, an interesting consistency relation holds between quantities that involve tensor perturbations. To start with, we will introduce the \textit{tensor-to-scalar ratio}
\begin{equation}
    \boxed{r\equiv \frac{\Delta^2_T(k)}{\Delta^2_s(k)}}\,,
    \label{eq:r}
\end{equation}
that gives the amplitude of the gravitational waves with respect to that of the scalar perturbations, at some fixed pivot scale $k$. From the definition \eq{PSfirst} and \eq{PhenomenologicalTensor} we get
\begin{equation}
   \boxed{r=16\eh}
   \label{eq:reps}
\end{equation}
or
\begin{equation}
    r=\frac{8}{\M}\left(\frac{\dot{\phi}}{H}\right)^2\,.
    \label{eq:rEvolution}
\end{equation}
Namely, $r$ depends on the time-evolution of the inflaton field. We eventually arrive at the important relation~\cite{Liddle:2000cg}
\begin{equation}
    \boxed{r=-8n_T}
    \label{eq:TensorProblem}
\end{equation}
at the lowest order in slow-roll parameters. \eq{TensorProblem} implies an almost scale-invariant slightly red-tilted spectrum. However this relation can be violated in many non-standard realizations of inflation such as in modified gravity theories~\citep{Baumann:2015xxa,Odintsov:2020ilr,Giare:2020plo,Oikonomou:2021kql,Odintsov:2022cbm}, in multi-fields inflationary models~\citep{Namba:2015gja,Peloso:2016gqs,Pi:2019ihn,Ozsoy:2020ccy}, or from trans-Planckian Physics~\citep{Ashoorioon:2014nta,Ashoorioon:2005ep}. Depending on the underlying phenomenology, the tensor tilt can range from being red ($n_{\rm T}<0$) to blue ($n_{\rm T}>0$), see e.g.~\citep{Stewart:2007fu, Mukohyama:2014gba,Giovannini:2015kfa,Giovannini:2018dob,Giovannini:2018nkt,Giovannini:2018zbf,Giare:2020vhn,Baumgart:2021ptt} and the references therein. As a result, constraining the tensor tilt (and in general the shape of the tensor spectrum) without any underlying assumption is crucial for testing new physics and the standard slow-roll scenario~\citep{Franciolini:2018ebs,DEramo:2019tit,Giare:2019snj,Caldwell:2018giq,Clarke:2020bil}.

From the definition \eq{r} with \eq{running} and \eq{PhenomenologicalTensor}, we can write 
\begin{equation}
    \boxed{\ln \Delta^2_{\rm T}(k)=\ln(r\,A_{\rm s}) + n_{\rm T}\,\ln(k/k_{\star}) + \alpha_{\rm T}\,\ln^2(k/k_{\star})}
\label{eq:PLtensorexpansion}
\end{equation}
The first parameter, \textit{i.e.} the tensor amplitude $A_{\rm t}\doteq r\,A_{\rm s}$, is currently constrained to\footnote{We recall that $A_{\rm s}\simeq 2.1\times 10^{-9}$ is the amplitude of primordial scalar perturbations~\citep{Planck:2018jri}.} $r<0.032$ at 95\%CL~\citep{Tristram:2021tvh} when \textit{Planck}~\citep{Planck:2020olo} and BK18~\citep{BICEP:2021xfz} datasets are combined, together with BAO~\citep{eBOSS:2020yzd} and CMB lensing~\citep{Planck:2018lbu}. Hopefully, in the upcoming decade, new CMB experiments such as LiteBIRD~\citep{LBIRD} and CMB-S4~\citep{CMB-S4} should reach a better sensitivity $r \sim 0.001$, possibly leading to the first detection of B-mode polarization.

\paragraph{\textit{Constraining the tensor tilt}} Relaxing the slow-roll consistency relation, the analysis of the CMB data only weakly constrains the tensor tilt to $-0.55<n_{\rm T}<2.54$ at 95\% CL~\citep{Akrami:2018odb}. However, important improvements in the upper limit can be achieved by exploiting other CMB-independent observables. For instance, along with B-modes polarization, primordial tensor fluctuations may contribute also to the stochastic background of gravitational waves (SGWB), the analogous of CMB for gravitational waves~\cite{Caprini_2018}. Interestingly, if the spectrum is enough blue-tilted, according to \eq{PLtensorexpansion} the inflationary contribution should be much amplified on scales of direct gravitational wave detection so that we can use data from ground-based interferometers such as LIGO and VIRGO to infer constraints on $n_{\rm T}$. These experiments set an upper bound on the fraction of the energy-density of the Universe in gravitational radiation $\Omega_{\rm GW}\lesssim 10^{-7}$ ~\citep{LIGO_SGWB-2017,LIGO_SGWB-2019} in the frequency range $f\in\left(20\,\rm{-}\,85.8\right)$ Hz (which corresponds to the wave-number range  $k_{\rm LV} \in \left(1.3\,\rm{-}\,5.5\right)\times 10^{16} \,\rm{Mpc}^{-1}$), leading to a more stringent upper limit  $n_T < 0.52$ at 95\% CL ~\citep{Akrami:2018odb}. While this approach is largely used in the literature, it should be noted that these bounds are obtained by extrapolating the relation \eq{PhenomenologicalTensor} on frequencies (those probed by GWs experiments) where it is not granted that the spectrum still follows a power-law behavior. Indeed, high wave-numbers $k$ correspond to modes that exit the horizon relatively close to the end of inflation where the spectrum may strongly depend on the higher-order terms in \eq{PLtensorexpansion}~\citep{Giare:2020vhn} and therefore on the specific form of the inflationary potential~\citep{Kinney:2021nje}, making it extremely difficult to derive reliable model-independent bounds on the tensor-tilt.

\begin{tcolorbox}[mybox]
Once we have accepted the fact that the seeds of perturbations are quantum fluctuations of the scalar field that has driven the accelerated expansion, and of the gravitational field, a prominent issue, still unsolved, arises: the \textit{Single Outcome problem} or \textit{Macro-objectivation problem}~\cite{Guzzetti:2016mkm}. The CMB radiation is an observable and then, according to quantum mechanics, it corresponds to a quantum operator. Thus, when we look at a CMB map we are considering the results of a measurement corresponding to a specific observable. Namely, the CMB state falls into an eigenvalue of the related observable. However, this means that, observing it today, the CMB perturbations get the value of the eigenvalue only at present time when we are making the measurements, since no observers existed before us. Furthermore, the same seeds give rise to the LLS of the Universe. Supposing their value is determined only today, our understanding of structure evolution will fail. How could they start growing  at early times? It has been pointed out~\cite{Grishchuk:1990bj} that inflationary perturbations evolve into highly squeezed quantum states on superhorizon scales because of an accelerated expansion, resulting in highly non-classical states. On the other hand, such a kind of quantum perturbations can be described as a realisation of a classical stochastic process in virtue of their large occupation number, so that we are justified to consider such quantities as classical. 
\end{tcolorbox}

\section{Post-inflationary evolution}
\label{sec:PIE}
As we have seen in the qualitative solutions in \eq{Lobello}, out of the horizon the amplitude of tensor perturbations is almost frozen. However, during the radiation and matter eras, they re-enter the causally connected space with the almost-scale-invariant power spectrum at the time of the first horizon crossing which occurred during inflation. Thus, Eq. (\ref{ICCROss}) provides an initial condition for the tensor modes produced during inflation. Once they are inside, they start oscillating with the amplitude damped by a factor $1/a$. In particular, they  acquire a time-dependence since the scale factor evolves as $a\sim \eta$ and $a\sim \eta^2$ during the radiation and matter dominance, respectively. To find the correct evolution of primordial gravitational waves, we take \eq{Lobello} and knowing that in MD and RD phases the scale factor has a power-law behaviour, we can write in the most generic way $a(\eta)=a_n\eta^n$, which covers both cases, and find the general solution
\begin{equation}
   \boxed{ h_\lambda(\boldsymbol{k},\eta)=\frac{A_\lambda(\boldsymbol{k})}{a(\eta)}\eta j_{n-1}(k\eta)+\frac{B_\lambda(\boldsymbol{k})}{a(\eta)}\eta Y_{n-1}(k\eta)}\,,
   \label{eq:NormalEvo}
\end{equation}
where $j_n(x)$ is the spherical Bessel function whilst $Y_n(x)$ is the spherical Bessel function of the second kind. $A_\lambda(\boldsymbol{k})$ and $B_\lambda(\boldsymbol{k})$ are dimensional constants to be established from the initial conditions. 

\subsubsection{Propagation of the primordial gravitational waves}
\eq{NormalEvo} clearly confirms the decaying behaviour of the inflationary modes in the expanding background, once they have crossed the horizon for the second time. Now, we want to choose the right initial conditions for $k\eta\ll 1$. For a Universe that is radiation-dominated, once the mode has re-entered the horizon, we find
\begin{equation}
    h_\lambda^{RD}(\boldsymbol{k},\eta)=h_{\infty}(k)j_0(k\eta),\quad \lambda=+,\times,
    \label{eq:RDS}
\end{equation}
where $h_{\infty}$ is the constant amplitude in \eq{HomeAlone} in the limit of exact de Sitter. It is valid for both polarizations. On the other hand, in a matter-dominated, Universe we obtain 
\begin{equation}
    h_\lambda^{MD}(\boldsymbol{k},\tau)=h_{\infty}(k)\frac{3j_1(k\eta)}{k\eta},\quad \lambda=+,\times.
    \label{eq:MDS}
\end{equation}
Looking at the dependence on $k$, these solutions tell us that tensor perturbations start oscillating with a damping factor greater for high frequency waves. To get the amplitude today, we should consider when the modes entered the horizon. If it is during the MD era, the solution in \eq{MDS} is enough. Nevertheless, if the modes entered during the RD era, neither of the above solutions apply. To find the result, we shall match the solution \eq{RDS} at the time of radiation-matter equality with the full solution valid in the matter era with free coefficients. That is~\cite{Caprini:2018mtu}

\begin{equation}
h^{MD,full}_\lambda(\boldsymbol{k},\eta)=\bar{A}_\lambda(\boldsymbol{k})\frac{j_{1}(k\eta)}{k\eta}+\bar{B}_\lambda(\boldsymbol{k})\frac{Y_{1}(k\eta)}{k\eta},
\end{equation}
where we have assumed an instantaneous transition. If we call $\eta_{eq}$ the moment in which we have the matter-radiation equality, and we define $x_{eq}=k\eta_{eq}$, the free coefficients read \cite{Watanabe:2006qe}:
\begin{gather}
    \bar{A}(k)=h_{inf}(k)\left[\frac32-\frac{\cos(2x_{eq})}{2}+\frac{\sin(2x_{eq})}{x_{eq}}\right],\\
    \bar{B}(k)=h_{inf}(k)\left[\frac{1}{x_{eq}}-x_{eq}-\frac{\cos(2x_{eq})}{x_{eq}}-\frac{\sin(2x_{eq})}{2}\right].
\end{gather}
Thus, from its definition \eq{TrF} we can now introduce the transfer function such that
\begin{equation}
    h_\lambda(\boldsymbol{k},\eta)=h_{inf}(k)T(k,\eta)\,,
   \label{eq:TransferHhH}
\end{equation}
where explicitly 
\begin{equation}
   \boxed{ T(k,\eta)=\left\{
    \begin{array}{cr}
    \frac{3j_1(k\eta)}{k\eta},& k<k_{eq}\\
    \frac{\bar{A}_\lambda(\boldsymbol{k})}{h_{inf}(k)}\frac{j_{1}(k\eta)}{k\eta}+\frac{\bar{B}_\lambda(\boldsymbol{k})}{h_{inf}(k)}\frac{Y_{1}(k\eta)}{k\eta}& k>k_{eq}
    \end{array}\right.}\,.
\end{equation}
At subhorizon scales $k\eta\gg 1$, and performing an oscillation-averaging procedure, we get \cite{Caprini:2018mtu}
\begin{equation}
\boxed{[T'(k,\eta)]^2\xrightarrow{k\eta\gg 1}\left\{
    \begin{array}{cr}
    \eta_{eq}^2/(2\eta^4),& k<k_{eq}\\
    9/(2\eta^4k^2)& k>k_{eq}
    \end{array}\right.}\,.
    \label{eq:forthe}
\end{equation}

\subsection{Energy density of gravitational waves}
\label{sec:PGWSradio}
We now want to focus on the identification of the gravitational wave energy density. In particular, we will set ourselves in the weak-field limit; namely, where the gravitational wave can be described as a spacetime ripple propagating on a fixed background:
\begin{equation}
    g_{\mu\nu}=g_{\mu\nu}^{(B)}+h_{\mu\nu}.
\end{equation}
A rather long computation \cite{Misner:1973prb} shows that the Ricci tensor can be explicitly written for the above metric as
\begin{equation}
R_{\mu\nu}=R^{(B)}_{\mu\nu}+R^{(1)}_{\mu\nu}(h)+R^{(2)}_{\mu\nu}(h)+\mathcal{O}(h^3).
\end{equation}
Being aware that the vacuum field equations are $R_{\mu\nu}=0$, we can write
\begin{equation}
    R^{(B)}_{\mu\nu}+\langle R^{(2)}_{\mu\nu}(h)\rangle=0
    \label{eq:Misner},
\end{equation}
where $R^{(1)}_{\mu\nu}(h)$ vanished because it is linear in the amplitude of the wave but the action of the waves to curve up the background is a non-linear phenomenon as the linearized theory shows no sign of it~\cite{Maggiore:2007ulw}. The first term in \eq{Misner} is free of ripples since it varies only on scales far larger than the reduced wavelength of the wave, $\lambda/2\pi$ (also called the \textit{coarse-grain viewpoint}), while the $\langle...\rangle$ in the second term indicates the average over several wavelengths which extracts the smooth contribution with respect to the coarse-graining scale. \eq{Misner} shows how the stress-energy in the waves creates the background curvature. We can write in vacuum~
\begin{equation}
    G^{(B)}_{\mu\nu}\equiv R^{(B)}_{\mu\nu}-\frac12 R^{(B)}g^{(B)}_{\mu\nu}=8\pi T^{(GW)}_{\mu\nu},
\end{equation}
where
\begin{equation}
    T^{(GW)}_{\mu\nu}\equiv -\frac{1}{8\pi}\left[\langle  R^{(2)}_{\mu\nu}(h)\rangle -\frac12 g^{(B)}_{\mu\nu}\langle  R^{(2)}_{\mu\nu}(h)\rangle\right].
\end{equation}
It is possible to show that, using the definition \eq{hbarr}, we obtain~\cite{Maggiore:2007ulw,Watanabe:2006qe,Misner:1973prb,Guzzetti:2016mkm}
\begin{equation}
    T^{(GW)}_{\mu\nu}=\frac{1}{32\pi}\langle \bar{h}_{\alpha\beta,\mu}\bar{h}^{\alpha\beta}_{,\nu}-\frac12 \bar{h}_{,\mu}\bar{h}_{,\nu}-2\bar{h}_{,\beta}^{\alpha\beta}\bar{h}_{\alpha(\mu,\nu)}\rangle.
\end{equation}
The latter relation, in the TT gauge (see \eq{tttt}), simplifies into \cite{Maggiore:2007ulw}
\begin{equation}
   \boxed{ T^{(GW)}_{\mu\nu}=\frac{1}{32\pi}\langle \bar{h}_{ij,\mu}\bar{h}^{ij}_{,\nu}\rangle}\,,
\end{equation}
which behaves like the other stress-energy tensor, \ie when $G^{(B)\mu\nu}_{,\nu}=0$ we also have $T^{(GW)\mu\nu}_{,\nu}=0$. Hence, from the definition of the stress-energy tensor on a FLRW background, the energy density reads
\begin{equation}
    \boxed{\rho_{GW}=\frac{1}{32\pi G a(\eta)^2}\langle h'_{ij}(\boldsymbol{x},\tau)h^{'ij}(\boldsymbol{x},\tau)\rangle}\,.
    \label{GWrho}
\end{equation}
We can also introduce the energy density per logarithmic frequency interval
\begin{equation}
    \boxed{\Omega_{GW}(k,\tau)\equiv \frac{1}{\rho_c}\frac{d\rho_{GW}}{d\ln{k}}}\,.
    \label{eq:OmegaGW}
\end{equation}

\subsection{Gravitational waves as extra radiation} 

GW have a radiation-like behavior and for this reason a background of GW acts as an additional radiation field in the Universe, contributing to the background expansion rate $a_0^4(\Omega_{GW}+\Omega_r)a^{-4}$. To constrain this additional value, we should look for any observable able to probe the evolution of the Universe. This is the case for BBN (see \sect{bBNPGW}), and CMB (see \sect{CMBPGW}). The constraints come from the amount of radiation possible in the Universe at that time, hence the GW energy density $\rho_{\text{GW}}(T)$ must not exceed the limits on the abundance of radiation during BBN and CMB decoupling. A constraint on the presence of `extra' radiation is usually expressed in terms of the effective number of relativistic species species $N_{\rm eff}$ defined in \eq{Neff}. To understand how this reference value is modified in presence of additional gravitational radiation, we focus on temperatures $T\gtrsim \mathcal O(1)$ MeV when the relativistic species in the Universe were electrons and their antiparticles, positrons $e^{\pm}$, neutrinos $\nu$ and photons $\gamma$. Including also the contributions of gravitons, the total amount of radiation will read \citep{Maggiore:1999vm}
\begin{equation}
\rho_{\rm rad} = \frac{\pi^2}{30} \left[ 2 \, T_{\gamma}^4 + \frac{7}{4}\, T_{e^{\pm}}^4 + \frac{7}{4} \,N_{\rm eff}\, T_{\nu}^4 + 2 \, T_{\rm GW}^4 \right]
\end{equation}
where the factor 2 in front of $T_{\rm GW}$ counts the two different polarization states $(+,\times)$ of tensor perturbations. Apart from the gravitons, all the other species were in thermal equilibrium and shared the same temperature: $T_{\gamma}=T_{e^{\pm}}=T_{\nu}$. Therefore it is straightforward to see that we can describe gravitational radiation as an  additional contribution to the effective number of relativistic species
\begin{equation}
\Delta N_{\rm eff}^{\rm GW} =  \frac {8}{7} \frac{T_{\rm GW}^4}{T_{\gamma}^4}=\left. \frac {8}{7} \frac{\rho_{\rm GW}}{\rho_{\gamma}}\right|_{\mathrm{T}_{\gamma}\gtrsim \mathcal O(1) \,\text{MeV}}
\end{equation}
To rescale this contribution to the present time, we must consider that after $T\gtrsim \mathcal O(1)\,\rm{MeV}$, as the Universe expands, the gravitational wave energy-density decays as $\rho_{\rm GW}\sim 1/a^4$, while, assuming entropy conservation, the CMB photon energy-density evolves as $\rho_{\gamma} \sim 1 /\left(a^{4} q_{*}^{4 / 3}\right)$ with $q_{*}$ the number of entropic degrees of freedom in \eq{qstaar}. Therefore, the present-day contribution will be given by
\begin{align}
\Delta N_{\rm eff}^{\rm GW} &=\left.\frac{8}{7}\left(\frac{q_{*}(T \gtrsim 1 \mathrm{MeV})}{q_{*}\left(T_{0}\right)}\right)^{\frac{4}{3}} \frac{\rho_{\mathrm{GW}}}{\rho_{\gamma}}\right|_{\rm Today}
\end{align}
with $q_{*}(T_0)\simeq 3.91$ the current number of entropic degrees of freedom. 
While the present CMB energy density $\rho_{\gamma}$ is accurately measured~\citep{Planck:2019nip,Planck:2018jri,Aghanim:2018eyx}, the present-day fraction of the energy budget of the Universe in gravitational radiation, can be easily computed by integrating the spectrum over all scales~\citep{Maggiore:1999vm,Boyle:2005se,Guzzetti:2016mkm}
\begin{equation}
\Omega_{\mathrm{GW}}=\frac{1}{12 H_{0}^{2}} \int \text{d} \ln k \, \Delta_{\rm T}(k) \, \dot{T}(\eta_0,k)^2   
\label{Eq:full_OmGW}
\end{equation}
where the contribution of each mode is weighted by the time derivative of the transfer function, from \eq{TransferHhH}. Using the limit in \eq{forthe} and the fact that the weakness of the gravitational interaction guarantees that the GW are decoupled from the rest of the universe, it appears that the energy density spectrum today of the tensor modes generated during inflation is flat in $k$ (assuming exact de Sitter) for modes that enter before the equivalence, and scaling as $k^{-2}$ for modes that enter during matter-dominated era. We can estimate the present time contribution at generic frequency $f=k/2\pi$ as~\citep{Akrami:2018odb,Bartolo:2016ami,Cabass:2015jwe,Stewart:2007fu,Graef:2018fzu,Liu:2015psa} 
\begin{equation}
\Omega_{\mathrm{GW}}(f) \simeq\frac{ \Delta_{\mathrm{T}}(f)}{24 z_{\mathrm{eq}}} 
\label{eq:ForcoOmegaGW}
\end{equation}
with $z_{\mathrm{eq}}\simeq 3400$ the redshift at equivalence and $\Delta_{\mathrm{T}}$ the spectrum of primordial tensor modes. By using \eq{ForcoOmegaGW}, putting everything together, we finally get~\citep{Maggiore:1999vm}
\begin{equation}
\Delta N_{\rm eff}^{\rm GW} \simeq \frac{h_0^2}{5.6\times10^{-6}}\left(\frac{1}{24\,z_{\rm eq}}\right) \int_{f_{\rm min}}^{f_{\rm max}} \frac{\mathrm{d}f}{f}\, \Delta_{\rm T}(f)
\label{eq:Negeff}
\end{equation}
recovering the standard result that gravitational waves contribute to the effective number of relativistic species through the logarithmic integral of their power spectrum over frequencies. 

\section{Relic radiation from primordial gravitational waves} 
\label{sec:bBNPGW}
According to \eq{Negeff} the contribution of inflationary tensor anisotropies to the effective number of relativistic degrees of freedom in the early Universe depends on the parametrization of primordial tensor spectrum. The common practice in literature is to assume a power-law tensor spectrum given by \eq{PhenomenologicalTensor} over the whole range of integration so that the integral \eq{Negeff} can be easily solved analytically:
\begin{align}
\Delta N_{\rm eff}^{\rm GW}& \simeq \frac{h_0^2}{5.6\times10^{-6}}\left(\frac{ r A_s}{24\,z_{\rm eq}}\right) \frac{1}{n_{\rm T}} \left[\left(\frac{f}{f_{\star}}\right)^{n_{\rm T}} \right]^{f_{\rm max}}_{f_{\rm min}}.
\label{eq:limit1}
\end{align}
Interestingly, a blue tensor tilt exponentially amplifies the GWs production on ultraviolet frequencies, contributing mostly in \eq{limit1}, possibly leading to a sizable $\Delta N_{\rm eff}$ from PGWs. This effect is commonly used in literature to bound blue-tilted models of inflation, leading to a limit $n_{\rm T}\lesssim 0.4$ that is more or less of the same order as those inferred by gravitational wave experiments ( see \textit{e.g.} Refs.\citep{Allen:1997ad,Smith:2006nka,Boyle:2007zx,Kuroyanagi:2014nba,Ben-Dayan:2019gll,Aich:2019obd,Cabass:2015jwe}), with several implications also for gravitational waves observations~\citep{Vagnozzi:2020gtf,Benetti:2021uea,Vagnozzi:2022qmc} and fundamental physics~\citep{Calcagni:2020tvw}.

\begin{tcolorbox}[mybox]
The contribution in \eq{Negeff} also depends on the frequency range $f\in[f_{\rm min}\,,\,f_{\rm max}]$ over which the integral runs. The choice of the frequency is quite debated. In particular, the infrared cutoff can be safely set to $f_{\rm min} = 10^{-10}\,\rm{Hz}$ which approximately corresponds to the size of the comoving horizon at the time of BBN~\citep{Cabass:2015jwe,Pritchard:2004qp,Smith:2006nka}. Conversely, the ultraviolet cutoff is more arbitrary. Being primordial gravitational waves produced during inflation, we expect an ultraviolet cutoff of the size of the horizon at the end of inflation~\citep{Meerburg:2015zua} (as PGWs with smaller wavelengths cannot be produced). Anyway, the size of the horizon at the end of inflation depends on the reheating temperature $T_{\rm RH}$ at the end of inflation. Assuming an almost GUT-scale inflation and an instant reheating we can set $T_{\rm RH}\sim 10^{15}\,\rm{GeV}$ which corresponds to $k_{\rm end}\sim 10^{23}\,\rm{Mpc}^{-1}$ and thus $f_{\rm max}\simeq 10^{8}\,\rm{Hz}$~\citep{Cabass:2015jwe}. Nevertheless, inflationary models with (very) lower reheating temperatures $T_{\rm RH}\sim 10^{10} - 100\, \rm{GeV}$ have been proposed in the literature (see e.g.~\citep{Kawasaki:1999na,Kawasaki:2000en,Giudice:2000dp,Giudice:2000ex,Hannestad:2004px,Khoury:2011ii,Hasegawa:2019jsa,Hasegawa:2020ctq,Carenza:2021ebx,Freese:2017ace,Litsa:2020rsm}) and, although such scenarios are typically not easy to realize, in these models the ultraviolet cutoff may be much smaller, limiting the high-frequency contributions in the integral \eq{Negeff}, see also Refs.~\citep{Vagnozzi:2020gtf,Benetti:2021uea}. 
\end{tcolorbox}

\subsubsection{Running of the Tensor Index}
The result in \eq{limit1} is derived assuming the tensor tilt to be exactly constant under the whole range of integration. Typically, in physical models of inflation where the tensor tilt can acquire such large positive values, it may also acquire a non-negligible scale dependence ~\citep{Giare:2020vhn,Giare:2020plo} which, at first order is parametrized by the running $\alpha_T$ (see \eq{running}). Hence, it is possible to extend the power low relation and use the one described in \eq{PLtensorexpansion} to solve \eq{Negeff}.
\begin{figure}[htbp]
   \centering
 	 \includegraphics[width=0.9\textwidth]{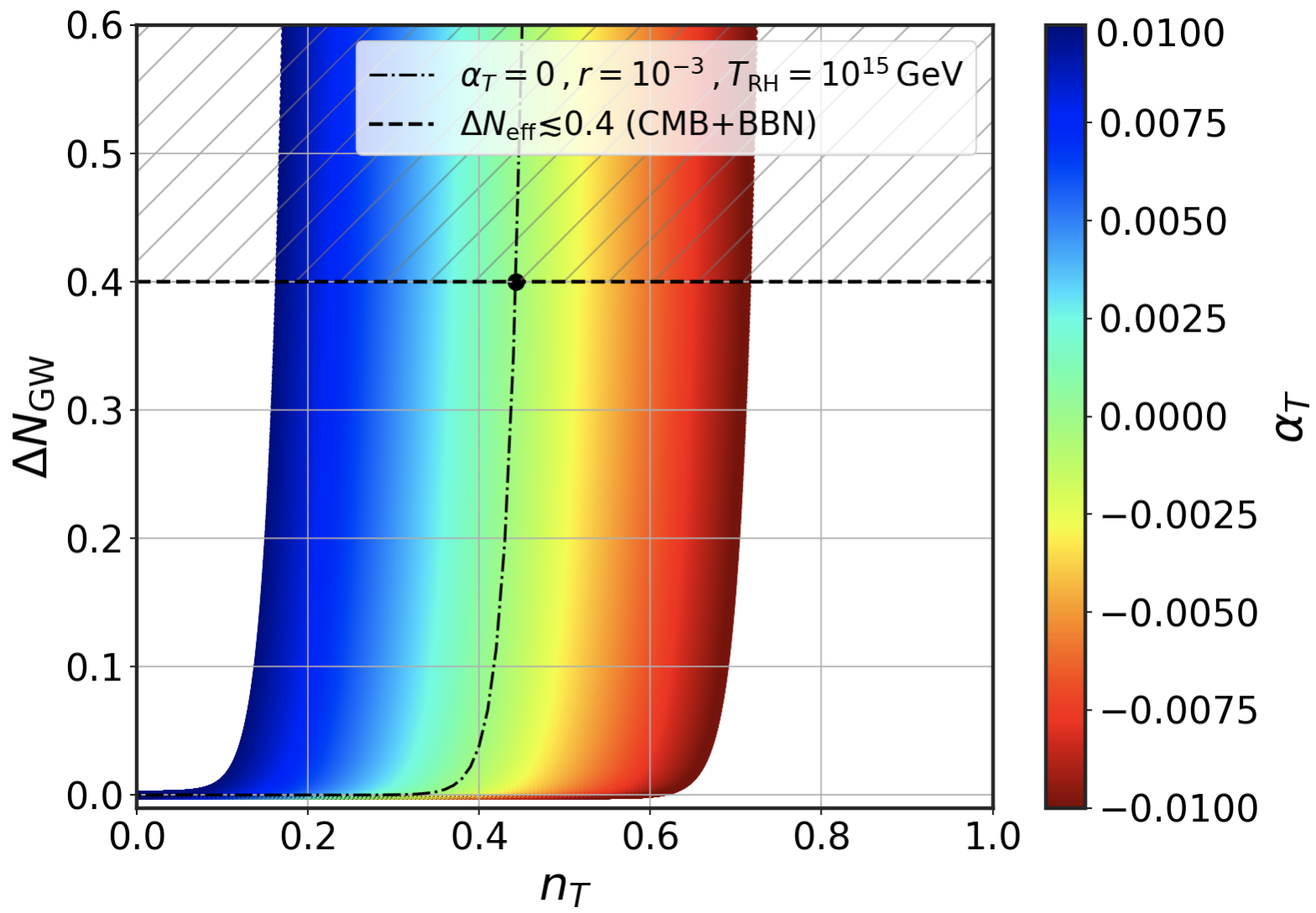}
 	 \caption[Inflationary tensor mode contribution to $N_{\rm eff}$]{\small Inflationary tensor mode contribution to the effective number of relativistic degrees of freedom as a function of the tensor tilt and its running $\alpha_{\rm T}$. The black dashed line represents the contribution for $\alpha_{\rm T}=0$ while the horizontal dashed line represents the limit on additional radiation from the BBN bounds.}
 	 \label{fig:figure1}
\end{figure}
In \autoref{fig:figure1}, it is possible to see the effect of a relatively small running of the tensor tilt on the calculation of $\Delta N_{\rm eff}^{\rm GW}$ finding that it can significantly change the results and so lead to a much tighter (relaxed) constraint on $n_{\rm T}$ represented by the horizontal dashed line in the figure. A positive (negative) $\alpha_{\rm T}$ amplifies (suppresses) the power spectrum on high frequency and its contributions in the integral \eq{Negeff}, providing another important clue that properly accounting for the ultraviolet behaviour of the tensor spectrum may be crucial in the calculation of $\Delta N_{\rm eff}^{\rm GW}$. In this regard, we notice that modes with frequency $f=k/2\pi$ will cross horizon $N_{k}$ e-folds before the end of inflation, where $N_{k}$ is given by \eq{Neff}. Hence, the "high frequencies" in the integral \eq{Negeff} correspond to tensor modes that exit the horizon extremely close to the end of inflation ($N_k\lesssim 2$ for $k\gtrsim 10^{21}\,\rm{Mpc^{-1}}$ and $T_{\rm RH}\sim 10^{15}$ GeV). This is precisely where, at least in the simplest inflationary scenarios, the potential decreases very rapidly to approach its minimum, and the slow-roll dynamics breaks down. It is not sure at all that a power-law parameterization (or even its next-to-leading order generalization) holds - even approximately - on such frequencies because the shape of the tensor spectrum will be strongly related to the shape of the inflationary potential~\cite{Kinney:2021nje,Giare:2020vhn}. As a result, the calculation of $\Delta N_{\rm eff}^{\rm GW}$ are largely sensitive to the underlying model.

\subsection{Current Bounds from Big Bang Nucleosynthesis}
\begin{figure}[h!]
	\centering
	\includegraphics[width=0.9\textwidth]{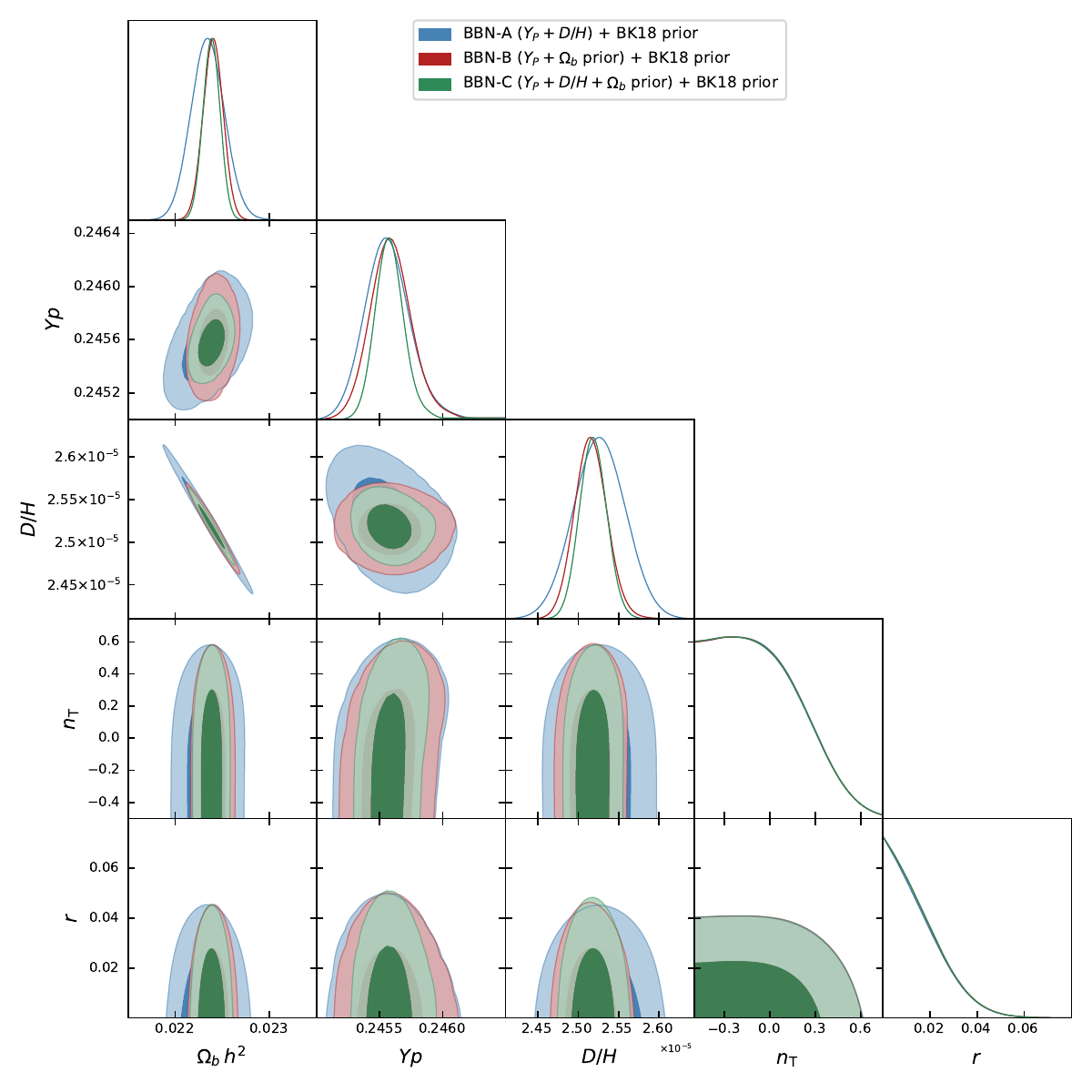}
	\caption[Probability posterior distributions for the most relevant cosmological parameters]{\small Two-dimensional 68\% and 95\% CL allowed regions and one-dimensional probability posterior distributions for the most relevant cosmological parameters obtained under the assumption of a power-law spectrum (see \eq{PhenomenologicalTensor}). The different colors refer to the different data combinations here considered for BBN analyses, see \tab{BBN}.}
	\label{fig:BBNBB}
\end{figure}
Before delving deeper into the model dependence of the tensor power spectrum in the calculation of $\Delta N_{\rm eff}^{\rm GW}$, it can be useful to update the observational constraints on blue-tilted models inflation resulting from the BBN epoch and quantifying how such results change with the parameterization of the primordial tensor spectrum. The BBN (see \sect{ThermalHistory}) explains the formation of the first light nuclei heavier than the lightest isotope of hydrogen by a solid understanding of the nuclear interactions involved in their production. It also provides a natural arena to test and constrain extensions to both cosmology and fundamental physics since any proposed model of the early Universe must be able to explain the abundances of light elements inferred by astrophysical and cosmological observations. The reason why the BBN constraining power can be applied to the analysis of blue-tilted models of inflation is quite straightforward: according to the Friedmann equation, additional gravitational radiation ($\Delta N_{\rm eff}^{\rm GW}$) will increase the expansion rate of the Universe $H(z)$. A faster expansion leads to a higher freeze-out temperature of the weak interactions, implying a higher fraction of primordial Helium and Deuterium, as well as a higher fraction of other primordial elements. This makes BBN an extremely powerful and quite general tool for constraining the total amount of relativistic species in the Universe, with several implications for physics beyond the Standard Model~\citep{Kawasaki:2004qu,Steigman:2007xt,Cyburt:2004yc,Sabti:2019mhn,DEramo:2022nvb}, the Neutrino flavour physics and inflationary cosmology. 

We begin our analysis by assuming a power-law primordial spectrum, parameterized by two quantities: the amplitude $r$ and the tilt $n_{\rm T}$. We randomly sample $N=10^6$ values of $r \in [0, 0.1]$ and $n_{\rm T} \in [-2, 2]$. For each of these points, we compute the contribution to the effective number of relativistic species $\Delta N_{\rm eff}^{\rm GW}$ as well as  the baryon energy density, in the range $\Omega_b h^2\in\left[0.020\,,\,0.025 \right]$ and create a grid in the plane $(\Delta N_{\rm eff}^{\rm GW} ,\Omega_b h^2)$. Then, we solve numerically the set of differential equations that regulate the BBN nuclear interactions in the primordial plasma~\citep{PitrouEtal2018,Pisanti:2007hk,Consiglio:2017pot,Gariazzo:2021iiu} using the code \textsc{PArthENoPE}~\cite{Gariazzo:2021iiu} and fixing the values of the neutron lifetime\footnote{The neutron lifetime is fixed to $\tau_n = 879.4$ s, corresponding to the latest measurement reported by the Particle Data Group ($\tau_n = 879.4 \pm 0.6 $ s)~\cite{ParticleDataGroup:2020ssz}}, for each point in the $(\Delta N_{\rm eff} ,\Omega_b h^2)$ plane the code computes the corresponding value of the primordial Helium fraction $Y_P$, the Deuterium abundance $D/H$ and all the other light element abundances, allowing direct comparison with observational data. In this regard, our baseline dataset for the BBN analyses consists of
\begin{itemize}
\item Two independent measurements of the primordial Helium \textit{fraction}, $Y_p = 0.2449 \pm 0.0040$~\citep{Aver:2015iza} and $Y_p = 0.2446\pm 0.0029$~\citep{Peimbert:2016bdg}. 
\item A percent determination of the primordial Deuterium abundance $D/H=\left(2.527\pm0.030\right)\cdot 10^{-5}$ based on six high precision and homogeneously analyzed $D/H$ measurements from~\citep{Cooke:2017cwo}.
\item The value of the baryon energy density parameter $\Omega_b\,h^2=0.0224 \pm 0.0001$ from the final 2018 Planck data release of temperature and polarization CMB angular power spectra~\citep{Aghanim:2018eyx}.
\item A prior on the tensor amplitude $r<0.037$ at 95\% CL coming from a combination of the final 2018 Planck data release of temperature and polarization CMB angular power spectra~\citep{Aghanim:2018eyx} and the B-modes 2018 likelihood from the Bicep Collaboration ~\citep{BICEP:2021xfz}.
\end{itemize}
We apply these priors on BBN abundances, re-weighting the contributions of the points using \textit{importance sampling}~\cite{DEramo:2022nvb}. Consequently,we obtain informative posterior distributions for the most interesting parameters to be inferred by observations. We summarize the results in \tab{BBN} while \fig{BBNBB} provides the marginalized posterior distributions of parameters.
\begin{table}
    \centering
    \renewcommand{\arraystretch}{1.5}
    \resizebox{0.8\textwidth}{!}{\begin{tabular}{|l|c|c|c|}
        \hline \hline
        \textbf{Parameter} & \textbf{BBN-A ($Y_p + D/H$)} & \textbf{BBN-B ($Y_p + \Omega_b h^2$)} & \textbf{BBN-C ($Y_p + D/H + \Omega_b h^2$)} \\
        \hline\hline
        $\Omega_{\rm b} h^2$ & $0.02234 \pm 0.00017$ & $0.02240 \pm 0.00010$ & $0.022382 \pm 0.000086$ \\
        $Y_p$ & $0.24558 \pm 0.00010$ & $0.24561 \pm 0.00010$ & $0.245591^{+0.000015}_{-0.000060}$ \\
        $(D/H)\cdot 10^{-5}$ & $2.527 \pm 0.030$ & $2.516 \pm 0.020$ & $2.519 \pm 0.016$ \\
        $\Delta N_{\rm eff}$ & $< 0.33 \, (<0.40)$ & $< 0.32 \, (<0.40)$ & $< 0.16 \, (<0.21)$ \\
        \hline \hline
    \end{tabular}}
    \caption[Results inferred from BBN primordial abundances]{\small Results inferred from BBN primordial abundances. The constraints on $\Omega_{\rm b} h^2$ , $Y_p$ and $10^{5}\cdot(D/H)$ are given at 68\%CL while the upper bounds on $\Delta N_{\rm eff}$ are given at 95\% CL (99\% CL).}
    \label{tab:BBN}
\end{table}
We start by using priors on $Y_p$ and $D/H$, alongside the tensor amplitude prior from B-modes. In this case the free parameters are $\Omega_b h^2$ and $n_{\rm T}$. We refer to this dataset as BBN-A. From it, we derive an upper limit on $\Delta N_{\rm eff}<0.3$ at 95\% CL, consistent with previous studies~\cite{Aver:2015iza,Peimbert:2016bdg,Cooke:2017cwo,DEramo:2022nvb,Giare:2021cqr,Aich:2019obd}. Assuming all this contribution to be made of primordial gravitational waves, we infer an upper limit on the tensor tilt $n_{\rm T}<0.3$ at 95\% CL, matching recent CMB-analyses \citep{Galloni:2022mok}. To test the robustness, we impose a prior on $\Omega_b h^2$ from Planck~\citep{Aghanim:2018eyx} alongside $Y_p$, labeling this dataset BBN-B. The free parameters are now $D/H$ and $n_{\rm T}$. The constraints on $\Delta N_{\rm eff}$ and $n_{\rm T}$ remain unchanged, with the $D/H$ value serving as a consistency check. Finally, combining all priors ($Y_p + D/H + \Omega_b h^2 + r$), labeled BBN-C, we find an improved constraint on $\Delta N_{\rm eff}<0.16$ at 95\% CL. However, the bound on $n_{\rm T}$ remains unchanged. This lack of improvement is explained by the behavior of $\Delta N_{\rm eff}$ as a function of $n_{\rm T}$, where for $n_{\rm T} \sim 0.4$, $\Delta N_{\rm eff}$ shows little sensitivity to changes in $n_{\rm T}$. It can be easily understood by looking at the black dashed line in \fig{figure1}: when we are close to $n_{\rm T}\sim 0.4$, the line in the plane $(n_{\rm T}\,,\,\Delta N_{\rm s})$ becomes almost horizontal. The results of the constraints on the tensor sector can be seen in \tab{BBN2}. 
\begin{table}
    \centering
    \renewcommand{\arraystretch}{1.5}
    \resizebox{0.8\textwidth}{!}{\begin{tabular}{|l|c|c|c|}
        \hline \hline
        \textbf{Parameter} & \textbf{BBN-A ($Y_p + D/H$)} & \textbf{BBN-B ($Y_p + \Omega_b h^2$)} & \textbf{BBN-C ($Y_p + D/H + \Omega_b h^2$)} \\
        \hline \hline
        $n_{\rm T}$ & $<0.324 \, (<0.376)$ & $<0.323 \, (<0.374)$ & $<0.32 \, (0.368)$ \\
        $r$ & $<0.037$ & $<0.037$ & $<0.037$ \\
        \hline \hline
    \end{tabular}}
    \caption[BBN results for the inflationary parameters]{\small The upper bounds are given at 95\% CL (99\% CL). These results for the inflationary parameters  are inferred under the parameterizations of the spectrum with a scale-independent tensor tilt. A BK18 prior ($r<0.037$ at 95\% CL) is assumed on the tensor amplitude.}
    \label{tab:BBN2}
\end{table}
Next, we explore the effect of allowing for running of the tensor tilt, $\alpha_{\rm T} \in [-0.2, 0.2]$. In this case, we use \eq{PLtensorexpansion} and summarize the results in \tab{BBN3}. 
\begin{figure}[h!]
	\centering
	\includegraphics[width=0.7\textwidth]{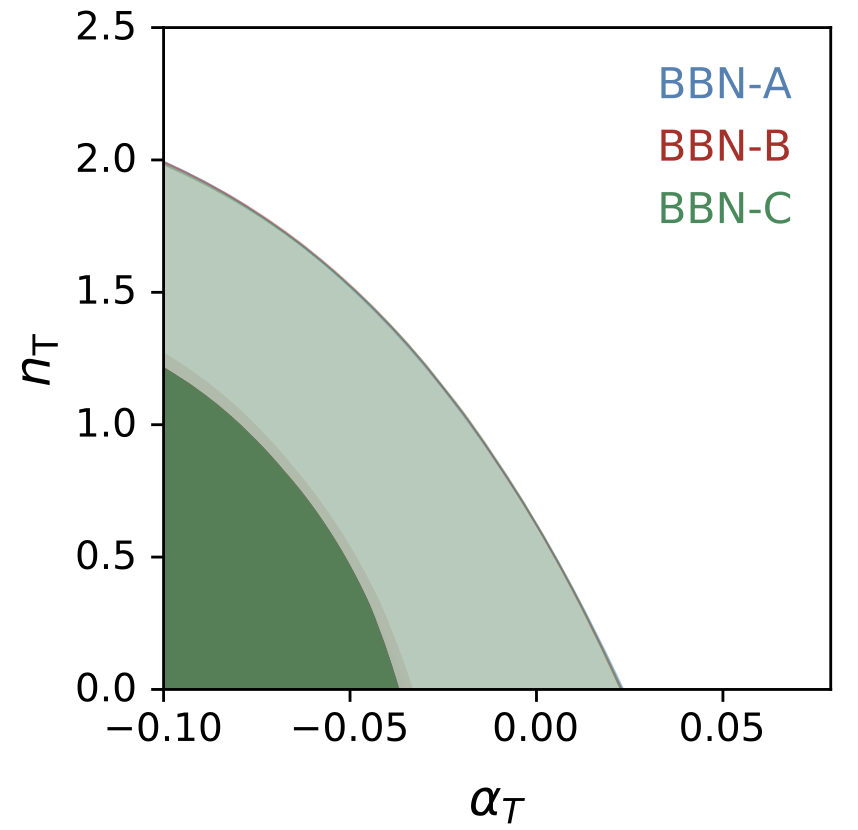}
	\caption[$(n_T,\alpha_T)$ plane]{\small 2D joint marginalized contours in the $(n_T,\alpha_T)$ plane obtained by allowing a non-vanishing running $\alpha_T$ to vary, see \ref{eq:PLtensorexpansion}.}
	\label{fig:BBNiiiBB}
\end{figure}
Both the bounds on additional radiation and light element abundances remain unchanged, as they do not depend on the tensor spectrum parameterization. However, the limit on $n_{\rm T}$ changes: with running allowed, we obtain $n_{\rm T} < 1.8$ at 95\% CL, while at 99\% CL the tilt is unbounded. This is due to a degeneracy between $n_{\rm T}$ and $\alpha_{\rm T}$ (see \fig{BBNiiiBB}), as a positive running amplifies the gravitational wave power on small scales, mimicking a larger scalar tilt. Conversely, negative running reduces gravitational wave power, relaxing the constraints on $n_{\rm T}$. These results confirm the importance of the parameterization choice when constraining blue-tilted inflationary models. For more details, see Appendix B in~\cite{Forconi:2}.
\begin{table}
    \centering
    \renewcommand{\arraystretch}{1.5}
    \resizebox{0.8\textwidth}{!}{\begin{tabular}{|l|c|c|c|}
        \hline \hline
        \textbf{Parameter} & \textbf{BBN-A ($Y_p + D/H$)} & \textbf{BBN-B ($Y_p + \Omega_b h^2$)} & \textbf{BBN-C ($Y_p + D/H + \Omega_b h^2$)} \\
        \hline \hline
        $n_{\rm T}$ & $<1.80 \, (\text{unc.})$ & $<1.80 \, (\text{unc.})$ & $<1.80 \, (\text{unc.})$ \\
        $r$ & $<0.037$ & $<0.037$ & $<0.037$ \\
        \hline \hline
    \end{tabular}}
    \caption[BBN results for the inflationary parameters for $\alpha_T\neq 0$]{\small The upper bounds are given at 95\% CL (99\% CL). These results for the inflationary parameters  are inferred under the parameterizations of the spectrum with a scale-dependent tensor tilt. A BK18 prior ($r<0.037$ at 95\% CL) is assumed on the tensor amplitude.}
    \label{tab:BBN3}
\end{table}

\subsection{Higher-order stochastic reconstruction}
\label{sec.3.3}

\begin{figure}[h!]
 	 \centering
 	 \includegraphics[width=0.9\textwidth]{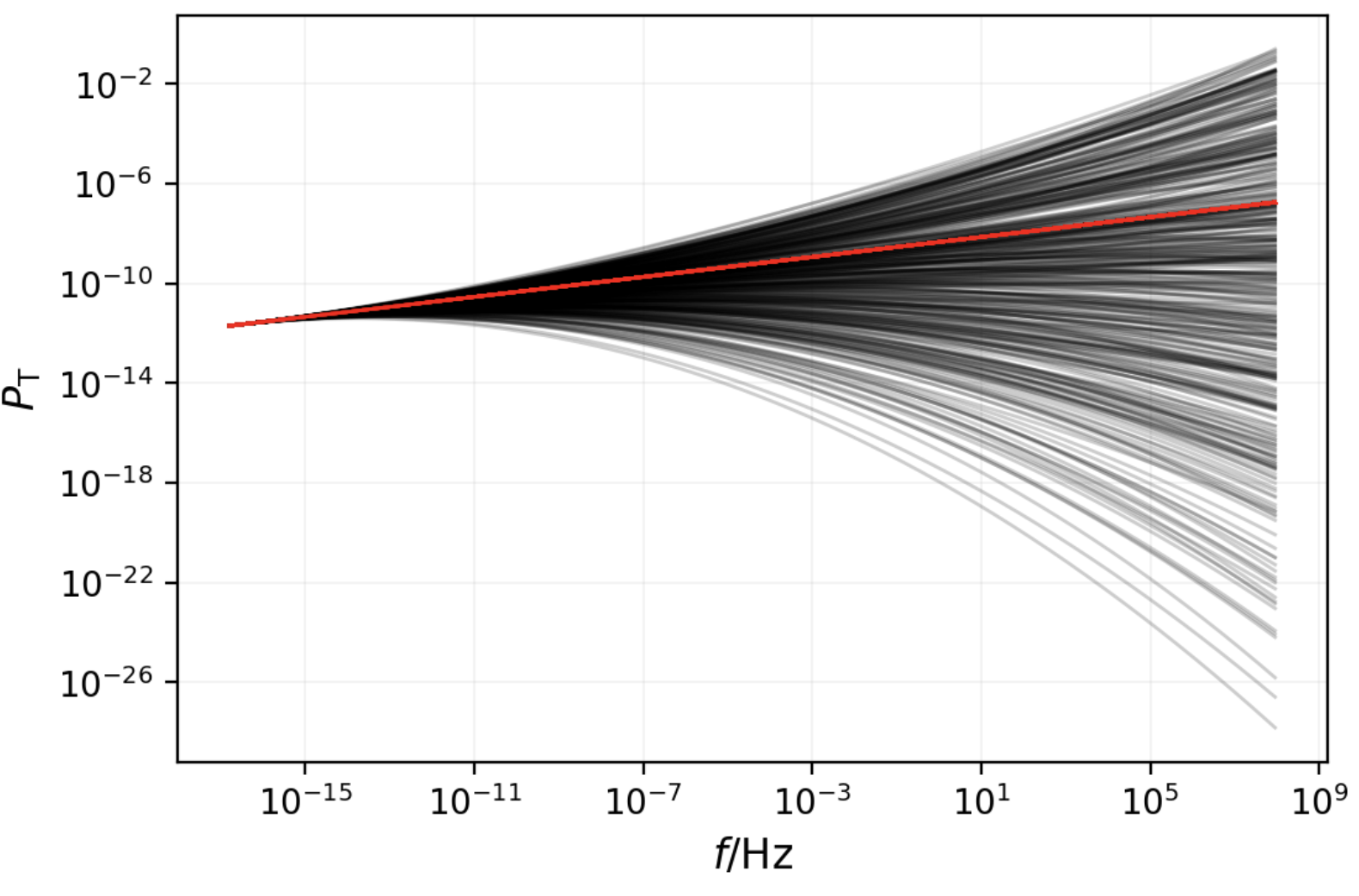}
 	 \caption[ Examples of randomly generated tensor spectra]{\small Examples of randomly generated tensor spectra (and the power-law extrapolation, red line).}
 	 \label{fig:figure2}
 \end{figure}

\begin{figure}[h!]
 	 \centering
 	 \includegraphics[width=0.9\textwidth]{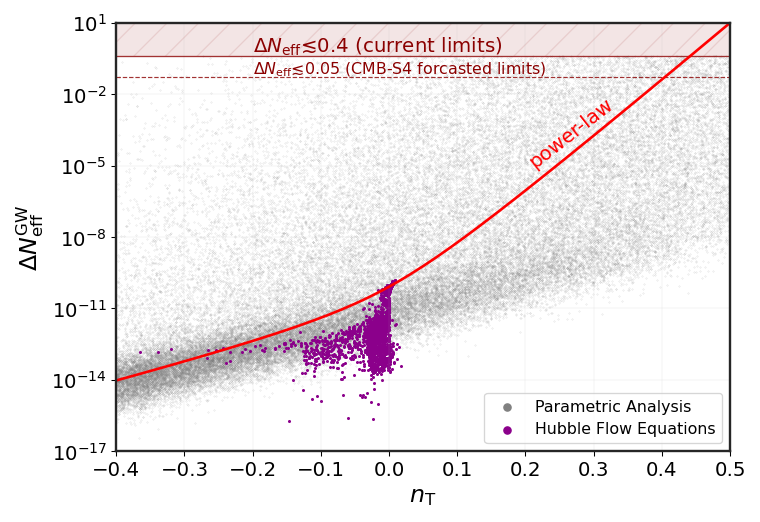}
 	 \caption[PGWs contribution to radiation energy-density]{\small Primordial Gravitational Wave contribution to radiation energy-density in the early Universe parametrized as a correction to the effective number of relativistic species ($\Delta N_{\rm eff}^{\rm GW}$). The red thick line represents the predictions for $\Delta N_{\rm eff}^{\rm GW}$ inferred by extrapolating a power-law parameterization for a scale-invariant tensor-tilt over all frequencies. The gray dots represent the results of the parametric analysis \eq{EXP}. The magenta points represent the observable predictions of an ensemble of physical models randomly realized within the framework of the EFT of inflation by means of a theoretical Monte Carlo. The horizontal red band (dashed line) represents the current (future forecasted) observational limit on radiation.}
    \label{fig:figure3}
\end{figure}

At the beginning of \sect{bBNPGW} we have seen how a simple additional term as the running, changes the bounds on PGWs coming from the BBN. A more general approach to the problem can be obtained by going beyond the first-order expansion in \eq{PLtensorexpansion} and expanding (the log of) the tensor spectrum as a series of powers 
\begin{align}
\ln \Delta_{\rm T}= \sum_{j=0}^{\infty} a_j (x-x_0)^j .
\label{eq:EXP}
\end{align}
If we choose the CMB frequency as the center of the expansion ($x_0=\ln f_{\star}$) the coefficients $a_j$ can be trivially related to (the derivatives of) the tensor spectrum evaluated at the CMB scales. In particular, the tensor amplitude and the tensor tilt are simply given by
\begin{equation}
a_0= \ln(rA_s), \quad a_1=\frac{d\ln \Delta_{\rm t}}{ d\ln f}\equiv n_{\rm T}
\label{eq:a0a1}
\end{equation}
while the higher order coefficients are related to the higher order derivatives of the spectrum (or the tensor tilt) as: 
\begin{equation}
\quad a_{j>1}=\frac{1}{j!} \frac{d^j\ln \Delta_{\rm T}}{ d\ln^j f} = \frac{1}{j!} \frac{d^{j-1} n_{\rm T}}{d\ln^{j-1} f}.
\label{eq:aj}
\end{equation}
Notice that if we stop the sum expansion at $j=1$ or $j=2$, we exactly recover \eq{PhenomenologicalTensor} or \eq{PLtensorexpansion} respectively. Therefore, including more and more terms in the sum will clearly guarantee a more accurate reconstruction of the tensor spectrum at $x \gg x_0$ since it employs also the other higher-order terms in the expansion. 
\begin{tcolorbox}[mybox]
If we want to adopt this parameterization in the integral \eq{Negeff}, we need to make sure that this sum will actually converge on the frequencies over which the integration runs. Although this depends on the specific model of inflation, in most models the tensor spectrum is a slow-evolving regular function of the frequency so that it is reasonable to expect a global convergence. For instance, the simplest slow-roll scenario is characterized by a hierarchy of parameters $n_{\rm t}=\mathcal O(\epsilon)$ and $d^{j} n_{\rm T} /d\ln^{j} f\lesssim \mathcal O(\epsilon^{j+1})$. Assuming such a scaling, the sum convergence can be easily proved by evaluating the radius of convergence
\begin{equation}
\frac{1}{R}\doteq\lim_{j\to \infty} \left|\frac{a_{j+1}}{a_j}\right| = \lim_{j\to \infty} \left|\frac{\mathcal O(\epsilon)}{j+1}\right|= 0.
\end{equation}
\end{tcolorbox}
In principle, we can adopt this parameterization to predict the value of the tensor spectrum at $x\gg x_0$. In practice, all the arbitrariness of the method is encapsulated into the coefficients $\{a_j\}$. Ultimately, fixing their values is equivalent to fixing a specific model of inflation. Here we sample different inflationary models by randomly varying the coefficients $\{a_j\}$ (for further details see~\cite{Forconi:2}). Examples of the spectra obtained within this method are provided in \fig{figure2}, together with a simple leading order power-law approximation (red line). Notice that we consider both negative and positive coefficients $\{a_j\}$, so that, on high frequencies, the spectrum can be either suppressed or amplified. Indeed, while in the simplest cases we expect suppression of power because of the rapid decrease of inflationary potential, in more elaborated scenarios it is in principle possible to build inflationary models with ultraviolet amplification of tensor perturbations~\citep{Oikonomou:2022ijs,Barrow:1993ad,Peng:2021zon,Ota:2022hvh,Odintsov:2022sdk,Baumgart:2021ptt}. In this latter case we may end up with large amounts of GW on the small scales as those probed by gravitational interferometers. Therefore, for all the simulated spectra, we also checked that the amplitude $\Delta^2_{\rm T}(k)$ remains consistent with the LIGO/VIRGO limit, keeping only the models able to satisfy observations. This is the reason why in \fig{figure2} we get much more suppressed spectra than amplified ones. From the same figure, we can also appreciate how the usual power-law parametrization is a precise approximation only at frequencies corresponding to the CMB scales (as required by construction) while important deviations are observed at higher frequencies, in spite of our efforts for keeping small the parameters $\{a_j\}$. 
Fixing the ultraviolet cutoff to $f_{\rm max}\simeq 10^{8}\,\rm{Hz}$, we numerically solve the integral \eq{Negeff} for all the different shapes of $\Delta^2_{\rm T}(f)$, thus computing the corresponding value of $\Delta N_{\rm eff}^{\rm GW}$. We ensure the computational relative error due to the numerical integration method to remain smaller than 1\%. In \autoref{fig:figure3} we show the results of our random analysis. Once again the red solid line represents the contribution $\Delta N_{\rm eff}^{\rm GW}$ obtained within the power-law parametrization corresponding to $\alpha_T=0$. Instead, the gray dots represent the values of $\Delta N_{\rm eff}^{\rm GW}$ obtained by the numerical integration method of the randomly obtained tensor spectra. 

This parametric analysis can be useful for pointing out potential limitations and weaknesses in current analyses, but a more reliable investigation of physical models of inflation and their respective contribution to the energy budget of the early Universe is needed. 

\subsection{Physical analysis}

In order to conduct a more reliable investigation of physical models of inflation, we use the Hubble flow equations in the EFT, as described in \sect{HFLEQ}. Taking into account all the operators in the quadratic effective action that induce tensor perturbations, one can derive the following leading order relation for the power-spectrum~\citep{Creminelli:2014wna,Noumi:2014zqa,Giare:2020vss}:
\begin{equation}
  \Delta_{\rm T}= \frac{1}{c_{\rm T}}\left(\frac{H^2}{\pi^2\,M_{\mathrm{pl}}^{2}} \right)
  \label{eq:PtCt}
\end{equation}
where $c_{\rm T}$ is the propagating speed of tensor modes that can be simply expressed in terms of $\bar{M}_3$ as $c_{\rm t}^{-2}=1 - \bar{M}_3^2 / M_{\mathrm{pl}}^{2}$ where $\bar{M}_3$ is defined in \eq{genaa}. In this case, it is straightforward to see, from its definition, that the tensor tilt acquires a further correction
\begin{equation}
    n_{\rm T}=-2\epsilon + \epsilon_{\rm T}
    \label{eq:ntt}
\end{equation}
where the evolution of the parameter
\begin{equation}
    \epsilon_{\rm T}= - \frac{\dot{c}_{\rm T}}{H\,c_{\rm T}}
    \label{eq:ctepsilont}
\end{equation}
is clearly governed by the system in \eq{system_Q}. It is also worth noting that \eq{TensorProblem} does not hold anymore. However, the tensor spectrum and its evolution are fully determined by the evolution of the background and the parameter $\epsilon_{\rm T}$. To test the observable predictions of a large number of stochastically generated models, we perform a theoretical Monte Carlo (see~\cite{Forconi:2} for more details):
\begin{itemize}
\item First and foremost, we truncated the hierarchy at $4$th-order. Then, we draw a suitable set of randomly chosen initial conditions for the background parameters.

\item Once the initial conditions are chosen,  we integrate \eq{HFE_background} forward in time for at most $\sim10^4$ e-folds of expansion. If we manage to get the end of inflation defined by the usual relation $\eh=1$, we store all the background parameters as functions of the number of e-folds $N$ before the end of inflation. Given a large number of repetitions ($\gtrsim 10^4$), approximately $90\%$ of the time the end of inflation is successfully reached. 

\item We then use the values reached by parameters at the end of inflation as new initial conditions at $N=0$ and perform a backward-in-time integration up to the e-folds when the primordial observables are evaluated ($N=60$). We reject all the results outside the range $0.94<n_s<0.98$, chosen conservatively around the Planck best-fit value, ending up with roughly $17\%$ of the total. We store the survived models and proceed to evolve $c_T$.   

\item  To solve the system of equations~\eq{system_Q} we need to specify some initial conditions for $\epsilon_T$ and the other high-order tensor parameters as well as perform the same consistency check as for the background.  Once that all the steps have been carried out, the model is either accepted or rejected. At the end of the process, only approximately $40\%$ of the attempts resolve in a successful inflation with a non-trivial tensor-speed sector.

\item Finally, we evolve the tensor spectrum dynamically from $N=60$ up to the end of inflation by means of the Hubble flow Equations. For each spectrum, we calculate the corresponding contribution to the energy density of the early Universe parametrized in terms of the effective number of relativistic degrees of freedom $\Delta N_{\rm eff}^{\rm GW}$. To do so, we evaluate the corresponding energy-density in gravitational waves $\Omega_{\rm GW}(f)$ by \eq{ForcoOmegaGW} and integrate it over frequency according to \eq{Negeff}.

\end{itemize}

Using this procedure, we are able to collect a sufficiently large ensemble of physical models ($\simeq 10.000$) which spans a reasonable range of possibilities, from realization with a canonical tensor-speed sector (\textit{i.e.}, $c_{\rm T}=1$  and $\epsilon_T=0$) to more general cases with time-dependent tensor parameters. 

\begin{tcolorbox}[mybox]
To achieve such an high number of models in the most direct and simple way some limitations are introduced. A first major restriction comes from limiting our analysis to a small subgroup of models with a fixed tensor amplitude $r\sim 0.001$ on the CMB scales. However that we do not expect this limitation to introduce a large bias on the frequency distribution of the values obtained for the tensor tilt. We are not particularly interested in studying the model frequency distribution, but rather in understanding whether models sharing similar parameters on the CMB scales may result into a significant different contribution to the energy budget of the early Universe because of their different evolutionary paths. Focusing only on models with the same $r$ at $N=60$ turns out to be particularly useful for this purpose since it ensures that the predictions for $\Delta N_{\rm eff}^{\rm GW}$ do not depend on the value of the tensor amplitude at the CMB scales. Furthermore, $r\sim 0.001$ is the declared target of future CMB-S4-like experiments~\cite{CMB-S4}. Therefore we believe it should be particularly interesting to understand what kind of physical models future surveys may be able to probe. A second minor limitation is introduced by taking only positive initial conditions for the parameter $\eh$, without considering models resulting from a background evolution with $\epsilon_{\rm in}<0$~\citep{Capurri:2020qgz}. This framework is quite general and can be applied also to more complicated scenarios where this possibility of negative values is viable, such as super-inflation models~\citep{Creminelli:2006xe,Gasperini:1992pa,Brustein:1995ah,Baldi:2005gk} or models with intermittent NEC violation~\citep{Cai:2020qpu,Cai:2022nqv}. It is important to acknowledge that this limitation can in fact result in a significant reduction in the number of models predicting a blue-shifted tensor tilt that our pipeline is able to investigate, see also \fig{figure3}. Despite this, our conclusions on $\Delta N_{\rm eff}^{\rm GW}$ cannot in any way rely on these exotic scenarios and we can safely exclude such models from the analysis without biasing the results.
\end{tcolorbox}

\begin{figure}[h!]
 	 \centering
 	 \includegraphics[scale=0.6]{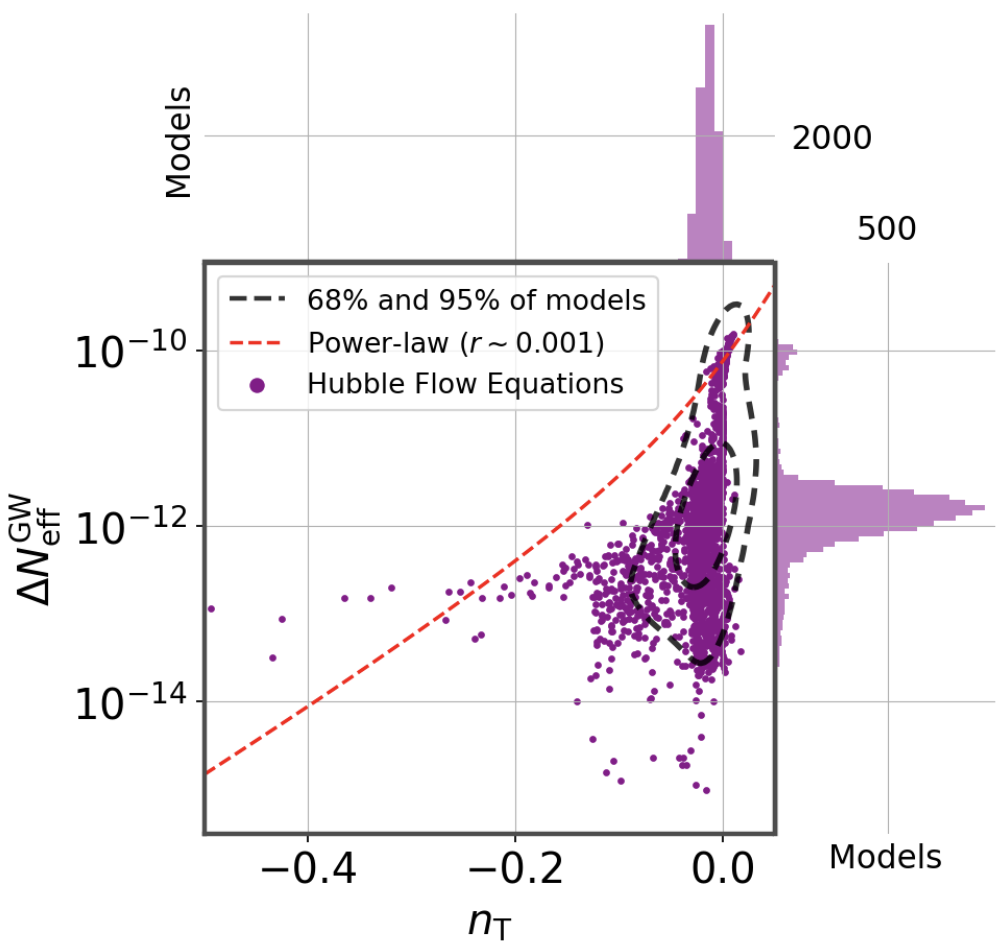}
 	 \caption[Observable predictions in the plane $(n_{\rm t}\,,\,\Delta N_{\rm eff}^{\rm GW})$]{\small Observable predictions in the plane $(n_{\rm t}\,,\,\Delta N_{\rm eff}^{\rm GW})$. The magenta dots represent the models realized within the Hubble Flow Equation method while the red dashed line represents the power-law prediction. The dashed black lines define the regions of the plane that contain the 68\% and 95\% of the total models and are calculated by marginalizing over the point frequency distribution (displayed by the two histograms on the axes).  }
 	 \label{fig:figure5}
\end{figure}

In \fig{figure3}, the results for $\Delta N_{\rm eff}^{\rm GW}$ from this approach are shown as dark magenta dots. \fig{figure5} zooms in on the $(n_{\rm T}, \Delta N_{\rm eff})$ plane, showing the distribution of physical models. The dashed black lines indicate the 68\% and 95\% regions, calculated by marginalizing over the point frequency (shown in the histograms). Most models show slightly negative tilt, as few blue-tilted models pass all consistency checks. Surviving models generally have canonical tensor speed ($c_{\rm T} = 1$, $\epsilon_{\rm T} = 0$) and respect the null energy condition ($\eh > 0$), following slow-roll consistency relations. This suggests that blue-tilted models satisfying all physical constraints (e.g. stability, causality, observational limits) are challenging to realize, with red-tilted models favored in Monte Carlo simulations. 
For blue-tilted models, only small $n_{\rm T}$ values are feasible, with the largest being $n_{\rm T} \simeq 0.08$. For red-tilted models, most have small tilt ($-0.1 < n_{\rm T} < 0$), though a few show $n_{\rm T} \lesssim -0.2$, which remain consistent with observational predictions. These models can arise from background and tensor speed evolution, particularly with $\epsilon_{\rm T} < 0$, implying $\dot{c}_{\rm T} > 0$ and an increasing tensor speed at CMB frequencies. Since the speed of gravitational interactions is not strictly constrained at these frequencies, such models remain viable, though a non-unitary $c_{\rm T}$ could raise concerns due to perturbative departures from General Relativity. However, these models are rare due to narrow initial condition constraints. An interesting observation in \fig{figure5} is a set of models following a near power-law behavior, with $n_{\rm T}$ close to zero, implying extremely slow-roll dynamics and a flat inflationary potential. While these models are not in the densest region, they still contribute to the 95\% region, producing a small peak in the $\Delta N_{\rm eff}$ histograms, as shown in \fig{figure5}.

\subsubsection{Observable Predictions: Relic Gravitational Radiation}

From \fig{figure5} we can appreciate that, while the histogram of $n_{\rm T}$ is very sharped and most models share similar values of the tensor tilt, the histogram of $\Delta N_{\rm eff}^{\rm GW}$ is instead much broader and the regions containing the 68\% and 95\% of models are almost vertical, spanning a quite large range of values $\Delta N_{\rm eff}^{\rm GW}\simeq 10^{-10}\, - \, \simeq 10^{-14}$. This means that models that share the same inflationary parameters on the CMB scales (i.e. the same amplitude $r$ and the same tilt $n_{\rm T}$) can easily result in a completely different contribution to $\Delta N_{\rm eff}^{\rm GW}$. The results of theoretical Monte Carlo suggest that extrapolating a power-law behavior over ultraviolet frequencies, in most cases leads to overestimating the gravitational wave contribution to the energy budget of the Universe. Nonetheless, we can observe a few models where the actual contribution to the effective number of relativistic species is larger than predicted by a power-law. While such points represent the vast minority of the models, it is still interesting to explain the physical reason underlyng this behavior. In particular, it is evident both from \fig{figure3} and from \fig{figure5} that this event is more frequent for those very few points that show a very red tensor tilt $n_{\rm T}\lesssim - 0.2$. On the other hand all the models obtained with the Hubble Flow Equation method give an extremely small $\Delta N_{\rm eff}^{\rm GW}$. The histogram of this parameter is in fact centred around values $\Delta N_{\rm eff}^{\rm GW}\sim 10^{-12}$, with a second small peak of models at $\Delta_{\rm eff}^{\rm GW}\sim 10^{-10}$, resulting from that class of models with an extremely slow evolution. These values are far away from the total amount of additional radiation allowed by data ($\Delta N_{\rm eff}^{\rm GW}\lesssim 0.3 - 0.4$) as well as from any current and future experimental sensitivity. 

At this point, a reasonable range of different scenarios and possibilities has been covered, consistently getting conclusive evidence that assuming a power-law spectrum over all scales can lead to a wrong estimation of the gravitational wave contribution in the early Universe. In light of this result, the calculation of $\Delta N_{\rm eff}^{\rm GW}$ proves to be remarkably model-dependent and more accurate analyses are needed before inferring any reliable conclusion on (blue-titled) inflationary models in light of the BBN bounds on additional radiation. 

\section{Cosmological constraints on slow-roll inflation}
\label{sec:CMBPGW}
As already anticipated, PGWs may imprints the CMB photon polarization, leading to a very distinctive signature in the B-modes spectrum on large angular scales~\cite{Guth:1980zm,Starobinsky:1980te,Linde:1981mu,Vilenkin:1983xq,Mukhanov:2005sc,Dodelson:2003ft,Weinberg:2008zzc,Martin:2013tda,Baumann:2009ds,Clarke:2020bil}. It has been demonstrated in \sect{perturbationss} and \sect{TensoTensoO} that, in the framework of single-field inflation with Einstein gravity, primordial scalar and tensor perturbations are expected to be (nearly) Gaussian and hence they can be described in terms of their two-point correlation functions and their primordial spectra as in \eq{PSscalar} and in \eq{PSTensor}. Since both scalar and tensor perturbations are sourced by the fluctuations of the inflaton field in an almost de Sitter background (see \chap{InfloBinglo}), we expect nearly but not exactly flat primordial spectra. As a matter of fact, the scalar tilt in \eq{nsns} and the tensor in \eq{ntnt} quantify the departure from the scale-invariant case and in this simplest scenario it is expected slightly tilted spectra because of the field evolution which breaks the de Sitter isometries, providing also a well-defined clock to measure the time to the end of inflation. However, inflation does not predict neither the precise values of the amplitudes nor those of the tilts, but they depend on the details of the inflationary dynamics which is clearly related to the precise shape of the potential. In fact, recalling \eq{nsnsepsilon} and \eq{ntepsH}, together with the relation between HSR and PSR, we see that we can also write
\begin{equation}
    n_s-1=2\eta_\Vl-6\epsilon_\Vl,\quad n_T=-2\epsilon_\Vl
    \label{eq:vepsveta}
\end{equation}
which implies that constraints on the spectral parameters can be translated into constraints on the inflationary potential (or the background dynamics) and vice-versa. Notice also that, within the power-law parametrization, both $n_{\rm s}$ and $n_{\rm T}$ are usually assumed to be scale invariant which implies that the runnings in \eq{runningss} and \eq{running} are set to zero and the higher order terms of the expansions in \eq{PLscalarexpansion} and \eq{PLtensorexpansion} are typically ignored. This is clearly an approximation and a further parametrization that includes also higher-order corrections could be considered \cite{Kuroyanagi:2011iw,Zarei:2014bta,Giare:2020vhn}. 

For this reason, even though current observations show a general agreement with the standard slow-roll predictions and many inflationary models proposed in literature can be ruled out, it should be noted that the missing evidence for tensor modes and, in general, the present day accuracy in data places only generic constraints on inflation that in many cases are obtained within the specific assumptions of the standard $\Lambda\rm{CDM}$ cosmological model (\textit{e.g.} an exactly flat background geometry, a vanishing scale dependence of the scalar and tensor tilt or even a negligible tensor amplitude).

\subsection{High order slow-roll framework}

Let us generalize the power spectrum expansion in \eq{PLscalarexpansion} one order higher
\begin{equation}
\ln\Delta^2_{\rm s}(k)=\ln A_{\mathrm{s}}+\left(n_{\mathrm{s}}-1\right) \ln\left(k / k_{*}\right) + \alpha_s \ln^2 \left(k / k_{*}\right)+\frac{\beta_s}{6} \ln^3 \left(k / k_{*}\right)
\label{eq:PhenomenologicalScalarsecond}
\end{equation}
where we have introduced a second term that takes into account the scale-dependence of the spectral index, the running of running $\beta_s$, defined as 
\begin{equation}
\beta_{\rm s}\doteq \left[\frac{d \alpha_s}{d\ln k}\right]_{k=k_*}\,.
\end{equation}
where the same pivot scale of $k_{*}=0.05\,\rm{Mpc}^{-1}$ has been adopted for both scalar and tensor perturbations. Notice that the running $\alpha_{\rm s}$ quantifies the rate of change of $n_{\rm s}$ per Hubble time ($d/d\ln k = 1/H\, d/dt$) while the running of running  $\beta_{\rm s}$ quantifies the rate of change of $\alpha_{\rm s}$ per Hubble time. These quantities are related to the shape of the inflationary potential and consequently to the underlying physics of inflation. Recalling the PSR parameters defined in \eq{evetav} and, using \eq{eVhierarchy}, we introduce two more parameters
\begin{equation}
\xi^2 _ { V } \doteq  M _ { \rm { pl } } ^ { 4 } \left(\frac { V _ { \phi } V _ { \phi \phi \phi } } { V ^ { 2 } }\right),\quad\varpi _ { V } ^ { 3 } \doteq M _ { \rm { pl } } ^ { 6 }\left(\frac { V _ { \phi } ^ { 2 } V _ { \phi \phi \phi \phi } } { V ^ { 3 } }\right)\,.
\label{eq:varpixi}
\end{equation}
These two additional parameters appear when we want, under the slow-roll assumption, express $\alpha_s$ and $\beta_s$ in terms of the PSR 
\begin{equation}
\alpha_{\rm s} =  16 \epsilon_{V} \eta _ { V } - 24 \epsilon _ { V } ^ { 2 } - 2 \xi _ { V } ^ { 2 }   
\label{eq:alpha_SR}
\end{equation}
\begin{equation}
\beta_{\rm s} =  - 192 \epsilon _ { V } ^ { 3 } + 192 \epsilon _ { V } ^ { 2 } \eta _ { V } - 32 \epsilon _ { V } \eta _ { V } ^ { 2 } - 24 \epsilon _ { V } \xi _ { V } ^ { 2 }  + 2 \eta _ { V } \xi _ { V } ^ { 2 } + 2 \varpi _ { V } ^ { 3 }
 \label{eq:beta_SR}
 \end{equation}
or, equivalently, in terms of the HSR, as 
\begin{equation}
\alpha_{\rm s} =-2\epsilon_\Hl\eta_\Hl-\eta_\Hl\xi^2_\Hl
\label{eq:alpha_s}
\end{equation}
\begin{equation}
\beta_{\rm s}=-2\epsilon_\Hl \eta_\Hl^2 -2\epsilon_\Hl\eta_\Hl\xi^2_\Hl-\eta_\Hl(\xi^2_\Hl)^2-\eta_\Hl\xi^2_\Hl\varpi^3_\Hl
 \end{equation}
 where $\xi^2_\Hl$ and $\varpi^3_\Hl$ are the equivalent of \eq{varpixi} but using \eq{Hierarchy}.

For the tensor spectrum we adopt a similar parametrization 
\begin{equation}
\ln\Delta^2_{\rm T}(k)=\ln\left(r\,A_{\mathrm{s}}\right)+n_{\mathrm{T}} \ln\left(k / k_{*}\right) + \alpha_{\rm T} \ln^2 \left(k / k_{*}\right)
+\frac{\beta_{\rm T}}{6} \ln^3 \left(k / k_{*}\right)
\label{eq:PhenomenologicalTensorsecond}
\end{equation} 
 with
\begin{equation}
\beta_{\rm T}\doteq \left[\frac{d\alpha_{\rm T}}{d\ln k}\right]_{k=k_*}\,.
\end{equation}
We can relate the higher order tensor runnings to the scalar ones by a set of slow-roll consistency relations. Indeed, under the assumption of slow roll inflation, a set of consistency relations among scalar and tensor parameters can be derived at any order~\cite{Martin:2013tda,Giare:2019snj}. It should be noted, however, that these relations can be violated in many nonstandard inflationary models, e.g. in presence of other spectator (rolling) fields~\cite{Mukohyama:2014gba,Namba:2015gja,Peloso:2016gqs,Giare:2020vhn,Ozsoy:2020ccy} or in modified gravity theories~\cite{Baumann:2015xxa,Giovannini:2015kfa,Giovannini:2018dob,Giovannini:2018nkt,Giovannini:2018zbf,Giare:2020vss,Giare:2020plo,Cicoli:2020bao} but it is not our case of study. In particular, the slow-roll consistency relations for the tensor running and running of running reads~\cite{Giare:2019snj}
\begin{equation}
\alpha_{\rm T}=\frac{r}{8}(n_s - 1) + \frac{r^2}{64},
\end{equation}
\begin{equation}
\beta_{\rm T}= \frac{r}{8}\left[ \alpha_{\rm s} - \left( n_{\rm s} -1 \right)^2\right] -\frac{3\,r^2}{64}\left(n_{\rm s} - 1\right) -\frac{r^3}{256}.
\end{equation} 
Therefore, given constraint on the scalar spectral index $n_{\rm s}$, its running $\alpha_{\rm s}$ and on the tensor-to-scalar ratio $r$, constraints can be derived on the tensor spectral index $n_{\rm T}$, its running $\alpha_{\rm T}$, its running of running $\beta_{\rm T}$.

Together with the standard $\Lambda$CDM parameters $\{\Omega_{\rm b}h^2$,$\Omega_{\rm c}h^2$, $\theta_{\rm{MC}}$,$\tau$,$\log(10^{10}A_{\rm s})$,$n_s\}$, we consider different combinations of the additional parameters $\alpha_{\rm s}$ and $\beta_{\rm s}$. Furthermore, we set the tensor-to-scalar ratio $r$ as a free parameter while we use the slow-roll consistency relation in \eq{TensorProblem} for the tensor tilt. Finally, we consider also the curvature density parameter $\Omega_k$ as an additional free parameter of the cosmological model. We explore the possibility of a nontrivial background geometry as a consistency check of the standard slow-roll paradigm. Indeed the vast majority of inflationary models predict flatness and constraints on the spatial curvature are an important test of this standard scenario. 

\subsection{Numerical analyses and datasets}
\label{sec:Foroni1Methods}

\begin{table}
	\begin{center}
		\renewcommand{\arraystretch}{1.5}
		\begin{tabular}{c@{\hspace{0. cm}}@{\hspace{1.5 cm}} c}
			\hline
			\textbf{Parameter}    & \textbf{Prior} \\
			\hline\hline
			$\Omega_{\rm b} h^2$         & $[0.005\,,\,0.1]$ \\
			$\Omega_{\rm c} h^2$     	 & $[0.001\,,\,0.99]$\\
			$100\,\theta_{\rm {MC}}$     & $[0.5\,,\,10]$ \\
			$\tau$                       & $[0.01\,,\,0.8]$\\
			$\ln(10^{10}A_{\rm s})$     & $[1.61\,,\,3.91]$ \\
			$n_{\rm s}$                  & $[0.8\,,\, 1.2]$ \\
			$\alpha_{\rm s}$             & $[-1\,,\, 1]$ \\ 
			$\beta_{\rm s}$              & $[-1\,,\, 1]$ \\
			$r$                          & $[0\,,\, 3]$ \\	
		    $\Omega_{\rm k}$             & $[-0.3\,,\,0.3]$\\
			\hline\hline
		\end{tabular}
		\caption{\small List of the parameter priors.}
		\label{tab.Priors}
	\end{center}
\end{table}

We perform MCMC (see \appx{DATa}) analyses using  the publicly available package \texttt{CosmoMC}~\cite{Lewis:2002ah,Lewis:2013hha} and computing the theoretical model described in the previous subsection with the latest version of the Boltzmann code \texttt{CAMB}~\cite{Lewis:1999bs,Howlett:2012mh}.
For all the different cosmological parameters we choose flat prior-distributions (unless otherwise stated), varying them uniformly in the conservative ranges listed in Table~\ref{tab.Priors}.
We explore the posteriors of our parameter space using the MCMC sampler developed for \texttt{CosmoMC} and tailored for parameter spaces with a speed hierarchy which also implements the "fast dragging" procedure~\cite{Neal:2005}. The convergence of the chains obtained with this procedure is tested using the Gelman-Rubin criterion with a threshold for chain convergence $R-1 \lesssim 0.02 $.

\subsubsection{Planck dataset}
Our baseline dataset consist of Planck 2018 temperature and polarization (TT TE EE) likelihood, which also includes low multipole data ($\ell <30$)~\cite{Aghanim:2019ame,Aghanim:2018eyx,Akrami:2018vks} (we refer to this combination as "Planck"). The reason why we consider both the high-multipole likelihood (which includes multipoles $30\lesssim \ell \lesssim 2500$ for the TT spectrum and $30\lesssim \ell \lesssim 2000$ for TE and EE spectra) and the "low-E" polarization likelihood (which covers the multipole range $2\le \ell \le 30$ for the EE spectrum) is for deriving constraints on all the cosmological parameters of the model. Furthermore, we include as dataset the Planck 2018 lensing likelihood~\cite{Aghanim:2018oex}, constructed from measurements of the power spectrum of the lensing potential (we refer to this dataset as "lensing"). Indeed the CMB photons that we measure today traversed almost the entire observable Universe and, along their paths, are deflected by gradients in the gravitational potentials associated with inhomogeneities in the Universe. This can cause a smoothing of the acoustic peaks and a conversion of E mode polarization into B-mode polarization. Therefore, the Planck lensing reconstruction, being the most significant detection of CMB lensing to date, is useful to improve the constraints on cosmological parameters, providing sensitivity above all on parameters that affect the late-time expansion and the background geometry. 

\subsubsection{BAO dataset}

While the Planck lensing measurements partially break the geometric degeneracy, it is well know that the inclusion of the baryon acoustic oscillation (BAO) measurements from galaxy surveys is a much more powerful way to break degeneracy in the geometrical sector. BAOs are the counterpart to the CMB acoustic peaks in the baryon distribution which remain imprinted also into the present-day matter distribution. Using the transverse BAOs information one can constrain the ratio between the comoving angular diameter distance ($D_M$) and the sound horizon ($r_d$) at the epoch when the baryon evolution becomes unaffected by coupling to photons. On the other hand, from the line-of-sight information we can constrain the quantity $H(z)\,r_d$. These two information can be combined together to constrain the acoustic-scale distance ratio $D_{V} / r_{\mathrm{d}} \doteq \left[c\, z\, D_{M}^{2}(z) H^{-1}(z)\right]^{1 / 3} / r_{\mathrm{d}}$. The acoustic scale measured by BAOs (at around $147$ Mpc), being much larger than the scale of virialized structures, makes the BAO measurements relatively simple geometric measurements insensitive to nonlinear physics, providing a robust geometrical test of cosmology. Here, in combination with the Planck data, we use the measurements of $D_{V} / r_{\mathrm{d}}$ from the 6dF survey at an effective redshift $z_{\rm eff} = 0.106$ \cite{Beutler:2011hx}, the SDSS Main Galaxy Sample at $z_{\rm eff}= 0.15$ \cite{Ross:2014qpa} and the final BOSS DR12 data with separate constraints on $H(z)\,r_{\rm d}$ and $D_M/r_{\rm d}$ in three correlated redshift bins at $z_{\rm eff} =\left[0.38\,,\,0.51\,,\,0.61\right]$ \cite{Alam:2016hwk} (we refer to this dataset as "BAO").

\subsubsection{B-mode dataset}

To improve also the constraints in the primordial tensor sector, we exploit the CMB B-modes power spectrum likelihood (cleaned from the foreground contamination) as released by Bicep2/Keck Array X Collaboration \cite{Ade:2018gkx} (we call it "BK15"). Indeed, it is well known that a satiable background of inflationary gravitational waves can produce B-modes polarization on large/intermediate angular scales where the cosmic variance is not very significant and gravitational lensing is not yet dominant. Notice however that the B-modes likelihood basically improves only the constraints on tensor modes. Therefore we include this dataset only when we analyze the tensor spectrum because interested in models with a satiable production of gravitational waves.

\subsubsection{Planck-independent CMB datasets}

Along with these combinations of datasets involving the Planck CMB measurements, we analyze also two other Planck-independent datasets. In particular we use the Atacama Cosmology Telescope DR4 likelihood~\cite{Hinshaw:2012aka} and the South Pole Telescope polarization measurements~\cite{Dutcher:2021vtw}. We combine both of them with WMAP 9-years observations data~\cite{Hinshaw:2012aka}. The reason is that the Atacama Cosmology Telescope has a minimum sensitivity in multipole of $600$ in TT, and $350$ in TE and EE, and so it lacks data around the first two acoustic peaks in the TT spectrum and the first full peak in TE/EE. Similarly, the South Pole Telescope measures only the TE and EE spectra over a range of multipoles $300\le \ell \le 1400$ for EE and $300 \le \ell \le 1700$ for TE. Therefore, the only way to obtain competitive Planck-independent measurements for all the cosmological parameters is to combine these two datasets with the public WMAP 9-year observations at intermediate scales ($2<\ell< 1200$ for TT and $\ell < 800$ for TE), as also done in~\cite{Dutcher:2021vtw,Aiola:2020azj} (we call these datasets "ACT+WMAP" and "SPT3G+WMAP"). Notice also that we use a Gaussian prior on $\tau = 0.065 \pm 0.015$ both for ACT+WMAP and for SPT3G+WMAP. Indeed, while our primary goal is to obtain a measurement of the cosmological (inflationary) parameters that is Planck-independent, neither ACT nor SPT-3G can constrain the optical depth at reionization $\tau$. Furthermore, there is evidence that WMAP large-scale polarization data ($2 < \ell < 23$ in TE spectrum) can be contaminated by dust, possibly affecting the WMAP bounds on $\tau$. For this reason in our analysis we exclude this multipoles range, using instead the conservative Gaussian prior on $\tau$ which is based on Planck measurements. This prior on $\tau$ is not expected to affect the constraints on the other cosmological parameters~\cite{Dutcher:2021vtw,Aiola:2020azj}.

\subsection{Running the scalar running}

\begin{table}
	\begin{center}
		\renewcommand{\arraystretch}{1.5}
		\resizebox{0.9\textwidth}{!}{\begin{tabular}{c c c c  c  c}
  	        \hline
			\textbf{Parameter} & \textbf{Planck18}  & \textbf{Planck18 + lensing}  & \textbf{Planck18 + BAO} & \textbf{ACTPol + WMAP} & \textbf{SPT3G+WMAP} \\
			\hline\hline
			$\Omega_{\rm b} h^2$ &$0.02235\pm 0.00017$&$0.02237\pm 0.00016$&$0.02243\pm 0.00015$ & $0.02195\pm 0.00025$ &$0.02251\pm 0.00025$\\
			$\Omega_{\rm c} h^2$ &$0.1207\pm 0.0015$&$0.1202\pm 0.0012$&$0.1195\pm 0.0010$ & $0.1190\pm 0.0029$ &$0.1139\pm 0.0032$\\
			$100\,\theta_{\rm {MC}}$ &$1.04085\pm 0.00031$&$1.04089\pm 0.00030$&$1.04100\pm 0.00028$& $1.04174\pm 0.00066$ &$1.03970\pm 0.00066$\\
			$\tau$   &$0.0575\pm 0.0086$&$0.0564\pm 0.0080$&$0.0590\pm 0.0087$& $0.061\pm 0.013$ &$0.063\pm 0.013$\\
			$\log(10^{10}A_{\rm S})$ &$3.053\pm 0.018$&$3.049\pm 0.015$&$3.053\pm 0.018$ & $3.051\pm 0.026$ &$3.037\pm 0.026$\\
			$n_s$ &$0.9612\pm 0.0054$&$0.9625\pm 0.0048$&$0.9645\pm 0.0045$ & $0.9680\pm 0.0082$ &$0.978\pm 0.011$\\
			$\alpha_s$ &$0.001\pm 0.010$&$0.002\pm 0.010$&$0.000\pm 0.010$ & $0.035\pm 0.012$ &$0.028\pm 0.017$\\
			$\beta_s$ &$0.012\pm 0.013$&$0.010\pm 0.013$&$0.009\pm 0.013$ & $0.035\pm 0.013$ &$0.023\pm 0.016$\\
			\hline 
			$\eta_V$ &$-0.0194^{+0.0027}_{-0.0026}$&$-0.0187^{+0.0025}_{-0.0023}$&$-0.0177^{+0.0021}_{-0.0022}$ & $-0.0160\pm 0.0041$ &$-0.0111\pm 0.0053$\\
			$\xi^2_V$ &$-0.0005\pm0.0050$&$-0.0008^{+0.0050}_{-0.0049}$&$-0.0001\pm0.0049$ & $-0.0174\pm 0.0058$ &$-0.0141\pm 0.0085$\\
			$\varpi _ { V } ^ { 3 }$ &$0.0058^{+0.0063}_{-0.0061}$&$0.0051^{+0.0062}_{-0.0061}$&$0.0046^{+0.0062}_{-0.0061}$ & $0.0172\pm 0.0064$ &$0.0115\pm 0.0078$\\
			$\eta_\Hl$ &$0.0388^{+0.0053}_{-0.0054}$&$0.0375^{+0.0047}_{-0.0049}$&$0.0355^{+0.0044}_{-0.0043}$ & $0.0320\pm 0.0082$ &$0.022\pm 0.011$\\
			$\xi^2_\Hl$ &$-0.02\pm0.26$&$-0.04^{+0.27}_{-0.26}$&$0.00\pm0.28$ & $<-0.02$ &$-$\\
			\hline	\hline
		\end{tabular}}
	\end{center}
	\caption[Results for $\Lambda\mathrm{CDM} + \alpha_s + \beta_s$]{\small Results for $\Lambda\mathrm{CDM} + \alpha_s + \beta_s$. The constraints on parameters are at $68\%$ CL, while upper bounds are at $95\%$ CL. The internal horizontal line divides the primary parameters of the cosmological model (those we directly sample in our MCMC analysis) from the derived parameters.}
	\label{tab:LCDM+runnings}
\end{table}

\begin{figure}[h!]
	\centering
	\includegraphics[width=0.9\textwidth]{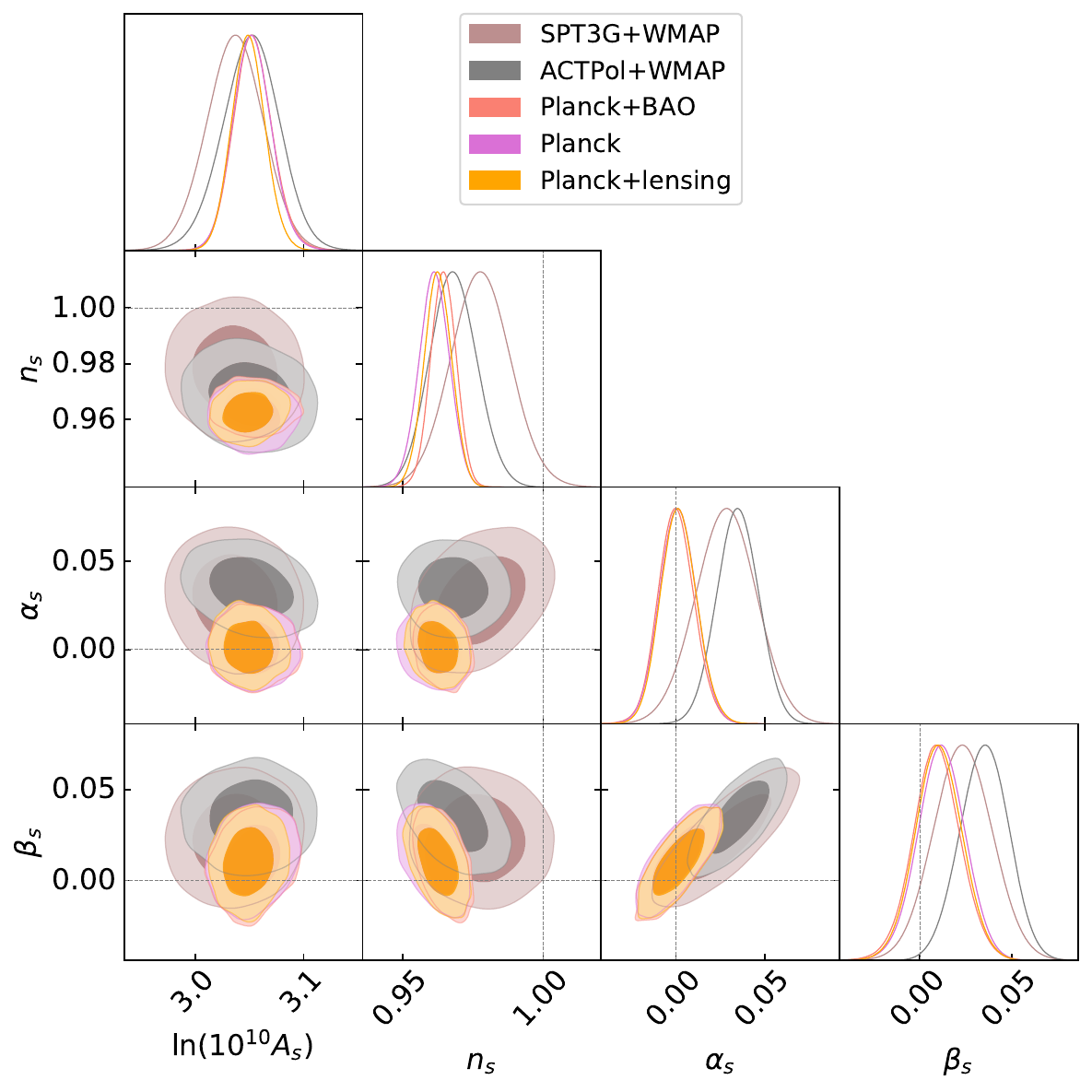}
	\caption[Marginalized 2D and 1D posteriors distributions for $\Lambda\mathrm{CDM} + \alpha_s + \beta_s$]{\small Marginalized 2D and 1D posteriors distributions for the $\Lambda\mathrm{CDM} + \alpha_s + \beta_s$ cosmological model obtained for the different combinations of the datasets listed in \sect{Foroni1Methods}. The dashed lines represent the case of vanishing inflationary parameters.}
	\label{fig:LCDM+runnings}
\end{figure}

We start analyzing an extended cosmological model which includes both the running of the scalar spectral index $\alpha_s$ and its running of running $\beta_{\rm s}$ as additional parameters. We refer to this model as $\Lambda\mathrm{CDM} +\alpha_{s}+\beta_{s}$. Notice that here we focus exclusively on the adiabatic scalar modes, parametrizing the scalar spectrum by \eq{PhenomenologicalScalarsecond} and assuming a negligible gravitational waves production. Assuming a negligible tensor amplitude $r=16\epsilon_\Vl\simeq 16\epsilon_\Hl\sim 0$ in terms of the slow-roll parameter means to consider $\epsilon_\Vl \simeq \epsilon_\Hl\sim 0$; i.e. negligibly small in the relations for the scalar tilt and its runnings. In \tab{LCDM+runnings} we summarize the results obtained for this model while in \fig{LCDM+runnings} we show the $68\%$ and $95\%$~CL contour plots for different parameters. 

From the Planck data we derive the constraints on the scalar tilt $n_s=0.9612\pm 0.0054$, on its running $\alpha_s=0.001\pm 0.010$, and on its running of running $\beta_s=0.012\pm 0.013$, all at 68\% CL\footnote{Unless otherwise stated, we always provide 68\% CL values for bounded parameters and $95\%$ CL for upper/lower bounds.}. The inclusion of the lensing spectrum and the BAO data does not change significantly the above mentioned constraints
and all these bounds are consistent with the case of vanishing runnings within one standard deviation, see also \fig{LCDM+runnings}. We compare the Planck results with other independent measurements derived using the different datasets listed in \sect{Foroni1Methods}. Considering the SPT-3G data combined with WMAP 9-years observations data, we get $\alpha_s=0.028\pm 0.017$ and $\beta_s=0.023\pm 0.016$, both and consistent with zero within 1.6 and 1.4 standard deviations, respectively. On the other hand, considering the ACTPol+WMAP data, we obtain a preference for nonvanishing running $\alpha_s=0.035\pm0.012$ and for a nonvanishing running of running $\beta_s=0.035\pm0.013$ at the level of 2.9$\sigma$ and 2.7$\sigma$, respectively. Interestingly, in both the cases positive values for the runnings are preferred with a statistical significance between about $1.7\,\sigma$ (SPT-3G+WMAP) and $2.9\,\sigma$ (ACTPol+WMAP). Notice also that, 
while both the ground based telescope measurements are in good agreement one with each other, they are in disagreement at more than $2\sigma$ with Planck regarding the value of the running $\alpha_s$, and in tension for the running of running $\beta_s$ (see also \fig{LCDM+runnings}). This tension indicates a difference coming from the high multipoles region, that can be an indication of small systematic errors unaccounted for, or physics beyond the standard models. In other words, the extended models considered in this paper recast the global tension between the datasets already present for a $\Lambda$CDM model~\citep{Handley:2020hdp} analysis.

Under the assumption of a negligible tensor amplitude, we derive constraints on the slow-roll parameters $\{\eta_V\,,\,\xi_V^2\,,\,\varpi_V^3\}$ and $\{\eta_\Hl,\xi^2_\Hl,\varpi^3_\Hl\}$.
Due to the Planck data evidence for a tilted scalar spectrum, we obtain nonzero slow-roll parameters $\eta_V=-0.0194^{+0.0027}_{-0.0026}$ or equivalently $\eta_\Hl=0.0388^{+0.0053}_{-0.0054}$. On the other hand, the missing evidence for scalar runnings only limits the parameter space allowed for higher-order slow-roll parameters to $\xi^2_V=-0.0005\pm0.0050$ and $\varpi^3_V=0.0058^{+0.0063}_{-0.0061}$ both consistent with zero within one standard deviation. Similarly, $\xi^2_\Hl=-0.02\pm 0.26$ while $\varpi^3_\Hl$ turns out to be unbounded, so we do not show it in the Table. Adding also lensing or BAO data to Planck, the constraints on the inflationary parameters do not change significantly, see also \tab{LCDM+runnings}.
Interestingly, considering the Atacama Cosmology Telescope DR4 likelihood combined with WMAP 9-years, the bounds on $\eta_V=-0.0160\pm0.0041$ and $\eta_\Hl=0.0320\pm0.0082$ remain basically unchanged with respect to the other datasets, while the preference for nonvanishing runnings is translated into the constraints on higher-order inflationary parameters $\xi_V^2=-0.0174\pm0.0058$ and $\varpi^3_V= 0.0172\pm0.0064$ (or equivalently $\xi^2_\Hl<-0.02$ at 95\% CL) that are all different from zero at more than 95\% CL. Finally, regarding the SPT3G+WMAP case, we find more than $1\sigma$ shift toward lower values of both $\eta_V=-0.0111\pm0.0053$ and $\eta_\Hl=0.022\pm0.011$, while we find $1\sigma$ preference for nonvanishing higher-order parameters $\xi_V^2=-0.0141\pm0.0085$ and $\varpi^3_V= 0.0115\pm0.0078$. For this dataset $\xi^2_\Hl$ is instead unconstrained.
 

\subsection{The tensor spectrum and slow-roll relations}

\begin{figure}[h!]
	\centering
	\includegraphics[width=1 \textwidth]{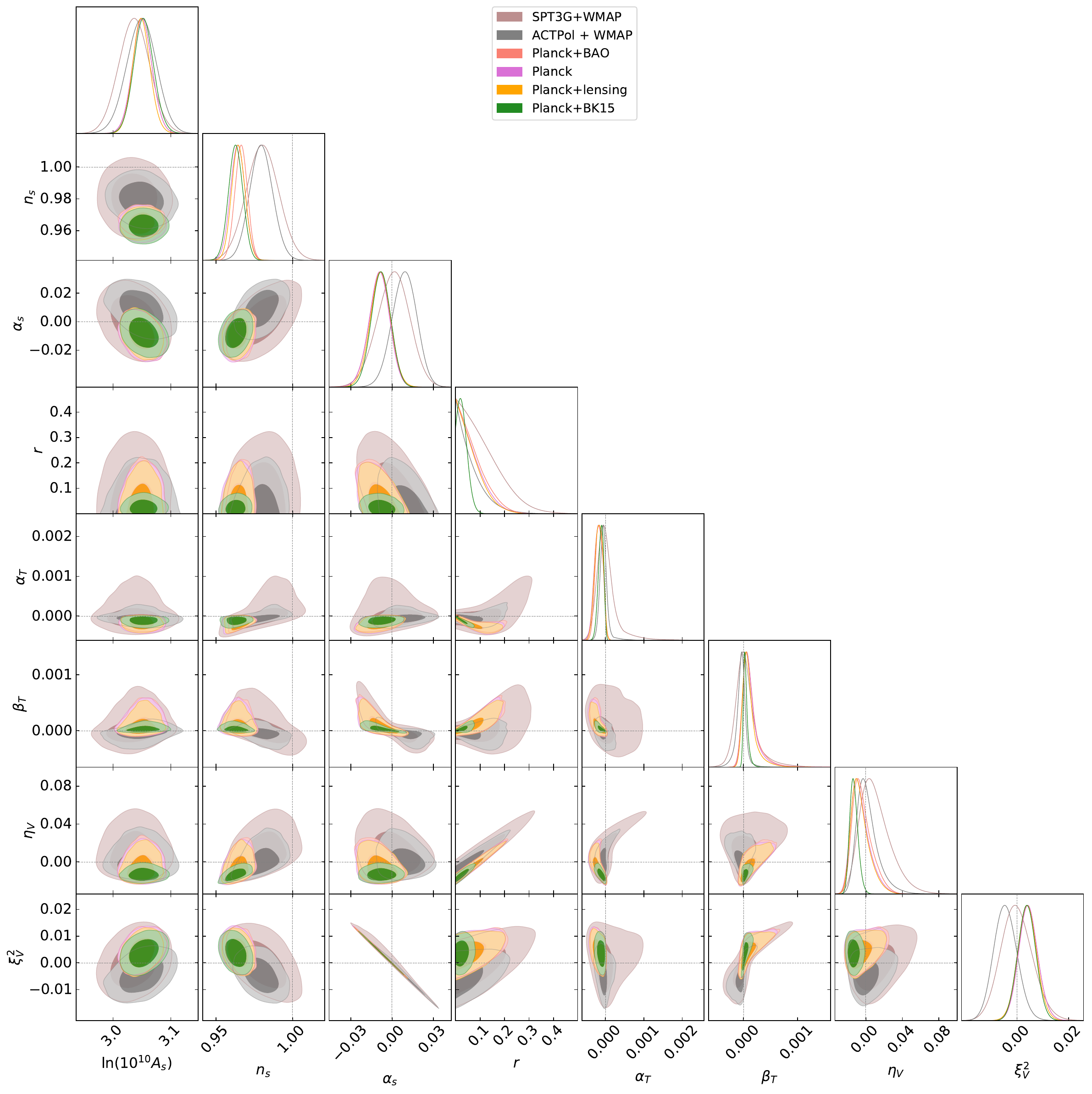}
	\caption[Marginalized 2D and 1D posteriors distributions for the $\Lambda\mathrm{CDM} + r+ \alpha_s$]{\small Marginalized 2D and 1D posteriors distributions for the $\Lambda\mathrm{CDM} + r+ \alpha_s$ cosmological model obtained for different combination of datasets listed in \sect{Foroni1Methods}. The dashed lines represent the case of vanishing inflationary parameters.}
	\label{fig:LCDM+r+nrun}
\end{figure}

\begin{table}
	\begin{center}
		\renewcommand{\arraystretch}{1.5}
		\resizebox{\textwidth}{!}{\begin{tabular}{c c c c c c c}
			\hline
			\textbf{Parameter} & \textbf{Planck18}  & \textbf{Planck18 + lensing}  & \textbf{Planck18 + BAO} & \textbf{ Planck18 + BK15} & \textbf{ACTPol + WMAP} & \textbf{SPT3G+WMAP} \\
			\hline\hline
			$\Omega_{\rm b} h^2$ &$0.02241\pm 0.00016$&$0.02242\pm 0.00015$&$0.02247\pm 0.00014$&$0.02239\pm 0.00015$ & $0.02234\pm 0.00022$ &$0.02273\pm 0.00024$\\
			$\Omega_{\rm c} h^2$ &$0.1202\pm 0.0014$&$0.1199\pm 0.0012$&$0.1193\pm 0.0010$&$0.1206\pm 0.0014$ & $0.1179\pm 0.0030$ &$0.1138\pm 0.0031$\\
			$100\,\theta_{\rm {MC}}$ &$1.04091\pm 0.00032$&$1.04093\pm 0.00030$&$1.04101\pm 0.00030$&$1.04087\pm 0.00031$ & $1.04186\pm 0.00065$ &$1.03978\pm 0.00067$\\
			$\tau$   &$0.0562\pm 0.0081$&$0.0560\pm 0.0076$&$0.0573\pm 0.0080$&$0.0570\pm 0.0083$& $0.058\pm 0.012$ &$0.060\pm 0.013$\\
			$\log(10^{10}A_{s})$ &$3.050\pm 0.017$&$3.049\pm 0.015$&$3.051\pm 0.017$&$3.053\pm 0.017$ & $3.049\pm 0.025$  &$3.037\pm 0.026$\\
			$n_s$ &$0.9642\pm 0.0047$&$0.9647\pm 0.0044$&$0.9665\pm 0.0041$&$0.9629\pm 0.0046$ & $0.9796\pm 0.0074$ &$0.980\pm 0.010$\\
			$\alpha_s$  &$-0.0094\pm 0.0074$&$-0.0084\pm 0.0073$&$-0.0091\pm 0.0075$&$-0.0080\pm 0.0069$ & $ 0.0090\pm 0.0087$  &$0.001\pm 0.012$\\
			$r$ &$<0.165 $&$<0.159$&$< 0.172$&$< 0.0658$ & $<0.176$ &$<0.260$\\
			\hline
			$n_T$ &$>-0.0206$&$>-0.0198$&$>-0.0215$&$>-0.0082$ & $>-0.022$ &$>-0.032$\\
			$\alpha_T$ &$\left(\,-18^{+12}_{-10}\,\right)\cdot 10^{-5}$&$\left(-17\pm 10\right)\cdot 10^{-5}$&$\left(\,-16.6^{+11}_{-9.5}\,\right)\cdot 10^{-5}$&$\left(\,-11.7^{+7.9}_{-5.9}\,\right)\cdot 10^{-5}$ & $\left(\,-4.2^{+6.9}_{-10}\,\right)\cdot 10^{-5}$  &$\left(\,3^{+13}_{-27}\,\right)\cdot 10^{-5}$\\
			$\beta_T$ &$\left(11.4^{+6.9}_{-15}\right)\cdot 10^{-5}$&$\left(9.96^{+6.1}_{-14}\right)\cdot 10^{-5}$&$\left(11.8^{+7.2}_{-16}\right)\cdot 10^{-5}$ &$\left(\,3.9^{+2.5}_{-4.8}\,\right)\cdot 10^{-5}$ & $\left(\,-4.4^{+8.1}_{-6.9}\,\right)\cdot 10^{-5}$  &$\left(\,5^{+12}_{-21}\,\right)\cdot 10^{-5}$\\
			$\epsilon_V\simeq\epsilon_\Hl$ &$< 0.0103$&$< 0.0099 $&$<0.0108$&$< 0.0041$ & $< 0.0110$ &$< 0.0163$\\
			$\eta_V$ &$-0.0058^{+0.0069}_{-0.012}$&$-0.0061^{+0.0066}_{-0.011}$&$-0.0039^{+0.0072}_{-0.012}$&$-0.0130^{+0.0038}_{-0.0050}$ & $0.0015^{+0.0074}_{-0.013}$  &$0.010^{+0.012}_{-0.019}$\\
			$\xi_V^2$ &$0.0044\pm 0.0037$&$0.0040\pm 0.0036$&$0.0043\pm 0.0038$&$0.0038\pm 0.0034$& $-0.0045\pm 0.0044$  &$-0.0001^{+0.0056}_{-0.0064}$\\
			$\eta_\Hl$ &$0.0277^{+0.0095}_{-0.0067}$&$0.0276^{+0.0091}_{-0.0062}$&$0.0250^{+0.0095}_{-0.0064}$&$0.0334\pm 0.0054$ & $ 0.0126^{+0.012}_{-0.0090}$ &$0.006^{+0.016}_{-0.013}$\\
			$\xi^2_\Hl$ &$-$&$0.37^{+0.26}_{-0.34}$&$0.62^{+0.16}_{-0.56}$&$0.24\pm 0.21$ & $-$ &$-$\\
			$V_{\rm inf}^{1/4}$ & $<2.04\times10^{16}\,\rm GeV$&$<2.01\times10^{16}\,\rm GeV$&$<2.06\times10^{16}\,\rm GeV$&$<1.62\times10^{16}\,\rm GeV$ & $<2.10\times10^{16}\,\rm GeV$ &$<2.31\times10^{16}\,\rm GeV$ \\
			\hline	\hline
		\end{tabular}}
	\end{center}
	\caption[Results for $\Lambda\mathrm{CDM} + r+ \alpha_s$]{\small Results for $\Lambda\mathrm{CDM} + r+ \alpha_s$. The constraints on parameters are at $68\%$ CL, while upper bounds are at $95\%$ CL. The internal horizontal line divides the primary parameters of the cosmological model (those we directly sample in our MCMC analysis) from the derived parameters (those we obtain from the others by the relations described in the text).}
	\label{tab:LCDM+r+nrun}
\end{table}
We now include as additional parameters the running of the scalar tilt $\alpha_s$ and the tensor amplitude $r$, fixing instead the scalar running of running to zero. We refer to this model as $\Lambda\mathrm{CDM}+\alpha_s+r$. In \tab{LCDM+r+nrun} we summarize the results obtained for this model while in Fig.~\ref{fig:LCDM+r+nrun} we show the $68\%$ and $95\%$~CL contour plots for different inflationary parameters.

For the scalar parameters, we see that the constraints on $n_{s}$ and $\alpha_s$ are slightly changed when replacing the running of running with the tensor-to-scalar ratio. This is due to the fact that, once the tensor amplitude varies, also the terms $\propto \epsilon_\Vl$ contribute in the slow roll relations in \eq{PhenomenologicalScalarsecond} and \eq{alpha_s}, modifying the correlation among the inflationary (scalar and tensor) parameters. Moreover since $\alpha_{s}$ and $\beta_s$ are strongly correlated for all the datasets (see also \fig{LCDM+runnings}) fixing $\beta_s=0$ produces a shift of $\alpha_s$ toward lower values. In particular, one can see that for the Planck data this shift is translated into a preference for negative values of $\alpha_s$ at the level of slightly more than $1\sigma$ even though the constraints on the running are always consistent with zero within two standard deviations. Notice also that these results remain unchanged when the lensing and BAO measurements are considered together with Planck. Furthermore, when the tensor amplitude can freely vary and the running of running is fixed to zero, also the ACTPol+WMAP and SPT3G+WMAP constraints on $\alpha_s$ shift toward lower values. This produces a reduction of $\alpha_s=0.0090\pm0.0087$ for ACTPol+WMAP, positive and larger than zero at slightly more than one standard deviation, and $\alpha_s=0.001\pm0.012$ for SPT3G+WMAP, completely in agreement with a vanishing scalar running. It should be noticed here that while SPT3G+WMAP is in agreement with Planck for the value of the running $\alpha_s$, ACTPol+WMAP is instead in tension at about $2\sigma$. As in the previous case the difference we see also in \fig{LCDM+r+nrun} is coming from the high multipole region.

As concerns the tensor spectrum, we see that its amplitude is constrained to be $r<0.165$ (at 95\% CL) by the Planck data alone while ACTPol+WMAP and SPT3G+WMAP give $r<0.176$ and $r<0.260$, respectively. A strong improvement in this upper bound is obtained including also the BK15 data that, combined with Planck, gives $r<0.0658$. Using the slow-roll relation between the tensor amplitude and the tensor tilt these upper bounds on the amplitude can be translated into a lower bounds on the (negative) tensor tilt, namely: $n_{\rm T}>-0.0206$ for the Planck data and $n_{\rm T}>-0.0082$ for Planck+BK15. Furthermore, in the slow-roll framework, any constraint to the tensor amplitude places also a constraint to the energy scale of inflation which reads
\begin{equation}
V^{1/4}_{\rm inf}=M_{\rm pl}\left(\frac{3}{2}\,\pi^2\,A_s\,r\right)^{1/4} \,\rm{GeV}. 
\label{eq:V1/4}
\end{equation}
Using the results in \tab{LCDM+r+nrun} from Planck data we derive $V^{1/4}_{\rm inf}<2.04\times 10^{16}\,\rm{GeV}$ while the inclusion of the BK15 data improves this upper bound to $V^{1/4}_{\rm inf}<1.62\times 10^{16}\,\rm{GeV}$.

Reversing the slow-roll relations for the scalar and tensor parameters, we derive constraints on the slow-roll parameters $\{\epsilon_\Vl\,,\eta_\Vl\,,\xi^2_\Vl \}$ that are related to the shape of the inflationary potential. In particular from Planck, we get $\epsilon_\Vl<0.0103$ while the improvement in the constraining power on the tensor amplitude due to the BK15 data is translated into the improved upper bound $\epsilon_\Vl<0.0041$. On the other hand, for $\eta_\Vl$ and $\xi_\Vl^2$ the Planck + BK15 data give $\eta_\Vl=-0.0130^{+0.0038}_{-0.0050}$ and $\xi_\Vl^2=0.0038\pm0.0034$, respectively, ruling out the null value at more than one standard deviation. On the contrary, ACTPol+WMAP finds $\eta_\Vl=0.0015^{+0.0074}_{-0.013}$ and $\xi_\Vl^2=-0.0045\pm0.0044$, always showing $1\sigma$ indication different from zero, but with an opposite sign with respect to Planck. In addition, SPT3G+WMAP prefer both the parameters $\eta_\Vl$ and $\xi_\Vl^2$ in agreement with the null value within the 68\% CL. Equivalently, we can derive constraints on the parameters $\{\eta_\Hl\,,\xi^2_\Hl\}$. For Planck + BK15 we obtain $\eta_\Hl = 0.0334\pm 0.0054$ and $\xi^2_\Hl = 0.24\pm 0.21$. Instead, the Atacama Cosmology Telescope and the South Pole Telescope data, even if they have larger experimental errors and lead to less constraining bounds, prefer $\eta_\Hl$ much lower than Planck, reducing the significance for a value different from zero, and $\xi^2_\Hl$ unconstrained.

Under the assumption of slow roll inflation, we see that the parameter space allowed for the (higher-order) tensor parameters in the slow-roll paradigm is strongly reduced since constraints on $r$ and the scalar spectrum are translated into constraints on tensor spectrum, see also \fig{LCDM+r+nrun}. In particular using the Planck+BK15 data we see that the results for the scalar parameters and the upper bound on the tensor amplitude, are translated into the constraints $\alpha_T=\left(\,-11.7^{+7.9}_{-5.9}\,\right)\cdot 10^{-5}$ and $\beta_T=\left(\,3.9^{+2.5}_{-4.8}\,\right)\cdot 10^{-5}$ for the tensor running and its running of running. It should be noted that these results are consistent with zero within less than two standard deviations and that, in any case, they are expected to be extremely small and therefore negligible in the slow-roll hierarchy. 
Similar results can be obtained also exploiting the Planck-independent measurements by ACTPol+WMAP and SPT3G+WMAP, see \tab{LCDM+r+nrun}. In particular, for these datasets the bounds on $\alpha_{\rm T}$ and $\beta_{\rm T}$ turn out to be less constraining with respect Planck(+BK15) because ACTPol and SPT3G in combination with WMAP have a smaller sensitivity both on the tensor amplitude and on scalar modes. However the higher-order corrections to the power-law spectrum of gravitational waves are always constrained to be extremely small by the slow-roll relations and, given also the large error bars of all the dataset, the bounds are all consistent with each other within 2 standard deviations. This leads to predict a scale invariant tensor tilt, unless corrections of order $|dn_{\rm T}/d\ln k|\lesssim 10^{-5}$.


\subsection{Implications for slow roll inflationary models}
\begin{figure}[h!]
	\centering
	\includegraphics[width=1 \textwidth]{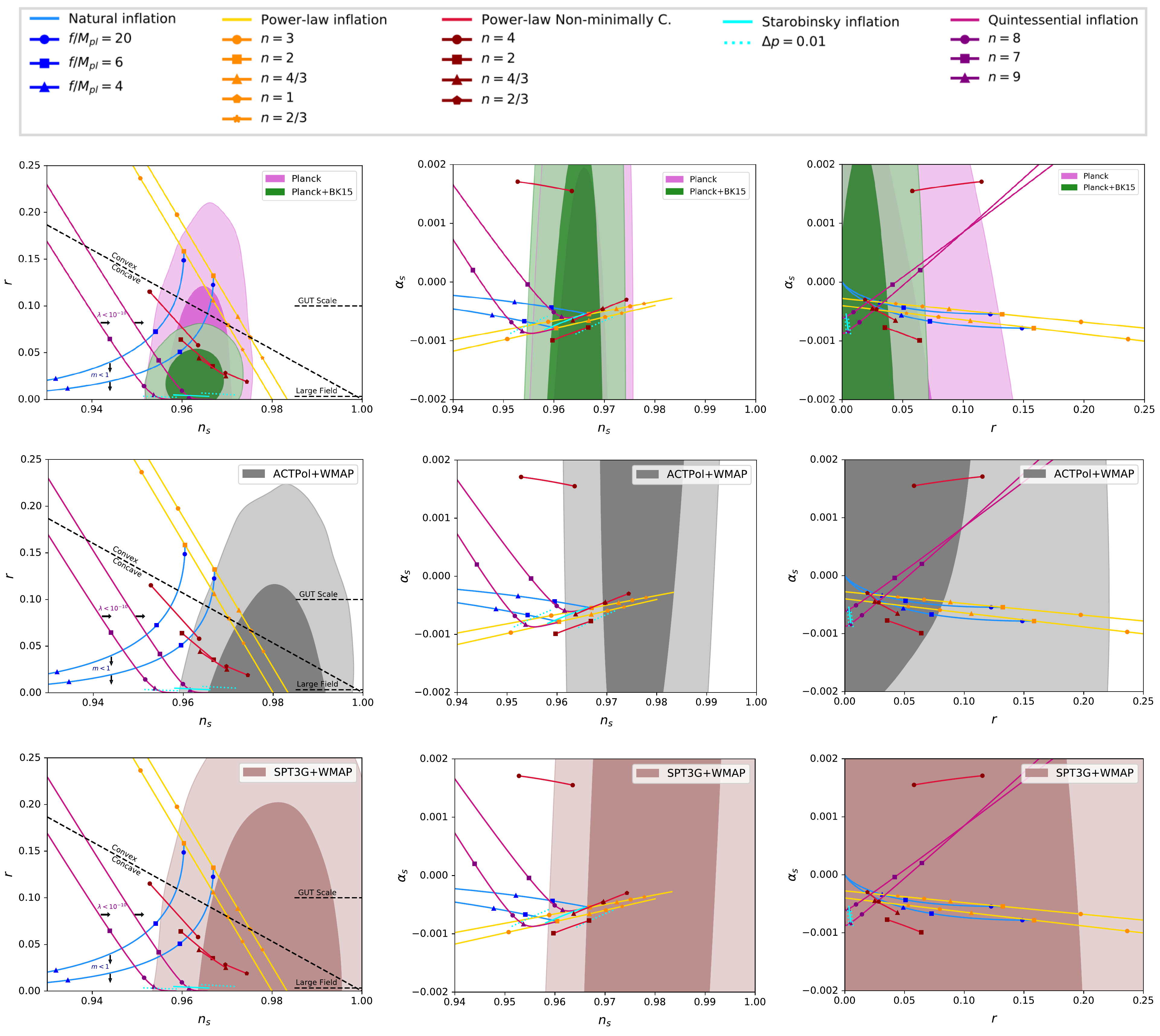}
	\caption[Marginalized joint 68\% and 95\% CL regions for $(n_s\,,\,r)$ , $(n_s\,,\,\alpha_s)$ and  $(r\,,\,\alpha_s)$]{\small Marginalized joint 68\% and 95\% CL regions for $(n_s\,,\,r)$ , $(n_s\,,\,\alpha_s)$ and  $(r\,,\,\alpha_s)$ from Planck(+BK15) (top panels), ACTPol+WMAP (middle panels) and SPT3G+WMAP (bottom panels) data. The marginalized contours can be compared to the theoretical predictions of some selected inflationary models opportunely described in the text.}
	\label{fig:Forconi1Models}
\end{figure}

Now, we shall focus on the constraints for a few selected models of slow-roll inflation. In particular, we compute the slow-roll parameters and consequently we predict the values of $n_s$, $\alpha_s$ and $r$ to first order in the slow-roll approximation (see \sect{SRAP}). We include an uncertainty in the number of \textit{e}-folds before the end of inflation of $50<N<60$ \cite{Akrami:2018odb}.  In \fig{Forconi1Models} we compare the theoretical predictions with the observational constraints obtained within the $\Lambda\mathrm{CDM}+r+\alpha_s$ cosmological model for the different datasets listed in \sect{Foroni1Methods}.

First, by noting that in a Universe dominated by the energy-density of the inflaton field during the slow-roll regime we have (see \eq{Ip})
\begin{equation}
    \dot{H}=-\frac{\Mpl}{2}\dot{\phi}^2=d^2N/dt^2,
\end{equation} 
one can relate the field excursion to the tensor amplitude by $\Delta\phi / M_{\rm pl}=\sqrt{r/8}\,N$ and using $N=50$ we set a lower bound
\begin{equation}
   \frac{\Delta\phi}{M_{\rm pl}}=1.01\left(\frac{r}{3.26\times10^{-3}}\right)^{\frac12}
\end{equation}
that is shown in \fig{Forconi1Models}. Note that both large and small field models are compatible with every dataset. Then, using \eq{V1/4}, we get an approximate limit for potentials that work on GUT scales finding that they are ruled out at 95\% CL by the combination Planck+BK15 even though they are still compatible with the other datasets, including ACTPol+WMAP and SPT3G+WMAP. 
This is again an indication of a tension between the Planck satellite results and the ground based telescopes measurements, that prefer a larger value for the scalar spectral index $n_s$ more consistent with a scale invariant spectrum $n_s=1$. This is not only a volume effect, due to the different constraining power of the experiments, but also an actual shift of the $n_s$ constraints coming from the power spectra damping tails.
Lastly, we determine whether the data are in agreement with a convex or a concave potential, being $r=-8/3 \,(n_s-1)$ the relation which defines the limit between the two different shapes. Due to the fact that B-modes polarization measurements are able to give more stringent constraints on tensor modes, in particular on $r$ that appears in the relation aforementioned, the BK15 data indicates that the potential should present a concave shape and exclude completely a convex one, whereas the other datasets are unable to give such a restriction and allow both shapes. 

We give below a concise review of some inflationary models studied in this work and the main results obtained by our analysis.

\paragraph{\textit{(Generalized) natural inflation}}: we  start from the general natural inflation \cite{Munoz:2014eqa}, which consider an axion model where a global $U(1)$ symmetry is spontaneously broken at scale $f$, with soft explicit symmetry breaking at a lower scale $\Lambda$; the inflaton field is the pseudo-Nambu-Goldstone boson~\cite{Adams:1992bn} (see \sect{NGB}). The potential reads
\begin{equation}
    V=2^{1-m}\Lambda^4\left[1+\cos{\frac{\phi}{f}}\right]^m.
    \label{NI}
\end{equation}
Fixing $m=1$ and recovering the natural inflation \cite{Freese:1990rb}, the parameters are
\begin{gather}
    n_s=1-\frac{1}{y}\left[\frac{1+2y^2(1+e^{-x})}{1+2y^2(1-e^{-x})}\right],\\
    \alpha_s=-\frac{4(2y^2+1)e^x}{y^2(-2y^2+(2y^2+1)e^x)^2},\\
    r=\frac{16e^{-x}}{1+2y^2(1-e^{-x})},
\end{gather}
where $x=N/y$ and $y=f/M_{\rm pl}$. Plotting the above quantities as functions of $f/M_{\rm pl}$ (blue curves in \fig{Forconi1Models}) we can see that this model in only compatible within one standard deviation for Planck and within two standard deviation with Planck+BK15. Anyway, relaxing the assumption $m=1$ and leaving $m$ a free parameter, the compatibility with Planck+BK15 increases as long as $m<1$. Given the tension present in the parameter space between the different experiments (as we can see from \fig{LCDM+r+nrun}), the model compatibility changes between the datasets. 
In fact, the South Pole Telescope data show only an agreement at 95\% CL for every N in the chosen interval, \ie both blue lines are in the lighter region of the dataset. Moreover, the shift toward high values of $n_s$ preferred by the Atacama Cosmology Telescope data basically excludes the (generalized) natural inflation from the 95\% CL contours. It should be stressed that the ground experiments (ACTPol and SPT-3G) are the ones responsible for the shift of the measurements and consequently changes the compatibility with the model, not WMAP 9-years \cite{2013ApJS..208...19H,Planck:2018nkj}.
Actually, the shift of the $n_s$ bounds is due to the high multipole region accurately constrained by the damping tail of the power spectra.
 
\paragraph{\textit{(Non minimally coupled) power-law inflation}}: by taking the limit $f\rightarrow \infty$ in Eq.\eqref{NI}, we recover the quadratic potential, a particular case of the general power-law inflation, represented as two yellow straight lines in \fig{Forconi1Models}, and described by the dominant term $\lambda_n\phi^n$. Values of the index $n=2/3,1,2$ have been obtained in string theory~\cite{Silverstein:2008sg,McAllister:2008hb,Dimopoulos:2005ac}. The spectral index, the scalar running and $r$ are simply
\begin{gather}
    n_s=1-\frac{2n+4}{n+4N},\\
    \alpha_s=-\frac{8(n+2)}{(n+4N)^2},\\
    r=\frac{16n}{n+4N}
\end{gather}
and we can see that there is no agreement when the B-modes BK15 observation are included, whereas we still have a consistency at 95\% CL with Planck alone, or up to within $1\sigma$ for ACTPol+WMAP and SPT3G+WMAP. Nevertheless, provided a nonminimal coupling with gravity, the simple power-law potential acquires a compatibility up to 68\% CL for some values of $n$ as shown by the red lines in \fig{Forconi1Models}. The coupling constant $\xi$ is chosen according to Ref.~\cite{Shokri:2019rfi} where the authors have made an analysis imposing this inflationary model at the beginning and using $\xi$ as a free parameter. For the sake of completeness the values are listed below
\begin{itemize}
    \item $\mathbf{n=4}$, with $\xi\simeq 0.0016$,
    \begin{gather}
        n_s=1-\frac{1}{N}(3-8\xi N),\\
        \alpha_s=\frac{1}{N^2}(-3+96\xi N-64\xi^2N^2),\\
        r=\frac{16}{N}(1-8\xi N).
    \end{gather}
    \item $\mathbf{n=2}$, with $\xi\simeq0.0015$,
    \begin{gather}
        n_s=1-\frac{2}{N}(1+\frac43\xi^2N^2),\\
        \alpha_s=\frac{2}{N^2}(-1+4\xi \alpha_sN-96\xi^2N^2),\\
        r=\frac{8}{N}(1-8\xi N).
    \end{gather}
    \item $\mathbf{n=\frac43}$, with $\xi\simeq 0.0011$,
    \begin{gather}
        n_s=1-\frac{1}{3N}(5+8\xi N),\\
        \alpha_s=\frac{5}{81N^2}(-27+48\xi N-704\xi^2 N^2),\\
        r=\frac{16}{9N}(3-32\xi N).
    \end{gather}
    \item $\mathbf{n=2/3}$, with $\xi\simeq 0.0007$,
    \begin{gather}
        n_s=1-\frac{4}{3N}(1+4\xi N),\\
        \alpha_s=\frac{4}{81N^2}(-27+84\xi N+464\xi^2 N^2),\\
        r=\frac{8}{9N}(3-40\xi N).
    \end{gather}
\end{itemize}
This model is consistent also with the ATCPol+WMAP and SPT3G+WMAP contours with the preference for higher values of the tensor tilt translated into slightly preferences for lower values of $n<2$, \textit{e.g.}, the one with $n=2/3$ acquires a compatibility of 68\% CL.

\paragraph{\textit{Quintessential inflation}}: in this scenario the early inflationary period and the late-time acceleration are combined. The potential in this case should be shallow at early times, i.e. satisfying the slow-roll conditions, and steep after. As the usual exponential model does not satisfy the observational constraints \cite{Geng:2017mic} a new parameter $n$ is added (\ref{Quint}) which also influences the steepness of $V(\phi)$, whose form is
\begin{equation}
    V=\Lambda e^{-\lambda \frac{\phi^n}{M^n_{pl}}},
    \label{Quint}
\end{equation}
with $n>1$. Imposing $\lambda\ll 1$ we end up with the large field inflation, called quintessential inflation \cite{Peebles:1998qn}. In this model, the parameters are
\begin{gather}
    n_s=1-\frac{2(n-1)}{(n-2)N}-\frac{[n(n-2)\lambda N]^{-\frac{2}{n-1}}}{(n-2)^2N^2},\label{eq:nsExpo}\\
    \alpha_s=-\frac{2(n-1)}{(n-2)N^2}+\frac{6(n-1)[n(n-2)\lambda N]^{-\frac{2}{n-2}}}{(n-2)^3 N^3},\\
    r=\frac{8[n(n-2)\lambda N]^{-\frac{2}{n-2}}}{(n-2)^2N^2}.
\end{gather}
Fixing $\lambda=10^{-10}$ and varying $n$, the purple curves in \fig{Forconi1Models} are drawn, showing, for example as reference, that $n=7$ is compatible with 95\% CL of Planck+BK15 with $N=60$ whereas it is not for $N=50$. A lower values of $\lambda$ move the curves to the right, increasing the inclination, whereas a higher value makes $n_s$ independent of it, as shown in \eq{nsExpo}. Concerning the other datasets, this model is in disagreement with ACTPol+WMAP data unless for significantly lower values of $\lambda$, and in tension with SPT3G+WMAP. Also in this case the different agreement of the models with the data is affected by the inconsistency between the datasets explored here.

\paragraph{\textit{Starobinsky-like inflation}}: lastly, we analyze the $R^2$ inflation~\cite{Starobinsky:1980te} which is characterized by adding higher curvature corrections ($R^2$) to the Hilbert action of gravity \eq{ActionInflatonMinimallycoupled}. This analysis comprehends also Higgs inflation~\cite{Bezrukov:2007ep} and universal attractors models~\cite{Kallosh:2013lkr} predictions since they are equal to the Starobinsky inflation \cite{Kehagias:2013mya}. The similarity is due to the fact that kinetic terms are negligible during the inflationary period and the slow-roll parameters differ from one another for $\sim 10^{-5}$ corrections which are still too small to be measured. However deviations from this scenario can be obtained considering different classes of inflationary models like $\alpha$-attractors \cite{Kallosh:2013tua}. The Starobinsky potential is
\begin{equation}
    V=\frac{M_{\rm pl}^2}{8}\lambda(1-e^{-\sqrt{\frac23}\frac{\phi}{M_{\rm pl}}})^2
\end{equation}
and the inflationary parameters are
\begin{gather}
    n_s=1-\frac{32N+24}{(4N-3)^2}\simeq 1-\frac{2}{N},\\
    \alpha_s=-\frac{64N(8N-13)}{(4N-3)^4}\simeq -\frac{2}{N^2},\\
    r=\frac{192}{(4N-3)^2}\simeq\frac{12}{N^2}.
\end{gather}
In this model the hierarchy of the parameters is $\xi\sim\epsilon\ll\eta\ll1$ instead of the more common $\xi\ll\eta\ll\epsilon\ll1$. Thus the value of $r$ is expected to be extremely small. In fact, we can see from the cyan line in \fig{Forconi1Models} that its smallness results in a compatibility within one standard deviation for all the datasets. Small deviation from this model, i.e. considering the term $R^p$ with $p\approx2$~\cite{Motohashi:2014tra}, worsen the agreement with Planck+BK15 as shown by the dotted lines which represent terms with $2+\Delta p$ where $\Delta p=0.01$. Considering also ACTPol+WMAP, we see that the model is excluded when $p$ is decreased, whereas for SPT3G+WMAP it is  still consistent within the 95\% CL contours. On the other hand, a bigger value of $p$ is completely in agreement with both datasets at 68\% CL.
\vspace{2mm}

The constraints on the slow roll inflationary models remain basically stable when $dn_s /d\ln k$ can freely vary in the sampling, see also the analogous discussion in~\cite{Akrami:2018odb} and also~\cite{Geng:2017mic,Wu:2018vuj,Renzi:2019ewp,Civiletti:2020fkm,Meza:2021xuq}. However, the tension~\cite{Handley:2020hdp} present between the cosmological datasets analysed in this work (i.e., Planck, ACTPol+WMAP and SPT3G+WMAP) produces different constraints on the inflationary parameter and consequently completely different results regarding the model compatibility, see also \fig{Forconi1Models}.


\subsection{Flatness as consistency check} 

In this section we study two different extensions of the standard cosmological model that both include the curvature parameter $\Omega_k$ as an additional parameter. In particular we first analyze the case $\Lambda\mathrm{CDM}+r+\Omega_k$ and then we add also the running of the scalar tilt, $\Lambda\mathrm{CDM}+r+\alpha_s+\Omega_k$. For both the models, we adopt the common power-law parameterization for the primordial spectra, assuming the usual slow-roll consistency relations to hold. Indeed, since the vast majority of inflationary models predict flatness, the constraints on the spatial curvature provide an important consistency check of this standard scenario, see also~\cite{Akrami:2018odb}.

\begin{table}
	\begin{center}
		\renewcommand{\arraystretch}{1.5}
		\resizebox{\textwidth}{!}{\begin{tabular}{c c c c c c c}
			\hline
			\textbf{Parameter} & \textbf{Planck18}  & \textbf{Planck18 + lensing}  & \textbf{Planck18 + BAO} & \textbf{Planck18 + BK15} & \textbf{ACTPol + WMAP} & \textbf{SPT3G+WMAP} \\
			\hline\hline
			$\Omega_{\rm b} h^2$ &$0.02263\pm 0.00018$&$0.02252\pm 0.00017$&$0.02241\pm 0.00015$&$0.02262\pm 0.00017$  &$0.02245\pm 0.00022$ &$0.02273\pm 0.00025$\\
			$\Omega_{\rm c} h^2$ &$0.1177\pm 0.0016$&$0.1181\pm 0.0015$&$0.1196\pm 0.0014$&$0.1179\pm 0.0015$ &$0.1184\pm 0.0030$ &$0.1141\pm 0.0033$\\
			$100\,\theta_{\rm {MC}}$ &$1.04120\pm 0.00033$&$1.04110\pm 0.00032$&$1.04097\pm 0.00031$&$1.04118\pm 0.00033$ &$1.04181\pm 0.00065$ &$1.03975\pm 0.00070$\\
			$\tau$   &$0.0480^{+0.0087}_{-0.0072}$&$0.0487^{+0.0085}_{-0.0075}$&$0.0554\pm 0.0080$&$0.0477^{+0.0086}_{-0.0072}$ &$0.059\pm 0.013$ &$0.060\pm 0.013$\\
			$\log(10^{10}A_{s})$ &$3.026^{+0.018}_{-0.015}$&$3.027^{+0.018}_{-0.016}$&$3.045^{+0.015}_{-0.017}$&$3.026\pm 0.018$ &$3.057\pm 0.027$ &$3.039\pm 0.026$ \\
			$n_s$ &$0.9728\pm 0.0052$&$0.9707\pm 0.0049$&$0.9671\pm 0.0046$&$0.9715\pm 0.0048$ &$0.9773\pm 0.0070$ &$0.9793\pm 0.0091$\\
			$r$ &$<0.170$&$<0.154$&$<0.120$&$<0.0613$ &$< 0.210$ &$< 0.259$\\
			$\Omega_k$&$-0.048^{+0.020}_{-0.016}$&$-0.0123^{+0.0072}_{-0.0063}$&$0.0007\pm 0.0020$&$-0.047^{+0.018}_{-0.015}$ &$-0.007^{+0.016}_{-0.012}$ &$0.0008^{+0.013}_{-0.0097}$\\
			\hline
			$n_T$ &$>-0.0212$&$>-0.0192$&$>-0.0150$&$>-0.0077$ &$>-0.0262$ &$>-0.0324$\\
			$\alpha_T$ &$\left(\,-10.8\pm 8.5\,\right)\cdot 10^{-5}$&$\left(-12\pm 7.8\right)\cdot 10^{-5}$&$\left(\,-12.7^{+9.5}_{-7.3}\,\right)\cdot 10^{-5}$&$\left(\,-7.5^{+5.6}_{-3.8}\,\right)\cdot 10^{-5}$ &$\left(\,-3.7^{+8}_{-16}\,\right)\cdot 10^{-5}$ &$\left(\,5^{+14}_{-31}\,\right)\cdot 10^{-5}$\\
			$\epsilon_V\simeq\epsilon_\Hl$ &$<0.0106$&$<0.0097$&$<0.0075$&$<0.0038$ &$< 0.0131$ &$< 0.0162$\\
			$\eta_V$ &$-0.0005^{+0.0081}_{-0.013}$&$-0.0033^{+0.0069}_{-0.012}$&$-0.0079^{+0.0053}_{-0.0091}$&$-0.0094^{+0.0038}_{-0.0049}$ &$0.005^{+0.010}_{-0.016}$ &$ 0.0096^{+0.013}_{-0.021}$\\
			$\eta_\Hl$ &$0.0184^{+0.011}_{-0.0080}$&$0.0217^{+0.0098}_{-0.0070}$&$0.0272^{+0.0081}_{-0.0058}$&$0.0252\pm 0.0055$ &$0.012^{+0.015}_{-0.010}$ &$0.007^{+0.019}_{-0.013}$\\
			$V_{\rm inf}^{1/4}$ & $<2.08\times10^{16}\,\rm GeV$&$<2.03\times10^{16}\,\rm GeV$&$<1.90\times10^{16}\,\rm GeV$&$<1.61\times10^{16}\,\rm GeV$ &$<2.19\times10^{16}\,\rm GeV$ &$<2.31\times10^{16}\,\rm GeV$\\
			$\Delta N_{\rm tot}$ &$63.55^{+0.30}_{-0.21}$&$-$&$-$&$63.31^{+0.31}_{-0.23}$ &$-$ &$-$\\
			$\Delta N(k_{\rm exit})$ &$1.55^{+0.30}_{-0.21}$&$-$&$-$&$1.31^{+0.31}_{-0.23}$ &$-$ &$-$\\
			\hline	\hline
		\end{tabular}}
	\end{center}
	\caption[Results for $\Lambda\rm CDM + r+ \Omega_k$]{\small Results for $\Lambda\rm CDM + r+ \Omega_k$. The constraints on parameters are at $68\%$ CL, while upper bounds are at $95\%$ CL. The internal horizontal line divides the primary parameters of the cosmological model (those we directly sample in our MCMC analysis) from the derived parameters (those we obtain from the others by the relations described in the text).}
	\label{tab:LCDM+r+omegak}
\end{table}

\begin{figure}[h!]
	\centering
	\includegraphics[width=0.8\textwidth]{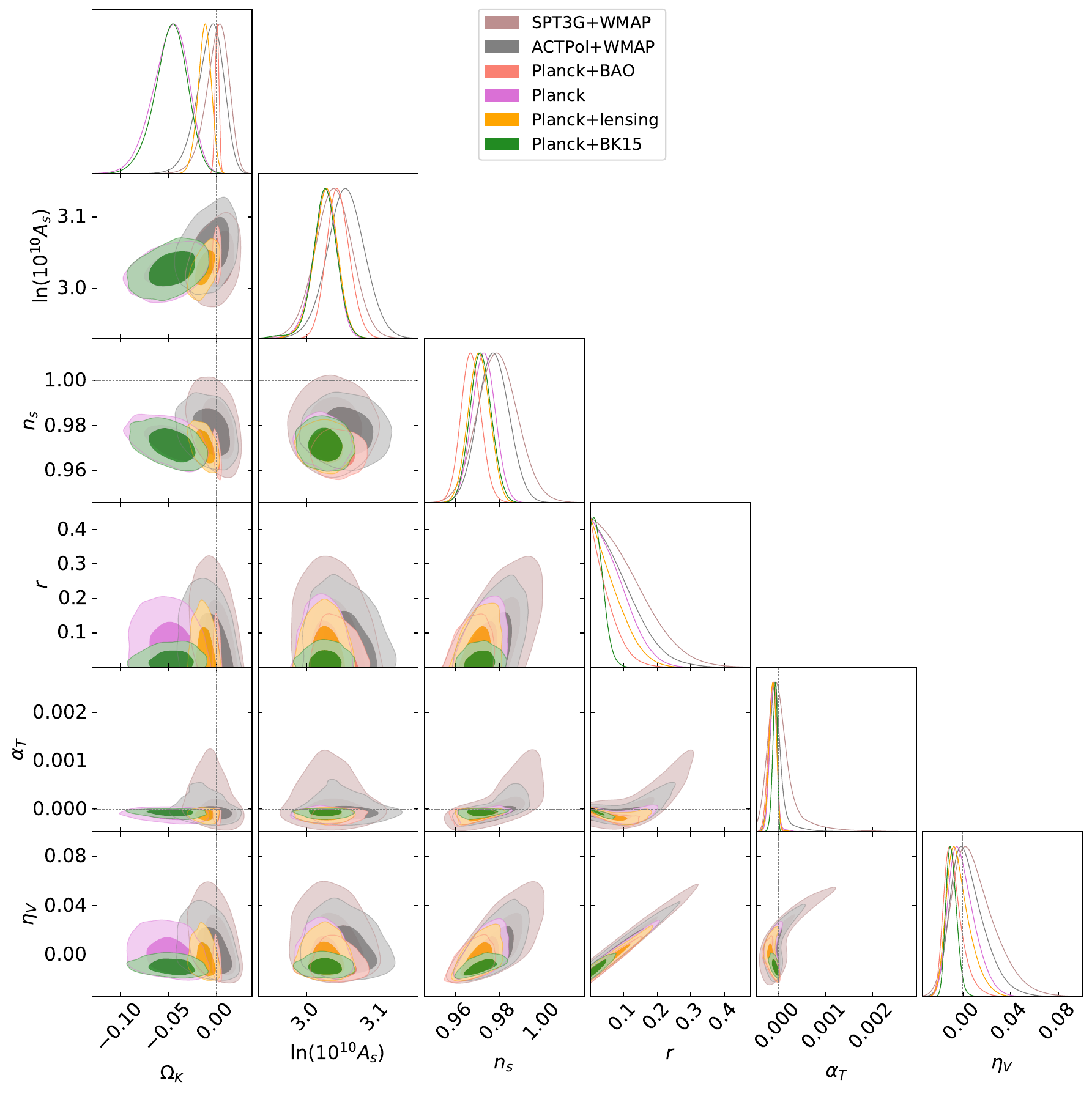}
	\caption[Marginalized 2D and 1D posteriors distributions for the $\Lambda\rm CDM + r+ \Omega_k$]{\small Marginalized 2D and 1D posteriors distributions for the $\Lambda\rm CDM + r+ \Omega_k$ cosmological model obtained for different combinations of the datasets listed in \sect{Foroni1Methods}. The dashed lines represent the  case of vanishing inflationary parameters and flat spacetime geometry.}
	\label{fig:LCDM+r+omegak}
\end{figure}

\begin{figure}[h!]
	\centering
	\includegraphics[width=1 \textwidth]{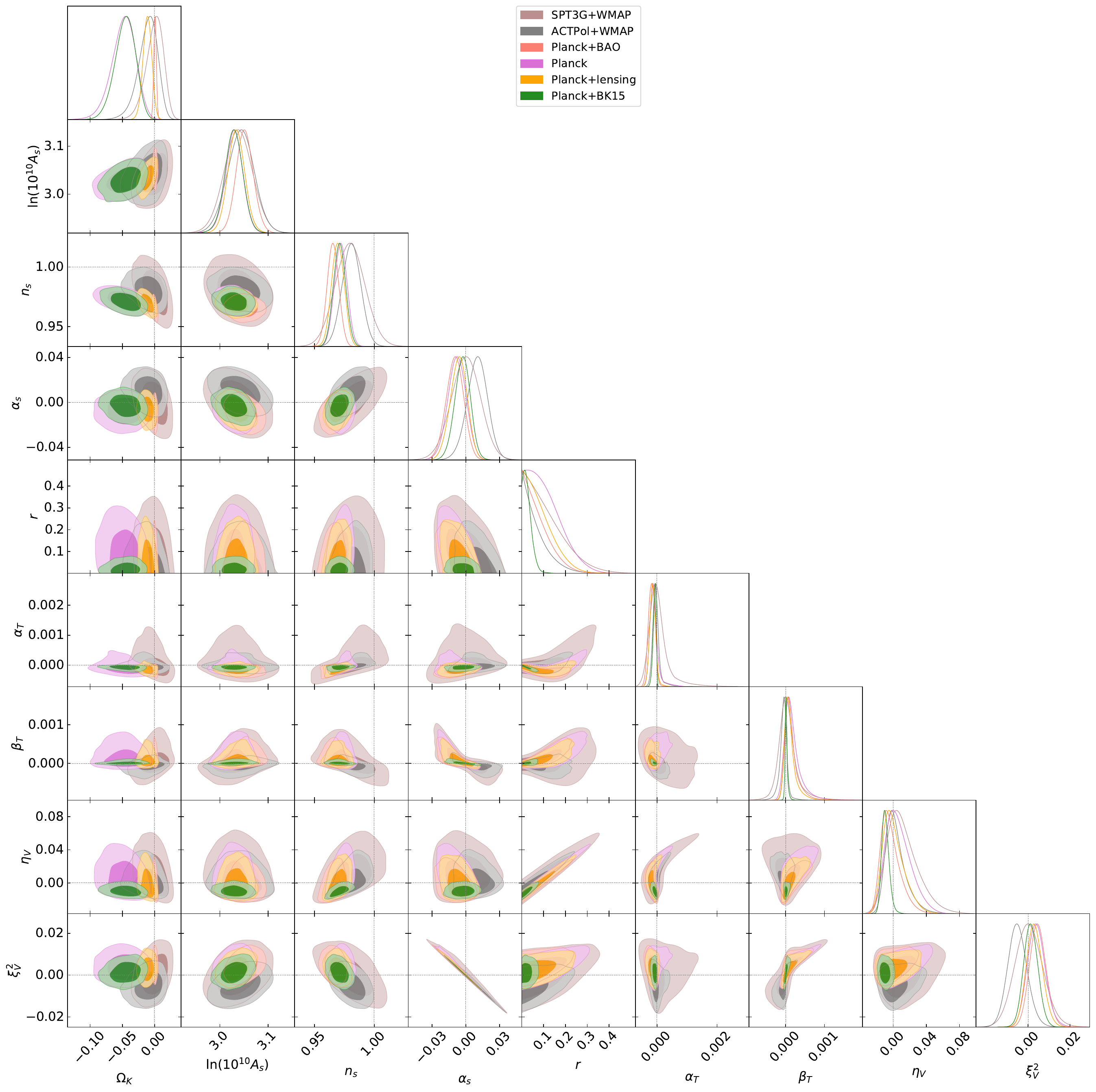}
	\caption[Marginalized 2D and 1D posteriors distributions for the $\Lambda\mathrm{CDM} + r+ \alpha_s+\Omega_k$]{\small Marginalized 2D and 1D posteriors distributions for the $\Lambda\mathrm{CDM} + r+ \alpha_s+\Omega_k$ cosmological model obtained for different combinations of the datasets listed in \sect{Foroni1Methods}. The dashed lines represent the  case of vanishing inflationary parameters and flat spacetime geometry.}
	\label{fig:LCDM+r+nrun+omegak}
\end{figure}

\tab{LCDM+r+omegak} summarizes the constraints derived for the model $\Lambda\rm{CDM}+r+\Omega_k$ and in \fig{LCDM+r+omegak} we show the $68\%$ and $95\%$~CL marginalized contours for different inflationary parameters in the same model. On the other hand, in \tab{LCDM+r+nrun+omegak} we present the results for the $\Lambda\mathrm{CDM}+r+\alpha_s+\Omega_k$ model showing in \fig{LCDM+r+nrun+omegak}  the $68\%$ and $95\%$~CL contours. 
\begin{table}
	\begin{center}
		\renewcommand{\arraystretch}{1.5}
		\resizebox{\textwidth}{!}{\begin{tabular}{c c c c c c c}
			\hline
			\textbf{Parameter} & \textbf{Planck18}  & \textbf{Planck18 + lensing}  & \textbf{Planck18 + BAO} & \textbf{Planck18 + BK15} & \textbf{ACTPol + WMAP} & \textbf{SPT3G+WMAP}\\
			\hline\hline
			$\Omega_{\rm b} h^2$ &$0.02268\pm0.00018$&$0.02255\pm0.00017$&$0.02245\pm0.0016$&$0.02263\pm0.00017$ &$0.02236\pm 0.00022$ &$0.02274\pm 0.00024$\\
			$\Omega_{\rm c} h^2$ &$0.1176\pm0.0016$&$0.1182\pm0.0015$&$0.1197\pm0.0015$&$0.1180\pm0.0015$ &$0.1171\pm 0.0032$ &$0.1141\pm 0.0038$\\
			$100\,\theta_{\rm {MC}}$ &$1.04121\pm0.00033$&$1.04110\pm0.00032$&$1.04097\pm0.00032$&$1.04118\pm0.00032$ &$1.04189\pm 0.00067$ &$1.03979\pm 0.00069$\\
			$\tau$   &$0.0491\pm0.0085$&$0.0514\pm0.0083$&$0.0573^{+0.0077}_{-0.0086}$&$0.0487\pm0.0086$ &$0.056^{+0.013}_{-0.012}$ &$0.060\pm 0.013$\\
			$\log(10^{10}A_{s})$ &$3.029\pm0.018$&$3.034\pm0.018$&$3.052\pm0.018$&$3.029\pm0.018$  &$3.043\pm 0.028$ &$3.038\pm 0.029$\\
			$n_s$ &$0.9720\pm0.0052$&$0.9696\pm0.0051$&$0.9655\pm0.0048$&$0.9710$ &$0.9810\pm 0.0077$ &$0.980\pm 0.012$\\
			$\alpha_s$  &$-0.0078\pm0.0080$&$-0.0064^{+0.0078}_{-0.0070}$&$-0.0097\pm0.0076$&$-0.0029\pm0.0068$ &$0.0102\pm 0.0090$ &$0.000\pm 0.013$\\
			$r$ &$<0.250$&$<0.205$&$<0.188$&$<0.0637$ &$<0.185$ &$< 0.282$\\
			$\Omega_k$&$-0.048^{+0.020}_{-0.016}$&$-0.0113\pm0.0066$&$0.0007\pm0.0020$&$-0.046^{+0.017}_{-0.014}$ &$-0.010^{+0.017}_{-0.011}$ &$0.000^{+0.015}_{-0.011}$\\
			\hline
			$n_T$ &$>-0.0312$&$>-0.0256$&$>-0.0235$&$>-0.0080$ &$>-0.0231$ &$>-0.0352$\\
			$\alpha_T$ &$\left(\,-9.6^{+8.4}_{-15}\,\right)\cdot 10^{-5}$&$\left(-13.6^{+8.8}_{-10}\right)\cdot 10^{-5}$&$\left(\,-17\pm 11\,\right)\cdot 10^{-5}$&$\left(\,-7.9^{+6.0}_{-3.9}\,\right)\cdot 10^{-5}$ &$\left(\,-2.5^{+2.7}_{-8.8}\,\right)\cdot 10^{-5}$ &$\left(\,5^{+16}_{-34}\,\right)\cdot 10^{-5}$\\
			$\beta_T$ &$\left(16^{+11}_{-22}\right)\cdot 10^{-5}$&$\left(10.6^{+7.3}_{-16}\right)\cdot 10^{-5}$&$\left(13.4^{+8}_{-18}\right)\cdot 10^{-5}$ &$\left(\,1.7^{+2.0}_{-3.4}\,\right)\cdot 10^{-5}$ &$\left(\,-5.8^{+9.3}_{-7.3}\,\right)\cdot 10^{-5}$ &$\left(\,6^{+16}_{-25}\,\right)\cdot 10^{-5}$\\
			$\epsilon_\Vl\simeq\epsilon_1$ &$<0.0156$&$< 0.0128 $&$<0.0118$&$<0.0040$ &$< 0.0116$ &$< 0.0176$\\
			$\eta_\Vl$ &$0.006^{+0.011}_{-0.018}$&$0.0003^{+0.0089}_{-0.015}$&$-0.0034^{+0.0079}_{-0.014}$&$-0.0095^{+0.0037}_{-0.0049}$ &$0.0030^{+0.0079}_{-0.014}$ &$0.011^{+0.013}_{-0.021}$\\
			$\xi_\Vl^2$ &$0.0040^{+0.0039}_{-0.0044}$&$0.0031^{+0.0035}_{-0.0040}$&$0.0046\pm 0.0038$&$0.0013\pm 0.0034$ &$-0.0050\pm 0.0046$ &$0.0004\pm0.0066$\\
			$\eta_\Hl$ &$0.0146^{+0.014}_{-0.0096}$&$0.0201^{+0.012}_{-0.0082}$&$0.0253^{+0.011}_{-0.0074}$&$ 0.0256\pm 0.0057$ &$0.0107^{+0.013}_{-0.0095}$ &$0.006^{+0.019}_{-0.016}$\\
			$\xi^2_\Hl$ &$-$&$-$&$-$&$0.10\pm 0.29$ &$-$ &$-$\\
			$V_{\rm inf}^{1/4}$ & $<2.3\times10^{16}\,\rm GeV$&$<2.2\times10^{16}\,\rm GeV$&$<2.1\times10^{16}\,\rm GeV$&$<1.6\times10^{16}\,\rm GeV$ &$<2.12\times10^{16}\,\rm GeV$ &$<2.35\times10^{16}\,\rm GeV$\\
			$\Delta N_{\rm tot}$ &$63.67^{+0.29}_{-0.21}$&$-$&$-$&$63.33^{+0.30}_{-0.22}$ &$-$ &$-$\\
			$\Delta N(k_{\rm exit})$ &$ 1.67^{+0.29}_{-0.21}$&$-$&$-$&$1.34^{+0.30}_{-0.22}$ &$-$ &$-$\\
			\hline	\hline
		\end{tabular}}
	\end{center}
	\caption[Results for $\Lambda\mathrm{CDM} + r+ \alpha_s+ \Omega_k$]{\small Results for $\Lambda\mathrm{CDM} + r+ \alpha_s+ \Omega_k$. The constraints on parameters are at $68\%$ CL, while upper bounds are at $95\%$ CL.The internal horizontal line divides the primary parameters of the cosmological model (those we directly sample in our MCMC analysis) from the derived parameters (those we obtain from the others by the relations described in the text).}
	\label{tab:LCDM+r+nrun+omegak}
\end{table}

For the inflationary parameters we see that in both the models, slightly higher values for scalar tilt are preferred with respect to the case without $\Omega_k$. In particular the Planck data gives $n_s=0.9720\pm 0.0052$ ($n_s = 0.9728\pm 0.0052$) when the running $\alpha_s$ is included (excluded). We can also appreciate that these constraints have $1\sigma$ shift toward higher values for the different datasets, including ACTPol+WMAP and SPT3G+WMAP. As concerns the scalar running, the bounds on $\alpha_s$ are consistent with those derived without considering $\Omega_k$, see also \tab{LCDM+r+nrun+omegak}. For the tensor amplitude, we see that, ignoring the scalar running, Planck data gives $r<0.170$ at 95\% CL while including $\alpha_s$ this bound is less stringent: $r<0.250$. Interestingly, for ACTPol+WMAP the upper bound $r<0.210$ becomes more stringent ($r<0.185$) including $\alpha_s$. We also confirm that the ACTPol+WMAP preference for a non-vanishing scalar running is reduced when the tensor amplitude can freely vary. A strong improvement in the constraining power is clearly obtained including also the B-modes BK15 likelihood and, in fact, including (excluding) the running, the combination Planck+BK15 gives $r<0.0637$ ($r<0.0613$). Also in this case the results appear to be stable and consistent with the case in which $\Omega_k$ is not varied. Using the slow-roll consistency relations among the inflationary parameters, we can appreciate how also in this case the parameters space allowed for the tensor spectrum is strongly constrained. On the other hand reversing the slow-roll relations for the scalar and tensor parameters, we can derive constraints on the slow-roll parameters $\{\epsilon_\Vl\,,\eta_\Vl\,,\xi^2_\Vl \}$. 
Exploiting the Planck+BK15 data, for the $\Lambda\rm{CDM}+r+\Omega_k$ model we obtain $\epsilon_\Vl<0.0038$ and $\eta_\Vl=-0.0094^{+0.0038}_{-0.0049}$ such results remain similar even if we let the scalar running $\alpha_s$ free to vary, in this scenario, however, we have also the result for the slow-roll parameter of the third order: $\xi_\Vl^2=0.0013\pm0.0034$. Considering the ACTPol+WMAP and SPT3G+WMAP datasets combination, we find instead both $\epsilon_\Vl$ and $\eta_\Vl$ in agreement with zero within the 68\% CL when the scalar running is fixed to zero or free to vary, while it appears $1\sigma$ indication for a negative $\xi_\Vl^2$ for ACTPol+WMAP in the $\Lambda\mathrm{CDM}+r+\alpha_s+\Omega_k$ model.
Equivalently, we can constrain the HSR parameters obtaining $\eta_\Hl=0.0256\pm 0.0057$ ($\eta_\Hl=0.0252\pm 0.0055$) and $\xi^2_\Hl=0.10\pm 0.29$ when $\alpha_s$ is considered (excluded) for Planck+BK15. This indication for the $\eta_\Hl$ parameter different from zero is reduced to more than $1\sigma$ for ACTPol+WMAP and disappears for SPT3G+WMAP. We would like to stress that all the results obtained analyzing the Planck 2018 data are in agreement with the ACTPol+WMAP and SPT3G+WMAP data within the 95\% CL.

Interestingly, as concerns the spatial curvature, the Planck preference for a closed universe~\cite{Aghanim:2018eyx,DiValentino:2019qzk,Handley:2019tkm,DiValentino:2020hov} is confirmed in both the scenarios, and slightly enforced when the BK15 data are combined together with Planck Data. Indeed in the extended parameter space of $\Lambda\rm{CDM}+r+\Omega_k$ we obtain $\Omega_k=-0.048^{+0.020}_{-0.016}$ for Planck and $\Omega_k=-0.047^{+0.018}_{-0.015}$ for Planck+BK15. Considering also the running of the scalar tilt as an additional parameter, the results are essentially unchanged. In any case, Planck and Planck+BK15 data prefer $\Omega_k<0$ at $2.4\,\sigma$ and $2.6\,\sigma$, respectively. Anyway, considering the lensing spectrum as measured by the Planck Collaboration the evidence for $\Omega_k\ne0$ is reduced to less then two standard deviation ($\Omega_k=-0.0123^{+0.0072}_{-0.0063}$ and $\Omega_k=-0.0113\pm 0.0066$ ignoring and considering $\alpha_s$, respectively). Finally, we have the indication for a spatially flat universe using also the BAO data ($\Omega_k=0.0007\pm0.0020$, for both the models), but this result should be considered with caution because these measurements are in strong disagreement with Planck when the curvature parameter is free to vary~\cite{DiValentino:2019qzk,Handley:2019tkm,DiValentino:2020hov}, so they cannot in principle be combined together. Similarly, exploiting the data from the Atacama Cosmology Telescope and the South Pole Telescope we do not find any evidence for $\Omega_k\ne 0$, with the constraints reading $\Omega_k=-0.007^{+0.016}_{-0.012}$ ($\Omega_k=-0.010^{+0.017}_{-0.011}$) for ACTPol+WMAP and $\Omega_k=-0.0008^{+0.013}_{-0.0097}$ ($\Omega_k=-0.000^{+0.015}_{-0.011}$) for SPT3G+WMAP when the running is excluded (included). It is important to stress here, that also in these extended scenario including a curvature free to vary the ACTPol+WMAP and SPT3G+WMAP dataset combinations show a tension with respect to the results obtained by Planck, as we can see in \fig{LCDM+r+omegak} and~\fig{LCDM+r+nrun+omegak}, always driven by the same effect discussed before. So, albeit the Universe is spatially flat or closed is still a very disputed issue, see also \cite{DiValentino:2020srs,Vagnozzi:2020zrh,Vagnozzi:2020dfn,Dhawan:2021mel,Efstathiou:2020wem,Yang:2020bpv}, in what follows we take into account the Planck(+BK15) preference for a closed cosmological spacetime, investigating the possible consequences for the slow-roll background dynamics. 

Inflation in a curved Universe has been largely discussed in literature, see e.g. Refs~\cite{Linde:1995xm,Linde:2003hc,Ratra:2017ezv,Bonga:2016iuf,Handley:2019anl,Bonga:2016cje,Ooba:2017ukj,Ellis:2001ym,Uzan:2003nk,Unger:2018oqo,Gordon:2020gel,Sloan:2019jyl,Motaharfar:2021gwi}. As a matter of fact, during inflation the spatial curvature is exponentially driven to flatness and so the only way to obtain an inflationary universe with $\Omega_k\ne0$ is to assume that it inflated only by a finite (small) number of \textit{e}-folds $\Delta N_{\rm tot}$. Furthermore, in a curved inflationary background, the power-law relations adopted in this work to compute the primordial spectra become disputed at low multipoles $\ell\lesssim 20$ and more reliable parameterizations should be considered~\cite{Ratra:2017ezv,Bonga:2016iuf,Handley:2019anl,Bonga:2016cje}. Anyway the differences are typically limited to low multipoles and the Planck estimation of cosmological parameters remains robust under the inclusion of positive spatial curvature~\cite{Bonga:2016iuf}. In what follows we therefore neglect these corrections and we provide constraints on the \textit{e}-fold of inflation compatible with Planck(+BK15) preference for a closed Universe. Indeed, in the case of a positive curvature, $\Omega_k<0$, assuming a slow-roll evolution and a reheating phase taking place just after the end of inflation ($\rho_{\rm reh}\simeq V_{\rm inf}$), the total of \textit{e}-fold can be estimated as~\cite{Ellis:2001ym,Uzan:2003nk} 
\begin{equation}
\Delta N_{\rm tot}\simeq \frac{1}{2}\log\left(\frac{\left(1+\delta_0-\Omega_{\rm r}\right)\mathcal R + \Omega_{\rm r} \mathcal R^2}{\delta_0}\right)
\end{equation}
with $\delta_0=\Omega_0-1$, $\Omega_{\rm r}\simeq 4\times 10^{-5}\,h^{-2}$ the radiation density parameter today~\cite{Aghanim:2018eyx} and 
\begin{equation}
\log \mathcal R \simeq 66 + \log\left(\frac{V^{1/4}_{\rm inf}}{10^{16}\,\rm{GeV} }\right).
\end{equation}

\begin{figure}[h!]
	\centering
	\includegraphics[width=0.85 \textwidth]{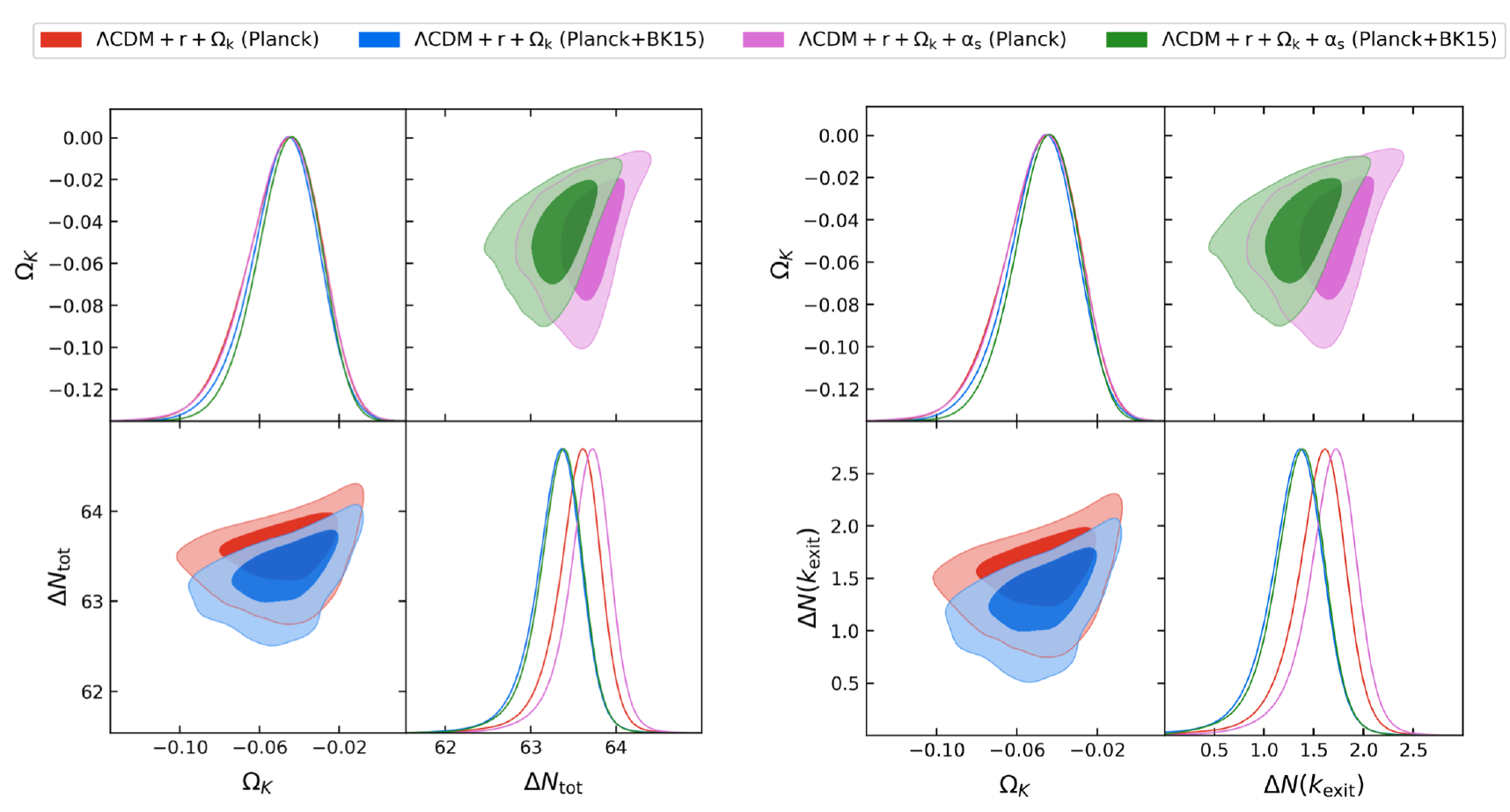}
	\caption[Marginalized 2D and 1D posteriors for the total number of \textit{e}-fold of inflation]{\small Marginalized 2D and 1D posteriors for the total number of \textit{e}-fold of inflation $\Delta N_{\rm tot}$ in a closed cosmological spacetime (left panel) and for the number of \textit{e}-fold before the largest observable scale exits the horizon during inflation $\Delta N(k_{\rm exit})$ (right panel).}
	\label{fig:Ntot_exit}
\end{figure}

In \fig{Ntot_exit}, we show the 68\% and 95\% CL marginalized contours for the total number of \textit{e}-fold of inflation compatible with Planck(+BK15) preference for a closed Universe. Within the $\Lambda\rm{CDM}+r+\Omega_k$ model, using only the Planck data, we obtain a maximum number of \textit{e}-fold $\Delta N_{\rm tot}=63.55^{+0.30}_{-0.21}$ at 68\% CL while including also the B-modes likelihood, for Planck+BK15 we get $\Delta N_{\rm tot}=63.31^{+0.31}_{-0.23}$ at 68\% CL. Including the scalar running in the sampling, the results remain almost unchanged, see also \tab{LCDM+r+nrun+omegak} and \fig{Ntot_exit}. This means that if the Planck(+BK15) evidence for a closed Universe will be confirmed by future measurements, one would need about 63 \textit{e}-fold of expansion while the total number of \textit{e}-folds in many physical models of inflation is typically extremely large, e.g. in power-law inflation one expects $\Delta N_{\rm tot}\sim 10^{12}$ \cite{Lyth:1998xn,Linde:2005ht}. This would strongly constrains the background dynamics before the largest observable scale exit the horizon, with important implications for the observed homogeneity in the cosmic microwave background. Indeed, assuming a standard slow roll inflation followed by a canonical reheating phase and supposing the Universe to be radiation-dominated from the end of reheating to the matter-radiation equality, the number of \textit{e}-folds between when the scale $k$ crosses the horizon and the end of inflation can be estimated as~\cite{Uzan:2003nk,Liddle:1993fq,Lidsey:1995np}
\begin{align}
\nonumber N(k)\simeq & 128 - \log \mathcal R-\log \left(\frac{k}{a_0\,H_0}\right) +2\log\left(\frac{V^{1/4}_{\rm inf}}{10^{16}\,\rm{GeV} }\right)\\& - \log\left(\frac{H_0}{100\,\rm{Km/s/Mpc}}\right) + \mathcal O \left(\log (V_k / V_{\rm inf})\right)
\end{align}
where, for a slow-roll dynamics, the effects of assuming  $V_k \simeq V_{\rm inf}$ are expected to be small for the scales of interest. By noting that the CMB roughly probes scales from 10 to $10^4$ Mpc, one can estimate the number of \textit{e}-fold before the largest observable scale in the Universe exits the horizon $\Delta N(k_{\rm exit})\simeq \Delta N_{\rm tot} - N(k_{\rm min})$. By noting that for the parameter space explored in this work $N(k_{\rm min})\simeq 61 - 62$, see also \cite{Uzan:2003nk}, from Planck(+BK15) data it follows that, within the $\Lambda\rm{CDM}+r+\Omega_k$ model, $\Delta N(k_{\rm exit})=1.55^{+0.30}_{-0.21}$ ($\Delta N(k_{\rm exit})=1.31^{+0.31}_{-0.23}$), while including also $\alpha_s$ we get $\Delta N(k_{\rm exit})=1.67^{+0.29}_{-0.21}$ ($\Delta N(k_{\rm exit})=1.34^{+0.30}_{-0.22}$), see also \fig{Ntot_exit}. Although the allowed number of \textit{e}-fold compatible with the constraints by structure formation (i.e. 50 - 60 \textit{e}-folds between the horizon exit and the end of inflation \cite{Akrami:2018odb}) are enough also to solve ‘flatness’ with an accuracy represented by the precision in $\Omega_k$ (a fine tuning of about 1\% is typically enough \cite{Linde:2003hc}), it should be also noted that the main difficulty for a successfully closed inflationary model is represented by homogeneity and isotropy. Indeed, in most of the models proposed in the literature, when the Universe does not inflate long enough to become flat, the density perturbations on the horizon scale are typically expected to be much larger than those observed, except for a specific class of models~\cite{Linde:2003hc}.

%% file: Chapters/Non-Gaussianity.tex
\chapter{Non-Gaussianity}
\label{ch:NGty}
In standard inflationary models, the primordial fluctuations that seeded the formation of cosmic structure are predicted to follow a nearly Gaussian distribution. However, small deviations from Gaussianity, referred to as non-Gaussianity (NG), provide a critical window into the physics of inflation and potential extensions beyond the simplest models. 

In this chapter, we explore the theoretical underpinnings of non-Gaussianity, beginning with the three-point correlation function, or bispectrum, which quantifies NG in the distribution of primordial fluctuations. We discuss how different inflationary models, such as quasi-single-field inflation and multi-field models, predict varying degrees of NG. These deviations offer a way to break degeneracies in the power spectrum and provide insights into the interactions and field content of the early Universe. We also examine the current observational bounds on NG, noting that while present measurements are consistent with single-field slow-roll models, future observations have the potential to reveal new physics beyond the standard model of inflation~\cite{Komatsu:2009kd,Bartolo:2004if,Fergusson:2006pr}. Even though current measurements of NG are within the expected prediction of the single-field slow-roll model, the current level of sensitivity cannot allow us to rule out alternative theories~\cite{Planck:2019kim}. For this reason, after giving a brief introduction on the bispectrum, we study the Quasi-single field inflationary models and the possible bias they can bring with a boosted trispectrum, to the cosmological parameters.

\section{Three-point function}
\label{eq:Ahan}

The deviation from pure Gaussian statistics is given mostly by the three-point function as it is usually much bigger than the others (but it is not always true, see \sect{QSFIFIA}). The advantage of using such a function is that, in the single field model, it can be explicitly calculated as a function of the slow-roll parameters\footnote{It turns out to be very small. The primordial fluctuations are Gaussian up to $10^{-6}$ which is beyond what we can measure in the near future} \cite{Maldacena:2002vr}. Furthermore, it potentially contains a lot of other information. The three-point function of $\mathcal{R}$ in momentum space is simply
\begin{equation}
    \boxed{\langle \mathcal{R}_{\boldsymbol{k}_1}\mathcal{R}_{\boldsymbol{k}_2}\mathcal{R}_{\boldsymbol{k}_3}\rangle=(2\pi)^3\delta(\boldsymbol{k}_1+\boldsymbol{k}_2,\boldsymbol{k}_3)\frac{(2\pi^2)}{k_1k_2k_3}B_{\mathcal{R}}(k_1,k_2,k_3)}\,,
    \label{eq:Shake}
\end{equation}
where the $\delta$, that ensures the momentum conservation, is a consequence of the homogeneity. The three wavevectors form a closed triangle and the strength of the signal depends on its shape~\cite{Babich:2004gb}. The function $B$ is called \textit{bispetrum} and depends only on the amplitude of $k_i$ due to isotropy and it is dimenisonless, thank to the factor $(k_1k_2k_3)^{-1}$. It is also real because the three-point function in position cannot change if we change sign to all coordinates \cite{Babich:2004gb}. We have said that the momenta form a triangle, so if we set the equilateral configuration, we can define the amplitude of NG~\cite{Baumann:2022mni} as
\begin{equation}
    f_{\rm NL}\equiv-\frac{5}{18}\frac{B_\mathcal{R}(k,k,k)}{\Delta_\mathcal{R}^4(k)}\,.
    \label{eq:nlfnl}
\end{equation}
Using this amplitude, a general bispectrum can be defined with the introduction of the the shape function $S(x_2,x_3)$ where $x_2\equiv k_2/k_1$ and  $x_3\equiv k_3/k_1$, obtaining
\begin{equation}
    B_\mathcal{R}(k_1,k_2,k_3)\equiv -\frac{18}{5}f_{\rm NL} \times S(x_2,x_3)\Delta^4_\mathcal{R}
\end{equation}

\paragraph{\textit{Local non-Gaussianity}}
NG arise also from self-interactive terms, therefore one way to make a  Gaussian field non-Gaussian is to add a square of itself~\cite{Gangui:1993tt,Komatsu:2001rj}
\begin{equation}
    \mathcal{R}(\boldsymbol{x})=\mathcal{R}_g(\boldsymbol{x})-\frac35 f^{loc}_{NL}[\mathcal{R}_g(\boldsymbol{x})^2-\langle \mathcal{R}_g(\boldsymbol{x})^2\rangle]\,.
    \label{eq:NLGU}
\end{equation}
In other words, we split the field into its linear Gaussian part $\mathcal{R}_g$ and its non-linear part which is the square of its local value minus the variance of its Gaussian part. We have used the superscript $loc$ as \eq{NLGU} describes at leading order the most generic form of NG which is local in real space. This form is expected for models where non-linearities develop outside the horizon \cite{Babich:2004gb}. This is the case for all the models in which the fluctuations of an additional light field, different from the inflaton, contribute to the curvature perturbations we observe. Experiment such as WMAP \cite{WMAP:2008lyn,Smith:2009jr} or Planck \cite{Planck:2019kim} puts a limit on the scalar variable $f^{loc}_{NL}$. The latest constrain is \cite{Planck:2019kim}
\begin{equation}
    f^{loc}_{\rm NL}=-0.9\pm 5.1\quad\text{with}\quad 68\%\quad\text{confidence level}.
    \label{eq:ConstraintNOnGauss}
\end{equation}
An amplitude $f_{NL}$ of order $100$ corresponds to a $0.1\%$ correction to $\mathcal{R}_g\sim 10^{-5}$ in \eq{NLGU}. The constraint \eq{ConstraintNOnGauss} implies that the CMB is highly Gaussian even if it has not be so. Nevertheless, if the single field slow-roll inflation is correct, then the observed Gaussianity is a rather natural consequence~\cite{Baumann:2009ds}. From \eq{NLGU} we can simply compute the Fourier transform
\begin{equation}
\mathcal{R}(\textbf{k})=\mathcal{R}_g(\textbf{k})-\frac35 f^{loc}_{NL}\(\int{\frac{d^3p}{(2\pi)^3}\mathcal{R}_g(\textbf{k}+\textbf{p})\mathcal{R}^*_g(\textbf{p})-(2\pi)^3\delta(\textbf{k})\langle \mathcal{R}_g(\boldsymbol{x})^2\rangle}\)
\end{equation}the bispectrum is
\begin{equation}
    B_{\mathcal{R}}(k_1,k_2,k_3)\propto (k_1k_2k_3)^{2}\[P_{\mathcal{R}}(k_1)P_{\mathcal{R}}(k_2)+P_{\mathcal{R}}(k_2)P_{\mathcal{R}}(k_3)+P_{\mathcal{R}}(k_3)P_{\mathcal{R}}(k_1)\].
    \label{eq:simply}
\end{equation}
If we consider a scale-invariant power spectrum we simplify \eq{simply} into
\begin{equation}
    B_{\mathcal{R}}(k_1,k_2,k_3)=\frac65 f_{\rm NL}^{\rm loc}\times S_{\rm local}\[\frac{1}{(k_1k_2)^2}+\frac{1}{(k_2k_3)^2}+\frac{1}{(k_3k_1)^2}\]\,
\end{equation}
with 
\begin{equation}
    S_{\rm local}(k_1,k_2,k_3)=\frac13\(\frac{k_3^2}{k_1k_2}+\frac{k_2^2}{k_1k_3}+\frac{k_1^2}{k_2k_3}\).
\end{equation}
As we previously mentioned, local NG have one wavevector smaller than the others. In fact the local NG bispectrum gives the most contribution in the limit $k_3\rightarrow0$, which consequently implies that the other two $k$ become equal and opposite, due to momentum conservation. This is called the \textit{squeezed limit}. The long wavelength freezes out much before the others and behaves as a background for their evolution. In this limit, the bispectrum is proportional to the power spectrum of the short and long wavelength modes
\begin{equation}
    \lim_{k_3\ll k_1\sim k_2} B_{\mathcal{R}}(k_1,k_2,k_3)=\frac{12}{5}f^{loc}_{NL}\times P_{\mathcal{R}}(k_1)P_{\mathcal{R}}(k_3).
    \label{eq:locshape}
\end{equation}

Being $k_3\rightarrow 0$ and $k_2\sim k_1=k_S$ where $S$ stands for short wavelength, we can see that~\cite{Baumann:2009ds}
\begin{equation}
    \langle \mathcal{R}_{k_1} \mathcal{R}_{k_2} \mathcal{R}_{k_3}\rangle \approx \langle(\mathcal{R}_{k_S})^2\mathcal{R}_{k_3}\rangle.
\end{equation}
The mode with longer wavelength freezes earlier and is already outside the horizon, acting as a background for the two short-wavelength modes. Consequently, the two point function $\langle\mathcal{R}_{k_1}\mathcal{R}_{k_2}\rangle$ depend linearly on the fluctuations of $\mathcal{R}_{k_3}$~\cite{Babich:2004gb}
\begin{equation}
    \langle \mathcal{R}_{\boldsymbol{k}_1}\mathcal{R}_{\boldsymbol{k}_2}\mathcal{R}_{\boldsymbol{k}_3}\rangle \propto P_{\mathcal{R}}(k_3) \frac{\partial}{\partial \mathcal{R}_{\boldsymbol{k}_3}}P_{\mathcal{R}}(k_S).
\end{equation}
Thus, we expect that any distribution will reduce to the local shape \eq{locshape}, if the derivative with respect to the background wave does not vanish. If we now assume the case for a single-field inflation, we obtain the interesting \textit{single-field consistency relation}~\cite{Maldacena:2002vr,Creminelli:2004yq,Cheung:2007sv,Babich:2004gb} 
\begin{equation}
\lim_{k_3\to0}B_{\mathcal{R}}(k_1,k_2,k_3) =(n_s-1)\frac{(k_1k_2k_3)^2}{(2\pi^2)^2}P_{\mathcal{R}}(k_S)P_{\mathcal{R}}(k_3).
\end{equation}
according to which, the single-field model of inflation in the squeezed limit have the bispectrum proportional to the scalar index so, being $n_s\sim 1$, as we have anticipated, the signal is naturally small and vanishes for perfectly scale-invariant perturbations. A detection of non-Gaussianity in this limit can therefore rule out single field inflation and give important information on extra fields~\cite{Arkani-Hamed:2015bza,Arkani-Hamed:2018kmz,Bernardeau:2002jy,Vernizzi:2006ve,Sasaki:2008uc,Byrnes:2006fr,Bernardeau:2001qr}.

\paragraph{\textit{Equilateral non-Gaussianity}} We have seen that the addition of a squared term gives local NG. If we now add high derivative corrections~\cite{Silverstein:2003hf,Chen:2006nt,Cheung:2007st}, e.g. $(\partial\phi)^4$, we end up with cubic terms such as $\dot{\mathcal{R}}^3$ and $\dot{\mathcal{R}}(\partial_i\mathcal{R})^2$, that gives a signal for the NG. Specifically, these terms are suppressed if an individual mode is outside the horizon, which implies that the maximum contribution is not in the squeezed limit, but in the \textit{equilateral configuration}, i.e. when all the modes are approximately the same. Also non-trivial speed of sound models produce a signal that peaks at equilateral configurations. The shape function for the equilateral NG depends on the specific terms in the cubic interactions but in general, for CMB data, the template used is~\cite{Baumann:2022mni,Creminelli:2005hu}
\begin{equation}
    S_{\rm equil}(k_1,k_2,k_3)=-\(\frac{k_1^2}{k_2k_3}+2\text{perms}\)+\(\frac{k_1}{k_2}+5\text{perm.}\)-2
\end{equation}
which is a good approximation to all forms of equilateral NG near the equilateral limit. Now that we have introduced this configuration, we can see that the bispectrum for slow-roll inflation can be written as a combination of the local and equilateral shape functions~\cite{Maldacena:2002vr,Baumann:2009ds}
\begin{equation}
    B_\mathcal{R}(k_1,k_2,k_3)\propto (4\eh -2\etah)S_{\rm local}(k_1,l_2,k_3)+\frac53\eh S_{\rm equil}(k_1,k_2,k_3)\,.
\end{equation}
The bispectrum peaks at squeezed triangles but has an amplitude that is suppressed by slow-roll parameters~\cite{Maldacena:2002vr} $f_{\rm NL}^{\rm SR}=\mathcal{O}(\eh,\etah)$.

The latest constrain is \cite{Planck:2019kim}
\begin{equation}
    f^{\rm equil}_{\rm NL}=-26\pm 47\quad\text{with}\quad 68\%\quad\text{confidence level}.
    \label{eq:ConstraintNOnGausseq}
\end{equation}
\begin{tcolorbox}[mybox]
In general, higher-derivatives terms appears in the model through a non-canonical kinetic term $P(X,\phi)$ where $X=(\partial_\mu\phi)^2$. These models have the general action as
\begin{equation}
    S=\frac12\int{d^4x\sqrt{-g}\[R-P(X,\phi)\]}
\end{equation}
and a non-trivial sound speed 
\begin{equation}
    c_s^2\equiv\frac{P_{,X}}{P_{,X}+2XP_{,XX}}
\end{equation}
In this case, the peaked signal in the equilateral configuration has an amplitude equal to 
\begin{equation}
    f_{\rm NL}^{\rm equil}=-\frac{35}{108}\(\frac{1}{c_s^2}-1\)+\frac{5}{81}\(\frac{1}{c_s^2}-1-2\Lambda\)
\end{equation}
with
\begin{equation}
    \Lambda=\frac{X^2P_{,XX}+\frac23X^3P_{,XXX}}{XP_{,X}+2X^2P_{,XX}}
\end{equation}
\end{tcolorbox}

\paragraph{\textit{Folded non-Gaussianities}} The idea of Bunch-Davis vacuum \eq{condition} is not the only possible configuration for the quantum fluctuations, in fact we can assume that the fluctuations could start from an excited state. This leads to NG that depends on the characteristic of the initial state~\cite{Chen:2006nt,Flauger:2013hra,Clough:2016ymm,Holman:2007na}. In this case, the maximum contribution is given by the \textit{folded configuration} which means that $k_1+k_2\sim k_3$.  The template used is~\cite{Chen:2010xka,Baumann:2022mni}
\begin{equation}
    S_{\rm folded}\propto k_1k_2k_3\frac{-k_1+k_2+k_3}{(k_c-k_1+k_2+k_3)^4}+2\text{perms}
\end{equation}
where $k_c$ is a cutoff that regulates the singularity in the folded limit.

Single-field inflationary models in general can produce equilateral or folded NG. In this parameter space, the contributions of different inflationary realiation is a linear combination of the euilateral shape with the \textit{orthogonal shape}, which is peaked both on equilateral and folded triangle configurations. The name is due to the fact that this shape is orthogonal to the equilateral one~\cite{Senatore:2009gt}. The constrain from CMB of the orthogonal amplitude is \cite{Planck:2019kim}
\begin{equation}
    f^{\rm ortho}_{\rm NL}=-38\pm 24\quad\text{with}\quad 68\%\quad\text{confidence level}.
    \label{eq:ConstraintNOnGaussort}
\end{equation}

\begin{tcolorbox}[mybox]
Can be useful to introduced another notation for the shape functions as done in~\cite{Fergusson:2008ra}. It is often found in literature and significantly compress the increasingly complex expression for the bispectra. We define
\begin{equation}
    K_p=\sum_i(k_i)^p,\quad K\equiv K_1
\end{equation}
\begin{equation}
    K_{pq}=\frac{1}{\Delta_{pq}}\sum_{i\neq j}(k_i)^p(k_j)^q\quad\text{with}\quad \Delta_{pq}=1+\delta_{pq},
\end{equation}
\begin{equation}
    K_{pqr}=\frac{1}{\Delta_{pqr}}\sum_{i\neq j\neq l}(k_i)^p(k_j)^q(k_l)^r\quad\text{with}\quad \Delta_{pqr}=\Delta_{pq}(\Delta_{qr}+\delta_{pr})
\end{equation}
and
\begin{equation}
    \widetilde{k}_{ip}=K_p-2(k_i)^p\quad\text{with}\quad\widetilde{k}_i=\widetilde{k}_{il}.
\end{equation}
 With the above definitions, we can write the shape function for the local
\begin{equation}
     \mathcal{S}_{local}(k_1,k_2,k_3)\propto \frac{K_3}{K_{111}},
\end{equation}
and the equilateral configuration
\begin{equation}
    \mathcal{S}_{\rm equil}(k_1,k_2,k_3)\propto \frac{\widetilde{k}_1\widetilde{k}_2\widetilde{k}_3}{K_{111}}.
\end{equation}
and eventually the folded case
\begin{equation}
    \mathcal{S}_{\rm folded}(k_1,k_2,k_3)\propto \frac{1}{K_{111}}(K_{12}-K_3)+4\frac{K_2}{(\widetilde{k}_1\widetilde{k}_2\widetilde{k}_3)^2}.
\end{equation}
\end{tcolorbox}

\section{Quasi-single field inflationary models}
\label{sec:QSFIFIA}
Quasi-single-field inflation~\cite{Chen:2009zp,Welling:2019bib,Noumi:2012vr} is a class of models which naturally emerges when considering
\begin{figure}[h!]
\centering
\includegraphics[width=0.7\textwidth]{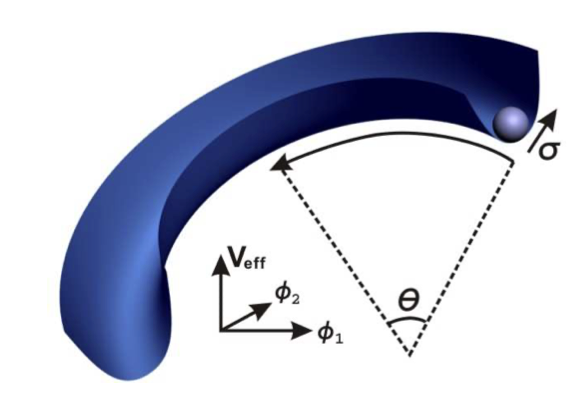}
  \caption[Model configuration]{\small This figure illustrates the model configuration. The $\sigma$ direction denotes the isocurvature direction, with mass typically of order $H$, and $\theta$ the inflationary direction. Picture taken from~\cite{Chen:2009zp}.}
  \label{fig:QSFI}
\end{figure}
UV completion. These models occupy a middle ground between single-field and multifield realizations of the inflationary theory. They are distinguished by a coupling between the inflaton and massive scalar fields. This interaction has the potential to produce large NG distinguishable from the single-field inflation~\cite{Chen:2009we} when the mass is of the order of the Hubble parameter. If the inflaton trajectory is straight, the predictions align with those of single-field inflation. On the other hand, once the trajectory turns, large NG can be generated. The mode in the tangential direction is called curvature mode and the perpendicular one, isocurvature mode. 
For simplicity, following~\cite{Chen:2009zp,Chen:2009we}, let us take a simple model, illustrated in \fig{QSFI}, characterised by the parameters that describes the turning trajectory, constant. Since we are moving along an arc, it is convenient to use polar coordinates when writing the action
\begin{equation}
    S=\int{d^4x\sqrt{-g}\[-\frac12(\hat{R}+\sigma)^2g^{\mu\nu}\partial_\mu\theta\partial_\nu\theta-\frac12g^{\mu\nu}\partial_\mu\sigma\partial_\nu\sigma-V_{\rm sr}(\theta)-V(\sigma)\]}\,,
\end{equation}
where $\theta$ represent the tangential direction, $\sigma$ the radial direction and $\hat{R}$ the radius of the arc. $V_{\rm sr}(\theta)$ is the SR potential and $V(\sigma)$ the potential that traps the isocurvaton at $\sigma_0$. The Friedmann equation and the continuity equation are modified as~\cite{Chen:2009we,Chen:2009zp}
\begin{equation}
    3\M H^2=\frac12 \hat{R}^2\dot{\theta}^2_0+V+V_{\rm sr},\quad -2\M \dot{H}=\hat{R}^2\dot{\theta}_0^2
\end{equation}
and the equation of motion are
\begin{equation}
    \sigma_0=\text{const},\quad V'(\sigma_0)=\hat{R}\dot{\theta}^2_0
    \label{eq:constsigma}
\end{equation}
together with
\begin{equation}
    \hat{R}^2\ddot{\theta}_0+3\hat{R}^2H\dot{\theta}_0+V'_{\rm sr}=0\,.
    \label{eq:InflR}
\end{equation}
$\hat{R}$ has been redefined so that $\hat{R}+\sigma_0\rightarrow \hat{R}$. If we perturb the fields $\delta\theta(\textbf{x},t)$ and $\delta\sigma(\textbf{x},t)$, we get 
we get~\cite{Chen:2009zp}
\begin{multline}
    \mathcal{L}_2=\frac{a^3}{2}\hat{R}^2\delta\dot{\theta}^2-\frac{a\hat{R}^2}{2}(\partial_i\delta\theta)^2-a^3\(\frac{V''_{\rm sr}}{2\hat{R}^2}-(3\epsilon_\Vl-\epsilon_\Vl^2+\epsilon_\Vl\eta_\Vl)H^2\)R^2\delta\theta^2+\frac{a^3}{2}\delta\dot{\sigma}^2\\
    -\frac{a}{2}(\partial_i\delta\sigma)^2-\frac{a^3}{2}(V''-\dot{\theta}^2_0)\delta\sigma^2
\end{multline} 
and
\begin{equation}
\delta\mathcal{L}_2=2a^3\hat{R}\dot{\theta}_0\delta\dot{\theta}\delta\sigma-2\epsilon_\Vl a^3\hat{R}\dot{\theta}_0H\delta\theta\delta\sigma,\quad \delta\mathcal{L}_3= -\frac{a^3}{6}V'''\delta\sigma^3+\dots
\label{eq:L2L3}
\end{equation}
where the PSR now reads
\begin{equation}
    \epsilon_\Vl=\frac{\M}{2}\(\frac{V'_{\rm sr}}{\hat{R}V_{\rm sr}}\)^2,\quad\eta_\Vl=2\M\[\(\frac{V'_{\rm sr}}{\hat{R}V_{\rm sr}}\)^2-\frac{V''_{\rm sr}}{\hat{R}^2V_{\rm sr}}\]\,.
\end{equation}
The perturbed Lagrangian has been divided into three part: $\mathcal{L}_2$ that describes two free fields, $\delta\mathcal{L}_2$ represents the coupling between them and $\delta\mathcal{L}_3$ which is the leading source for the three-point function for $\delta\theta$. We need to impose the condition $\lvert V'''\rvert/H\ll1$ so that the cubic term is sub-dominant with respect to the second.  Hereafter, for simplicity, we ignore the gravity perturbations and terms proportional to the PSR parameters. We are allowed to do that being the potential $V$ the main isocurvature contribution. Defining the conjugate momentum $\pi$ as in \eq{conjugatepi}, and working in the in-in formalism~\cite{Weinberg:2005vy}, we can construct the Hamiltonian as the sum of a kinetimatic part $\mathcal{H}_0$ and interaction part $\mathcal{H}_I$. Specifically
\begin{equation}
    \mathcal{H}_0=a^3\[\frac12 \hat{R}^2\delta\dot{\theta}_I^2+\frac{\hat{R}}{2a^2}(\partial_i\delta\theta_I)^2+\frac12\delta\dot{\sigma}_I^2+\frac{1}{2a^2}(\partial_i\delta\sigma_I)^2+\frac12(V''+7\dot{\theta}^2_0)\delta\sigma_I^2\]
\end{equation}
and
\begin{equation}
\mathcal{H}^I_2=-2\hat{R}\dot{\theta}_0a^3\delta\sigma_I\delta\dot{\theta}_I,\quad\mathcal{H}^I_3=\frac16V'''a^3\delta\sigma_I^3
\label{eq:H2H3I}
\end{equation}
where the perturbations are written in the interaction pictures and it has been used the relation $\delta\dot{\theta}^I=\partial\mathcal{H}_0/\partial(\delta\pi^I_\theta$ and equivalent for $\sigma$. From the kinetic part, we can see that $\sigma$ are massive modes with effective mass $m^2=V''+7\dot{\theta}^2_0$. We follow the procedure in \sect{perturbationss} and we quantize $\delta\theta^I_\textbf{k}$ and $\delta\sigma^I_\textbf{k}$. For the mode function $u_\textbf{k}$ for the curvature modes, we have the same solution \eq{Rising} coming from \eq{MS} in the SR approximation. On the other hand, for the modes function of the isocurvature modes have the equation of motion written as 
\begin{equation}
    v''_\textbf{k}-\frac2\eta v'_\textbf{k}+k^2v_\textbf{k}+\frac{m^2}{H^2\eta^2}v_\textbf{k}=0
\end{equation}
whose solution is~\cite{Chen:2009we,Chen:2009zp}
\begin{equation}
    v_\textbf{k}=-ie^{i(\nu+\frac12)\frac\pi2}\frac{\sqrt{\pi}}{2}H(-\eta)^{\frac32}H_\nu^{(1)}(-k\eta)
    \label{eq:nunu}
\end{equation}
for $m^2/H^2\leq 9/4$ and $\nu=\sqrt{9/4-m^2/H^2}$ whereas
\begin{equation}
    v_\textbf{k}=-ie^{-\hat{\nu}\frac\pi2+i\frac{\pi}{4}}\frac{\sqrt{\pi}}{2}H(-\eta)^{\frac32}H_{i\hat{\nu}}^{(1)}(-k\eta)
    \label{eq:nunubar}
\end{equation}
for $m^2/H^2> 9/4$ and $\hat{\nu}=\sqrt{m^2/H^2-9/4}$. If $k\gg H,m$ then we recover the Bunch-Davis vacuum \eq{BDmf}. It is interesting to see that after horizon exit \eq{nunu} decays as $(-\eta)^{-\nu+3/2}$, while \eq{nunubar} as $(-\eta)^{3/2}$ with an additional oscillatory factor $\eta^{\pm\bar{\nu}}$~\cite{Chen:2009we,Chen:2009zp}. This means that the perturbations for massive fields eventually roll back to zero. When the mass increases, we transit from an over-damped oscillator to an under-damped oscillator and the contribution in real space of isocurvaton to the curvature correlation is suppressed by $e^{-m/H}$. Because of that, we can ignore the under-damped case and concentrate on the case of $0\leq \nu < 3/2$. 

\subsubsection{Turning trajectories imprints}

Because of the presence of isocurvature modes, the power spectrum contains additional terms with respect to \eq{PSscalar}.  The transition from the isocurvature mode to the curvature mode is regulated by
\begin{equation}
    H^I_2=-2\hat{R}\dot{\theta}_0a^3\int{\frac{d^3\textbf{k}}{(2\pi)^3}\delta\sigma^I_\textbf{k}\delta\dot{\theta}^I_{-\textbf{k}}}
\end{equation}
As $\theta$ is our inflaton, we can use \eq{Rflz} and find the power spectrum as
\begin{equation}
    \Delta_{\mathcal{R}}=\frac{H^4}{4\pi^2\hat{R}^2\dot{\theta}^2_0}\[1+8\mathcal{C}\(\frac{\dot{\theta}_0}{H}\)^2\]
\end{equation}
and
\begin{equation}
    n_s-1=-2\epsilon_\Hl-\eta_\Hl+8\mathcal{C}\eta_\Hl\(\frac{\dot{\theta}_0}{H}\)^2
\end{equation}
It is evident that, in order to maintain a perturbative regime, we need to impose $\dot{\theta}_0^2/H^2\ll 1$. This condition, however, makes the power spectrum not relevant to constraints this class of model. The interesting signature are evident at higher-order correlation function.

\subsubsection{Isocurvaton self-interaction}

The third term in \eq{H2H3I}, desribes the self-interaction of $\sigma$. The interaction Hamiltonian is simply the integral
\begin{equation}
    \frac16V'''a^3\int{\frac{d^3\textbf{p}}{(2\pi)^3}\frac{d^3\textbf{q}}{(2\pi)^3}\delta\sigma^I_{\textbf{p}}(t)\delta\sigma^I_{\textbf{q}}(t)\delta\sigma^I_{-\textbf{p}-\textbf{q}}(t)}
\end{equation}

See~\cite{Chen:2009zp} for detailed computation of the bispectra. Qualitatively, the bispectrum lies in the intermediate shape forms. If $\nu$ is large, the shape shift towards locality whereas for small $\nu$ the shapes resemble the equilateral one. Despite that, it is possible to define the amplitude $f^{\rm int}_{\rm NL}$ for intermidiate shape so that it matches the usual $f^{\rm loc}_{\rm NL}$ in \eq{nlfnl}. It reads
\begin{equation}
    f^{\rm int}_{\rm NL}=\alpha(\nu)\Delta^{-1}_{\mathcal{R}}\(\frac{\dot{\theta}_0}{H}\)^3\(-\frac{V'''}{H}\)
    \label{eq:fnltris}
\end{equation}
where $\alpha(\nu)$ is a numerical coefficient that can be very large until it blows up if $\nu\rightarrow 3/2$. This divergence is due to the imposition of constant turn trajectory.

The three point correlation function in the squeezed limit (see \eq{locshape}), is~\cite{Chen:2009we,Chen:2009zp} 
\begin{equation}
    \langle \mathcal{R}(\textbf{k}_1)\mathcal{R}(\textbf{k}_2)\mathcal{R}(\textbf{k}_3)\rangle \xrightarrow[]{k_3\ll k_1=k_2} s(\nu)\frac{4\dot{\theta}_0^3V'''}{3H\hat{R}^3}\frac{1}{k_1^{\frac72-\nu}k_2k_3^{\frac32+\nu}}(2\pi)^3\delta^3(\Sigma_i \textbf{k}_i)
    \label{eq:accra}
\end{equation}
with $s(\nu)$ function of integral over real space of the Hankel function~\cite{Chen:2009zp}. When $\nu$ approaches 0 the Hankel function change form of the expansion and the NG signal change shape. In the denominator in \eq{accra}, we see that the squeezed limit $(\sim k_3^{-\frac32-\nu})$ lies between the equilateral $(\sim k_3^{-1})$ and local $(\sim k_3^{-3})$ bispectrum. For this reason these shapes are called \textit{intermediate shapes}. As we have previously stated, equilateral NG are generated when the horizon crossing happens at the same time for all the modes as opposed to local bispectrum which is generated when the modes have already crossed the horizon. In the quasi-single field inflation, the isocurvaton mass can vary around $\mathcal{O}(H)$. This means, as we have seen in \eq{nunu} and \eq{nunubar}, for heavier field ($\nu<1/2$) the amplitude decays faster after horizon exit, and hence, it mimics the equilateral behavior. Lighetr field, on the other hand, decays slower. So the conversion from isocurvature to curvature mode is still coninuing after horizon crossing, and the nature become increasingly local. In the limiting case of $\nu=3/2$ they coincide. In general, $f^{\rm int}_{\rm NL}$ is small and we cannot generate large NG~\cite{Salopek:1990jq}, due to the presence of $V'''$ which often becomes a SR potential. The peculiarity of this model can be seen going one step further in the non-linear theory.

\subsubsection{Trispectrum}

A noteworthy feature of perturbations in this regime is the fact that the trispectrum amplitude $\tau_{NL}$ is boosted with respect to the bispectrum amplitude $f_{NL}$~\cite{Baumann:2012bc,Baumann:2011nk,Assassi:2012zq,Chen:2009we,Chen:2009zp}. The interacting Hamiltonian terms  that contribute to the trispectrum are proportional to $V'''$ and $V''''$. The amplitude of the trispetra for a given shape is denoted with $\tau_{\rm NL}$. In the regular tetrahedron limit~\cite{Chen:2009bc}
\begin{equation}
    \langle\mathcal{R}^4\rangle\rightarrow(2\pi)^9\Delta^{6}_{\rm \mathcal{R}}\delta^3(\sum_i\textbf{k}_i)\frac{1}{k^9_1}\tau_{\rm NL}\,.
\end{equation}
The contibution from $V''''$ comes from the expansion to the fifth order of the correlation function. To have an approximate value for the amiplitute of the trispectrum, from the turning trajctory  we have a contribution
\begin{equation}
    \tau_{\rm NL}^{\rm CI}\sim\Delta^{-2}_{\mathcal{R}}\(\frac{\dot{\theta}}{H}\)^4V''''
\end{equation}
with CI denoting the contact-interaction. In our approximation, this term is small. However, we have another contribution coming from the scalar-exchange. The vertices of the interaction contribute with the third derivative of the potential. This has a great consequence because
\begin{equation}
    \tau_{\rm NL}^{\rm SE}\sim\Delta^{-2}_\mathcal{R}\(\frac{\dot{\theta}}{H}\)^4\(\frac{V'''}{H}\)^2\sim \(\frac{H}{\dot{\theta}}\)^2f^2_{\rm NL}
    \label{eq:trisfnl}
\end{equation}
where we have compared the result with \eq{fnltris}. If we have a slow turning trajectory, $(\dot{\theta}/H)^2\ll1$, we get $\tau_{\rm NL}\gg f_{\rm}^2$. This implies that a large bispectra may be a better probe for quasi-single inflation than bispectra. To be more specific, also $\tau_{\rm NL}^{\rm CI}$ can contribute non-negligibly but it strictly depends on the details of the potential $V$. 

The dominant contribution is coming from the \textit{collapsed limit} of the four point function. The collapsed limit of an N-point function corresponds to the limit where one internal momentum is smaller than all the external ones (the internal momentum is the vectorial sum of M external momenta). The Suyama-Yamaguchi inequality~\cite{Suyama:2007bg} puts a constrain from below on the \textit{collapsed limit} of the four-point function by the amplitude in the \textit{squeezed limit} of the three-point function. More precisely, $\tau_{NL}\gtrsim (\frac65 f_{NL})^2$. This inequality is particularly sensitive to the presence of multiple sources; if only one source is considered, it reaches saturation. In the collapsed limit, we have
\begin{equation}
\langle\mathcal{R}_{\textbf{k}_1}\mathcal{R}_{\textbf{k}_2}\mathcal{R}_{\textbf{k}_3}\mathcal{R}_{\textbf{k}_4}\rangle\xrightarrow[]{k_{12}\rightarrow 0}4\tau_{\rm NL}(\nu)\Delta^2_{\mathcal{R}}(k_1)\Delta^2_{\mathcal{R}}(k_3)\Delta^2_{\mathcal{R}}(k_{12})\(\frac{k_{12}}{\sqrt{k_1k_3}}\)^{3-2\nu}
\label{eq:trispectrumspaghetti}
\end{equation}

To better study the possible implications of such large NG coming from the trispectrum, we can work within the Super-$\Lambda$CDM framework~\cite{Adhikari:2022moo,Adhikari:2019fvb,Forconi:4}, an extension of the $\lcdm$ model where NG are parametrized by an additional parameter in the angular power spectrum due to super sample signal (i.e. Super). Precedent works found the existence of possible NG at 95\% CL that is increased at more than 3$\sigma$ when the curvature of our Universe is included in the picture. If the \slcdm proves to be accurate, it could profoundly impact current cosmological constraints that rely on the $\lcdm$ assumption. Specifically, in the following sections, we aim to examine the potential consequences for the current cosmological constraints in the neutrino sector. Such an inquiry is highly pertinent at this time, as direct constraints of comparable precision regarding parameters like the neutrino mass are beginning to emerge from ground-based laboratories (see e.g.~\cite{Capozzi:2021fjo}).

\begin{tcolorbox}[mybox]

The covariance matrix of the two-point harmonic space observables gets contributions from the measurement noise and the sample variance due to the incomplete sampling of Fourier modes caused by the finiteness of the survey volume. In harmonic space, it is composed of three terms: the Gaussian (G), the connected non-Gaussian (cNG) and the super-sample (SSC) covariance. The first represents the covariance of the observables if the statistical distribution of the corresponding underlying field was perfectly Gaussian. The second and third arise because of the non-Gaussian coupling of the different Fourier modes of the fields. Specifically, the cNG term describes mode coupling within the survey volumes~\cite{Krause:2016jvl,Barreira:2018jgd} whereas the latter term describes the effects of the coupling of modes respectively larger and smaller than the survey typical linear size $L=\sqrt[3]{V}$ (being $V$ the survey volume)~\cite{Hu:2002we,Takada:2013wfa,Hamilton:2005dx,Rimes:2005dz,Bayer:2022nws}.

The SSC describes an effect that  sets a natural scale for the minimum wavenumber $k$ which can be sampled by any real survey. In fact, Fourier modes with $k<L^{-1}$ simply cannot be accommodated by the survey volume. However, the nonlinear mode coupling intertwines the evolution of these “super-survey”, or “soft” modes with the evolution of “sub-survey”, or “hard” modes. The net result of these two factors is a modulation of the observables within the survey volume by an unobservable background perturbation which biases our measurements. The modulation induced by the super-survey modes is an equivalent to a change in the background density of the observed region. This is accounted for as an additional term in the data covariance matrix, which becomes non-diagonal in $(k_1,k_2)$ because the different modes do not evolve independently.
    
\end{tcolorbox}

\section{Super-$\Lambda$CDM}
\label{sec:PSOGFNSN}

By introducing $\epsilon=3/2-\nu$ into \eq{trispectrumspaghetti} and performing the shift $n_s\rightarrow n_s+\epsilon$ for $k_{1,3}$ and $n_s\rightarrow n_s-2\epsilon$ for $k_{12}$, we can derive the full-sky expression for the NG covariance for the contribution of \eq{trispectrumspaghetti} (see~\cite{Adhikari:2019fvb} and the references therein)
\begin{equation}
    \mathcal{C}_{NG}=\frac9\pi\tau_{NL}(\epsilon)C^{SW}_{L=0}(n_s-2\epsilon)C_\ell(n_s+\epsilon)C_{\ell'}(n_s+\epsilon)
    \label{eq:covNG}
\end{equation}
where $C_\ell$'s are the lensed harmonics~\cite{2012JCAP...03..011P}  and $C_L^{SW}$ is the Sachs-Wolfe angular power, defined as
\begin{equation}
    C^{SW}_{L}(n_s-2\epsilon)=\frac{4\pi A_s}{9(k_0r_\star)^a}\frac{\sqrt{\pi}\Gamma(1-\frac{a}{2})\Gamma(L+\frac{a}{2})}{4\Gamma(\frac32-\frac{a}{2})\Gamma(2+L-\frac{a}{2})}
    \label{eq:CSWL}
\end{equation}
where $a=n_s-2\epsilon-1$, $0<a<2$ and $r_\star$ is the comoving distance to the last scattering surface. 

If we compute the \textit{soft limits} by  splitting the fields into short and long modes ($\theta=\theta_L+\theta_S$ and $\sigma=\sigma_L+\sigma_S$)~\cite{Assassi:2012zq} it becomes more evident that the interaction couples long and short modes. We can take advantage of this fact and include the trispectrum contribution to the covariance matrix using the Super-sample method~\cite{Li:2014jr,Manzotti:2014wca}. Even though they are not directly observable, modes larger than the survey scale affect the evolution of sub-sample modes and consequently affecting the power spectrum. Rather than considering this impact as an additional source of noise, we treat it as an extra parameter with the same analysis pipeline. This approach allows us to examine the response of the power spectrum alongside other parameters, streamlining the data analysis process. As the effect of this parameter should be equal to a noise contribution, its mean value has to be zero, whereas its variance should be set in a way that \eq{covNG} is recovered. Therefore, in our case of study, taking under consideration the trispectrum consistency condition for super sample-signal~\cite{Li:2014sga,Takada:2013wfa}, the power spectrum can be modified as follows:
\begin{equation}
    C^{m}_\ell=C_\ell-A_0C_\ell(n_s+\epsilon)
    \label{eq:Cl-A0}
\end{equation}
where $m$ stands for the power spectrum measured in the presence of the NG whereas $C_\ell$ represents the CMB power spectrum for a realization without super-sample signal coupling. $A_0$ is our additional parameter which quantifies the contribution of the trispectrum and, to recover $\mathcal{C}_{\rm NG}$, is defined as
\begin{equation}
    \langle A_0^2\rangle= \frac{9}\pi \tau_{NL}(\epsilon)C^{SW}_{L=0}(n_s-2\epsilon)\,.
\end{equation}

It is important to underline that the latest constraints on $\tau_{NL}$~\cite{Marzouk:2022utf} are referred to the local trispectrum for multifield inflation (\ie $\epsilon=0$). For a vanishing $\epsilon$, the Sachs-Wolfe angular power spectrum computed at $L=0$ diverges for small modes, and, therefore, it cannot be directly related to our constraints of $A_0$. 

Taking into account the framework we have outlined, we can promote the $\Lambda$CDM model to the \slcdm model, i.e., to the usual six cosmological parameters, we add $\epsilon$, originating from the quasi-single-field framework \eq{trispectrumspaghetti}, and $A_0$ introduced within the Super-sample approach. This extension of the standard model, initially introduced in~\cite{Adhikari:2022moo,Adhikari:2019fvb}, provides a more convenient picture to study the NG. The response of the CMB power spectrum in the Super$\Lambda$CDM scenario was already addressed in~\cite{Adhikari:2019fvb}; it has been investigated using the temperature power spectrum from Planck 2015 release \cite{Planck:2015bpv} also in conjunction with Pantheon type Ia supernovae~\cite{Pan-STARRS1:2017jku} and Baryon Acoustic Oscillations data \cite{Beutler:2011hx,Ross:2014qpa,Gil-Marin:2015nqa}. Additionally, in~\cite{Adhikari:2022moo}, the curvature of our Universe was included and constrained the resulting scenario using both polarization and temperature spectra from Planck 2018~\cite{Planck:2018vyg,Planck:2019nip}, Pantheon type Ia supernovae~\cite{Pan-STARRS1:2017jku} and SH0ES \cite{Riess:2020fzl}. Our purpose is to add new constraints to the parameters by using different datasets. We also want to explore possible correlations among NG and neutrino physics.

\subsection{Extensions in Neutrino Physics} 
\label{sec:methodForconi4}

To take into account the response parametrized in \eq{Cl-A0}, we have modified our theoretical Boltzman solver code \texttt{CAMB}~\cite{Lewis:1999bs,Howlett:2012mh} to generate the theoretical angular power spectrum
\begin{equation}
    C_\ell\rightarrow C_\ell+A_0C_\ell(n_s+\epsilon)\,.
    \label{eq:CAMBRelation}
\end{equation}

As anticipated, we studied the \slcdm model setting constraints to its eight cosmological parameters, as well as extensions in the neutrino sector. The neutrino sector is usually parametrize using combinations of three quantity: 
\begin{itemize}
\item The effective number of neutrino species in the Universe ($N_{\rm eff}$ see \eq{Neff})). We have seen that if neutrino decoupling had been instantaneous, we would have $N_{\rm eff}=3$ but due to non-instantaneous decoupling we observe $N_{\rm eff}\approx 3.04$~\cite{Akita:2020szl,Froustey:2020mcq,Bennett:2020zkv,Cielo:2023bqp}. Any additional relativistic particle produced before recombination can be treated as an additional contribution to this number and even primordial gravitational waves (see \sect{PGWSradio}). Therefore, observing a $\Delta N_{\rm eff}\neq 0$ can be a hint of new physics. Conversely, smaller values suggest a lower-temperature reheating~\cite{deSalas:2015glj} than expected in the $\lcdm$ universe. Notably, as the radiation energy density $\rho_r$ is proportional to the effective number of neutrinos, different values of $N_{\rm eff}$ modify the sound horizon at recombination. In particular, larger values decrease the horizon and, consequently, require higher values of $H_0$ (and $\sigma_8$) potentially moving towards late-time $H_0$ measurements. However, we cannot take this possibility with pure optimism, as the increased value of the $\sigma_8$ parameter exacerbates tensions with large-scale structure data~\cite{DiValentino:2020vvd}.
\item The sum of neutrino masses (see e.g.~\cite{Lesgourgues:2006nd}) $\sum{m_\nu}$. The Planck $\Lambda$CDM base model assumes a normal mass hierarchy, with the minimal mass $\sum{m_\nu}=0.06$ eV. However, it is worth noting that the case of the smallest mass splitting does not determine the value, and $\sum{m_\nu}>0.06$ eV remains a plausible possibility. On the other hand, an inverted hierarchy increases the lower bound to be  $\sum{m_\nu}>0.1$ eV; thus, a stringent upper bound can exclude the latter scenario. In general, this extension can be considered the best motivated as laboratory experiments confirm that at least two neutrinos are massive~\cite{DeSalas:2018rby,Gariazzo:2023joe}. Also, cosmological probes provide constraints on the sum of neutrino masses (see, e.g.,~\cite{Lattanzi:2017ubx,Tanseri:2022zfe,DiValentino:2023fei,DiValentino:2021hoh,RoyChoudhury:2018gay,RoyChoudhury:2019hls,Craig:2024tky}). The effect of massive neutrinos is to suppress power on scales smaller than their free-streaming scale, which can be related to the reduction of lensing potential. Therefore, due to the \textit{lensing anomaly}, caution is advised when interpreting Planck results, as they might yield an overly strong upper limit~\cite{Capozzi:2021fjo,DiValentino:2021imh,diValentino:2022njd}. Additionally, increasing the neutrino mass intensifies the cosmological tension, as it leads to lower values of $H_0$.
\item The effective mass of sterile neutrinos~\cite{Asaka:2005an,Mastrototaro:2021wzl,Asadi:2022nj,Gelmini:2019deq} $m^{\rm eff}_{\nu,\rm sterile}$. For example, if the sterile neutrinos were to thermalize with the same temperature as active neutrinos, we should expect $N_{\rm eff}\approx 4$. However, to maintain generality, we can equally consider an arbitrary temperature $T_s$ or a distribution proportional to the active-sterile neutrino mixing angle~\cite{Dodelson:1993je}. In this case, a relationship between the effective mass and the physical mass of sterile neutrinos can be derived through $N_{\rm eff}$. For instance, within the context of a thermally distributed scenario, we have $m_{\nu,\rm sterile}=(\Delta N_{\rm eff})^{3/4}m^{\rm thermal}_{\rm sterile}$. We can see that for small values of the effective number of relativistic species, the physical mass $m^{\rm thermal}_{\rm sterile}$ increases. Consequently, neutrinos become nonrelativistic before recombination. A limit to the physical mass is required in order to leave out the cases where sterile neutrinos can be considered a candidate for warm and cold dark matter~\cite{Planck:2013pxb}. Specifically, we set $m^{\rm thermal}_{\rm sterile}<10$ eV, as done by the Planck Collaboration~\cite{Planck:2018vyg}.
\end{itemize}

For all the different cosmological parameters, we choose flat-prior distributions, varying them uniformly in the conservative ranges listed in \tab{Priors}. 
\begin{table}
	\begin{center}
		\renewcommand{\arraystretch}{1.5}
		\begin{tabular}{c@{\hspace{0. cm}}@{\hspace{1.5 cm}} c}
			\hline
			\textbf{Parameter}    & \textbf{Prior} \\
			\hline\hline
			$\Omega_{\rm b} h^2$         & $[0.005\,,\,0.1]$ \\
			$\Omega_{\rm c} h^2$     	 & $[0.001\,,\,0.99]$\\
			$100\,\theta_{\rm {MC}}$     & $[0.5\,,\,10]$ \\
			$\tau$                       & $[0.01\,,\,0.8]$\\
			$\log(10^{10}A_{\rm S})$     & $[1.61\,,\,3.91]$ \\
			$n_{\rm s}$                  & $[0.8\,,\, 1.2]$ \\
			\hline
            $A_0$                        & $[-0.6\,,\,0.6]$\\
            $\epsilon$                 & $[-1\,,0]$\\
            $N_{\rm eff}$                & $[1\,,\,5]$\\
            $m^{\rm eff}_{\nu,\rm sterile}$ [eV]        & $[0\,,\,3]$\\
            $\Sigma m_\nu$ [eV]                & $[0\,,\,2]$\\
            \hline\hline
		\end{tabular}
		\caption[List of the parameter priors]{\small List of the parameter priors. The cutoff for $\epsilon$ is due to the fact that at $\epsilon=0$ we leave the quasi-single-field model and obtain the trispectrum for a multifield inflationary model. 
  }
		\label{tab:Priors}
	\end{center}
\end{table}
Then, for each model, we perform MCMC analyses using the publicly available package \texttt{Cobaya}~\cite{Torrado:2020dgo} and computing the theory with our modified version of \texttt{CAMB}~\cite{Lewis:1999bs,2012JCAP} according to \eq{CAMBRelation}. We explore the posteriors of our parameter space using the MCMC sampler developed for \texttt{CosmoMC}~\cite{Lewis:2002ah,Lewis:2013hha} and tailored for parameter spaces with a speed hierarchy which also implements the "fast dragging" procedure~\cite{neal2005taking}. The convergence of the chains obtained with this procedure is tested using the Gelman-Rubin criterion~\cite{Gelman:1992zz}, and we choose as a threshold for chain convergence $R-1\lesssim 0.02$.  The likelihoods we used are the following:
\begin{itemize}
    \item CMB temperature and polarization power spectra from the legacy Planck release~\cite{Planck:2018vyg,Planck:2019nip} with  \textit{plik}TTTEEE+lowl+lowE, which we will call from now on, Planck.
    
    \item Lensing Planck 2018 likelihood~\cite{Planck:2018lb}, reconstructed from the measurements of the power spectrum of the lensing potential. We refer to this dataset as just Lensing.
    
    \item Baryon Acoustic Oscillations (BAO) measurements extracted from data from the 6dFGS~\cite{Beutler:2011hx}, SDSS MGS~\cite{Ross:2014qpa}, BOSS DR12~\cite{Alam:2016hwk} and eBOSS DR16~\cite{eBOSS:2020yzd} surveys. 
    We call this dataset BAO. 
    
    \item Pantheon sample which consists of $1048$ type Ia supernovae measurements spanning the redshift range $0.01<z<2.3$~\cite{Pan-STARRS1:2017jku}.
\end{itemize}

\subsection{Cosmological Bounds} 
\label{sec:resultsNG}

\begin{figure}[h!]
	\centering
	\includegraphics[width=0.9\textwidth]{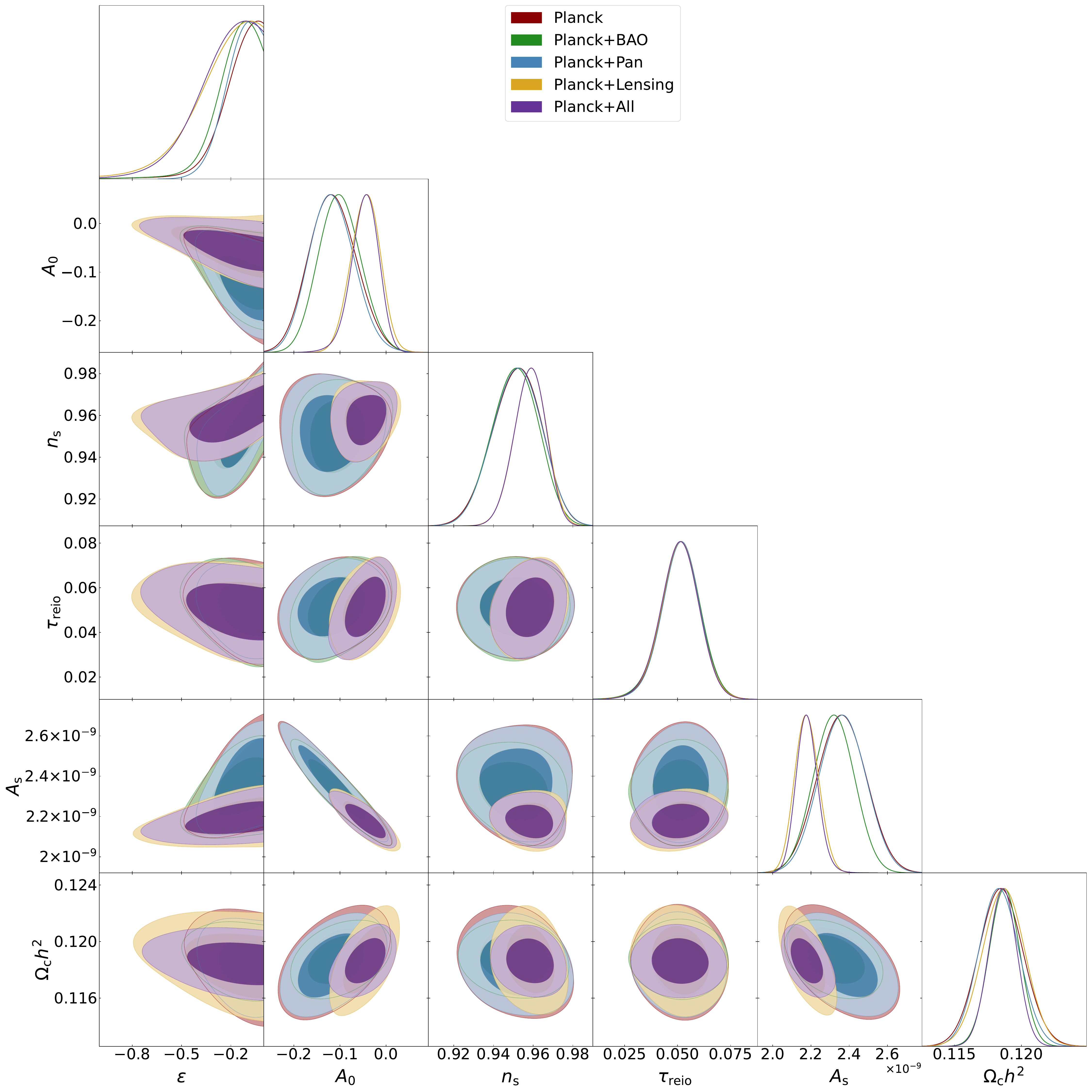}
	\caption{\small Marginalized 2D and 1D posterior distributions for the \slcdm.}
	\label{fig:lcdm}
\end{figure}

The full set  of tables and figures for all combinations of parameters can be found in~\cite{Forconi:4}. Here, we list only the full result of the constraints on \slcdm in \tab{lcdm} and \fig{lcdm} and for Super-$\Lambda$CDM$+\nnu+\mnu+\meff$ in \tab{nnumeffmnu} and \fig{nnumeffmnu}.
\begin{table}[h!]
	\begin{center}
		\renewcommand{\arraystretch}{1.5}
		\resizebox{0.9\textwidth}{!}{\begin{tabular}{c c c c  c  c}
  	        \hline
			\textbf{Parameter} & \textbf{Planck}  & \textbf{Planck+Lensing}  & \textbf{Planck+BAO} & \textbf{Planck+Pantheon} & \textbf{Planck+All} \\
			\hline\hline
			$\Omega_\mathrm{b} h^2$ & $0.02258\pm 0.00017$& $0.02250\pm 0.00016$& $0.02255\pm 0.00015$& $0.02259\pm 0.00017$& $0.02251\pm 0.00014$\\
			$\Omega_\mathrm{c} h^2$ & $0.1185\pm 0.0016$&$0.1187\pm 0.0015$& $0.1187\pm 0.0011$& $0.1183\pm 0.0014$& $0.11861\pm 0.00099$\\
			$100\theta_\mathrm{MC}$ & $1.04110\pm 0.00033$&$1.04105 \pm 0.00033$&$1.04107\pm 0.00029$&$1.04112\pm 0.00032$&$1.04106\pm 0.00029$\\
            $\tau_\mathrm{reio}$ & $0.0516\pm 0.0086$&$0.0512\pm 0.0085$&$0.0514\pm 0.0087$&$0.0515\pm 0.0084$&$0.0510\pm 0.0085$\\
			$\log(10^{10} A_\mathrm{s})$ & $3.160\pm 0.050$&$3.081\pm 0.027$&$3.143\pm 0.043$&$3.162\pm 0.048$&$3.081\pm 0.024$\\
            $n_\mathrm{s}$ & $0.952\pm 0.012$&$0.9586\pm 0.0076$&$0.951\pm 0.011$&$0.952\pm 0.011$&$0.9584\pm 0.0075$\\
			\hline	
            $H_0$ [km s$^{-1}$ Mpc$^{-1}$] & $68.13\pm 0.71$&$67.99\pm 0.68$&$68.01\pm 0.49$&$68.20\pm 0.65$&$68.01\pm 0.45$\\
            $\sigma_8$ & $0.849\pm 0.019$&$0.8190\pm 0.0085$&$0.843\pm 0.016$&$0.849\pm 0.018$&$0.8190^{+0.0084}_{-0.0095}$\\
            $S_8$ & $0.857\pm 0.020$&$ 0.828\pm 0.013$&$ 0.852\pm 0.017$&$ 0.856\pm 0.019$&$ 0.828\pm 0.011$\\
            $A_{\rm 0}$& $-0.116\pm 0.048$&$-0.045\pm 0.029$&$-0.101\pm 0.043$&$-0.118\pm 0.045$&$-0.046^{+0.031}_{-0.027}$\\
            $\epsilon$ &$> -0.369$&$> -0.605$&$> -0.384$&$> -0.325$&$> -0.573$\\
\hline\hline    
		\end{tabular}}
	\end{center}
	\caption[Results for \slcdm]{\small Results for \slcdm. The constraints on parameters are at $68\%$ CL, while upper bounds are at $95\%$ CL.}
	\label{tab:lcdm}
\end{table}
First of all, it should be stressed that, as pointed out in~\cite{Adhikari:2019fvb,Adhikari:2022moo}, the BAO and Lensing datasets have to be taken \textit{cum grano salis}\footnote{From Latin, it translates into "with a pinch of salt" and means that care is needed.} as theoretical prediction for the Super-$\Lambda$CDM model is still under development. However, using an agnostic approach, we include both in our exploration of the parameter spaces. On the other hand, supernovae data in the Pantheon measurements are not affected by the primordial NG and can be safely included in our discussion. The label Planck+All  refers to the combination of all the datasets (``All'' means ``Lensing+BAO+Pantheon''). We present the results quoting lower bounds at $95\%$ CL and constraints at $68\%$ CL, if not otherwise stated. For the sake of simplicity we refer to $\meff$ as $\mef$.
\subsubsection{Super-parameters}

In presenting the constraints on $A_0$, we should bear in mind that $A_0=0$ corresponds to the standard $\Lambda$CDM scenario with no contribution from NG at large scale. Therefore, any deviation of $A_0$ from its null value will go in favor of Super-$\Lambda$CDM. Polarization did not change much the constraints obtained in~\cite{Adhikari:2019fvb}. In fact, with only TT+$\tau$ prior (with the Planck 2015 release), $A_0=-0.15^{+0.14}_{-0.13}$ at \ncl~\cite{Adhikari:2019fvb} which increases when combined with distance ladder $H_0$ and SNe Ia Pantheon leading to  $A_0=-0.21^{+0.12}_{-0.13}$ at \ncl~\cite{Adhikari:2019fvb}. In our case, for Planck, we obtain $A_0=-0.12\pm0.10$ and for Planck+Pantheon, $A_0=-0.118^{+0.095}_{-0.092}$, both at \ncl. Despite the absolute value being lower than the previous analysis~\cite{Adhikari:2019fvb}, it is better constrained. On the other hand, if we allow the possibility of different relativistic species in the Universe, by allowing $\nnu$ to vary, the mean value shifts toward $0$. For example, with Planck only we have  $A_{\rm 0} = -0.109^{+0.095}_{-0.098}$ at \ncl. The highest absolute value is obtained when we promote as free parameters not only $\nnu$ but also $\mef$ and $\mnu$. In particular,  for Planck only, we get $A_{\rm 0} = -0.16 \pm 0.11$ at \ncl. 
Interestingly enough, combinations of Planck data with Lensing or BAO leads to the least solid evidence for NG. For example, Planck+Lensing predicts the most compatible value with zero for $A_0$, leading to 
$A_{\rm 0} = -0.040^{+0.055}_{-0.060}$
at \ncl with $\nnu$ as free parameter (i.e. for the scenario Super-$\Lambda$CDM$+\nnu$). The latter result hints that NG are negligible at less than $2\sigma$. Nonetheless, apart from the Lensing combination, our predictions for a negative non-zero $A_0$ lies within $\sim 2.5\sigma$. The lensing dataset affects greatly the Planck+All combination, and, therefore, $A_0$ is shifted toward zero, thus increasing the compatibility with $\Lambda$CDM with respect to other datasets. In \slcdm, $A_0$ negatively correlates with the primordial amplitude $A_s$ and there is a hint of correlation with $\Omega_c h^2$, as can be seen in the triangular plot in \fig{lcdm}. When considering extensions to the neutrino sector, the degeneracy with $A_s$ persists (and is even more pronounced, e.g., see \fig{nnumeffmnu}), while the correlation with $\Omega_c h^2$ weakens or it is completely absent (see, for example, \fig{nnumeffmnu}).
\begin{table}[h!]
	\begin{center}
		\renewcommand{\arraystretch}{1.5}
		\resizebox{0.9\textwidth}{!}{\begin{tabular}{c c c c  c  c}
  	        \hline
			\textbf{Parameter} & \textbf{Planck}  & \textbf{Planck+Lensing}  & \textbf{Planck+BAO} & \textbf{Planck+Pantheon} & \textbf{Planck+All} \\
			\hline\hline
$\Omega_\mathrm{b} h^2$&$0.02268\pm 0.00021$&$0.02255\pm 0.00019$&$0.02273\pm 0.00018$&$0.02273\pm 0.00019$&$0.02263\pm 0.00017$\\
$\Omega_\mathrm{c} h^2$&$ 0.1186^{+0.0038}_{-0.0032}$&$0.1191^{+0.0035}_{-0.0029}$&$0.1177^{+0.0035}_{-0.0030}$&$0.1180^{+0.0038}_{-0.0032}$&$0.1182^{+0.0030}_{-0.0025}$\\
$100\theta_\mathrm{MC} $&$ 1.04083\pm 0.00038$&$1.04081\pm 0.00037$&$1.04100\pm 0.00034$&$1.04094\pm 0.00036$&$1.04098\pm 0.00033$\\
$\tau_\mathrm{reio} $&$0.0507\pm 0.0084$&$0.0507\pm 0.0087$&$0.0514\pm 0.0088$&$0.0515\pm 0.0086$&$0.0515\pm 0.0085$\\
$\log(10^{10} A_\mathrm{s})$&$3.210\pm 0.059$&$3.117^{+0.039}_{-0.044}$&$3.199\pm 0.057$&$3.204\pm 0.056$&$3.109\pm 0.034$\\
$n_\mathrm{s}$&$ 0.952\pm 0.014$&$0.9583\pm 0.0099$&$ 0.955\pm 0.013$&$0.955\pm 0.014$&$0.9612\pm 0.0093$\\
\hline
$H_0$ [km s$^{-1}$ Mpc$^{-1}$] &$66.6^{+1.9}_{-1.7}$&$66.7\pm 1.5$&$68.18\pm 0.73$&$67.8\pm 1.2$&$68.18\pm 0.67$\\
$\sigma_8$&$ 0.787\pm 0.044$&$0.771\pm 0.030$&$0.822\pm 0.025$&$0.813\pm 0.035$&$0.800^{+0.020}_{-0.018}$\\
$S_8$ &$0.828\pm 0.029$&$ 0.809^{+0.022}_{-0.019}$&$ 0.835\pm 0.023$&$ 0.834\pm 0.028$&$ 0.812^{+0.019}_{-0.017}$\\
$A_{\rm 0} $&$-0.156\pm 0.052$&$-0.076\pm 0.041$&$-0.148\pm 0.051$&$ -0.151\pm 0.050$&$ -0.070\pm 0.034$\\
$\epsilon$&$ > -0.264$&$ > -0.457$&$ > -0.277$&$ > -0.268$&$> -0.431$\\
$N_{\rm eff}$ &$3.21^{+0.11}_{-0.15}$&$3.177^{+0.088}_{-0.13}$&$3.168^{+0.083}_{-0.12}$&$3.188^{+0.095}_{-0.14}$&$3.150^{+0.072}_{-0.11}$\\
$\Sigma m_\nu $ [eV]&$< 0.593$&$ < 0.483$&$ < 0.234$&$< 0.335$&$< 0.198$\\
$m_{\rm eff}$ [eV]&$ < 0.909$&$< 0.695$&$< 0.689$&$< 0.790$&$ < 0.550$\\
\hline\hline
		\end{tabular}}
	\end{center}
	\caption[Results for \slcdm+$N_{\rm eff}+m_{\rm eff}+\Sigma m_\nu$]{\small Results for \slcdm+$N_{\rm eff}+m_{\rm eff}+\Sigma m_\nu$. The constraints on parameters are at $68\%$ CL, while upper bounds are at $95\%$ CL.}
	\label{tab:nnumeffmnu}
\end{table}

Concerning the second Super-parameter $\epsilon$,
it has to be underlined that the usual trispectrum for multifield inflationary model is recovered for $\epsilon=0$. That is, we impose a cut-off at $0$ as indicated in \tab{Priors}. 
In our analysis, we did not find corroborating evidence for a non-zero value of $\epsilon$. Instead, we put lower bounds at \ncl which are generally more relaxed than the previous analysis, where it was found $\epsilon>-0.320$ (for TT+$\tau$ prior) and $\epsilon>-0.200$ when late-time measurements were included~\cite{Adhikari:2019fvb}. Highest absolute values are possible when the lensing data are included. In fact, for \slcdm+$\nnu$, we have relaxed bounds, such as $\epsilon>-0.692$ and $\epsilon>-0.388$ for Planck+Lensing and Planck+Pantheon, respectively. In all triangular plots, it is evident that Planck, Planck+Pantheon and Planck+BAO exhibit a certain correlation between $n_s$ and $\epsilon$, as we might expect due to the shift in the spectral index in \eq{CAMBRelation}. Conversely, the datasets, Planck+Lensing and Planck+All display a less pronounced degeneracy.

\subsubsection{Neutrino parameters}
Hints for additional massless particles are described by the parameter $N_{\rm eff}$ whose standard value is approximately $N_{\rm eff}\approx 3.044$. 
In our analysis, when we allow $N_{\rm eff}$ to vary, we see similar constraints as in the simple extension $\Lambda$CDM$+N_{\rm eff}$~\cite{Planck:2018vyg}
\begin{figure}[h!]
\centering
    \includegraphics[width=0.8\textwidth]{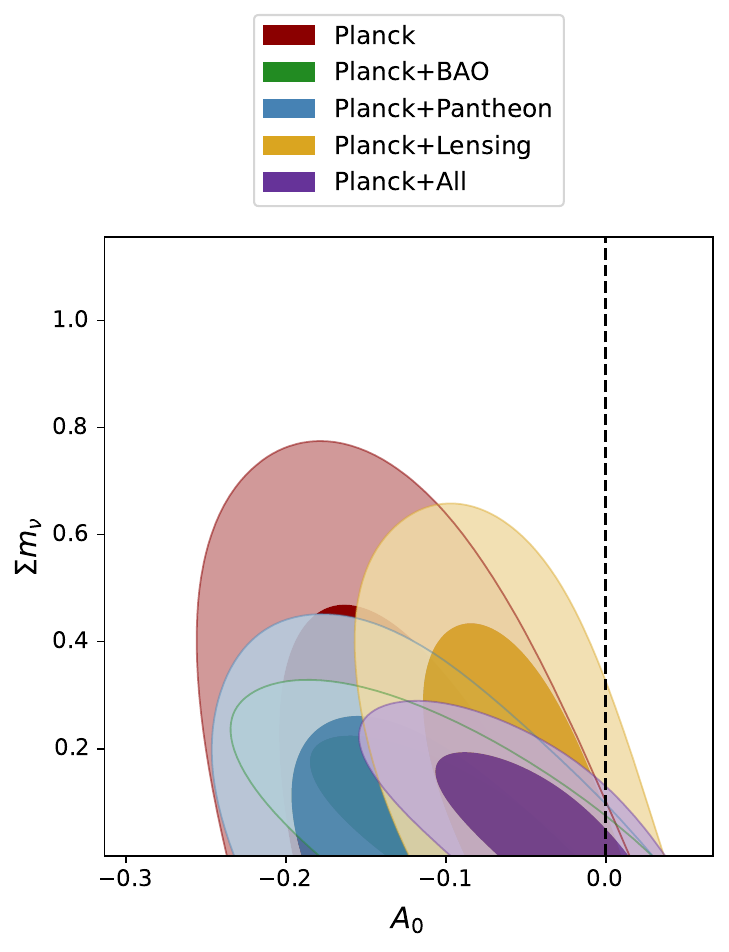}
  \caption{\small Marginalized 2D posterior distributions for the \slcdm+$\mnu$.}
  \label{fig:A0mnu}
\end{figure}
for Planck only, which is $N_{\rm eff}=2.92\pm0.19$ at \scl, whereas in \slcdm+$\nnu$, $N_{\rm eff}=2.93\pm0.21$ at \scl. Therefore,Planck alone does not alter the constraints on $N_{\rm eff}$. 
The addition of BAO induces a preference for smaller values of $N_{\rm eff}$ in the \slcdm scenario leading to $N_{\rm eff}=3.01\pm0.18$ (at \scl for $\Lambda$CDM$+N_{\rm eff}$) and $N_{\rm eff}=2.91\pm0.18$ (at \scl for \slcdm+$\nnu$). 
Thus, the standard value remains within the $1\sigma$ range but deviation from it of more than $20\%$ is allowed at $95\%$~CL. In a similar way, when the sum of neutrino masses $\mnu$ is let free to vary, analogous consideration can be drawn where for Planck alone,$N_{\rm eff}=2.91\pm0.19$ at \scl (for $\Lambda$CDM$+\nnu+\mnu$)~\cite{Planck:2018vyg}
and $N_{\rm eff}=2.94 \pm 0.22$ at 68\% CL for the \slcdm$+\nnu+\mnu$  model. 
Conversely, as soon as the possibility of a sterile neutrino is added to the model instead of $\mnu$, we obtain higher values for all datasets but 
the standard value still lies within the 95\% CL. Likewise, if we consider all three neutrino parameters in our model, i.e. for the model Super-$\Lambda$CDM$+\nnu+\mnu+\mef$, we have a preference for a higher effective number of relativistic species. In fact, when all the datasets are combined, we obtain $N_{\rm eff}=3.150^{+0.072}_{-0.11}$ at \scl (for Super-$\Lambda$CDM$+\nnu+\mnu+\mef$) while for Planck alone, $N_{\rm eff}=3.21^{+0.11}_{-0.15}$ at \scl (for Super-$\Lambda$CDM$+\nnu+\mnu+\mef$). Notably, these values are higher than the predictions of the $\Lambda$CDM model. For the same datasets (Planck+All and Planck), at \scl, we obtain $N_{\rm eff}=3.133^{+0.061}_{-0.099}$ and $N_{\rm eff}=3.156^{+0.074}_{-0.12}$ respectively. 

The possibility of ruling out the inverted mass hierarchy is far from being reached in this analysis. Nonetheless, it is worth highlighting that the constraints are considerably weaker compared to those of the $\Lambda$CDM model. It was the case for $N_{\rm eff}$ but it is more evident with the sum of neutrino masses. The reason for the weaker constraints is not only due to a volume effect, but a genuinely negative correlation between $A_0$ and $\Sigma m_\nu$, as we can see in the triangular plots and in \fig{A0mnu}. In other words, given that there is a slight indication for $A_0<0$ coming out from our analysis, this translates to more room for massive neutrinos. Fixing $A_0=0$ could, therefore, bias the constraints on the total neutrino mass on being too strong. For example, in \slcdm with Planck only and Planck+BAO we have $\Sigma m_\nu<0.624\ev$ and $\Sigma m_\nu<0.266\ev$ at 95\% CL, respectively, while the $\lcdm$ predictions are $\Sigma m_\nu<0.257\ev$ and $\mnu<0.126\ev$ at 95\% CL. Because of the absence of noticeable correlation with the possibility of massless relics (i.e. $\Delta N_{\rm eff}\neq 0$), when $N_{\rm eff}$ is added to the set, the constraints remain almost unchanged. If also the sterile neutrino is included in the picture, the upper bounds are slightly stronger as, for example, for Planck only we have $\Sigma m_\nu<0.593\ev$ at 95\% CL. These results preserve the weaknesses of the constraining power of \slcdm compared to $\lcdm$ (as for Planck only we have, $\Sigma m_\nu<0.352\ev$ at 95\% CL). Additionally, it is noteworthy that the constraints on the neutrino mass sum get more relaxed when the analysis is limited to early Universe physics, in contrast to scenarios confined to low-redshift cosmology. For example, in comparison with our result for \slcdm+$\mnu$ using only Planck data, in~\cite{Giare:2023aix} they found an increase of $~21\%$ in the constraints, when lensing, BAO and type Ia Supernovae data are combined, without relying on the CMB.

Similar results are obtained for $\mef$. In the scenario where massive sterile neutrinos are combined with the standard active neutrinos, we exclude the possibility of considering these neutrinos as candidates for cold dark matter particles. To be specific, we apply a cutoff at thermal masses greater than $10$~eV. Again, we obtain weaker constraints with respect to $\lcdm$~\cite{Planck:2018vyg}. For example, in $\lcdm+\nnu+\mef$, Planck alone gives $\mef<0.753\ev$ at 95\% CL and it increases to $\mef<0.893\ev$ at 95\% CL considering \slcdm. In \slcdm, the upper bound is stronger when Planck is combined with Pantheon, i.e. $\mef<0.801\ev$ at 95\% CL. If we add also the sum of neutrino masses, the bound is again stronger not only for Planck+Pantheon $\mef<0.790\ev$ at 95\% CL, but also for other datasets except for Planck alone, where the limit relaxes to $\mef<0.909\ev$ at 95\% CL. For comparison, Planck alone in $\Lambda$CDM$+\mef+\nnu+\mnu$ at \ncl gives $\mef<0.339\ev$.
This is an interesting observation in this context which dictates that the bounds on the neutrino mass obtained in \slcdm are relaxed compared to the $\Lambda$CDM paradigm. 
\begin{figure}[h!]
	\centering
	\includegraphics[width=0.9 \textwidth]{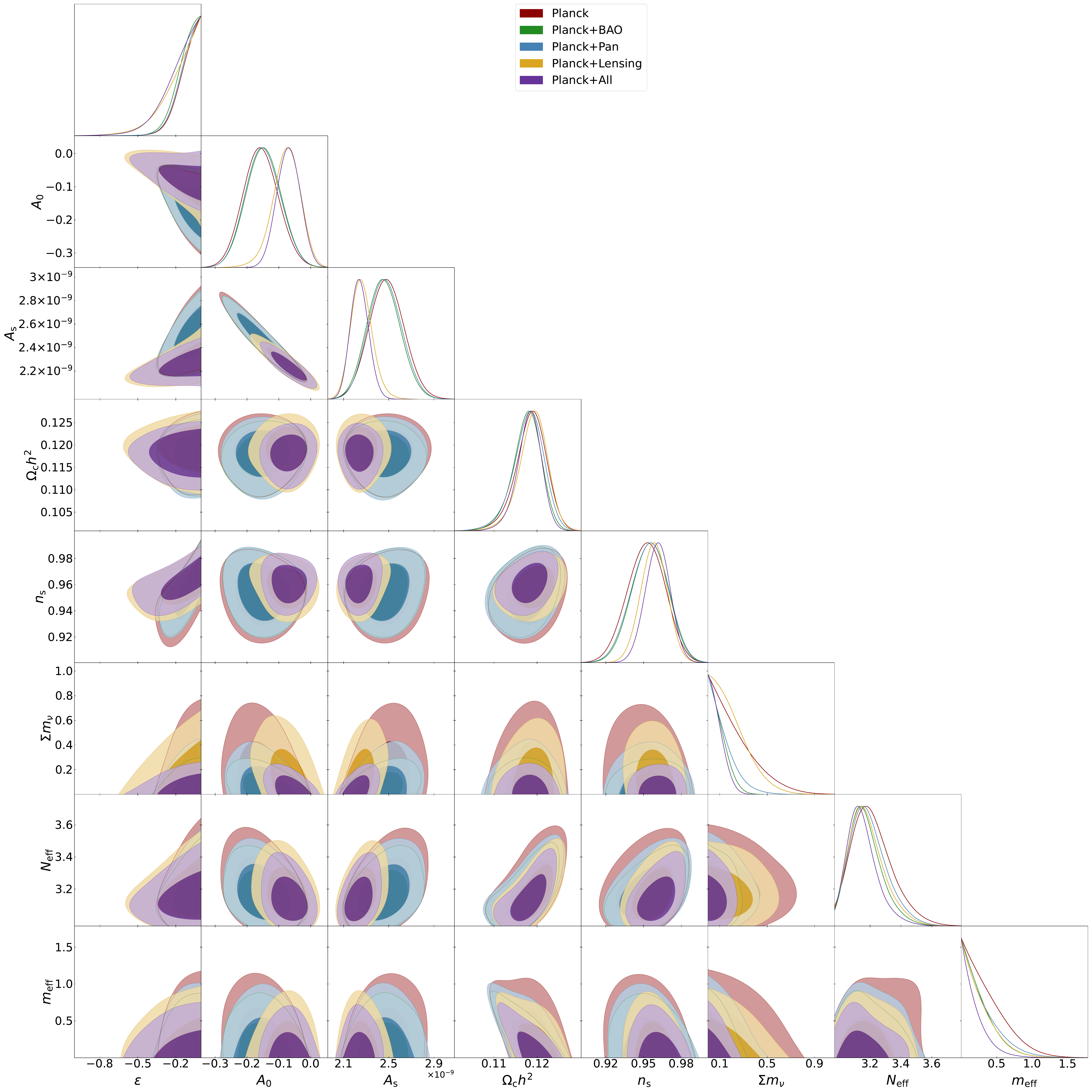}
	\caption{\small Marginalized 2D and 1D posterior distributions for the \slcdm$+ N_{\rm eff}+m_{\rm eff}+\Sigma m_\nu$.}
	\label{fig:nnumeffmnu}
\end{figure}

\subsubsection{$H_0$ and $S_8$}
Concerning the cosmological tensions, we can see that, adding the Super-parameter $A_0$,  the value of $H_0$  increases a bit with respect to the $\Lambda$CDM cosmology. In $\Lambda$CDM we have $H_0=67.27\pm 0.60 \,\mathrm{km}\, \mathrm{s}^{-1}\,\mathrm{Mpc}^{-1}$ at \scl that becomes $H_0=68.13\pm0.71 \, \mathrm{km}\, \,\mathrm{s}^{-1}\, \mathrm{Mpc}^{-1}$ at \scl in \slcdm.  
It is, however, smaller than the one found in the previous work with CMB temperature data only~\cite{Adhikari:2019fvb}. A similar increase with the inclusion of the Super-parameters is seen also within the extension of the neutrino phenomenology. For instance with $\nnu$ we have $H_0=66.4\pm1.4 \,\mathrm{km}\, \mathrm{s}^{-1}\,\mathrm{Mpc}^{-1}$ (at \scl) for $\lcdm+\nnu$  that changes to  $H_0=67.3 \pm 1.7 \mathrm{km}\, \mathrm{s}^{-1}\,\mathrm{Mpc}^{-1}$ (at \scl) in \slcdm +$\nnu$. 
Moving into \slcdm tends to ease the tension with late-time measurements, which, however, are still well beyond $3\sigma$. 

If we let the effective mass of sterile neutrinos free to vary, we see that $A_0$ increases and $H_0$ increases accordingly,  while the $S_8$ values decrease. For example, with Planck only data the \scl constraint on $H_0$ is $H_0=67.97 \pm 0.97\; \mathrm{km}\, \mathrm{s}^{-1}\,\mathrm{Mpc}^{-1}$. On the other hand, if we consider the sum of the neutrino masses, due to the decrease in its expectation values as well as its inverse proportionality with $H_0$, we obtain a lower value for $H_0$. For instance, we have $H_0=67.0^{+1.2}_{-0.97}\, \mathrm{km}\, \mathrm{s}^{-1}\,\mathrm{Mpc}^{-1}$ and $H_0=66.5^{+2.0}_{-1.7}\, \mathrm{km}\, \mathrm{s}^{-1}\,\mathrm{Mpc}^{-1} $  both at \scl for Planck only $\Lambda$CDM+$\mnu$ and \slcdm+$\mnu$ respectively.
If we promote the effective number of relativistic species as a parameter of the theory, i.e. \slcdm+$\mnu$+$\nnu$, with $A_0=-0.133\pm0.056$, $H_0$ decreases as much as (with Planck only) $H_0=65.8\pm2.2 \, \mathrm{km}\, \mathrm{s}^{-1}\,\mathrm{Mpc}^{-1}$ at \scl.
The latter value is greater than the one corresponding to $A_0=0$ (which at \ncl is $66.1^{+3.5}_{-3.6} \, \mathrm{km}\, \mathrm{s}^{-1}\,\mathrm{Mpc}^{-1}$.
When all three neutrino parameters are included in the analysis (i.e., for the scenario Super-$\Lambda$CDM$+\nnu+\mnu+\mef$), we get $H_0=68.18\pm 0.73 \, \mathrm{km}\, \mathrm{s}^{-1} \mathrm{Mpc}^{-1}$ at 68\% CL for Planck+BAO and $H_0 = 67.8\pm1.2 \,\mathrm{km}\, \mathrm{s}^{-1} \mathrm{Mpc}^{-1}$ at 68\% CL for Planck+Pantheon. The corresponding values for the same model with null $A_0$, at \scl, are $H_0=68.0^{+0.67}_{-0.81}\, \mathrm{km}\, \mathrm{s}^{-1} \mathrm{Mpc}^{-1}$ and $H_0 = 67.53\pm0.97 \,\mathrm{km}\, \mathrm{s}^{-1} \mathrm{Mpc}^{-1}$.
Thus, looking at the estimated values of $H_0$ in \slcdm and its various extensions where the maximum mean value of $H_0$ is $\sim 68.33\; \mathrm{km}\, \mathrm{s}^{-1} \mathrm{Mpc}^{-1}$  (obtained in Super-$\Lambda$CDM$+\nnu+\mef$) with uncertainties less than $1 \,\mathrm{km}\, \mathrm{s}^{-1} \mathrm{Mpc}^{-1}$, one can conclude that neither \slcdm, nor its extensions, are efficient in resolving the $H_0$ tension between Planck~\cite{Planck:2018vyg} and SH0ES~\cite{Riess:2021jrx}. On the other hand, focusing on the 
estimated values of $S_8$ in \slcdm and its various extensions, we see that  compared to the $\Lambda$CDM-based Planck's estimation ($S_8 =  0.834 \pm 0.016$)~\cite{Planck:2018vyg}, the values of $S_8$ do not make any dramatic changes except in the case for Planck+Lensing for which a mild reduction in the $S_8$ parameter is observed for Super-$\Lambda$CDM$+\nnu+\mef$ ($S_8 = 0.809^{+0.023}_{-0.020}$ at 68\% CL),  Super-$\Lambda$CDM$+\nnu+\mnu+\mef$ ($S_8 = 0.809^{+0.022}_{-0.019}$ at 68\% CL). However, for other datasets, $S_8$ takes $\Lambda$CDM-like values or even higher. 
Therefore, it is clear that easing tension on $S_8$ in these scenarios  does not seem promising at least according to the current level of sensitivity of the astronomical data, even if for a fair comparison we should analyze the weak lensing data only assuming the Super-$\Lambda$CDM model as well.  
\begin{figure}[h!]
    \centering
    \begin{subfigure}[b]{0.48\textwidth}
        \centering
        \includegraphics[width=\textwidth]{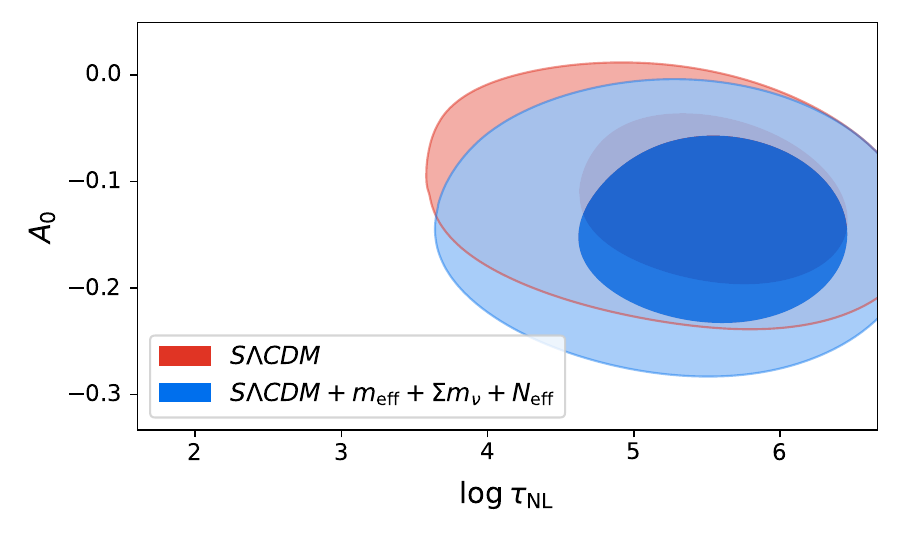}
        \label{fig:a0-vs-tau}
    \end{subfigure}
    \hfill
    \begin{subfigure}[b]{0.48\textwidth}
        \centering
        \includegraphics[width=\textwidth]{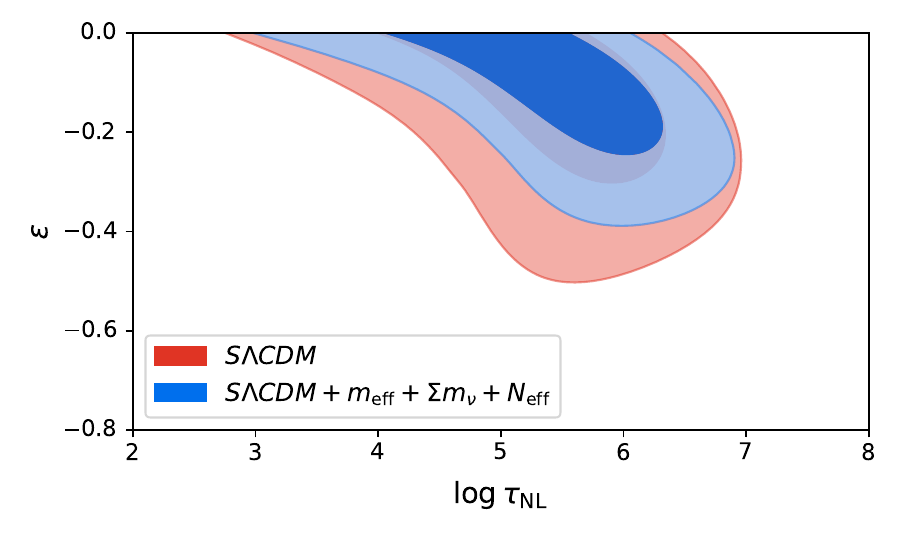}
        \label{fig:eps-vs-tau}
    \end{subfigure}
    \caption[$\tau_{\rm NL}$]{\small $\tau_{\rm NL}$ is weakly constrained as we set only a lower bound for $\epsilon$. Here, only the Planck dataset has been used.}
    \label{fig:taunl}
\end{figure}

\subsubsection{Trispectrum Constraint}

In this section, we study how our findings translate into constraints for $\tau_{\rm NL}$ and how the results change when we impose a conservative upper limit on the trispectrum. The trispectrum's shape for quasi-single-field inflation scenario occupies the \textit{intermediate}-shape ground, primarily due to its dependence on $\epsilon$. Our analysis establishes only a lower bound on $\epsilon$, resulting in relatively weak constraints on the amplitude of $\tau_{\rm NL}$. Specifically, using only the Planck likelihood, at \ncl, we find $log{\tau_{\rm NL}}=5.4^{+1.3}_{-1.5}$ for \slcdm and $log \tau_{\rm NL}=5.5^{+1.2}_{-1.4}$ for \slcdm+$m_{\rm eff}$+$\Sigma m_\nu$+$ N_{\rm eff}$. These results are illustrated in \fig{taunl}. 

\begin{figure}[h!]
\centering
\includegraphics[width=0.9\textwidth]{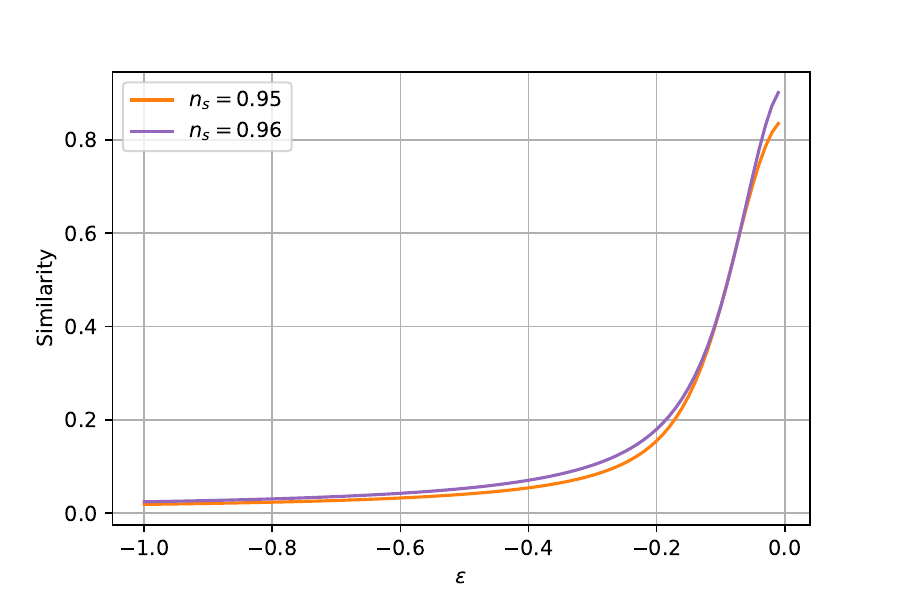}
  \caption[Cosine similarity]{\small  Cosine similarity between \eq{trispectrumspaghetti} and the standard template. When $\epsilon\sim0$ we have a near overlap: The \textit{intermediate} shape tends to the local one, where the bounds are applicable.}
  \label{fig:similarity}
\end{figure}

\begin{figure}[h!]
    \centering
    \begin{subfigure}[b]{0.48\textwidth}
        \centering
        \includegraphics[width=\textwidth]{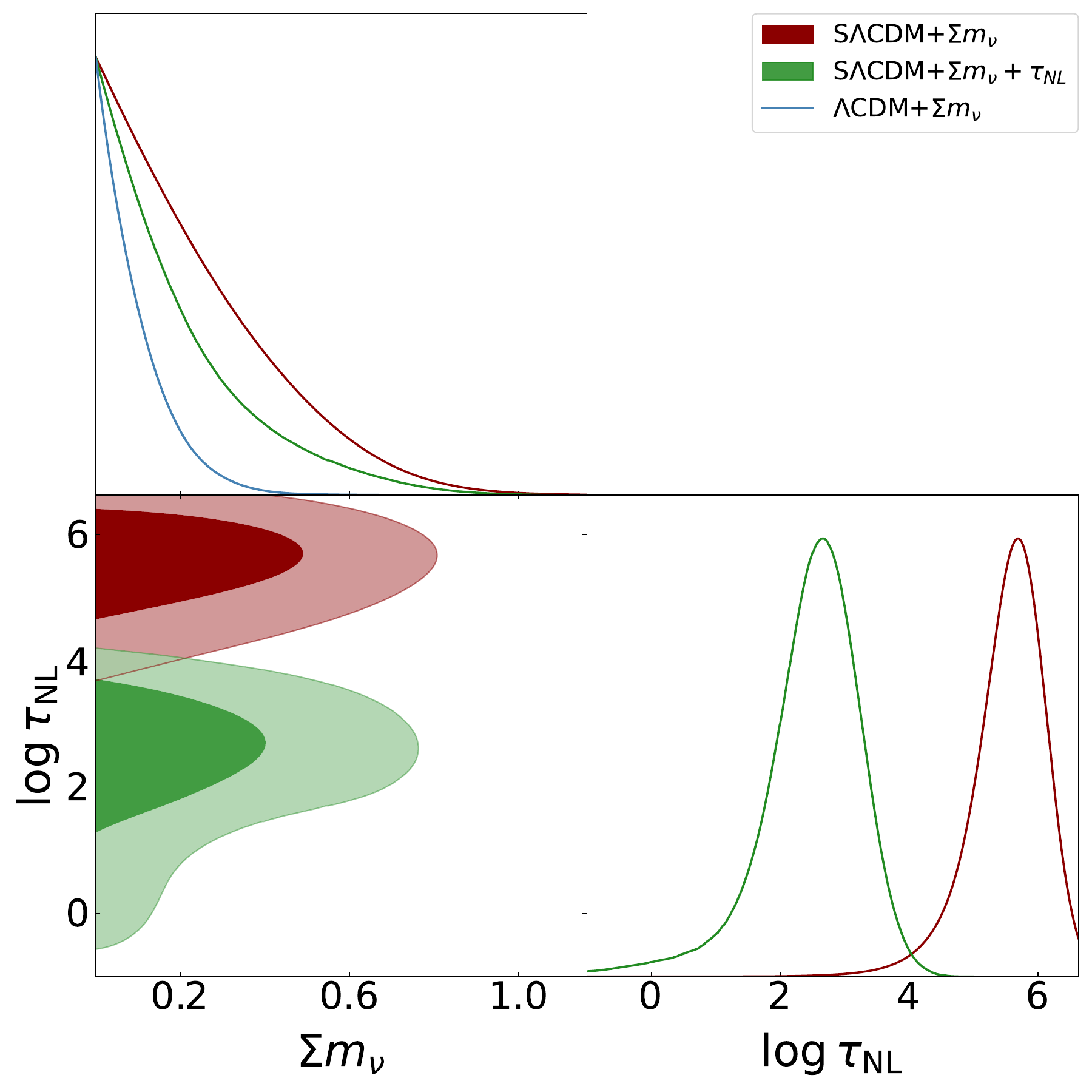}
        \caption{\small $\Sigma m_\nu$}
        \label{fig:1}
    \end{subfigure}
    \hfill
    \begin{subfigure}[b]{0.48\textwidth}
        \centering
        \includegraphics[width=\textwidth]{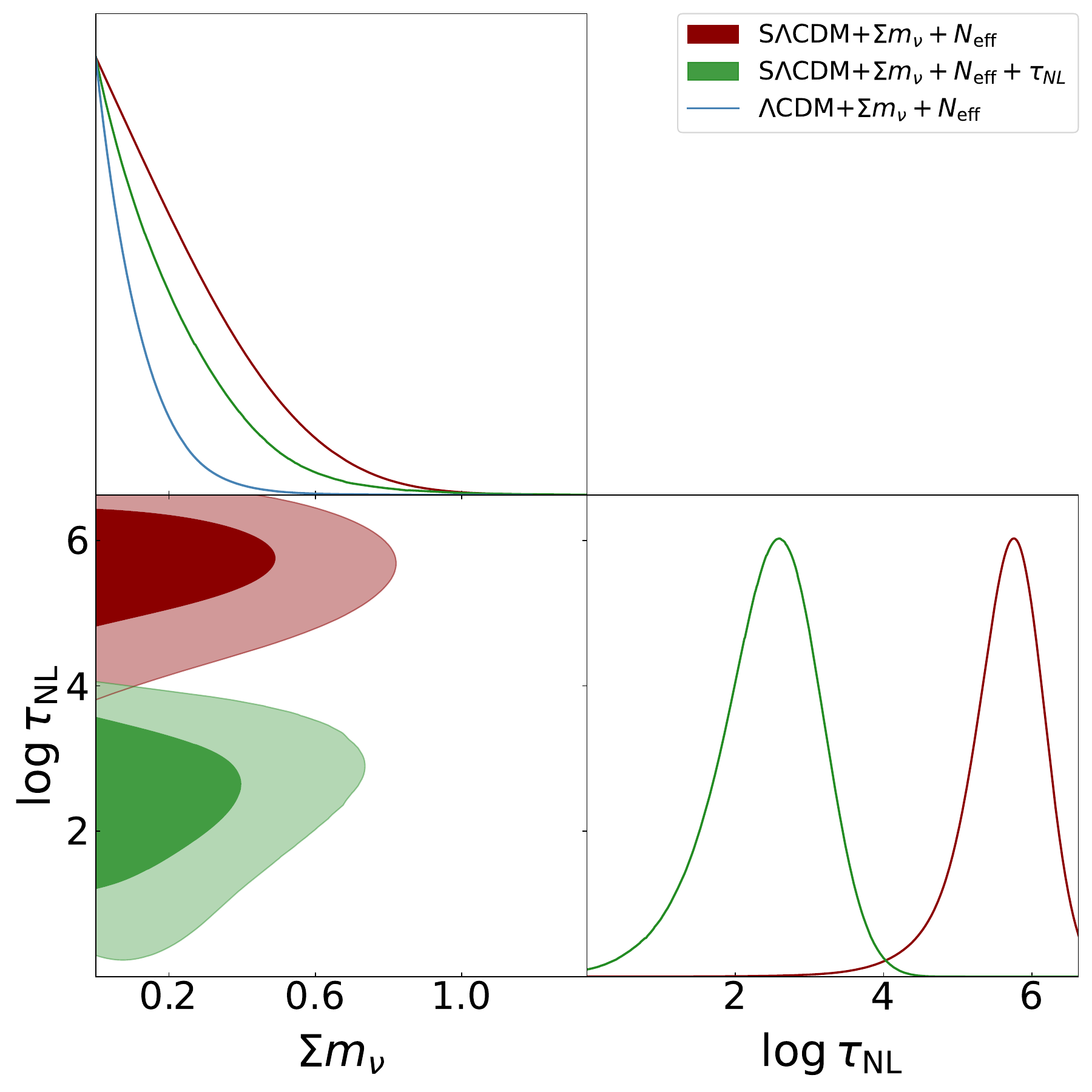}
        \caption{\small $\Sigma m_\nu+N_{\rm eff}$}
        \label{fig:2}
    \end{subfigure}
    \vspace{0.3cm}
    \begin{subfigure}[b]{0.66\textwidth}
        \centering
        \includegraphics[width=\textwidth]{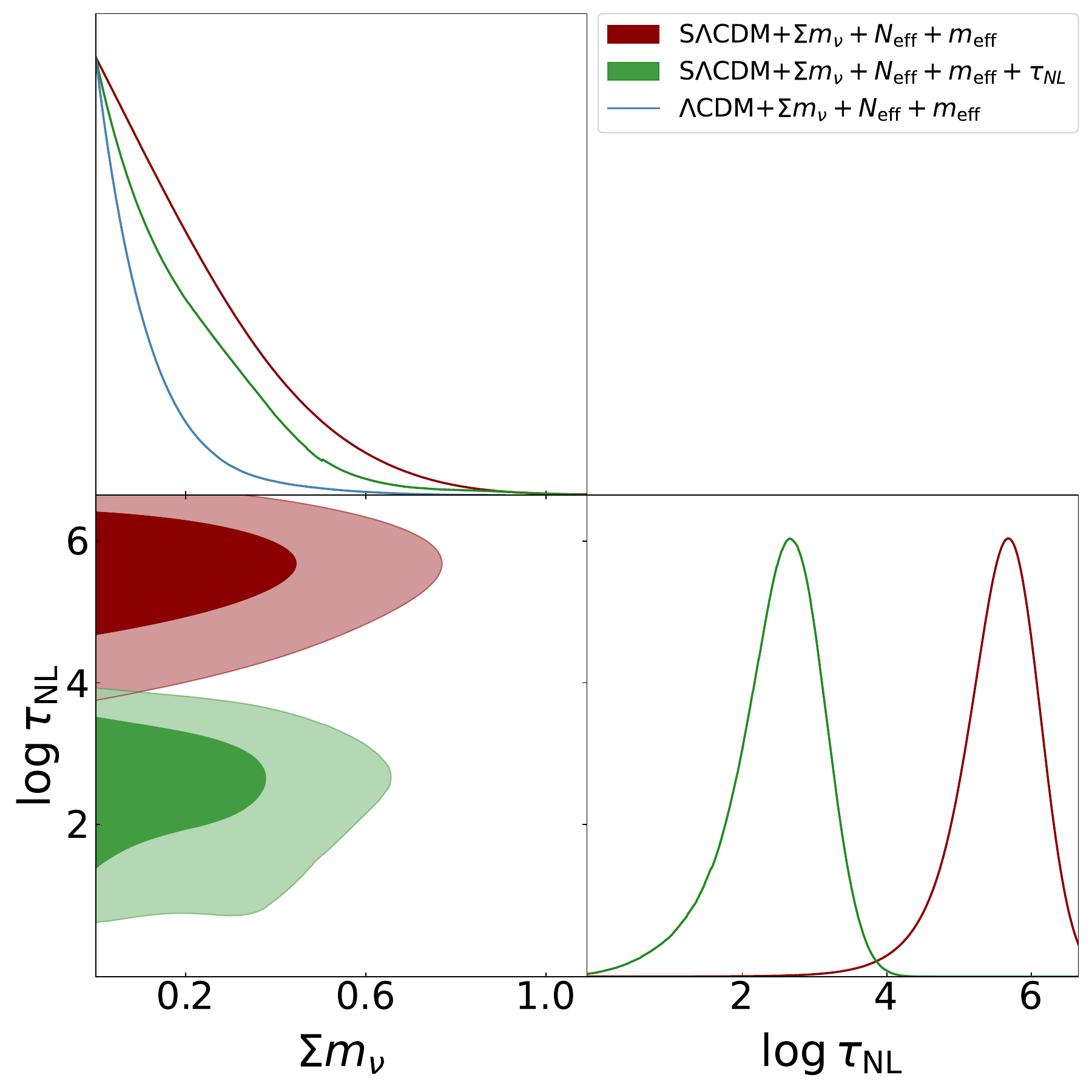}
        \caption{\small $\Sigma m_\nu+N_{\rm eff}+m_{\rm eff}$}
        \label{fig:3}
    \end{subfigure}
    \caption[Constraints on $\Sigma m_\nu$ including the prior on $\tau_{\rm NL}$]{\small Constraints on $\Sigma m_\nu$ for the Planck dataset. Comparing the $\Lambda$CDM case with Super$\Lambda$CDM (S$\Lambda$CDM), including ($+\tau_{NL}$) and not including the prior on the local trispectrum~\cite{Marzouk:2022utf}, we can see that, despite being tighter (from left to right, $~8\%$, $~15\%$, and $~18\%$ more constraining) when we apply the trispectrum bound, the constraints on the sum of neutrino mass are still more relaxed than the case for $A_0=0$.}
    \label{fig:mnu_tauprior}
\end{figure}

Previous studies~\cite{Marzouk:2022utf} have established a constraint on $\tau_{\mathrm{NL}}$, setting a limit of $\tau_{NL}<1700$ \ncl, which is 2 orders of magnitude smaller than our central value but consistent with models inside our \ncl contour. To check if these bounds apply to our model, we compute the similarity between the trispectrum shape for our quasi-single field inflation (QSFI) in \eq{trispectrumspaghetti} and the standard template (ST)
\begin{equation}
    \vec{T}_{\rm ST}(k_1,k_2,k_3,k_4)=\tau_{\rm NL}P_\mathcal{R}(k_1)P_\mathcal{R}(k_2)P_\mathcal{R}(k_3)P_\mathcal{R}(k_4)
\end{equation}
The trispectrum is calculated using a grid of $k\in[10^{-4};1]$ and fixing the primordial parameters. Specifically, the similarity corresponds to the value of the cosine:
\begin{equation}
    \text{Similarity}(\epsilon) = \frac{\vec{T}_{\rm QSFI} \cdot \vec{T}_{\rm ST}}{\|\vec{T}_{\rm QSFI}\| \|\vec{T}_{\rm ST}\|}\,.
\end{equation}
It strongly depends on $\epsilon$, as shown in \fig{similarity}. For $\epsilon\sim0$, we recover a good alignment with the standard form. However, for values within our lower bounds, we cannot exclude cases of low similarity, where the constraints are not relevant.
As qualitatively demonstrated in Fig. 1 in~\cite{Adhikari:2019fvb}, and confirmed in \fig{taunl},  when $\epsilon$ approaches zero (allowing the parameter $a$, as defined in \eq{CSWL}, to approach nearly zero), the current constraint on $\tau_{\mathrm{NL}}$ can still produce a variance consistent with our results. Additionally, the triangular plots show that the constraints on the sum of neutrino masses are less stringent for smaller $\epsilon$. Therefore, it is interesting to examine what happens to our conclusion on the possible bias on $\Sigma m_\nu$ when tight constraints on $\tau_{\rm NL}$ are applied. Motivated by this consideration, we assumed a similarity of approximately $1$, indicating a nearly perfect alignment with the local shape, making the constraints relevant. Imposing, therefore, an external prior on $\tau_{\rm NL}<1700$ at \ncl we have re-evaluated our constraints. Studying all cases where $\Sigma m_\nu$ is a free parameter, we see that, although the upper bound for $\Sigma m_\nu$ is lower, it is still not as stringent as in the scenario where NG are not considered. For example, if we let only $\Sigma m_\nu$ vary, the bound  $\Sigma m_\nu<0.617$~eV shifts to $\Sigma m_\nu<0.569$~eV as we apply the constraints on the trispectrum. These results can be seen in \fig{mnu_tauprior} and are listed in \tab{mnu_tau}. We can conclude that the intriguing possibility of a bias in constraining the total neutrino mass persists even when we assume a tight prior on the trispectrum.
\begin{table}[htp]
    \centering
    \begin{tabular}{lccc}
        \toprule
        Model & Super-$\Lambda$CDM & Super-$\Lambda$CDM$+\tau_{\rm NL}$ & $\Lambda$CDM \\
        \midrule
        $\Sigma m_\nu$ & $< 0.617$\,eV & $< 0.569$\,eV & $< 0.257$\,eV \\
        $\Sigma m_\nu + N_{\rm eff}$ & $< 0.627$\,eV & $< 0.539$\,eV & $< 0.305$\,eV \\
        $\Sigma m_\nu + N_{\rm eff} + m_{\rm eff}$ & $< 0.589$\,eV & $< 0.478$\,eV & $< 0.352$\,eV \\
        \bottomrule
    \end{tabular}
        \caption[Constraints on $\Sigma m_\nu$]{\small Constraints on $\Sigma m_\nu$ [eV] at \ncl for $\Lambda$CDM, \slcdm with the $\tau_{\rm NL}$ prior ($+\tau_{\rm NL}$) and without. We can see that the constraints of neutrino mass, when the similarity is assumed to be nearly one, are still more relaxed than the case where we neglect NG. These values are obtained using Planck dataset alone.}
        \label{tab:mnu_tau}
\end{table}
We emphasize that a direct and precise analysis of the trispectrum is beyond the scope of this work. However, the flexibility in the model permits exploration within the bounds of a minimal $\epsilon$ value. Similarly, constraints on the trispectrum's contribution~\cite{Agarwal:2013qta,Green:2023uyz} to scale-dependent bias~\cite{Baumann:2012bc} can be bypassed by exploiting the condition where $a$ approaches zero. Since we have not imposed strict constraints on $\epsilon$, we plan to address this aspect in more detail in future work. Finally, it is important to note that, although existing bounds on nonlocal shapes~\cite{Smith:2015uia} exist, they do not directly apply to our specific model framework.

%% file: Chapters/NonLinear.tex
\chapter{Massive Galaxies at High Redshift}
\label{ch:Chahahah}
It is extremely important to test the $\lcdm$ paradigm at different epochs and scales. One of the main challenges in testing the $\Lambda$CDM model was the lack of observations for objects at high redshift ($z\sim 10$). Current mainstream probes such as BAO and SNe Ia are unable to provide direct observations at these high redshifts, making it challenging to study the accuracy of the $\Lambda$CDM model around these epochs. This is a significant issue since it is during the early stages of the Universe's evolution that the main structures of the Universe are formed.

Recently, the James Webb Space Telescope (JWST) has observed galaxies at high redshifts, offering a glimpse into these \textit{dark ages}. These observations have revealed a population of surprisingly massive galaxy candidates (see e.g.~\cite{Santini_GLASS_Mstar,Castellano_GLASS_hiz,Finkelstein_CEERS_I1,Naidu_UVLF_2022,Treu:2022iti,Harikane_UVLFs,PerezGonzalez_CEERS, Papovich_CEERS_IV}) with stellar masses of the order of $M \ge 10^{10.5} M_{\odot}$. In a recent study \cite{2023Natu} it has been pointed out that the JWST data indicates a higher cumulative stellar mass density (CSMD) in the redshift range $7 < z < 11$ than predicted by the $\Lambda$CDM model, leading to several speculations that question the validity of the former~\cite{Boylan-Kolchin:2022kae,Biagetti:2022ode,Bouwens:2022gqg,xiao2023massive,2023arXiv231015284V,2023arXiv231107483W,2024ApJ...961...37C,Desprez:2023pif,Qin:2023rtf}. Nevertheless, the origin of the discrepancy could rely on other factors. One possibility to consider is that there may be inaccuracies in measuring the properties of galaxies. In fact, the stellar properties of the considered samples of massive high-redshift galaxies are derived from fitting a template of spectral energy distributions (SEDs) to the emissions in different photometric bands. The present lack of complete spectroscopic data raises the possibility of ambiguity in distinguishing between early star-forming galaxies and quiescent galaxies at lower redshifts, around $z \sim 5$. Additionally, the measured stellar masses strongly depend on the assumed Initial Mass Function (IMF). However, recent comparisons of photometric and spectroscopic redshifts in overlapping samples of galaxies with both measurements solidify the evidence for a high space density of bright galaxies at $z\gtrsim 8$ compared to theoretical model predictions~\cite{2023arXiv230405378A,2023ApJ...951L..22A}. More severe uncertainties affect the measurements of stellar mass, which are derived assuming a Salpeter IMF. While adopting other universal forms for the IMF based on low-redshift conditions would not change (or may even amplify) the masses derived from the current measurements, the star formation process can be significantly different at high redshifts, resulting in a top-heavy IMF. In particular, the increase in gas temperatures in star-forming, high-redshift galaxies (also contributed by the heating due to CMB photons) could lead to a greater contribution of massive stars to the galactic light, resulting in significantly lower values for the stellar masses (a factor of 3-10, with the exact value depending on the assumed gas temperature) compared to those measured by Labbe' et al. (2022) \cite{2022arXiv220807879S}. However, recent studies have, for the first time, exploited the unique depth and resolution properties of JWST to perform spatially resolved SED fitting of galaxies in the SMACS 0723 JWST ERO Field, using six NIRCam imaging bands spanning the wavelength range 0.8–5 $\mu$m \cite{Gimenez-Arteaga:2022ubw}. This approach unveiled the presence of older stellar populations that are only distinguishable in the spatially resolved maps, since in the observations with single aperture photometry they are overshadowed by the younger ($10 Myr$)  population. As a result, the more accurate estimates of stellar masses derived from the resolved analyses are significantly larger (up to a factor of 10) than those obtained from the single-aperture photometry previously employed in the analysis of high-redshift galaxies observed with JWST. This would further strengthen the tension with LCDM predictions. A second potential explanation is that the limited JWST observations thus far (covering an area of approximately $38$ square arcminutes) may be a highly atypical and unusually dense region of the Universe. This hypothesis can be tested through upcoming JWST surveys like COSMOS-Web.

If neither of the aforementioned possibilities can account for the discrepancies between the JWST results and the $\Lambda$CDM model, it may be necessary to consider modifications to the model itself. Alternatives models within the dark matter and dark energy sector have been tested (for example, see ~\cite{Menci:2022wia,Wang:2022jvx,Gong:2022qjx,Zhitnitsky:2023znn,Dayal:2023nwi,Maio:2022lzg,Domenech:2023afs}), as well as Primordial Black Holes solutions~\cite{Liu:2022bvr,Yuan:2023bvh,Hutsi:2022fzw}, Cosmic strings~\cite{Jiao:2023wcn}, and large scale-dependence Non-Gaussianity~\cite{Biagetti:2022ode}. But also solutions within the $\Lambda$CDM paradigm have been studied~\cite{Haslbauer:2022vnq}, such as modifications of the primordial power spectrum~\cite{Parashari:2023cui} or using Extreme Value Statistics~\cite{Lovell:2}. Unfortunately, none of the extensions considered until present are capable of providing a compelling solution to the JWST stellar mass density observations.

In \sect{PolJWST} we investigate a possible explanation, not explored
yet in the literature, i.e. the presence of possible, unknown, systematics in the Planck data. In \sect{JDE}, on the other hand, we assume no systematic issue behind these preliminary findings and we test whether these emerging anomalies could be somehow connected to other long-standing cosmological puzzles, such as the Hubble tension. All these issues might hint at a shared limitation in our current comprehension of the Universe, motivating the need to consider alternative theoretical scenarios. Before presenting the results, in \sect{NLC} a brief discussion on non-linear clustering is given, as well as important statistical functions are introduced, fundamental for the following JWST tension exploration

\section{Non-linear clustering}
\label{sec:NLC}
To better understand the formation of small-scale structures, a non-linear treatment is needed. However, it is not trivial because, for instance, we cannot treat each Fourier mode independently and natural NG arises. For this reason the non-linearized analysis is tackled using numerical simulations. Nevertheless, it is instructive to introduce high symmetries to the model and compute the evolution analytically~\cite {Gunn:1972sv} in the so-called \textit{spherical collapse}. Being the structure formation placed at late time, let us consider a flat MD universe whose average density is $\bar{\rho}(t)$. If we compress a region of radius $R_i$ within a sphere of radius $r_i$, conserving the mass, we have
\begin{equation}
\rho_i=\frac{\bar{\rho}_iR^3_i}{r_i^3}\equiv\bar{\rho}_i(1+\delta_i)
\end{equation}
with $\delta$ the density contrast $\delta\rho/\bar{\rho}$. Between $r_i$ and $R_i$ the is a gap with no matter. The degree of symmetries in the model allow to consider this shells as independently-evolving, up to the \textit{shell-crossing}. The mass shell evolves following Newtonian gravity and, according to the conservation of energy, we have
\begin{equation}
    \frac12\dot{r}^2-\frac{GM(r)}{r}=E.
    \label{eq:NIfog}
\end{equation}
During the evolution, the mass and energy are constant. If $r(t_i)\geq R_i$ then the $M(r)$ is the same prior the perturbation therefore nothing changes in the normal evolution. Inside the overdensity shell, we have $E<0$. The solution of \eq{NIfog} in parametric form is
\begin{equation}
    r(\theta)=\frac{GM}{2|E|}(1-\cos{\theta})=A(1-\cos{\theta}),\quad t(\theta)=\frac{GM}{(2|E|)^{3/2}}(\theta-\sin{\theta})=B(1-\cos{\theta})
    \label{eq:Hope}
\end{equation}
with $A^3=GMB^2$. We want to apply the solution \eq{Hope} to the edge of the over-dense region $(r(t_i)=r_i$. At early times ($\theta\ll 1$), $\cos\approx 1-\theta^2/2$ and $\sin \approx \theta -\theta^3/6$ therefore
\begin{equation}
    r(t)\approx\frac{A}{2}\(\frac{6}{B}\)^{\frac23}t^{\frac23},
\end{equation}
which evolves exactly as the background density in MD (see \tab{Table}). However, the overdensity will slow down the evolution until a turning point ($\theta_{\rm turn}=\pi$). Then it will reverse the motion and collpase in a time $\theta_{\rm col}=2\pi$. We know that the background density evolves like $t^{-2}$ and the perturbation density is simply the mass over volume. We can straightforwardly compute the ratio of these quantities, using \eq{Hope} and find
\begin{equation}
    \frac{\rho}{\bar{\rho}}=1+\delta=\frac92\frac{(\theta-\sin{\theta})^2}{(1-\cos{\theta})^3}\,.
\end{equation}
At the beginning, we are in the \textit{linear regime},  characterised by the (still) smallness of $\delta$. If we go to the second order of the Taylor expansion ($\theta-\sin{\theta}\approx\theta^3/6-\theta^5/120$ and $1-\cos{\theta}\approx \theta^2/2-\theta^4/24$. This gives 
\begin{equation}
    \delta\approx \frac{3}{20}\theta^2=\frac3{20}\(\frac{6}{B}\)^{2/3}t^{2/3}\equiv\delta_{\rm lin}(t)\propto a
    \label{eq:lin}
\end{equation}
that recovers the result found in \eq{mmdom}. However, when the contrast is
\begin{equation}
    \delta(\theta_{\rm turn})=\frac{9\pi^2}{16}-1\approx 4.55
    \label{eq:turnturn}
\end{equation}
we have the \textit{turn-around}. It is more than four times the value one would extrapolate maintaining a linear treatment $\delta_{\rm lin}(t_{\rm turn})\approx 1.06$~\cite{Baumann:2022mni}. Eventually, the region collapse $\delta(\theta_{\rm col}=2\pi)=\infty$. Being $t_{\rm col}=2t_{\rm turn}$, the linear extrapolation gives $\delta(t_{\rm col})=1.686$. An implication of this would be that, working within linear perturbation theory, when $\delta_c\equiv\delta_{\rm lin}\approx 1.686$, then it should interpreted as a complete collapse. 

\subsubsection{Halos}
The fact that $\delta(t_{\rm col})=\infty$ implies that the assumption of spherical collapse eventually breaks down. In fact, infinity density indicates the presence of black-holes. But our Universe is not dominated by black-holes. In fact, the matter, during collpase, eventually stops and find an equilibrium configuration which is called violent relaxation or \textit{virialisation} because it obeys the virial theorem. This is a \textit{dark matter halo}, in which the galaxies are embedded due to baryon clustering.

The virial theorem relates the average kinetic energy $T$ with the average potential $V$ by
\begin{equation}
    T=-\frac12V\,.
    \label{eq:vir}
\end{equation}

The kinetic energy of the perturbation is $\dot{r}^2/2$ and the potential $V=-GM/r$. At turn around $T=0$ and therefore 
\begin{equation}
    V_{\rm turn}=-\frac{GM}{r_{\rm turn}}\,.
\end{equation}
If the total energy $E$ is conserved, as e have assumed, using \eq{vir} we eventually obtain after virialisation 
\begin{equation}
    \frac12V_{\rm vir}=V_{\rm turn}\longrightarrow r_{\rm vir}=\frac12r_{\rm turn}
\end{equation}
which implies $\rho_{\rm vir}=8\rho_{\rm turn}$. If we take the virialisation time to coincide with collapse time, we now that $t_{\rm vir}=2t_{\rm turn}$ and the background energy has diluted by a factor 4, therefore
\begin{equation}
    \delta_{\rm vir}=32\frac{\rho_{\rm turn}}{\bar{\rho}_{\rm turn}}-1\,.
\end{equation}
Using \eq{turnturn}, we get $\delta_{\rm vir}\approx 177$.
\begin{tcolorbox}[mybox]
    It is interesting to see what happens if we add the contribution of a cosmological constant in our computations. The presence of $\Lambda$ modifies \eq{NIfog} 
    \begin{equation}
        \frac12\dot{r}^2-\frac{GM}{r}-\frac16\Lambda r^2=E\,.
        \label{eq:fogLambda}
    \end{equation}
    In order for the over-dense region to turn around and collapse, $\dot{r}$ must vanish. It can happen only if exist a solution $r(t)$ for
    \begin{equation}
        \frac16\Lambda r^3-|E|r+GM=0\,.
        \label{eq:labradoodle}
    \end{equation}
    If $r>0$, then \eq{labradoodle} has a solution only if $\sqrt{\Lambda}<(3B)^{-1}$ with $B$ defined in \eq{Hope}. Using \eq{lin}, then we can relate the bound on $\Lambda$ to the perturbation
    \begin{equation}
        \Lambda<0.01\frac{\delta^3}{t^2}\,.
    \end{equation}
    This implies that, if we want gravitational collapse to occur and galaxies to form then there is an upper bound on the cosmological constant, which depends on the strength of the initial perturbations. To quantify this for our Universe, we use $\delta\approx 10^{-3}$ at the time of last-scattering $t\approx 10^{13}s$. This gives $\Lambda<10^{-37}s^{-2}$ which implies $\rho_{\Lambda}<10^{10}\text{eV}m^{-3}$~\cite{Baumann:2022mni}.
\end{tcolorbox}

\subsubsection{Mass function}

The density contrast implicitly depend on $\textbf{x}$ which is a specific point in space. However, in general, to maintain a certain homogeneity in our Universe, a corse-grained description is more needed. To this end, we introduce a smoothing of the density field, that is, we remove contributions below a certain scale $R$. In other word, this filtered field $\delta_{\rm R}$ is the average density in a volume of size $R^3$. Whenever $\delta_{\rm R}>\delta_c$, we declare a halo of size $R$. The smoothing process is a convolution with a \textit{window function} $W(\textbf{x};R)$
\begin{equation}
    \delta_{\rm R}(\textbf{x})\equiv\delta(\textbf{x};R)=\int{d^3x'W(\textbf{x}-\textbf{x}';R)\delta(\textbf{x}')}\,.
\end{equation}
In Fourier space, we have
\begin{multline}
\delta(\textbf{k};R)=\int{d^3xd^3x'e^{i\textbf{k}\cdot\textbf{x}}W(\textbf{x}-\textbf{x}';R)\delta(\textbf{x}')}=\\
\int{d^3xd^3x'e^{i\textbf{k}\cdot(\textbf{x}-\textbf{x}')}W(\textbf{x}-\textbf{x}';R)e^{i\textbf{k}\cdot\textbf{x}'}\delta(\textbf{x}')}=W(\textbf{k};R)\delta(\textbf{k})\,.
\end{multline}
Some common choice for window functions are
\begin{itemize}
    \item The \textit{Spherical Top-Hat} 
    \begin{equation}
        W(\textbf{x},R)=\frac1V\times\left\{\begin{array}{cc}
            1 & |\textbf{x}|\leq R \\
            0 & |\textbf{x}|> R
        \end{array}\right.\,,\quad V=\frac{4\pi}{3}R^3\,.
    \end{equation}
    This window has a sharp cut-off in real space whereas in Fourier space
    \begin{equation}
        W(\textbf{k},R)=\frac{3}{(kR)^3}\[\sin{kR}-kR\cos{kR}\].
        \label{eq:THfu}
    \end{equation}
    \item The \textit{Sharp \textit{k} Filter} where the sharp cut-off takes place in Fourier space
    \begin{equation}
        W(kR)=\left\{\begin{array}{cc}
            1 & kR\leq 1 \\
            0 & kR>1
        \end{array}\right.
    \end{equation}
    where we used the propriety $W(\textbf{k};R)=W(kR)$. In real space, the window becomes
    \begin{equation}
        W(\textbf{x},R)=\frac{1}{2\pi^2r^3}\[\sin{r/R}-\frac{r}{R}\cos{r/R}\]
    \end{equation}
    with $|\textbf{x}|\equiv r$.
    \item The \textit{Gaussian} is described in real space as
    \begin{equation}
        W(r,R)=\frac{1}{(2\pi)^{3/2}R^3}e^{-\frac{r^2}{2R^2}}
    \end{equation}
    and
    \begin{equation}
        W(kR)=e^{-\frac{k^2R^2}{2}}\,.
    \end{equation} 
\end{itemize}
In general, as a mass $M=(4\pi/3)R^3\bar{\rho}$ with $\bar{\rho}$ the average density is associated to a radius $R$, $\delta_{\rm R}$ is often called $\delta_{\rm M}$.

Once we have the window function, we can compute the distribution of masses contained within a sphere of radius $R$. The average mass inside the sphere is
\begin{equation}
    \bar{M}(R)=\int{d^3xW(\textbf{x};R}\bar{\rho}(\textbf{x})=\frac{4\pi R^3\bar{\rho}}{3}\gamma,\quad\left\{\begin{array}{cc}
        1 & \text{Top-Hat} \\
        9\pi/2 & \text{k Filter}\\
        3\sqrt{\pi/2} & \text{Gaussian}
    \end{array}\right.
\end{equation}
The variance in the mass distribution is then 
\begin{equation}
    \sigma^2(M)=\langle \delta^2(\textbf{x}; R)\rangle=\int{d^3x'd^3x''W(\textbf{x}-\textbf{x}';R)W(\textbf{x}-\textbf{x}'';R)\langle(\delta(\textbf{x}')\delta(\textbf{x}'')\rangle}
\end{equation}
and, recalling the power spectrum $P(k)$ in \eq{xiR}, we are left with
\begin{equation}
    \sigma^2(M)=\frac1{2\pi^2}\int{dk k^2W(kR)P(k)}\,.
    \label{eq:VarianceM}
\end{equation}

Let us now compute the number of halos in a given mass range. The number of halos of mass \textbf{M} at a position \textbf{x} and time $t$ is $n_h(t,\textbf{x},M)$. If we assume that the smoothed density is a Gaussian random field, the probability that a region of space has an overdensity $\delta_M$ is
\begin{equation}
    \mathbb{P}(\delta_M)=\frac{1}{\sqrt{2\pi\sigma^2(M)}}\text{exp}\[-\frac12\frac{\delta_M^2}{\sigma^2(M)}\]. 
\end{equation}
Consequently, the probability for a region to exceed the density threshold $\delta_c$ is~\cite{Press:1973iz}
\begin{equation}
    \mathbb{P}(\delta_M>\delta_c)=\int_{\delta_c}^\infty d\delta_M\mathbb{P}(\delta_M)=\int_\nu^{\infty}dxe^{-x^2/2}=\frac12\text{erfc}\(\frac{\nu}{\sqrt{2}}\),
    \label{eq:PShechter}
\end{equation}
where $\nu(M)\equiv\delta_c/\sigma(M)$ is called \textit{peak height}. Being $\sigma$ a decreasing function of $M$, the small-scale fluctuations are the first to collapse giving the typical \textit{bottom up} type of structure formation. It should be noted that with \eq{PShechter} we are taking into account only overdense regions whereas also underdense region can become part of halos if enclosed in larger overdense region. Therefore we can multiply \eq{PShechter} by a factor $2$~\cite{Press:1973iz} that comes from the broader extension of \textit{excursion set theory}~\cite{Bond:1990iw}.

The probability that a halo formed in the range $\[M,M+dM\]$ is $-d\mathbb{P}/dM$. Then the \textit{mass function} which is the abundance of halos of mass $M$, is
\begin{equation}
    \boxed{\frac{d\bar{n}_h}{dM}=-\frac{\bar{\rho}}{M}\frac{d\mathbb{P}}{dM}=-f_{\rm PS}(\nu)\frac{\bar{\rho}}{M^2}\frac{d\ln{\sigma}}{d\ln{M}}}
    \label{eq:massfunc}
\end{equation}
with $\bar{n}_h$ the mean value of the number of halos, $\bar{\rho}/M$, instead is the maximum number density of halos of mass $M$ in a region of mass density $\bar{\rho}$. $f_{\rm PS}(\nu)$ is the \textit{halo multiplicity}~\cite{Press:1973iz} 
\begin{equation}
    f_{\rm PS}(\nu)=\sqrt{\frac{2}{\pi}}\nu\text{exp}\[-\frac{\nu^2}{2}\]\,.
    \label{eq:hakab}
\end{equation}
Qualitatively, \eq{massfunc} behaves as a power law if  $\nu$ is small while for large masses (large $\nu$) it has an exponential fall-off. \eq{massfunc} remarkably captures the correct shape and overall normalizaiton of the mass function. However, N-body simulations disagrees quantitatively with the predictions. Many more precise functions have been developed (see e.g.~\cite{Zentner:2006vw,Cooray:2002dia,Sheth:1999mn}). 

\section{JWST and Planck’s CMB polarization}
\label{sec:PolJWST}
 The Planck data is widely regarded as one of the most reliable datasets in cosmology today. However, the measurement of large-scale polarization, in particular, has proven to be extremely challenging, as evidenced by the various constraints on the optical depth reported in different data releases (see e.g. \cite{Planck:2016mks}). The optical depth $\tau$ is directly linked to the amplitude of CMB polarization on large angular scales. For instance, the value of this parameter ranged from $\tau=0.089^{+0.012}_{-0.014}$ in the 2013 data release \cite{Planck:2013pxb} to $\tau=0.054^{+0.081}_{-0.071}$ in the 2018 data release (both at the $68\%$ confidence level). This $2.5$-sigma shift clearly highlights the difficulties in accurately determining the large-scale CMB polarization. Furthermore, the $TE$ spectrum at multipoles $ \ell<30$ has not been utilized for the extraction of cosmological parameters due to the existence of systematic uncertainties. The $TE$ spectra indeed exhibit excess variance compared to simulations at low multipoles, particularly at $\ell = 5$ and $\ell = 18$ and $\ell = 19$, for reasons that are not yet understood. At the very same time, Planck polarization measurements at high angular scales have also been plagued by systematics, such as temperature-to-polarization leakage and uncertainties in polarization efficiencies. As clearly stated in~\cite{Planck:2018vyg}, one should avoid over-interpreting the Planck polarization results, considering the sensitivity of those to small changes in specific choices and assumptions made in the data analyses. It is also worth noting that several physical effects exist that could alter the amplitude and shape of CMB polarization spectra, such as magnetic fields, interactions with pseudoscalar fields (Chern-Simons coupling), and axion-like particles (see e.g. \cite{Harari:1992ea, Pogosian:2019jbt,Lue:1998mq,Fujita:2020aqt,Finelli:2008jv}). Moreover, the modelling of the reionization process can in principle also affect the constraints from Planck on $\tau$ (see e.g. \cite{Mortonson:2007tb}), even if this is claimed to be small in the analysis of~\cite{Planck:2018vyg}. Taking a more conservative approach, we demonstrate that by excluding the constraints from CMB polarization data, a higher value of the $\sigma_8$ parameter becomes more compatible with the Planck data, leading to better agreement with the JWST observations. It should be noted that this solution is not entirely satisfactory as it exacerbates the tension with cosmic shear surveys, which favour a lower value of the $S_8$ parameter. Nevertheless, we show that this approach significantly alleviates the Hubble tension instead, which is currently one of the most challenging issues in cosmology.

\subsection{Methodology}
\label{sec:CSMD}

Our main observable is the \textit{Cumulative Stellar Mass Density} (CSMD) given by
\begin{equation}
    \rho_\star(\bar{M})=\epsilon f_b\int^{z_2}_{z_1}\int^{\infty}_{\bar{M}}\frac{dn_h}{dM}MdM\frac{dV}{V(z_1,z_2)}~,
    \label{eq:CSMD}
\end{equation}
where $f_b=\Omega_b/\Omega_m$ is the cosmic baryon fraction, $\epsilon$ is the efficiency of converting baryons into stars, $V$ is the comoving volume of the Universe between redshift $z_2$ and $z_1$, given by $4/3\pi [R_2^3-R_1^3]$, $R_2$ 
and $R_1$ being the comoving radius at respective redshifts. The mass function $dn_h/dM$ is given in \eq{massfunc} where $\bar{\rho}$ is the comoving matter density of the Universe ($\rho_m = \Omega_m\rho_{crit}$) and $\rho_{crit}$ is critical density of Universe, $\rho_{crit}= 2.8\times10^{11} h^2 \msolar Mpc^{-3}$. In what follows, instead of using \eq{hakab} as halo multiplicity, we make use of the Sheth-Tormen alternative~\cite{Sheth:1999mn,Sheth:1999su}
\begin{equation}
    F(\nu)=\nu f_{\rm ST}(\nu)=A\sqrt{\frac{2a}{\pi}}\nu\left(1+\frac{1}{\bar{\nu}^p}\right)e^{-\frac{\bar{\nu}^2}{2}}\,,
    \label{eq:STMF}
\end{equation}
with $A=0.3222$, $a=0.707$, $p=0.6$. $\bar{\nu}=\sqrt{a}\nu$ and $\nu=\frac{\delta_c}{\sigma(R)}$. The parameter $\delta_c$ is the already encountered value of linear density contrast at the time of the collapse of non-linear density and $\sigma(R,z)$ is the variance of linear density field smoothed at scale R = (3M/ 4$\pi \bar{\rho})^{1/3}$ shown in \eq{VarianceM}. We assume here a Top-Hat window function \eq{THfu}. Though it has been confirmed by high-quality N-body simulations that several phenomenological fitting mass functions may work better than ST Mass function \cite{basilakos,bhattacharya}, we choose to work within the ST formalism, as it is theoretically motivated in terms of the collapse of halos~\cite{Maggiore_2010,Achitouv_2012}; additionally, it has been exhaustively tested by N-body simulations for different dark energy models taking different priors for $\Omega_m$ and $\Omega_{\Lambda}$~\cite{despali_2015}. Various studies confirm that the ST mass function works universally as a function of redshift ($z<10$) and cosmology with 20 percent expected error bounds~\cite{reed_2006,despali_2015}. Nevertheless, analyses of future  JWST data may need to consider a more sophisticated fit. 

\subsubsection{Dataset}
For the Planck data, we consider the CMB temperature and polarization power spectra from the final Planck release~\cite{Planck:2018vyg,Planck:2019nip} and we split it into four parts (see \fig{PlotAngular}). The full Planck Temperature and Polarization anisotropies power spectra with Planck TTTEEE+low$\ell$+low$E$ and without the large angular scale polarization (Planck TTTEEE+low$\ell$). The full Temperature anisotropies spectra (Planck TT+low$\ell$) and then with only the small scale anisotropy (Planck TT). For the case of JWST's CSMD, we consider separately two sets of datapoints in two redshift bins respectively as taken from~\cite{2023Natu}. For the redshift bin $7<z<8.5$ we consider:
\begin{itemize}
    \item $\log_{10}\rho^{\rm obs}_*(M_1)=5.90\pm0.35$ at 
 $log_{10}(M_1)=10.1$ 
 \item $\log_{10}\rho^{\rm obs}_\star(M_2)=5.70\pm0.65$ at $log_{10}(M_2)=10.8$ 
 \end{itemize}
while for the redshift bin $8.5<z<10$ we exploit the following data points:
\begin{itemize}
    \item $\log_{10}\rho^{\rm obs}_\star(M_1)=5.7\pm0.40$ at 
 $log_{10}(M_1)=9.7$ 
 \item $\log_{10}\rho^{\rm obs}_\star(M_2)=5.40\pm0.65$ at $log_{10}(M_2)=10.4$ 
 \end{itemize}
 
We assumed a Log Normal distribution for $\rho^{\rm obs}$, suggested by the symmetry of the lower and upper error bars under this choice. We also opt for this approach given the susceptibility of current measurements to substantial systematic errors and the exploratory character of our paper, which seeks to propose potential solutions rather than assert them. Notice that the observational data points are model dependent because the comoving volume for $\rho^{obs}$ has been computed assuming the best-fit $\Lambda$CDM model to Planck TTTEEE+low$E$+lensing CMB measurements ($h=0.6732$, $\Omega_m=0.3158$, $n_s=0.96605$, $\sigma_8=0.8120$, see ~\cite{2023Natu}). Therefore, to use these datapoints to analyze cosmologies different from the minimal $\Lambda$CDM  picture we need to rescale them properly by means of the comoving volume $V_C$ of the models under consideration. Given a cosmological model defined by set of cosmological parameters $\bar{p}$, we need to rescale the previous datapoints as 
\begin{equation}
\rho^{obs}(M_i,\bar p)=\frac{{V_C^{Planck}}{V_C(\bar p)}}\rho^{obs}_*(M_i)\,.
\end{equation}
A similar rescaling is also applied using the square of the luminosity distance, due to the usage of the flux in the derivations of both $\rho^{obs}$ and the masses $M_i$ and, as before, the underlying assumption of the best-fit Planck $\Lambda$CDM model.

For our analyses, we perform a MCMC using the publicly available package \texttt{Cobaya} ~\cite{Torrado:2020dgo} and computing the theorical predictions exploiting the latest version of the publicly available software \texttt{CAMB}~\cite{Lewis:1999bs,2012JCAP}. We explore the posteriors of our parameter space using the MCMC sampler developed for CosmoMC~\cite{Lewis:2002ah,Lewis:2013hha} and tailored for parameter spaces with a speed hierarchy which also implements the "fast dragging" procedure~\cite{neal2005taking}. The convergence of the chains obtained with this procedure is tested using the Gelman-Rubin criterion~\cite{Gelman:1992zz} and we choose as a threshold for chain convergence $R-1\lesssim 0.02$. Once the chains have converged, we compute the CMSD for each model, i.e. for each of the parameter combination explored by the MCMC analysis. In particular, we set the cosmology by varying the cosmological parameters $\{\Omega_b,\Omega_c,\theta_s,\tau\}$ and the inflationary parameters $\{n_s,A_s\}$. Afterwards, we use \texttt{CAMB} to compute the primordial power spectrum and subsequently the variance in \eq{VarianceM}. Then, we take its derivative and estimate the Halo mass function in \eq{massfunc}. Eventually, by double integrating over mass and redshift, \eq{CSMD}, we finally obtain $\rho_\star(M)$.

For each model in the chain, we include the JWST's CSMD data using  the (simple) $\chi^2$ function:
\begin{equation}
    \chi^2_{\textrm{JWST}}(\bar{p}) =  \sum^2_{i=1} \left[ \frac{
    \log_{10}{\rho^{\textrm{th}}(M_i,\bar{p})}- \log_{10}{\rho^{\textrm{obs}}(M_i,\bar{p})}}{\sigma_i} \right]^2~,
    \label{eq:chi2JWST}
\end{equation}
where $\rho^{th}$ is the CSMD presented in \eq{CSMD} for the different models explored and for an assumed valued of $\epsilon$, and $\sigma_i^2$ is the variance for each $\log_{10}\rho^{obs}(M_i,\bar{p})$ value \footnote{It is worth noting that one could assume free variation in $\epsilon$ and utilize the combined Planck+JWST data to impose constraints on this parameter. However, due to the potential existence of significant systematics, we opt for a more conservative approach. We choose to analyze only a few cases for this parameter, with a greater emphasis on understanding its variation rather than seeking stringent constraints on it.}. We then obtain the updated constraints on the cosmological parameters by re-weighting the MCMC chains, i.e. performing an importance sample, using the package \texttt{getdist}.

\subsection{Results}
\tab{Tablechi2} presents the main results from our analyses, which is the goodness of fit of the $\Lambda$CDM models preferred by the Planck experiment compared to those from a combined Planck+JWST analysis. We consider three possible values for the efficiency parameter $\epsilon$ (see \eq{CSMD}): $\epsilon=0.1$, in agreement with current local observations; $\epsilon=0.2$, in agreement with high-redshift simulations; and a larger value of $\epsilon=0.32$, which serves as a conservative upper limit~\cite{Tacchella:2018qny}.
\begin{table}[htb]                                
\begin{center} 
\begin{tabular}{|l|c|c|}                                              
\hline\hline                                                          
Dataset & $\Delta \chi^2$ & $\Delta \chi^2$ \\ 
& $7\le z \le 8.5$ & $8.5 \le z \le 10$\\ 
\hline
\hline
$\epsilon=0.1$&&\\
\hline
Planck TT +JWST& $20.14$& $25.71$\\
Planck TT+Low$\ell$+JWST&$21.61$&$28.11$\\
Planck TTTEEE+Low$\ell$+JWST&$26.30$&$33.03$\\
Planck TTTEEE+Low$\ell$+LowE+JWST&$42.20$&$52.71$\\
\hline
\hline
$\epsilon=0.2$&&\\
\hline
Planck TT +JWST& $5.69$& $7.49$\\
Planck TT+Low$\ell$+JWST&$5.49$&$7.65$\\
Planck TTTEEE+Low$\ell$+JWST&$6.42$&$10.30$\\
Planck TTTEEE+Low$\ell$+LowE+JWST&$11.87$&$17.76$\\
\hline
\hline
$\epsilon=0.32$&&\\
\hline
Planck TT +JWST& $2.11$& $2.53$\\
Planck TT+Low$\ell$+JWST&$2.34$&$3.06$\\
Planck TTTEEE+Low$\ell$+JWST&$2.17$&$2.88$\\
Planck TTTEEE+Low$\ell$+LowE+JWST&$3.68$&$6.19$\\
\hline
\end{tabular}
\caption[$\Delta \chi^2$ between the best fit model]{\small $\Delta \chi^2$ between the best fit model in the corresponding Planck and Planck+JWST chains.}\label{tab:Tablechi2}       
\end{center}                                                          
\end{table}                                                          
\paragraph{\textit{Local limit}} When $\epsilon=0.1$, there is a strong incompatibility between the Planck CMB angular spectra measurements (even temperature-only) and the CMSD derived from JWST observations. A $\Delta \chi^2 \sim 10.6$ for two degrees of freedom, corresponding to the inclusion of the two JWST data points, already exceeds the $99.5\%$ confidence level (C.L.), indicating a high level of incompatibility between the datasets. Assuming that both datasets and the theory are correct, this result strongly suggests a value of $\epsilon > 0.1$. This is not surprising, as simulations predict larger values for $\epsilon$ at higher redshifts~\cite{Tacchella:2018qny}.

\paragraph{\textit{High redshift limit}} On the other hand, when assuming a value of $\epsilon=0.2$, there is a much better compatibility with the Planck temperature data alone, with a $\Delta \chi ^2 \le 5.9$ below the $95\%$ CL exclusion threshold for two degrees of freedom. Although the comparison at higher redshifts leads to a poorer fit, it is still reasonable since $\Delta \chi ^2 \le 7.4$ corresponds to an exclusion below the $97.5\%$ CL In the case of the low-redshift bin data, the inclusion of polarization data at small angular scales slightly increases the exclusion to just over $95\%$ CL However, the addition of Planck's large angular scale polarization ($LowE$) significantly raises the $\Delta \chi^2$ and excludes any compatibility below the $99.5\%$ CL.
In conclusion, when $\epsilon=0.2$, a reasonable compatibility is achieved between the Planck temperature-only dataset and the CMSD derived from JWST. If this result is confirmed in the future, it could hint towards a systematic issue with the measurements of large-scale polarization data from Planck.

\paragraph{\textit{Conservative limit}} If we consider the case $\epsilon =0.32$, we observe that both datasets are relatively well-matched, with only the full Planck dataset exhibiting a tension above $95\%$ C.L. However, given the potential presence of various systematics in both datasets, this tension should not be regarded as a serious concern. This result may come as a surprise, as an efficiency of approximately $\epsilon=0.32$ is not inherently impossible, and, considering the errors associated with the CMSD, this finding is reasonable. In conclusion, if the large-scale polarization measured by Planck is accurate and the $\Lambda$CDM model is valid, the Planck vs JWST controversy can be resolved by a significant increase in the efficiency parameter $\epsilon>0.32$~\footnote{In~\cite{Boylan-Kolchin:2022kae} the effect of $\epsilon$ is also explored. However, a full MCMC analysis with different possible data combinations, as the one presented here, was missing in the literature.}.

\subsubsection{Cosmological parameters}
\begin{figure}[h!]
	\centering
	\includegraphics[width=0.47 \textwidth]{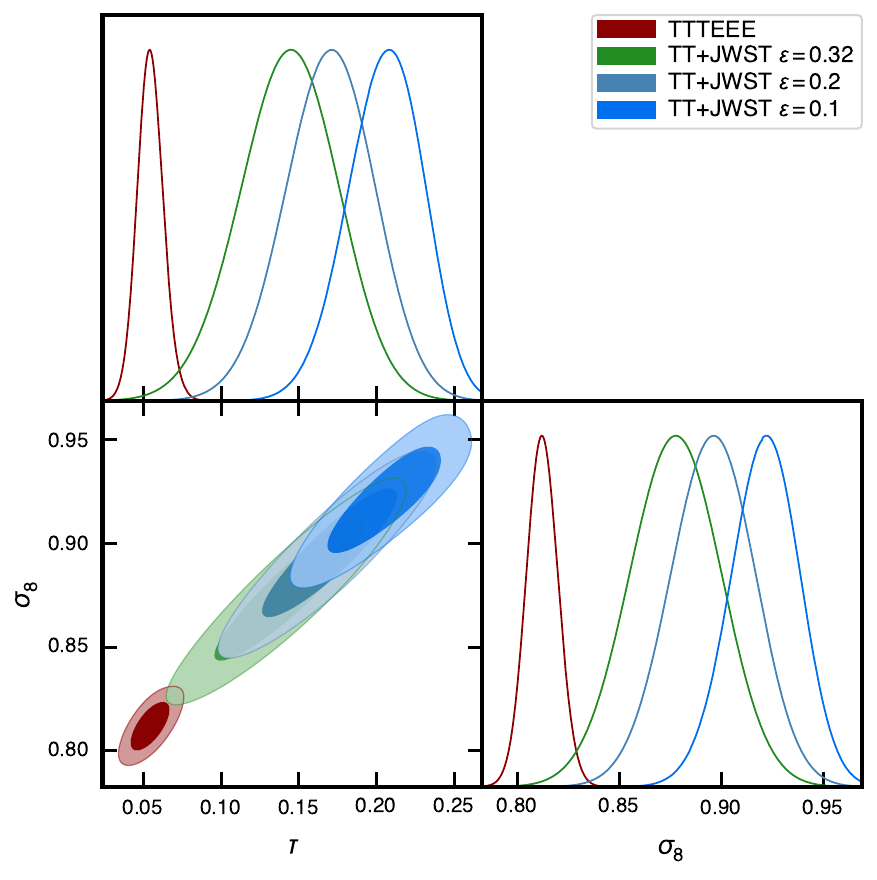}
 	\includegraphics[width=0.47 \textwidth]{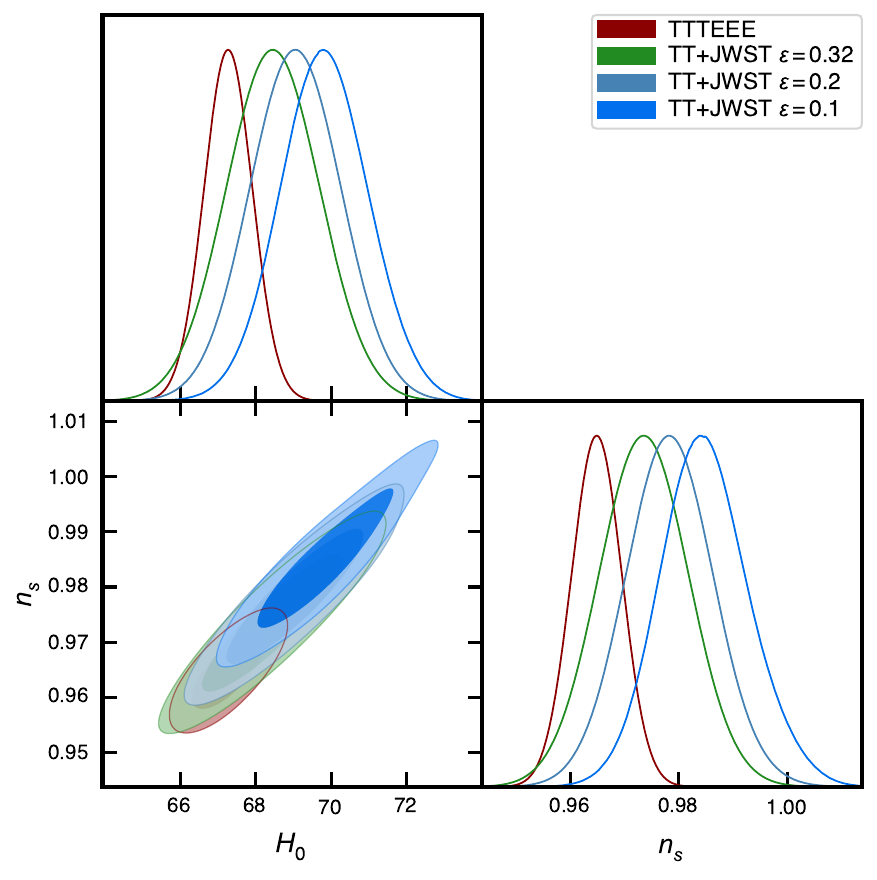}
	\caption[Marginalized 2D and 1D posterior distributions under the
 assumption of a $\Lambda$CDM]{\small Marginalized 2D and 1D posterior distributions under the
 assumption of a $\Lambda$CDM model for the full Planck dataset and Planck TT+JWST in the range  $8.5\le z \le 10$. Constraints are reported in terms of the $\sigma_8$ and $\tau$ parameters (Left Panel) and in terms of the $n_S$ and $H_0$ parameters.} 
	\label{fig:fig1Fo}
\end{figure}

Given the higher level of compatibility, it is interesting now to examine the constraints on cosmological parameters obtained from a combined analysis of Planck TT+JWST data and compare them with a standard analysis that utilizes the full Planck dataset (temperature and polarization). \fig{fig1Fo} illustrates the comparison between a combined analysis of Planck TT+JWST data to a standard analysis that utilizes the full Planck dataset (temperature and polarization)

\paragraph{\textit{Clustering power}} In the left panel of \fig{fig1Fo}, we observe that the combined Planck TT+JWST analysis shifts the values of the $\sigma_8$ parameter towards higher values. This shift is made possible by the fact that the optical depth $\tau$ remains unconstrained without the inclusion of large-scale polarization data from Planck.
The 2D marginalized plot demonstrates that the full Planck dataset and the Planck TT+JWST dataset exhibit significant tension when $\epsilon \le 0.2$ and marginal compatibility when $\epsilon =0.32$. This tension can be quantified further by examining the constraints on the $\sigma_8$ parameter. For $\epsilon=0.2$, the Planck TT+JWST analysis yields a constraint of $\sigma_8=0.895\pm0.019$ at $68 \%$ C.L., which deviates from the $\sigma_8=0.8120\pm0.0073$ constraint derived from the full Planck dataset by $4.7$ standard deviations (see also the posteriors for $\sigma_8$ in the left panel of \fig{fig1Fo}). Such high value of the $\sigma_8$ parameter is certainly in tension with current determinations from cosmic shear data that actually prefer an even lower value ( see e.g. \cite{DES:2021vln,DES:2022ygi,Heymans:2020gsg,KiDS:2020ghu}). Lyman-$\alpha$ forest data, hovewer, shows generally an higher value for $\sigma_8\sim0.9$ (see e.g. \cite{2022MNRAS.515..857E}). 

\paragraph{\textit{Hubble parameter}} Interestingly, as shown in the right panel of \fig{fig1Fo}, the Planck TT+JWST dataset also favors a higher value for the Hubble constant, with $H_0 = 68.8\pm 1.1$ in the range $7\le z \le8.5$ and
$H_0 = 69.0\pm 1.1$ in the range  $8.5\le z \le 10$ at $68 \%$ C.L. for $\epsilon=0.2$. The exclusion of the Planck large-scale CMB polarization not only potentially reconciles the Planck and JWST results but also brings them into agreement with current local measurements of the Hubble constant within a range of $3$ standard deviations. However, abandoning the $lowE$ polarization does not completely resolve the tension, but rather reduces it to a level that is more consistent with statistical fluctuations. It is evident that if there is a systematic issue in the $LowE$ polarization data, it can hinder the accurate identification of a theoretical solution to the Hubble tension.

\paragraph{\textit{Reionization}} It is also crucial to quantify the discrepancy in the optical depth $\tau$. Using the Planck TT+JWST data under the assumption of $\epsilon=0.2$ and considering the higher redshift bin, we obtained $\tau=0.17\pm0.027$, which is approximately $4$ standard deviations away from the full Planck result. This may seem highly significant, but a similar internal discrepancy between the TT and TTTEEE results is also observed in the case of $\Lambda$CDM, as seen in the more recent PR4 NPIPE analysis (\cite{Rosenberg:2022sdy}, see Table 1 in that paper). Therefore, a fluctuation of this magnitude is not entirely unexpected, considering the presence of similar internal discrepancies in the Planck data.

The future Litebird satellite \cite{LiteBIRD:2020khw,Sugai:2020pjw}, scheduled for launching around 2030, holds great promise for accurately measuring the large-scale CMB polarization. This will provide a crucial test of the Planck results and shed further light on the tension observed between the full Planck dataset and the Planck TT+JWST dataset. By obtaining precise measurements of the large-scale CMB polarization, Litebird has the potential to validate or challenge the exclusion of Planck's large-scale CMB polarization in the combined analysis. Additionally, Euclid satellite \cite{EUCLID:2011zbd}, is expected to provide valuable insights into the $S_8$ tension and the nature of dark matter clustering. If the Euclid mission confirms the $S_8$ tension observed between the full Planck dataset and the current cosmic shear data, it could also exclude the possibility of a dark matter clustering scenario where $\sigma_8 \sim 0.9$.

\section{JWST and dark energy sector}
\label{sec:JDE}
Although it is certainly premature to draw any definitive conclusions from these preliminary observations, if neither of the aforementioned possibilities can account for the discrepancies between the JWST results and the theoretical predictions of a baseline $\Lambda$CDM cosmology, it may be necessary to consider modifications to the model itself~\cite{Adil:2023ara,Menci:2022wia,Wang:2022jvx,Gong:2022qjx,Zhitnitsky:2023znn,Dayal:2023nwi,Maio:2022lzg,Domenech:2023afs,Liu:2022bvr,Yuan:2023bvh,Hutsi:2022fzw,Jiao:2023wcn,Biagetti:2022ode,Menci:2024rbq,Parashari:2023cui,Lovell:2,Haslbauer:2022vnq,Gandolfi:2022bcm,Lovyagin:2022kxl,Wang:2023ros,Hirano:2023auh,Paraskevas:2023itu} or to the galaxy formation process~\cite{Qin:2023rtf,Pallottini:2023yqg,Ferrara:2022dqw,Pacucci:2023oci,2023arXiv231105030M}. In this section, we take a step forward in this direction by testing alternative models where galaxy evolution could be notably different than in $\Lambda$CDM. In particular, we consider extensions related to the dark energy sector of the theory as possible phenomenological alternatives to explain the JWST preliminary findings. We test Early Dark Energy (EDE) and Interacting Dark Energy (IDE) cosmologies as both these extended scenarios, featuring modifications in the growth of structure, might predict a different evolution of perturbations, potentially resulting in the formation of more massive galaxies. We demonstrate that while EDE emerges as an excellent candidate to explain (at least partially) the unexpected JWST preference for larger stellar mass densities, IDE is generally disfavored by JWST measurements, despite yielding higher values of matter clustering parameters $\sigma_8$ and $S_8$.

\subsection{Extended Dark Energy Scenarios}
\label{sec:plumino}
Despite the undeniable uncertainties surrounding the preliminary findings from JWST, one might wonder whether these new emerging anomalies could be somehow linked to other well-known longstanding problems in cosmology, such as the Hubble tension. This raises the question of whether they could both originate from a common issue related to our current theoretical understanding of the Universe and, on a broader scale, what kind of beyond-$\Lambda$CDM phenomenology (if any) can increase the present-day expansion rate of the Universe while also leading to a higher cumulative stellar mass density at earlier times. At first glance, this question can even appear misplaced, as these observations are often believed to imply an older universe compared to the $\Lambda$CDM predictions. Since the age of the Universe is roughly $\propto 1/H_{0}$, this would suggest that increasing the Hubble constant could worsen the discrepancy with the observations released by JWST. However, structure formation is influenced by the evolution of primordial density perturbations and the underlying cosmology. Models addressing the Hubble tension often propose modifications at both the background level and in perturbation dynamics. This could allow for comparable or greater structure formation in a younger Universe (various N-body simulations of extension to $\Lambda$CDM show such variations, e.g.~\cite{Maio:2006zs,Baldi:2008ay,Baldi:2012ky,Barreira:2013xea,Maio:2014qwa,Adamek:2017uiq}). Furthermore, in beyond-$\Lambda$CDM models different correlations among cosmological parameters can shift their fitting values. These effects may significantly impact parameters related to structure formation, such as the matter density $\Omega_m$, and the other matter clustering parameters $\sigma_8$ and $S_8 = \sigma_8 \cdot (\Omega_m/0.3)^{1/2}$. These correlations are crucial as they could affect the amplitude and shape of the matter power spectrum.
\begin{figure}[h!]
\centering
\includegraphics[width=0.86\textwidth]{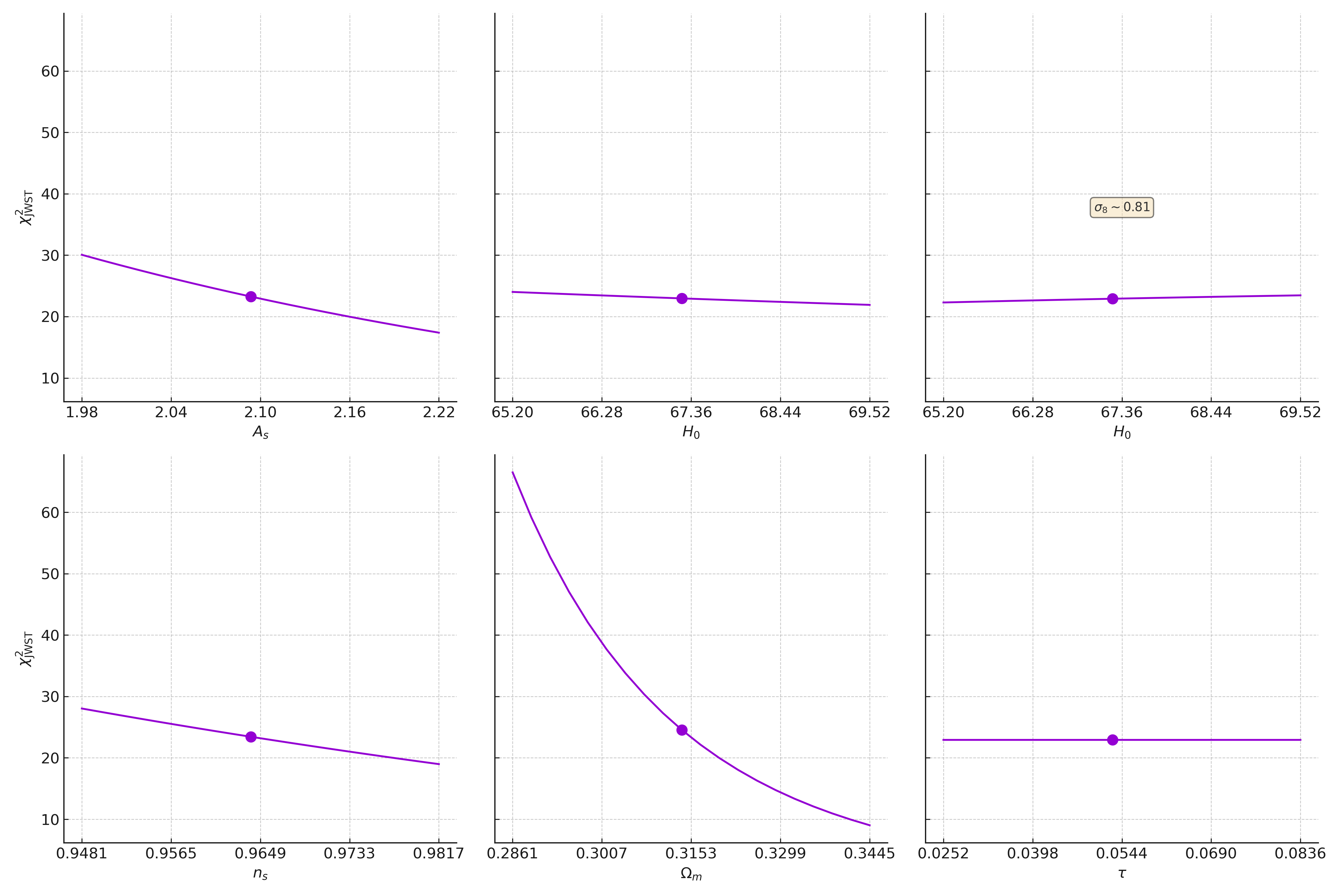}
\caption[Changes in $\chi^2_{\rm JWST}$ when varying a single parameter]{\small Changes in $\chi^2_{\rm JWST}$ when varying a single parameter while keeping the others fixed to the Planck $\Lambda$CDM best-fit values (bold points in the figure): $\Omega_bh^2=0.02238$, $\Omega_ch^2=0.1201$, $H_0=67.32$ km/s/Mpc, $n_s=0.9659$, $A_s\times10^9=2.1$, $\tau=0.0543$, and $\Omega_mh^2=0.143$. In the top panel's third plot, when $H_0$ is free to vary, $\sigma_8$ is kept fixed by rescaling $A_s$ accordingly.}
\label{fig:Chi2_params}
\end{figure}

\subsubsection{Correlation with Standard Parameters}
An exercise certainly useful for understanding which kind of phenomenology could hit two targets with one arrow -- increasing $H_0$ and aligning more closely with JWST --  is  breaking down the problem into smaller parts. In particular, we focus on the baseline $\Lambda$CDM model fixing all parameters to the best-fit values provided by Planck and altering each one individually within a 4-standard-deviation range. Through this analysis, focusing on the JWST likelihood ($\chi^2_{\rm JWST}$), we identify physical adjustments for better consistency with observations. From \fig{Chi2_params} we can derive a quite significant amount of information:

\begin{itemize}
    \item First and foremost, we observe that the parameter on which $\chi^2_{\rm JWST}$ is most sensitive is the matter density parameter $\Omega_m$. In particular, a larger fraction of matter in the Universe will considerably improve the quality of the fit to JWST observations by facilitating structure formation. \footnote{For similar discussions involving quasars at high redshifts, see, e.g.~\cite{Risaliti:2018reu,Yang:2019vgk}.}
    \item Secondly, we can clearly note that increasing the amplitude of the primordial perturbations $A_s$ or considering a larger tilt $n_s$ results in a significant reduction in $\chi^2_{\rm JWST}$. This is in line with previous findings in \sect{PolJWST}, where it was argued that relaxing the Planck constraints on polarization, which in turns allows $\tau$ to reach considerably higher values~\cite{Giare:2023ejv}, can substantially improve the agreement between JWST and CMB data\footnote{Since there exists a well-known degeneracy relation $A_s e^{-2\tau}$, high values of $\tau$ can be compensated by higher values of $A_s$}. Similarly, larger $n_s$ can substantially increase the power in the matter power spectrum on small scales, also facilitating more structure to form.
    \item Finally, concerning $H_0$, we observe that a straightforward increase in the value of this parameter while simultaneously keeping $\sigma_8$ constant worsens the fit to JWST data. This aligns with the argument that, when fixing structure formation parameters, a younger Universe reduces the number of structures that can form. However, the impact of $H_0$ on $\chi^2_{\rm JWST}$ is relatively small and it proves how the Hubble constant plays only a partial role in a more complex interplay of various parameters.
\end{itemize}

Summarizing these results, from a phenomenological standpoint, an effective model to increase the value of $H_0$ and improve the agreement with preliminary JWST data should predict a higher spectral index, along with a greater quantity of matter in the Universe and possibly higher values of $\sigma_8$ and $S_8$. On one side, this phenomenology is common in proposals aimed at resolving the Hubble tension by introducing new physical components that act before recombination. For this reason, we explore extensions related to the early Universe and, as a case study, analyse EDE cosmology. On the other hand, larger values of matter clustering parameters $\sigma_8$ and $S_8$ can also be achieved within late-time solutions of the Hubble tension that attempt to modify physics after recombination, influencing the value of $H_0$ derived from the angular distance from the CMB. Therefore, in the spirit of not leaving anything untried, we also test IDE cosmologies where both the growth of perturbations and the matter clustering are significantly different than in $\Lambda$CDM.

\subsubsection{Early Dark Energy}
Early Dark Energy models are a natural hypothesis of dark energy, see e.g.~\cite{Wetterich:2004pv,Doran:2006kp,Hollenstein:2009ph,Calabrese:2010uf,Calabrese:2011hg,Calabrese:2011hg,Pettorino:2013ia,Archidiacono:2014msa,Poulin:2023lkg,Poulin:2018dzj,Poulin:2018zxs,Smith:2019ihp,Niedermann:2019olb,Niedermann:2020dwg, Murgia:2020ryi,Ye:2020btb,Klypin:2020tud,Hill:2020osr,Herold:2021ksg,Herold:2022iib,Reeves:2022aoi,Jiang:2022uyg,Simon:2022adh,Smith:2022hwi,Kamionkowski:2022pkx,Niedermann:2023ssr,Cruz:2023lmn,Eskilt:2023nxm,Smith:2023oop,Sharma:2023kzr,Efstathiou:2023fbn,Gsponer:2023wpm,Goldstein:2023gnw}. Deviating from the traditional cosmological constant framework, EDE models  account for a non-negligible contribution from dark energy in the early Universe. In addition, these EDE models can be based on a generic dark energy fluids which are inhomogeneous. Their density and pressure vary over time, leading to a non-static equation of state. The phenomenological analyses
of these inhomogeneous dark energy models usually require additional dark energy clustering parameters, the dark energy effective sound speed and the dark energy anisotropic stress.  The effective sound speed determines the clustering properties of dark energy and consequently it affects the growth of matter density fluctuations. Therefore, in principle, its presence could be revealed in large scale structure observations. The growth of perturbations can also be affected by the anisotropic stress contributions  which lead to a damping in the velocity perturbations. Recently, EDE models have garnered significant attention, particularly due to their potential role in addressing some of the aforementioned cosmological tensions~\cite{Kamionkowski:2022pkx,Poulin:2023lkg,Abdalla:2022yfr}. Our analysis will concentrate on the EDE implementation detailed in~\cite{Hill:2020osr}. This model proposes that, in the early Universe, a light scalar field deviates from its potential minimum and, constrained by Hubble friction, is functionally similar to a cosmological constant. As soon as, at some particular redshift $z_\star$, the Hubble parameter reduces to be less than the mass of the field, the scalar field rolls down its potential and begins to oscillate about the minimum. To avoid spoiling late-time
cosmology, the vacuum energy must redshift away quicker than matter (i.e. faster than $a^{-3}$), and the field should behave as a subdominant component. A typical set of parameters used in this model is: the fractional contribution to the total energy density of the Universe, $f_{\rm EDE}(z) \equiv \rho_{\rm EDE}(z)/\rho_{\rm tot}(z)$ evaluated at the critical redshift $z_c$ at which it reaches the maximum value, and $\theta_i$, which is the parameter that usually describes the initial field displacement. This particular behavior implies a larger amount of energy-density in the early Universe (just prior to recombination), a reduction of the sound horizon and, consequently, a larger value of the Hubble constant inferred by CMB observations. This is the reason why EDE models have been proposed as a possible solution to the Hubble constant tension.

\subsubsection{Interacting Dark Energy}
Interacting Dark Energy models describe a phenomenological scenario where the dark fluids of the Universe interact with each other by allowing a transfer of energy and/or momentum between them, see e.g.~\cite{Valiviita:2008iv,Gavela:2009cy,Salvatelli:2014zta,Sola:2016jky,DiValentino:2017iww,Kumar:2017dnp,Wang:2016lxa,SolaPeracaula:2017esw,Sola:2017znb,Gomez-Valent:2018nib,Martinelli:2019dau,Yang:2019uog,DiValentino:2019ffd,Pan:2019jqh,Kumar:2019wfs,Yang:2018euj,Escudero:2015yka,Kumar:2016zpg,Murgia:2016ccp,Pourtsidou:2016ico,Yang:2018ubt,Barros:2018efl,Yang:2019uzo,Pan:2019gop,DiValentino:2019jae,DiValentino:2020leo,Yao:2020pji,Lucca:2020zjb,DiValentino:2020kpf,Gomez-Valent:2020mqn,Yang:2020uga,Yao:2020hkw,Pan:2020bur,DiValentino:2020vnx,Hogg:2020rdp,SolaPeracaula:2021gxi,Lucca:2021dxo,Kumar:2021eev,Yang:2021hxg,Gao:2021xnk,Yang:2021oxc,Lucca:2021eqy,Halder:2021jiv,Kaneta:2022kjj,Gariazzo:2021qtg,Nunes:2021zzi,Yang:2022csz,Nunes:2022bhn,Goh:2022gxo,Gomez-Valent:2022bku,vanderWesthuizen:2023hcl,Zhai:2023yny,Bernui:2023byc,deCruzPerez:2023wzd,Escamilla:2023shf}. Instead, the other components of the Universe (such as radiation and baryons) remain unaffected. The background evolution for dark energy and dark matter is modified, as the continuity equations for the single component present an interaction function $Q$ whose sign governs the energy-momentum flow. A negative value of the interaction rate, $Q < 0$, implies a transfer of energy and/or momentum from pressureless dark matter to dark energy, while the opposite, refers to an energy-momentum flow from the dark energy sector to the dark matter one. In order to solve the background evolution, one would need a specific interaction function $Q$. Depending on such a function, it can be solved either analytically or numerically, together with the equation for the Hubble rate evolution. In what follows we shall use the well-known interaction rate~\cite{He:2008si,Valiviita:2008iv,Gavela:2009cy,Gavela:2010tm,Honorez:2010rr}:    
\begin{eqnarray}
Q =  \xi {\cal H} \rho_{\rm de}\,,
\label{eq:coupling}
\end{eqnarray}
where $\xi$ is a dimensionless coupling parameter. The equation governing the evolution of the density perturbations for the dark sector can be found in~\cite{Valiviita:2008iv,Gavela:2009cy,Gavela:2010tm}. IDE models may suffer from instabilities in the perturbation evolution~\cite{Valiviita:2008iv,He:2008si}. Our analysis adheres to the criterion of~\cite{Gavela:2009cy} in terms of the so-called \emph{doom factor} 
\begin{equation}
\textbf{d}= \frac{Q}{3\mathcal{H}(1+w) \rho_{\rm de}}~,
\end{equation}
which is required to be negative in order to avoid instabilities. Consequently this stability condition for our case is translated into a stable parameter space in which $(1+w)$ and $\xi$ must have opposite sign~\cite{Gavela:2009cy}. Therefore, in the phantom regime in which $(1+w)$ is a negative quantity the dimensionless coupling $\xi$ must be positive. On the other hand, in the quintessence region $\xi$ must be negative. For earlier studies, see~\cite{Valiviita:2008iv,He:2008si,Jackson:2009mz,Gavela:2010tm,Clemson:2011an,Li:2014eha,Li:2014cee,Guo:2017hea,Zhang:2017ize,Guo:2018gyo,Yang:2018euj,Dai:2019vif}. 

We conclude this section with a final remark: even if the interaction scenario considered here is a pure  phenomenological model, some studies have shown that using a multi-scalar field action, the coupling function can be derived~\cite{Pan:2020zza}. Therefore, the interaction model of the form given by \eq{coupling} also benefits from a solid theoretical motivation. 

\subsection{Methodology}

To explore the extended dark energy scenarios in relation to the JWST observations, we strictly follow the methodology detailed in \sect{CSMD}. We compute the predicted CSMD given by \eq{CSMD}. For our analysis, we opt for a conservative approach and fix $\epsilon=0.2$ following~\cite{Tacchella:2018qny}. However, as suggested by~\cite{Tacchella:2018qny}, in principle star formation efficiency can be a function of the halo mass and further adjustments to star formation physics might be needed for more precise computations~\cite{2023arXiv231114804C}. Nonetheless, for the mass scale we are working with, $\epsilon$ can only vary smoothly as a function of mass, following a power law. Therefore, given the short mass range we are using in our analyses, this does not change in a significant way our main conclusions. Notice also that we consider the cosmic baryon fraction instead of computing the baryon evolution in different halos~\cite{Allen:2004cd,Vikhlinin:2005mp,Kravtsov:2005ab,Borgani:2009cd,Mantz:2014xba,Maio:2014qwa,Panchal:2024dcl}\footnote{ Notice that, with the baryon fraction $f_b$ playing the role of a multiplicative factor in front of the integral \eqref{eq:CSMD}, if $f_b$ increases, the theoretical predictions are proportionally pushed towards the observed data points. Just for reference, fixing $f_b$ as large as $f_b = 0.23$, we find an expected decrease in the $\chi^2$ of about 55\%.}. All these methodological choices and simplifications are widely used in the literature and allow us to present conservative and credible results that, without sacrificing generality, can be directly compared with similar works following the same approach. Furthermore, we employ the  Sheth-Tormen (ST) Halo mass function \cite{Sheth:1999mn,Sheth:1999su} as in \eq{STMF}.

\paragraph{\textit{Galaxy Observations}} Regarding the JWST observations, given their preliminary nature and the wide research interest they have generated, a multitude of datasets obtained following different methodologies have been released. Consequently, a choice of which dataset to use becomes necessary. In this study, we consider four independent datasets summarized in \autoref{tab:obs_JWST} and listed below: Notice that many additional measurements have been released, see, e.g.~\cite{Bouwens_2023,xiao2023massive}. In future, these measurements could potentially serve as independent tests of our preliminary findings.
\begin{table}[h]
    \centering
    \renewcommand{\arraystretch}{1.5}
    \begin{tabular}{|c|c|c|c|}
        \hline
        $z$ & $\ln{\left(\rho/M_\odot Mpc^{-3}\right)}$ & $\ln{\left(M/M_\odot\right)}$ & Dataset \\
        \hline
        \multirow{2}{*}{$7<z<8.5$} & $5.893\pm0.345$ & $10.030$ & \multirow{4}{*}{CEERS~\cite{2023Natu}} \\
         & $5.676\pm0.652$ & $10.75$  & \\
         \cline{1-3}
        \multirow{2}{*}{$8.5<z<10$} & $5.709\pm0.386$ & $9.704$ & \\
         & $5.386\pm0.653$ & $10.408$  & \\
        \hline
        \hline
        $3.5<z<4.5$ & $7.00^{+0.14}_{-0.16}$ &$10.48\pm0.15$  & \multirow{5}{*}{HUDF $\&$ UDS~\cite{Navarro-Carrera:2023ytd}} \\
         \cline{1-3}
         $4.5<z<5.5$ & $6.79^{+0.20}_{-0.28}$ & $10.45\pm0.27$  & \\
         \cline{1-3}
        $5.5<z<6.5$ & $6.67^{+0.21}_{-0.23}$ & $10.33\pm0.36$ & \\
        \cline{1-3}
        $6.5<z<7.5$ & $6.51^{+0.42}_{-0.60}$ & $10.68\pm0.79$  & \\
         \cline{1-3}
        $7.5<z<8.5$ & $5.75^{+0.59}_{-0.1.10}$ &[10.70] & \\
        \hline
        \hline
        $6.9<z<8.5$ & $5.07\pm0.52$ &\multirow{2}{*}{$7.2< \ln{\left(M/M_\odot\right)} <9.3$}  & \multirow{2}{*}{GLASS~\cite{Santini_GLASS_Mstar}} \\
        \cline{1-2}
        $3.5<z<4.5$ & $4.52\pm0.65$ &  &  \\
        \hline
        \hline
        & $\ln{\left(n(>M_{\rm halo}\right)}$ & $\ln{\left(M_{halo}/M_\odot\right)}$ & \\
         \cline{2-3}
        \multirow{3}{*}{$5<z<6$} & \multirow{3}{*}{$-5.52^{+0.69}_{-0.58}$} & $12.88^{+0.11}_{-0.13}$ & \multirow{3}{*}{FRESCO~\cite{xiao2023massive}} \\
        \cline{3-3}
         &  & $12.68^{+0.23}_{-0.17}$  & \\
         \cline{3-3}
         &  & $12.54^{+0.17}_{-0.18}$  & \\
         \hline
        
    \end{tabular}
    \caption[Observational points for JWST]{\small Observational points for JWST for the four different dataset. The values here must be rescaled by the corresponding comoving volume and luminosity distance for the Planck bestfit $\Lambda$CDM model.}
    \label{tab:obs_JWST}
\end{table}

\begin{itemize}
\item Confirmed CSMD measurements taken from~\cite{2023Natu}. This dataset was already used in \sect{PolJWST}. We find this dataset particularly well-suited for our analysis as it provides explicit constraints on the CSMD that are directly linked to cosmological structure formation. For this dataset, the errors reported in~\autoref{tab:obs_JWST} are assumed to follow a Log Normal distribution. We refer to this dataset as \textit{JWST-CEERS}.  

\item Five observational CSMD coming from the photometric data of JWST coverage of the UKIDSS Ultra Deep Survey (UDS) and Hubble Ultra Deep Field (HUDF)~\cite{Navarro-Carrera:2023ytd}. The errors for HUDF $\&$ UDS displayed in~\autoref{tab:obs_JWST} are conservatively taken as the maximum value, while the stellar masses are set to be equal to $M=10^8M_\odot$ as suggested in Table 6 of~\cite{Navarro-Carrera:2023ytd}. We refer to this dataset as \textit{JWST-HUDF\,$\&$UDS}. 

\item Two values of the observational CSMD coming from optical data in the GLASS-ERS 1324 program~\cite{Santini_GLASS_Mstar}. The stellar mass for GLASS datapoints is set to the average between the mass interval reported in~\autoref{tab:obs_JWST} where we refer to this dataset as \textit{JWST-GLASS}.

\item JWST FRESCO NIRCam/grism survey~\cite{xiao2023massive}. This dataset spans an area of 124 arcmin$^2$, covering a survey volume of approximately $1.2 \times 10^6$ Mpc$^3$ within the redshift range $z \in [5,9]$. We refer to it as \textit{JWST-FRESCO}. Taking the \textit{JWST-FRESCO} data at face value (and barring any potential systematic errors), we consider 3 obscured galaxies located within densely dusty regions, with redshifts in the range $5 \lesssim z \lesssim 6$. Referring to Fig. 3 of~\cite{xiao2023massive}, we can see that these galaxies show exceptionally extreme properties such as dark matter halo masses of $\log{\left( M_{\mathrm{halo}} / M_\odot \right)} = 12.88^{+0.11}_{-0.13}$, $12.68^{+0.23}_{-0.17}$, and $12.54^{+0.17}_{-0.18}$. Notice also that the quantity measured by \textit{JWST-FRESCO} is the cumulative comoving number density of dark matter halos, not the CSMD. For any given model of cosmology, the cumulative comoving number density of dark matter halos can be computed as
\begin{equation}
    n(>M_{\mathrm{halo}}) =  \int_{z_1}^{z_2} \int_{M_{\rm{halo}}}^{\infty} \frac{dn_h}{dM} \, dM,
    \label{eq:CNDM}
\end{equation}
where the integrand in \eq{CNDM} is defined by \eq{massfunc}. Regarding the error for the FRESCO dataset, we use the approach of Poissonian approximation for small numbers of observed events as done in~\cite{Menci:2022wia, Menci:2024rbq}. 
\end{itemize}

\begin{tcolorbox}[mybox]
The Poissonian approximation for small numbers of observed events was firstly introduced in Refs.~\cite{1986ApJ...303..336G,Ebeling:2003tf}. At its core, the methodology relies on approximations to the exact Poissonian confidence limits for small numbers of observed events (that in our case is $3$). More quantitatively, we approximate the true Poissonian upper limit, by means of Eq.(4) of~\cite{Ebeling:2003tf}, that, for 3 events, reads
\begin{equation}
\Delta \log n_{\rm upper} = 4 \left[ \frac{35}{36} + \frac{S}{6} + 4^{c(S)} b(S)\right]^{3},
\end{equation}
where we fix $S\simeq 1.645$ which corresponds to choosing a 95\%CL interval uncertainty. For Gaussian statistics (i.e., a normal probability distribution) the desired CL is related to $S$ by 
\begin{equation}
\text{CL}(S) = \frac{1}{\sqrt{2\pi}} \int_{-\infty}^{S} e^{-t^2 /2} \, dt,
\end{equation}
see also the third column of Tab. 3 in~\cite{1986ApJ...303..336G}. Notice also that, once $S$ is fixed, $c(S)$ and $b(S)$ are numerical coefficients that can be easily calculated by using Eqs.(6)-(7) in~\cite{Ebeling:2003tf}. Similarly, adopting Eq.(11) of~\cite{Ebeling:2003tf}, we estimate the lower limit on the error bar as
\begin{equation}
\Delta  \log  n_{\rm lower} = 4 \left[\frac{26}{27} + \frac{S}{3\sqrt{3}} + 3^{\gamma(S)} \beta(S) + \delta(S) \sin\left(\frac{10\pi}{13}\right)\right]^3
\end{equation}
where, as usual, $S\simeq 1.645$ and $\beta(S)$, $\gamma(S)$ and $\delta(S)$ are given by Eqs.(9), (10) and (12) of~\cite{Ebeling:2003tf}, respectively. We stress that this statistical methodology, while accurate, necessarily introduces an additional layer of approximation. For this reason, we remain conservative proving the uncertainties on $\log{n(>M_{\rm halo})}$ at 95\% CL (corresponding to fixing $S \simeq 1.645$).
\end{tcolorbox}
As already done in the previous computations, in \eq{CSMD} $V(z_1,z_2) = \frac{4}{3} \pi \left[R^3(z_2) - R^3(z_1)\right]$ represents the model-dependent comoving volume of the Universe between redshifts $z_1$ and $z_2$. Hence, the values of $\rho^{\text{obs}}$ given in \tab{obs_JWST} need to be appropriately rescaled before interfacing them with the predicted values for dark energy scenarios. A similar rescaling has to be applied, using the squared ratios of the luminosity distances too.

\paragraph{\textit{Numerical Analysis}} For our analyses, we perform a MCMC using the publicly available package \texttt{Cobaya}~\cite{Torrado:2020dgo} and generate theoretical predictions exploiting with a modified version of the software \texttt{CLASS}~\cite{2011arXiv1104.2932L,2011JCAP07034B} to address the IDE scenario, while, for EDE, the publicly available software \texttt{CLASS EDE},\footnote{\url{https://github.com/mwt5345/class_ede}} which solves the evolution of the background and of the perturbations in the presence of a scalar field by means of the KG equation. We investigate the posterior distributions of our parameter space through the MCMC sampler developed for \texttt{CosmoMC}~\cite{Lewis:2002ah,Lewis:2013hha} and tailored for parameter spaces with a speed hierarchy which also implements the ”fast dragging” procedure~\cite{neal2005taking}.  The likelihood used for the MCMC analysis are:
\begin{itemize}
\item CMB temperature and polarization power spectra from the legacy Planck release~\cite{Planck:2018vyg,Planck:2019nip} with  \textit{plik} TTTEEE+low-$\ell$+lowE.
\item Lensing Planck 2018 likelihood~\cite{Planck:2018lb}, reconstructed from measurements of the power spectrum of the lensing potential.
\end{itemize}
In the following discussion, we will refer to the combinations of these two datasets simply as CMB. The convergence of the chains obtained with this procedure is assessed using the Gelman-Rubin criterion~\cite{Gelman:1992zz} setting a convergence threshold at $R-1\lesssim 0.02$.

Once the chains have converged,\footnote{The converged chains are taken with 50$\%$ of burn-in.} we follow the same procedure as in \sect{CSMD}, namely we compute the CMSD and we calculate the $\chi^2_{\rm JWST}$ defined in \eq{chi2JWST} for various datasets listed in \autoref{tab:obs_JWST}. We then obtain the updated constraints on the cosmological parameters by re-weighting the MCMC chains, i.e. performing an importance sample, using the package \texttt{getdist}.

\subsection{Results for Early Dark Energy}

We start by examining EDE. For this model, we summarize the best-fit values related to the most relevant parameters in \tab{EDE} for CEERS and \tab{EDE_Fresco} for FRESCO. For the other dataset, the tables can be found in~\cite{Forconi:5}.
\begin{table}
	\begin{center}
		\renewcommand{\arraystretch}{1.5}
		\begin{tabular}{|c|  c | c |c |c| }
  	        \hline
			\textbf{Parameter} &\textbf{CMB}  &  & \textbf{JWST-CEERS}  & \textbf{CMB+JWST}\\
        \hline
             \multirow{3}{*}[+2ex]{$n_s$}&\multirow{3}{*}[+2ex]{$0.981$}&$z_{\rm low}$&$0.997$&$0.981$\\
			 &&$z_{\rm high}$&$0.997$&$0.981$\\
        \hline
             \multirow{3}{*}[+2ex]{$H_0$}&\multirow{3}{*}[+2ex]{$69.45$}&$z_{\rm low}$&$72.60$&$69.45$\\
			 &&$z_{\rm high}$&$72.60$&$69.45$\\
        \hline
             \multirow{3}{*}[+2ex]{$\sigma_8$}&\multirow{3}{*}[+2ex]{$0.8273$}&$z_{\rm low}$&$0.854$&$0.8273$\\
			 &&$z_{\rm high}$&$0.854$&$0.8273$\\
        \hline
             \multirow{3}{*}[+2ex]{$\tau$}&\multirow{3}{*}[+2ex]{$0.0575$}&$z_{\rm low}$&$0.0497$&$0.05753$\\
			 &&$z_{\rm high}$&$0.0497$&$0.05753$\\
        \hline
             \multirow{3}{*}[+2ex]{$\Omega_m$}&\multirow{3}{*}[+2ex]{$0.307$}&$z_{\rm low}$&$0.304$&$0.307$\\
			 &&$z_{\rm high}$&$0.304$&$0.307$\\
        \hline
             \multirow{3}{*}[+2ex]{$f_{\rm EDE}$}&\multirow{3}{*}[+2ex]{$0.0628$}&$z_{\rm low}$&$0.151$&$0.0628$\\
			 &&$z_{\rm high}$&$0.151$&$0.0628$\\
        \hline
            \multirow{3}{*}[+2ex]{$\chi^2$}&\multirow{3}{*}[+2ex]{$2772$}&$z_{\rm low}$&$5.75$&$2782.76$ $(2772+10.76)$\\
			 &&$z_{\rm high}$&$7.99$&$2787.34$ $(2772+15.34)$\\
        \hline
\end{tabular}
	\end{center}
	\caption[Results for EDE]{\small Results for EDE. We provide the best-fit values of cosmological parameters, namely the combination that minimizes the $\chi^2$ of the fit to the CMB data alone ($\chi^2_{\rm CMB}$), \textit{JWST-CEERS} data alone ($\chi^2_{\rm JWST-CEERS}$), and CMB+JWST-CEERS data ($\chi^2_{\rm CMB+JWST}$).}
	\label{tab:EDE}
\end{table}
\begin{figure}[h!]
	\centering
	\includegraphics[width=0.95\textwidth]{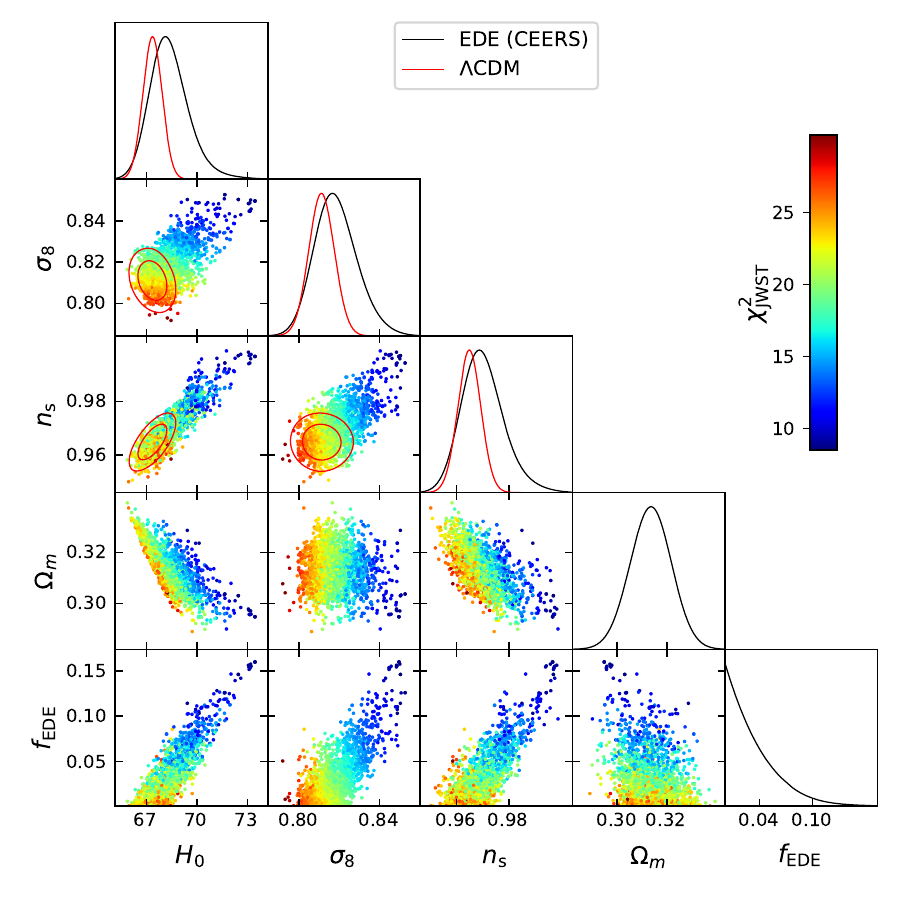}
	\caption[Correlations among the most relevant parameters of EDE]{\small Triangular plot showing the distribution of points and the correlations among the most relevant parameters of EDE. The color map refers to the value of $\chi^2_{\rm JWST}$ so that the color pattern in the figure represents the direction towards which one needs to move in the parameter space to improve the fit to JWST data.}
	\label{fig:EDE_LCDM}
\end{figure}
First and foremost, we examine the best-fit values obtained by exclusively considering CMB measurements from the Planck satellite (indicated as CMB in the table). In this case, the values reported in \tab{EDE} and \tab{EDE_Fresco}, simply represent the combination of cosmological parameters for which $\chi^2_{\text{CMB}}$ ($=2772$) acquires its minimum value among those obtained within the MCMC analysis. It is worth noting that we retrieve results widely documented in the literature. In particular, the Planck data, while not showing any substantial preference for a non-vanishing fraction of EDE, produce a best-fit value of $f_{\text{EDE}}=0.06$. This leads to a present-day expansion rate of the Universe $H_0=69.45$ km/s/Mpc, which is generally higher than the best-fit value obtained for this parameter within the standard cosmological model. Another point that is worth emphasizing is that for the inflationary spectral index we get a best-fit value $n_s=0.981$, confirming once more the trend of EDE models in predicting a spectrum of primordial perturbations closer to the scale-invariant case than what is observed in $\Lambda$CDM and predicted by the most typical inflationary potentials~\cite{DiValentino:2018zjj,Ye:2021nej,Ye:2022efx,Jiang:2022uyg, Jiang:2022qlj,Takahashi:2021bti,Jiang:2023bsz,Peng:2023bik}
.\footnote{For other discussions surrounding the value of this parameter and the agreement among the results of different CMB probes, see, e.g.~\cite{Forconi:1,ACT:2020gnv,Handley:2020hdp,DiValentino:2022rdg,Giare:2022rvg,Giare:2023wzl,DiValentino:2022oon,Calderon:2023obf,Giare:2023xoc}.} As a second step, following the methodology outlined in the previous section, for each combination of parameters in the MCMC chains (i.e., for each collected model), we calculate the $\chi^2$ against the four different JWST datasets listed in \tab{obs_JWST}. 

\subsubsection{EDE in Light of CEERS Measurements}
We start discussing the results obtained re-weighting the chains in light of $\chi^2_{\rm JWST-CEERS}$ resulting from the \textit{JWST-CEERS}  dataset. This dataset has been recently analyzed in many similar studies and allows us for direct comparison with the existing findings in the literature~\cite{Boylan-Kolchin:2022kae,Menci:2022wia}. In this case, we summarize the results in \tab{EDE}, distinguishing between the low ($z_{\rm low}$) and high ($z_{\rm high}$) redshift bins (also see \tab{obs_JWST}). Similar to the CMB analysis,  we present the specific combination of parameters that minimizes $\chi^2_{\rm JWST-CEERS}$. Furthermore, in \fig{EDE_LCDM} we provide a triangular plot showing the distribution of sampled models and the correlations among different parameters, together with a color-map representing the value of $\chi^2_{\rm{JWST-CEERS}}$. For the sake of comparison, in the same figure, we also depict the predictions of $\Lambda$CDM. A few intriguing conclusions can be derived from both \tab{EDE} and \fig{EDE_LCDM}. Firstly, there are no significant differences between the results obtained for the high and low redshift bins. Secondly, as for the best-fit values of cosmological parameters, we now find a pronounced preference for a non-vanishing fraction of EDE, $f_{\rm EDE}=0.151$. We also get higher $\sigma_8=0.85$ and observe the same trend towards higher values of inflationary spectral index $n_s=0.997$, now essentially consistent with a Harrison-Zel’dovich spectrum. As pointed out in \sect{plumino}, this is exactly the kind of phenomenology one needs to increase the agreement with JWST data. Therefore, not surprisingly, the minimum value of $\chi^2_{\rm JWST-CEERS}$ for both the high ($\chi^2_{\rm{JWST-CEERS}}=7.99$) and low ($\chi^2_{\rm{JWST-CEERS}}=5.75$) redshift bins are significantly better than what we get in $\Lambda$CDM (where $\chi_{\rm JWST-CEERS}^2\sim 17$, see \tab{Tablechi2}). This suggests that EDE stands as a valid phenomenological alternative to explain (at least partially) the preliminary measurements released by \textit{JWST-CEERS}. Furthermore, regarding $H_0$, the \textit{JWST-CEERS} best-fit value reads $H_0=72.60$ km/s/Mpc. Therefore, not only within the context of EDE we can improve the agreement between the theoretical predictions of the model and the \textit{JWST-CEERS} data, but to achieve this, we move through the parameter space in the same direction needed to solve the Hubble tension, as well. This is also clearly confirmed by the color pattern in \fig{EDE_LCDM}, underscoring that it is indeed possible to address both issues within the same theoretical framework.

Finally, always in \tab{EDE}, we present the results inferred by summing up the $\chi^2$ values of CMB and \textit{JWST-CEERS}. We observe that the combination of parameters that minimizes the total $\chi^2_{\text{CMB+JWST}}$ is the same as that minimizing the fit to only the Planck data $\chi^2_{\text{CMB}}$. At first glance, this implies that the cost of improving the fit to the \textit{JWST-CEERS} data is an overall deterioration in the fit to the CMB. On the other hand, such deterioration is entirely expected, given the strong preference of Planck data for a $\Lambda$CDM cosmology and the general disagreement between \textit{JWST-CEERS} data and the latter. As extensively documented in the literature and confirmed by our analysis, individual Planck data do not provide clear evidence in favor of an EDE cosmology. In any case, the best-fit parameters suggest an inclination towards a model where the fraction of EDE remains modest and well below $f_{\rm EDE}\lesssim 0.1$. In contrast, reconciling \textit{JWST-CEERS} with an EDE cosmology would require an EDE fraction $f_{\rm EDE}\gtrsim 0.1$. Forcing such an EDE fraction into the model would source significant effects in the CMB spectra that can only be partially compensated by the observed shift in the fitting values of other cosmological parameters. Just to provide an illustrative example, the increase in the expansion rate of the Universe before recombination due to a substantial EDE component leads to a significant reduction in the value of the sound horizon at the combination, forcing the value of $H_0$ in the direction of SH0ES, which is certainly not the direction favored by CMB data. Additionally, since EDE does not alter the physics of post-recombination, a higher $H_0$ implies a lower angular diameter distance from the CMB, $D_A$. In turn, this leads to a shift in the wavenumber associated with the damping tail $k_D$, as these two parameters are related by the relationship $\ell_D \sim k_D D_A$, where the multipole $\ell_D$ is also fixed by CMB measurements. In an attempt to maintain a good fit to the damping scale, the value of $n_s$ is shifted towards a scale-invariant primordial spectrum (i.e., $n_s \to 1$) which certainly improves the agreement between \textit{JWST-CEERS} and EDE but is again highly disfavored by Planck (by over $8\sigma$ in $\Lambda$CDM). Overall, all these effects and shifts in the fitting values of cosmological parameters seem to favor \textit{JWST-CEERS} observations. However, although they partially compensate for each other, they still remain somewhat disfavored based solely on CMB data, leading to a deterioration in the fit. 
\begin{table}
	\begin{center}
		\renewcommand{\arraystretch}{1.5}
		\begin{tabular}{|c|  c  |c |c| }
  	        \hline
			\textbf{Parameter} &\textbf{CMB}  & \textbf{JWST-FRESCO}  & \textbf{CMB+JWST}\\
        \hline
             $n_s$&$0.981$&$0.997$&$0.978$\\
        \hline
             $H_0$&$69.45$&$72.60$&$69.88$\\
        \hline
             $\sigma_8$&$0.8273$&$0.854$&$0.8267$\\
        \hline
            $\tau$&$0.0575$&$0.0497$&$0.0510$\\
        \hline
             $\Omega_m$&$0.307$&$0.304$&$0.312$\\
        \hline
             $f_{\rm EDE}$&$0.0628$&$0.151$&$0.0853$\\
        \hline
            $\chi_{\rm FRESCO}^2$&$2772$&$21.42$&$2804$ $(2774+30)$\\
        \hline
\end{tabular}
	\end{center}
	\caption[Results for EDE with FRESCO]{\small Results for EDE. We provide the best-fit values of cosmological parameters, namely the combination that minimizes the $\chi^2$ of the fit to the CMB data alone ($\chi^2_{\rm CMB}$), \textit{JWST-FRESCO} data alone ($\chi^2_{\rm JWST-FRESCO}$), and CMB+\textit{JWST-FRESCO} data ($\chi^2_{\rm CMB+JWST}$).}
	\label{tab:EDE_Fresco}
\end{table}

\subsubsection{EDE in Light of FRESCO Measurements}
\begin{figure}[h!]
	\centering
	\includegraphics[width=0.95\textwidth]{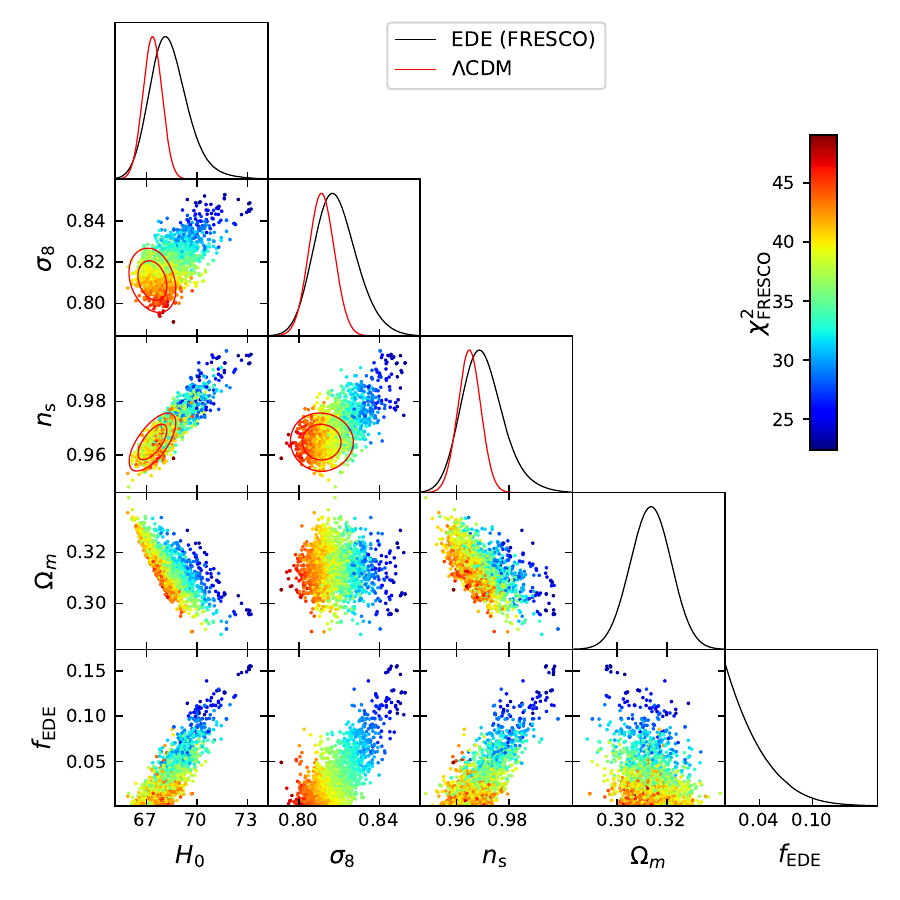}
\caption[correlations among the most relevant parameters of EDE with FRESCO]{\small Triangular plot showing the distribution of points and the correlations among the most relevant parameters of EDE. The color map refers to the value of $\chi^2_{\rm FRESCO}$ so that the color pattern in the figure represents the direction towards which one needs to move in the parameter space to improve the fit to JWST FRESCO data.}
	\label{fig:EDE_LCDM_FRESCO}
\end{figure}

When analyzing the other JWST datasets listed in \tab{obs_JWST}, all the conclusions we have drawn so far remain mostly true. For instance, by comparing the results obtained for \textit{JWST-CEERS} in \fig{EDE_LCDM} with those obtained for \textit{JWST-FRESCO} in \fig{EDE_LCDM_FRESCO}, at first glance, we can spot the very same color patterns, indicating that a non-vanishing fraction of EDE could, in principle, help to reduce $\chi^2_{\rm FRESCO}$ while also yielding higher values of $H_0$. However, paying closer attention to the colorbar scale, we observe that as we approach the limit $f_{\rm EDE}\to 0$ moving towards the $\Lambda$CDM cosmology, we get $\chi^2_{\rm JWST-FRESCO}\sim 50$ for 3 data points. This value can be reduced all the way down to $\min(\chi^2_{\rm JWST-FRESCO})\sim 21.42$ when $f_{\rm EDE}\sim 0.15$ and $H_0\sim73$ km/s/Mpc, as seen in \tab{EDE_Fresco}. On one hand, this lends weight to the idea that EDE could potentially pave the way to partially explaining more massive galaxies and higher values of $H_0$. On the other hand, it is important to note a nearly threefold increase in $\chi^2_{\mathrm{JWST-FRESCO}}$ compared to the results obtained for \textit{JWST-CEERS}. Taking the large $\chi^2$ at face value, we must draw the conclusion that the \textit{JWST-FRESCO} dataset remains in strong disagreement with the theoretical predictions of the standard cosmological model and, to a lesser extent, with EDE as well. 

Tables and figures for \textit{JWST-GLASS} and \textit{JWST-{HUD$\&$UDS}} can be found in~\cite{Forconi:5}, and it is possible to see that a similar conclusion can be drawn. As a result, we conclude that while EDE represents a phenomenological possibility to partially address the JWST data, it falls short of being exhaustive in fully addressing issues, leaving the quest for a more comprehensive solution wide open. Having that said, it is worth keeping in mind some caveats surrounding the joint analyses. For instance, the total $\chi^2_{\text{CMB+JWST}}$ is obtained by considering the sum of $\chi^2$ for each sampled model in the MCMC chains \textit{afterward} and not through a joint analysis of the two experiments from the outset. Additionally, only CMB data are taken into account in the MCMC analysis, which we know do not favor high values of $f_{\text{EDE}}$ and $H_0$. Considering other datasets, such as the measurements of the expansion rate provided by the SH0ES collaboration, could lead to significantly different results in terms of the $\chi^2$ analysis, as typically pointed out by the EDE community (see for example the discussion on page 25 of~\cite{Poulin:2023lkg}). Hence, a full joint likelihood analysis of all these datasets (which is beyond the aim of this work) is needed before deriving any definitive conclusions on this matter.

\subsection{Results for Interacting Dark Energy}

We now move to the study of IDE. In this case, we consider three different models: the usual IDE cosmology with a fixed dark energy equation of state $w \simeq -1$, and $w$IDE models where the equation of state parameter $w$ is free to vary, although limited either in quintessence ($w > -1$) or phantom ($w < -1$) regime. For the sake of simplicity, in this subsection, we present \textit{only} the results obtained from \textit{JWST-CEERS}. Several well-motivated reasons underpin this decision. Firstly, no significant differences emerged in the results for EDE when analyzing the four different JWST datasets listed in \tab{obs_JWST}. Overall, all these observations converge on anomalous galaxies that are more massive than predicted by the standard cosmological model. Therefore, no significant disparities are anticipated when analyzing the same four datasets across the various IDE models proposed in this section. Yet another motivation involves noting that addressing these JWST anomalous observations requires a somewhat clear beyond-$\Lambda$CDM phenomenology that none of the three IDE models proposed here can offer. To streamline the analysis and emphasize this point, we focus on \textit{JWST-CEERS}, which will provide the phenomenological guidelines applicable directly to all datasets not explicitly mentioned, without exception. 
\begin{figure}[h!]
	\centering
	\includegraphics[width=0.95\textwidth]{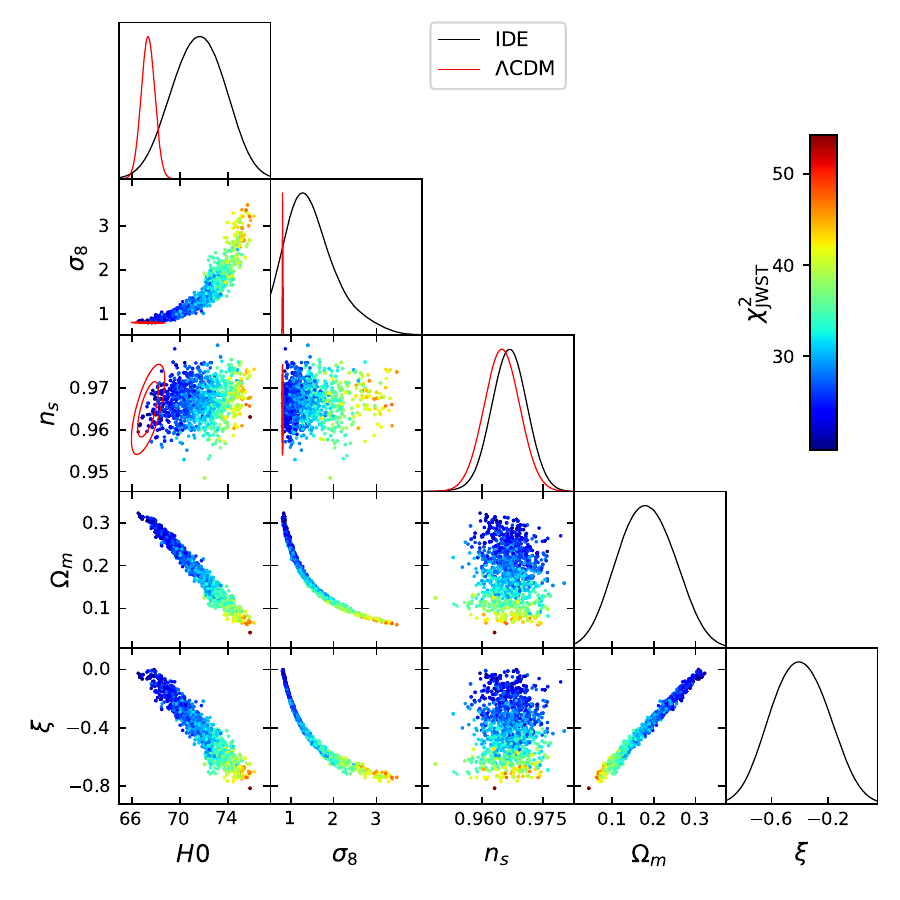}
	\caption[Correlations among the most relevant parameters of IDE]{\small Triangular plot showing the distribution of points and the correlations among the most relevant parameters of IDE, when $w$ is fixed to $w\simeq -1$. The color map refers to the value of $\chi^2_{\rm JWST}$ so that the color pattern in the figure represents the direction towards which one needs to move in the parameter space to improve the fit to JWST data.}
	\label{fig:IDE_LCDM}
\end{figure}

\paragraph{\textit{IDE}} \tab{IDE} displays the results for the IDE model with a fixed dark energy equation of state. Similar to EDE, we consider three different combinations of data: CMB, \textit{JWST-CEERS}, and CMB+\textit{JWST-CEERS}. In the table, we always show the combination of parameters that minimizes the $\chi^2$ for these three datasets. When focusing solely on the Planck CMB data, we note that the best-fit value for the parameter encapsulating new physics, i.e. the coupling $\xi$, reads $\xi=-0.28$. This suggests a quite significant transfer of energy-momentum from the dark matter sector to the dark energy sector. As widely documented in the literature, such a transfer of energy-momentum leads to a higher present-day expansion rate of the Universe, whose best-fit value reads $H_0=70.98$ km/s/Mpc (significantly higher than in the standard cosmological model). Furthermore, while we do not observe significant differences in the value of the spectral index, we notice a tendency toward higher values of $\sigma_8=1.09$. This makes the model potentially interesting for JWST. Nevertheless, the results we obtain from \textit{JWST-CEERS} data seem to indicate precisely the opposite. In contrast to the CMB fit (which  prefers $\xi < 0$), when considering only the \textit{JWST-CEERS} likelihood, the coupling parameter $\xi$ tends towards $\xi \to 0$. 
\begin{table}[h!]
	\begin{center}
		\renewcommand{\arraystretch}{1.5}
		\begin{tabular}{|c|  c | c |c |c| }
  	        \hline
			\textbf{Parameter} &\textbf{CMB}  &  & \textbf{JWST-CEERS}  & \textbf{CMB+JWST}\\
        \hline
             \multirow{3}{*}[+2ex]{$n_s$}&\multirow{3}{*}[+2ex]{$0.971$}&$z_{\rm low}$&$0.968$&$0.963$\\
			 &&$z_{\rm high}$&$0.968$&$0.961$\\
        \hline
             \multirow{3}{*}[+2ex]{$H_0$}&\multirow{3}{*}[+2ex]{$70.98$}&$z_{\rm low}$&$67.43$&$68.27$\\
			 &&$z_{\rm high}$&$67.43$&$66.69$\\
        \hline
             \multirow{3}{*}[+2ex]{$\sigma_8$}&\multirow{3}{*}[+2ex]{$1.09$}&$z_{\rm low}$&$0.878$&$0.89$\\
			 &&$z_{\rm high}$&$0.878$&$0.85$\\
        \hline
             \multirow{3}{*}[+2ex]{$\tau$}&\multirow{3}{*}[+2ex]{$0.057$}&$z_{\rm low}$&$0.0609$&$0.057$\\
			 &&$z_{\rm high}$&$0.0609$&$0.051$\\
        \hline
             \multirow{3}{*}[+2ex]{$\Omega_m$}&\multirow{3}{*}[+2ex]{$0.214$}&$z_{\rm low}$&$0.301$&$0.315$\\
			 &&$z_{\rm high}$&0.301&0.285\\
        \hline
             \multirow{3}{*}[+2ex]{$\xi$}&\multirow{3}{*}[+2ex]{$-0.28$}&$z_{\rm low}$&$-0.0734$&$-0.098$\\
			 &&$z_{\rm high}$&$-0.073$&$-0.053$\\
        \hline
             \multirow{3}{*}[+2ex]{$\chi^2$}&\multirow{3}{*}[+2ex]{$2781$}&$z_{\rm low}$&$12.50$&$2800.79$ $(2785+15.79)$\\
			 &&$z_{\rm high}$&$18.22$&$2809.29$ $(2788+21.29)$\\
        \hline
\end{tabular}
	\end{center}
	\caption[Results for IDE]{\small Results for IDE, with the dark energy equation of state fixed to $w\simeq -1$. We provide the best-fit values of cosmological parameters, namely the combination that minimizes the $\chi^2$ of the fit to the CMB data alone ($\chi^2_{\rm CMB}$), \textit{JWST-CEERS} data alone ($\chi^2_{\rm JWST-CEERS}$), and CMB+JWST data ($\chi^2_{\rm CMB+JWST}$).}
 \label{tab:IDE}
\end{table}
Consequently, we lose any ability to increase the present-day expansion rate, getting a best fit value $H_0 = 67.43$ km/s/Mpc. \fig{IDE_LCDM} further reinforces our conclusions: it is not $n_s$ but $\Omega_m$ that now assumes a critical role. When moving in the direction $\xi < 0$, the matter density undergoes a significant decrease due to the energy transfer from dark matter to dark energy. This leads to a substantial increase in the value of $\sigma_8$ to compensate for the reduced $\Omega_m$. However, as illustrated by the color pattern in \fig{IDE_LCDM}, in order to minimize $\chi^2_{\rm JWST-CEERS}$, it becomes necessary to revert to the $\Lambda$CDM framework by preventing such energy transfer, essentially moving towards $\xi\to0$. This behaviour is further supported by comparing the best-fit values of $\Omega_m$ ($\sigma_8$) for CMB and JWST: while in the former case $\Omega_m = 0.214$ ($\sigma_8=1.09$), for \textit{JWST-CEERS}, we get back to more typical values $\Omega_m = 0.301$ ($\sigma_8=0.878$). Consequently, this model fails to provide a satisfactory fit to the \textit{JWST-CEERS} observations.
\begin{table}[h!]
	\begin{center}
		\renewcommand{\arraystretch}{1.5}
		\begin{tabular}{|c|  c | c |c |c| }
  	        \hline
			\textbf{Parameter} &\textbf{CMB}  &  & \textbf{JWST-CEERS}  & \textbf{CMB+JWST}\\
        \hline
             \multirow{3}{*}[+2ex]{$n_s$}&\multirow{3}{*}[+2ex]{$0.9615$}&$z_{\rm low}$&$0.967$&$0.9614$\\
			 &&$z_{\rm high}$&$0.967$&$0.9614$\\
        \hline
             \multirow{3}{*}[+2ex]{$H_0$}&\multirow{3}{*}[+2ex]{$67.35$}&$z_{\rm low}$&$65.44$&$64.92$\\
			 &&$z_{\rm high}$&$65.44$&$64.92$\\
        \hline
             \multirow{3}{*}[+2ex]{$\sigma_8$}&\multirow{3}{*}[+2ex]{$0.9342$}&$z_{\rm low}$&$0.839$&$0.871$\\
			 &&$z_{\rm high}$&$0.839$&$0.871$\\
        \hline
             \multirow{3}{*}[+2ex]{$\tau$}&\multirow{3}{*}[+2ex]{$0.05742$}&$z_{\rm low}$&$0.0686$&$0.055$\\
			 &&$z_{\rm high}$&$0.0686$&$0.055$\\
        \hline
             \multirow{3}{*}[+2ex]{$\Omega_m$}&\multirow{3}{*}[+2ex]{$0.274$}&$z_{\rm low}$&$0.326$&$0.314$\\
			 &&$z_{\rm high}$&$0.326$&$0.314$\\
        \hline
             \multirow{3}{*}[+2ex]{$\xi$}&\multirow{3}{*}[+2ex]{$-0.1743$}&$z_{\rm low}$&$-0.045$&$-0.115$\\
			 &&$z_{\rm high}$&$-0.045$&$-0.115$\\
        \hline
            \multirow{3}{*}[+2ex]{$w$}&\multirow{3}{*}[+2ex]{$-0.9483$}&$z_{\rm low}$&$-0.923$&$-0.90$\\
			 &&$z_{\rm high}$&$-0.923$&$-0.90$\\
        \hline
             \multirow{3}{*}[+2ex]{$\chi^2$}&\multirow{3}{*}[+2ex]{$2774$}&$z_{\rm low}$&$11.94$&$2790.89$ $(2776+14.89)$\\
			 &&$z_{\rm high}$&$17.5$&$2797.63$ $(2778+21.63)$\\
        \hline
\end{tabular}
	\end{center}
	\caption[Results for $w$IDE]{\small Results for $w$IDE, with $w>-1$ free to vary in the quintessence regime. We provide the best-fit values of cosmological parameters, namely the combination that minimizes the $\chi^2$ of the fit to the CMB data alone ($\chi^2_{\rm CMB}$), \textit{JWST-CEERS} data alone ($\chi^2_{\rm JWST-CEERS}$), and CMB+JWST data ($\chi^2_{\rm CMB+JWST}$).}
	\label{tab:nwIDE}
\end{table}

\begin{figure}[h!]
	\centering
	\includegraphics[width=0.95\textwidth]{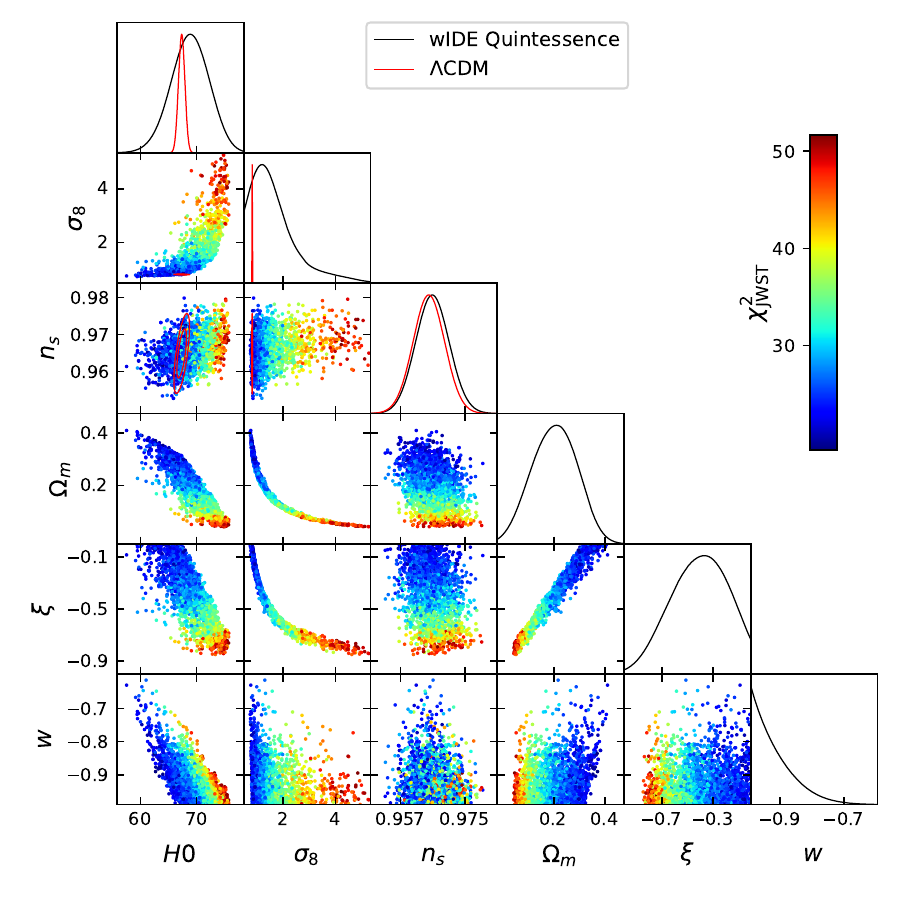}
	\caption[Correlations among the most relevant parameters of $w$IDE]{\small Triangular plot showing the distribution of points and the correlations among the most relevant parameters of $w$IDE, when $w$ is free to vary in the quintessence region $w> -1$. The color map refers to the value of $\chi^2_{\rm JWST}$ so that the color pattern in the figure represents the direction towards which one needs to move in the parameter space to improve the fit to JWST data.}
	\label{fig:nwIDE_LCDM}
\end{figure}

\paragraph{\textit{Quintessence Regime}} The best-fit values of cosmological parameters for the $w$IDE model with $w > -1$ (i.e., confined to the quintessence regime) are displayed in \autoref{tab:nwIDE}. Qualitatively, these results mirror those previously obtained for $w = -1$. There are no differences between the high and low redshift bins, and there is no preference for $\xi \neq 0$ from \textit{JWST-CEERS} data. The behaviors of the parameters depicted in \fig{nwIDE_LCDM} clearly indicate that introducing a coupling while leaving $w$ free to vary does not improve the fit to \textit{JWST-CEERS} observations. Once more, the reason behind this phenomenon is the decrease in matter density resulting from the energy flow within the interacting model.

\begin{figure}[h!]
	\centering
	\includegraphics[width=0.95\textwidth]{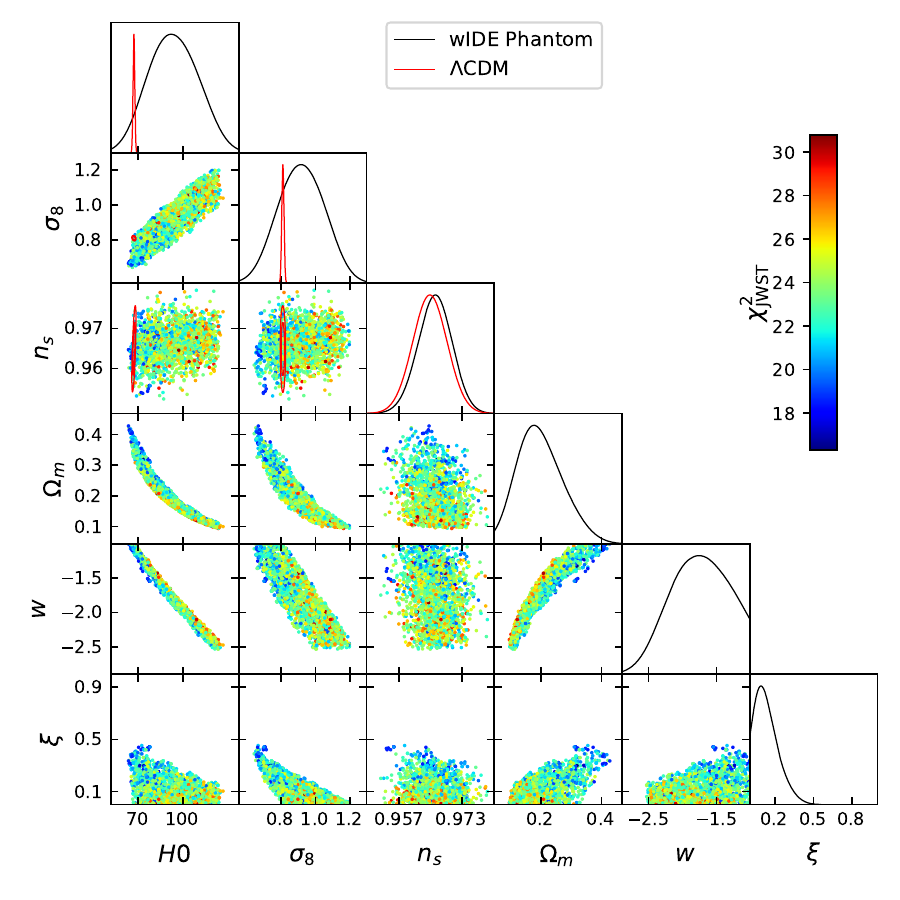}
	\caption[correlations among the most relevant parameters of $w$IDE in the phantom region]{\small Triangular plot showing the distribution of points and the correlations among the most relevant parameters of $w$IDE, when $w$ is free to vary in the phantom region $w< -1$. The color map refers to the value of $\chi^2_{\rm JWST}$ so that the color pattern in the figure represents the direction towards which one needs to move in the parameter space to improve the fit to JWST data.}
	\label{fig:pwIDE_LCDM}
\end{figure}

\paragraph{\textit{Phantom Regime}} The situation becomes somewhat more intricate when we turn to the case where the dark energy equation of state is confined to the phantom regime $w<-1$. In this case, the best-fit values of parameters are summarized in \tab{pwIDE} for the usual combinations of datasets. When considering only the best-fit values from the CMB, we find the well-documented Planck preference for a phantom equation of state~\cite{Yang:2021flj,Escamilla:2023oce}, with the best-fit value reading $w=-2.04$. Given the well-known degeneracy between the effects produced by a phantom $w$ and increasing the present-day expansion rate of the Universe, interacting phantom models can provide a much larger value of $H_0$~\cite{Yang:2022csz} and, in fact, we obtain a best-fit value $H_0=103.8$ km/s/Mpc. This essentially indicates that, without including datasets at lower redshifts, breaking this line of degeneracy proves challenging and values of $H_0$ in line with those measured by SH0ES collaboration can always be reintroduced by considering a sufficiently phantom dark energy component. In addition, due to a combination of correlations among different parameters such as $H_0$, $w$, and $\Omega_m$, we observe that for the latter parameter, the best-fit value reads $\Omega_m=0.139$ and is compensated by a substantial increase in $\sigma_8=1.026$. In light of these effects on parameters governing the matter clustering in the Universe, the question of whether (and to what extent) a phantom $w$IDE model could effectively contribute to explaining the anomalies observed in JWST remains a topic of debate. Looking at the brighter side, when we consider the fit to \textit{JWST-CEERS} data, we observed that both for the high ($\chi^2_{\rm JWST-CEERS}=16.3$) and low ($\chi^2_{\rm JWST-CEERS}=10.8$) redshift bins, $\chi^2_{\rm JWST-CEERS}$ slightly improves compared to the standard cosmological model. Furthermore, the \textit{JWST-CEERS} data seem to point toward a non-zero coupling $\xi\sim 0.2 - 0.3$, which is substantially higher than that preferred by the Planck data (whose best-fit value is $\xi\simeq0.05$).
\begin{table}[h!]
	\begin{center}
		\renewcommand{\arraystretch}{1.5}
		\begin{tabular}{|c|  c | c |c |c| }
  	        \hline
			\textbf{Parameter} &\textbf{CMB}  &  & \textbf{JWST-CEERS}  & \textbf{CMB+JWST}\\
        \hline
             \multirow{3}{*}[+2ex]{$n_s$}&\multirow{3}{*}[+2ex]{$0.9663$}&$z_{\rm low}$&$0.963$&$0.969$\\
			 &&$z_{\rm high}$&$0.967$&$0.969$\\
        \hline
             \multirow{3}{*}[+2ex]{$H_0$}&\multirow{3}{*}[+2ex]{$103.8$}&$z_{\rm low}$&$67.10$&$75.62$\\
			 &&$z_{\rm high}$&$73.88$&$75.62$\\
        \hline
             \multirow{3}{*}[+2ex]{$\sigma_8$}&\multirow{3}{*}[+2ex]{$1.026$}&$z_{\rm low}$&$0.685$&$0.760$\\
			 &&$z_{\rm high}$&$0.739$&$0.760$\\
        \hline
             \multirow{3}{*}[+2ex]{$\tau$}&\multirow{3}{*}[+2ex]{$0.05087$}&$z_{\rm low}$&$0.0612$&$0.0617$\\
			 &&$z_{\rm high}$&$0.0644$&$0.0617$\\
        \hline
             \multirow{3}{*}[+2ex]{$\Omega_m$}&\multirow{3}{*}[+2ex]{$0.139$}&$z_{\rm low}$&$0.396$&$0.292$\\
			 &&$z_{\rm high}$&$0.319$&$0.292$\\
        \hline
             \multirow{3}{*}[+2ex]{$\xi$}&\multirow{3}{*}[+2ex]{$0.05235$}&$z_{\rm low}$&$0.374$&$0.229$\\
			 &&$z_{\rm high}$&$0.299$&$0.229$\\
        \hline
            \multirow{3}{*}[+2ex]{$w$}&\multirow{3}{*}[+2ex]{$-2.0436$}&$z_{\rm low}$&$-1.149$&$-1.33$\\
			 &&$z_{\rm high}$&$-1.33$&$-1.33$\\
        \hline
             \multirow{3}{*}[+2ex]{$\chi^2$}&\multirow{3}{*}[+2ex]{$2767$}&$z_{\rm low}$&$10.8$&$2783.90$ $(2771+12.90)$\\
			 &&$z_{\rm high}$&$16.3$&$2790.15$ $(2771+19.15)$\\
        \hline
\end{tabular}
	\end{center}
	\caption[Results for $w$IDE with $w<-1$]{\small Results for $w$IDE with $w<-1$ free to vary in the phantom regime. We provide the best-fit values of cosmological parameters, namely the combination that minimizes the $\chi^2$ of the fit to the CMB data alone ($\chi^2_{\rm CMB}$), \textit{JWST-CEERS} data alone ($\chi^2_{\rm JWST-CEERS}$), and CMB+JWST data ($\chi^2_{\rm CMB+JWST}$).}
	\label{tab:pwIDE}
\end{table}
The reason underlying this preference is that to ensure the stability of the perturbations, $\xi$ is now required to be positive. This results in a shift of the energy flow from the dark energy sector to the dark matter one. Consequently, increasing the value of the coupling means injecting more power into the matter sector, facilitating structure formation, and improving the \textit{JWST-CEERS} fit. However, looking at \fig{pwIDE_LCDM}, where, as usual, we plot the correlations among different parameters together with a color-map representing the value of $\chi^2_{\rm JWST-CEERS}$, it becomes really difficult to identify a color pattern representing the direction in which we need to move in the parameter space to improve the fit to \textit{JWST-CEERS} data. Despite this, with a good degree of imagination, we can speculate that by moving towards larger values of the coupling $\xi\to 0.4$ (corresponding to other values of $\Omega_m\to 0.4$), the value of $\chi^2_{\rm JWST-CEERS}$ seems to undergo a general improvement for the highlighted reasons, and its value becomes as good as the $\Lambda$CDM one. Interestingly, looking again at \fig{pwIDE_LCDM}, we note that the value of the Hubble rate corresponding to such a coupling is $H_0\sim 70$ km/s/Mpc, i.e. close to the result provided by SH0ES collaboration.

Therefore, in the context of IDE cosmology, the only potential scenario in which we can simultaneously slightly improve the agreement between the model predictions and JWST data while obtaining values of $H_0$ in line with local distance ladder measurements seems to involve considering a phantom component of the equation of state of dark energy. That being said, the ability of this model to address these two issues remains somewhat limited, above all when compared to the competing EDE solutions discussed in the previous section.

%% file: Chapters/Conclusions.tex
\label{conclusionnn}

In this thesis we outlined the main ideas of the standard model of cosmology together with the derivation of the major equations on which the framework lies upon. Moreover, we studied the dynamics of the leading character of this wonderful play that is the history and evolution of our Universe. It allowed us to start from primordial seeds of quantum fluctuations of the inflationary scalar field and arrive to massive galaxies in the Dark Ages of our Universe, passing through the statistical proprieties of the observations, the implications of an effective field theory approach, while also trying to adventure in the plethora of possible alternatives to the standard model, both from a theoretical and a phenomenological point of view. It has been possible by leveraging precision cosmological observations to constrain new physics in the early Universe. By analyzing data from the Cosmic Microwave Background, large-scale structures, high-redshift galaxies observed by the James Webb Space Telescope, and primordial gravitational waves, we have investigated the implications of these extensions on inflationary dynamics, the effective number of relativistic species, structure formation, and neutrino physics.

In \textbf{Section \ref{sec:bBNPGW}} we revisit the calculation of the inflationary gravitational wave contribution to the radiation energy-density in the early Universe. Behaving as additional radiation, primordial gravitational waves may in fact increase the effective number of relativistic species ($N_{\rm eff}$) by a further correction that depends on the integrated energy-density in gravitational radiation over all scales which, in turns, affect the BBN epoch. This effect is particularly relevant, because it is commonly used to infer stringent bounds on the additional radiation energy-density and, in its turn, to constrain (blue-titled) models of inflation. We study how (much) different parameterizations of the tensor spectrum impact on the final predictions of $\Delta N_{\rm eff}^{\rm GW}$. We perform parametric analysis by expanding the spectrum in full generality as a sum of powers and randomly collecting $10^{6}$ different shapes of the spectrum able to satisfy all the observational constraints, consistently towards all cosmological epochs and scales. The results prove that relaxing that assumption of power-law spectrum \textit{on high frequencies}, the value of the tensor tilt becomes basically uncorrelated with $\Delta N_{\rm eff}^{\rm GW}$ so that models with the same $n_{\rm T}$ can contribute very differently to the energy budget of the Universe. Additionally, we work within the framework of the EFT of inflation and follow the methodology of the Hubble Flow Equation. We solve numerically the truncated system of the Hubble Flow Equations for a set of suitably defined initial conditions and we produce an ensemble of very general physical models studying the resulting observable predictions. We showed that in most cases extrapolating a power-law behavior over 24 orders of magnitude in frequency leads to overestimating the power in gravitational waves, above all on the ultraviolet frequencies that are the most relevant in the calculation. As a result, the predicted relic energy-density in gravitational wave can be ultimately incorrect. 

In \textbf{Section \ref{sec:CMBPGW}} we provide new updated constraints on slow roll inflation analyzing different extended scenarios beyond the $\Lambda\rm{CDM}$ cosmological model. Together with the usual six standard parameters, we simultaneously vary different combinations of additional parameters, including the running of the scalar spectral index $\alpha_s\doteq dn_s/d\log k$, its running of running $\beta_s\doteq d\alpha_s / d\log k$, the tensor amplitude $r\doteq A_T / A_s$ and the spatial curvature $\Omega_k$. As concerns the spectrum of primordial scalar perturbations, analyzing the different combinations of the Planck, lensing, BAO and BK15 data, we find no evidence for a scalar running or a running of running. On the other hand, analyzing the ACTPol+WMAP data we find a preference for nonzero $\alpha_s$ and $\beta_s$ at the level of $2.9\sigma$ and $2.7\sigma$, respectively (\textit{i.e.}, both at about 99\% CL). Anyway, such a preference is reduced when the running of running is replaced by tensor amplitude in the model. Regarding the spectrum of inflationary gravitational waves, we provide different upper bounds on the tensor amplitude, with $r<0.0658$ at 95\% CL our most constraining bound for Planck+BK15 data at the pivot scale $k_{\star}=0.05\rm{Mpc}^{-1}$. This result remains stable in almost all the models considered in the work, implying for the inflationary energy scale $V^{1/4}_{\rm inf}\lesssim 2 \times 10^{16}\,\rm{GeV}$. Furthermore, given the constraints on the tensor spectrum and the upper limits on the tensor amplitude, we show that the slow-roll consistency relations strongly reduce the parameter space allowed for the tensor spectrum, basically predicting a scale-independent tensor tilt unless corrections of order $d n_{T}/d\log k \lesssim 10^{-5}$, for all the datasets. Always using the slow-roll relation, we provide constraints on the slow-roll parameters $\{\epsilon_\Vl \, , \, \eta_\Vl \, , \, \xi_\Vl^2 \, , \, \varpi^3_\Vl\}$ that define the shape of the inflationary potential or similarly on the parameters $\{\eh,\etah,\xi^2_\Hl,\varpi^3_\Hl\}$ that define the dynamics of the background expansion. We then compare the theoretical predictions of some selected inflationary models- The extensions of the standard model considered in this paper recast the global tension between the datasets already present for a $\Lambda$CDM model analysis on a difference between the inflationary parameters. Finally, we vary the spatial curvature parameter $\Omega_k$ in realistic inflationary models that include both a tensor amplitude and a scalar running. Since the vast majority of inflationary models predict flatness, the constraints on the spatial curvature provide an important consistency check of this standard scenario. Interestingly, analyzing Planck(+BK15) data we instead find a preference for a closed cosmological spacetime at $2.4\,\sigma$ ($2.6\,\sigma$), while no relevant evidence is obtained adding also lensing and BAO to Planck or analyzing the Atacama Cosmology Telescope and the South Pole Telescope data. Maintaining an agnostic perspective on the spatial geometry, we investigate the possible consequences of a curved cosmological spacetime for the inflationary slow-roll background dynamics.

In \textbf{Section \ref{sec:PSOGFNSN}} we promote the $\lcdm$ model to the \slcdm with the introduction of two additional parameters: $\epsilon$, which comes from the inflationary theory, and $A_0$, a noise parameter with zero average that mimics the NG covariance contribution according to the Super-sample signal. With this setup, we  modified the theoretical code \texttt{CAMB} and explored the parameter space allowing the neutrino sector of the Universe.  We have not found any correlation between NG and $\nnu$ or $\mef$, but the indication for a negative correlation of $A_0$ with $\mnu$ is signaled. In particular, as our analysis suggests a slight indication of $A_0<0$, this implies greater leeway for massive neutrinos. We have then assumed that, despite being in the \textit{intermediate} shape, there is an alignment with the local shape with a similarity$\sim1$. Then, the trispectrum constraint found in literature can be applied. We saw that, albeit being more stringent, we still obtain more relaxed upper bounds with respect to the case when NG are not considered. Setting $A_0=0$ might consequently skew the constraints on the total neutrino mass, making them overly stringent. 

In \textbf{Section \ref{sec:PolJWST}} we conducted a joint analysis of the latest observations of the CSMD from high redshift galaxies measured by the JWST satellite and the constraints on CMB anisotropies from the Planck experiment\footnote{Being present numerous assumptions involved and the potential for systematic uncertainties in the data, the result presented should be considered valuable mainly for future analyses that will use less premature data.}.
We find that the efficiency of star production from baryons, $\epsilon$, plays a crucial role in our analysis to the point that the tension between the full Planck dataset and the JWST dataset can potentially be eased by increasing the efficiency parameter $\epsilon$. When $\epsilon=0.32$, the tension becomes below $3$ standard deviations, offering a possible resolution to the discrepancy with higher efficiencies. However, we found that if the Planck large angular scale EE data is omitted, there is a much better agreement between the Planck temperature data alone and the CSMD from JWST even for a lower efficiency of $\epsilon=0.2$. The combined analysis of Planck TT+JWST data indicates a shift towards higher values of the $\sigma_8$ parameter compared to the standard analysis using the full Planck dataset. This shift is possible thanks to the exclusion of large-scale polarization data. For $\epsilon=0.2$, the Planck TT+JWST analysis yields constraints on the $\sigma_8$ parameter that deviate from the constraints derived from the full Planck dataset by approximately $4$ standard deviations. Interestingly, the Planck TT+JWST analysis also provides a higher value for the Hubble constant of $H_0 = 69.0\pm 1.1$ at $68 \%$ C.L., suggesting a potential reconciliation with current local measurements. It is also important to note that since the reionization of the universe is primarily caused by the energetic radiation emitted by the first generation of stars and galaxies, the higher the redshift of massive galaxies, the higher the expected redshift of reionization. Consequently, this would result in a larger polarization signal, which contrasts with Planck's observations. Considering the challenges associated with such measurements, one could argue that a possible (even partial) solution to the current tensions lies in a systematic issue with the Planck polarization data at low multipoles.

Finally, in \textbf{Section \ref{sec:JDE}} we changed the perspective when tackling the issue of Massive galaxies observed by JWST. We took aside the hypothesis of any systematic in the observed data and we explore the possibility that these emerging anomalies could be somehow connected to other long-standing cosmological puzzles, such as the Hubble tension. We took the four independent JWST datasets at face value, we first study the correlation between the JWST likelihood and the fundamental six-$\Lambda$CDM parameters and we notice that it is sensitive to a phenomenology that is very common in models featuring new physics in the dark energy sector (both at early and late times). Inspired by this idea, we explore whether in the context of Early Dark Energy or Interacting Dark Energy, the JWST findings could be explained. We find that EDE constitutes an excellent candidate. Not only we can improve the agreement between the theoretical predictions of the model and the JWST data, but to achieve this, we move through the parameter space in the same direction needed to solve the Hubble tension. This underscores that it is indeed possible to address both issues within the same theoretical framework. Conversely, $w$IDE models featuring a dark energy equation of state $w \geq -1$ are generally disfavored from JWST, despite yielding higher values of matter clustering parameter $\sigma_8$. This is due to the energy flow from the dark matter sector to the dark energy one, implying a smaller $\Omega_m$. On the other hand, when the equation of state is confined to the phantom regime $w < -1$, the situation becomes somewhat more intricate. Whether, and to what extent, a phantom $w$IDE model could effectively contribute to explaining the anomalies observed in JWST findings remains an open question. The energy-momentum dynamics and parameter degeneracies can lead to a significant increase in the matter component, which in turn can slightly improve the agreement between the model predictions and the JWST data while also yielding values of $H_0$ in line with local distance ladder measurements. However, the ability of this model to address these two issues simultaneously remains limited compared to EDE.

Collectively, this thesis demonstrates how precision cosmology can be employed to probe and constrain new physics beyond the standard $\Lambda$CDM model. By exploring extensions in inflationary dynamics, the early Universe's radiation content, structure formation, and primordial non-Gaussianities, we have provided deeper insights into the mechanisms that shaped the early Universe. Our work highlights the power of cosmic observables in testing fundamental physics and guiding the development of more comprehensive cosmological models to accommodate emerging observational data.

%% file: Appendices_More/Acknowledgment.tex
\noindent First and foremost, I would like to express my deep gratitude to my PhD supervisor, \textbf{Alessandro Melchiorri}, for granting me the independence to explore my academic curiosity throughout my doctoral journey. Our collaboration began during my Master’s thesis, and over the years, his trust and guidance have allowed me to grow as a researcher. His willingness to let me pursue my own ideas has been invaluable in shaping the scientist I am today.
\vspace{2mm}


\noindent I am immensely grateful to \textbf{Eleonora Di Valentino}, who has been as a mentor to me, since my Master’s studies. Despite not being your official student, you have always made time to listen to my doubts and concerns, no matter how busy you were. You have been a constant source of encouragement and you have offered valuable advice, whether it concern specific physics issues or the broader existential questions that often accompany a PhD student. Your dedication to my academic and personal growth has left an indelible mark on my path as a researcher.
\vspace{2mm}

\noindent Equally, if not more, I wish to express my deepest gratitude to \textbf{William Giarè}. Words may fall short in capturing the profound influence you have had on my development as a physicist. Your door was always open, and your patience and availability made the 2,125 kilometres between us feel non-existent. You have been both a guide and a constant source of encouragement and your ability to teach with such clarity and understanding is something I am truly grateful for. I genuinely believe you are one of the finest physicists I know, and your dedication to research, combined with your humility, is inspiring and will lead you to even greater achievements. From our very first email exchange in 2021, I gained not only a colleague but a true friend. I cherish our friendship as much as our professional collaboration, and I sincerely hope that we can continue working together for many years. Thank you for everything, you have truly been indispensable.
\vspace{2mm}

\noindent I would also like to thank \textbf{Carsten van de Bruck}, who welcomed me warmly during my visits in Sheffield. Your enthusiasm led to an engaging journey through neutrino cosmology, one I hope we can continue to explore. Our mutual love for Hans Zimmer's music was a fun bonus during our work together. My thanks extend as well to Enrico, Yujia, Mariaveronica, Elsa, Rita, Steffen, Andrea and all the members of the \textbf{CRAG research group}, whose warmth and stimulating discussions made all my visits to Sheffield both fun and academically enriching.
\vspace{2mm}


\noindent The first year of a PhD is often filled with uncertainty, and I am deeply grateful to \textbf{Salvatore Capozziello}, whose authority and availability helped ease this transition. His guidance, along with \textbf{Micol Benetti}’s kindness and patience, made that period much more manageable. Micol, you have not only been a great collaborator but also a friend who listened and offered support, both in physics and beyond. Your willingness to help and your understanding made a significant difference during the challenging times of my PhD. It has been a real pleasure working with both of you, and I sincerely hope that we can continue collaborating in the future.
\vspace{2mm}


\noindent I am also thankful to \textbf{Ruchika} for the many ideas, discussions, and collaborative work we shared. Her contributions to this thesis have been invaluable, and I deeply appreciate her insights and support throughout this process.
\vspace{2mm}


\noindent Additionally, I extend my thanks to \textbf{David Mota} for welcoming me into his research group and sharing truly fascinating ideas. I also wish to thank \textbf{Signe Riemer-Sorensen} for her support, and invaluable lessons on machine learning. I hope that we can continue our collaboration and achieve the goals we have set together.

%% file: Appendices_More/HarmonicOscillator.tex
\chapter{Quantization Methods}
\label{sjsdjk}
\subsubsection{Quantization of a One-Dimensional Harmonic Oscillator}
\label{app:HO1D}
Fixing the potential as
\begin{equation}
    V(x,t)=\frac12 m^2\omega^2(t)x^2,
\end{equation} 
we specialize in the case of a harmonic oscillator. The equation of motion takes the form
\begin{equation}
    \ddot{x}+\omega^2(t)x=0\,.
    \label{eq:HOEquation}
\end{equation}
The latter equation is both second-order and linear. Thus, the solution is determined by two hermitian operators\footnote{The hermiticity ensures that their spectrum is real as it should be for the classical correspondent is a real number.} $x(0)$ and $\dot{x}(0)$ and is linear as well. We shall now introduce a complex function $f(t)$ which satisfies \eq{HOEquation} and expand the operator $x$ in terms of creation and annihilation operators (which are time-independent)
\begin{equation}
    x(t)=f(t)a+f^*(t)a^\dagger.
    \label{eq:boxeda}
\end{equation}
In this form, the commutator \eq{HOcanonical2} gives
\begin{equation}
    \langle f,f\rangle[a,a^\dagger]=1,
\end{equation}
where the bracket notation is defined by
\begin{equation}
    \langle f,g\rangle\equiv\frac{im}{\hbar}(f^*g_{,t}-f^*_{,t}g).
    \label{eq:HOBracket}
\end{equation}
If \eq{HOEquation} holds both for $f$ and $g$, then the bracket \eq{HOBracket} is independent of time $t$. Without loss of generality, we assume from now on that the solution $f$ is chosen so that the real number $\langle f,f \rangle$ is positive. Once we have imposed such a condition, we can re-scale it 
\begin{equation}
    \langle f,f\rangle=1.
    \label{eq:HONOrmalization}
\end{equation}
This result brings us to
\begin{equation}
    \boxed{[a,a^\dagger]=1}\,,
    \label{eq:HOcommutatora}
\end{equation}
that is the standard relation for raising and lowering operators of a harmonic oscillator. We can pluck out the annihilation and creation operators from
\begin{gather}
    \langle f,x\rangle=\frac{im}{\hbar}(f^*\dot{f}a+\cancel{f^*\dot{f}^*a^\dagger}-\dot{f}^*fa-\cancel{\dot{f}^*f^*a^\dagger})=\langle f, f\rangle a=a,
    \label{eq:HOannhilator}\\
    \langle f^*,x\rangle=\frac{im}{\hbar}(\cancel{f\dot{f}a}+f\dot{f}^*a^\dagger-\cancel{\dot{f}fa}-\dot{f}f^*a^\dagger)=-\langle f, f\rangle a^\dagger=-a^\dagger.
\end{gather}
We can now define the vacuum state $\ket{0}$ satisfying the condition
\begin{equation}
   a\ket{0}=0.
   \label{eq:HOVacuum}
\end{equation}
Excited states of the system are created by repeated application of the creation operators
\begin{equation}
   \ket{n}\equiv\frac{1}{\sqrt{n!}}(a^\dagger)^n\ket{0},
   \label{eq:HONstates}
\end{equation}
where the normalized states are eigenstates of the number operator $N=a^\dagger a$ with eigenvalue $n$
\begin{equation}
    N\ket{n}=n\ket{n}.
\end{equation}The span of all these states defines a Hilbert space of \textit{excitations} above the vacuum state. 

We have not determined the mode function $f$ uniquely as we have only imposed \eq{HONOrmalization}. As a result, the vacuum state is not fixed: a change in $f(t)$ could be accompanied by a change in $a$ that keeps the solution $x(t)$ unchanged (see \eq{HOannhilator}). If we consider only time-dependent frequencies $\omega(t)$,  there is in fact no unique notation of the vacuum. On the other hand, if we focus on the special case of a constant frequency $\omega(t)=\omega$, the energy is conserved and we can make a particular choice of $f$ in order to make the vacuum state the ground state of the Hamiltonian. We write down the relation
\begin{equation}
    H=\frac12 m\dot{x}^2+\frac12m\omega^2x^2,
    \label{eq:HOHamiltonian}
\end{equation}
\eq{HOHamiltonian} represents the Hamiltonian of the system which, for a general mode function $f(t)$, becomes
\begin{equation}
    H=\frac12m\left[(\dot{f}^2+\omega^2 f^2)aa+(\dot{f}^2+\omega^2f^2)^*a^\dagger a^\dagger+(\lvert\dot{f}\rvert^2+\omega^2\lvert f\rvert^2)(aa^\dagger+a^\dagger a)\right].
\end{equation}
If we now use \eq{HOVacuum} and \eq{HOcommutatora}, we find the action of the Hamiltonian operator on the vacuum state
\begin{equation}
    H\ket{0}=\frac12m(\dot{f}^2+\omega^2f^2)^*a^\dagger a^\dagger\ket{0}+(\lvert\dot{f}\rvert^2+\omega^2\lvert f\rvert^2)\ket{0}.
\end{equation}
As we want $\ket{0}$ be the eigenstate of $H$, the  first term must vanish, which requires
\begin{equation}
   \boxed{ \dot{f}=\pm i\omega f}\,,
   \label{eq:HOffixed}
\end{equation}
and thus,
\begin{equation}
    \langle f,f\rangle= \mp \frac{2m\omega}{\hbar}\lvert f \rvert^2.
\end{equation}
Since we have asked the bracket to be a positive number, we select from \eq{HOffixed} the minus sign. If we now impose \eq{HONOrmalization}, we obtain what is called the \textit{positive frequency solution}
\begin{equation}
    \boxed{f(t)=\sqrt{\frac{\hbar}{2m\omega}}e^{-i\omega t}}\,.
    \label{eq:HOPositivefrequency}
\end{equation}
With $f$ given by \eq{HOPositivefrequency}, the Hamiltonian becomes
\begin{equation}
    H=\frac12\hbar\omega(aa^\dagger+a^\dagger a)=\hbar\omega\left(N+\frac12\right);
    \label{eq:HHO}
\end{equation}
the minimum energy state is the one with $N=0$, and \textit{zero point energy} $\hbar\omega/2$. This is just the state $\ket{0}$ annihilated by $a$. If we used any other function than \eq{HOPositivefrequency}, the state annihilated by $a$ would differ from the ground state of the oscillator. As last consideration, although the mean value of the position is zero in the ground state, the means of its square is
\begin{equation}
    \bra{0}x^2\ket{0}\equiv\bra{0}x^\dagger x\ket{0}=\bra{0}(f^*a^\dagger+fa)(fa+f^*a^\dagger)\ket{0}=\lvert f(\omega,t)\rvert^2\bra{0}[a,a^\dagger]\ket{0}
\end{equation}
and hence, \textit{the zero-point fluctuations} of the position in the vacuum state are identified as the square of the mode function
\begin{equation}
    \boxed{\langle \lvert x\rvert^2\rangle=\lvert f(\omega,t)\rvert^2=\frac{\hbar}{2m\omega}}\,.
    \label{eq:HOFluctuations}
\end{equation}

\subsubsection{Quantization of a Scalar Field in Curved Spacetime}
\label{app:qsfcs}
Let us see how to quantize a scalar field in curved spacetime~\cite{Jacobson:2003vx}. For simplicity, for the rest of the chapter we will use the notation $\M=1$.
In an arbitrary dimension $D$, the action of a generic scalar field $\phi$ is
\begin{equation}
    S=\int{d^Dx\sqrt{-g}\frac12(\partial^\mu\phi\partial_\mu\phi-(m^2+\xi R)\phi^2)},
\end{equation}
and the equation of motion is
\begin{equation}
    (\Box +m^2+\xi R)\phi=0.
    \label{eq:EOMQUaQUa}
\end{equation}
\begin{tcolorbox}[mybox]
If the coupling $\xi$ vanishes, we restrict ourselves to the case of \textit{minimal coupling} and the equation of motion turns to be the Klein-Gordon equation. If also the mass vanishes, we have a \textit{massless minimally coupled scalar}. Furthermore, we have the special case of \textit{conformal coupling} with 
\begin{equation}
    m=0,\qquad \xi=\frac{D-2}{4(D-1)}.
    \label{eq:conformalCoupling}
\end{equation}
Let us briefly introduce it. If we make a conformal transformation
\begin{equation}
    \widetilde{g}_{\mu\nu}=\Omega^2(x)g_{\mu\nu},
\end{equation}
we also have
\begin{gather}
    \widetilde{g}^{\mu\nu}=\Omega^{-2}(x)g^{\mu\nu},\\
    \widetilde{R}=\Omega^{-2}(x)\left(R-2(D-1)\Box\ln{\Omega}-(D-1)(D-2)g^{\alpha\beta}(\ln{\Omega})_{,\alpha}(\ln{\Omega})_{,\beta}\right).
\end{gather}
With $D=2$, the action is invariant without any change of the scalar field $S(\phi,g)=S(\phi,\widetilde{g})$. In any other dimension, the kinetic term is invariant only if (the scalar field has dimension $[l]^{\frac{2-D}{2}}$).
\begin{equation}
    \Omega(x)=\Omega=constant,\qquad \widetilde{\phi}=\Omega^{\frac{2-D}{2}}\phi.
\end{equation}
On the other hand, if $\Omega$ is not constant, it can be shown that the action is invariant only if $\xi$ is chosen as in \eq{conformalCoupling}. In this case we have $S=(\phi,g)=S(\widetilde{\phi},\widetilde{g})$.
 \end{tcolorbox}

We now promote $\phi$ and $p$ to hermitian operators and we impose the canonical commutation relation
 \begin{equation}
     [\phi(x'),p(y')]=i\hbar\delta^{D-1}(x',y').
     \label{eq:commutatorQua}
 \end{equation}
 We introduce the \textit{Klein-Gordon inner product}, \ie a conserved bracket of two complex solutions to the scalar wave equation
 \begin{equation}
     \langle f,g\rangle=\int_{\Sigma}{d\Sigma_\mu j^\mu},\quad j^\mu(f,g)=\frac{i}{\hbar}\sqrt{-g}g^{\mu\nu}\left(f^*g_{,\nu}-f^*_{,\nu}g\right),
     \label{eq:KGinnerProduct}
 \end{equation}
which is independent of the space-like surface $\Sigma$ and satisfies
\begin{equation}
    \langle f,g\rangle^*=-\langle f^*,g^*\rangle=\langle g,f\rangle,\qquad \langle f,f^*\rangle=0,
\end{equation}
from which we impose
\begin{equation}
    a(f)=\langle f,\phi\rangle.
    \label{eq:AnnhiliatorScalar}
\end{equation}
As both $\phi$ and $f$ satisfy the wave equation, $a(f)$ is well-defined and independent of the surface on which the bracket is evaluated. From \eq{AnnhiliatorScalar} it follows
\begin{equation}
    a^\dagger(f)=-a(f^*).
\end{equation}
Using the relation in \eq{commutatorQua}, we obtain
\begin{equation}
    [a(f),a^\dagger(g)]=\langle f, g\rangle,\quad [a(f),a(g)]=-\langle f, g^*\rangle\quad\text{and}\quad [a^\dagger(f),a^\dagger(g)]=-\langle f^*,g\rangle.
    \label{eq:ConditionssFock}
\end{equation}
Eventually, if $f$ is a positive norm solution with unit norm $\langle f,f\rangle=1$, then the creator and annihilator operators satisfy the usual commutation relation for a harmonic oscillator (see \appx{HO1D}). We now call $\ket{\psi}$ the normalized state which satisfies
\begin{equation}
    a(f)\ket{\psi}=0
\end{equation}
but it is not completely specified. The normalized eigenstate of the number operator $N(f)=a^\dagger(f)a(f)$ with eigenvalue $n$, is
\begin{equation}
    \ket{n,\psi}=\frac{1}{\sqrt{n!}}(a^\dagger(f))^n\ket{\Psi}.
    \label{eq:FOck}
\end{equation}
However, we have still not defined a Hilbert space. In the span of all the states, \eq{FOck} defines only a Fock space of $n$-\textit{particle excitations} above the state $\ket{\psi}$. In order to construct the full Hilbert space, we shall find a decomposition of the space of complex solutions to the wave equation $\mathcal{S}$ into 
\begin{equation}
    \mathcal{S}=\mathcal{S}_p\oplus \mathcal{S}^*_p,
    \label{eq:directSS}
\end{equation}
which is the direct sum of both a positive norm and its complex conjugate subspaces, such that
\begin{equation}
    \langle f,f,\rangle >0 \quad \forall f\in \mathcal{S}_p.
\end{equation}
Namely, each $f$ in $\mathcal{S}_p$ can be scaled in its own harmonic oscillator sub-algebra, and
    \begin{equation}
    \langle f,g^*\rangle =0 \quad \forall f,g\in\mathcal{S}_p,
    \label{eq:NhNial}
\end{equation}
which implies, according to \eq{ConditionssFock}, that 
\begin{equation}
    [a(f),a(g)]=[a^\dagger(f),a^\dagger(g)]=0.
\end{equation}
Then, we can define the total Hilbert space for the field theory as the space of finite norm sums of possibly infinitely many states. Its form is~\cite{Jacobson:2003vx}
\begin{equation}
    a^\dagger(f_1)...a^\dagger(f_n)\ket{0},\quad\text{with}\quad a(f)\ket{0}=0\quad \forall f, f_n\in\mathcal{S}_p.
    \label{eq:FockVacuum}
\end{equation}
The state $\ket{0}$ is the \textit{Fock vacuum} and depends on the decomposition \eq{directSS} and in general is not the ground state. The representation of the field operator on this Fock space is hermitian and satisfies the canonical commutation relations.

\subsubsection{Special Case of Flat Spacetime}
Let us now apply what we have found so far to the case of a massive scalar field in a flat spacetime. A natural decomposition of the space solutions \eq{directSS} is the positive and negative frequency with respect to a Minkowski time translation. Besides, the corresponding Fock vacuum \eq{FockVacuum} is the ground state. As we are in an infinite volume of space, plane wave solutions are not normalizable. Thus, we will consider periodic boundary conditions so that space becomes a large three-dimensional torus with circumferences $L$ and volume $V=L^3$. The allowed wave vectors are $\boldsymbol{k}=(2\pi/L)\boldsymbol{n}$, where the components of $\boldsymbol{n}$ are integers. 

In \eq{EOMQUaQUa} $\Box$ becomes the flat d'Alambertian and $R=0$. A complete set of modes to this equation is given by
\begin{equation}
    f_{\boldsymbol{k}}(t,\boldsymbol{x})=\sqrt{\frac{\hbar}{2V\omega(\boldsymbol{k})}}e^{-i\omega(\boldsymbol{k})t}e^{i\boldsymbol{k}\vdot\boldsymbol{x}},
\end{equation}
where
\begin{equation}
    \omega(\boldsymbol{k})=\sqrt{\boldsymbol{k}^2+m^2}.
\end{equation}
The bracket satisfies:
\begin{gather}
    \langle f_{\boldsymbol{k}}, f_{\boldsymbol{l}}\rangle=\delta_{\boldsymbol{k,l}},\\
    \langle f^*_{\boldsymbol{k}}, f^*_{\boldsymbol{l}}\rangle=-\delta_{\boldsymbol{k,l}},\\
    \langle f_{\boldsymbol{k}}, f^*_{\boldsymbol{l}}\rangle=0.
\end{gather}

So, we have an orthogonal decomposition of the solution space into positive norm solutions and their conjugates as in \eq{NhNial}, with $\mathcal{S}_p$ the space spanned by the positive frequency modes $f_{\boldsymbol{k}}$. As in the general case, we get a Fock space representation. We proceed to define the annihilator operator associated to $f_{\boldsymbol{k}}$,
\begin{equation}
    a_{\boldsymbol{k}}=\langle f_{\boldsymbol{k}},\phi\rangle,
\end{equation}
from which we can decompose the field operator
\begin{equation}
    \phi=\sum_{\boldsymbol{k}}{\left(f_{\boldsymbol{k}}a_{\boldsymbol{k}}+f^*_{\boldsymbol{k}}a^\dagger_{\boldsymbol{k}}\right)}.
\end{equation}
All the $f_{\boldsymbol{k}}$ have positive frequency, the Hamiltonian is a sum over the contribution from each $\boldsymbol{k}$ value and define the vacuum state
\begin{equation}
    a_{\boldsymbol{k}}\ket{0}=0
    \label{eq:007}
\end{equation}
as the ground state for all $\boldsymbol{k}$. The states
\begin{equation}
    a^\dagger_{\boldsymbol{k}}\ket{0}
\end{equation}
have momentum $\hbar\boldsymbol{k}$, energy $\hbar\omega(\boldsymbol{k})$ and are interpreted as single particle states. Although the field Fourier component $\phi_{\boldsymbol{k}}=f_{\boldsymbol{k}}a_{\boldsymbol{k}}+f^*_{\boldsymbol{k}}a^\dagger_{\boldsymbol{k}}$ has zero mean in the vacuum state, in the same way as the harmonic oscillator (see \appx{HO1D}), the zero-point fluctuations are characterized by
\begin{equation}
\bra{0}\phi^\dagger_{\boldsymbol{k}}\phi_{\boldsymbol{k}}\ket{0}=\lvert f_{-\boldsymbol{k}}\rvert^2=\frac{\hbar}{2V\omega(\boldsymbol{k})}\,.
\end{equation}

%% file: Appendices_More/Gravitons.tex
\chapter{Gravitons}
\label{ewjkq}
Looking at \eq{Lobello} we see that, once we have introduced the variable $\omega_k$ such that
\begin{equation}
    f_k''+\omega_k+\omega^2_k(\eta)f_k=0,
\end{equation}
the equation describes an harmonic oscillator with time-dependent frequency (see for example \appx{HO1D}), whose energy per mode is described by
\begin{equation}
    H_k\equiv \frac12(\lvert f'_k\rvert^2+\omega^2_k\lvert f_k\rvert^2)
\end{equation}
 When $\omega^2_k(\eta)$ varies in time adiabatically, namely
\begin{equation}
    \omega_k'\ll\omega_k^2\quad\text{for}\quad \omega^2_k>0,
\end{equation}
we can associate an occupation number $n_k$ to each mode $k$ so that $\lvert \Delta \boldsymbol{k}\rvert^3n_k$ represents the number density of quanta, called \textit{gravitons}, with momenta $[\boldsymbol{k},\boldsymbol{k}+\Delta\boldsymbol{k}]$. The occupation number is given by the ratio of the energy per mode $E_k$, that is analogous to the eigenvalue of $H$ in the one-dimensional harmonic oscillator, and $\omega_k$
\begin{equation}
    n_k+\frac12\equiv \frac{1}{2\omega_k}(\lvert f'_k\rvert^2+\omega^2_k\lvert f_k\rvert^2).
    \label{eq:Occupationnumber}
\end{equation}
Here, we should substitute $f_k$ to its expression according to the scale. At sub-horizon scales, we have \eq{SubSUbBUS} and consequently $n_k=0$, as it should for vacuum in flat spacetime. Nevertheless, the frequency is red shifting and at some point it becomes too small to keep up with Hubble expansion. At that point, while the frequency goes to zero, the state remains trapped in the vacuum state of the would-be harmonic oscillator with frequency $k\sim a_\star H_\star$. Under the condition of adiabatically variation for $\omega_k$, the occupation number to each mode is an adiabatic invariant. However, the inflationary expansion violates both conditions, resulting in an abundant production of gravitons as the modes leave the Hubble radius. Strictly speaking, the occupation number is not well-defined during inflation as there is no adiabatic conservation. We evaluate it just after inflation assuming an instantaneous transition into a power law expansion era. Therefore, we consider $a(\eta)\propto \eta^p$ and for simplicity the approximate result in de Sitter stage. For $k\ll H_\star a_\star$, from \eq{HomeAlone} and the definition \eq{Lobello} we can write explicitly the terms into \eq{Occupationnumber} and find 
\begin{gather}
    n_k\sim \left(\frac{H_k}{H_\star}\right)^2\left(\frac{a_\star H_\star}{k}\right)^3\quad\text{if}\quad p\neq 1\\
     n_k\sim \left(\frac{H_k}{H_\star}\right)^2\left(\frac{a_\star H_\star}{k}\right)^4\quad\text{if}\quad p= 1
\end{gather}
which corresponds, respectively, to the matter-dominated and the radiation-dominated phase, according to \tab{Table}. Thus, super-Hubble modes exhibit a very large occupancy $n_k\gg1$, as it corresponds to a large ensemble of gravitons. The original quantum nature of the tensor perturbations is lost (due to the time evolution leading to squeezing), but reflected in the stochastic nature of the emerging effectively classical field distribution \cite{Polarski:1995jg}. We can also check indeed that the wave function is semiclassical. We should see if the $\phi$-length scale over which the amplitude of the wave function changes is much longer than the $\phi$-length scale over which the phase changes. It is possible to show that~\cite{Senatore:2016aui}
\begin{equation}
    \Delta\phi_{AV}\sim \frac{H}{k^{\frac32}}\frac{1}{V^{\frac12}},\quad \Delta\phi_{PV}\sim \frac{H}{k^{\frac32}}\frac{1}{V^{\frac12}}\left(\frac{k}{aH}\right)^{\frac12} 
\end{equation}
where $PV$ and $AV$ stand for amplitude variation and phase variation, respectively, and we have reintroduced the potential $V$. So their ratio is
\begin{equation}
    \frac{\Delta\phi_{PV}}{\Delta\phi_{AV}}\sim \left(\frac{k}{aH}\right)^{\frac12} \rightarrow 0.
\end{equation}
This means that the system approaches a classical stochastical description on super-horizon scales. 

In general, taken a background of stochastic gravitational waves whose source is in the early Universe, the amplitude of the tensor perturbation is a random variable which can be characterised only statistically by means of ensamble average. We always have the one-observable-universe problem; namely, we invoke the hypothesis of ergodicity and the fair sample hypothesis. However, besides the usual requirement of an almost homogeneous and isotropic Universe, we shall add two more assumptions:
\begin{enumerate}
    \item The initial conditions of the gravitational wave generating process are the same at every point in space.
    \item The gravitational wave source fulfills causality and operates at a time when the typical size of a region of causal contact in the Universe was smaller than the causal horizon today.
\end{enumerate}
Under these conditions, the signal of the gravitational waves from the early Universe takes the form of a stochastic background. Obviously, such conditions are not satisfied in the inflationary stage as the causal horizon grows exponentially. Nevertheless, the primordial gravitational waves' signal is still a \textit{stochastic background}~\cite{Allen:1996vm}. The reason resides in the intrinsic quantum nature of the generating process as we have demonstrated above.

%% file: Appendices_More/DataAnalysis.tex
\chapter{Data Analysis} 
\label{app:DATa}

There are two different ways of defining what probability is; if we define it as the long-term frequency of an event occurring in repeated independent trials, we are assuming a \textit{frequentist} interpretation. This point of view depends on what the experimenter thinks about the probability of data that have not been observed. On the other hand, if we think the probability as a measure of the degree of belief about a proposition, we are adopting the \textit{Bayesian} method. Bayesian statistics only deals with the data that were actually observed~\cite{Trotta:2008qt} and is particularly useful in situations where repeated trials are not feasible. For this reason, the latter interpretation is the most used in cosmology and is the one we are going to focus on, in this Appendix. First and foremost, let us introduce the fundamental principles of statistical inference~\cite{D_Agostini_2003,Trotta:2017wnx,Trotta:2008qt}.

\subsubsection{Elementary Notions}

We call $P(A)$ the probability of the statement $A$. If $A$ is conditioned to a certain information $I$, we write $P(A|I)$ where $I$ is assumed to be true\footnote{For example, the probability ($P$) to find a job ($A$) with a degree in physics ($I$).}. The fundamental rules to bear in mind when dealing with probabilities are the \textit{sum rule}, which states that the probability of $A$ plus the probability of its complement $\bar{A}$ is always equal to one, and the \textit{product rule} that defines the joint probability of two events $A$ and $B$ as the product of the probability of $B$ and the conditional probability of $A$, given $B$ (always under the conditional on information $I$)
\begin{equation}
P(A|I) + P(\bar{A}|I) = 1\,,\quad P(A, B|I) = P(A | B,I) P(B|I)\,.
\label{eq:sumprod}
\end{equation}
From these two relations, we can see that the probability of $B$ alone is easily obtained as\footnote{From the sum rule, $\sum_{A}P(A|B,I)=1$} 
\begin{equation}
    P(B|I)=\sum_A{P(A,B|I)}\,.
\end{equation}
Using the fact that the product rule is symmetric, we can now write down the most important (yet so simple) theorem, which defines the Bayesian method, and which is the ground base for all the data analysis: the \textit{Bayes Theorem}
\begin{equation}
P(A |B,I) = \frac{P(B|A,I) P(A|I)}{P(B|I)}.
\label{eq:BTab}
\end{equation}

Before further exploring the implications of the Bayes theorem, we should briefly digress, introducing the case of $A$ replaced by a continuous random variable $x$. The probability distribution for $x$ is called the probability density function (pdf), $p(x)$, and $p(x')dx'$ represents the probability that the random variable $x$ falls within the interval $[x', x'+dx']$. The cumulative distribution function (cdf) for a continuous random variable is defined as
\begin{equation}
P(x) = \int_{-\infty}^{x} p(y) dy.
\end{equation}
Given a distribution, two important proprieties are the expectation value $\langle\cdot\rangle$ (which controls the location of the distribution) and the variance $\text{Var}(\cdot)$ (which controls how much the distribution is spread out)
\begin{equation}
    \langle x\rangle \equiv \int{x'p(x')dx'},\quad \text{Var}(x)\equiv \langle x^2\rangle-\langle x\rangle^2=\int{x^{'2}p(x')dx'}-\(\int{x'p(x')dx'}\)^2\,.
\end{equation}
An important result in probability theory is the \textit{Central Limit Theorem:} Let $ x_1, x_2, \dots, x_N $ be independent random variables with finite means $\mu_i$ and variances $\sigma_i^2$. The normalized sum
\begin{equation}
Y = \sum_{i=1}^{N} \frac{(x_i - \mu_i)}{\sqrt{\sigma_i^2}}
\end{equation}
will converge to a Gaussian distribution with mean $0$ and variance $1$ as $N$ increases.

\begin{tcolorbox}[mybox]
The Gaussian pdf is a continuous distribution with mean $\mu$ and standard deviation $\sigma$ given by
\begin{equation}
    p(x|\mu,\sigma)=\frac{1}{\sqrt{2\pi\sigma^2}}e^{-\frac12\frac{(x-\mu)^2}{\sigma^2}}
\end{equation}
the expectation value is $\mu$ and the variance $\sigma^2$. The probability content of a Gaussian of standard deviation $\sigma$ for a given symmetric interval around the mean of width $k\sigma$ on each side is given by
\begin{equation}
    P(\mu-k\sigma<x<\mu+k\sigma)=\text{erf}\(\frac{k}{\sqrt{2}}\)
\end{equation}
with
\begin{equation}
    \text{erf}(x)=\frac{2}{\sqrt{\pi}}\int^x_0e^{-y^2}dy
\end{equation}
Some commonly used values are $k=1$ which gives $0.683$ probability of content and $k=2$ with $0.954$ probability of content.

Let us now define the random variable $\chi^2$ as the sum of the squares of n standardised independent identically distributed Gaussian random variables $x_1,x_2,\dots,x_N$ 
\begin{equation}
    \chi^2=\sum_{i=1}^{N}(x_i-\mu)^2/\sigma^2
\end{equation}
the $\chi^2$ follows the Chi-Square distribution with $N$ degrees of freedom
\begin{equation}
    p(\chi^2)=\frac{1}{\Gamma(\frac{N}{2})2^{\frac{N}{2}}}\(\chi^2\)^{\frac{N}{2}-1}e^{-\frac12\chi^2}
\end{equation}
with $\langle x \rangle =N$ and $Var(x)=2N$
\end{tcolorbox}

In the right-hand side of \eq{BTab}, we see how given the evidence $B$ we can update the probability of a hypothesis $A$. For our purpose, it is straightforward to identify $A$ as the parameters $\theta$ for our theoretical model, and $B$, the observed data $d$, so that we end up with  
\begin{equation}
\boxed{p(\theta | d) = \frac{p(d|\theta) p(\theta)}{p(d)}}
\label{eq:BT}
\end{equation}
where we omit $I$ which, as we see below, can identify different underlying models in the model comparison case. In \eq{BT}, $p(\theta |d)$ is the posterior probability, which represents our degree of belief about the parameters $\theta$, after we observed $d$. This is what we are really looking for: the probability distribution of the parameters given that we observed the data $d$. $p(d|\theta)$ is also called the likelihood $\mathcal{L}(\theta)$,  the probability of observing the data given the parameters. It is crucial to understand that the likelihood is not a probability distribution over $\theta$, but rather a function of $\theta$  given the observed data. The \textit{ Maximum Likelihood} (ML) Principle  asserts that the best estimate of the parameter $\theta$ is the one that maximizes the likelihood function $\theta_{\text{ML}} = \text{max}_{\theta} \mathcal{L}(\theta)$. To find the \textit{Maximum Likelihood Estimator} (MLE), one typically takes the derivative of the log-likelihood function with respect to the parameter $\theta$ and sets it to zero, whereas the second derivative is then checked to be less than zero to ensure that it corresponds to a maximum. If we Taylor expand $\ln \mathcal{L}$ around its maximum, we find
\begin{equation}
\mathcal{L}(\theta) \approx \mathcal{L}(\theta_{\text{ML}}) \exp \left( -\frac{(\theta - \theta_{\text{ML}})^2}{2 \Sigma_{\theta}^2} \right),
\label{eq:MLp}
\end{equation}
where $\Sigma_{\theta}^2$ is the variance of the estimator, given by:
\begin{equation}
\Sigma_{\theta}^2 = \left[ - \frac{\partial^2 \ln \mathcal{L}(\theta)}{\partial \theta^2} \Big|_{\theta_{\text{ML}}} \right]^{-1}.
\end{equation}
Therefore, the likelihood function can be approximated to second order as a Gaussian around the ML value which, in turns, implies we can express the uncertainity around the ML value as $\Sigma_\theta$. From this approximation we can identify the \textit{Fisher Matrix}~\cite{Fisher:1935,Tegmark:1996bz} as
\begin{equation}
    \boxed{F_{ij}\equiv-\frac{\partial^2\ln{\mcal{L}(\theta)}}{\partial\theta_i\partial\theta_j}\Big|_{\theta_{\rm ML}},\quad \Sigma^2_\theta=\frac{1}{F_{ii}}}\,.
\end{equation}
One of the many propriety of the Fisher matrix is the Cramér-Rao inequality~\cite{Kendall:1967} where we have that the variance of any unbiased estimate of a parameter must exceed the reciprocal of the diagonal element of the Fisher matrix $\langle\hat{\theta}_i^2\rangle\geq\Sigma^2_\theta$. Finally, the factor $p(\theta)$ in \eq{BT} is called the prior and it represents our prior knowledge on the parameters. We are expecting the posterior to be independent of it. It is common in literature to choose the prior to be flat $p(\theta)=(\theta_{\rm max}-\theta_{\rm min})^{-1}$ for $\theta\in\[\theta_{\rm min},\theta_{\rm max}\]$. In this case, the posterior becomes functionally identical to the likelihood up to a proportionality constant. Flat priors do not come without caveats, for example, a flat prior on a parameter $\theta$ does not correspond to a flat prior on a non–linear function of that parameter, $f(\theta)$. In fact, the prior on $f(\theta)$ becomes informative and is proportional to the Jacobian for the transformation. Furthermore, if we are ignorant about the scale of a quantity $\theta$, it can be shown~\cite{Jaynes:2003jaq} that the appropriate prior is flat on $\ln{\theta}$, which gives equal weight to all orders of magnitude. This prior corresponds to $p(\theta)\propto \theta^{-1}$ and is called Jeffreys’ prior. The fact that the posterior contains prior knowledge on the parameters makes the \textit{Maximum a Posteriori} (MAP) Principle a better method to compute the best parameters, where instead of looking for the parameter that makes the derivatives of the likelihood zero, we are looking for $\theta_{MP}$ such that $\partial p(\theta|d)/\partial \theta=0$. For flat uninformative priors MAP and ML are identical~\cite{Maltoni:2003cu,Gariazzo:2023joe}. From \eq{BT}, if we want the posterior for one parameter, we \textit{marginalize} (i.e. integrating over the uninteresting parameters)
\begin{equation}
    p(\theta_1|d)\propto \int{\mathcal{L}(\theta)p(\theta)d\theta_1d\theta_2\dots d\theta_N}\,.
    \label{eq:marginalize}
\end{equation}
Another way to "get rid" of a parameter, instead of marginalizing, is to have the posterior \textit{maximized} with repsect to one parameter. This happens when we fix the parameter at its MAP.

The denominator in \eq{BT} is a normalization factor that will play a crucial rol for model comparison. The posterior is a probability distribution of the parameters, therefore it has to be normialized to unity
\begin{equation}
    \int{p(\theta|d)d\theta}=1=\frac{\int{p(d|\theta)p(\theta)d\theta}}{p(d)}\longrightarrow \int{p(d|\theta)p(\theta)d\theta}=p(d)\,.\label{eq:BEv}
\end{equation}
 
To summarize the steps, in data analysis it is important to
\begin{enumerate}
    \item Define the parameters for our model we want to use to describe the observation.
    \item Use our theoretical knowledge to assign a prior to the parameters.
    \item Build a likelihood function that depends on the experiment and usually contains nuisance parameters.
    \item Obtain the posterior distribution
\end{enumerate}

The last point, is not easy to achieve. Usually analytical solutions do not exist and we need to rely on numerical simulations, in particular in the form of \textit{Monte Carlo Markov Chain} (MCMC) techniques (e.g.~\cite{Trotta:2017wnx,Verde:2009tu,Speagle:2019ffr})

\subsubsection{Monte Carlo Markov Chain}
\label{sec:MCMC}
A Markov chain is a sequence $\{0,\dots, t,\dots \}$ of samples, which in our case could be the set of parameters $\{\theta^{(0)},\dots,\theta^{(t)},\dots\}$ that describes the model, such that the probability of the $t+1$ element in the chain only depends on the value of the $t$ element. The sequence of generated points takes a a kind of random walk in parameter space. Two important proprieties are:
\begin{itemize}
    \item The proportionality of the samples density to the posterior pdf
    \item The chain converge to a stationary state where successive elements of the chain are samples from the target distribution, in our case the posterior $p(\theta|d)$.
\end{itemize}
Samples are generated through the  \textit{transition probability} $T(\theta^{(t)},\theta^{(t+1)})$, which gives the probability of moving from point $\theta^{(t)}$ to point $\theta^{(t+1)}$ in parameter space. A sufficient condition to obtain a Markov Chain is that the transition probability satisfies the \textit{detailed balance condition}
\begin{equation}  \frac{T(\theta^{(t)},\theta^{(t+1)})}{T(\theta^{(t+1)},\theta^{(t)})}=\frac{p(\theta^{(t+1)}|d)}{p(\theta^{(t)}|d)}\,.
\end{equation}
When the chain converged, the estimates of expectation value of any function of the parameters is
\begin{equation}
        \langle f(\theta)\rangle\approx\int{p(\theta|d)f(\theta) d\theta}=\frac{1}{M}\sum^{M-1}_{t=0}f(\theta^{(t)})
        \label{eq:espvalue}
\end{equation}
In \eq{espvalue} $t=0$ do not correspond to the initial starting point of the chain but to the new starting point, after having discarded the \textit{burn-in} period (usually $30\%$ of the chain). This is because at the beginning the sampling process do not follow the equilibrium distribution. It happens qualitatively when the chain has moved in the neighborhood of the posterior peak and consequently the curve of the log posterior as a function of the step number flattens. Furthermore, an important issue is to understadn when \eq{espvalue} is sufficiently accurate. A useful diagnostic tool is the \textit{Gelman and Rubin criterion}~\cite{Gelman:1992zz}; in short, this method compare several sequences from a set of random starting points and check if they are indistinguishable. This tests either convergence and poor mixing.
\begin{tcolorbox}[mybox]
Let us consider $M$ chains with $2N$ elements with a burn-in of $50\%$ (see e.g.~\cite{Verde:2007wf,Verde:2009tu,Heavens:2009nx}). $y$ denotes a chain element. Therefore, the mean of the chain is
\begin{equation}
    \bar{y}^j=\frac{1}{N}\sum_{i=1}^Ny^j_i
\end{equation}
where $j$ represents the j-th chain. The mean of the distribution is
\begin{equation}
    \bar{y}=\frac{1}{NM}\sum_{j=1}^{M}\sum_{i=1}^{N}y^j_i
\end{equation}
The variances between chains is
\begin{equation}
    B_n=\frac{1}{M-1}\sum_{j=1}^{M}(\bar{y}^j-\bar{y})^2
\end{equation}
and within a chain
\begin{equation}
    W=\frac{1}{M(N-1)}\sum_{j=1}^{M}\sum_{i=1}^{N}(y^j_i-\bar{y}^j)2
\end{equation}
the quantity
\begin{equation}
    \boxed{R=\frac{\frac{N-1}{N}W+B_n\(1+\frac1M\)}{W}}
    \label{eq:GR02}
\end{equation}
is the ratio of two estimates of the variance in the target distribution: the numerator is an estimate of the variance that is unbiased if the distribution is stationary, but is otherwise an overestimate. The denominator is an underestimate of the variance of the target distribution if the individual sequences did not have time to converge.
The convergence of the Markov chain is then monitored by recording the quantity $R$ for all the parameters and running the simulations until the values for $R$ are always $<0.02$~\cite{Gelman:1992zz}
\end{tcolorbox}
As a last remark, it is common to \textit{thinning} the chain, which means to select only one sample every $k$. This ensure a certain degree of independence between the samples from the posterior. The auto-correlation is a good measure of the number of steps required before the chain has “forgotten” its previous state~\cite{Dunkley:2004sv}. 

The $1$–dimensional marginal probability for the j–th element of $\theta$, $\theta_j$, is easily found from \eq{marginalize} and \eq{BT}
\begin{equation}
    p(\theta_1|d)=\int{p(\theta|d)d\theta_2\dots d\theta_n}.
\end{equation}

The widely used algorithm to perform MCMC analysis is the \textit{Metropolis-Hastings algorithm}~\cite{Metropolis:1953am,Hastings:1970aa}. It works as follows:
\begin{enumerate}
    \item We start from a random point $\theta^{(0)}$ with posterior $p_0\equiv p(\theta^{(0)}|d)$
    \item The next point $\theta^{(c)}$ is drawn from the proposal distribution $q(\theta^{(0)},\theta^{(c)})$ which can be a Gaussian of fixed width $\sigma$, centered around the current point.
    \item We evaluate the posterior at the new proposed point $p_c$ and compute the acceptance probability
    \begin{equation}
        \alpha=\text{min}\(\frac{p_c}{p_0},1\)
        \label{eq:acceptance}
    \end{equation}
    \item We generate a random number $r$ from the uniform distribution $[0,1)$ and accept the new step if $r<\alpha$ and rejecting otherwise. From \eq{acceptance} we see that whenever the next proposed value for the parameters has a larger posterior than the previous one, the step is always taken. 
    \item If the candidate point is accepted, we add it to the chain and move there. Otherwise stay at the old point (which is thus counted twice in the chain).
\end{enumerate}
By construction, the algorithm does not depend on the normalization constant, since what matters is the ratio of the pdf’s. From step $2$, it is easily to understand that the proposal distribution $q$ is crucial for an efficient exploration of the parameter space. If it is too small, then the algorithm spends too much time locally and takes a long time to explore the target distribution. The target density hardly changes and we end up with a case of poor mixing. If instead the scale of $q$ is too large, the chain gets stuck because the points are always rejected being sampled in places where the target density is low\footnote{Sometimes, the best course of action is to run an MCMC in order to get a covariance matrix. Then use it as the multivariate Gaussian proposal distribution for the new MCMC.}.

\begin{tcolorbox}[mybox]
Sometimes after an MCMC analysis, it is interesting to add a new dataset to the model. In that case,it is possible to do an \textit{importance sampling}. If $p$ is the posterior with the new dataset and $q$ the previous posterior, we can get the expectation value under $p$ by computing the one under $q$ but re-weighting it with the factor $p/q$~\cite{Trotta:2017wnx}. Given $f$ a generic function, we can write
\begin{equation}
    \langle f(\theta)\rangle \approx\frac{1}{M}\frac{\sum_{t=0}^{M-1} w_t f(\theta^{(t)})}{\sum_{t=0}^{M-1} w_t}
\end{equation}
where $w_t=p(\theta^{(t)})/q(\theta^{(t)})$ are the importance sampling weights and $\theta^{(t)}\sim q(\theta)$
\end{tcolorbox}

\subsubsection{Models comparison}

The simplest way to compare models is the frequentist approach. We need to simply evaluate the goodness of fit computing the $\chi^2$ statistics of the model. The best model is the one with the best statistics (combination of lowest $\chi^2$ and number of degrees of freedom). However this is too rough and we need to improve. Firstly because we re lacking priors, secondly because it penalize models with parameters unconstrained by the data (which might be a observational problem and not an issue of the theory). Hence, we need to improve our method of model comparison. 

In our context, a model $\mscr{M}$ is simply the combination of a set of parameters $\theta$ and of their prior distribution, $p(\theta|\mscr{M})$. To find the best model to fit of the data is pathological with respect to the complexity of the model. In fact, with a sufficient high number of parameters it is possible to fit almost every observed data. Therefore, it is important to quantify the necessity, for a model, to not only fit the data, but to ensure that all parameters are "essential". Which means \textit{Pluralitas non est ponenda sine neccesitate}\footnote{Franciscan monk William of Ockham (ca. 1285-1349)} or, in other words, the simplest theory compatible with the available evidence ought to be preferred. this is called the \textit{Occam's razor principle}. 

Let us introduce with another light the quantity presented in \eq{BEv}. It is an important tool to quantify the model performance and is called called \textit{Bayesian evidence}, here we explicitly write the model conditionality
\begin{equation}
\boxed{p(d|\mscr{M})\equiv\int_{\Omega_\mscr{M}}{p(d|\theta,\mscr{M})p(\theta|\mscr{M})d\theta}}
    \label{eq:BayesianEv}
\end{equation}
\eq{BayesianEv} is the average of the likelihood $p(d|\theta,\mscr{M})$ under the prior for a specific model choice. Also, the conditional information $I$ in \eq{BTab} in this case is our model $\mscr{M}$. The posterior probability for a model, given the data, is obtained simply using \ref{eq:BT} $p(\mscr{M}|d)\propto p(\mscr{M})p(d|\mscr{M})$ (sometimes called as \textit{doubt}~\cite{Starkman:2008py,March_2010}). It can give some insight looking at the analytical expression for the evidence when the the likelihood and the prior can be approximated by Gaussian distributions and the likelihood has its bestfit parameters $\theta_{ML}$ and $\sigma_{ML}$, and the prior with means $\theta_{P}$ and variances $\sigma_{P}$~\cite{Amendola:2015ksp}. It is 
\begin{equation}
    p(d)=\mathcal{L}(\theta_{ML})\frac{\sigma_{MP}}{\sigma_P}\text{exp}\{-\frac12\[\(\frac{\theta_{ML}}{\sigma_{ML}}\)^2+\(\frac{\theta_{p}}{\sigma_{p}}\)^2-\(\frac{\theta_{MAP}}{\sigma_{MAP}}\)^2\]\}
    \label{eq:evian}
\end{equation}
with 
\begin{equation}  \theta_{MAP}=\frac{\sigma_{ML}^2\theta_P+\sigma_{P}^2\theta_{ML}}{\sigma_{ML}^2+\sigma_{P}^2}\,,\quad \sigma_{MAP}=\frac{\sigma_{ML}^2+\sigma_{P}^2}{\sigma_{ML}^2+\sigma_{P}^2}
\end{equation}
\eq{evian} is composed by three factors. The likelihood computed at ML and express how well the model fits the data. Then we have the ratio of variances which is a ratio of parameter volumes because the variance tells us how much parameter space available we have for the parameters and therefor tells us how the parameter volume changes from the prior to the posterior. Also, since the ratio is smaller than one for each parameter, then adding more parameters penalizes the evidence, following the Occam's razor principle. On the other hand, unconstrained parameters have $\sigma_{ML}\gg\sigma_P$ so the ratio is close to unity. In the third factor, we see that if $\theta_{MAP}$ differ from the prior mean or ML parameters the evidence is penalized 

If we have two different models $\mscr{M}_0$ and $\mscr{M}_1$ and we want to check if $\mscr{M}_1$ is better at describing the data, we can use the Bayes theorem, introducing a prior assigned for each model, and we can introduce the \textit{Bayes factor}~\cite{Kass:1995loi,Marshall:2004zd}
\begin{equation}
    \frac{p(\mscr{M}_0|d)}{p(\mscr{M}_1|d)}=\frac{p(d|\mscr{M}_0)}{p(d|\mscr{M}_1)}\frac{p(\mscr{M}_0)}{p(\mscr{M}_1)}=B_{01}\frac{p(\mscr{M}_0)}{p(\mscr{M}_1)}
    \label{eq:BFactor}
\end{equation}
where usually $p(\mscr{M}_0)=p(\mscr{M}_1)$
According to our previous discussion, this tool is better than the simple $\chi^2$ statistics because, from the \eq{BayesianEv}, we see that if a certain parameter $\bar{\theta}$ is poorly constrained, then the likelihood is almost constant when varying $\bar{\theta}$. This implies that, being the prior factorizable in almost all the cases, the integral decouples and gives unity, not entering into the evidence integral and therefore the Bayesian factor. This is an improvement of the $\chi^2$ method because now, unconstrained parameters are rightly related to poor measurements rather than penalising the model.

If the Bayes factor $B_{01}$ is grater than one, it suggests that the original model is favored and we should not switch to $\mscr{M}_1$. In the other hand, if it is smaller than one, we have no evidenceto favor any model. To better quantify the strength of the evidence, we can use the modified~\cite{Trotta:2008qt} Jefferys' scale~\cite{Jeffreys:1939xee}
\begin{itemize}
    \item $\lvert\ln{B_{01}}\rvert<1$ is ranked as \textit{inconclusive} and $\mscr{M}_0$ has a posterior probability $<0.750$
    \item $\lvert\ln{B_{01}}\rvert=1.0$ is ranked as \textit{weak} and $\mscr{M}_0$ has a posterior probability $0.750$
    \item $\lvert\ln{B_{01}}\rvert=2.5$ is ranked as \textit{moderate} and $\mscr{M}_0$ has a posterior probability $0.923$
    \item $\ln{B_{01}}\rvert=5$ is ranked as \textit{strong} and $\mscr{M}_0$ has a posterior probability $0.993$
\end{itemize}

\begin{tcolorbox}[mybox]
Following~\cite{Trotta:2008qt} we can consider $\mscr{M}_0$ as a model that predict $\theta=0$ with no free parameter, whereas $\mscr{M}_1$ assigns to $\theta$ a Gaussian prior distribution with $\mu=0$ and $\sigma^2=\Sigma^2$. If we perform a measurement of $\theta$ that is described by a Gaussian likelihood whose maximum value lying $\lambda$ standard deviations away from $0$, then the Bayes factor between these two models is
\begin{equation}
    B_{01}=\sqrt{1+\(\frac{\Sigma}{\sigma}\)^2}\text{exp}\(-\frac{\lambda^2}{2\(1+\(\frac{\sigma}{\Sigma}\)^2\)}\)
\end{equation}
where $\sigma$ is the standard deviation of the Gaussian likelihood. If the $\theta_{ML}$ lays distant from $0$ thend the exponential dominates and the Bayes factor is small, favouring $\mscr{M}_1$. However, if $\lambda\lesssim 1$ and $\sigma/\Sigma\ll 1$, where the latter condition implies that the likelihood is more sharply peaked than the prior, then we have that the Bayes factor increases giving strong evidence in favour of the simpler model $\mscr{M}_0$. However, in the event that $\sigma\gg\Sigma$, we have a likelihood which is less informative than the prior, that implies $B_{01}\sim 1$ and the data do not change our relative belif in the two models. 
\end{tcolorbox}

If $\mscr{M}_1$ is simply a nested model of $\mscr{M}_0$, in the sense that $\mscr{M}_1$ include new parameters with respect to the original model, the Bayes factor approximates to the Savage–Dickey density ratio~\cite{Trotta:2005ar,Verdineli:1995} 
\begin{equation}
p(\theta',\theta|\mscr{M}_1)=p(\theta'|\mscr{M}_1)p(\theta|\mscr{M}_0)
\end{equation}
and the Bayes factor can be written as
\begin{equation}
    B^{\rm SD}_{01}=\frac{p(\theta'|d,\mscr{M}_1)}{p(\theta'|\mscr{M}_1)}\Big |_{\theta'=0}
\end{equation}
The numerator is simply the marginal posterior under the more complex model evaluated at the simpler model’s parameter value, while the denominator is the prior density of the more complex model evaluated at the same point. 

If the extra parameters are phenomenological or, in general, we do not have enough knowledge to assign a prior, an interesting procedure is to choose the prior on the new parameters in order to maximise the probability of the new model, given the data. If, even under this best case scenario, the more complex model is not significantly more probable than the simpler model, then one can confidently say that the data does not support the addition of the new parameters, without worrying that some other choice of prior will make the new model more probable~\cite{Berger:1987,Berger:1987v2,Sellke:2001}.

With the advent of precision cosmology, we cannot rely on the simple equation that complexity means higher degrees of freedom of the model and therefore we need to penalize it. Also, we have the ambiguity from evidence with unmeasured parameters: we are unable to discern whether they are  unconstrained or if they improve the quality just enough to give the same evidence. For this reason, we can reverse the point of view and start investigating what is the number of parameters that the data can support~\cite{Spiegelhalter:2002}. The information gained when upgrading the prior to the posterior can be quantified using the \textit{Kullback-Leibler (KL) divergence}~\cite{Kullback:1951zyt}
\begin{equation}
\mcal{D}_{\text{KL}} \equiv \int p(\theta|d, \mscr{M}) \ln \frac{p(\theta|d, \mscr{M})}{p(\theta|\mscr{M})} d\theta=\left\langle \ln \frac{p(\theta|d, \mscr{M})}{p(\theta|\mscr{M})}\right\rangle_{\text{p}}
\label{eq:DKL}
\end{equation}
where "p" indicates the average over the posterior. It is evident from \eq{DKL} that there is a strong prior dependency~\cite{Handley:2019wlz}. From the definition of the KL divergence, we can introduce the \textit{Shannon information} $\mcal{I}(\theta)$~\cite{Shannon:1948dpw} which is defined as $\mcal{I}(\theta)\equiv\ln\(p(\theta|d, \mscr{M})/p(\theta|\mscr{M})\)$. The Shannon information represents the amount of information gained in natural bits about $\theta$ when moving from the prior to the posterior. For indpeendent parameters we have $\mcal{I}(\theta_1,\theta_2)=\mcal{I}(\theta_1)+\mcal{I}(\theta_2)$ that is transported to the KL divergence, being the latter a linear function of the former. Using \eq{BT} and the fact that $\int{p(\theta|d,\mscr{M})d\theta}=0$ , we have
\begin{equation}
    \mcal{D}_{\text{KL}} = -\ln{p(d|\mscr{M})} + \int p(\theta|d, \mscr{M}) \ln p(d|\theta,\mscr{M}) d\theta.
\end{equation}
(used for examples in~\cite{Seehars:2014ora,Nicola:2018rcd,Li:2024hjf,Hosoya:2004nh,Grandis:2015qaa}). 

The information of the prior-posterior compression gained with $\mcal{D}_{\rm KL}$ is lacking to any indication on which parameters are providing us with information. To this end, we ought to introduce a new quantity. The number of parameters effectively constrained by the data can be measured using the \textit{Bayesian model complexity}~\cite{Gelman:1992zz,Spiegelhalter:2002} $\mcal{C}_b\equiv2\mcal{D}_{\text{KL}}$ (a less stable but more interesting quantity is the \textit{Bayesian model dimensionality}\cite{Handley:2019pqx}). Unlike the KL divergence, now the Bayesian model complexity is weakly prior dependent With this quantity, if $p(d|\mscr{M}_0) \approx p(d|\mscr{M}_1)$, we are now able to distinguish~\cite{Trotta:2008qt}:

\begin{enumerate}
    \item  $\mcal{C}_b(\mscr{M}_1) > \mcal{C}_b(\mscr{M}_0)$: the quality of the data is sufficient to measure the additional parameters of the more complicated model $\mscr{M}_1$, but they do not improve its evidence by much. We should prefer model $\mscr{M}_0$, with fewer parameters.
    \item $\mcal{C}_b(\mscr{M}_1) \approx \mcal{C}_b(\mscr{M}_0)$: both models have comparable evidence and the effective number of parameters is about the same. In this case, the data is not good enough to measure the additional parameters of the more complicated model (given the choice of prior), and we cannot draw any conclusions as to whether the extra parameter is needed.
\end{enumerate}

\begin{tcolorbox}[mybox]

There are other information criteria that have been widely used in several astrophysical and cosmological contexts~\cite{Liddle:2004nh,Liddle:2007fy,Trotta:2008qt,Trotta:2017wnx}
\begin{itemize}
    \item The \textit{Akaike Information Criterion (AIC)~\cite{Akaike:1974}} is an essentially frequentist criterion that sets the penalty term equal to twice the number of free parameters in the model, $(k)$
    \begin{equation}
    \text{AIC} \equiv -2 \ln \mcal{L}_{\text{max}} + 2k
    \label{eq:AIC}
    \end{equation}
    The derivation of the AIC follows from an approximate minimization of the KL divergence between the true model distribution and the distribution being fitted to the data.
    \item The \textit{Bayesian Information Criterion (BIC)}~\cite{Schwarz:1978}: follows from a Gaussian approximation to the Bayesian evidence in the limit of large sample size ($N$ is the number of data points):
    \begin{equation}
    \text{BIC} \equiv -2 \ln \mcal{L}_{\text{max}} + k \ln N
    \end{equation}
    The best model is again the one that minimizes the BIC. With respect to \eq{AIC}, BIC penalizes more the free parameters.
    \item The \textit{Deviance Information Criterion (DIC)}~\cite{Spiegelhalter:2002} can be written as
    \begin{equation}
    \text{DIC} \equiv -2 \mcal{D}_{\text{KL}} + 2\mcal{C}_b.
    \label{eq:DIC}
    \end{equation}
    In the limit of well–constrained parameters, the AIC is recovered from \eq{DIC}, but the DIC has the advantage of accounting for unconstrained directions in parameter space. 
\end{itemize}
However, great care must be used when dealing with these criteria for a large number of data points. They assume implicitly that all free parameters are well constrained by the data. But in general, the number of free parameters might not be a good representation of the actual number of effective parameters. 
\end{tcolorbox}

\subsubsection{Dataset Comparison}
Suppose now that we fix a specific model $\mscr{M}$ and that the data are coming from two independent datasets we call $A$ and $B$\footnote{New parameters are possibly added also when we combine independent dataset, because of new nuisance or more constraining power.}. If $A$ is described by a set of parameters $\theta_A$ and $B$ by $\theta_B$, then the total evidence can be written as the product of the individual evidences (because of the likelihood, the posterior do not factorize). Therefore, we can test the confidence in the ability to combine the datasets with the \textit{robustness}~\cite{Amendola:2012wc,Handley:2019wlz,Marshall:2004zd} 
\begin{equation}
    R=\frac{p(A,B)}{p(A)p(B)}
    \label{eq:R}
\end{equation}
where $p(A,B)=\int{p(A|\theta)p(B|\theta)p(\theta)d\theta}$. The numerator represents the case where both datasets are explained by the same parameters within the model, while the denominator allows each data set to be explained by different parameters. The robustness is a ratio between evidences similarly to the Bayes factor \eq{BFactor} but within a specific model in the case of testing the compatiblity between datasets, therefore the same considerations apply~\cite{Handley:2019wlz}. If $R\gg 1(R\ll 1)$ we can interpret it as both datasets consistent (inconsistent). Just as a reminder, if the datasets are not compatible with one another, since only the likelihoods can be multiplied, we should not expect $R=1$ in this scenario. $R$ is strongly prior-dependent for shared parameters that are constrained, in fact narowing the priors means decreasing the value of $R$. Hence, it must be handled with great care because with an ill-defined prior, $R$ can indicate an agreement between two dataset that are not (but the other way around is not possible due to the one-directional increasing of volume effects). For a better interpretation of $R$~\cite{Handley:2019wlz,Raveri:2018wln,Amendola:2012wc}, we can use the definition of conditional probability \eq{sumprod} and see that it can be written as $R=p(A|B)/p(A)$ and thus, we can say that $R$ represents the relative confidence that we have in dataset $A$ in light of knowing $B$, compared to the confidence in $A$ alone. From this point of view, it comes straightforward that $R>1$ means that $B$ strengthened our confidence in $A$. 

We can now introduce two other important quanitties for dataset comparison. If we take the logarithm of \eq{R}, we have
\begin{equation}
    \boxed{\log{S}=\log{R}-\log{I}}
\end{equation}
where $S$ is the \textit{Suspiciousness}~\cite{Lemos:2019txn,Handley:2019wlz} and $I$ is the \textit{Information ratio}. $I$ contains the proportionality of the prior and is defined as $\log{I}=\mcal{D}_{A}+\mcal{D}_{B}-\mcal{D}_{AB}$ therefore $S$ is prior independent~\cite{DES:2020hen} and depends only on the actual mismatch between the posteriors. If the posteriors are such that we may approximate them in the cosmological parameters by a Gaussian (in a broader sense~\cite{Handley:2019wlz,DES:2020hen}), with means and covariance $\mu$ and $\sigma$, then the suspiciousness follows the $\chi^2_d$ distribution where $d$ is the effective number of degrees of freedom constrained by both datasets~\cite{Handley:2020hdp,Handley:2019wlz,DiValentino:2022rdg,Gariazzo:2023joe}:
\begin{equation}
    \log S = \frac{d}{2}-\frac{\chi^2}{2}
\end{equation}
with 
\begin{equation}
    \chi^2=(\mu_A-\mu_B)(\sigma_A+\sigma_B)^{-1}(\mu_A-\mu_B)
\end{equation}
For an easier interpretation of the result we can use the inverse cumulative $\chi^2$ distribution and obtain the tension probability
\begin{equation}
    p=\int^{\infty}_{\chi^2}\frac{x^{d/2-1}e^{-x/2}}{2^{d/2}\Gamma(d/2)}dx\,.
\end{equation}
the expression
\begin{equation}
    \sigma(p)=\sqrt{2}\text{erfc}^{-1}(1-p)
\end{equation}
represents the relationship between a given cumulative probability $p$ and its corresponding number of standard deviations $\sigma$ in a normal distribution. So a $3\sigma$ deviation corresponds to $p<0.3\%$ and points towards a strong tension. So the probability tension quantifies the probability of the observed tension occurring by chance. 
\begin{tcolorbox}[mybox]
We are observing a parameter value, $\theta$, that we are expecting to follow a distribution under the null hypothesis $H_0: \theta=\theta_0$ and we want to perform some test statistics $T(X)$ comparing $H_0$ with an alternative $H_1: \theta \neq \theta_0$. For this purpose, we define the \textit{p-value}~\cite{Berger:1987,Berger:2002,Sellke:2001}. The p-value (or observed significance level) is given by the probability under $H_0$, that $T$ achieves values as extremes or more extremes that have been observed:
\begin{equation}
    p_V=p(T(X)\geq T^{\rm obs}|H_0)
\end{equation}
\end{tcolorbox}

If we want to understand how \textit{strong} is the compromise we need to ask, when joining two datasets instead of taking them independently, we want the goodness-of-fit test~\cite{DES:2020hen,Raveri:2018wln}. This test evaluates the cost of explaining datasets with the same parameter values and is quantified by the estimator:
\begin{equation}
    \mcal{Q}_{\rm DMAP}=2\ln{\mcal{L}_A(\hat{\theta}_{\rm A})}+2\ln{\mcal{L}_B(\hat{\theta}_{\rm B})}-2\ln{\mcal{L}_{A+B}(\hat{\theta}_{\rm AB})}
\end{equation} 
Here $\hat{\theta}_A$ denotes the parameter values which “best” describe dataset $A$. In the frequentist statistics, the parameters are taken such as they maximize the likelihood sand therefore $\mcal{Q}_{\rm DMAP}\equiv\chi^2$. On the other hand, in the context of Bayesian analysis, $\hat{\theta}_A$ are the parameters that maximize the posterior. As we have seen, for flat uninformative priors MAP and ML are identical. The test statistic $Q_{DMAP}$ is widely used in cosmology (see e.g.~\cite{Schoneberg:2021qvd,DES:2020hen,Cruz:2023lmn}). When the likelihoods and posteriors are Gaussian $\mcal{Q}_{\rm DMAP}$ is $\chi^2$ distributed~\cite{Raveri:2018wln,DES:2020hen}. The goodness-of-fit is expected to degrade by one for each measured parameter, and indicates tension if the decrease is higher. Only the parameters that are constrained by the data over the prior can contribute to a tension since prior-constrained parameters cannot be optimized to improve the data fit. In the limits where the prior is uninformative or fully informative $\mcal{Q}_{\rm DMAP}$ is the likelihood expression for parameter shifts discussed in the previous sections and its statistical significance should match the one obtained with parameter-shift techniques.